\documentclass[School=Warsaw]{Dissertate}

\begin{document}

\title{Real-time optimal quantum control for atomic magnetometers with decoherence}
\author{J\'ulia Amor\'os Binefa}
\firstadvisor{Prof. Konrad Banaszek}
\secondadvisor{Dr. Jan Ko\l{}ody\'nski}
\firstreviewer{Prof. Marco G. Genoni}
\secondreviewer{Prof. M\v{a}d\v{a}lin Gu\c{t}\v{a}}
\thirdreviewer{Prof. \L{}ukasz Cywi\'nski}

\degree{Doctor of Philosophy}
\field{Physics}
\degreeyear{2025}
\degreemonthsub{June}
\degreemonthdef{October}
\department{Physics}

\maketitle
\copyrightpage

	\phantomsection
	\addcontentsline{toc}{chapter}{Abstract}
	\newpage
	\pagenumbering{roman}
	\setcounter{page}{3}
	\pagestyle{fancy}
	\renewcommand{\headrulewidth}{0.0pt}
	\vspace*{35pt}
	\begin{center}
	\scshape Abstract \\ \rm
	\end{center}

Quantum entanglement can enhance the sensitivity of atomic sensors to static or slowly varying fields. But many important applications in fields such as medicine or navigation require tracking fast or transient signals. This presents unique challenges, one of them being that the potential benefits of entanglement in such cases are still not fully understood. To investigate this, we apply concepts from continuous quantum measurements and estimation theory to optical atomic magnetometers, aiming to accurately model these devices, interpret their measurement data, control their dynamics, and achieve optimal sensitivity. 

Quantifying this optimal performance requires determining a fundamental quantum limit on sensitivity. The above bound imposed by noise is derived and shown to scale at best linearly with the sensing time and number of atoms $N$, ruling out any super-classical scaling. Moreover, this quantum limit is independent of the initial state, measurement, estimator, and measurement-based feedback, and depends only on the decoherence model and strength of the field fluctuations. Thus, finding an estimator that attains this limit proves a given sensing strategy optimal.  

To approach this bound, we develop a quantum dynamical model scalable w.r.t. $N$, based on a co-moving Gaussian approximation of the stochastic master equation, which includes both measurement backaction and decoherence. This enables the construction of a real-time estimation and control architecture that integrates an extended Kalman filter (EKF) with a linear quadratic regulator (LQR). 

By simulating the magnetometer with our model and our proposed EKF+LQR strategy, we show that quantum-limited tracking of constant and fluctuating fields is within reach of current atomic magnetometers. Strikingly, our sensing strategy can also track biological relevant signals, such as heartbeat-like waveforms. It can furthermore be used to drive the atomic ensemble into an entangled state, even when the measurement record is used for feedback but afterwards discarded.

	\vspace*{\fill}
	\newpage \lhead{} \rhead{}
	\cfoot{\thepage}

	\phantomsection
	\addcontentsline{toc}{chapter}{Streszczenie}
	\newpage
	\pagenumbering{roman}
	\setcounter{page}{3}
	\pagestyle{fancy}
	\renewcommand{\headrulewidth}{0.0pt}
	\vspace*{35pt}
	\begin{center}
	\scshape Streszczenie \\ \rm
	\end{center}

Splątanie kwantowe może zwiększyć czułość sensorów atomowych na pola statyczne lub wolnozmienne. Jednak wiele istotnych zastosowań w dziedzinach takich jak medycyna czy nawigacja wymaga śledzenia sygnałów szybko lub nagle zmieniających się. Wiąże się to ze szczególnymi wyzwaniami, biorąc pod uwagę brak wystarczającego zrozumienia potencjalnych korzyści ze splątania w takich przypadkach. Aby to zbadać, w pracy doktorskiej zastosowano narzędzia z zakresu kwantowych pomiarów ciągłych i teorii estymacji do optycznych magnetometrów atomowych, w celu dokładnego modelowania tych urządzeń, interpretacji danych pomiarowych, kontroli ich dynamiki oraz osiągnięcia optymalnej czułości.

Określenie tej optymalnej wydajności wymaga wyznaczenia fundamentalnej kwantowej granicy czułości. Powyższa granica, zdeterminowana przez szum, którą wyznaczamy w pracy, w najlepszym wypadku skaluje się liniowo z czasem pomiaru i liczbą atomów $N$, co wyklucza jakiekolwiek skalowanie lepsze niż klasyczne. Co więcej, ta kwantowa granica jest niezależna od stanu początkowego, pomiaru, estymatora oraz sprzężenia zwrotnego opartego na pomiarze i zależy wyłącznie od modelu dekoherencji oraz siły fluktuacji pola. Zatem znalezienie estymatora osiągającego tę granicę jest dowodem na optymalność danej strategii pomiarowej.

Aby zbliżyć się do tej granicy, opracowujemy kwantowy model dynamiczny dobrze skalujący się z $N$, oparty na współporuszającym się przybliżeniu Gaussowskim kwantowego równania "master", który uwzględnia zarówno działanie pomiaru, jak i dekoherencję. Umożliwia to stworzenie architektury estymacji i kontroli w czasie rzeczywistym, integrującej rozszerzony filtr Kalmana (EKF) z regulatorem liniowo-kwadratowym (LQR).

Poprzez symulację magnetometru z wykorzystaniem naszego modelu oraz proponowanej strategii EKF+LQR pokazujemy, że kwantowo-ograniczone śledzenie zarówno pól stałych, jak i fluktuujących, jest w zasięgu obecnych magnetometrów atomowych. Co istotne, nasza strategia pomiarowa umożliwia także śledzenie sygnałów istotnych biologicznie, takich jak sygnały podobne do tych generowanych przez serce, ale może być po prostu użyta do wprowadzenia układu atomowego w stan splątany, także kiedy dane pomiarowe są wywołane przez sprzężenie zwrotne ale później zapominane.

	\vspace*{\fill}
	\newpage \lhead{} \rhead{}
	\cfoot{\thepage}

\setcounter{tocdepth}{2}
\setcounter{secnumdepth}{4}

	\newpage 
        \thispagestyle{fancy} 
        \vspace*{\fill}
	\scshape 
        \noindent 

To my parents, Rosaura and Josep Maria.
	\vspace*{\fill} 
        \newpage
        \rm

	\chapter*{Acknowledgments}
	\noindent

First and foremost, I would like to thank my supervisor, Dr.~Janek  Ko\l{}ody\'nski, for his guidance and support throughout my PhD. Somehow, the vision he shared in our  first meeting turned out to be pretty much what this thesis became, a rare thing! His positivity, even when results were uncertain, helped me stay motivated, and he never hesitated to dive into the nitty-gritty when we were stuck. I also deeply thank him for his efforts in teaching me how to write, proofreading my thesis, encouraging me to present at conferences, and ensuring consistent funding throughout my PhD.

I would also like to thank my other PhD supervisor, Prof.~Konrad Banaszek, for his support in facilitating my doctoral studies as well as his efforts as the head of the QOT center in attracting funding. 

I am also especially grateful to Dr.~Francesco Albarelli, Prof.~Marco G. Genoni and Dr.~Matteo A. C. Rossi for their openness and generosity. Although we never formally collaborated, they helped me understand their code, answered my questions and shared their excellent notes on continuous measurements, which clarified lots of important concepts in my mind. 

Another special thanks goes to our collaborator, Prof.~Morgan W. Mitchell, for his invaluable expertise and guidance in applying the model and techniques developed in this thesis to experimentally realistic regimes. His insights were crucial in motivating our work and in bringing the theory closer to experiments.

I would also like to express my sincere gratitude to Prof.~Marco G. Genoni, Prof.~M\v{a}d\v{a}lin Gu\c{t}\v{a} and Prof.~\L{}ukasz Cywi\'nski for their careful and thorough review of this thesis, as well as for their very insightful comments, which I have incorporated into this revised version.

Finally, I would like to thank my friends and colleagues: Dr.~Lewis Clark, Mateusz Molenda, Dr.~Rahul Deshpande, and Dr.~Aditya Jain for reading parts of this thesis and providing valuable feedback. I am also thankful to Adam Parafiańczuk and Marcin Koźbiał for their help in translating the abstract into Polish.

	\vspace*{\fill} \newpage


	\chapter*{List of Publications}
	\noindent
	\begin{itemize}
    \item[] J. Amor\'{o}s-Binefa and J. Ko\l{}ody\'nski, \textit{``Noisy atomic magnetometry in real time,''} \href{https://iopscience.iop.org/article/10.1088/1367-2630/ac3b71}{New J. Phys. 23, 123030 (2021).}
    \item[] J. Amor\'{o}s-Binefa and J. Ko\l{}ody\'nski, \textit{``Noisy atomic magnetometry with Kalman filtering and measurement-based feedback,''} \href{https://link.aps.org/doi/10.1103/k7nk-lrwd}{PRX Quantum 6, 030331 (2025)}.
    \item[] J. Amor\'{o}s-Binefa, M. W. Mitchell, and J. Ko\l{}ody\'nski, \textit{``Tracking time-varying signals with quantum-enhanced atomic magnetometers,''} \href{https://arxiv.org/abs/2503.14793}{arXiv:2503.14793 (2025)}.
\end{itemize}
	\vspace*{\fill} \newpage

\tableofcontents
\listoffigures
\listoftables

	\chapter*{Abbreviations}
	\noindent
	
\begin{tabular}{ll}
  \textbf{aMSE} & \textbf{a}veraged \textbf{M}ean \textbf{S}quared \textbf{E}rror \\
  \textbf{BCRB} & \textbf{B}ayesian \textbf{C}ram\'{e}r-\textbf{R}ao \textbf{B}ound \\
  \textbf{BI} & \textbf{B}ayesian \textbf{I}nformation \\
  \textbf{CoG} & \textbf{Co}-moving \textbf{G}aussian approximation \\
  \textbf{CRB} & \textbf{C}ram\'er-\textbf{R}ao \textbf{B}ound \\
  \textbf{CPTP} &  \textbf{C}ompletely \textbf{P}ositive and \textbf{T}race-\textbf{P}reserving map \\
  \textbf{CS} & \textbf{C}lassically-\textbf{S}imulated limit \\
  \textbf{CSS} & \textbf{C}oherent \textbf{S}pin \textbf{S}tate \\
  \textbf{DP} & \textbf{D}ynamical \textbf{P}rogramming \\
  \textbf{EKF} & \textbf{E}xtended \textbf{K}alman \textbf{F}ilter \\
  \textbf{EM} & \textbf{E}uler-\textbf{M}aruyama method \\
  \textbf{FI} & classical \textbf{F}isher \textbf{I}nformation \\
  \textbf{GKSL} & \textbf{G}orini-\textbf{K}osakowski-\textbf{S}udarshan-\textbf{L}indblad form \\
  \textbf{HJB} & \textbf{H}amilton-\textbf{J}acobi-\textbf{B}ellman equation \\
  \textbf{KF} & \textbf{K}alman \textbf{F}ilter \\
  \textbf{LG} & \textbf{L}inear and \textbf{G}aussian \\
  \textbf{LQG} & \textbf{L}inear-\textbf{Q}uadratic-\textbf{G}aussian control \\
  \textbf{LQR} & \textbf{L}inear-\textbf{Q}uadratic \textbf{R}egulator \\
  \textbf{MAP} & \textbf{M}aximum \textbf{A}-\textbf{P}osteriori estimator \\
  \textbf{MCG} & \textbf{M}agneto\textbf{C}ardio\textbf{G}ram \\
  \textbf{ML} & \textbf{M}aximum \textbf{L}ikelihood estimator \\
  \textbf{MMSE} & \textbf{M}inimum \textbf{M}ean \textbf{S}quared \textbf{E}rror estimator \\
  \textbf{MSE} & \textbf{M}ean \textbf{S}quared \textbf{E}rror \\
  \textbf{OU} & \textbf{O}rnstein-\textbf{U}hlenbeck process \\
  \textbf{PDF} & \textbf{P}robability \textbf{D}ensity \textbf{F}unction \\
  \textbf{PMF} & \textbf{P}robability \textbf{M}ass \textbf{F}unction \\
  \textbf{POVM} & \textbf{P}ositive \textbf{O}perator-\textbf{V}alued \textbf{M}easure \\
  \textbf{RWA} & \textbf{R}otating-\textbf{W}ave \textbf{A}pproximation \\
  \textbf{SDE} & \textbf{S}tochastic \textbf{D}ifferential \textbf{E}quation \\
  \textbf{SME} & \textbf{S}tochastic \textbf{M}aster \textbf{E}quation \\
  \textbf{SQL} & \textbf{S}tandard \textbf{Q}uantum \textbf{L}imit \\
  \textbf{SS} & \textbf{S}teady-\textbf{S}tate \\
  \textbf{VdP} & \textbf{V}an \textbf{d}er \textbf{P}ol oscillator \\
    \textbf{w.r.t.} & \textbf{w}ith \textbf{r}espect \textbf{t}o
\end{tabular}

	\vspace*{\fill} \newpage
	\setcounter{page}{1}
	\pagenumbering{arabic}

\onehalfspacing

\setcounter{chapter}{-1}  
\chapter{Introduction}
\label{introduction}

\newthought{Metrology --- the science of measurement ---} relies on classical estimation theory to infer unknown parameters from measurement data and quantify the uncertainty of these estimates. Perhaps most crucially, it also tells us how to identify optimal estimators that minimize this uncertainty, such as those that saturate the Cram\'{e}r–Rao bound \cite{Kay1993,Van-Trees}.

Even though classical estimation theory does not require the measurement outcomes to be probabilistic, we still use probabilities to express our uncertainty about the true value of a parameter. This uncertainty arises not from the theory itself, but from practical limitations such as environmental noise, sensor imperfections, or incomplete information about the system. In contrast, the randomness observed in measurements of quantum states is not due to external factors, but rather an intrinsic feature of quantum systems arising from the postulates of quantum mechanics. Even in ideal, noiseless conditions, the outcomes of quantum measurements are fundamentally stochastic, with their associated likelihoods dictated by Born's rule~\cite{bornsrule}. Quantum systems, moreover, have non-classical features such as coherence and entanglement, which can be harnessed to reduce measurement uncertainty beyond classical limits \cite{Caves1981,Wineland1992,Wineland1994,Bollinger1996}. This reduction in uncertainty improves estimation accuracy, which typically scales with the number of resources (e.g. the number of particles used, $N$). In classical strategies using uncorrelated particles, the estimation error scales as $1/\sqrt{N}$, known as the Standard Quantum Limit (SQL). However, quantum-enhanced strategies can achieve more favorable scalings, such as $1/N$, typically referred to as the Heisenberg Limit (HL)~\cite{DAriano2001,Giovannetti2001_other,Giovannetti2004,Giovannetti2006_other,Pezze2018RMP}. Even though numerous experiments have demonstrated surpassing the SQL \cite{Kuzmich2000,Wasilewski2010,Shah2010,Koschorreck2010,Sewell2012}, decoherence and other noise sources can significantly degrade entanglement and prevent achieving super-classical scalings \cite{Escher2011,Demkowicz2012}. 

Much early work in quantum metrology focused on repeated measurements of step-wise protocols \cite{Giovannetti2006_other}. However, many real-world sensing tasks require tracking time-varying signals~\cite{mcg_paper,Bison2009,jensen_magnetocardiography_2018,Kim2019,Yang2021,Canciani2020,Canciani2022}, where repeated measurements are not possible and the processes of preparation, evolution, detection and estimation are occurring simultaneously. In such cases, sensors must operate in one shot and in real time. To achieve this, one needs both practical real-time estimation and control strategies, as well as an open-system description of continuous-time quantum dynamics that include measurement backaction.

An estimation strategy naturally suited to continuous monitoring is Bayesian filtering, which continuously updates probabilistic estimates as new, noisy data becomes available \cite{crassidis2011optimal,sarkka2013bayesian}. This real-time process involves two steps: (1) a predictive step, relying on a dynamical model of the system, and (2) a measurement update step, which refines the estimate based on incoming measurement outcomes~\cite{sarkka2013bayesian}. A classical example is the Kalman filter (KF)~\cite{kalman_new_1960,kalman_new_1961}, optimal for linear systems with Gaussian noise~\cite{sarkka2013bayesian}. For nonlinear quantum systems such as atomic sensors, methods like the extended Kalman filter (EKF) are better suited \cite{crassidis2011optimal,simon2006}.

Since Bayesian filtering relies on a model, it is essential to have a reliable description of both the system and the measurement process to implement any algorithm effectively \cite{crassidis2011optimal,sarkka2013bayesian}. In continuously-monitored quantum sensors, this requires modeling the measurement backaction, i.e., how measurements disturb the system and thus shape its subsequent evolution. The formal framework for this is the theory of continuous quantum measurements~\cite{Davies1976,Srinivas1981,carmichael_book,Srinivas1996}, which describes the evolution of the quantum state conditioned on the measurement outcomes~\cite{Belavkin1989,Wiseman1993,handel_modelling_2005}. In this setting, the dynamics are governed by a stochastic master equation (SME) that incorporates backaction either in the form of quantum jumps (e.g., in photon counting) or as diffusive noise (e.g., in homodyne detection)~\cite{Wiseman1993,Wiseman_thesis,wiseman2010quantum}.

Combining Bayesian filtering with this framework allows for real-time tracking while correctly accounting for measurement backaction. This has been achieved in experiments with Gaussian systems such as mechanical oscillators~\cite{Iwasawa2013,wieczorek_optimal_2015, Wilson2015, Rossi2018, rossi_observing_2019,Magrini2021}, which have the advantage that both the conditional and unconditional distributions remain Gaussian.

Atomic sensors, particularly optical atomic magnetometers, can also benefit from these ideas. These devices rival state-of-the-art superconducting quantum interference devices (SQUIDs) in sensitivity, yet operate without the need for cryogenic cooling~\cite{Kominis2003,clarke2004squid} and can potentially be miniaturized to chip-scale sizes~\cite{kitching_chip-scale_2018}. For spin-precession sensors, theoretical models have been proposed that assume Gaussianity~\cite{Geremia2003, Madsen2004, Molmer2004, Albarelli2017}, or resort to brute-force numerics for low atomic numbers~\cite{rossi_noisy_2020}. In experiments, however, spin-precession sensors are large atomic ensembles ($N \sim 10^6 - 10^{15}$) that naturally evolve toward non-Gaussian unconditional distributions~\cite{budker_optical_2007}, and to date Gaussian
models have been successfully applied only when ignoring~\cite{Jimenez2018} or evading~\cite{Kong2020, Troullinou2021, Troullinou2023} the measurement backaction. This motivates the need for a realistic nonlinear model that captures both the measurement backaction and the decoherence effects in a way that is scalable to systems with large atom numbers.

To address this, we propose an approximate model referred to as the ``co-moving Gaussian'' picture. This model reduces the complexity of the full quantum dynamics while still accounting for dephasing and the measurement backaction that creates spin squeezing \cite{Amoros-Binefa2024,Amoros-Binefa2025}. This model not only enables the design of an estimation and feedback scheme by combining an EKF with a linear quadratic regulator (LQR) \cite{crassidis2011optimal,Stockton2004}, but also allows us to simulate realistic atomic magnetometry experiments. Its accuracy is validated by comparing it against the exact simulation of the SME for moderate-sized systems \cite{Amoros-Binefa2024}. 

Once the model is validated, the next step is to asses the performance of the sensor when tracking fluctuating fields by establishing fundamental limits on the estimation error. This is achieved by deriving a lower bound on the Bayesian Cram\'{e}r-Rao bound (BCRB) \cite{Van-Trees,Fritsche2014}, which we refer to as classically-simulated (CS) limit or quantum limit~\cite{matsumoto_metric_2010,Demkowicz2012,Amoros-Binefa2021,Amoros-Binefa2024,Amoros-Binefa2025}. Notably, this limit depends only on the dephasing rate and the strength of field fluctuations, and scales at best linearly with the atom number and sensing time, thereby precluding any possibility of surpassing the SQL \cite{Amoros-Binefa2021,Amoros-Binefa2024,Amoros-Binefa2025}. Remarkably, our magnetometry setup attains this quantum limit, confirming that the entire sensing protocol is optimal under the given noise conditions, regardless of the initial state, measurement, estimator, or measurement-based feedback \cite{Amoros-Binefa2025}. Beyond demonstrating optimal real-time sensing of fluctuating signals, we also apply our method to track signals commonly found in medicine and biology, such as magneto-cardiograms (MCG)~\cite{jensen_magnetocardiography_2018,Amoros-Binefa2025}, which requires filtering to extract heartbeat-like signals from noisy backgrounds. 

Finally, this framework can also be used for quantum state preparation \cite{Amoros-Binefa2024}. Specifically, we show that the LQR not only enhances estimation accuracy but also steers the atomic ensemble into an entangled state without the need to store past measurement data. This makes the protocol a practical method for preparing entangled states in real time.

\newthought{This thesis is organized into six chapters.} The first three chapters cover background material that is important for understanding the rest of the work. In particular, the first part of \chapref{chap:pre} reviews important topics like probability theory, stochastic processes and stochastic calculus. The second part of \chapref{chap:pre} introduces basic concepts of quantum mechanics, including angular momentum operators, the Wigner function and spin squeezing. Then, \chapref{chap:bayesian} presents an in-depth discussion of Bayesian filtering, with complete treatments of the KF, EKF and LQR. In \chapref{chap:cm}, we derive the SME for both photodetection and homodyne measurement. While these chapters aim to keep the thesis self-contained, my background in quantum physics inevitably influences what I consider to be introductory material, so readers with different expertise may find some sections more familiar than others. If you are already comfortable with the topics described above, feel free to skip ahead to  \chapref{chap:bounds} and \chapref{chap:model}, where the main results and contributions are discussed. In \chapref{chap:bounds}, we derive the quantum limit or classically-simulated limit on the estimation error via lower-bounding the BCRB. Finally, \chapref{chap:model} applies these theoretical and numerical tools to atomic magnetometry. We introduce the “co-moving Gaussian” model to capture both the measurement backaction and decoherence in large atomic ensembles, and then put forward a complete estimation and control protocol. This protocol consists of an EKF combined with LQR, which achieves optimal performance and even enables real-time entanglement preparation. Finally, \chapref{chap:conclusions} summarizes the results and discusses potential directions for future research.

\chapter{Preliminaries} \label{chap:pre}

\newthought{The main focus of this thesis} is the tracking of quantities that vary randomly over time. In many complex systems, such as those in physics, finance, or biology, the parameters of interested are not static but fluctuate continuously due to inherent randomness and external disturbances. These quantities are referred to as \emph{stochastic processes}, and to capture their dynamics, it is essential to combine differential equations with probability theory.  

\newthought{In the first part of this opening chapter}, we begin by reviewing a bit of probability theory: random variables, probability density functions, expectation and covariances and the Gaussian distribution. Next, we explain in \secref{sec:stochastic_process} what a stochastic process is through some simple examples, and introduce in \secref{sec:stochastic_calculus} the basics of stochastic calculus with topics like It\^{o}'s Lemma, stochastic differential equations and their corresponding numerical methods. 

In the second part of this chapter, i.e. \secref{sec:fundamentals_qm},  we briefly introduce a few concepts related to the dynamics and properties of quantum systems, mostly for later reference. These include the position and momentum operators, the dynamics of open quantum systems, angular momentum operators and coherent spin states. We also cover the Wigner quasiprobability distribution in \secref{sec:Wigner_quasiprobability} and how to map the Wigner function onto a sphere in \secref{sec:Wigner_on_Bloch}, followed by a brief explanation of spin squeezing in \secref{sec:spin-squeezing_intro}. 

\section{Fundamental concepts of probability theory}
\subsection{Random variables}

In probability theory, a \emph{random variable} assigns numerical values to the outcomes of a random process, each with an associated likelihood. Formally, it is a function mapping a sample space $\Omega$ to a subset of the real numbers $\Real$. For example, when flipping a coin, the sample space $\Omega$ representing all possible outcomes of this experiment is $\Omega = \{\text{Heads},\text{Tails}\}$. Now, let $\mrm{X}$\footnote{In mathematics, random variables are typically denoted by a capital letter (e.g. $\mrm{X}$), and their possible values, a.k.a. realizations, by the corresponding lowercase letter ($x$). We will follow this convention here but use lowercase for both in later chapters.} be a random variable that assigns numerical values to these outcomes, e.g.:
\begin{align}
    \mrm{X}(\text{Heads}) = 1 \quad\quad \text{and} \quad \quad \mrm{X}(\text{Tails}) = 0,
\end{align}
where $x = 0$ and $x = 1$ are \emph{realizations} of the random variable $\mrm{X}$. Random variables like the one above, or others such as the outcome of a dice roll, take on specific, countable values. Therefore, these are referred to as \emph{discrete} random variables. Additionally, each possible outcome has an associated probability, the probability mass function (PMF), and these probabilities must sum to 1.

\begin{definition}[Probability mass function] For a discrete random variable $\mrm{X}$, let the PMF be a function $p \, : \, \Real \to [0,1]$  that gives the probability of the random variable $\mrm{X}$ taking the specific value $x$, i.e.
    \begin{align}
        p(x) = \Prob{\mrm{X} = x},
    \end{align}
    where $p(x) \geq 0 \; \forall x$ and the sum of all probabilities must be equal to 1:
    \begin{align}
        \sum_{x\in \mathcal{X}} p(x) = 1,
    \end{align}
    with $\mathcal{X}$ corresponding to the set of all the possible realizations $x$ of the random variable $X$.
    \label{def:PMF}
\end{definition} 
For example, if $\mrm{X}$ represents the outcome of rolling a fair six-sided dice, the PMF is:
\begin{align}
    p(x) = \frac{1}{6} \quad \text{for} \quad x \in \mathcal{X} = \Omega = \{1,2,3,4,5,6\}.
\end{align}

In contrast, a \emph{continuous random variable} takes values from a continuous range, such as time, position or temperature. Since a continuous random variable can have infinitely many values, the probability of it taking any specific value is zero. Therefore, probabilities are determined over intervals using the probability density function (PDF). The total probability for all values (the area under the PDF curve) must also sum to 1.

\begin{definition}[Probability density function] The PDF of a continuous-valued random variable $\mrm{X}$ is denoted as $p(x)$. Its integration over an interval $[a,b]$ yields the probability of $\,\mrm{X} \in [a,b]$, i.e.
    \begin{equation}
        \Prob{a \leq \mrm{X} \leq b} = \int_a^b p(x) \dd x,
    \end{equation}
    and thus, relating the PDF to the PMF. Additionally, the PDF must satisfy the following conditions
    \begin{equation}
        p(x) \geq 0 \; \forall x, \;\; \text{and} \;\; \int p(x) \, \dd x = 1. \nonumber
    \end{equation}
    \label{def:PDF}
\end{definition} 

\subsection{Multivariate random variables}

Additionally, the definitions of PMF and PDF can be extended also to \emph{multivariate random variables}, also referred to as \emph{random vectors}. Formally, a multivariate random variable is a column vector $\pmb{\mrm{X}} = (\mrm{X}_1,\dots,\mrm{X}_n)^\Trans$, whose components are random variables. 

\begin{definition}[Multivariate PMF] For a multivariate discrete random variable $\mrm{X}$, let the PMF be a function $p \, : \, \Real^n \to [0,1]$  that gives the probability of the random vector $\pmb{\mrm{X}}$ taking the specific value $\pmb{x}$, i.e.
    \begin{align}
        p(\pmb{x}) = \Prob{\pmb{\mrm{X}} = \pmb{x}}
    \end{align}
    where $p(\pmb{x}) \geq 0 \; \forall \pmb{x}$ and the sum of all probabilities must be equal to 1:
    \begin{align}
        \sum_{\pmb{x}\in \pmb{\mathcal{X}}} p(\pmb{x}) = 1
    \end{align}
    with $\pmb{\mathcal{X}}$ corresponding to the set of all the elements $\pmb{x}$ can take.
    \label{def:multivariate_PMF}
\end{definition}

\begin{definition}[Multivariate PDF] \label{def:multivariate_pdf} The multivariate PDF of a continuous-valued random vector $\pmb{\mrm{X}}$ is denoted as $p(\pmb{x})$. Its integration over a domain $\pmb{\Zeta}$ in the $n$-dimensional space of the values of $\,\pmb{\mrm{X}}$ yields the probability of            $\,\pmb{\mrm{X}} \in \pmb{\Zeta}$, i.e:
    \begin{equation}
        \Prob{\pmb{\mrm{X}}\in\pmb{\Zeta}} = \int_{\pmb{\Zeta}} p(\pmb{x}) \, \D \pmb{x},
    \end{equation}
    where $\int \D \pmb{x}$ is shorthand for $\int_{-\infty}^{\infty} \cdots \int_{-\infty}^{\infty} \dd x_1 \dots \dd x_n$. Just like for the single-variable PDF, its multivariate counterpart must satisfy the following conditions
    \begin{equation}
        p(\pmb{x}) \geq 0 \; \forall \pmb{x}, \;\; \text{and} \;\; \int p(\pmb{x}) \, \D \pmb{x} = 1. \nonumber
    \end{equation}
    \label{def:multivariate_PDF}
\end{definition} 

\subsection{Joint probability functions}

When working with more than one random variable, their combined behavior is described by \emph{joint probability functions}, which capture the probability of two or more random variables taking specific values simultaneously. 

\begin{definition}[Joint PMF] For discrete random variables, the joint behavior of $\,\mrm{X}$ and $\mrm{Y}$ is described by the joint PMF, which yields the probability of $\,\mrm{X}$ taking the value $x$ and $\mrm{Y}$ taking the value $y$, simultaneously:
    \begin{equation}
        p(x,y) = \Prob{\mrm{X}=x, \mrm{Y}=y}.
    \end{equation}
\end{definition}

\begin{definition}[Joint PDF] For continuous random variables, the joint behavior of random variables $\mrm{X}$ and $\mrm{Y}$ is described by the joint PDF $p(x,y)$, which provides the relative likelihood of $\,\mrm{X}$ taking a value near $x$ and $\mrm{Y}$ taking a value near $y$:
    \begin{align}
        \Prob{a\leq \mrm{X} \leq b,c\leq \mrm{Y} \leq d} = \int_a^b \int_c^d p(x,y) \, \dd y \, \dd x.
    \end{align}
\end{definition}

Both definitions can be easily extended to multivariate random variables. 

\subsection{Independent and dependent random variables}

Two random variables are \emph{independent} if neither of their probability functions depends on the value of the other variable. Their definition is formalized through the use of joint probability functions. In particular:
\begin{definition}[Independent discrete random variables] Let $\mrm{X}$ and $\mrm{Y}$ be two discrete random variables, with ranges $\mathcal{X}$ and $\mathcal{Y}$, respectively. We say that $\mrm{X}$ and $\mrm{Y}$ are independent if for all $x\in\mathcal{X}$ and $y\in\mathcal{Y}$, the joint PMF satisfies
    \begin{align}
        \Prob{\mrm{X} = x, \mrm{Y} = y} = \Prob{\mrm{X}=x} \cdot \Prob{\mrm{Y}=y} \;\;\;\; \forall \, x,y
    \end{align}
    In other words, for every pair of possible outcomes $x$ and $y$, the probability that $\mrm{X}$ takes the value $x$ and $\mrm{Y}$ takes the value $y$ is the product of the individual probabilities of those events. That can be extended to multivariate discrete random variables, as
    \begin{align}
        \Prob{\mrm{X}_1 = x_1,\dots,\mrm{X}_n = x_n} = \Prob{\mrm{X}_1 = x_1} \cdot \ldots \cdot \Prob{\mrm{X}_n = x_n}, \;\;\;\; \forall \, x_1,\dots,x_n.
    \end{align}
\end{definition}

Additionally, a similar definition applies to independent continuous random variables, where independence is characterized by the joint PDF factoring into the product of their respective probability functions. Specifically,
\begin{definition}[Independent continuous random variables] Let $\mrm{X}$ and $\mrm{Y}$ be two continuous random variables with joint PDF $p(x,y)$. We say that $\mrm{X}$ and $\mrm{Y}$ are independent if the joint PDF can be written as a product of the marginal densities:
    \begin{align}
        p(x,y) = p(x) \cdot p(y),
    \end{align}
    where 
    \begin{align}
        p(x) = \int_{-\infty}^\infty p(x,y) \dd y, \quad p(y) = \int_{-\infty}^\infty p(x,y) \dd x.
    \end{align}
    Equivalently, for all real $a\leq b$ and $c \leq d$ s.t. $x\in[a,b]$ and $y\in[c,d]$,
    \begin{align}
        &\Prob{\mrm{X}\in[a,b],\mrm{Y}\in[c,d]} = \int_a^b \int_c^d p(x,y) \, \dd y\, \dd x = \int_a^b \int_c^d p(x) p(y) \, \dd y \, \dd x \nonumber \\
        &\quad= \left(\int_a^b p(x) \, \dd x\right)\left(\int_c^d p(y) \, \dd y \right) = \Prob{\mrm{X}\in[a,b]} \cdot \Prob{\mrm{Y}\in[c,d]}.
    \end{align}
    In other words, the probability that $\mrm{X}$ falls in the interval $[a,b]$ as well as $\mrm{Y}$ falls in $[c,d]$ is the product of the individual probabilities of these events. Just like in the discrete case, this definition can also be extended to multivariate continuous random variables, as
    \begin{align}
        p(\pmb{x}) = p(x_1,\dots,x_n) = p(x_1) \cdot \ldots \cdot p(x_n), \;\;\;\; \forall \, x_1,\dots,x_n.
    \end{align}
\end{definition}

On the other hand, two random variables are \emph{dependent} when they have a joint probability density that cannot be factored into the product of their respective probability functions. Unlike independent random variables, where knowing the outcome of one variable tells us nothing about the other, the value of one dependent variable is influenced by the value of the other.

For continuous random variables, this dependency is best described through the use of the   \emph{conditional PDFs} and \emph{marginal PDFs}. Marginal PDFs are isolated densities of a single random variable where we have removed the influence of all the other variables. All these functions are central to the understanding of Bayesian filtering and estimation. 

For that reason, let us start by defining the probability density for $\mrm{X}$ if we were to know nothing about $\mrm{Y}$, i.e. the marginal of $\mrm{X}$, through the law of total probability. Namely,
\begin{property}[Marginal PDF / The law of total probability / Sum rule] \label{propty:sum_rule}
    Let $\mrm{X}$ and $\mrm{Y}$ be two random variables, such that $\mrm{Y} \in \mathcal{Y}$, then the law of total probability states that the marginal PDF reads as:
    \begin{equation} \label{eq:sum_rule}
        p(x) = \int_{y\,\in \,\mathcal{Y}} p(x,y) \dd y. 
    \end{equation}
\end{property}
Thus, marginalizing over $\mrm{Y}$ is simply integrating over all possible values of $\mrm{Y}$ such that we are left with a PDF that depends only on $\mrm{X}$. 

Now we have all the tools needed to define the conditional PDF, which in turn yields a very important property in probability theory: the product rule. A conditional distribution is the probability distribution of one random variable given that another has taken a specific value. It is derived from their joint distribution, since $p(x,y)$ with $y=a$ gives the relative probability for $x$ given that $\mrm{Y} = a$. However, $p(x,y=a)$ is not normalized and therefore, we have to divide it by the integral over all values of $x$, i.e., the marginal of $\mrm{Y}$. 

\begin{property}[Conditional PDF / Product rule] \label{propty:prod_rule} 
    Let $\mrm{X}$ and $\mrm{Y}$ be two random variables with the joint PDF, $ p(x,y)$. The conditional density function, $p(x|y)$, which reads as the probability density of $x$ given $y$, is defined as
    \begin{equation} \label{eq:cond_def_1}
        p(x|y) = \frac{p(x,y)}{\int_{-\infty}^{\infty} p(x,y) \, \dd x},
    \end{equation}
    which, since the denominator of \eqnref{eq:cond_def_1} is the marginal of $\mrm{Y}$, it can be rewritten as
    \begin{equation}
        p(x|y) = \frac{p(x,y)}{p(y)}.
    \end{equation}
    The definition of the conditional density function yields the product rule:
    \begin{equation} \label{eq:prod_rule}
        p(x,y) = p(x|y) p(y).
    \end{equation}    
\end{property}
Note that $p(x|y)$ should be interpreted as a function of only $x$ since $y$ is a known value, i.e. we have no uncertainty in $y$.

\subsection{Expectation and covariance}

One of the most important distributions for both linear and nonlinear filtering is the Gaussian distribution, which can be described using only the mean vector (or expected value) and the covariance matrix of the Gaussian random variable. And if our results are not Gaussian, we anyway tend to approximate them using the mean and covariance of this non-Gaussian distribution, even though these do not fully capture all the facets of our actual distribution.
Therefore, let us first define what is an expected value and a covariance matrix.

\begin{definition}[Expected value / mean] The expected value (i.e. mean) of a continuous random variable $\mrm{X}$ is defined as
    \begin{equation} \label{eq:mean_def}
        \EE{\mrm{X}} \coloneqq \int_{-\infty}^\infty x \; p(x) \; \dd x.
    \end{equation}
\end{definition}

The expectation of $\mrm{X}$ is the integral of $x$ weighted by the probability density of $x$. Thus, values of $\mrm{X}$ with high probability density will influence our expected values more than values of $\mrm{X}$ with low probability density. The expected value of $\mrm{X}$ is also sometimes referred to as the first moment of $\mrm{X}$. We can further define higher order moments of $\mrm{X}$, such as the variance of $\mrm{X}$ as:
\begin{definition}[Variance] The variance of a random variable $\mrm{X}$ is the expected value     of the squared deviation of $\mrm{X}$ from the mean $\EE{\mrm{X}}$:
    \begin{equation} \label{eq:variance_def}
        \mrm{V}[\mrm{X}] \coloneqq \EE{(\mrm{X} - \EE{\mrm{X}})^2} = \EE{\mrm{X}^2} - \EE{\mrm{X}}^2 .
    \end{equation}
\end{definition}

Additionally, the definition of the mean and other moments can also be extended to random vectors. Specifically, the multivariate mean will read as:
\begin{definition}[Multivariate mean]
    A random vector $\pmb{\mrm{X}} = [\mrm{X}_1,\mrm{X}_2,\ldots,\mrm{X}_n]^\Trans$ has an expected value (mean) given by
    \begin{equation}
        \EE{\pmb{\mrm{X}}} \coloneqq \int \pmb{x} \; p(\pmb{x}) \; \D \pmb{x},
    \end{equation}
    where $p(\pmb{x})$ is the multivariate PDF of \defref{def:multivariate_pdf}. 
\end{definition}
We can also extend the definition of the variance to multivariate random variables by defining the covariance matrix of $\pmb{\mrm{X}}$. For a random vector $\pmb{\mrm{X}} = (\mrm{X}_1,\mrm{X}_2,\ldots,\mrm{X}_n)^\Trans$, the covariance matrix encodes how each pair of variables $\mrm{X}_i$ and $\mrm{X}_j$ vary together, revealing patterns of correlation and dependencies across the components of $\pmb{\mrm{X}}$. Formally, the covariance matrix is defined as:
\begin{definition}[Covariance matrix]
    Let $\pmb{\mrm{X}}$ be a random vector. Then, the covariance matrix is
    \begin{equation}
        \cov[\pmb{\mrm{X}}] \coloneqq \EE{(\pmb{\mrm{X}} - \EE{\pmb{\mrm{X}}})(\pmb{\mrm{X}} - \EE{\pmb{\mrm{X}}})^{\!\Trans\;}}.
    \end{equation}
\end{definition}
We can view the factor $(\pmb{\mrm{X}} - \EE{\pmb{\mrm{X}}})$ as the distance between the random vector $\pmb{\mrm{X}}$ and its mean. That is, how much does it spread around its mean value. Additionally, note that the diagonal elements of $\cov[\pmb{\mrm{X}}]$, i.e. $\cov[\pmb{\mrm{X}}]_{ii}$ are the variances of each individual variable $\mrm{X}_i$, while the off-diagonal elements indicate the correlation between different variables.

\subsection{Law of large numbers}

Sometimes we are interested in finding the expected value of a random value, but an analytical solution is not available. Perhaps we are not able to solve the involved integral explicitly, or more commonly, because the exact form of the underlying distribution is not known. However, if instead we have access to a large number of samples from the random variable, we can use the law of large numbers and estimate the expected value numerically. 

In particular, the law of large numbers states that, as the number of independent and identically distributed samples of a random variable increases, the sample average converges to the true expected value of that variable. In other words:

\begin{theorem}[The law of large numbers] \label{def:law_large_numbers} 
    Let $\mrm{X}_1, \; \mrm{X}_2, \ldots, \mrm{X}_n$ be a sequence of independent and identically distributed random variables. In other words, they are all distributed according to the same PDF, $p(x)$, and thus, have a finite expected value $\mu = \EE{\mrm{X}_i} \; \forall i = 1,\dots,n$. Then,
    \begin{equation}
        \lim_{n \rightarrow \infty} \bar{\mrm{X}}_n = \lim_{n \rightarrow \infty} \frac{1}{n} \sum_{i=1}^n \mrm{X}_i = \mu.
    \end{equation}
    Or, in other words, the sample average $\bar{X}_n$ converges almost surely\footnote{$\Prob{\lim_{n \rightarrow \infty} \bar{\mrm{X}}_n = \mu} = 1$} to the expected value $\mu$.
\end{theorem}

\subsection{The multivariate Gaussian distribution}

Let $\pmb{\mrm{X}} \sim \Gauss(\pmb{x}|\pmb{\mu},\pmb{\Sigma})$ denote a Gaussian random vector drawn from a multivariate PDF with mean $\pmb{\mu}$ and covariance $\pmb{\Sigma}$. The PDF of $\pmb{x}$ is known as a Gaussian distribution and defined as:
\begin{definition}[Multivariate Gaussian distribution] A Gaussian random vector $\pmb{\mrm{X}} \in \Real^n$ has a probability density
    \begin{equation} \label{eq:multivariate_Gaussian}
        \Gauss(\pmb{x}|\,\pmb{\mu},\pmb{\Sigma}) = \frac{1}{(2\pi)^{n/2} |\pmb{\Sigma}|^{1/2}} \exp{\left(-\frac{1}{2} (\pmb{x} - \pmb{\mu})^\Trans \pmb{\Sigma}^{-1} (\pmb{x} - \pmb{\mu}) \right)}
    \end{equation}
    where $|\;\cdot\;|$ denotes the determinant, $\pmb{\mu} = \EE{\pmb{\mrm{X}}}$ is the mean of $\pmb{\mrm{X}}$ and $\pmb{\Sigma}$ represents the covariance of $\,\pmb{\mrm{X}}$, i.e. $\pmb{\Sigma} \coloneqq \cov[\pmb{\mrm{X}}] =\EE{(\pmb{\mrm{X}} - \EE{\pmb{\mrm{X}}})(\pmb{\mrm{X}} - \EE{\pmb{\mrm{X}}})^{\!\Trans\;}}$. 
\end{definition}
Just as in the case of a Gaussian random variable, the PDF of a Gaussian random vector comes completely determined by its mean and covariance matrix. Furthermore, one of the most useful properties of Gaussian random variables is their linear combination property. 
\begin{property}[Linearity of Gaussian random vectors]
    Let $\pmb{\mrm{X}} \sim \Gauss(\pmb{x}|\,\pmb{\mu}_x, \pmb{\Sigma}_x)$ be a Gaussian random vector, as well as  $\pmb{\mrm{Y}} \sim \Gauss(\pmb{y}|\,\pmb{\mu}_y,\pmb{\Sigma}_y)$. Then, a linear combination of $\pmb{\mrm{X}}$ and $\pmb{\mrm{Y}}$ yields another Gaussian random vector $\pmb{\mrm{Z}}$:
    \begin{equation}
        \pmb{\mrm{Z}} = \pmb{A}\pmb{\mrm{X}} + \pmb{B}\pmb{\mrm{Y}},
    \end{equation}
    where $\pmb{\mrm{Z}}$ has a mean and covariance:
    \begin{align}
        \pmb{\mu}_z &= \EE{\pmb{A}\pmb{\mrm{X}} + \pmb{B}\pmb{\mrm{Y}}} = \pmb{A}\pmb{\mu}_x + \pmb{B}\pmb{\mu}_y, \\
        \pmb{\Sigma}_z &= \cov\!\left[\pmb{A}\pmb{\mrm{X}} + \pmb{B}\pmb{\mrm{Y}}\right] = \cov\!\left[\pmb{A}\pmb{\mrm{X}}\right] + \cov\!\left[\pmb{B}\pmb{\mrm{Y}}\right] = \pmb{A}\pmb{\Sigma}_x\pmb{A}^\Trans + \pmb{B}\pmb{\Sigma}_y\pmb{B}^\Trans\!,
    \end{align}
    with $\pmb{A}$ and $\pmb{B}$ being some deterministic matrices specifying the linear combination of $\pmb{\mrm{X}}$ and $\pmb{\mrm{Y}}$ that gives $\pmb{\mrm{Z}}$. 
\end{property}

\section{Stochastic processes} \label{sec:stochastic_process}

A \emph{stochastic process} is defined as a collection of random variables
\begin{equation}
    \{\mrm{X}(t), \; \text{where} \; t \in T\},
\end{equation}
where each random variable is associated with a distinct point $t$ in a parameter space $T$. The parameter space $T$, often interpreted as time, can be any set index, though it is commonly a subset of the real numbers $\Real$. 

We can sample the stochastic process $\mrm{X}(t)$ at times $t_0,t_1,\ldots,t_n$, ordered such that: $t_0 \leq t_1 \leq \ldots \leq t_n$. The evaluation of the stochastic process at these times yields a sequence of random variables $\pmb{\mrm{X}}_{0:n} = \{\mrm{X}_0,\mrm{X}_1,\ldots,\mrm{X}_n\}$ with a joint probability density:
\begin{equation} \label{eq:joint_PDF_formarkov}
    p(\pmb{x}_{0:n}) = p(x_0,x_1,\ldots,x_n),
\end{equation}
which can be related to the conditional PDF through the product rule:
\begin{equation} \label{eq:probability_x_0:n}
   p(\pmb{x}_{0:n}) =p(\pmb{x}_{0:n\shortminus1},x_n) = p(x_{n}|\pmb{x}_{0:n\shortminus1}) p(\pmb{x}_{0:n\shortminus1}).
\end{equation}

\subsection{Markov process} \label{sec:markov_process}

A stochastic process is called a \emph{Markov process} if it satisfies the Markov property, which states that the conditional probability distribution of future states of the process depends only upon the knowledge of the most recent realization, not on the sequence of events that preceded it. 
\begin{property}[Markov property] \label{propty:markov} 
    A sequence of random variables $\pmb{\mrm{X}}_{0:n} = \{\mrm{X}_k\}_{k = 0, 1, \dots, n}$ form a Markov sequence (or Markov chain) if $\mrm{X}_k$ given $\mrm{X}_{k\shortminus1}$ is independent of all the other variables up to time $k\!\shortminus\!1$:
    \begin{equation} \label{eq:Markov_def}
        p(x_k|\pmb{x}_{0:k\shortminus1}) = p(x_k|x_{k\shortminus1}).
    \end{equation}
\end{property}
It follows that the form given in \eqnref{eq:probability_x_0:n} can be simplified as:
\begin{align}
    p(\pmb{x}_{0:n}) = p(x_0,x_1,\ldots,x_n) &\underset{\ref{eq:prod_rule}}{=} p(x_n|\pmb{x}_{0:n\shortminus1}) p(\pmb{x}_{0:n\shortminus1}) \nonumber \\
    &\underset{\ref{eq:Markov_def}}{=} p(x_n|x_{n\shortminus1}) p(\pmb{x}_{0:n\shortminus2},x_{n\shortminus 1})  \nonumber \\
    &\underset{\ref{eq:prod_rule}}{=}  p(x_n|x_{n\shortminus1}) p(x_{n\shortminus1}|\pmb{x}_{0:n\shortminus2}) p(\pmb{x}_{0:n\shortminus2}) \nonumber \\
    &= p(x_n|x_{n\shortminus1}) p(x_{n\shortminus1}|x_{n\shortminus2}) \cdot \ldots \cdot p(x_1|x_0) p(x_0), \label{eq:Markov_form_PDF}
\end{align}
provided that the stochastic process $\mrm{X}(t)$ is evaluated at times $t_0,t_1,\ldots,t_n$ fulfilling $t_0 \leq t_1 \leq \ldots \leq t_n$.

\begin{definition}[Chapman-Kolmogorov Equation] \label{def:Chapman_Kolmogorov_def}
The Chapman-Kolmogorov equation expresses the marginal transition probability between two states of a Markov process in terms of an intermediate state. Using the Markov property and the product rule, it takes the form
\begin{align}
    p(x_k|x_{k\shortminus2}) = \int \dd x_{k\shortminus1} \, p(x_k|x_{k\shortminus1}) \, p(x_{k\shortminus1} | x_{k\shortminus2}),
    \label{eq:Chapman_Kolmogorov_def}
\end{align}
where the conditional independence $ p(x_k | x_{k\shortminus1}, x_{k\shortminus2}) = p(x_k | x_{k\shortminus1})$ follows from the Markov assumption.
\end{definition}

\begin{myproof}
    The statement above can be quickly shown using the sum rule, the Markov assumption and the product rule:
    \begin{align}
    p(x_k|x_{k\shortminus2}) &\underset{\ref{eq:sum_rule}}{=} \int \dd x_{k\shortminus1} \, p(x_k,x_{k\shortminus1}|x_{k\shortminus2}) \underset{\ref{eq:prod_rule}}{=} \int \dd x_{k\shortminus1} \, p(x_k|x_{k\shortminus1},x_{k\shortminus2}) p(x_{k\shortminus1}|x_{k\shortminus2}) \nonumber \\
    &\underset{\ref{eq:Markov_def}}{=} \int \dd x_{k\shortminus1} \, p(x_k|x_{k\shortminus1}) p(x_{k\shortminus1}|x_{k\shortminus2}).
\end{align}
\end{myproof}

\subsection{Poisson process} \label{sec:Poisson_process}

\begin{definition}[Poisson process] Let $\{\N(t), t \in [0,\infty) \}$ be a stochastic process that counts the number of events occurring up to time $t$. The process $\N(t)$ is said to be a Poisson process with rate $\lambda > 0$ if the following conditions are satisfied:

\begin{enumerate}
    \item \textbf{Initial condition:} The process starts with zero events at time zero, i.e. $\N(0) = 0$.
    \item \textbf{The process has independent increments:} For any $ 0 \leq s < t $, the number of events occurring in the interval $ [s, t] $ is given by $\Delta \N(t-s) \coloneqq \N(t) - \N(s) $, which is independent of the number of events occurring prior to $ s $.
    \item \textbf{The process has stationary increments:} For any $ 0 \leq s < t $, the number of events in $ [s, t] $ depends only on the length of the interval $ t - s $, and not on the specific values of $ s $ and $ t $.
    \item \textbf{Continuity in probability:} For any $\epsilon > 0$ and $t\geq 0$, it holds that
    \begin{equation}
        \lim_{\Dt\to0} \Prob{|\N(t+\Dt)-\N(t)|>\epsilon} = 0.
    \end{equation}
    In other words, the Poisson process $\N(t)$ has almost surely continuous trajectories. Crucially, this condition still allows the infinitesimal increments of $\N(t)$ to be finite, i.e. for the process to jump. 
    \item  \textbf{Poisson distribution for the increments:} For each $ t \geq 0 $, the number of events in $ [t, t+\Dt] $ follows a distribution with rates:
    \begin{align} \label{eq:conditons_DeltaN}
        &\Prob{\Delta \N(\Dt) \coloneqq \N(t + \Dt) - \N(t) = 1} = \lambda \Dt + \littleo(\Dt), \\
        &\Prob{\Delta \N(\Dt) \coloneqq \N(t +  \Dt) - \N(t) = 0} = 1 - \lambda  \Dt + \littleo(\Dt), \\
        &\Prob{\Delta \N(\Dt) \coloneqq \N(t +  \Dt) - \N(t) > 1} = \littleo(\Dt),
    \end{align}
    where the probability of a jump occurring between $t$ and $t+\Dt$ vanishes as $\Dt \to 0$. Here, the little-o notation $\littleo(\Dt)$\footnote{The notation $\littleo(\Dt)$ means that the additional terms vanish faster than $\Dt$, i.e., they become negligible as $\Dt \to 0$. This differs from Big-O notation, which describes the scaling behavior up to a constant factor. Formally: 
    \begin{equation}
        f(\Dt) = \begin{cases}
            \littleo(\Dt^k) \quad \Longleftrightarrow \quad \lim_{\Dt \to 0} \dfrac{f(\Dt)}{\Dt^k} = 0\\
            \bigO(\Dt^k) \quad \Longleftrightarrow \quad \lim_{\Dt \to 0} \dfrac{f(\Dt)}{\Dt^k} = \text{const}
        \end{cases}
    \end{equation}} represents a function that fulfills $ \lim_{ \Dt \rightarrow 0} \frac{\littleo(\Dt)}{ \Dt} = 0 $. 
\end{enumerate}
\end{definition}

Therefore, if we now take the limit of $\Dt \to 0$, we can define an infinitesimal increment:
\begin{equation} \label{eq:dN_def}
    \dN \coloneqq \N(t +  \dt) - \N(t), 
\end{equation}
that counts the number of events that occur in the interval $ [t, t+ \dt] $. Namely,
\begin{align} \label{eq:conditions_dN}
    &\Prob{\dN = 1} = \lambda \dt + \littleo(\dt), \\
    &\Prob{\dN = 0} = 1 - \lambda \dt + \littleo(\dt), \\
    &\Prob{\dN > 1} = \littleo(\dt).
\end{align}
In other words, in an infinitesimal interval $[t,t+\dt]$, there can be either one jump with a vanishing probability of $\lambda \dt$, or no jumps, with a probability $1-\lambda \dt$. Therefore, since $\dN$ can only have values $0$ or $1$, it follows that 
\begin{equation} \label{eq:dN2=dN}
    \dN\tinyspace^2 = \dN. 
\end{equation}
Additionally, both the expectation and variance of this increment, $ \dN $, are equal to $ \lambda  \dt $:
\begin{align} \label{eq:mean_variance_Poisson}
   &\EE{\dN} = \sum_{k=0}^1 k \; \Prob{\dN = k} =  \lambda  \dt + \littleo(\dt), \\
   &\var{\dN} = \EE{\dN\tinyspace^2} - \EE{\dN}^2 = \sum_{k=0}^1 k^2 \; \Prob{\dN = k} - (\lambda \dt)^2 = \lambda  \dt +  \littleo(\dt).
\end{align}
The distribution above can be shown to be a Poisson distribution by discretizing a total time evolution $[0,T]$ by time steps $\Dt$ s.t. $n = t/\Dt$ and calculating the probability that $k$ jumps occur in $k$ intervals in a total of $n$  intervals. The number of different combinations of $k$ elements from a total group of $k \leq n$ elements is given by: 
\begin{equation}
    \binom{n}{k} = \frac{n!}{k!(n-k)!}.
\end{equation}
Therefore, in the limit of $\Dt \to 0$, or equivalently, $n\to\infty$, we get
\begin{align}
    \Prob{(\N(t) = k} &= \lim_{n\to\infty} \frac{n!}{k!(n-k)!} (\lambda \Dt)^k (1-\lambda \Dt)^{n-k} \nonumber \\
    &= \lim_{n\to\infty} \frac{n!}{(n-k)! \, n^k} \frac{(\lambda t)^k (1-\lambda \tinyspace t/n)^n}{k! (1-\lambda \tinyspace t/n)^k} \nonumber \\
    &= \frac{(\lambda t)^k \ee^{-\lambda t}}{k!}, \;\;\; \text{where} \;\;\; k = 0,1,2,\ldots
\end{align}
where in the last step we have used the following identities:
\begin{align}
    &\lim_{n\to\infty} (1-\lambda\tinyspace t/n)^n = \ee^{-\lambda t}, \\
    &\lim_{n\to\infty} (1-\lambda \tinyspace t/n)^k = 1, \\
    &\lim_{n\to\infty} \frac{n!}{(n-k)!n^k} = \lim_{n\to\infty} \frac{n}{n}\frac{n-1}{n}\cdot \ldots \cdot \frac{n-k+1}{n} = \lim_{n\to\infty} \frac{n^k}{n^k} = 1.
\end{align}

As stated in the third condition of the definition of a Poisson process $\N(t)$, the process has stationary increments. Therefore, the increment $\Delta \N(\Dt) = \N(t+\Dt) - \N(t)$ has the same distribution as $\N(\Dt)$. Then,
\begin{align}
    \Prob{\Delta \N = k} &= \Prob{(\N(t+\Dt) - \N(t)) = k} = \Prob{\N(\Dt) = k} \nonumber \\
    &= \frac{(\lambda \Dt)^k}{k!} \ee^{-\lambda \Dt} \;\;\; \mrm{for} \;\;\; k = 0,1,2,\dots .
\end{align}

So to sum up, the Poisson increment, $\Delta \N(\Dt) \sim \Pois(\lambda \Dt)$, is drawn from a Poisson distribution with occurrences $k$, which in principle can be $0, 1, 2,$ and so on. However, as $\Dt \rightarrow 0$, $\dN = \lim_{\Dt\to 0} \Delta \N(\Dt)$, and the probability of $\dN$ taking values greater than $1$ becomes negligible. In other words, $\dN$ is only either $0$ or $1$ almost surely\footnote{An event happens \emph{almost surely} when it happens with probability 1.}, i.e. $\Prob{\dN \in \{0,1\}} = 1$. 

These properties describe a discrete process where events occur one at a time, and with the time between events following an exponential distribution with the parameter $ \lambda $.

So far we have considered the case of a homogeneous Poisson process. Namely, a Poisson process whose rate $\lambda$ is constant in time. However, in future chapters we will require Poisson processes whose rate $\lambda$ is actually time-dependent, i.e. $\lambda(t)$. For that reason, let us introduce next \emph{non-homogeneous} Poisson processes, which fulfill the same conditions of a homogeneous Poisson process with the exception of a constant rate $\lambda$ and stationary increments. 

\begin{definition}[Non-homogeneous Poisson process] \label{def:non-hom_Poisson}
    Let $\N(t) \in [0,\infty)$ be a non-homogeneous Poisson processes with rate $\lambda(t) > 0$ that is locally integrable if it fulfills:
    \begin{enumerate}
        \item \textbf{Initial condition:} The process starts with $\N(0) = 0$.
        \item \textbf{The process has independent increments:} For any $0 \leq s < t$, the number of events occurring in the interval $[s,t]$ is given by $\Delta \N (t-s) \coloneqq \N(t) - \N(s)$, which is independent of the number of events prior to $s$. 
        \item \textbf{The process is almost surely continuous} over its trajectories.
        \item \textbf{Poisson distribution of the increments:} For any $t \geq 0$, the number of events in $[t,t+\Delta t]$ follow a distribution with rates:
        \begin{align} \label{eq:conditons_DeltaN_nonhom}
        &\Prob{\Delta \N(\Dt) \coloneqq \N(t + \Dt) - \N(t) = 1} = \lambda(t) \Dt + \littleo(\Dt), \\
        &\Prob{\Delta \N(\Dt) \coloneqq \N(t +  \Dt) - \N(t) = 0} = 1 - \lambda(t)  \Dt + \littleo(\Dt), \\
        &\Prob{\Delta \N(\Dt) \coloneqq \N(t +  \Dt) - \N(t) > 1} = \littleo(\Dt),
    \end{align}
    where $\littleo(\Dt)$ is the little-o notation fulfilling $ \lim_{ \Dt \rightarrow 0} \frac{\littleo(\Dt)}{ \Dt} = 0 $. 
    \end{enumerate}
\end{definition}

Another equivalent definition of a non-homogeneous Poisson process contains points 1-3 from \defref{def:non-hom_Poisson} and also that the probability of having $k$ events in the interval $[t, t+\Delta t]$ is:
\begin{align}
    \Prob{\Delta \N(\Dt) = k} = \Prob{\N(t+\Delta t) - \N(t) = k} = \frac{\Lambda(t,t+\Dt)^k \ee^{-\Lambda(t,t+\Dt)}}{k!}, 
\end{align}
where $k = 0,1,2,\dots$ and the Poisson rate is
\begin{equation}
    \Lambda(t,t+\Dt) = \int_t^{t+\Dt} \lambda(\tau) \dd \tau.
\end{equation}

\subsection{Wiener process} \label{sec:Wiener_process}

\begin{definition}[Wiener process] Let $\{\W(t), t \in [0,\infty)\}$ be a stochastic process representing the continuous evolution of a random variable over time. The process $\W(t)$ is said to be a Wiener process if it satisfies the following conditions:
\begin{enumerate}
    \item \textbf{Initial condition:} $\W(0) = 0$ almost surely.
    \item \textbf{Independent increments:} For any $0 \leq s < t$, the increment $\Delta\!\W(t-s) \coloneqq \W(t) - \W(s)$ is independent of the process history up to time $s$. Thus, evolution over any interval depends only on that interval, independent of past values. 
    \item \textbf{Stationary increments:} For any $0 \leq s < t$, the increment $\W(t) - \W(s)$ depends only on the length of the interval $t - s$ and not on the specific values $s$ and $t$. This means that the distribution of $\Delta\!\W(t-s) = \W(t) - \W(s)$ is identical to that of $\W(t-s)$, i.e. the increment is just another Wiener process. 
    \item \textbf{Continuity in probability:} The process $\W(t)$ is almost surely continuous in $t$. Specifically, for any $\epsilon > 0$ and $t\geq 0$, it holds that
    \begin{equation}
        \lim_{\Dt\to0} \Prob{|\W(t+\Dt)-\W(t)|>\epsilon} = 0
    \end{equation}
    \item \textbf{Normally distributed increments:} For any $0 \leq s < t$, the increment $\W(t) - \W(s)$ is normally distributed with mean zero and variance $t-s$. Specifically, 
    \begin{align} \label{eq:W_def_1} 
        \Delta\!\W(t-s) \coloneqq \W(t) - \W(s) \sim \Gauss(0, t-s), 
    \end{align} 
    where $\Gauss(\,\cdot\; ,\,\cdot\,)$ is a Gaussian distribution, introduced in \eqnref{eq:multivariate_Gaussian}, with first and second moments:
    \begin{align}
        &\EE{\Delta\!\W(t-s)} = 0, \\
        &\EE{\Delta\!\W(t-s)^2} = t-s. \label{eq:EdW^2=t-s}
    \end{align}
\end{enumerate}
\end{definition}

Thus, if we take the limit $\Dt \to 0$, we can define an infinitesimal increment
\begin{equation}
    \dW \coloneqq \W(t+\dt) - \W(t) \sim \Gauss(0,\dt),
\end{equation}
which represents the change in the Wiener process over the infinitesimal interval $[t,t+\dt]$ and has mean zero and variance $\dt$:
\begin{align}
    \EE{\dW} &= 0, \\
    \mrm{V}[\dW] &= \EE{\dW^2} = \dt.
\end{align}

The paths of the Wiener process are continuous but nowhere differentiable. In other words, although $\W(t)$ has continuous sample paths, it has no well-defined slope at any point.

\section{Stochastic calculus with Gaussian noise} \label{sec:stochastic_calculus}
Stochastic calculus is a branch of mathematics that extends traditional calculus to handle systems driven by randomness. It addresses how to integrate and differentiate functions of stochastic processes, such as $f(\mrm{X}(t))$, with respect to other stochastic processes. One possible way of doing that is with the It\^{o} integral, which integrates a given stochastic process $\mrm{X}(t)$ with respect to the Wiener process $\W(t)$:
\begin{equation}
    \int_0^T \mrm{X}(t) \dW.
\end{equation}
Since $\W(t)$ has (random) continuous but nowhere differentiable paths, classical calculus techniques such as Riemann-Stieltjes integration is not applicable. 

\subsection{It\^{o} integral} \label{sec:ito_integral}

To define the It\^{o} integral, let us consider a function $f\left(t,\mrm{X}(t)\right)$ that depends on a time parameter $t$ and some stochastic process $\mrm{X}(t)$ up to time $t$. It follows that the function $f\left(t,\mrm{X}(t)\right)$ is in itself another stochastic process, and we have to further assume that $f\left(t,\mrm{X}(t)\right)$ is a  \emph{non-anticipating} function (a.k.a. \emph{non-anticipating} process or \emph{adapted} process), i.e. $f\left(t,\mrm{X}(t)\right)$ is independent of the behavior of $\mrm{X}(t)$ and the Wiener process $\W(t)$ in the future of $t$. More rigorously:
\begin{definition}[Non-anticipating function] \label{def:non-anticipating_f}
A function $f(t, \mrm{X}(t))$ is said to be non-anticipating (or adapted) if, $\forall \tau > t$, it is statistically independent of the future increment $\W(\tau) - \W(t)$ of the Wiener process.
\end{definition}

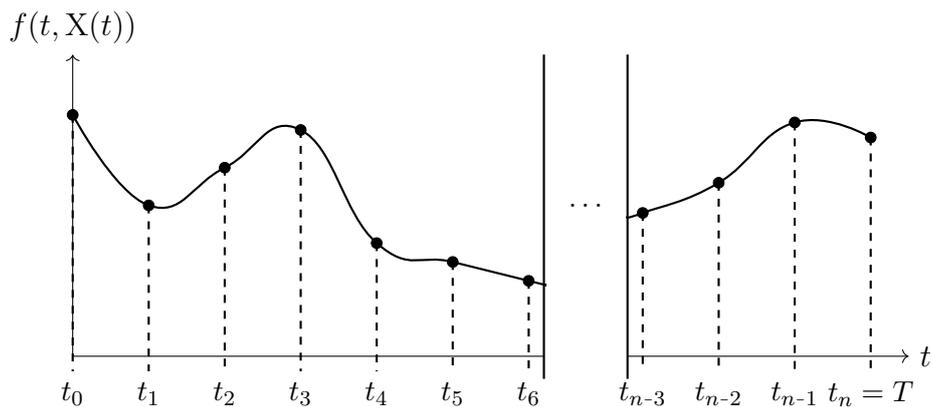
\begin{figure}[htbp]
\begin{center}
    \begin{tikzpicture}
    \draw[-] (0, 0) -- (6.2, 0);
    \draw[->] (7.3, 0) -- (11, 0) node[right] {$t$};
    \draw[->] (0, 0) -- (0, 4) node[above] {$f(t,\mrm{X}(t))$};
    
    \draw[thick] plot[smooth, tension=0.8] coordinates {(0,3.2) (1, 2) (2, 2.5) (3,3) (4, 1.5) (5, 1.25) (6, 1) (6.2,0.95)};

    \draw[thick] plot[smooth, tension=0.8] coordinates {(7.3,1.85) (7.5,1.9) (8.5,2.3) (9.5,3.1) (10.5,2.9)};
    
    \filldraw[black] (0, 3.2) circle (2pt) node[above] {};
    \filldraw[black] (1, 2) circle (2pt) node[above] {};
    \filldraw[black] (2, 2.5) circle (2pt) node[above] {};
    \filldraw[black] (3, 3) circle (2pt) node[above] {};
    \filldraw[black] (4, 1.5) circle (2pt) node[above] {};
    \filldraw[black] (5, 1.25) circle (2pt) node[above] {};
    \filldraw[black] (6, 1) circle (2pt) node[above] {};
    \filldraw[black] (7.5, 1.9) circle (2pt) node[above] {};
    \filldraw[black] (8.5, 2.3) circle (2pt) node[above] {};
    \filldraw[black] (9.5, 3.1) circle (2pt) node[above] {};
    \filldraw[black] (10.5, 2.9) circle (2pt) node[above] {};

    \draw[thick] (6.2, -0.3) -- (6.2, 4);
    \draw[thick] (7.3, -0.3) -- (7.3, 4);
    \node at (6.78, 2) {\small $\dots$};

    \draw[thick, dashed] (0, 3.2) -- (0, -0.2);
    \node at (0, -0.5) {\small $t_0$};
    \draw[thick, dashed] (1,2) -- (1,-0.2);
    \node at (1, -0.5) {\small $t_1$};
    \draw[thick, dashed] (2,2.5) -- (2,-0.2);
    \node at (2, -0.5) {\small $t_2$};
    \draw[thick, dashed] (3,3) -- (3,-0.2);
    \node at (3, -0.5) {\small $t_3$};
    \draw[thick, dashed] (4,1.5) -- (4,-0.2);
    \node at (4, -0.5) {\small $t_4$};
    \draw[thick, dashed] (5,1.25) -- (5,-0.2);
    \node at (5, -0.5) {\small $t_5$};
    \draw[thick, dashed] (6,1) -- (6,-0.2);
    \node at (6, -0.5) {\small $t_6$};
    \draw[thick, dashed] (7.5,1.9) -- (7.5,-0.2);
    \node at (7.5, -0.5) {\small $t_{n\shortminus3}$};
    \draw[thick, dashed] (8.5,2.3) -- (8.5,-0.2);
    \node at (8.5, -0.5) {\small $t_{n\shortminus2}$};
    \draw[thick, dashed] (9.5,3.1) -- (9.5,-0.2);
    \node at (9.5, -0.5) {\small $t_{n\shortminus1}$};
    \draw[thick, dashed] (10.5,2.9) -- (10.5,-0.2);
    \node at (10.5, -0.5) {\small $t_n=T$};
\end{tikzpicture}
\end{center}
\caption[Discrete evaluation of the function $f(t,\mrm{X}(t))$]{\textbf{Discrete evaluation of the function $f(t,\mrm{X}(t))$.} Visual representation of the partitioning of the time interval $[0,T\,]$ to evaluate the function $f(t,\mrm{X}(t))$ and define the It\^{o} integral of $f(t,\mrm{X}(t))$ with respect to the noise process $\W(t)$.}
\label{fig:discretization_ito_int}
\end{figure}

Additionally, $f\left(t,\mrm{X}(t)\right)$ cannot grow too quickly since the average of its squared value must be finite. Then,

\begin{definition}[The It\^{o} integral]
    The It\^{o} integral of $f\left(t,\mrm{X}(t)\right)$ with respect to the Wiener process $\W(t)$ over an interval $[0,T\,]$ is defined as
    \begin{equation} \label{eq:ito_int_def}
        S = \int_0^T f\left(t,\mrm{X}(t)\right) \dW \coloneqq \underset{n\to\infty}{\mslim} \; S_n
    \end{equation}
    where $\mslim$ denotes the mean-squared limit of the approximating sum $S_n$:
    \begin{equation} \label{eq:Riemannlike_S_n}
        S_n = \sum_{i = 1}^n f\left(t_{i\shortminus1},\mrm{X}(t_{i\shortminus1})\right) \Delta\! \W(\Dt_i),
    \end{equation}
    with the increment $\Delta\! \W(\Dt_i)$ given as $\Delta \!\W(\Dt_i) \coloneqq \W(t_i) - \W(t_{i\shortminus1})$ and $\{t_i\}_{i=0}^n$ being a partition of the interval $[0,T\,]$ with $0 = t_0 \leq t_1 \leq \dots \leq t_{n\shortminus1} \leq t_n = T$.    
\end{definition}
Note that the limit in \eqnref{eq:ito_int_def} is taken in the mean square sense. Therefore, let us define what we mean by that:
\begin{definition}[Mean-Squared Limit]
Let $\{S_n\}_{n=1}^\infty$ be a sequence of random variables and let $S$ be another random variable. We say that $S_n$ converges to $S$ in the mean-squared sense, and write
\begin{equation}
    \underset{n\to\infty}{\mslim} \; S_n = S,
\end{equation}
if
\begin{equation} \label{eq:ms_limit}
    \lim_{n\to\infty} \EE{(S_n - S)^2} = 0,
\end{equation}
where $\EE{(S_n - S)^2}$ is what we call in statistics the mean squared error (MSE), which will be defined in more detail in the next chapter. 
\end{definition}
This type of convergence means that the approximating sum $S_n$ will, on average, stay close to the It\^{o} integral value $S$ as we refine the partition, without requiring every individual path of the process $S_n$ to converge exactly (pathwise) to the integral $S$. This might remind the reader of how, in statistics, we seek an estimator for a true quantity that minimizes the MSE. In that case, we easily recognize that even a good estimator may produce a poor estimate in a particular realization. What matters is that such deviations become increasingly rare on average as the approximation improves.

Besides the It\^{o} integral, there are other ways to define a stochastic integral w.r.t. the Wiener process. Another approach, known as the \emph{Stratonovich} integral, is defined as
\begin{equation} \label{eq:Strat_int}
    \Sint^T_{\!\!\!0} f\left(t,\mrm{X}(t)\right) \dW \coloneqq \underset{n\to\infty}{\mslim} \sum_{i=1}^n \, f\left(\frac{t_{i\shortminus1}+t_{i}}{2},\frac{\mrm{X}(t_i)+\mrm{X}(t_{i\shortminus1})}{2}\right) \Delta\! \W(\Dt_i)
\end{equation}
where we superimpose $\mathcal{S}$ over the integral symbol to highlight that this integral is different from the It\^{o} integral.

\begin{myexample} \textbf{(It\^{o} integral of a Wiener process):} 
    Let us compute 
    \begin{equation}
        \int_0^T \W(t)\, \dW,
    \end{equation}
    where, for this stochastic integral, $f\left(t,\mrm{X}(t)\right) = f\left(t,\W(t)\right) = \W(t)$. Recall the definition of an It\^{o} integral as given in \eqnref{eq:ito_int_def}. In our case, $S_n$ is of the form:
    \begin{equation}
        S_n = \sum_{i=1}^n \W(t_{i\shortminus1}) \Delta\!\W(\Dt_i)
    \end{equation}
    where $\Delta\!\W(\Dt_i) \coloneqq \W(t_i) - \W(t_{i\shortminus1})$. To simplify the sum, we can expand the summand as:
    \begin{align}
        \W(t_{i\shortminus1}) \Delta\!\W(\Dt_i) = \frac{1}{2} \left(\W(t_i)^2 - \W(t_{i\shortminus1})^2 - \Delta\!\W(\Dt_i)^2\right),
    \end{align}
    which makes evaluating the sum easier. In particular, substituting this identity into the sum $S_n$ yields:
    \begin{align}
        S_n = \frac{1}{2} \sum_{i=1}^n \left( \W(t_i)^2 - \W(t_{i\shortminus1})^2 \right) - \frac{1}{2} \sum_{i=1}^n \left( \Delta\!\W(\Dt_i)^2 \right).
    \end{align}
    Next, note that the first part of the sum $S_n$ forms a telescopic sum. This means that all intermediate terms cancel out, leaving only the boundary terms:
    \begin{align}
        \sum_{i=1}^n \left( \W(t_i)^2 - \W(t_{i\shortminus1})^2 \right) = \W(T\,)^2 - \W(0)^2 = \W(T\,)^2,
    \end{align}
    where in the last step we have used the initial condition of a Wiener process, $\W(0) = 0$ (see \secref{sec:Wiener_process}). Therefore, we can now rewrite $S_n$ as:
    \begin{align}
        S_n = \frac{1}{2} \left(\W(T\,)^2 - \sum_{i=1}^n  \Delta\!\W(\Dt_i)^2 \right),
    \end{align}
    and hence computing the mean-squared limit of $S_n$
    \begin{equation}
        S = \underset{n\to\infty}{\mslim} \; S_n = \frac{1}{2} \left(\W(T\,)^2 - \underset{n\to\infty}{\mslim} \sum_{i=1}^n  \Delta\!\W(\Dt_i)^2 \right) 
    \end{equation}
    reduces to computing the mean-squared limit of the sum of $\Delta\!\W(\Dt_i)^2$:
    \begin{equation}
        I = \W(T\,)^2 - 2S =  \underset{n\to\infty}{\mslim} \, I_n = \underset{n\to\infty}{\mslim} \sum_{i=1}^n  \Delta\!\W(\Dt_i)^2.
    \end{equation}
    In other words, we have now to find the It\^{o} integral $I$ s.t. 
    \begin{equation} \label{eq:def_mslimit_for_I}
        \lim_{n\to\infty} \EE{(I_n - I\,)^2} = 0
    \end{equation}
    is fulfilled. Our approach to tackle this problem is to give an ansatz to the solution of the It\^{o} integral $I$ and then check that it satisfies the mean-squared limit definition in \eqnref{eq:def_mslimit_for_I}. 
    In particular, we choose the following ansatz:
    \begin{equation}
        I = T,
    \end{equation}
    which is simply the mean of $I_n$. Namely,
    \begin{align}
        \EE{I_n} = \EE{\sum_{i=1}^n  \Delta\!\W(\Dt_i)^2} = \sum_{i=1}^n  \EE{\Delta\!\W(\Dt_i)^2} \underset{\eqref{eq:EdW^2=t-s}}{=} \sum_{i=1}^n (t_i - t_{i\shortminus1}) = T,
    \end{align}
    where in the last step we have again a telescopic sum. Now that we know the reason behind our ``inspired'' ansatz, we just need to check whether $I_n$ converges to $T$ in the mean-squared sense:
    \begin{equation}
        \lim_{n\to\infty} \EE{(I_n - I\,)^2} = \lim_{n\to\infty} \EE{\left(\sum_{i=1}^n \Delta\!\W(\Dt_i)^2 - T\right)^{\,\!\!\!\!2}\,}.
    \end{equation}
    The first step is to expand the squared sum in the limit above:
    \begin{align}
        &\left(\sum_{i=1}^n \Delta\!\W(\Dt_i)^2 - T\right)^{\,\!\!\!\!2} = \left(\sum_{i=1}^n \Delta\!\W(\Dt_i)^2\right)^{\,\!\!\!\!2} \! - 2T\sum_{i=1}^n \Delta\!\W(\Dt_i)^2 + T^{\,2} \nonumber \\
        &\quad= \sum_{i=1}^n \Delta\!\W(\Dt_i)^4 + 2 \sum_{i=1}^n \sum_{j=1}^{i\shortminus 1} \Delta\!\W(\Dt_i)^2 \, \Delta\!\W(\Dt_j)^2- 2T\sum_{i=1}^n \Delta\!\W(\Dt_i)^2 + T^{\,2}, 
    \end{align}
    by employing the following identity:
    \begin{align} \label{eq:identity_sumx2}
        \left( \!\sum_{i=1}^n x_i \!\right)^{\,\!\!\!\!2} = \sum_{i=1}^n x_i^2 + 2 \sum_{i=1}^n \sum_{j=1}^{i-1} x_i \, x_j.
    \end{align}
    Then, we can take its expected value:
    \begin{align}
        \EE{\left(\sum_{i=1}^n \Delta\!\W(\Dt_i)^2 - T\right)^{\,\!\!\!\!2}\,} &= \sum_{i=1}^n \EE{\Delta\!\W(\Dt_i)^4} + 2 \sum_{i=1}^n \sum_{j=1}^{i\shortminus 1} \EE{\Delta\!\W(\Dt_i)^2\, \Delta\!\W(\Dt_j)^2} \nonumber \\
        &\quad - 2T\sum_{i=1}^n \EE{\Delta\!\W(\Dt_i)^2} + T^{\,2},
    \end{align}
    and use the properties of the Wiener process (see \secref{sec:Wiener_process}) to calculate the higher moments of $\Delta\!\W(\Dt_i)$ appearing above:
    \begin{align}
        &\EE{\Delta\!\W(\Dt_i)^4} = 3 \Dt_i^2 = 3(t_i-t_{i\shortminus1})^2, \\
        &\EE{\Delta\!\W(\Dt_i)^2 \Delta\!\W(\Dt_j)^2} = \EE{\Delta\!\W(\Dt_i)^2}\mrm{E}\big[\Delta\!\W(\Dt_j)^2\big] = \Dt_i \, \Dt_j,
    \end{align}
    where the last expression uses the property of independent increments, since $i>j \quad \forall \; i,j$. 
    
    Therefore, 
    \begin{align}
        \EE{\left(\sum_{i=1}^n \Delta\!\W(\Dt_i)^2 - T\right)^{\,\!\!\!\!2}\,} &= 3 \sum_{i=1}^n (t_i-t_{i\shortminus1})^2 + 2 \sum_{i=1}^n \sum_{j=1}^{i\shortminus 1} (t_i-t_{i\shortminus1}) (t_j-t_{j\shortminus1}) \nonumber \\
        &\quad - 2T\sum_{i=1}^n (t_i-t_{i\shortminus1}) + T^{\,2}. \label{eq:middle_WdW_E_term}
    \end{align}
    If we now simplify the telescopic sums
    \begin{equation}
        \sum_{i=1}^n (t_i-t_{i\shortminus1}) = T,
    \end{equation}
    and employ the identity specified in \eqnref{eq:identity_sumx2} to roll back the term
    \begin{align}
        2 \sum_{i=1}^n \sum_{j=1}^{i\shortminus 1} (t_i-t_{i\shortminus1}) (t_j-t_{j\shortminus1}) &= \left(\sum_{i=1}^n (t_i-t_{i\shortminus1}) \right)^{\,\!\!\!2} - \sum_{i=1}^n (t_i-t_{i\shortminus1})^2 \nonumber \\
        &= T^{\,2} - \sum_{i=1}^n (t_i-t_{i\shortminus1})^2, 
    \end{align}
    we can finally substitute all that into \eqnref{eq:middle_WdW_E_term} and take its limit to verify that $I = T$. Namely,
    \begin{align}
        \lim_{n\to\infty} \EE{\left(\sum_{i=1}^n \Delta\!\W(\Dt_i)^2 - T\right)^{\,\!\!\!\!2}\,} = 2 \lim_{n\to\infty} \sum_{i=1}^n (t_i-t_{i\shortminus1})^2 = 0.
    \end{align}
    Hence, it follows that
    \begin{align}
        I = \underset{n\to\infty}{\mslim} \; I_n = \underset{n\to\infty}{\mslim} \sum_{i=1}^n \Delta\!\W(\Dt_i)^2 = T,
    \end{align}
    and thus,
    \begin{align}
        S &= \int_0^T \W(t) \dW = \underset{n\to\infty}{\mslim} \sum_{i=1}^n \W(t_{i\shortminus1}) \Delta\!\W(\Dt_i) = \frac{1}{2} \left(\W(T\,)^2 - \underset{n\to\infty}{\mslim} \;I_n \right) \nonumber \\
        &= \frac{1}{2} \left(\W(T\,)^2 - T \, \right). 
    \end{align}
\end{myexample}

\subsection{$\dW^2 = \dt$ and $\dW^n = 0$ for $n > 2$}

The differential of a Wiener process, denoted as $\dW$, behaves differently from ordinary differentials. In particular, a key result in stochastic calculus is that the square of $\dW$ is equal to the differential of time, $\dt$, while higher powers of $\dW$ vanish. More precisely, we want to prove that 
\begin{equation}
    (\dW)^{2+N} = \begin{cases}
        \dt \quad &\text{for} \quad N = 0, \\
        0 \quad &\text{for} \quad N > 0.
    \end{cases}
\end{equation}
To do so, we have to use the It\^{o} integral and show
\begin{align}
    \int_0^T f\left(t,\mrm{X}(t)\right) (\dW)^{2+N} &\coloneqq \underset{n\to\infty}{\mslim} \sum_{i=1}^n f\left(t_{i\shortminus1},\mrm{X}(t_{i\shortminus1})\right) \Delta \!\W(\Dt_i)^{2+N} \nonumber \\
    &= 
    \begin{cases}
        \int_0^T f\left(t,\mrm{X}(t)\right) \dt \quad &\text{for} \quad N = 0, \\
        0 \quad &\text{for} \quad N > 0
    \end{cases}.
\end{align}

To tackle this proof, we need to divide it into two: first show that $\dW^2 = \dt$ and then, demonstrate that $\dW^{2+N} = 0$ for $N > 2$. 

\begin{myproof}
    To prove that $\dW^2 = \dt$ we have to employ the definition of the It\^{o} integral and show:
    \begin{equation} \label{eq:ms_lim_for_dW2}
        \underset{n\to\infty}{\mslim} \; \sum_{i=1}^n \, f\left(t_{i\shortminus1},\mrm{X}(t_{i\shortminus1})\right) \Delta \!\W(\Dt_i)^{2} = \int_0^T f\left(t,\mrm{X}(t)\right) \dt.
    \end{equation}
    However, to understand this limit, we must recall the definition of convergence in the mean-square sense, as introduced in \eqnref{eq:ms_limit}. This means that, for the sequence of approximating sums $S_n$ and the limiting value $S$, we require
    \begin{equation}
        \lim_{n\to\infty} \EE{(S_n - S)^2} = 0,
    \end{equation}
    where we define
    \begin{equation} \label{eq:def_Sn_for_dW2}
        S_n = \sum_{i=1}^n \, f\left(t_{i\shortminus1},\mrm{X}(t_{i\shortminus1})\right) \Delta \!\W(\Dt_i)^{2},
    \end{equation}
    and
    \begin{equation} \label{eq:def_S_for_dW2}
        S = \int_0^T f\left(t,\mrm{X}(t)\right) \dt = \lim_{n\to\infty} \sum_{i=1}^n \,f\left(t_{i\shortminus1},\mrm{X}(t_{i\shortminus1})\right) \Dt_i.
    \end{equation} 
    Therefore, if we substitute the expressions for $S_n$ and $S$, we get:
    \begin{align}
        &\lim_{n\to\infty} \EE{\left(S_n - S\right)^2} = \lim_{n\to\infty} \EE{\left( \sum_{i=1}^n  \, f\left(t_{i\shortminus1},\mrm{X}(t_{i\shortminus1})\right)  (\Delta \!\W(\Dt_i)^2 - \Dt_i) \right)^{\!\!\!2}\,}. 
    \end{align}
    Note that the squared sum in the limit above can be expanded according to the identity in \eqnref{eq:identity_sumx2}, i.e.: 
     \begin{align}
        &\EE{\left( \sum_{i=1}^n  \, f\left(t_{i\shortminus1},\mrm{X}(t_{i\shortminus1})\right)  (\Delta \!\W(\Dt_i)^2 - \Dt_i) \right)^{\!\!\!2}\,} =  \sum_{i=1}^n \EE{\,f_{i\shortminus1}^{\;2}} \EE{\left(\Delta \!\W(\Dt_i)^2 - \Dt_i\right)^2} \nonumber \\
        &+ 2 \sum_{i=1}^n \sum_{j=1}^{i-1} \EE{\,f_{i\shortminus1} \; f_{j\shortminus1}} \left(\EE{\Delta\!\W(\Dt_i)^2} - \Dt_i\right)\big(\mrm{E}[\Delta \!\W(\Dt_j)^2] - \Dt_j\big), \label{eq:intermediate_dW2_proof}
    \end{align}
    where we simplify the notation by writing $f_{i\shortminus1}$ instead of $f\left(t_{i\shortminus1},\mrm{X}(t_{i\shortminus1})\right)$. Above we also use that $f_{i\shortminus1} = f\left(t_{i\shortminus1},\mrm{X}(t_{i\shortminus1})\right)$ is independent of $\Delta\!\W(\Dt_i) = \W(t_i) - \W(t_{i\shortminus1})$. This follows from:
    \begin{enumerate}
        \item The function $f\,$ being a non-anticipating function, i.e. it does not depend on future values (recall \defref{def:non-anticipating_f}).
        \item The property of stationary increments, fundamental to the definition of L\'{e}vy processes, and in particular, Wiener processes (if needed, refer back to \secref{sec:Wiener_process}).
    \end{enumerate}
    Even though $\Delta\!\W(\Dt_i)$ is defined in terms of $\W(t_{i\shortminus1})$, its distribution is independent of $\W(t_{i\shortminus1})$ because it is determined solely by the interval $\Dt_i$. Namely, recall from \secref{sec:Wiener_process} that $\Delta\!\W(\Dt_i) = \W(t_i) - \W(t_{i\shortminus1}) = \W(\Dt_i) \sim \Gauss(0,\Dt_i)$. This lack of dependency on the actual path of $\W(t)$ up to $t_{i\shortminus1}$ makes $\Delta\!\W(\Dt_i)$ statistically independent of $\W(t_{i\shortminus1})$ despite being defined as $\Delta\!\W(\Dt_i) \coloneqq \W(t_i) - \W(t_{i\shortminus1})$. Therefore, even if $f$ was a function of the process $\W(t)$, i.e. $f\left(t,\W(t)\right)$, or of a process $\mrm{X}(t)$ somehow correlated with $\W(t)$, $f_{i\shortminus1}$ is still independent of $\Delta\!\W(\Dt_i)$. Hence,
    \begin{align}
        &\EE{\;f_{i\shortminus1}^{\;2} \left(\Delta \!\W(\Dt_i)^2 - \Dt_i\right)^2} = \EE{\,f_{i\shortminus1}^{\;2}} \EE{\left(\Delta \!\W(\Dt_i)^2 - \Dt_i\right)^2}, \\
        &\EE{\,f_{i\shortminus1} \; f_{j\shortminus1} \left(\Delta\!\W(\Dt_i)^2 - \Dt_i\right)\big(\Delta \!\W(\Dt_j)^2 - \Dt_j\big) } =  \\
        &\quad =\EE{\,f_{i\shortminus1} \; f_{j\shortminus1}} \left(\EE{\Delta\!\W(\Dt_i)^2} - \Dt_i\right)\big(\mrm{E}[\Delta \!\W(\Dt_j)^2] - \Dt_j\big), \;\;\; \text{since} \;\;\; i>j \;\;\; \forall \, i,j. \nonumber
    \end{align}
    Note that in the last expression, we have used that 
    \begin{equation}
        \EE{\Delta\!\W(\Dt_i)\Delta\!\W(\Dt_j)} = \EE{\Delta\!\W(\Dt_i)}\EE{\Delta\!\W(\Dt_j)} \quad \text{since} \quad i>j,
    \end{equation}
    which holds because the Wiener increments of different time steps are independent (again, see \secref{sec:Wiener_process}). Furthermore, given that each increment $\Delta\!\W(\Dt_i)$ is normally distributed, then
    \begin{align}
        &\EE{\Delta\!\W(\Dt_i)^2} = \Dt_i, \\
        &\EE{\left(\Delta\!\W(\Dt_i)^2 - \Dt_i\right)^2} = \EE{\Delta\!\W(\Dt_i)^4 - 2\Delta\!\W(\Dt_i)^2\Dt_i + \Dt_i^2} \nonumber \\
        &\quad\quad\quad\quad\quad\quad\quad\quad\;\;= 3 \Dt_i^2 - 2\Dt_i^2 + \Dt_i^2 = 2\Dt_i^2.
    \end{align}
    When we apply these results to \eqnref{eq:intermediate_dW2_proof} and take the limit of $n \to \infty$, we obtain
    \begin{align}
        \lim_{n\to\infty} \EE{\left(S_n - S\right)^2} = \lim_{n\to\infty} 2 \sum_{i=1}^n \EE{f_{i\shortminus1}^{\;2}} \Dt_i^2 = 0,
    \end{align}
    since $\Dt_i$ is squared. In other words, the average mean squared error of the difference between the ``Riemann-sum'' approximation $S_n$ and the limiting stochastic integral $S$ converges to zero as the partition is refined. Thus, we have shown that
    \begin{align}
        \int_0^T \!\! f\left(t,\mrm{X}(t)\right) \!(\dW)^{2} \!\coloneqq \underset{n\to\infty}{\mslim} \! \sum_{i=1}^n \, f\left(t_{i\shortminus1},\mrm{X}(t_{i\shortminus1})\right) \Delta \!\W(\Dt_i)^{2} = \!\int_0^T \!\!f\left(t,\mrm{X}(t)\right) \dt,
    \end{align}
    and hence
    \begin{align}
        \dW^2 = \dt.
    \end{align}
\end{myproof}
The proofs of $(\dW)^{2+N} = 0$ when $N>2$ and $\dW \dt = 0$ follow very similar steps to the previous one, and are provided in \appref{ap:dW^2+N=0_proof} and  \appref{ap:dWdt=0_proof}. All these results are only valid for the It\^{o} integral, since we have used that $\Delta \! \W_i$ is independent of the non-anticipating function $f_{i\shortminus1}$. However, the integrand in the Stratonovich integral is evaluated at the midpoint of the interval, i.e. $f_{i\shortminus1} = f\left(\frac{1}{2} (t_{i\shortminus1}+t_{i}),\frac{1}{2}(\mrm{X}(t_i)+\mrm{X}(t_{i\shortminus1}))\right)$, while the increment remains defined as $\Delta\!\W(\Dt_i) = \W(t_i) - \W(t_{i\shortminus1})$. Hence, $\Delta\!\W(\Dt_i)$ and $f_{i\shortminus1}$ are not necessarily independent even though $f$ is non-anticipating. Thus, $\dW^2 = dt$ does not hold in Stratonovich calculus~\cite{Gardiner1985_book_stochastic}. 

\subsection{It\^{o}'s lemma} \label{sec:itos_lemma}

Let us now introduce It\^{o}'s lemma, the stochastic analog of the chain rule. It tells us how to differentiate functions of stochastic processes and is essential for formulating and solving stochastic differential equations (SDEs).

\begin{lem}[It\^{o}'s lemma]\label{lem:itos_lemma}
    Let $\W(t)$ be a Wiener process and consider a function $f\left(t,\W(t)\right)$, twice differentiable w.r.t. $\W(t)$ and once differentiable with respect to $t$. Then, It\^{o}'s lemma states:
    \begin{align}
        \dd f\left(t,\W(t)\right) = \left(\frac{\partial f}{\partial t} + \frac{1}{2} \frac{\partial^2 f}{\partial \W^2} \right) \! \dt + \frac{\partial f}{\partial \W} \dW,
    \end{align}
    where $\partial f/\partial \W$ and $\partial^2 f/\partial \W^2$ are the first and second partial derivatives of $f\left(t,\W(t)\right)$ w.r.t. $\W(t)$, evaluated at $(t,\W(t))$.
\end{lem}
In this form, It\^{o}'s lemma shows that the differential of $f\left(t,\W(t)\right)$ has an additional term $\sfrac{1}{2} \, \partial^2 f/\partial \W^2$ that accounts for the quadratic variation of the Wiener process $\W(t)$.
\begin{myproof}
    Consider the function $f(t+\Dt,\W(t+\Dt))$, where $\Dt$ is a finite time-step s.t. $\Dt > 0$. Additionally, let us define the Wiener increment over this small interval $\Dt$ as $\Delta\!\W(\Dt) \coloneqq \W(t+\Dt)-\W(t)$. Then, if we expand $f(t+\Dt,\W(t+\Dt))$ using the Taylor series to second order in $W(t)$ and first in $t$, we get:
    \begin{align}
        \!\!f(t\!+\!\Dt,\!\W(t\!+\!\Dt)) \approx f\left(t,\!\W(t)\right) + \frac{\partial f}{\partial t} \Dt + \frac{\partial f}{\partial \W} \Delta\!\W(\Dt) + \frac{1}{2}\frac{\partial^2 f}{\partial \W^2} (\Delta\!\W(\Dt))^2,
    \end{align}
    where higher-order terms in $\Dt$ and $\Delta\!\W(\Dt)$ are ignored since they will vanish as $\Dt\to 0$. 
    
    The key difference from ordinary calculus comes from the term including $(\Delta\!\W(\Dt))^2$, since as shown in the previous section, $\dW^2 = \dt$ in the limit of $\Dt \to0$. 
    
    Therefore, by now rearranging the terms above and taking the limit of $\Dt \to 0$, we get
    \begin{align}
        \dd f\left(t,\W(t)\right) &= 
        \left(\frac{\partial f}{\partial t} + \frac{1}{2} \frac{\partial^2 f}{\partial \W^2} \right) \!
        \dt + \frac{\partial f}{\partial \W} \dW. \label{eq:Ito_lemma}
    \end{align}
\end{myproof}

Additionally, there is also a more general form for It\^{o}'s lemma, providing the differential of a function $f\left(t,\mrm{X}(t)\right)$, where $\mrm{X}(t)$ is a general stochastic process. Namely,

\begin{lem}[General It\^{o}'s lemma] Let $\mrm{X}(t)$ be an It\^{o} process and let $f\left(t,\mrm{X}(t)\right)$ be a twice-differentiable function of time and the process $\mrm{X}(t)$. Then, It\^{o}'s lemma states:
\begin{align} \label{eq:general_Ito_lemma}
    \dd  f\left(t,\mrm{X}(t)\right) =
        \frac{\partial f}{\partial t} \dt + \frac{\partial f}{\partial \mrm{X}} \dd \mrm{X} + \frac{1}{2} \frac{\partial^2 f}{\partial \mrm{X}^2} \dd \mrm{X} \dd \mrm{X},
\end{align}
where $\dd \mrm{X} \dd \mrm{X}$ represents the quadratic variation of $\mrm{X}(t)$. For $\mrm{X}(t) = \W(t)$, the quadratic variation is $\dt$.
\end{lem}

\begin{myexample} \textbf{(How to apply the It\^{o} lemma (I)):} 
    Let us consider now $f\left(t,\W(t)\right) = \mu \, t + \sigma \W(t)$, where $\mu, \; \sigma$ are constant parameters. If now we evaluate the differential $\dd f\left(t,\W(t)\right)$ using the It\^{o} form, we get:
    \begin{align} \label{eq:differential_f_1}
        \dd f\left(t,\W(t)\right) = \mu \, \dt + \sigma \, \dW.
    \end{align}
    Note that a function of a stochastic process is simply another stochastic process, and thus, we can rename the function of $\W(t)$ as $\mrm{X}(t) \coloneqq f\left(t,\W(t)\right)$. Then, we can view the It\^{o} differential in \eqnref{eq:differential_f_1} as a stochastic differential equation of the process $\mrm{X}(t)$. Namely,
    \begin{equation} \label{eq:example_SDE}
        \dd \mrm{X}(t) = \mu \, \dt + \sigma \, \dW.
    \end{equation}
\end{myexample}

\begin{myexample} \textbf{(How to apply the It\^{o} lemma (II)):} 
    In this second example, let us consider a different function: $f\left(t,\W(t)\right) = \W(t)^2$. If we apply It\^{o}'s lemma to $f\left(t,\W(t)\right)$, we obtain
    \begin{align}
        \dd f\left(t,\W(t)\right) = \dd \left(\W(t)^2 \right) = 0 \cdot \dt + 2\W(t) \dW + \frac{1}{2} 2 \dt = \dt + 2\W(t) \dW,
    \end{align}
    since $\partial f / \partial t = 0$ because $f$ does not depend on $t$, and 
    \begin{align}
        &\frac{\partial f}{\partial \W} = 2 \W(t), \\
        &\frac{\partial^2 f}{\partial \W^2} = 2.
    \end{align}
    This result shows how the second derivative term in It\^{o}'s lemma influences the differential of functions that are at least quadratic in $\W(t)$.
\end{myexample}

\subsection{It\^{o}-Leibniz product rule}

In ordinary calculus, the Leibniz product rule provides the derivative of a product of two or more functions. For stochastic processes, however, the corresponding rule for differentiating the product of two stochastic processes --- or two functions of stochastic processes, which are themselves stochastic processes too --- is known as the It\^{o}-Leibniz product rule.

\begin{rulesec} \textbf{(It\^{o}-Leibniz product rule):} Let $\mrm{X}(t)$ and $\mrm{Y}(t)$ be two It\^{o} processes. Then, the differential of the product $\mrm{Z}(t) = \mrm{X}(t) \mrm{Y}(t)$ is given by:
\begin{equation}
    \dd \mrm{Z}(t) = \dd\left(\mrm{X}(t) \mrm{Y}(t)\right) = \mrm{X}(t) \dd \mrm{Y}(t) + \mrm{Y}(t) \dd \mrm{X}(t) + \dd \mrm{X}(t) \dd \mrm{Y}(t).
\end{equation}
\end{rulesec}
\begin{myproof}
    To derive the It\^{o}-Leibniz product rule, we can start considering the following increment of $\mrm{Z}(t) = \mrm{X}(t) \mrm{Y}(t)$:
    \begin{align}
        \Delta \mrm{Z}(\Dt) = \mrm{Z}(t+\Dt) - \mrm{Z}(t) = \mrm{X}(t+\Dt)\mrm{Y}(t+\Dt) - \mrm{X}(t)\mrm{Y}(t), 
    \end{align}
    and then add and subtract the term $\mrm{X}(t)\mrm{Y}(t+\Dt)$:
    \begin{align}
        \Delta \mrm{Z}(\Dt) &=  \mrm{X}(t+\Dt)\mrm{Y}(t+\Dt) + \mrm{X}(t)\mrm{Y}(t+\Dt) - \mrm{X}(t)\mrm{Y}(t+\Dt) - \mrm{X}(t)\mrm{Y}(t) \nonumber \\
        &= \mrm{Y}(t+\Dt) \left(\mrm{X}(t+\Dt)-\mrm{X}(t)\right) + \mrm{X}(t) \left(\mrm{Y}(t+\Dt)-\mrm{Y}(t)\right).
    \end{align}
    Next, we again add and subtract another term, $\mrm{Y}(t) \left(\mrm{X}(t+\Dt)-\mrm{X}(t)\right)$:
    \begin{align}
        \Delta \mrm{Z}(\Dt) &= \mrm{Y}(t+\Dt) \left(\mrm{X}(t+\Dt)-\mrm{X}(t)\right) - \mrm{Y}(t) \left(\mrm{X}(t+\Dt)-\mrm{X}(t)\right) \nonumber \\
        &+ \mrm{Y}(t) \left(\mrm{X}(t+\Dt)-\mrm{X}(t)\right) + \mrm{X}(t) \left(\mrm{Y}(t+\Dt)-\mrm{Y}(t)\right) \nonumber \\
        &= \left(\mrm{Y}(t+\Dt) - \mrm{Y}(t)\right) \left(\mrm{X}(t+\Dt)-\mrm{X}(t)\right) \nonumber \\
        &+ \mrm{Y}(t) \left(\mrm{X}(t+\Dt)-\mrm{X}(t)\right) + \mrm{X}(t) \left(\mrm{Y}(t+\Dt)-\mrm{Y}(t)\right)  \nonumber \\
        &= \Delta \mrm{Y}(\Dt) \Delta \mrm{X}(\Dt) + \mrm{Y}(t) \Delta \mrm{X}(\Dt) + \mrm{X}(t) \Delta\mrm{Y}(\Dt). 
    \end{align}
    If we now take the limit of $\Dt\to 0$, we derive the It\^{o}-Leibniz product rule:
    \begin{align}
        \dd \mrm{Z}(t) = \mrm{X}(t) \dd \mrm{Y}(t) + \mrm{Y}(t) \dd \mrm{X} (t)+ \dd \mrm{Y}(t) \dd \mrm{X}(t).
    \end{align}
\end{myproof}

\subsection{Stochastic differential equations} \label{sec:SDEs}

Stochastic differential equations are ordinary differential equations with the right hand side perturbed with a random term. To mathematically define SDEs, we use It\^{o} notation. An example of a simple SDE is given in \eqnref{eq:example_SDE}, which its form in It\^{o} notation is derived using It\^{o}'s lemma. More generally though, a differential equation perturbed by white Gaussian noise\footnote{White Gaussian noise is a random signal with a flat power spectral density and Gaussian amplitude distribution; it can be informally thought of as the derivative of the Wiener process. For a more in-depth explanation, see \secref{sec:white_noise}.} can be written in It\^{o} form as:
\begin{equation} \label{eq:general_form_SDE}
    \dd \mrm{X}(t) = \mu(\mrm{X}(t),t) \dt + \sigma(\mrm{X}(t),t) \dW,
\end{equation}
where $\mrm{X}(t)$ is the stochastic process we aim to solve for, i.e. the state of the system at time $t$, and $\dW$ is the Wiener differential. The terms $\mu(\mrm{X}(t),t)$ and $\sigma(\mrm{X}(t),t)$ are the drift and diffusion term, representing the deterministic and random parts of the evolution, respectively. 

Solutions to SDEs can be generally classified into two categories: analytical, with closed-form solutions, and numerical, which approximate the evolution of $\mrm{X}(t)$ through discretized steps. A simple example of an analytical closed-form solution is given by the SDE in \eqnref{eq:example_SDE}, which has the solution $\mrm{X}(t) = \mu \, t + \sigma \W(t)$.

\begin{myexample} \textbf{(Geometric Brownian motion solution):}
    Another well-known example is the geometric Brownian motion satisfying the following SDE:
    \begin{equation} \label{eq:gBm}
        \dd \mrm{X}(t) = \mu \, \mrm{X}(t) \dt + \sigma \, \mrm{X}(t)  \dW,
    \end{equation}
    which has the solution
    \begin{equation}
        \mrm{X}(t) = \mrm{X}(0) \exp{\left\{\!\left(\!\mu - \frac{1}{2}\sigma^2 \!\right) \! t + \sigma \W(t)\!\right\}},
    \end{equation}
    where $\mrm{X}(0)$ is the initial value of the process $\mrm{X}(t)$.    
\end{myexample}
\begin{myproof}
    To find the solution of the geometric Brownian motion defined by \eqnref{eq:gBm}, let us compute the differential of the natural logarithm of $\mrm{X}(t)$. In particular, since $\ln \!\left( \mrm{X}(t) \right)$ is a function of the process $\mrm{X}(t)$, i.e. $f\left(t,\mrm{X}(t)\right)$, we have to use It\^{o}'s lemma \eqref{eq:general_Ito_lemma} to compute the differential of $f\left(t,\mrm{X}(t)\right) = \ln \!\left( \mrm{X}(t) \right)$. Namely, 
    \begin{align}
        \dd \left(\ln \!\left( \mrm{X}(t) \right)\right) = \frac{1}{2}  \left( \!\frac{\partial^2}{\partial \mrm{X}^2} \ln \!\left( \mrm{X}(t) \right)\!\right)\!\dd \mrm{X} \, \dd \mrm{X} +  \left( \!\frac{\partial}{\partial \mrm{X}} \ln \!\left( \mrm{X}(t) \right)\!\right)\!\dd \mrm{X},
    \end{align}
    where $\dd \mrm{X} \, \dd \mrm{X}$ is the quadratic variation of $\mrm{X}(t)$:
    \begin{equation}
        \dd \mrm{X} \, \dd \mrm{X} = \mu^2 \mrm{X}^2(t) \dt^2 + 2\sigma \mu  \mrm{X}^2(t) \, \dW \dt + \sigma^2 \mrm{X}^2(t) \, \dW^2 = \sigma^2\mrm{X}^2(t) \, \dt + \bigO{(\dt^{\sfrac{3}{2}})}.
    \end{equation}
    Therefore, we can write the differential of $\ln \!\left( \mrm{X}(t) \right)$ as
    \begin{align}
        \dd \left(\ln \!\left( \mrm{X}(t) \right)\right) &= -\frac{1}{2} \frac{1}{\mrm{X}^2(t)} \sigma^2 \mrm{X}^2 (t) \dt + \frac{\dd \mrm{X}}{\mrm{X}(t)} = -\frac{1}{2} \sigma^2 \dt + \frac{\dd \mrm{X}}{\mrm{X}(t)} = \nonumber \\
        &= -\frac{1}{2} \sigma^2 \dt + \mu \, \dt + \sigma \dW,
    \end{align}
    where in the last step we used \eqnref{eq:gBm}. Hence, by now integrating from $0$ to $t$, we get
    \begin{equation}
        \ln \!\left( \mrm{X}(t) \right) - \ln \!\left( \mrm{X}(0) \right) = \ln \!\left( \!\frac{\mrm{X}(t)}{\mrm{X}(0)} \!\right)= \left(\!\mu - \frac{1}{2} \sigma^2\!\right) \! t + \sigma \W(t), 
    \end{equation}
    which, by exponentiating its both sides, yields an analytical form for $\mrm{X}(t)$:
    \begin{equation}
        \mrm{X}(t) = \mrm{X}(0) \exp{\left\{\left(\!\mu - \frac{1}{2} \sigma^2\!\right) \! t + \sigma \W(t) \right\}}.
    \end{equation}
\end{myproof}

\subsection{Ornstein-Uhlenbeck process} \label{sec:OUP_intro}

The Ornstein-Uhlenbeck (OU) process is another process with Gaussian noise that models mean-reverting behavior, i.e. the system has a tendency to return to an equilibrium state. The dynamics of the OU process $\mrm{X}(t)$ are governed by the following SDE:
\begin{equation} \label{eq:OUP_form}
    \dd \mrm{X}(t) = -\theta \, (\mrm{X}(t) - \mu) \dt  + \sigma \, \dW,
\end{equation}
where $\theta > 0$, $\mu$ and $\sigma$ are constants and $\dW$ is the Wiener differential. The deterministic term $-\theta \, \mrm{X}(t)$ pulls the process back towards a long-term mean, $\mu$, while the stochastic term $\sigma \dW$ introduces the random kicks. 

Just like in the example of the geometric Brownian motion, the OU process (here we assume $\mu = 0$) also has a formal solution:
\begin{align}
    \mrm{X}(t) = \mrm{X}(0) \ee^{-\theta t} + \sigma \int_0^t \ee^{-\theta(t-\tau)} \dW(\tau).
\end{align}
\begin{myproof}
    Define a function of the stochastic process $\mrm{X}(t)$ as $f\left(\mrm{X}(t),t\right) \coloneqq \mrm{X}(t) \ee^{\theta t}$, such that when computing its differential with It\^{o}'s lemma we get:
    \begin{equation}
        \dd \left(\mrm{X}(t) \ee^{\theta t} \right) = \theta \mrm{X}(t) \ee^{\theta t} \dt + \ee^{\theta t} \dd \mrm{X}(t) = \sigma \, \dW(t) \ee^{\theta t}.
    \end{equation}
    Hence, integrating both sides from $0$ to $t$ yields:
    \begin{equation}
        \mrm{X}(t) \ee^{\theta t} = \mrm{X}(0) + \sigma \int_0^t \ee^{\theta \tau} \dW(\tau),
    \end{equation}
    which can be simplified by taking the exponential term to the r.h.s. to reveal the formal solution of the process:
    \begin{equation}
        \mrm{X}(t) = \mrm{X}(0) \ee^{-\theta t} + \sigma \int_0^t \ee^{-\theta(t-\tau)} \, \dW(\tau).
    \end{equation}
\end{myproof}

From this expression, one can derive the mean and the covariance of the process:
\begin{align}
    \EE{\mrm{X}(t)} &= \mrm{X}(0) \ee^{-\theta t}, \label{eq:mean_OUP} \\
    \cov[\mrm{X}(t)\mrm{X}(s)] &= \EE{(\mrm{X}(t)-\EE{\mrm{X}(t)})(\mrm{X}(s)-\EE{\mrm{X}(s)})} \nonumber \\
    &= \sigma^2 \ee^{-\theta(s+t)}\EE{\int_0^t \ee^{\theta \tau} \, \dW(\tau) \int_0^s \ee^{\theta \upsilon} \, \dW(\upsilon)} \nonumber \\
    &= \sigma^2 \ee^{-\theta(s+t)} \int_0^t \int_0^s \ee^{\theta \tau} \ee^{\theta \upsilon} \, \EE{\dW(\tau) \, \dW(\upsilon)} \nonumber \\
    &= \sigma^2 \ee^{-\theta(s+t)} \int_0^t \ee^{\theta \tau} \left[ \int_0^s \ee^{\theta \upsilon} \, \delta(\tau-\upsilon) \dd \upsilon \right] \dd\tau \nonumber \\
    &=\sigma^2 \ee^{-\theta(s+t)} \int_0^{\trm{min}(t,\,s)} \ee^{2\theta \tau}  \dd\tau \nonumber \\
    &=\sigma^2 \ee^{-\theta(s+t)} \frac{1}{2\theta}\left( \ee^{2\theta \, \trm{min}(t,\,s)} - 1\right) = \frac{\sigma^2}{2\theta} \left(\ee^{-\theta|t-s|} - \ee^{-\theta(s+t)} \right) \label{eq:covariance_OUP}
\end{align}
where $\trm{min}(t,s)$ appears because $\ee^{\theta \tau}$ is non-zero only when $0 \leq \tau \leq s$. It follows that the variance is:
\begin{align} \label{eq:variance_OUP}
    \mrm{V}[\mrm{X}(t)] = \cov[\mrm{X}(t)\mrm{X}(t)] = \frac{\sigma^2}{2\theta} \left(1 - \ee^{-2\theta t}\right),
\end{align}
which for small $\theta$ can be approximated as
\begin{align} \label{eq:var_OUP_smallchi}
    \mrm{V}[\mrm{X}(t)] \approx \sigma^2 t.
\end{align}

The OU process can also be described in terms of a PDF $p(x,t)$, which describes the likelihood of the process $\mrm{X}(t)$ being in a state $x$ at a time $t$. This probability density evolves according to the Fokker-Planck equation, a partial differential equation that governs the time evolution of the probability distribution for a stochastic process. For the zero-mean ($\mu = 0$) OU process, the Fokker-Planck equation is given by:
\begin{equation}
    \frac{\partial p}{\partial t} = \theta \frac{\partial}{\partial  x} (x \, p) + \frac{\sigma^2}{2} \frac{\partial^2 p}{\partial x^2}.
\end{equation}

Solving the Fokker-Planck equation with the initial condition $p(x,t_0) = \delta(x-x_0)$ yields the following transition probability function:
\begin{equation} \label{eq:transition_probability}
    p(x,t|x_0, t_0) = \sqrt{\frac{\theta}{\pi \sigma^2 (1-\e^{-2\theta\,(t-t_0)})}} \exp{\left(-\frac{\theta}{\sigma^2} \frac{\left(x-x_0 \e^{-\theta \,(t-t_0)} \right)^2}{1-\e^{-2\theta\,(t-t_0)}}\right)},
\end{equation}
where $t>t_0$, $x = \mrm{X}(t)$ and $x_0 = \mrm{X}(t_0)$. Note that the mean and variance of the OU process $\mrm{X}(t)$, given in \eqnref{eq:mean_OUP} and \eqnref{eq:variance_OUP}, respectively, can be easily inferred from the Gaussian form of the transition probability by setting the starting time $t_0 = 0$ in \eqnref{eq:transition_probability}.

\subsection{Numerical methods for SDEs}

For cases where closed-form solutions of SDEs are not attainable, numerical methods provide a practical way to approximate the behavior of stochastic processes over time. Numerical solutions to SDEs rely on time discretization and iterative computation of approximate values of the stochastic process. One of the most widely used methods for numerically solving SDEs is the \emph{Euler-Maruyama} (EM) method, a straightforward extension of the classical Euler method for deterministic differential equations. 

\subsubsection{The Euler-Maruyama method}

Consider a general SDE of the form specified in \eqnref{eq:general_form_SDE}. Namely,
\begin{equation}
    \dd \mrm{X}(t) = \mu\left(\mrm{X}(t),t\right) \dt + \sigma\left(\mrm{X}(t),t\right) \dW,
\end{equation}
where $\mrm{X}(t)$ is the stochastic process for which we want to numerically solve the equation above over a time interval $[0,T\,]$. To implement the EM method, the time interval $[0,T\,]$ is divided into $n$ discrete time steps of equal size $\Dt = T/n$. Let $t_k = k\Dt$ denote the discrete time points, with $k = 0,1,2,\ldots,n$. Then, the approximate solution at each time step, $\mrm{X}[k] \coloneqq \mrm{X}(t_k)$, is computed iteratively using the update formula:
\begin{align}
    \mrm{X}[k+1] = \mrm{X}[k] + \mu\left(t_k,\mrm{X}[k]\right) \Dt + \sigma\left(t_k,\mrm{X}[k]\right) \Delta\!\W_k,
\end{align}
where $\Delta\!\W_k \coloneqq \W(t_{k+1}) - \W(t_k)$ is the increment of the Wiener process over the time step $\Dt$, which is simulated by drawing from a Gaussian distribution with mean zero and variance $\Dt$:
\begin{equation}
    \Delta\!\W_k \sim \Gauss(0,\Dt).
\end{equation}

The EM method is a first-order method in both the time-step $\Dt$ and the Wiener increment $\Delta\!\W_k$. Specifically, it converges strongly with order $1/2$, meaning that the expected error between the true solution of $\mrm{X}$ and the numerical solution scales as $\bigO{(\sqrt{\Dt})}$. For practical purposes, this implies that reducing size of the time step improves the accuracy of the method, but its precision is limited. Higher-order methods exist but they are often more complex and computationally expensive. In particular, when the drift and/or diffusion coefficients exhibit high nonlinearity or stiffness, then more sophisticated methods, such as the Milstein method or higher-order Runge-Kutta methods for SDEs may be necessary to achieve better accuracy. 

\subsection{White noise}\label{sec:white_noise}

Ordinary differential equations can be extended to describe the dynamics of stochastic processes by adding a white noise term, such as:
\begin{equation}
    \frac{\dd x}{\dt} = \mu \, x + w(t),
\end{equation}
where $w(t)$\footnote{In general, this noise in the   Langevin equation does not need to be white \cite{Gardiner1985_book_stochastic}.} represents the white noise, also sometimes referred as a Langevin term. Even though we previously mentioned that white Gaussian noise $w(t)$ can be informally viewed as the formal derivative of a Wiener process, we have yet to rigorously define what we mean by white noise.

White noise processes differ significantly between discrete and continuous time, particularly in how they are defined and interpreted. In the discrete-time setting, white noise has an intuitive definition, involving sequences of uncorrelated (or independent) random variables with finite variance. In contrast, continuous-time white noise cannot be defined as a standard stochastic process, requiring more complicated mathematical machinery. Thus, we begin by introducing the simpler notion of discrete white noise and then move to continuous time. 

\begin{definition}[Discrete white noise] \label{def:discrete_wn}
    Let $\{\pmb{q}_k\}_{k\in\mathbb{Z}}$ be a real‑valued discrete‑time stochastic process. We say that it is a wide‑sense white noise process if it satisfies~\cite{hayes1996statistical}:
    \begin{align}
      &\EE{\pmb{q}_k} = 0, \\
      &\cov[\pmb{q}_k, \pmb{q}_j]
      = \EE{\pmb{q}_k \, \pmb{q}_j} = \pmb{Q}_k\,\delta_{kj},
    \end{align}
    where $\delta_{kj}$ is the Kronecker delta and $\pmb{Q}_k>0$ is the variance of $\pmb{q}_k$.  If, in addition, the random variables in $\{\pmb{q}_k\}_{k\in\mathbb{Z}}$ are mutually independent, i.e., 
    \begin{align}
        p(\pmb{q}_1,\dots,\pmb{q}_k) = p(\pmb{q}_1) \cdot \dots \cdot p(\pmb{q}_k)
    \end{align}
    then $\{\pmb{q}_k\}_{k\in\mathbb{Z}}$ is also strict‑sense white noise. Finally, when each element $\pmb{q}_k$ of the white process is Gaussian, one speaks of a discrete Gaussian white noise:
    \begin{align}
         \pmb{q}_k \;\sim\;\mathcal{N}(0,\,\pmb{Q}_k).
    \end{align}
\end{definition}

Unlike in discrete time, a continuous-time white noise process cannot be defined as a standard stochastic process that when evaluated point-wise yields a random variable with a finite variance. Instead, continuous white noise is treated as a \emph{generalized} stochastic process, defined only through its action on test functions via integration:

\begin{definition}[Continuous white noise] \label{def:cont_wn}
    A (generalized) continuous-time stochastic process $\{\pmb{w}(t), t \in [0,\infty)\}$ taking values in $\mathbb{R}^n$ is called white noise if~\cite{papoulis2002probability}:
         \begin{align}
          &\EE{\pmb{w}(t)} = 0, \\
          &\cov[\pmb{w}(t), \pmb{w}(s)]
          = \EE{\pmb{w}(t) \pmb{w}(s)} = \pmb{Q}(t)\,\delta(t-s), \label{eq:covariance_cont_wn}
        \end{align}
    where $\delta(\,\cdot\,)$ is the Dirac delta and $\pmb{Q}(t)$ is the (continuous) covariance matrix, also known as the spectral density matrix \cite{sarkka2013bayesian}.
\end{definition}

The term ``white'' in white noise refers to its flat power spectral density $S_w(\omega)$ over an infinite frequency range. To visualize this, consider the case of a white noise with a constant covariance, i.e. $Q(t) = Q$. Its power spectral density is given by:
\begin{equation}
    S_w(\omega) = \int_{-\infty}^\infty  \EE{\pmb{w}(t) \pmb{w}(t+\tau)} \ee^{\ii \omega \tau} \dd \tau = Q.
\end{equation}
Thus, this constant power spectral density across an infinite bandwidth implies that the generalized process has an infinite variance in the time domain. This is reflected by the Dirac delta $\delta(0)$ in the covariance function of \eqnref{eq:covariance_cont_wn} when $t = s$. Due to this infinite variance, $\pmb{w}(t)$ cannot be treated as a standard stochastic process that can be evaluated point-wise yielding random variables with finite variance. Instead, it is referred to as a \emph{generalized} process, just like the Dirac delta is called a generalized function: i.e. it is defined by multiplying it with a test function and then integrating:
\begin{align}
    \!\!\int_{-\infty}^\infty \!\!\!\!\! f(t) \delta(t) \dd t = f(0)  \Longleftrightarrow \int_{-\infty}^\infty  \!\!\!\!\! \pmb{f}^{\;\Trans}\!(t) \pmb{w}(t) \dd t = \brkt{\pmb{f},\pmb{w}} \;\; \text{is a well-defined random vector.} 
\end{align}

In the case where the white noise is also Gaussian, it can be formally related to a Wiener process $\{\W(t), t \in [0,\infty)\}$ (introduced in \secref{sec:Wiener_process}) as:
\begin{equation}
    w(t) = \frac{\dd\W(t)}{\dt},
\end{equation}
in a distributional sense. This means that for any suitable test function
\begin{equation}
    \int f(t) w(t) \dt = \int f(t) \dW,
\end{equation}
where the right-hand side is an It\^{o} integral (see \secref{sec:ito_integral} for more on It\^{o} integrals).

Another important point we should discuss is how to discretize continuous white noise. Since $w(t)$ is a generalized stochastic process whose properties are only well-defined under integration with a test function $f(t)$, we must utilize this and set $f(t) = 1$ in order to generate discrete random variables. Namely, over an interval $[t_{k\shortminus1},t_k)$ of length $\Dt$, the discretized white noise reads as
\begin{equation}
    \pmb{q}_k = \frac{1}{\Dt}\int_{t_{k\shortminus1}}^{t_k} \pmb{w}(\tau) \dd \tau.
\end{equation}
In the case where $\pmb{w}(t)$ is a continuous \emph{Gaussian} white noise, then its integral over the interval $[t_{k\shortminus1},t_k)$ is simply the Wiener increment $\Delta\!\W_k$:
\begin{equation}
    \Delta\!\W_k = \W(t_k) - \W(t_{k\shortminus1}) = \int_{t_{k\shortminus1}}^{t_k} \dW_\tau =  \int_{t_{k\shortminus1}}^{t_k} \pmb{w}(\tau) \dd \tau,
\end{equation}
such that
\begin{equation}
    \pmb{q}_k = \frac{\Delta\!\W_k}{\Dt} = \frac{1}{\Dt} \int_{t_{k\shortminus1}}^{t_k} \pmb{w}(\tau) \dd \tau, \label{eq:discretization_white_noise}
\end{equation}
where if the covariance of the continuous white Gaussian process is $\pmb{Q}$, then this discretized white noise process is drawn from a Gaussian distribution with mean zero and covariance $\pmb{Q}/\Dt$:
\begin{equation}
    \pmb{q}_k \sim \Gauss(0,\pmb{Q}/\Dt).
\end{equation}

\subsection{Discretization of a continuous linear and Gaussian system} \label{sec:discretization_cont_model}

A common class of stochastic dynamical models are continuous-time linear Gaussian (LG) systems. These describe the evolution of a state vector and a measurement output through linear dynamics driven by Gaussian noise. Because of this, the system is fully characterized by its means and covariances, which makes it analytically tractable and widely applicable in areas ranging from physics and engineering to finance and biology. 

Even though LG models are often written as SDEs in a continuous-time formulation, many real-world applications rely on digital computation. It is therefore necessary to derive a discrete-time equivalent of the continuous LG model that preserves its statistical properties. Below, we discuss how such a discretization can be obtained.

\begin{prop}[Discretization of a continuous‐time LG system]\label{prop:disc_LG}
    Let $\pmb{x}(t)$ be the state of the system, $\pmb{u}(t)$ the control input, and $\pmb{y}(t)$ the measurement, and let $\pmb{w}(t)$ and $\pmb{v}(t)$ be zero-mean white Gaussian noises with covariances:
    \begin{align}
        \EE{\pmb{w}(t) \pmb{w}(s)^\Trans} &= \pmb{Q}(t) \delta(t-s), \label{eq:cov_Q_ch1} \\
        \EE{\pmb{v}(t) \pmb{v}(s)^\Trans} &= \pmb{R}(t) \delta(t-s), \label{eq:cov_R_ch1}\\
        \EE{\pmb{w}(t) \pmb{v}(s)^\Trans} &= 0. \label{eq:uncorrel_S_zero_ch1}
    \end{align}
    Consider the continuous-time LG model:
    \begin{align} 
        \dot{\pmb{x}}(t) &= \pmb{F}(t) \, \pmb{x}(t) + \pmb{B}(t) \,\pmb{u}(t) + \pmb{G}(t) \, \pmb{w}(t), \label{eq:cont_system_model_L&G_ch1}\\
        \pmb{y}(t) &= \pmb{H}(t) \, \pmb{x}(t) + \pmb{v}(t). \label{eq:cont_meas_model_L&G_ch1}
    \end{align}
    Then, by discretizing the system above, we get:
    \begin{align}
        \pmb{x}_k &= \pmb{A}_{k\shortminus1} \, \pmb{x}_{k\shortminus1} +  \pmb{B}_{k\shortminus1} \, \pmb{u}_{k\shortminus1} + \pmb{G}_{k\shortminus1} \, \pmb{q}_{k\shortminus1}, \label{eq:discrete_system_L&G_ch1} \\
        \pmb{y}_k &= \pmb{H}_k \, \pmb{x}_k + \pmb{r}_k, \label{eq:discrete_meas_L&G_ch1}
    \end{align}
    where $\pmb{q}_{k-1}\sim\Gauss(0,\pmb{Q}_{k-1})$, $\pmb{r}_k\sim\Gauss(0,\pmb{R}_k)$, $\EE{\pmb{q}_{k-1} \, \pmb{r}_k^\Trans}=0$, and
    \[
    \begin{array}{lll}
    \pmb{A}_{k\shortminus1} = \I + \pmb{F}(t_{k\shortminus1})\,\Dt,
      & \pmb{B}_{k\shortminus1} = \pmb{B}(t_{k\shortminus1})\,\Dt,
      & \pmb{G}_{k\shortminus1} = \pmb{G}(t_{k\shortminus1})\,\Dt, \\[1ex]
    \pmb{H}_{k\shortminus1} = \pmb{H}(t_{k\shortminus1}),
      & \pmb{Q}_{k\shortminus1} = \dfrac{\pmb{Q}(t_{k\shortminus1})}{\Dt},
      & \pmb{R}_{k\shortminus1} = \dfrac{\pmb{R}(t_{k\shortminus1})}{\Dt}.
    \end{array}\]
\end{prop}

\begin{figure}[htbp]
\begin{center}
    \begin{tikzpicture}
    \draw[->] (0, 0) -- (7, 0) node[right] {time};
    \draw[->] (0, 0) -- (0, 4) node[above] {signal};
    
    \draw[thick] plot[smooth, tension=0.8] coordinates {(0,0.5) (1, 2) (2, 2.5) (3,3) (4, 1.5) (5, 1.25) (6, 1)};
    
    \draw[solid, line width = 0.5mm] (0, 0.5) -- (1, 0.5);
    \draw[solid, line width = 0.5mm] (1, 2) -- (2, 2);
    \draw[solid, line width = 0.5mm] (2, 2.5) -- (3, 2.5);
    \draw[solid, line width = 0.5mm] (3, 3) -- (4, 3);
    \draw[solid, line width = 0.5mm] (4,1.5) -- (5, 1.5);
    \draw[solid, line width = 0.5mm] (5,1.25) -- (6, 1.25);

    \filldraw[black] (0, 0.5) circle (2pt) node[above] {};
    \filldraw[black] (1, 2) circle (2pt) node[above] {};
    \filldraw[black] (2, 2.5) circle (2pt) node[above] {};
    \filldraw[black] (3, 3) circle (2pt) node[above] {};
    \filldraw[black] (4, 1.5) circle (2pt) node[above] {};
    \filldraw[black] (5, 1.25) circle (2pt) node[above] {};
    \filldraw[black] (6, 1) circle (2pt) node[above] {};

    \draw[thick, dashed] (0, 0.5) -- (1, 0.5) -- (1,2) -- (2,2) -- (2,2.5) -- (3,2.5) -- (3, 3) -- (4, 3) -- (4, 1.5) -- (5, 1.5) -- (5, 1.25) -- (6, 1.25) -- (6, 1);

    \node at (2, 1.2) {\small continuous};
    \draw[->] plot[smooth, tension=0.8] coordinates {(2.9, 1.2) (3.3,1.3) (3.6, 1.9)};
    \node at (5, 2.5) {\small sampled};
    \draw[->] plot[smooth, tension=0.8] coordinates {(5.7, 2.5) (6,2) (5.8, 1.31)};
\end{tikzpicture}
\end{center}
\caption[A continuous signal sampled using the zero-order hold assumption]{\textbf{A continuous signal sampled using the zero-order hold assumption.} To discretize a signal using zero-order hold, the time axis is divided in increments of $\Dt$ in order to evaluate the function at these steps: $k\Dt$, with $k \in \mathbb{Z}$. The signal is then further assumed to maintain a constant value $f(k\Dt)$ from time $k \Dt$ to time $(k+1)\Dt$. }
\label{fig:discretization}
\end{figure}
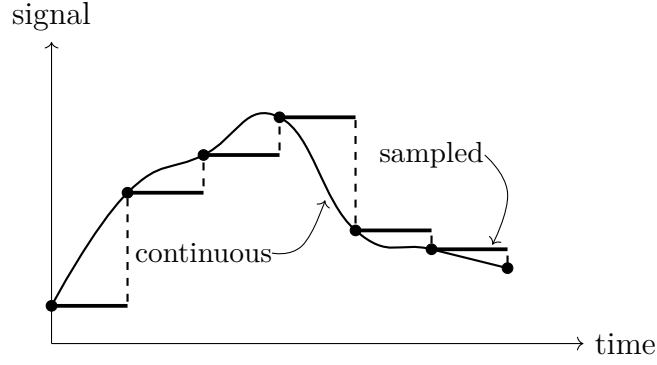

\begin{myproof}
    Let us establish the equivalence between the continuous LG system and its discrete counterpart. The first step is to apply \theoremref{theorem:sol_inhom_diff} to solve  \eqnref{eq:cont_system_model_L&G_ch1}, focusing on the interval between time points $t_{k\shortminus1} = (k\!\shortminus\!1)\Dt$ and $t_k = k\Dt$:
\begin{equation} \label{eq:sol_with_phi_general_timestep}
    \pmb{x}(t_k) = \pmb{\Phi}(t_k,t_{k\shortminus1}) \, \pmb{x}(t_{k\shortminus1}) + \int_{t_{k\shortminus1}}^{t_k} \pmb{\Phi}(t_k,\tau) \pmb{B}(\tau) \pmb{u}(\tau) \dd \tau + \int_{t_{k\shortminus1}}^{t_k}  \pmb{\Phi}(t_k,\tau) \pmb{G}(\tau) \pmb{w}(\tau) \dd \tau,
\end{equation}
where $\pmb{\Phi}(t_k,t_{k\shortminus1})$ is the state-transition matrix that satisfies
\begin{equation} \label{eq:def_ode_for_phi}
    \frac{\dd \pmb{\Phi}(t_k,t_{k\shortminus1})}{\dd t} = \pmb{F}(t) \pmb{\Phi}(t_k,t_{k\shortminus1}), \;\;\;\;\; \pmb{\Phi}(t_{k\shortminus1},t_{k\shortminus1}) = \I.
\end{equation}
and fulfills the following properties for all $t_{k\shortminus1} \leq t_k \leq T$:
\begin{align}
    \pmb{\Phi}(t_{k\shortminus1},t_k) = \pmb{\Phi}^{-1}(t_k,t_{k\shortminus1}) , \label{eq:phi_inv_main}\\
    \pmb{\Phi}(T,t_{k\shortminus1}) = \pmb{\Phi}(T,t_k) \pmb{\Phi}(t_k,t_{k\shortminus1}) \label{eq:phi_concat_main}.
\end{align}
To now find an expression for the transition matrix of the state, $\pmb{\Phi}(t_k,t_{k\shortminus1})$, we apply the zero-order hold approximation. In other words, we assume a small enough time-step $\Dt$ during which, as depicted in \figref{fig:discretization}, each deterministic continuous signal in \eqnref{eq:cont_system_model_L&G_ch1} and \eqnref{eq:cont_meas_model_L&G_ch1} is constant within the time-step $\Dt$:
\begin{align}
    \pmb{F}(t) \approx \pmb{F}(t_{k\shortminus1}), \;\; 
    \pmb{B}(t) \approx \pmb{B}(t_{k\shortminus1}), \;\;
    \pmb{G}(t) \approx \pmb{G}(t_{k\shortminus1}), \;\;
    \pmb{H}(t) \approx \pmb{H}(t_{k\shortminus1}), \;\;
    \pmb{u}(t) \approx \pmb{u}(t_{k\shortminus1}), \nonumber
\end{align}
with the only exception being the zero-mean Gaussian noise processes $\pmb{w}(t)$ and $\pmb{v}(t)$, as explained in \secref{sec:white_noise}. Then, the solution to \eqnref{eq:def_ode_for_phi} is simply 
\begin{equation}
    \pmb{\Phi}(t_k,t_{k\shortminus1}) = \e^{\pmb{F}(t_{k\shortminus1})(t_k-t_{k\shortminus1})} = \e^{\pmb{F}(t_{k\shortminus1})\Dt},
\end{equation}
and \eqnref{eq:sol_with_phi_general_timestep} becomes
\begin{align}
    \!\!\!&\pmb{x}(t_k) \!=\! \e^{\pmb{F}(t_{k\shortminus1})\Dt} \pmb{x}(t_{k\shortminus1}) \!+\! 
    \left[\int_{t_{k\shortminus1}}^{t_k} \!\!\!\! \e^{\pmb{F}(t_{k\shortminus1})(t_k\shortminus\tau)} \pmb{B}(\tau) \pmb{u}(\tau) \dd \tau\right]  \!+\! \int_{t_{k\shortminus1}}^{t_k} \!\!\!\! \e^{\pmb{F}(t_{k\shortminus1})(t_k\shortminus\tau)} \pmb{G}(\tau) \pmb{w}(\tau) \dd \tau  \\
    \!&=\! \e^{\pmb{F}(t_{k\shortminus1})\Dt} \pmb{x}(t_{k\shortminus1}) \!+\!\!
    \left[\int_{t_{k\shortminus1}}^{t_k} \!\!\!\!\! \e^{\pmb{F}(t_{k\shortminus1})(t_k\shortminus\tau)} \dd \tau\right] \!\!\pmb{B}(t_{k\shortminus1}) \pmb{u}(t_{k\shortminus1}) \!+\!\! \int_{t_{k\shortminus1}}^{t_k} \!\!\!\!\! \e^{\pmb{F}(t_{k\shortminus1})(t_k\shortminus\tau)} \pmb{G}(t_{k\shortminus1}) \pmb{w}(\tau) \dd \tau. \label{eq:continuous_to_discrete_xk}
\end{align}
If we now compare \eqnref{eq:continuous_to_discrete_xk} with \eqnref{eq:discrete_system_L&G_ch1}, we can identify:
\begin{align}
     \pmb{x}_k &= \pmb{x}(t_k), \;\; \pmb{u}_{k\shortminus1} = \pmb{u}(t_{k\shortminus1}), \;\; \pmb{B}_{k\shortminus1} = \left[\int_{t_{k\shortminus1}}^{t_k} \!\!\!\! \e^{\pmb{F}(t_{k\shortminus1})(t_k\shortminus\tau)} \dd \tau\right] \!\!\pmb{B}(t_{k\shortminus1}), \\
     \pmb{A}_{k\shortminus1} &= \e^{\pmb{F}(t_{k\shortminus1})\Dt}, \;\;
     \pmb{G}_{k\shortminus1} \, \pmb{q}_{k\shortminus1} = \int_{t_{k\shortminus1}}^{t_k} \!\!\!\! \e^{\pmb{F}(t_{k\shortminus1})(t_k\shortminus\tau)} \pmb{G}(t_{k\shortminus1}) \pmb{w}(\tau) \dd \tau, \label{eq:Gq_to_continuous}
\end{align}
where we carefully maintain the order of matrix and vector operations, as they are not generally commutative. The last three terms can be further simplified by noting that the continuous equations are equivalent to their discrete counterparts only in the limit when the time-step $\Dt$ approaches zero (i.e., as $t_k \rightarrow t_{k\shortminus1}$). Consequently, we can assume the time-step $\Dt$ to be infinitesimally small and only keep the terms up to first-order in $\Dt$:
\begin{align}
    \pmb{B}_{k\shortminus1} &= \left[\int_{t_{k\shortminus1}}^{t_k} \!\!\!\! \e^{\pmb{F}(t_{k\shortminus1})(t_k\shortminus\tau)} \dd \tau\right] \!\!\pmb{B}(t_{k\shortminus1}) \approx \footnotemark[1] \Dt \, \pmb{B}(t_{k\shortminus1}), \\
    \pmb{A}_{k\shortminus1} &= \e^{\pmb{F}(t_{k\shortminus1})\Dt} \approx \I + \pmb{F}(t_{k\shortminus1})\Dt, \\
    \pmb{G}_{k\shortminus1} \, \pmb{q}_{k\shortminus1} &= \int_{t_{k\shortminus1}}^{t_k} \!\!\!\! \e^{\pmb{F}(t_{k\shortminus1})(t_k\shortminus\tau)} \pmb{G}(t_{k\shortminus1}) \pmb{w}(\tau) \dd \tau \approx \footnotemark[2]  \pmb{G}(t_{k\shortminus1}) \!\!\! \int_{t_{k\shortminus1}}^{t_k} \!\!\! \pmb{w}(\tau) \dd \tau .
\end{align}
\footnotetext[1]{$\int_0^{\Dt} \!\!  f(t) \dt \approx f(0) \Dt, \;\; \text{when} \;\; \Dt \rightarrow 0$} 
\footnotetext[2]{$ \int_0^{\Dt} \!\! f(t) w(t) \dt \approx f(0) \!\! \int_0^{\Dt} \!\!\!\! w(t) \dt, \;\; \text{when} \;\; \Dt \rightarrow 0$} 

From the last equation we can further identify that
\begin{align} \label{eq:process_noise_discrete_w_continuous}
    \pmb{G}_{k\shortminus1} = \pmb{G}(t_{k\shortminus1}) \Dt, \;\;\;\;\; \text{and} \;\;\;\;\; \pmb{q}_{k\shortminus1} = \frac{1}{\Dt} \int_{t_{k\shortminus1}}^{t_k} \!\!\! \pmb{w}(\tau) \dd \tau,
\end{align}
where the discrete white noise sample $\pmb{q}_k$ comes from averaging the continuous white noise over the interval $[t_{k\shortminus1},t_k)$, as explained in \eqnref{eq:discretization_white_noise}. Note that unlike for the state, control term and other functions in \eqnref{eq:sol_with_phi_general_timestep}, we do not approximate the noise terms $\pmb{w}(t)$ and $\pmb{v}(t)$ using zero-order hold. In other words, we do not sample these continuous noises and approximate them with their value at the beginning of the interval, like in \figref{fig:discretization}. The main reason behind this lays in the definition of continuous white Gaussian noise, which as discussed in \secref{sec:white_noise}, has a delta-correlated covariance. This simply means that the noise is completely uncorrelated from one time instant to another, but that it has infinite variance. If we try to approximate the process noise with the zero-order hold, i.e. $\pmb{q}_{k\shortminus1} \approx \pmb{w}(t_{k\shortminus1})$, and apply the definition of the process covariance of \eqnref{eq:cov_Q_ch1}:
\begin{align}
    \pmb{Q}_k = \EE{\pmb{q}_k \, \pmb{q}_k^\Trans} = \EE{\pmb{w}(t_k) \pmb{w}^\Trans\!(t_k)} = \pmb{Q}(t_k) \delta(0),
\end{align}
we obtain an unbounded discrete process noise covariance. This clearly illustrates why direct sampling of the noise, as required by the zero-order hold, is mathematically problematic. If instead we discretize the continuous process noise by averaging it over the infinitesimally small time-step $\Dt$, as suggested by \eqnref{eq:process_noise_discrete_w_continuous}, we retrieve a mathematically sound expression relating $\pmb{Q}_k$ and $\pmb{Q}(t)$, which yields a well-behaved process:
\begin{align}
    \pmb{Q}_k &= \EE{\pmb{q}_k \,\pmb{q}_k^\Trans} = \frac{1}{\Dt^2} \!\! \int_{t_{k}}^{t_{k+1}} \!\!\! \int_{t_{k}}^{t_{k+1}} \!\!\! \EE{\pmb{w}(\tau) \pmb{w}^\Trans\!(s)} \dd \tau \, \dd s = \frac{1}{\Dt^2} \!\!\int_{t_{k}}^{t_{k+1}} \!\!\! \int_{t_{k}}^{t_{k+1}} \!\!\! \pmb{Q}(\tau) \delta(\tau\!\shortminus\!s) \dd \tau \, \dd s  \nonumber \\
    &=\frac{1}{\Dt^2} \int_{t_{k}}^{t_{k+1}} \!\! \pmb{Q}(\tau) \dd \tau \approx \frac{1}{\Dt^2} \, \pmb{Q}(t_{k}) \Dt = \frac{\pmb{Q}(t_{k})}{\Dt},
\end{align}
where we have now applied zero-order hold to approximate $\pmb{Q}(t) \approx \pmb{Q}(t_k)$ inside the integral of $\pmb{Q}(t)$ over an infinitesimal time-step $\Dt$ in the interval $[t_k,t_{k+1})$. It then follows that
\begin{equation} \label{eq:discr_to_cont_GQG}
    \pmb{G}_k \pmb{Q}_k \pmb{G}_k^\Trans = \Dt \, \pmb{G}(t_k)\pmb{Q}(t_k)\pmb{G}(t_k)^\Trans.
\end{equation}
We can apply the same reasoning to the measurement process noise, since $\pmb{v}(t)$ is a continuous Gaussian white noise. Namely, the discretized measurement noise is obtained by integrating $\pmb{v}(t)$ over the interval $\Dt$:
\begin{equation}
    \pmb{r}_k = \frac{1}{\Dt} \int_{t_{k\shortminus1}}^{t_k} \pmb{v}(\tau) \dd \tau.
\end{equation}
Then, since $\pmb{y}(t)$ follows \eqnref{eq:cont_meas_model_L&G_ch1} and by applying the zero-hold assumption for the state $\pmb{x}(t)$ and measurement matrix $\pmb{H}(t)$, we get
\begin{align}
    \pmb{y}_{k} &\coloneqq \frac{1}{\Dt} \int_{t_{k}}^{t_{k+1}} \pmb{y}(t) \dt = \frac{1}{\Dt} \int_{t_{k}}^{t_{k+1}} \left(\pmb{H}(t) \pmb{x}(t) + \pmb{v}(t) \right) \dt \\
    &\approx \pmb{H}(t_k) \pmb{x}(t_k) + \frac{1}{\Dt} \int_{t_{k}}^{t_{k+1}} \!\! \pmb{v}(t) \dt,
\end{align}
where the relation between the discrete and continuous measurement covariance, $\pmb{R}_k$ and $\pmb{R}(t_k)$, is:
\begin{align}
    \pmb{R}_k &= \EE{\pmb{r}_k\pmb{r}_k^\Trans} = \frac{1}{\Dt^2} \!\!\int_{t_{k}}^{t_{k+1}}\!\!\int_{t_{k}}^{t_{k+1}} \!\!\EE{\pmb{v}(t) \pmb{v}^\Trans(\tau)} \dt \, \dd \tau = \frac{1}{\Dt^2} \!\! \int_{t_{k}}^{t_{k+1}} \!\! \pmb{R}(t) \dt = \frac{\pmb{R}(t_k)}{\Dt}. 
\end{align}
\end{myproof}

\begin{table}[H]
    \centering
    \renewcommand{\arraystretch}{1.75}
    \setlength{\extrarowheight}{0pt} 
    \begin{tabular}{|>{\centering\arraybackslash}m{4cm}|>{\centering\arraybackslash}m{4cm}|}
    \hline
        $\pmb{x}_k = \pmb{x}(t_k)$\vspace{.1pt} & $\pmb{u}_k = \pmb{u}(t_k)$\vspace{.1pt} \\
    \hline
        $\pmb{A}_k = \I + \pmb{F}(t_k)\Dt$ \vspace{.1pt} & $\pmb{B}_k = \pmb{B}(t_k)\Dt$ \vspace{.1pt}\\
    \hline
        $\pmb{H}_k = \pmb{H}(t_k)$ \vspace{.1pt} & $\pmb{G}_k = \pmb{G}(t_k) \Dt$ \vspace{.1pt} \\
    \hline
        $\pmb{r}_k = \frac{1}{\Dt} \int_{t_{k\shortminus1}}^{t_k} \pmb{v}(\tau) \dd \tau$ \vspace{1pt} & $\pmb{q}_k = \frac{1}{\Dt} \int_{t_{k\shortminus1}}^{t_k} \pmb{w}(\tau) \dd \tau$ \vspace{1pt} \\
    \hline
        $\pmb{R}_k = \frac{\pmb{R}(t_k)}{\Dt}$ & $\pmb{Q}_k = \frac{\pmb{Q}(t_k)}{\Dt}$ \\
    \hline
    \end{tabular}
    \caption[Table summarizing how to discretize a continuous linear and Gaussian system]{\textbf{Table summarizing how to discretize a continuous linear and Gaussian system.} This table details how each vector, white noise component, and matrix defining the continuous LG system of \eqnsref{eq:cont_system_model_L&G_ch1}{eq:cont_meas_model_L&G_ch1} is transformed to obtain the corresponding elements of the discrete LG system given in \eqnsref{eq:discrete_system_L&G_ch1}{eq:discrete_meas_L&G_ch1}.}
    \label{tab:summary_discretization}
\end{table}

\section{(some) Fundamentals of quantum mechanics} \label{sec:fundamentals_qm}
\subsection{Position and momentum operators}
The position and momentum operators, $\position$ and $\momentum$, can be written in terms of creation and anihilation operators as
\begin{equation} \label{eq:pos&mom_a&adagger}
    \position = \sqrt{\frac{1}{2}} \left(\creat +  \anihil \right), \;\;\;\;\; \momentum = i \sqrt{\frac{1}{2}} \left( \creat - \anihil \right),
\end{equation}
These operators satisfy the canonical commutation relation $[\position,\momentum] = \ii$, which follows directly from the algebra $[\anihil,\creat] = 1$.

\subsubsection{Ground wavefunction in the position eigenbasis of the Harmonic Oscillator} \label{subsubsec:GW_HO}

The ground-state wavefunction of the quantum harmonic oscillator in the position eigenbasis is a Gaussian function:
\begin{equation}
    \psi_0(x) = \braket{x}{0} = \frac{1}{\pi^{1/4}} \ee^{-x^2/2},
\end{equation}
which has been normalized, as required from a wavefunction. 

\begin{myproof}   
From \eqnref{eq:pos&mom_a&adagger}, we can write the annihilation and creation operators w.r.t. the position and momentum operators as
\begin{align}
    \anihil = \frac{1}{\sqrt{2}} (\position + i \momentum), \quad \text{and} \quad \creat = \frac{1}{\sqrt{2}} (\position - i\momentum).
\end{align}
As standard, applying $\anihil$ to the ground state yields zero, 
\begin{equation}
    \anihil \ket{0} = 0.
\end{equation}
Thus, applying $\bra{x}$ to the equation above, where $\bra{x}$ is an eigenstate of $\position$, will also yield zero:
\begin{equation}
    \braketop{x}{\anihil}{0} = \frac{1}{\sqrt{2}} \braketop{x}{(\position + i \momentum )}{0} = 0.
\end{equation}
By using that the $\momentum$ operator represented in the position basis can be written as 
\begin{equation}
    \braketop{x}{\momentum}{0} = -\ii \frac{\dd}{\dd x} \braket{x}{0},
\end{equation}
we find a first-order linear differential equation for the wavefunction $\psi_0(x) = \braket{x}{0}\,$:
\begin{equation}
    \braketop{x}{\position}{0} + \ii \braketop{x}{\momentum}{0} = x \, \braket{x}{0} + \frac{\dd}{\dd x} \braket{x}{0} = 0,
\end{equation}
which has a solution 
\begin{equation}
    \braket{x}{0} = A \e^{-x^2/2},
\end{equation}
with a normalization constant $A = 1/\pi^{1/4}$, since
\begin{equation}
    1 = \int_{-\infty}^\infty \braketsquared{x}{0} \, \dd x = \int_{-\infty}^\infty |A|^2 \e^{-x^2} \, \dd x = |A|^2 \sqrt{\pi}.
\end{equation}
Therefore, the wavefunction of the ground state in the position eigenbasis reads as:
\begin{equation} \label{eq:ground_state_wf_ho}
    \braket{x}{0} = \frac{1}{\pi^{1/4}} \e^{-x^2/2}.
\end{equation}
\end{myproof}

\subsection{Dynamics of open quantum systems}

\subsubsection{The Gorini–Kossakowski–Sudarshan–Lindblad generator} \label{sec:Lindbladian}

A quantum system interacting with a Markovian environment is governed by a master equation of the form
\begin{equation} \label{eq:master_equation}
    \frac{d\rho}{ \dt} = \Lin \, \rho,
\end{equation}
where the Gorini–Kossakowski–Sudarshan–Lindblad (GKSL) generator of the evolution \cite{GKS1976,Lindblad1976} reads as:
\begin{equation} \label{eq:Linbladian_form}
    \Lin \rho = -i [\Ham , \rho] + \sum_{k = 1}^K \D[\LinOp_k] \rho.
\end{equation}
Here, $\Ham$ is a Hermitian operator representing the Hamiltonian of the system and $\{ \LinOp_j\}$ are a collection of operators, often referred to as Lindblad operators, that characterize the various irreversible processes via the (dissipative) superoperator:
\begin{equation} \label{eq:dissipation_superoperator}
    \D[ \GenOp ] \rho = \GenOp \rho \GenOp^\dagger - \frac{1}{2} (\GenOp^\dagger \GenOp \rho + \rho \GenOp^\dagger \GenOp).
\end{equation}
The formal solution for a general master equation \eqref{eq:master_equation} with a time-dependent Lindblad form can be written as
\begin{equation}
    \rho(t) = \LinN_t \, \rho(0)
\end{equation}
where the superoperator $\LinN_t$ is defined as
\begin{equation} \label{eq:time-dep_N}
    \LinN_t = \TimeOrd \left\{\e^{\int_0^t \Lin_\tau d\tau} \right\},
\end{equation}
with $\TimeOrd \left\{\exp{\{\;\cdot\;\}} \right\}$ denoting the time-ordered exponential. In the special case when $\Lin$ is time-independent, the solution is a semigroup:
\begin{equation} \label{eq:time-indep_N}
    \LinN_t = \e^{\Lin t}.
\end{equation}

\subsection{Angular momentum}

The three \emph{total} angular momentum operators $\Jx$, $\Jy$ and $\Jz$ obey the commutation relation
\begin{equation}
    [\Jindex{j},\Jindex{k}] = i \tinyspace \epsilon_{jkl} \tinyspace \Jindex{l}
\end{equation}
where $\epsilon_{jkl}$ is the Levi-Civita symbol and the indeces $i,j,k = x,y,z$. In other words, 
\begin{align}
    [\Jx,\Jy] &= \ii \Jz \\
    [\Jy,\Jz] &= \ii \Jx \\
    [\Jz,\Jx] &= \ii \Jy. 
\end{align}
The collection of these angular momentum operators form a total angular momentum \emph{vector} operator:
\begin{equation}
    \Jqvec{J} = (\Jx,\Jy,\Jz)^\Trans,
\end{equation}
whose squared magnitude defines another operator:
\begin{equation}
    \Jsq \coloneqq \Jx^2+\Jy^2+\Jz^2.
\end{equation}
The operators $\Jsq$ and $\Jz$ have a common eigenbasis with eigenvectors labeled with quantum numbers $j$ and $m$:
\begin{align}
    \Jsq \ket{\,j,m} &= j(j+1)\ket{\,j,m}, \nonumber \\
    \Jz \ket{\,j,m} &= m \ket{\,j,m}, \nonumber 
\end{align}
where $j \in \{0,\frac{1}{2},1,\frac{3}{2},2,\dots\}$, and for a given $j$, the value of $m$ ranges as $m = - j, - j~+~1,$ $\dots, j~-~1,j$, i.e. from $-j$ to $j$ in steps of unity \cite{Biedenharn_Louck_1984}. Using the operators $\Jx$ and $\Jy$, we define the ladder operators for the angular momentum operators as:
\begin{align}
    \Jpm = \Jx \pm i \Jy,
\end{align}
which fulfill $\Jp = \Jm^{\dagger}$. These operators act on the eigenstates $\ket{\,j,m}$ by raising or lowering the eigenvalue $m$ by one unit:
\begin{align}
    \Jpm \ket{\,j,m} = \sqrt{j(j+1) - m(m\pm 1)} \ket{\,j,m \pm 1}.    
\end{align}

If $\Jqvec{J}$ is the angular momentum vector of an ensemble of $N$ two-level systems, then $j \leq N/2$ and $-j \leq m \leq j$, with $j$ and $m$ integer or half and $j_{\trm{min}} = 0,1/2$ depending whether there is an even or odd number of two-level systems, i.e. whether $N$ is even or odd \cite{Shammah2018}. Then, the collective angular momentum operators are defined as the sum over the individual contributions:
\begin{equation} \label{eq:collective_angular_momentum}
    \Jx = \frac{1}{2} \sum_{k=1}^N \Paulix^{(k)}, \quad \Jy = \frac{1}{2} \sum_{k=1}^N \Pauliy^{(k)}, \quad \Jz = \frac{1}{2} \sum_{k=1}^N \Pauliz^{(k)},
\end{equation}
where the angular momentum operators for each individual two-level system are  $\sx^{(k)} = \frac{1}{2} \Paulix^{(k)}$, $\sy^{(k)}~=~\frac{1}{2}\Pauliy^{(k)}$ and $\sz^{(k)} = \frac{1}{2}\Pauliz^{(k)}$. The collective operators obey the same commutation relations and eigenvalue equations as those for a single system while encapsulating the macroscopic spin properties of the entire ensemble.

\subsection{Coherent spin state} \label{sec:CSS_main}

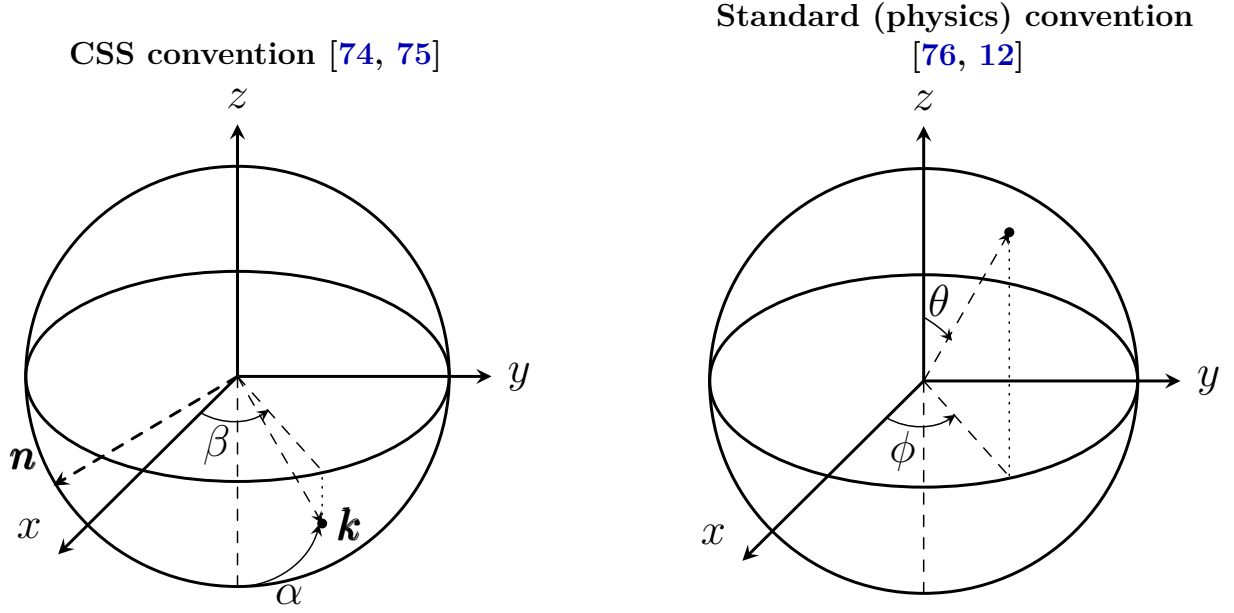
\begin{figure}[ht!]
    \centering
    \begin{subfigure}{0.45\textwidth}
        \centering
        \textbf{\hspace{-0.75cm} CSS convention \cite{Arecchi1972,Dowling1994}}
        \vspace{0.2cm}
        \resizebox{\textwidth}{!}{%
    \begin{tikzpicture}[scale=2, line cap=round, line join=round, >=stealth]
        \draw[thick, ->] (0,0,0) -- (0,0,2.2) node[above left] {$x$};
        \draw[thick, ->] (0,0,0) -- (1.2,0,0) node[right] {$y$};
        \draw[thick, ->] (0,0,0) -- (0,1.2,0) node[above] {$z$};
        
        \draw[dashed] (1,0,0) arc[start angle=0, end angle=60, radius=1];
        \draw[dashed, ->] (0,0,0) -- (0.4,-0.7,0) node[right] {$\pmb{k}$}; 
        \draw[dashed] (0,0,0) -- (0.4,-0.45,0); 
        \draw[dotted] (0.4,-0.7,0) -- (0.4,-0.45,0); 
        \draw[dashed] (0,0,0) -- (0,-1,0); 
        \draw[dashed, thick, ->] (0,0,0) -- (-0.35,0,1.35) node[above left] {$\pmb{n}$}; 
        
        \draw[thick] (0,0,0) circle (1); 
        \draw[thick] (0,0,0) ellipse (1 and 0.5); 
        
        \draw[->] (0,-1,0) arc[start angle=270, end angle=345, radius=0.4] node[midway, below] {$\alpha$};
        
        \draw[->] (0,0,0.45) arc[start angle=240, end angle=305, radius=0.3] node[midway, below, yshift=3pt, xshift=-5pt] {$\beta$};
        
        \filldraw (0.4,-0.7,0) circle (0.02) node[above right] {$ $}; 
    \end{tikzpicture}
    \vspace{0.5em}
    }
    \end{subfigure}
    \hfill 
    \begin{subfigure}{0.45\textwidth}
        \centering
        \textbf{\hspace{-0.5cm} Standard (physics) convention \cite{Schmied2011,Pezze2018RMP}}
        \resizebox{\textwidth}{!}{%
    \begin{tikzpicture}[scale=2, line cap=round, line join=round, >=stealth]
        \draw[thick, ->] (0,0,0) -- (0,0,2.2) node[above left] {$x$};
        \draw[thick, ->] (0,0,0) -- (1.2,0,0) node[right] {$y$};
        \draw[thick, ->] (0,0,0) -- (0,1.2,0) node[above] {$z$};
        
        \draw[dashed] (1,0,0) arc[start angle=0, end angle=60, radius=1];
        \draw[dashed, ->] (0,0,0) -- (0.4,0.7,0); 
        \draw[dashed] (0,0,0) -- (0.4,-0.45,0); 
        \draw[dotted] (0.4,0.7,0) -- (0.4,-0.45,0); 
        \draw[dashed] (0,0,0) -- (0,-1,0); 
        
        \draw[thick] (0,0,0) circle (1); 
        \draw[thick] (0,0,0) ellipse (1 and 0.5); 
        
        \draw[->] (0,0.3,0) arc[start angle=60, end angle=40, radius=0.5] node[midway, above, yshift=-1pt] {$\theta$};
        
        \draw[->] (0,0,0.45) arc[start angle=240, end angle=305, radius=0.3] node[midway, below, yshift=3pt, xshift=-5pt] {$\phi$};
        
        \filldraw (0.4,0.7,0) circle (0.02); 
    \end{tikzpicture}
    }
    \vspace{0.01em}
    \end{subfigure}

    \caption[Spherical coordinates conventions]{\textbf{Spherical coordinates conventions.} The choice of rotation angles $\beta$ and $\alpha$ (see left sketch) define the rotation operator $R(\alpha,\beta)$ and thus, the axis along which the CSS is aligned (along $\pmb{k}$). Given that the rotation operator around axis $\pmb{n}$ is applied to the ground state of the ensemble $\ket{j,-j}$, by convention centered around the south pole, the rotation angle $\alpha$ is measured off the south pole \cite{Arecchi1972,Dowling1994}. Thus, the standard spherical coordinates parametrizing a sphere, $(\theta,\phi)$ (see right), relate to $(\alpha,\beta)$ as $\theta = \pi - \alpha$ and $\phi = \beta$ \cite{Dowling1994}. When we discuss Wigner functions of CSS mapped onto the Bloch sphere, it is important to note that $(\alpha,\beta)$ simply define the direction of the CSS, whereas $(\theta,\phi)$ map the Wigner function onto the 3D sphere.}
    \label{fig:angles_for_CSS}
\end{figure}

As might be deduced from the name, coherent spin states \cite{Radcliffe1971,Arecchi1972} (also known as atomic coherent states) are simply an extension of the coherent states of a field first introduced by Glauber. A coherent spin state (CSS) is a product state, i.e. the tensor product of $N$ qubits, e.g. spin-$1/2$, all aligned in the same direction $\pmb{k} = \left(\sin{\alpha} \cos{\beta},\sin{\alpha} \sin{\beta},\cos{\alpha}\right)$ (see \figref{fig:angles_for_CSS}):
\begin{align}
    \ket{\alpha,\beta} = \bigotimes_{k=1}^N \left[ \cos{\frac{\alpha}{2}} \ket{0}_k + \ee^{\ii \beta} \sin{\frac{\alpha}{2}} \ket{1}_k \right].
\end{align}

As discussed in detail in \refcite{Arecchi1972}, a CSS can also be written in terms of the angular momentum basis by rotating the ground state $\ket{j,-j}$ by an angle $\alpha$ about an axis $\pmb{n} = (\sin{\beta},-\cos{\beta},0)$ with the operator
\begin{equation} \label{eq:rotation_op_forCSS_1}
    R(\alpha,\beta) = \ee^{-\ii \alpha \left(\Jx \sin{\beta} - \Jy \cos{\beta} \right)} = \ee^{\xi \Jp - \xi^* \Jm},
\end{equation}
where $\xi = \frac{1}{2} \alpha \ee^{-\ii \beta}$. Namely,
\begin{equation} \label{eq:def_CSS_1}
    \ket{\alpha,\beta} \coloneqq R(\alpha,\beta) \ket{j,-j}.
\end{equation}
In \refcite{Arecchi1972} they further show how to rewrite \eqnref{eq:rotation_op_forCSS_1} as
\begin{equation}
    R(\alpha,\beta) = R(\eta) = \ee^{\eta \Jp} \ee^{\ln{\left(1+|\eta|^2\right)}\Jz} \ee^{-\eta^* \Jm},
\end{equation}
where
\begin{equation} \label{eq:def_eta_CSS}
    \eta = \ee^{-\ii \beta} \tan{\frac{\alpha}{2}}.
\end{equation}
By now applying this form of the rotation operator to the ground state $\ket{j,-j}$, as indicated in \eqnref{eq:def_CSS_1}, we get
\begin{equation} \label{eq:coh_state_deriv_1}
    \ket{\eta} \coloneqq R(\eta) \ket{j,-j} = \left(-\frac{1}{1+|\eta|^2}\right)^j \ee^{\eta \Jp} \ket{j,-j},
\end{equation}
since
\begin{align}
    \ee^{-\eta^* \Jm} \ket{j,-j} &= \I \ket{j,-j} \\
    \ee^{\ln{\left(1+|\eta|^2\right)}\Jz} \ket{j,-j} &= \ee^{\ln{\left(1+|\eta|^2\right)}(-j)} \ket{j,-j} = \ee^{\ln{\left(1+|\eta|^2\right)^{-j}}} \ket{j,-j} \nonumber \\
    &= \left(\frac{1}{1+|\eta|^2}\right)^{j} \ket{j,-j}.
\end{align}

Next, we expand the exponent in \eqnref{eq:coh_state_deriv_1} as
\begin{align}
    \ket{\eta} \coloneqq R(\eta) \ket{j,-j} = \left(\frac{1}{1+|\eta|^2}\right)^j \sum_{k=0}^{2j} \frac{\eta^{\,k}}{k!} \Jp^{\,k} \ket{j,-j},
\end{align}
where the power series expansion of $\ee^{\eta \Jp}$ terminates at $k=2j$, because $\Jp^k \ket{j,j} = 0$ for $k>0$ and thus, $\Jp^k \ket{j,-j} = 0$ for $k>2j$. By shifting $k$ by $j$ and redefining it as $k = m + j$, the summation becomes
\begin{align}
    \ket{\eta} = \left(\frac{1}{1+|\eta|^2}\right)^j \sum_{m = -j}^{j} \frac{\eta^{\,m+j}}{(m+j)!} \Jp^{\,m+j} \ket{j,-j}.
\end{align}
By then applying the relation derived in \propref{prop:exp_jm_asJpfromj-j}:
\begin{equation}
    \binom{2j}{m+j}^{1/2} \ket{j,m} = \frac{1}{(m+j)!} \Jp^{m+j} \ket{j,-j},
\end{equation}
we get the form of a CSS in the eigenbasis of the angular momentum 
\begin{align}
    \ket{\eta} = (1+|\eta|^2)^{-j} \sum_{m=-j}^j \binom{2j}{m+j}^{1/2} \eta^{m+j} \ket{j,m}, \ \ \ \ \eta \in \Complex,
\end{align}
where $\eta = \tan{\frac{\alpha}{2}} e^{-i\beta}$ as given in \eqnref{eq:def_eta_CSS}. Assuming the CSS of the spin-1/2 ensemble points in the $x$-direction s.t. the angles $\alpha$ and $\beta$ have values:
\begin{equation}
    j = \frac{N}{2}, \ \ \ \ \alpha = \frac{\pi}{2}, \ \ \ \ \text{and} \ \ \ \ \beta = 0,
\end{equation}
then, the value of $\eta$ is 1 and the CSS, aligned along $x$, reads as:
\begin{equation} \label{eq:CSS_along_x_jmbasis}
    \ket{\eta} = \frac{1}{2^{N/2}} \sum_{m = -N/2}^{N/2} \binom{N}{\frac{N}{2} + m}^{1/2} \ket{\frac{N}{2},m}.
\end{equation}

A CSS of that form has the following mean and variance:
\begin{align}
    \braketavg{\;\Jqvec{J} \,}_{\trm{CSS}_x} &= \trace{\rho_{\trm{CSS}_x} \; \Jqvec{J}} = \left(\braketavg{\Jx}_{\trm{CSS}_x} , \; \braketavg{\Jy}_{\trm{CSS}_x} , \; \braketavg{\Jz}_{\trm{CSS}_x}\right)^{\!\!\Trans} = \left(\frac{N}{2},\;0,\;0\right)^{\!\!\!\Trans} \\
    \braketavg{\Delta^2\Jqvec{J}\,}_{\trm{CSS}_x} &= \left(\Delta^{\!2}\braketavg{\Jx}_{\trm{CSS}_x} , \; \Delta^{\!2}\braketavg{\Jy}_{\trm{CSS}_x} , \; \Delta^{\!2}\braketavg{\Jz}_{\trm{CSS}_x}\right)^{\!\!\Trans} = \left(0,\;\frac{N}{4},\;\frac{N}{4}\right)^{\!\!\!\Trans}
\end{align}
with $\braketavg{\Delta^2\Jindex{i}} = \braketavg{\Jindex{i}^2} - \braketavg{\Jindex{i}}^2$, where $i = x, \ y,$ or $z$. The values for the means and variances are calculated in \secref{sec:CSS_m_v}.

\subsection{The Wigner quasiprobability distribution} \label{sec:Wigner_quasiprobability}

\begin{definition}[Wigner distribution]
    Let $\rho$ be a mixed state, and $x$ and $p$ a pair of conjugate variables representing position and momentum. Then, the Wigner distribution is defined as:
    \begin{equation}
        \Wigner{\rho}(x,p) \coloneqq \frac{1}{\pi} \int_{-\infty}^{\infty} \braketop{x-y}{\rho}{x+y} \e^{2 \ii p y} \dd y,
    \end{equation}
    where we assume $\hbar = 1$. Equivalently, for a pure state $\ket{\psi}$, it can be written as:
    \begin{equation}
        \Wigner{\psi}(x,p) \coloneqq \frac{1}{\pi} \int_{-\infty}^{\infty} \psi^*(x+y) \psi(x-y) \ee^{2 i p y} \dd y.
    \end{equation}
\end{definition}

The Wigner function fulfills the following properties:
\begin{enumerate}
    \item $\Wigner{\rho}(x,p)$ is a $\Real$-valued function.
    \item It may take on negative values. For continuous variables, this is often interpreted as a sign of nonclassical behavior \cite{Pezze2018RMP}.
    \item It provides proper marginal distributions:
        \begin{align}
            \braketop{x}{\rho}{x} &= \int \dd p \, \Wigner{\rho}(x,p), \\
            \braketop{p}{\rho}{p} &= \int \dd x \, \Wigner{\rho}(x,p). 
        \end{align}
    \item The state overlap of two pure states $\ket{\psi}$ and $\ket{\phi}$ is calculated as
    \begin{equation}
        \braketsquared{\psi}{\phi} = 2\pi \int_{-\infty}^\infty \dd x \int_{-\infty}^\infty \dd p \, \Wigner{\psi}(x,p)\Wigner{\phi}(x,p)
    \end{equation}
    \item Operator averages are calculated as
    \begin{align}
        \braketavg{\GenOp} = \trace{\rho \, \GenOp} = 2\pi \int_{-\infty}^\infty \dd x \int_{-\infty}^\infty \dd p \Wigner{\rho}(x,p) \; \Wigner{\GenOp} (x,p),
    \end{align}
    where $\Wigner{\GenOp} (x,p)$ is the Wigner function of the operator $\GenOp$.
\end{enumerate}

\begin{myexample} \textbf{(Wigner distribution of the vacuum state):}
    Let us explicitly derive the Wigner quasiprobability distribution for the vacuum state $\ket{0}$. Its wavefunction in the position eigenbasis, $\psi_0(x)$, is:
    \begin{equation}
        \braket{x}{0} = \psi_0(x) = \frac{1}{\pi^{1/4}} e^{-x^2/2},
    \end{equation}
    as derived in \secref{subsubsec:GW_HO}. Therefore, its Wigner function can be written as:
    \begin{align}
        \Wigner{0}(x,p) = \frac{1}{\pi} \int_{-\infty}^\infty \dd y \; \ee^{2\ii p y} \frac{1}{\sqrt{\pi}} \ee^{-(x-y)^2/2} \ee^{-(x+y)^2/2} .
    \end{align}
    The sum of the exponents of the wavefunctions yields:
    \begin{equation}
        -\frac{(x-y)^2}{2} - \frac{(x+y)^2}{2} = -x^2 - y^2.
    \end{equation}
    Therefore,
    \begin{equation}
        \Wigner{0}(x,p) = \frac{1}{\pi^{\sfrac{3}{2}}} \ee^{-x^2} \int_{-\infty}^\infty \dd y \; \ee^{-y^2} \ee^{2\ii p y}.
    \end{equation}
    Next we can evaluate the integral above by noting it is a standard Fourier transform of a Gaussian:
    \begin{equation}
        \int_{-\infty}^{\infty} \dd y \; \ee^{-y^2} \ee^{2\ii p y} = \mathcal{F}_y \!\left[\ee^{-y^2}\right]\!(p) = \sqrt{\pi} \ee^{-p^2}.
    \end{equation}
    Thus, the Wigner function of the vacuum state is
    \begin{equation}
        \Wigner{0}(x,p) = \frac{1}{\pi} \ee^{x^2+p^2}.
    \end{equation}
    Furthermore, note that the Wigner distribution for the vacuum state $\Wigner{0}(x,p)$ is invariant under rotations in the phase space, since the variables $x$ and $p$ define a radial distance $r^2 = x^2 + p^2$. herefore, if we define new operators $\position_\phi$ and $\momentum_\phi$ rotated in phase space by $\phi$ through the following transformations:
    \begin{align}
        \position_\phi &= \position \cos{\phi} + \momentum \sin{\phi}, \\
        \momentum_\phi &= - \position \sin{\phi} + \momentum \cos{\phi},
    \end{align}
    we can show that the wavefunction of the rotated eigenket $\ket{x_\phi}$ is simply
    \begin{equation} \label{eq:rotated_GW_HO}
        \braket{x_\phi}{0} = \psi_0(x_\phi) = \frac{1}{\pi^{1/4}} \ee^{-x_\phi^2/2},
    \end{equation}
    since
    \begin{equation}
        \braketsquared{x_\phi}{0} = \int \dd p_\phi \Wigner{0}(x_\phi,p_\phi) = \frac{1}{\sqrt{\pi}} \ee^{-x_\phi^2},
    \end{equation}
    because $x^2 + p^2 = x_\phi^2 + p_\phi^2$, i.e. the Wigner function of the vacuum state remains invariant under rotations in the phase space. 
\end{myexample}

\subsection{The Wigner function on a sphere} \label{sec:Wigner_on_Bloch}

Computing the Wigner distribution and mapping it into the Bloch sphere is very useful for visualizing quantum states and how different operations affect them, specially in the case of atomic ensembles \cite{Pezze2018RMP,Dowling1994,Schmied2011}. The Wigner quasiprobability distribution, mapped onto the generalized Bloch sphere, is parametrized using the standard spherical convention with angles $\theta$ and $\phi$ \cite{Pezze2018RMP} (see right in \figref{fig:angles_for_CSS}):
\begin{align}
    \Wigner{\rho}(\theta,\phi) = \sqrt{\frac{N+1}{4\pi}} \sum_{k = 0}^N \sum_{q=-k}^{k} \rho_{kq} Y_k^{\;q}(\theta,\phi)
    \label{eq:Wigner_quasi_BlochSphere}
\end{align}
where $Y_k^{\;q}(\theta,\phi)$ are the complex spherical harmonics \cite{Khersonskii1988}. These functions are defined on the surface of a unit sphere, mapping $Y_k^{\;q} : \mrm{S}^2 \rightarrow \Complex$, where $\mrm{S}^2$ is the 2D surface of the sphere. Spherical harmonics provide a complete, orthonormal basis for square-integrable functions defined on $\mrm{S}^2$. As a result, each function on the surface of a sphere can be written as a weighted sum of these spherical harmonics. In a way, spherical harmonics generalize the Fourier series from periodic functions on a circle ($\mrm{S}^1$) to functions on a sphere ($\mrm{S}^2$). While the Fourier series decomposes a periodic function into a sum of sines and cosines, spherical harmonics extend this concepts to two dimensions by employing both azimuthal ($\phi$) and polar ($\theta$) angular dependencies. Here, $\theta$ and $\phi$ follow the standard convention, with $\theta$ being the polar angle measured off the $+z$-axis \cite{Dowling1994,Schmied2011} (see right in \figref{fig:angles_for_CSS}). 

The coefficients $\rho_{kq}$ of the spherical harmonic decomposition of the Wigner function, 
\begin{equation}
    \rho_{kq} = \sum_{m_1,m_2 = -J}^J \rho_{m_1,m_2} t_{kq}^{m_1 m_2},
\end{equation}
are determined by the part of the density matrix supported by the \emph{totally symmetric subspace}, in particular, its elements $\rho_{m_1,m_2} \coloneqq \bra{\,J,m_1} \rho \ket{\,J,m_2}$, written in the angular momentum basis for the maximal total spin $\,J=N/2$, as well as the coefficients $t_{kq}^{m_1 m_2}$ dictating the transformation from the Dicke space to the $k$-space~\cite{Schmied2011,Agarwal1981}:
\begin{equation}
    t_{kq}^{m_1 m_2} \coloneqq (-1)^{\,J-m_1-q} \brkt{\,J, m_1 ; J, -m_2 | k,q},
\end{equation}
where $\brkt{\,J, m_1 ; J, -m_2 | k,q}$ are the Clebsch-Gordan coefficients.
Note that the exact density matrix is needed to generate the Wigner quasiprobability distribution.

\subsection{Spin squeezing} \label{sec:spin-squeezing_intro}

Spin-squeezed states are defined as states in which the variance of one collective spin component is reduced below that of a CSS, at the expense of increased variance along an orthogonal direction. Consider an ensemble of $N$ two-level atoms described by collective spin operators. The Wineland spin squeezing parameter in \refcite{Wineland1992,Wineland1994} is defined as 
\begin{equation}
    \xi^{\,2}(t) \coloneqq
    \!\frac{\mrm{V}_{\!\perp\!}(t)}{\brkt{\Jindex{s}(t)}^2}
     \left(\frac{\mrm{V}_{\!\perp}^\CSS}{\brkt{\Jindex{s}}_\CSS^2} \right)^{-1} = \frac{N\;\mrm{V}_{\!\perp\!}(t)}{\brkt{\Jindex{s}(t)}^2},
    \label{eq:def_Wineland_par}
\end{equation}
where $s$ is the mean spin direction, and $\perp$ is the perpendicular direction, along which the ensemble is being squeezed. For an ideal CSS where all atoms are uncorrelated, $\xi^{\,2}(t) = 1$. Hence, when $\xi^{\,2}(t) < 1$, the state is (metrologically) spin squeezed. It is important to note that the definition of spin squeezing is not unique. Another widely used definition, introduced by Kitagawa and Ueda \cite{Kitagaba_Ueda_SSS}, is inspired by photon squeezing and focuses on the minimum variance of a spin component orthogonal to the mean spin direction:
\begin{equation}
\xi_S^{\,2} = \frac{4 \min_{\perp} (\Delta \Jindex{\perp})^2}{N},
\label{eq:def_Kitagawa_par}
\end{equation}
where just like before, $\xi_S^{\,2} < 1$ defines the spin squeezing condition. This parameter quantifies the reduction in quantum noise along a particular spin direction and is directly related to the metrological spin squeezing parameter through the following inequality:
\begin{equation}
\xi_S^{\,2} \leq \xi^{\,2}.
\label{eq:relation_S_R}
\end{equation}
In other words, metrological spin squeezing ($\xi^{\,2}<1$) implies spin squeezing in the sense of Kitagawa and Ueda ($\xi_S^{\,2} < 1$), but the converse is not necessarily true. That is, while a state with reduced spin variance ($\xi_S^{\,2} < 1$) may exhibit quantum correlations, it does not automatically guarantee an improvement in metrological sensitivity. 

\chapter{Bayesian estimation and control} \label{chap:bayesian}

\newthought{Bayesian inference} is a statistical framework in which parameters are treated as random variables~\cite{Van-Trees,crassidis2011optimal}. In this approach, probabilities quantify our uncertainty or beliefs about these parameters or hypothesis, and are systematically updated as new data becomes available~\cite{Van-Trees,crassidis2011optimal,sarkka2013bayesian}. That contrasts with the frequentist approach, where parameters are constant and probabilities are interpreted as the proportion of outcomes in a large number of repeated identical experiments~\cite{Kay1993,Van-Trees}. How to update our knowledge by incorporating new observations is described by the Bayes' theorem\cite{Bayes1763,sarkka2013bayesian}.  It combines prior information, represented by a \emph{prior} distribution, with the statistical model of the observations, given by the \emph{likelihood} distribution, to produce an updated probability known as the \emph{posterior} distribution: 
\begin{equation}
    p(\theta|y) \propto p(y|\theta)p(\theta) \;\; \longleftrightarrow \;\; \text{posterior} \propto \text{likelihood} \, \cdot \, \text{prior},
\end{equation}
where $\theta$ is the parameter, and $y$ represents the observed data.  

Bayesian inference is especially well suited for problems such as optimal filtering, where the parameters of the system or state evolve over time, requiring continuous updates as new observations are performed~\cite{sarkka2013bayesian}. Filtering involves estimating the time-varying state of a system using noisy observations when the true state is not directly observable. This estimation process is typically implemented in two stages: a prediction step, which uses a model of the dynamics of the system to propagate in time the state distribution, and an update step, which refines this estimation based on the latest observation \cite{crassidis2011optimal,sarkka2013bayesian}. This recursive process ensures that the state estimate improves as more data becomes available. The Kalman filter (KF) is a well-known example, providing an optimal solution when the system is linear and the noise is Gaussian~\cite{kalman_new_1960,kalman_new_1961,Kay1993,crassidis2011optimal,sarkka2013bayesian}. For nonlinear or non-Gaussian scenarios, extensions like the extended Kalman filter (EKF) are used, which linearize the system dynamics around the current estimate to handle mild nonlinearities\cite{simon2006,crassidis2011optimal}. In more challenging situations, which are beyond the scope of this thesis, more general methods such as particle filtering can be employed, using a set of particles to approximate the posterior distribution~\cite{sarkka2013bayesian}. 

Besides estimating parameters in real time, we are often also interested in how to optimally steer the system into a state more advantageous for metrology, e.g. more sensitive to the parameter of interest or more robust to noise. To define what we mean by optimal control, we must first specify a control cost, i.e. a function that balances two competing objectives: (1) achieving the desired control task with high accuracy, and (2) minimizing the cost or effort required by the actuator~\cite{kolosov2020optimal}. This cost function typically includes adjustable parameters, or ``knobs'', that allow us to trade off estimation performance against control effort\cite{crassidis2011optimal,kolosov2020optimal,Stockton2004}. The optimal control law is then the one that minimizes this total cost. For linear and Gaussian (LG) systems, the optimal control is given by the linear-quadratic Gaussian (LQG) controller, which combines a KF with a linear-quadratic regulator (LQR) \cite{crassidis2011optimal,kolosov2020optimal}. A complete and rigorous proof on why LQG is optimal requires knowledge of dynamical programming and the broader field of optimal control. To avoid delving into that but still provide convincing arguments, we present simpler --- albeit less rigorous --- proofs that show how both LQR and LQG minimize their respective control costs\cite{crassidis2011optimal,kolosov2020optimal}.

The results and derivations presented in this chapter are not original contributions. Rather, they are re-derivations and explanations in my own words, inspired by and adapted from well-established treatments in the literature, particularly: \citet{sarkka2013bayesian}, \citet{crassidis2011optimal}, \citet{simon2006}, and \citet{kolosov2020optimal}. Much of the material on Bayesian filtering and the KF is drawn from \citet{sarkka2013bayesian} and \citet{crassidis2011optimal}; the EKF section closely follows the treatment in \citet{simon2006}; and the discussion on control, including the LQR and LQG controller, is based primarily on \citet{crassidis2011optimal} and \citet{kolosov2020optimal}.

\newthought{This chapter} is organized as follows: we begin in \secref{sec:bayesian_statistics} with a brief overview of Bayesian inference, introducing Bayes’ rule and common estimators such as the maximum likelihood estimator, the maximum a-priori (MAP) estimator, and the minimum mean squared error (MMSE) estimator. In \secref{sec:fisher_info}, we define the Fisher information and introduce the Bayesian Cramér-Rao bound (BCRB), a lower bound on the average mean squared error (aMSE). We then move to Bayesian filtering in \secref{sec:bayesian_filtering}, where we highlight the importance of modeling both the system dynamics and the measurement process in \secref{sec:state_space_model}. These models define the probability distributions used to track the state of the system over time, and they form the foundation of the two key steps in any Bayesian filter: the prediction and measurement update, which are detailed in \secref{sec:bayesian_optimal_filtering}.

In \secref{sec:uncorr_discrete_KF}, we derive the discrete-time KF under the assumption of uncorrelated system and measurement noise. This is the simplest form of the KF and a particularly clear example of how Bayesian filtering works when the system is linear and all noise is Gaussian. It provides closed-form equations for the estimate and its covariance, which correspond to the mean and variance of the (Gaussian) posterior distributions. This eliminates the need to track full probability distributions explicitly since updating the mean and covariance is sufficient. Moreover, because the mean of a Gaussian posterior is the MMSE estimator, the KF yields the optimal estimate in this setting. 

In \secref{sec:uncorr_continuous_KF}, we extend the KF to the continuous-time setting. To do so, we had address already in \secref{sec:discretization_cont_model} how to discretize LG systems with white Gaussian noise, which is not a conventional Gaussian random variable but the formal derivative of a Wiener process. Properly discretizing this noise to avoid pathological behavior requires integrating it over a finite time step, which yields a Wiener increment. With this clear, we then take the continuous-time limit in \secref{sec:uncorr_continuous_KF} to arrive at the continuous-time KF. The same steps are later followed in \secref{sec:correlated_KF}, where we extend the KF to handle correlated process and measurement noise. This is the form necessary for continuously monitored atomic magnetometers, where correlations between the system and the measurement naturally arise due to measurement back-action. Since the systems we are interested in are often nonlinear, we derive the EKF in \secref{sec:EKF_theory} by linearizing the dynamics around the current estimate. Finally, in \secref{sec:control}, we turn to optimal control. In \secref{sec:LQR}, we study the LQR for systems where the state is directly accessible. We then combine this with state estimation using the KF to derive the LQG controller in \secref{sec:LQG_controller}, which handles noisy measurements and provides optimal feedback control.

\section{Introduction to Bayesian inference}
\subsection{Bayesian statistics} \label{sec:bayesian_statistics}

Suppose we want to use Bayesian inference to estimate some quantity of interest, $\theta$, given a measurement $y$. 
Any Bayesian method has three key steps:
\begin{enumerate}
    \item \textbf{Modelling.} Model what we know about $\theta$ even before making any measurements (using a prior $p(\theta)$), as well as how the measurements $y$ relate to $\theta$ (using a probability density $p(y|\theta)$, also known as a likelihood function). 
    \item \textbf{Measurement update.} Combine what we know before (the prior) with our measurement (i.e., with the likelihood $p(y|\theta)$) in order to get a function of $\theta$, i.e. the posterior distribution $p(\theta|y)$, which takes into account our updated knowledge on $\theta$ after performing a measurement $y$.
    \item \textbf{Decision making.} We can now combine what we know about $\theta$, i.e., the posterior $p(\theta|y)$, with a cost function, in order to perform an \emph{optimal decision}. 
\end{enumerate}

\subsection{Prior and likelihood}

As hinted at in the previous section, Bayesian methods rely on three important components: the prior, the likelihood and the posterior. For now, let us define and discuss the role of the prior and the likelihood when analyzing an unknown parameter $\theta \in \Theta$ based on some observed data $y$. The posterior, which requires introducing Bayes' rule, will be discussed in the subsequent section.

\begin{definition}[The likelihood function]
    The first key assumption in any Bayesian model is that the observed data, $y$, given that the parameter is $\theta$, follows the distribution 
\begin{equation}
    y \sim p(y|\theta),
\end{equation}
which is referred to as the likelihood. The model given by the likelihood $p(y|\theta)$ describes how we expect our observations $y$ to behave when we know that the true value of the parameter of interest is $\theta$. This distribution is often easier to compute than the probability of $\theta$ conditioned w.r.t. $y$. Since $y$ is observed, $p(y|\theta)$ can be viewed as a function of $\theta$,
\begin{equation}
    I(\theta|y) = p(y|\theta),
\end{equation}
where $I(\theta|y)$ is also called the likelihood function. Note that the likelihood is not a PDF w.r.t. $\theta$, but w.r.t. $y$. In other words, if we integrate $p(y|\theta)$ w.r.t. $\theta \in (-\infty,\infty)$, it does not yield one, unlike when integrating w.r.t. $y \in (-\infty,\infty)$.  
\end{definition}

\begin{definition} [Prior] Another fundamental component of many Bayesian methods is the prior distribution, where the term ``prior'' simply means earlier or before:
    \begin{equation}
        p(\theta).   
    \end{equation}
    This probability distribution expresses our a-priori knowledge concerning the parameter $\theta$, i.e., the knowledge held before any data $y$ is taken into account.
\end{definition}

\subsection{Bayes' rule and the posterior}

The Bayes' rule is the cornerstone of any Bayesian method because it enables us to compute the posterior $p(\theta|y)$ from the likelihood $p(y|\theta)$ and $p(\theta)$. It straightforwardly follows from \propertyref{propty:prod_rule} and \propertyref{propty:sum_rule}. Specifically, the product rule implies the following equivalence:
 \begin{equation}
     p(\theta,y) = p(\theta|y) p(y) = p(y|\theta) p(\theta).
 \end{equation}
 Then, we can simply write the posterior as:
 \begin{equation}
     p(\theta|y) = \frac{p(y|\theta) p(\theta)}{p(y)},
 \end{equation}
 where $p(\theta)$ is our initial knowledge of the parameter,  $p(y|\theta)$ is given by the model and $p(y)$ is the marginal PDF given in \propertyref{propty:sum_rule}, i.e. $p(y) = \int p(y|\theta) p(\theta) \dd \theta$.

Thus, by employing Bayes' theorem as derived above to combine the prior and likelihood, we get the \emph{posterior}, $p(\theta|y)$, one of the main goals of Bayesian statistics.

\begin{definition}[Posterior] The posterior distribution describes what we know about $\theta$ after observing $y$, and can be easily derived from the likelihood and prior using Bayes' rule:
    \begin{equation}
        p(\theta|y) = \frac{p(y|\theta) p(\theta)}{p(y)} \propto I(\theta|y) p(\theta).
    \end{equation}
   
\end{definition}

\subsection{Bayesian decision theory}

Next comes the question of how we can use the posterior, $p(\theta|y)$, to make decisions. Typically, in Bayesian statistics, the goal is to minimize some function known as the cost, $C(\theta,\alpha)$, where $\theta$ is the quantity of interest and $\alpha$ denotes the decision. Since we do not know the true value of $\theta$, we cannot directly minimize the actual cost $C(\theta,\alpha)$. Instead, we consider the expected cost under our current beliefs about $\theta$, which are encoded in the posterior distribution $p(\theta|y)$. 
Therefore, an optimal decision $\alpha$ is one that minimizes the average cost we would expect to incur, weighted by the probabilities given by the posterior:
\begin{equation} \label{eq:posterior_exp_cost}
    \alpha_{\mrm{opt}} = \text{arg} \, \underset{\alpha}{\text{min}} \;  \E{C(\theta,\alpha)}{p(\theta|y)}  = \text{arg} \, \underset{\alpha}{\text{min}} \; \int_\Theta C(\theta,\alpha) p(\theta|y) \dd \theta.
\end{equation}
where the expectation is taken w.r.t. to the posterior $p(\theta|y)$, for which $y$ is given and $\theta \in \Theta$ is a random variable. Note that this process follows all the steps outlined in \secref{sec:bayesian_statistics}.

\subsubsection{Cost functions for estimation problems} \label{sec:quadratic_cost}

Let us now introduce two commonly used cost functions for estimation problems and then derive their respective estimators. In the context of estimation, choosing a decision $\alpha$ corresponds to selecting a specific value as our best guess for the unknown parameter $\theta$. Therefore, the decision $\alpha$ is simply our estimator of $\theta$, denoted from now on as $\esttheta$. Importantly, the estimator $\esttheta$ is a function of the observed measurements $y$, reflecting how our estimate depends on the data collected. However, not every choice of estimator $\esttheta(y)$ is guaranteed to be optimal. When an estimator is optimal with respect to a chosen cost function, we will denote it with a subscript, such as $\esttheta_{\mrm{opt}}(y)$ for a general optimal estimator, or $\esttheta_{\mrm{MMSE}}(y)$ specifically when it is the optimal estimator minimizing the mean squared error (MSE). 

\paragraph*{Quadratic cost and the MMSE estimate} 
One of the most common cost functions in estimation theory is the quadratic cost or the \emph{squared error}:
\begin{equation}
    C(\theta,\esttheta\,) = (\theta - \esttheta(y))^\Trans (\theta - \esttheta(y)).
\end{equation}

Its corresponding optimal estimator is known as the minimum mean squared error (MMSE) estimator. To derive it, we need to minimize the posterior expected cost as described in \eqnref{eq:posterior_exp_cost} for the case of a quadratic cost:
\begin{align}
    \esttheta_{\mrm{MMSE}}(y) &= \text{arg} \, \underset{\esttheta}{\text{min}} \;  \E{C(\theta,\esttheta\,)}{p(\theta|y)} = \text{arg} \, \underset{\esttheta}{\text{min}} \;  \E{(\theta - \esttheta(y))^\Trans(\theta - \esttheta(y) )}{p(\theta|y)} \\
    &= \text{arg} \, \underset{\esttheta}{\text{min}} \;  \int_\Theta (\theta - \esttheta(y))^\Trans(\theta - \esttheta(y)) \, p(\theta|y) \; \dd \theta.
\end{align}
Note that the quantity we are minimizing, i.e. the posterior expected quadratic cost, is precisely the \emph{mean squared error} (MSE), a term likely more familiar to the reader. We can formally define the MSE as follows:
\begin{definition}[Mean squared error] The MSE of an estimator $\esttheta(y)$ for a parameter $\theta \in \Theta$, given observations $y$, is defined as
    \begin{equation}
        \E{(\theta - \esttheta(y) \,)^\Trans(\theta - \esttheta(y) \,)}{p(\theta|y)} =  \int_\Theta (\theta - \esttheta(y))^\Trans(\theta - \esttheta(y)) \, p(\theta|y) \; \dd \theta
    \end{equation}
    Here, the ``mean'' (or average) is taken with respect to the posterior distribution, $p(\theta|y)$. 
\end{definition}
Later, we will discuss the \emph{average mean squared error} (aMSE), which refers to the squared error or quadratic cost averaged with respect to the joint PDF of the parameter and the data.

\begin{figure}
    \centering
    \begin{tikzpicture}
    
    \begin{axis}[
        no markers,
        domain=0:8,
        samples=500,
        axis lines*=middle,
        xlabel={$\theta$},
        ylabel={PDF},
        xlabel style={at={(0.95,-0.05)}, anchor=west}, 
        ylabel style={at={(-0.01,0.9)}, anchor=south}, 
        height=8cm,
        width=12cm,
        xtick=\empty,
        ytick=\empty,
        enlargelimits=false,
        clip=false,
        axis on top,
        grid = major,
        legend pos=north east,
        legend cell align={left},
        legend style={draw=none},
        axis line style={-Stealth}
    ]
    
    \addplot [blue, ultra thick] {1.1*exp(-2 * (x-2)^2)/1.88 + 0.3*exp(-3 * (x-5)^2)/1.88};
    \addlegendentry{Likelihood}
    
    \addplot [orange, ultra thick] {exp(-3* (x-3.5)^2)/1.52465 + 0.2*exp(-0.5 * (x-6)^2)/1.52465};
    \addlegendentry{Posterior}
    
    \draw[dashed, blue] (axis cs:2,0) -- (axis cs:2,1.1/1.88) node[above, pos=1] {};
    \fill[blue] (axis cs:2,1.1/1.88) circle (3pt);
    \node at (axis cs:2,0) [below] {$\esttheta_{\mrm{ML}}$};
    
    \draw[dashed, orange] (axis cs:3.5,0) -- (axis cs:3.5,1/1.52465) node[above, pos=1] {};
    \fill[orange] (axis cs:3.5,1/1.52465) circle (3pt);
    \node at (axis cs:3.5,0) [below] {$\esttheta_{\mrm{MAP}}$};
    
    \draw[dashed, black] (axis cs:4.32203,0) -- (axis cs:4.32203,0.11848) node[above, pos=1] {};
    \fill[black] (axis cs:4.32203,0.11848) circle (3pt);
    \node at (axis cs:4.32203,0) [below] {$\esttheta_{\mrm{MMSE}}$};
    
    \end{axis}
    
    \end{tikzpicture}
    \caption[Comparison of the ML, MAP and MMSE estimates]{\textbf{Comparison of the ML, MAP and MMSE estimates.} Visual representation of the maximum likelihood estimator ($\esttheta_{\mrm{ML}}(y)$, blue dot), the maximum a-priori estimator ($\esttheta_{\mrm{MAP}}(y)$, yellow dot) and the minimum mean squared error estimator ($\esttheta_{\mrm{MMSE}}(y)$, black dot).}
    \label{fig:visualization_of_different_estimators}
\end{figure}
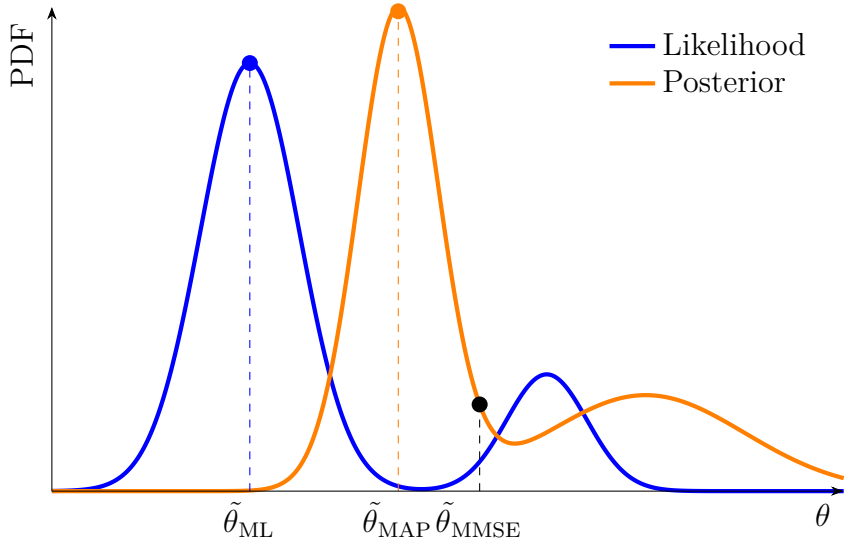

The optimal estimatior that solves the above minimization problem can be shown to be the mean of the posterior, which is therefore referred to as the MMSE estimate. Namely,
\begin{equation} \label{eq:MMSE_estimator}
    \esttheta_{\mrm{MMSE}}(y) = \text{arg} \, \underset{\esttheta}{\text{min}} \;  \E{(\theta - \esttheta(y) )^\Trans(\theta - \esttheta(y) )}{p(\theta|y)} = \bartheta,
\end{equation}
where $\bartheta$ denotes the mean of $\theta$ with respect to the posterior:
\begin{equation} \label{eq:mean_of_the_posterior}
    \bartheta = \int_\Theta \theta \; p(\theta|y) \dd \theta.
\end{equation}
To visualize the MMSE estimator, see \figref{fig:visualization_of_different_estimators}, where it is marked as a black dot.
\begin{myproof}
    Let us show that the optimal estimator of $\theta$ is the mean of the posterior. To do so, let us first rewrite the cost function $C(\theta,\esttheta\,)$ as follows:
    \begin{align}
        C(\theta,\esttheta\,) = (\theta - \esttheta(y) )^\Trans(\theta - \esttheta(y) ) = (\theta - \bartheta + \bartheta - \esttheta(y))^\Trans(\underbrace{\theta -\bartheta}_{\text{zero mean}} + \! \underbrace{\bartheta - \esttheta(y)}_{\text{deterministic}}),
    \end{align}
    where we have added and subtracted the mean of the posterior $\bartheta$. That splits the factor $\theta - \esttheta(y)$ into two terms, one still probabilistic but with zero mean, $\theta - \bartheta$, and another completely deterministic, $\bartheta - \esttheta(y)$. If we now expand the quadratic term and take the expectation value w.r.t. to the posterior $p(\theta|y)$, we obtain:
    \begin{align}
        \E{C(\theta,\esttheta\,)}{p(\theta|y)} &= \E{(\theta - \bartheta)^\Trans(\theta - \bartheta)}{p(\theta|y)} + \overbrace{\E{(\theta - \bartheta)^\Trans}{p(\theta|y)}}^{0}(\bartheta - \esttheta(y)) \nonumber \\
        &+ (\bartheta - \esttheta(y))^\Trans \underbrace{\E{(\theta - \bartheta)}{p(\theta|y)}}_{0} + \E{(\bartheta - \esttheta(y))^\Trans(\bartheta - \esttheta(y))}{p(\theta|y)} \nonumber \\
        &= \E{(\theta - \bartheta)^\Trans(\theta - \bartheta)}{p(\theta|y)} + (\bartheta - \esttheta(y))^\Trans(\bartheta - \esttheta(y)). 
    \end{align}
    Thus, it follows that the argument of $\esttheta(y)$ that minimizes the quantity above can only be $\bartheta$. Therefore, the optimal estimator that minimizes the expectation of the squared error w.r.t. the posterior, $p(\theta|y)$, is the mean of the posterior, i.e. the MMSE estimator given in \eqnref{eq:MMSE_estimator}. 
\end{myproof}

\paragraph*{The 0-1 cost and the MAP estimator}

Let us now consider the following cost function:
\begin{equation}
    C(\theta,\esttheta\,) = -\delta(\theta - \esttheta(y)\,),
\end{equation}
where $\delta(\,\cdot\,)$ represents the Dirac delta. This cost function, also referred to as the 0-1 loss function, picks only the exact value of $\theta$ and disregards all the other values of $\esttheta(y)$. It is an idealized cost function, since it penalizes any deviation from the true value equally and maximally. 

To find the optimal estimator for this cost function, we compute the expected value of $C(\theta,\esttheta\,)$ w.r.t. the posterior:
\begin{align}
    \E{C(\theta,\esttheta\,)}{p(\theta|y)} = -\int_{\theta\in\Real^n} \delta(\theta-\esttheta(y)\,) p(\theta|y) \dd \theta = -p(\theta|y) |_{\theta = \esttheta},
\end{align}
where in the last step we have used the Dirac delta property: $\int_{-\infty}^\infty f(x) \delta(x-\esttheta(y)) \dd x = f(\esttheta(y))$. If now we minimize over all possible arguments of $\esttheta(y)$, we get the optimal estimator $\esttheta(y)$ :
\begin{equation}
    \esttheta_{\mrm{MAP}}(y) = \text{arg} \; \underset{\esttheta}{\text{min}} \left[-p(\theta|y) |_{\theta = \esttheta}\right] = \text{arg} \; \underset{\theta}{\text{max}} \; p(\theta|y). 
\end{equation}
As shown by a yellow dot in \figref{fig:visualization_of_different_estimators}, the optimal estimator for the 0-1 cost is the $\theta$ that maximizes the posterior distribution, which we call the maximum a-posteriori (MAP) estimator.

While the Bayesian approach allows for the incorporation of prior information through $p(\theta)$, sometimes prior knowledge is not available or we ignore it altogether. In that case, the goal is to directly maximize the likelihood function $p(y|\theta)$. This leads to the maximum likelihood (ML) estimator:
\begin{equation}
    \esttheta_{\mrm{ML}}(y) = \text{arg} \; \underset{\theta}{\text{max}} \; p(y|\theta),
\end{equation}
which is marked in \figref{fig:visualization_of_different_estimators} with a blue dot, i.e. the maximum of the likelihood. The ML estimator can be seen as a method that focuses entirely on the data, with no consideration of prior beliefs. 

\section{Fisher information and estimation theory} \label{sec:fisher_info}

Having established the foundations of classical Bayesian estimation, we now turn our attention to a quantity crucial for assessing the quality of an estimator: the Fisher information (FI). Besides quantifying the amount of information that a dataset contains about an unknown parameter, the FI also lower bounds the variance of unbiased estimators, serving as an estimation benchmark. This bound, referred to as Cram\'{e}r-Rao bound (CRB), can be extended to a Bayesian setting by incorporating prior information. Known as the Bayesian Cram\'{e}r-Rao bound (BCRB), it establishes a lower bound on the average mean squared error (aMSE) or average quadratic cost of all (biased or unbiased) estimators, given a well-behaved prior. 

\subsection{Fisher information} \label{sec:Fisher_info}

An essential question in parameter estimation is: how much information about the unknown parameter $\theta$ is contained in the data $y$? The FI provides an answer by quantifying the sensitivity of the likelihood function $p(y|\theta)$ with respect to changes in $\theta$. Intuitively, if the likelihood varies strongly with $\theta$, then the data $y$ carries a lot of information about $\theta$, and accurate estimation is possible. Conversely, if $p(y|\theta)$ is nearly flat in $\theta$, then estimation will be fundamentally imprecise. 

\begin{definition}[Fisher information] \label{def:FI}
Let $y$ be a random variable representing an observation drawn from the likelihood distribution $p(y|\theta)$ depending on an unknown parameter $\theta$. The Fisher information (FI) associated with $p(y|\theta)$ is defined as~\cite{Fisher1922,Van-Trees}
\begin{align}
    \Fisher[p(y|\theta)] &= \int \dd y \frac{1}{p(y|\theta)} \left(\frac{\partial p(y|\theta)}{\partial \theta }\right)^2  \label{eq:FI_def1} \\
    &= \E{\left(\frac{\partial}{\partial \theta} \ln p(y|\theta)\right)^2}{p(y|\theta)}  \label{eq:FI_def2} \\
    &= - \E{\frac{\partial^2}{\partial \theta^2} \ln p(y|\theta)}{p(y|\theta)},  \label{eq:FI_def3}
\end{align}
where the equivalence between \eqref{eq:FI_def1} and \eqref{eq:FI_def3} holds under regularity conditions ensuring interchangeability of differentiation and integration.
\end{definition}

\subsection{Cram\'er-Rao bound}

Knowing how much information the data carries about $\theta$ is useful, but the natural follow-up question is: what does this imply for the achievable precision of an estimator?
The CRB provides such a link by establishing a fundamental lower bound on the MSE of any locally unbiased estimator. It formalizes the intuition that no matter how cleverly we process the data, we cannot achieve an accuracy beyond the inverse of the Fisher information.

\begin{prop}[Cram\'{e}r-Rao bound] \label{prop:CRB}
Let $y \sim p(y|\theta)$ be observations drawn from a regular likelihood, i.e.
\begin{align}
    \int \dd y \frac{\partial}{\partial \theta} \, p(y|\theta) = 0,
\end{align}
and let $\esttheta(y)$ be an estimator of $\theta$ that is locally unbiased, i.e.
\begin{align}
    \int \esttheta(y) \frac{\partial}{\partial\theta}\, p(y|\theta) \dd y = 1. 
\end{align}
Then the MSE of $\esttheta$ satisfies the Cram\'{e}r-Rao inequality
\begin{align}
    \E{\Delta^{\!2} \esttheta\,}{p(y|\theta)} \geq \frac{1}{\Fisher\,[p(y|\theta)]},
\end{align}
where $\Fisher\,[p(y|\theta)]$ denotes the Fisher information given in \eqnsref{eq:FI_def1}{eq:FI_def3} and $\Delta^{\!2} \esttheta$ is the squared error:
\begin{equation}
    \Delta^{\!2} \esttheta \coloneqq (\theta - \esttheta(y) )^2.
\end{equation}
\end{prop}

\subsection{Bayesian Cram\'er-Rao bound} \label{sec:BCRB_section}

In the Bayesian framework, the parameter $\theta$ is treated as a random variable with a prior distribution $p(\theta)$, specified before observing any data. Therefore, an optimal estimator in this setting should not only minimize the MSE with respect to the posterior, but by its average performance across all possible values of $\theta$, weighted according to the prior distribution $p(\theta)$. To reflect this, we must define the average mean squared error (aMSE):
\begin{definition}[Average mean squared error] Let $\theta \in \Theta$ be a parameter, $\esttheta$ its estimate and $y$ the observed data with a joint PDF $p(\theta,y)$. The aMSE of the estimator $\esttheta$ is defined as
    \begin{align}
        \EE{\Delta^{\!2} \esttheta \,} \coloneqq \E{(\theta - \esttheta(y))^2}{p(\theta, y)} &= \int_Y \int_\Theta (\theta - \esttheta(y))^2 \, p(\theta, y) \; \dd \theta \, \dd y \\ 
        &= \int_\Theta \dd \theta \; p(\theta) \int_Y \dd y \; p(y|\theta)  (\theta - \esttheta(y))^2 .
    \end{align} \label{def:aMSE_def}
\end{definition}
The aMSE can be lower bounded by different classes of Bayesian bounds, which are equivalent when working in the LG regime \cite{Fritsche2014}. In this thesis, we focus on the marginal unconditional BCRB \cite{bobrovsky1987,Fritsche2014}.

\begin{prop}[Marginal unconditional Bayesian Cram\'er-Rao bound]
    The aMSE of any estimator $\esttheta$ is bounded as
    \begin{align} \label{eq:BCRB_ch1}
        \EE{\Delta^{\!2} \esttheta \,} \geq \frac{1}{J_B},
    \end{align}
    where $J_B$ is the Bayesian information (BI)~\cite{Van-Trees},
    \begin{align} \label{eq:BI}
        J_B = \E{\left(\partial_{\theta} \log p(\theta,y) \right)^2}{p(\theta,y)}.
    \end{align}
\end{prop}

The BI can be split into two terms, $\,J_B = J_P + J_M$. The first term, $J_P$, represents the contribution of our prior knowledge about $\theta$, 
\begin{align} \label{eq:ap_J_P}
    J_P = \Fisher[p(\theta)] = \E{\left(\partial_{\theta} \log p(\theta) \right)^2}{p(\theta)}.
\end{align}
The second term, namely the contribution of the measurement records, or $J_M$, can be understood as averaging the FI of the likelihood over the prior distribution, i.e., 
\begin{align} \label{eq:ap_J_M}
    J_M &= \E{\left( \partial_{\theta} \log p(y,\theta) \right)^2}{p(\theta,y)} = \\
    &= \int d\theta \; p(\theta) \, \Fisher[p(y|\theta)],
\end{align}
with
\begin{equation} \label{eq:ap_Fisher_pyB}
    \Fisher[p(y|\theta)]=\E{\left( \partial_{\theta} \log p(y|\theta) \right)^2}{p(y|\theta)}
\end{equation}
being the FI of the likelihood $p(y|\theta)$ of observing a measurement trajectory $y$ given that the parameter to estimate has a value $\theta$, as introduced in \eqnsref{eq:FI_def1}{eq:FI_def3}.

\section{Bayesian filtering} \label{sec:bayesian_filtering}

Bayesian filtering provides a recursive framework for estimating the state parameters of a system based on indirect, noisy measurements over time. To formulate this problem, we begin by discretizing time s.t. $t = k \Dt$, where $\Dt$ is the time step and $k\in\mathbb{N}$. With this discretization, the evolution of parameters and observations can be written as follows:
\begin{itemize}
    \item $\pmb{x}_k$: the state vector at time $k$, representing the unknown parameters we aim to estimate.
    \item $\pmb{y}_k$: the measurement vector at time $k$. 
\end{itemize}
All the state parameters and measurements \emph{up to} time $k$ can be collected into:
\begin{itemize}
    \item $\pmb{x}_{0:k} = \{\pmb{x}_0,\pmb{x}_1,\dots,\pmb{x}_k\}$: the state trajectory up to time $k$,
    \item $\pmb{y}_{0:k} = \{\pmb{y}_0,\pmb{y}_1,\dots,\pmb{y}_k\}$: the measurement trajectory up to time $k$.
\end{itemize}

\subsection{State space model} \label{sec:state_space_model}

An essential step to formulate any filtering problem is to model the system and measurement by creating a state space model. These models describe how the state vector evolves recursively, as well as how it relates to the observations. In particular, for a state vector $\pmb{x}_k$ and a measurement vector $\pmb{y}_k$, where $k$ denotes time, their respective dynamics can be modeled with either equations or probability distributions:
\begin{align}
    &\pmb{x}_k = \pmb{f}_{k\shortminus1}[\pmb{x}_{k\shortminus1},\pmb{u}_{k\shortminus1},\pmb{q}_{k\shortminus1}] &&\hspace{-50pt}\Longleftrightarrow&&\hspace{-50pt} p(\pmb{x}_{k}|\pmb{x}_{0:k\shortminus1},\pmb{y}_{0:k};\pmb{u}_{0:k-1})\label{eq:system_model} \\
    &\pmb{y}_k = \pmb{h}_k[\pmb{x}_k,\pmb{r}_k] &&\hspace{-50pt}\Longleftrightarrow&&\hspace{-50pt} p(\pmb{y}_k|\pmb{x}_{0:k},\pmb{y}_{0:k}) \label{eq:meas_model}
\end{align}
where $\pmb{u}_k$ is an external known control vector. Furthermore, we assume the state noise $\pmb{q}_k$ and the measurement noise $\pmb{r}_k$ to be zero-mean white noise vectors, with known probability distributions which are not necessarily Gaussian (see \secref{sec:white_noise}).

The initial value for the state vector is drawn from a prior distribution $\pmb{x}_0 \sim p(\pmb{x}_0)$. The first equation describes how the system evolves in time and updates the state vector from time $k\!\shortminus\!1$ to time $k$. The measurement model given by \eqnref{eq:meas_model} relates the observation vector to the state vector and the measurement noise.

Thus far, we have made no assumptions regarding the noise, besides it being white. In what follows, we assume that the process and measurement noise are uncorrelated. This is an essential condition to ensure that the state follows a Markov evolution and that the measurements are conditionally independent of past measurements and states. We will later revisit the case of correlated noise and discuss how to de-correlate it when necessary.
 
\subsubsection{Uncorrelated process and measurement noise} \label{sec:uncorr_process_and_meas}

From the system and measurement model in \eqnsref{eq:system_model}{eq:meas_model}, it follows that the state vector $\pmb{x}_k$ depends solely on the previous state $\pmb{x}_{k\shortminus1}$ and control input $\pmb{u}_{k\shortminus1}$, while the measurement $\pmb{y}_k$ depends only on the current state $\pmb{x}_k$. Assuming that the process noise $\pmb{q}_{k\shortminus1}$ and measurement noise $\pmb{r}_k$ are white (i.e., temporally uncorrelated), as well as mutually uncorrelated, there is no mechanism through which they can introduce dependencies between $\pmb{x}_k$ and earlier states $\pmb{x}_{0:k\shortminus2}$ or past measurements $\pmb{y}_{0:k\shortminus1}$. Thus, the probability distributions of \eqnsref{eq:system_model}{eq:meas_model} can be simplified to
\begin{align}
    &\pmb{x}_k = \pmb{f}_{k\shortminus1}[\pmb{x}_{k\shortminus1},\pmb{u}_{k\shortminus1},\pmb{q}_{k\shortminus1}] &&\hspace{-70pt}\Longleftrightarrow&&\hspace{-90pt} p(\pmb{x}_{k}|\pmb{x}_{k\shortminus1};\pmb{u}_{k\shortminus1}) \label{eq:system_eqv} \\
    &\pmb{y}_k = \pmb{h}_k[\pmb{x}_k,\pmb{r}_k] &&\hspace{-70pt}\Longleftrightarrow&&\hspace{-90pt} p(\pmb{y}_k|\pmb{x}_k) \label{eq:meas_eqv}\\
    &\text{when} \;\;\; \cov\left[\pmb{q}_k, \,\pmb{r}_s\right] = 0 \;\; \forall \; k,s.
\end{align}
Note that \eqnref{eq:system_eqv} and \eqnref{eq:meas_eqv} fulfill two very important properties: the Markovianity of states and the conditional independence of measurements. In particular, the system function $\pmb{f}_{k\shortminus1}$ in \eqnref{eq:system_model} depends only on $\pmb{x}_{k\shortminus1}$ and $\pmb{u}_{k\shortminus1}$, and not on any measurement $\pmb{y}_{0:k}$ (not even indirectly since the system and measurement noise are uncorrelated). Hence, the transition probability $p(\pmb{x}_{k}|\pmb{x}_{0:k\shortminus1},\pmb{y}_{0:k};\pmb{u}_{0:k\shortminus1})$ in \eqnref{eq:system_model} is conditionally dependent only on $\pmb{x}_{k\shortminus1}$ and $\pmb{u}_{k\shortminus1}$, i.e.  $p(\pmb{x}_k|\pmb{x}_{k\shortminus1};\pmb{u}_{k\shortminus1})$, which means that \eqnref{eq:system_eqv} obeys the Markov property defined in \propertyref{propty:markov}.

Additionally, the likelihood in \eqnref{eq:meas_eqv} is conditionally dependent only on $\pmb{x}_k$, and independent on other measurements and state vectors. Therefore, it fulfills the following property:

\begin{property}[Conditional independence of measurements] \label{propty:cond_indep} 
    The current measurement $\pmb{y}_k$ given the current state $\pmb{x}_k$ is conditionally independent of the measurement and state histories:
    \begin{equation}
        p(\pmb{y}_k|\pmb{x}_{0:k},\pmb{y}_{0:k\shortminus1}) = p(\pmb{y}_k|\pmb{x}_k).
    \end{equation}
\end{property}

Moving forward we will drop the dependence of the transition probability on the control law $\pmb{u}_{k\shortminus1}$ for brevity.

\subsection{Bayesian optimal filtering} \label{sec:bayesian_optimal_filtering}

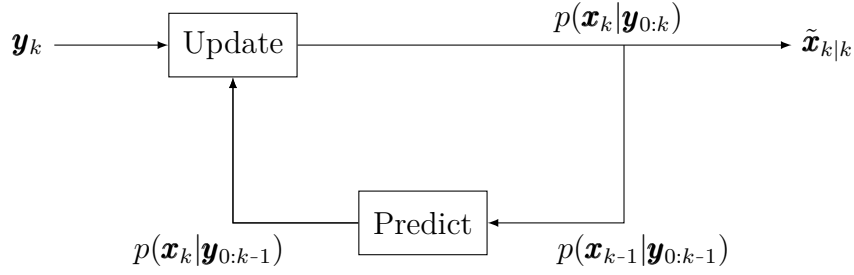
\begin{figure}[htbp]
\begin{center}
    \begin{tikzpicture}[auto, node distance=2cm,>=latex]
    \tikzstyle{block} = [rectangle, draw, fill=white, 
    text centered, minimum height=2em, minimum width=4em]
    \tikzstyle{input} = [coordinate]
    
    \node [name=y_k] (y_k) {$\pmb{y}_k$};
    \node [block, right=1.5cm of y_k] (update) {Update};
    \node [block, below=1.5cm of update, xshift = 2.5cm] (predict) {Predict};
    \node [coordinate, right=6.5cm of update] (output) {};

    \coordinate (midpoint) at ($(update)!0.7!(output)$);

    \draw [->] (y_k) -- (update);
    \draw [->] (update) -- node [above, pos = 0.65] {$p(\pmb{x}_k | \pmb{y}_{0:k})$} node [right, pos = 1] {$\est{\pmb{x}}_{k|k}$} (output);

    \draw [->] (predict) -| node [pos=0.25, anchor=north east] {$p(\pmb{x}_k | \pmb{y}_{0:k\shortminus1})$} (update);
    \draw [->] (predict) -| node [pos=-0.75, anchor=north west] {$p(\pmb{x}_{k\shortminus1} | \pmb{y}_{0:k\shortminus1})$} (update);

    \draw [->] (midpoint) |- (predict);
\end{tikzpicture}
\end{center}
\caption[A scheme illustrating the recursive algorithm of a Bayesian filter]{\textbf{A scheme illustrating the recursive algorithm of a Bayesian filter.} In the ``Predict'' box, a model is used to propagate the \emph{update PDF} from step $k\!\shortminus\!1$ to the \emph{prediction PDF} $p(\pmb{x}_k| \pmb{y}_{0:k\shortminus1})$. Then, in the ``Update'' box, a new measurement $\pmb{y}_k$ is incorporated into the algorithm to update the \emph{prediction PDF} into a new \emph{update PDF} for time $k$: $p(\pmb{x}_k|\pmb{y}_{0:k})$. Finally, before repeating this process again, an estimate at time $k$ is computed from the just-updated posterior: $\est{\pmb{x}}_{k|k}$.}
\label{fig:recursive_Bayes_filter}
\end{figure}

The goal of Bayesian optimal filtering is to compute the posterior density of the state $\pmb{x}_k$ given the history of measurements up to time $k$, i.e. $p(\pmb{x}_k|\pmb{y}_{0:k})$. To do so, we can either use a brute force approach where we find the posterior by simply applying Bayes' rule and then marginalizing w.r.t. $\pmb{x}_{0:k\shortminus1}$, or use a more efficient recursive approach whose complexity does not grow with $k$. In particular, we can recursively compute the posterior density at time $k$, i.e. $p(\pmb{x}_{k}|\pmb{y}_{0:k})$, from the posterior density at the previous time, $p(\pmb{x}_{k\shortminus1}|\pmb{y}_{0:k\shortminus1})$, as summarized in \figref{fig:recursive_Bayes_filter}. First, we assume that in the previous step, $k\!\shortminus\!1$, we have computed $p(\pmb{x}_{k\shortminus1}| \pmb{y}_{0:k\shortminus1})$. Next, we take this density, which summarizes our knowledge of the state $\pmb{x}_{k\shortminus1}$ given the measurement trajectory $\pmb{y}_{0:k\shortminus1}$, and use it to predict the state at time $k$ given the measurement trajectory up to time $k\!\shortminus\!1$. Namely, we use the system or process model to predict the probability density $p(\pmb{x}_k | \pmb{y}_{0:k\shortminus1})$. Then, we update this probability distribution using the measurement model, which gives us the current measurement $\pmb{y}_k$, and allows us to derive the posterior for the state $\pmb{x}_k$: $p(\pmb{x}_k|\pmb{y}_{0:k})$. From this posterior, we can then compute the relevant quantities of interest, such as the estimator of $\pmb{x}_k$ that minimizes the quadratic cost or MSE.

Let us now write a detailed step-by-step guide on how to recursively update the Bayesian filter:
\begin{enumerate}[label=(\roman*)]
    \item \textbf{Initialize.} The first step of the recursion is the prior $p(\pmb{x}_0)$.
    \item \textbf{Predict.} Compute $p(\pmb{x}_k|\pmb{y}_{0:k\shortminus1})$ from $p(\pmb{x}_{k\shortminus1}| \pmb{y}_{0:k\shortminus1})$. Namely,
    \begin{align}
        p(\pmb{x}_k| \pmb{y}_{0:k\shortminus1}) &\underset{\ref{propty:sum_rule}}{=} \int p(\pmb{x}_k,\pmb{x}_{k\shortminus1}| \pmb{y}_{0:k\shortminus1}) \dd \pmb{x}_{k\shortminus1} \label{eq:marginal_of_xk_from_xk,xk-1} \\
        &\underset{\ref{propty:prod_rule}}{=} \int p(\pmb{x}_k| \pmb{x}_{k\shortminus1}, \pmb{y}_{0:k\shortminus1}) p(\pmb{x}_{k\shortminus1}|\pmb{y}_{0:k\shortminus1}) \dd \pmb{x}_{k\shortminus1}. 
    \end{align}
    However, if now we recall \propertyref{propty:markov}, the probability density $p(\pmb{x}_k| \pmb{x}_{k\shortminus1}, \pmb{y}_{0:k\shortminus1})$ can be simplified to $p(\pmb{x}_k|\pmb{x}_{k\shortminus1})$ because the current state $\pmb{x}_k$ is conditionally independent of the measurements $\pmb{y}_{0:k\shortminus1}$. Thus, we get
    \begin{equation} \label{eq:original_prediction_PDF}
         \text{prediction : } p(\pmb{x}_k| \pmb{y}_{0:k\shortminus1}) = \int p(\pmb{x}_k|\pmb{x}_{k\shortminus1}) p(\pmb{x}_{k\shortminus1}|\pmb{y}_{0:k\shortminus1}) \dd \pmb{x}_{k\shortminus1},
    \end{equation}
    which is also known as the Chapman-Kolmogorov equation, introduced in \defref{def:Chapman_Kolmogorov_def}.
    \item \textbf{Update.} Compute $p(\pmb{x}_k|\pmb{y}_{0:k})$ from $p(\pmb{x}_k|\pmb{y}_{0:k\shortminus1})$. To do so, we want to update our knowledge about $\pmb{x}_k$ using the new measurement $\pmb{y}_k$:
    \begin{align}
        p(\pmb{x}_k|\pmb{y}_{0:k}) = p(\pmb{x}_k|\pmb{y}_k,\pmb{y}_{0:k\shortminus1}) = \frac{p(\pmb{y}_k|\pmb{x}_k,\pmb{y}_{0:k\shortminus1}) p(\pmb{x}_k|\pmb{y}_{0:k\shortminus1})}{p(\pmb{y}_k|\pmb{y}_{0:k\shortminus1})}.
    \end{align}
    This equation can be further simplified if we recall that our measurement model fulfills \propertyref{propty:cond_indep}: the likelihood of measuring $\pmb{y}_k$ is given only by $\pmb{x}_k$ and independent of previous measurements. Namely,
    \begin{equation}
        p(\pmb{y}_k|\pmb{x}_k,\pmb{y}_{0:k\shortminus1}) = p(\pmb{y}_k|\pmb{x}_k).
    \end{equation}
    Therefore,
    \begin{equation} \label{eq:original_update_PDF}
        \text{update : } p(\pmb{x}_k|\pmb{y}_{0:k}) = \frac{1}{p(\pmb{y}_k|\pmb{y}_{0:k\shortminus1})} p(\pmb{y}_k|\pmb{x}_k) p(\pmb{x}_k|\pmb{y}_{0:k\shortminus1}),
    \end{equation}
    where $p(\pmb{x}_k|\pmb{y}_{0:k\shortminus1})$ is the prediction PDF, and $p(\pmb{y}_k|\pmb{y}_{0:k\shortminus1})$ is the likelihood of detecting $\pmb{y}_k$ given a past measurement trajectory $\pmb{y}_{0:k\shortminus1}$:
    \begin{equation}
        p(\pmb{y}_k|\pmb{y}_{0:k\shortminus1}) = \int \dd \pmb{x}_k \, p(\pmb{y}_k,\pmb{x}_k|\pmb{y}_{0:k\shortminus1}) = \int \dd \pmb{x}_k \, p(\pmb{y}_k|\pmb{x}_k) p(\pmb{x}_k|\pmb{y}_{0:k\shortminus1}).
    \end{equation}
\end{enumerate}

Note that these expressions are general and provide a recursive solution to any filtering problem. In the case of LG systems, these equations can be solved exactly. When the system is nonlinear and/or non-Gaussian, good approximations can also be found.

\section{The discrete uncorrelated Kalman filter} \label{sec:uncorr_discrete_KF}

The Kalman filter (KF) is one of the few filters that, under certain modeling assumptions, yields an exact solution. Additionally, it is also the foundation of more advanced filters that provide approximate solutions to the Bayesian filtering equations, enabling them to handle more general models.

The starting assumption to find a close form solution to the Bayesian filtering equations is that the system and measurement models are both LG. Namely,
\begin{align}
    \pmb{x}_k &= \pmb{A}_{k\shortminus1} \, \pmb{x}_{k\shortminus1} +  \pmb{B}_{k\shortminus1} \, \pmb{u}_{k\shortminus1} + \pmb{G}_{k\shortminus1} \, \pmb{q}_{k\shortminus1}, \label{eq:system_L&G} \\
    \pmb{y}_k &= \pmb{H}_k \, \pmb{x}_k + \pmb{r}_k, \label{eq:meas_L&G}
\end{align}
where $\pmb{x}_k$ is the state vector, $\pmb{y}_k$ the measurement. Assuming the model to be Gaussian implies that the process noise $\pmb{q}_{k\shortminus1} \sim \Gauss(0,\pmb{Q}_{k\shortminus1})$ and the measurement noise $\pmb{r}_k \sim \Gauss(0,\pmb{R}_k)$ are both mutually uncorrelated white Gaussian noises with zero mean and covariances $\pmb{Q}_{k\shortminus1}$ and $\pmb{R}_k$, respectively.  The dependence of the model and the measurement on the state is clearly linear, and it is guided by a transition matrix $\pmb{A}_{k\shortminus1}$ and by a measurement model matrix $\pmb{H}_k$, respectively. Then, the model described by \eqnsref{eq:system_L&G}{eq:meas_L&G} can be equivalently written in probabilistic terms as
\begin{align} \label{eq:L&G_model_syst}
    &\pmb{x}_k \sim p(\pmb{x}_k|\pmb{x}_{k\shortminus1}) = \Gauss(\pmb{x}_k| \pmb{A}_{k\shortminus1} \pmb{x}_{k\shortminus1}+\pmb{B}_{k\shortminus1} \pmb{u}_{k\shortminus1}, \pmb{G}_{k\shortminus1}\pmb{Q}_{k\shortminus1}\pmb{G}_{k\shortminus1}^\Trans), \\
    &\pmb{y}_k \sim p(\pmb{y}_k|\pmb{x}_k) = \Gauss(\pmb{y}_k|\pmb{H}_k\,\pmb{x}_k,\pmb{R}_k), \label{eq:L&G_model_meas}
\end{align}
where the notation $\Gauss(\pmb{x}|\,\pmb{\mu},\pmb{\Sigma})$ simply denotes a multivariate Gaussian PDF of a state $\pmb{x}$ with mean $\pmb{\mu}$ and covariance $\pmb{\Sigma}$. 

The KF recursively computes the prediction and update probability densities of \eqnref{eq:original_prediction_PDF} and \eqnref{eq:original_update_PDF}, which now, given that our model is LG, will also be Gaussian:
\begin{align}
    \text{prediction} \hspace{-50pt}&&\Longrightarrow&&\hspace{-50pt} p(\pmb{x}_k|\pmb{y}_{0:k\shortminus1}) &\coloneqq \Gauss(\pmb{x}_k|\est{\pmb{x}}_{k|k\shortminus1},\pmb{\Sigma}_{k|k\shortminus1}) \label{eq:PDF_predict}\\
    \text{update} \hspace{-50pt}&&\Longrightarrow&&\hspace{-50pt} p(\pmb{x}_k|\pmb{y}_{0:k}) &\coloneqq \Gauss(\pmb{x}_k|\est{\pmb{x}}_{k|k},\pmb{\Sigma}_{k|k}) \label{eq:PDF_update}
\end{align}
where $\est{\pmb{x}}_{k|k\shortminus1}$ and $\est{\pmb{x}}_{k|k}$ are their respective means, and $\pmb{\Sigma}_{k|k\shortminus1}$ and $\pmb{\Sigma}_{k|k}$, their covariances. Additionally, given that $\est{\pmb{x}}_{k|k\shortminus1}$ and $\est{\pmb{x}}_{k|k}$ are the means of the prediction and update \emph{posterior} PDFs, they are also the optimal estimators of $\pmb{x}_k$ that minimize the quadratic cost or MSE, as shown in \secref{sec:quadratic_cost}. Namely,
\begin{align}
    \est{\pmb{x}}_{k|k\shortminus1} &= \int \pmb{x}_k \; p(\pmb{x}_k|\pmb{y}_{0:k\shortminus1}) \dd \pmb{x}_k &&\!\!\!\!\!\!\!\!\!\!\!\!\!\!\!\!\!\!\!\!\!\!\!\!\!\!\!\!= \quad \text{ \emph{a-priori} estimate}, \\
    \est{\pmb{x}}_{k|k} &= \int \pmb{x}_k \; p(\pmb{x}_k|\pmb{y}_{0:k}) \dd \pmb{x}_k  &&\!\!\!\!\!\!\!\!\!\!\!\!\!\!\!\!\!\!\!\!\!\!\!\!\!\!\!\!= \quad \text{ \emph{a-posteriori} estimate}.
\end{align}
Note that both $\est{\pmb{x}}_{k|k\shortminus1}$ and $\est{\pmb{x}}_{k|k}$ are estimates of the same quantity, i.e. $\pmb{x}_k$, with the crucial difference that the \emph{a-posteriori} estimate incorporates the latest measurement, $\pmb{y}_k$, while the \emph{a-priori} estimate does not. Therefore, it is to be expected that the \emph{a-posteriori} estimate is more accurate than the \emph{a-priori} one. 
    
In \secref{sec:bayesian_optimal_filtering} we discussed how the posterior distribution of $\pmb{x}_k$  conditioned on measurements up to time $k$: $p(\pmb{x}_k|\pmb{y}_{0:k})$, can be computed recursively starting from $p(\pmb{x}_0)$ using recursive Bayesian optimal filtering. In the case of LG systems, we will show that the posterior, a.k.a. update density of \eqnref{eq:PDF_update}, remains fully Gaussian at all times. Then, given that Gaussian distributions are uniquely determined by their mean and covariance, the filtering problem reduces to recursively updating the moments of \eqnsref{eq:PDF_predict}{eq:PDF_update}:
\begin{enumerate}[label=(\roman*)]
    \item The \textbf{Prediction step} is defined by the following equations
    \begin{align}
        \est{\pmb{x}}_{k|k\shortminus1} &= \pmb{A}_{k\shortminus1} \, \est{\pmb{x}}_{k\shortminus1|k\shortminus1} + \pmb{B}_{k\shortminus1} \pmb{u}_{k\shortminus1}, \label{eq:predict_mean} \\
        \pmb{\Sigma}_{k|k\shortminus1} &= \pmb{A}_{k\shortminus1} \, \pmb{\Sigma}_{k\shortminus1|k\shortminus1} \, \pmb{A}^\Trans_{k\shortminus1} + \pmb{G}_{k\shortminus1}\pmb{Q}_{k\shortminus1}\pmb{G}_{k\shortminus1}^\Trans .\label{eq:predict_cov}
    \end{align}
    \item The \textbf{Update step} is given by
    \begin{align}
        \est{\pmb{x}}_{k|k} &= \est{\pmb{x}}_{k|k\shortminus1} + \pmb{K}_k (\pmb{y}_k - \pmb{H}_k \est{\pmb{x}}_{k|k\shortminus1}), \label{eq:update_mean} \\ 
        \pmb{\Sigma}_{k|k} &= \left(\I - \pmb{K}_k \pmb{H}_k \right) \pmb{\Sigma}_{k|k\shortminus1}, \label{eq:update_cov}
    \end{align}
    where
    \begin{align}
        \pmb{K}_k &= \pmb{\Sigma}_{k|k\shortminus1} \pmb{H}^\Trans_k \pmb{T}_k^{-1}, \label{eq:Kk_wT}\\
        \pmb{T}_k &= \pmb{H}_k \pmb{\Sigma}_{k|k\shortminus1} \pmb{H}_k^\Trans + \pmb{R}_k. \label{eq:Tk_wH}
    \end{align}
Furthermore, let us highlight that $\pmb{T}_k$ is not just a handy definition to shorten our notation but actually is the \emph{innovation covariance}. Namely,
\begin{equation} \label{eq:TkE}
    \pmb{T}_k = \EE{(\pmb{y}_k - \est{\pmb{y}}_k)(\pmb{y}_k - \est{\pmb{y}}_k)^\Trans},
\end{equation}
where $\pmb{y}_k - \est{\pmb{y}}_k$ is the   innovation, which is the difference between the measurement and its prediction 
\begin{equation}
    \est{\pmb{y}}_k = \pmb{H}_k \est{\pmb{x}}_{k|k\shortminus1}.
\end{equation}
The relationship between \eqnref{eq:Tk_wH} and \eqnref{eq:TkE} can be established by:
\begin{align}
    \pmb{T}_k &= \EE{(\pmb{H}_k \pmb{x}_k + \pmb{r}_k - \pmb{H}_k \est{\pmb{x}}_{k|k\shortminus1})(\pmb{H}_k \pmb{x}_k + \pmb{r}_k - \pmb{H}_k \est{\pmb{x}}_{k|k\shortminus1})^\Trans} \\
    &= \EE{(\pmb{H}_k (\pmb{x}_k - \est{\pmb{x}}_{k|k\shortminus1}) + \pmb{r}_k )(\pmb{H}_k (\pmb{x}_k - \est{\pmb{x}}_{k|k\shortminus1}) + \pmb{r}_k)^\Trans} \\
    &= \pmb{H}_k \, \EE{(\pmb{x}_k - \est{\pmb{x}}_{k|k\shortminus1})(\pmb{x}_k - \est{\pmb{x}}_{k|k\shortminus1}))^\Trans} \! \pmb{H}_k^\Trans + \EE{\pmb{r}_k\pmb{r}_k^\Trans} \\
    &= \pmb{H}_k \pmb{\Sigma}_k \pmb{H}_k^\Trans + \pmb{R}_k,
\end{align}
where the cross terms vanish due to the assumption that the state prediction error and the measurement noise are uncorrelated:
$\EE{(\pmb{x}_k - \est{\pmb{x}}_{k|k\shortminus1})\pmb{r}_k^\Trans} = \EE{\pmb{r}_k(\pmb{x}_k - \est{\pmb{x}}_{k|k\shortminus1})^\Trans} = 0$. Additionally, by taking \eqnref{eq:update_cov}: 
\begin{equation}
    \pmb{\Sigma}_{k|k\shortminus1} - \pmb{\Sigma}_{k|k} = \pmb{K}_k \pmb{H}_k \pmb{\Sigma}_{k|k\shortminus1},
\end{equation}
and substituting it into \eqnref{eq:Kk_wT} and \eqnref{eq:Tk_wH}, one can rewrite the Kalman gain:
\begin{align}
    \pmb{\Sigma}_{k|k\shortminus1} \pmb{H}_k^\Trans = \pmb{K}_k \pmb{T}_k = \pmb{K}_k (\pmb{H}_k \pmb{\Sigma}_{k|k\shortminus1} \pmb{H}_k^\Trans + \pmb{R}_k) = (\pmb{\Sigma}_{k|k\shortminus1} - \pmb{\Sigma}_{k|k}) \pmb{H}_k^\Trans + \pmb{K}_k \pmb{R}_k, 
\end{align}
such that
\begin{equation}
    \pmb{K}_k = \pmb{\Sigma}_{k|k} \pmb{H}_k^\Trans \pmb{R}_k^{-1}.
\end{equation}
\end{enumerate}

Let us now show how to derive the prediction and update equations through inductive hypothesis. 

\begin{myproof} The outline of the proof is as follows:  we assume that the posterior at time $k\!\shortminus\!1$ holds, and then show that both the prediction and update posteriors are also Gaussians with means and covariances following the equations given above. 
    \begin{enumerate}[label=(\roman*)]
        \item \textbf{Predict.}
            To find an expression for \eqnref{eq:PDF_predict}, we assume that the step $k\!\shortminus\!1$ holds, i.e. the posterior at time $k\!\shortminus\!1$ is a Gaussian PDF of the form:
            \begin{equation} \label{eq:PDF_Gaussian_step_k-1}
                p(\pmb{x}_{k\shortminus1}|\pmb{y}_{0:k\shortminus1}) = \Gauss(\pmb{x}_{k\shortminus1}|\est{\pmb{x}}_{k\shortminus1|k\shortminus1},\pmb{\Sigma}_{k\shortminus1|k\shortminus1}),
            \end{equation}
            Then, we follow the steps of \eqnref{eq:marginal_of_xk_from_xk,xk-1} to find an expression for the prediction PDF:
            \begin{align} \label{eq:derivation_predict_PDF}
                &p(\pmb{x}_k| \pmb{y}_{0:k\shortminus1}) \underset{\ref{propty:sum_rule}}{=} \int p(\pmb{x}_k,\pmb{x}_{k\shortminus1}| \pmb{y}_{0:k\shortminus1}) \dd \pmb{x}_{k\shortminus1} \nonumber \\
                &\underset{\ref{propty:prod_rule}}{=} \int p(\pmb{x}_k| \pmb{x}_{k\shortminus1}, \pmb{y}_{0:k\shortminus1}) p(\pmb{x}_{k\shortminus1}|\pmb{y}_{0:k\shortminus1}) \dd \pmb{x}_{k\shortminus1} \nonumber \\
                &\underset{\ref{propty:markov}}{=}  \int p(\pmb{x}_k|\pmb{x}_{k\shortminus1}) p(\pmb{x}_{k\shortminus1}|\pmb{y}_{0:k\shortminus1}) \dd \pmb{x}_{k\shortminus1} \nonumber \\
                &= \int \Gauss(\pmb{x}_k| \pmb{A}_{k\shortminus1} \pmb{x}_{k\shortminus1}+\pmb{B}_{k\shortminus1} \pmb{u}_{k\shortminus1}, \pmb{G}_{k\shortminus1}\pmb{Q}_{k\shortminus1}\pmb{G}_{k\shortminus1}^\Trans) \Gauss(\pmb{x}_{k\shortminus1}|\est{\pmb{x}}_{k\shortminus1|k\shortminus1},\pmb{\Sigma}_{k\shortminus1|k\shortminus1}) \dd \pmb{x}_{k\shortminus1} \nonumber \\ 
                &\underset{\ref{lem:joint_distribution}}{=} \int \! \Gauss\!\left( \!\!\begin{pmatrix} \pmb{x}_{k\shortminus1} \nonumber \\ \pmb{x}_k \end{pmatrix} \middle| \,
                \pmb{m}_{k|k\shortminus1} ,
                \pmb{P}_{k|k\shortminus1}
                \right) \! \dd \pmb{x}_{k\shortminus1} \nonumber \\
                &\underset{\ref{lem:cond_PDF_Gaussian_var}}{=} \Gauss(\pmb{x}_k|\pmb{A}_{k\shortminus1} \est{\pmb{x}}_{k\shortminus1|k\shortminus1} \!+\! \pmb{B}_{k-1} \pmb{u}_{k-1}, \pmb{A}_{k\shortminus1} \pmb{\Sigma}_{k\shortminus1|k\shortminus1} \pmb{A}_{k\shortminus1}^\Trans + \pmb{G}_{k\shortminus1}\pmb{Q}_{k\shortminus1}\pmb{G}_{k\shortminus1}^\Trans )
            \end{align}
            where in the last step we computed the marginal of the joint probability density, with mean and covariance
            \begin{align}
                \pmb{m}_{k|k\shortminus1} &\!=\! \begin{pmatrix} \est{\pmb{x}}_{k\shortminus1|k\shortminus1} \nonumber \\ \pmb{A}_{k\shortminus1} \est{\pmb{x}}_{k\shortminus1|k\shortminus1} \!+\! \pmb{B}_{k-1} \pmb{u}_{k-1}\end{pmatrix}\!\!, \\
                \pmb{P}_{k|k\shortminus1} &\!=\! \begin{pmatrix} 
                    \pmb{\Sigma}_{k\shortminus1|k\shortminus1} & \pmb{\Sigma}_{k\shortminus1|k\shortminus1} \pmb{A}_{k\shortminus1}^\Trans \\ 
                    \pmb{A}_{k\shortminus1} \pmb{\Sigma}_{k\shortminus1|k\shortminus1} & \pmb{A}_{k\shortminus1} \pmb{\Sigma}_{k\shortminus1|k\shortminus1} \pmb{A}_{k\shortminus1}^\Trans \!+\! \pmb{G}_{k\shortminus1}\pmb{Q}_{k\shortminus1}\pmb{G}_{k\shortminus1}^\Trans
                \end{pmatrix}.
            \end{align}
            Therefore, the prediction density at time $k$ is also a Gaussian:
            \begin{align}
                 \!\!p(\pmb{x}_k| \pmb{y}_{0:k\shortminus1}) &\coloneqq \Gauss(\pmb{x}_k|\est{\pmb{x}}_{k|k\shortminus1},\pmb{\Sigma}_{k|k\shortminus1}) \\
                 &= \Gauss(\pmb{x}_k|\pmb{A}_{k\shortminus1} \est{\pmb{x}}_{k\shortminus1|k\shortminus1} \!+\! \pmb{B}_{k-1} \pmb{u}_{k-1}, \pmb{A}_{k\shortminus1} \pmb{\Sigma}_{k\shortminus1|k\shortminus1} \pmb{A}_{k\shortminus1}^\Trans + \pmb{G}_{k\shortminus1}\pmb{Q}_{k\shortminus1}\pmb{G}_{k\shortminus1}^\Trans ),
            \end{align}
            such that
            \begin{align}
                \est{\pmb{x}}_{k|k\shortminus1} &= \pmb{A}_{k\shortminus1} \est{\pmb{x}}_{k\shortminus1|k\shortminus1} \!+\! \pmb{B}_{k-1} \pmb{u}_{k-1},\\
                \pmb{\Sigma}_{k|k\shortminus1} &= \pmb{A}_{k\shortminus1} \pmb{\Sigma}_{k\shortminus1|k\shortminus1} \pmb{A}_{k\shortminus1}^\Trans + \pmb{G}_{k\shortminus1}\pmb{Q}_{k\shortminus1}\pmb{G}_{k\shortminus1}^\Trans.
            \end{align}
        \item \textbf{Update.} The update density of \eqnref{eq:PDF_update} can be derived by first  employing  \lemref{lem:joint_distribution} to find an expression for the joint PDF $p(\pmb{x}_k,\pmb{y}_k|\pmb{y}_{0:k\shortminus1})$. Then, using \lemref{lem:cond_PDF_Gaussian_var}, one can derive the conditional density $\pmb{x}_k|\pmb{y}_k \sim p(\pmb{x}_k|\pmb{y}_k,\pmb{y}_{0:k\shortminus1})$. The joint PDF, conditioned on the measurement outcomes up to time $k\!\shortminus\!1$, is:
        \begin{align}
            &p(\pmb{x}_k,\pmb{y}_k|\pmb{y}_{0:k\shortminus1}) = p(\pmb{y}_k|\pmb{x}_k,\pmb{y}_{0:k\shortminus1}) p(\pmb{x}_k|\pmb{y}_{0:k\shortminus1}) \underset{\ref{propty:cond_indep}}{=} p(\pmb{y}_k|\pmb{x}_k) p(\pmb{x}_k|\pmb{y}_{0:k\shortminus1}) \nonumber \\
            &= \Gauss(\pmb{y}_k|\pmb{H}_k\pmb{x}_k,\pmb{R}_k) \Gauss(\pmb{x}_k|\est{\pmb{x}}_{k|k\shortminus1},\pmb{\Sigma}_{k|k\shortminus1}) \nonumber \\
            &=  \Gauss\!\left( \!\!
            \begin{pmatrix} \pmb{x}_k \\ \pmb{y}_k \end{pmatrix} 
            \middle|
            \begin{pmatrix} \est{\pmb{x}}_{k|k\shortminus1} \\ \pmb{H}_k \est{\pmb{x}}_{k|k\shortminus1} \end{pmatrix}
            \!,\! 
            \begin{pmatrix} 
            \pmb{\Sigma}_{k|k\shortminus1} & \pmb{\Sigma}_{k|k\shortminus1} \pmb{H}_k^\Trans \\ 
            \pmb{H}_k \pmb{\Sigma}_{k|k\shortminus1} & \pmb{H}_k \pmb{\Sigma}_{k|k\shortminus1} \pmb{H}_k^\Trans + \pmb{R}_k
            \end{pmatrix} \!\!
            \right) \label{eq:joint_p(x,y)},
        \end{align}
        where in the last step we have applied \lemref{lem:joint_distribution}. Now, we can compute the conditional PDF $p(\pmb{x}_k|\pmb{y}_{0:k})$ from \eqnref{eq:joint_p(x,y)} using \lemref{lem:cond_PDF_Gaussian_var}:
        \begin{align}
            p(\pmb{x}_k|\pmb{y}_{0:k}) &= p(\pmb{x}_k|\pmb{y}_k,\pmb{y}_{0:k\shortminus1}) = \Gauss(\pmb{x}_k| \est{\pmb{x}}_{k|k},\pmb{\Sigma}_{k|k}), \label{eq:update_PDF}\\
            \est{\pmb{x}}_{k|k} &= \est{\pmb{x}}_{k|k\shortminus1} + \pmb{K}_k (\pmb{y}_k - \pmb{H}_k \,\est{\pmb{x}}_{k|k\shortminus1}), \\
            \pmb{\Sigma}_{k|k} &= \pmb{\Sigma}_{k|k\shortminus1} - \pmb{K}_k\pmb{T}_k\,\pmb{K}_k^\Trans \label{eq:cov_update_intermediate}
        \end{align}
        where
        \begin{align}
            \pmb{T}_k &= \pmb{H}_k \pmb{\Sigma}_{k|k\shortminus1} \pmb{H}_k^\Trans + \pmb{R}_k, \\
            \pmb{K}_k &= \pmb{\Sigma}_{k|k\shortminus1} \pmb{H}_k^\Trans \pmb{T}_k^{-1}. 
        \end{align}
        Note that $\pmb{T}_k = \pmb{T}_k^{\,\Trans}$ and $(\pmb{T}_k^{\,-1})^\Trans = (\pmb{T}_k^{\,\Trans})^{-1}$. It follows from \eqnref{eq:cov_update_intermediate} that the covariance update equation can also be written as
        \begin{align}
            \pmb{\Sigma}_{k|k} &= \pmb{\Sigma}_{k|k\shortminus1} - \pmb{K}_k\pmb{T}_k( \pmb{\Sigma}_{k|k\shortminus1} \pmb{H}_k^\Trans \pmb{T}_k^{\,-1})^\Trans =  \pmb{\Sigma}_{k|k\shortminus1} - \pmb{K}_k\pmb{T}_k \pmb{T}_k^{\,-1} \pmb{H}_k \pmb{\Sigma}_{k|k\shortminus1} \nonumber \\
            &= (\I - \pmb{K}_k\pmb{H}_k) \pmb{\Sigma}_{k|k\shortminus1} 
        \end{align}
    \end{enumerate}    
\end{myproof}

\begin{table}[h!]
\begin{center}
\renewcommand{\arraystretch}{2.5} 
\setlength{\extrarowheight}{0pt} 
\begin{tabular}{|>{\centering\arraybackslash}m{2.5cm}|>{\centering\arraybackslash}m{10cm}|}
\hline
\textbf{Model} \vspace{1pt}& 
\makecell[c]{$\pmb{x}_{k} = \pmb{A}_{k\shortminus1} \pmb{x}_{k-1} + \pmb{B}_{k\shortminus1} \pmb{u}_{k\shortminus1} + \pmb{G}_{k\shortminus1} \pmb{q}_{k\shortminus1}, \quad \pmb{q}_k \sim \Gauss(0, \pmb{Q}_k)$ \\ $\pmb{y}_k = \pmb{H}_k \pmb{x}_k + \pmb{r}_k, \quad \pmb{r}_k \sim \Gauss(0, \pmb{R}_k)$} \vspace{1pt}\\
\hline
\textbf{Initialize} \vspace{1pt}& 
\makecell[c]{$\est{\pmb{x}}(t_0) = \est{\pmb{x}}_{0|0}$ \\ $\pmb{\Sigma}_{0|0} = E \left\{ (\tilde{\pmb{x}}_{0|0} - \pmb{x}_0) (\tilde{\pmb{x}}_{0|0} - \pmb{x}_0)^T \right\}$} \vspace{1pt} \\
\hline 
\textbf{Predict} \vspace{1pt}& 
\makecell[c]{$\est{\pmb{x}}_{k|k\shortminus1} = \pmb{A}_{k\shortminus1} \est{\pmb{x}}_{k\shortminus1|k\shortminus1} + \pmb{B}_{k\shortminus1} \pmb{u}_{k\shortminus1}$ \\ $\pmb{\Sigma}_{k|k\shortminus1} = \pmb{A}_{k\shortminus1} \pmb{\Sigma}_{k\shortminus1|k\shortminus1} \pmb{A}_{k\shortminus1}^T + \pmb{G}_{k\shortminus1}\pmb{Q}_{k\shortminus1}\pmb{G}_{k\shortminus1}^\Trans$} \vspace{1pt}\\
\hline
\textbf{Gain} \vspace{1pt}& 
$\pmb{K}_k = \pmb{\Sigma}_{k|k\shortminus1} \pmb{H}_k^T \left( \pmb{H}_k \pmb{\Sigma}_{k|k\shortminus1} \pmb{H}_k^T + \pmb{R}_k \right)^{-1}$ \vspace{1pt}\\
\hline
\textbf{Update} \vspace{1pt} & 
\makecell[c]{$\est{\pmb{x}}_{k|k} = \est{\pmb{x}}_{k|k\shortminus1} + \pmb{K}_k \left( \pmb{y}_k - \pmb{H}_k \, \est{\pmb{x}}_{k|k\shortminus1} \right)$ \\ $\pmb{\Sigma}_{k|k} = \left( \I - \pmb{K}_k \pmb{H}_k \right) \pmb{\Sigma}_{k|k\shortminus1}$} \vspace{1pt}\\
\hline
\end{tabular}
\caption[Summary of the discrete uncorrelated Kalman filter]{Summary of the discrete uncorrelated Kalman filter.}
\label{tab:summary_discrete_uncorrelated_KF}
\end{center}
\end{table}

\section{The continuous uncorrelated Kalman filter} \label{sec:uncorr_continuous_KF}

To transform the discrete KF to a continuous one, we have to take the limit of $\Dt \rightarrow 0$. Therefore, we do not need anymore to distinguish between predict and update steps and can combine their equations for both the estimator and the covariance, summarized in \tabref{tab:summary_discrete_uncorrelated_KF}, into a joint \emph{a-priori} recursive form of the KF:
\begin{align}
    \est{\pmb{x}}_{k+1} &= \pmb{A}_k\est{\pmb{x}}_{k} + \pmb{B}_k\pmb{u}_k + \pmb{A}_k \pmb{K}_k \left(\pmb{y}_k - \pmb{H}_k \est{\pmb{x}}_{k}\right) , \label{eq:merged_est_eq} \\
    \pmb{K}_k &= \pmb{\Sigma}_{k} \pmb{H}_k^T \left( \pmb{H}_k \pmb{\Sigma}_{k} \pmb{H}_k^T + \pmb{R}_k \right)^{-1}\!\!\!\!, \label{eq:kalman_gain}\\
    \pmb{\Sigma}_{k+1} &= \pmb{A}_{k} \pmb{\Sigma}_{k} \pmb{A}_{k}^T - \pmb{A}_{k} \pmb{K}_k \pmb{H}_k \pmb{\Sigma}_{k} \pmb{A}_{k}^T + \pmb{G}_{k}\pmb{Q}_{k}\pmb{G}_{k}^\Trans. \label{eq:merged_cov_eq}
\end{align}

First, let us derive the expression for the continuous Kalman gain starting from its discrete version in \eqnref{eq:kalman_gain} and substituting the first-order approximations for $\pmb{H}_k$ and $\pmb{R}_k$ summarized in \tabref{tab:summary_discretization}:
\begin{align}
    \pmb{K}_k = \pmb{\Sigma}_{k} \pmb{H}_k^\Trans \left( \pmb{H}_k \pmb{\Sigma}_{k} \pmb{H}_k^\Trans + \pmb{R}_k \right)^{-1} \!\! = \pmb{\Sigma}_{k} \, \pmb{H}^{\Trans}\!(t_k) \!\! \left( \pmb{H}(t_k) \pmb{\Sigma}_{k} \, \pmb{H}^{\Trans}\!(t_k) + \frac{\pmb{R}(t_k)}{\Dt} \right)^{-1},
\end{align}
which, since  $\pmb{R}(t_k)/\Dt \gg \pmb{H}(t_k) \pmb{\Sigma}_{k} \, \pmb{H}^{\Trans}\!(t_k)$ when $\Dt \to 0$, then it can be simplified to
\begin{align}
    \pmb{K}_k = \pmb{\Sigma}_{k} \, \pmb{H}^{\Trans}\!(t_k) \pmb{R}(t_k)^{-1} \Dt.
\end{align}
By now dividing the discrete Kalman gain $\pmb{K}_k$ by $\Dt$, we define the continuous Kalman gain as
\begin{equation} \label{eq:discr_to_cont_KG}
    \pmb{K}(t_k) \coloneqq \frac{\pmb{K}_k}{\Dt},
\end{equation}
which by taking the limit of $\Dt \rightarrow 0$, yields the following expression:
\begin{equation} \label{eq:uncorr_KG}
    \pmb{K}(t) = \pmb{\Sigma}(t) \, \pmb{H}^{\Trans}(t) \pmb{R}^{-1}(t).
\end{equation}

Next, we turn to the derivation of the Riccati equation, i.e. the continuous differential equation for the covariance. The first step is to substitute the first-order approximation of $\pmb{A}_k = \left(\I + \pmb{F}(t_k)\Dt\right)$ from \propref{prop:disc_LG}, and note that both terms $\pmb{G}_k \pmb{Q}_k \pmb{G}_k^\Trans$ and $\pmb{K}_k$ are of order $\Dt$, as indicated in \eqnref{eq:discr_to_cont_GQG} and \eqnref{eq:discr_to_cont_KG}, respectively. Then,
\begin{align}
    \pmb{\Sigma}_{k+1} &=  \left(\I + \pmb{F}(t_k)\Dt\right) \pmb{\Sigma}_{k} \! \left(\I + \pmb{F}^{\,\Trans}\!(t_k)\Dt\right) \! -  \left(\I + \pmb{F}(t_k)\Dt\right) \pmb{K}(t_k) \Dt \, \pmb{H}(t_k) \pmb{\Sigma}_{k} \! \left(\I + \pmb{F}^{\,\Trans}\!(t_k)\Dt\right) \nonumber \\
    &+ \pmb{G}(t_{k})\pmb{Q}(t_{k})\pmb{G}^\Trans(t_{k})\Dt = \pmb{\Sigma}_k + \pmb{F}(t_k) \pmb{\Sigma}_k \Dt + \pmb{\Sigma}_k \, \pmb{F}^{\,\Trans}\!(t_k) \Dt - \pmb{K}(t_k) \pmb{H}(t_k) \pmb{\Sigma}_k \Dt \nonumber \\
    &+ \pmb{G}(t_{k})\pmb{Q}(t_{k})\pmb{G}^\Trans(t_{k})\Dt.
\end{align}
By now rearranging terms and taking the limit of $\Dt \rightarrow 0$, we retrieve a  Riccati equation for the covariance:
\begin{align}
    \lim_{\Dt \rightarrow 0} \!\frac{\pmb{\Sigma}_{k+1} \!-\! \pmb{\Sigma}_k}{\Dt} = \frac{\dd \pmb{\Sigma}(t)}{\dt} &=  \pmb{F}(t) \pmb{\Sigma}(t) \!+\! \pmb{\Sigma}(t) \, \pmb{F}^{\,\Trans}\!(t)  \!-\! \pmb{K}(t) \pmb{H}(t) \pmb{\Sigma}(t) \!+\! \pmb{G}(t)\pmb{Q}(t)\pmb{G}^\Trans\!(t) \nonumber \\
    &=  \pmb{F}(t) \pmb{\Sigma}(t) \!+\! \pmb{\Sigma}(t) \, \pmb{F}^{\,\Trans}\!(t)  \!-\! \pmb{K}(t) \pmb{R}(t) \pmb{K}(t) \!+\! \pmb{G}(t)\pmb{Q}(t)\pmb{G}^\Trans\!(t).
\end{align}
Note that the Riccati equation is quadratic w.r.t. the covariance matrix $\pmb{\Sigma}(t)$, which becomes evident when substituting the Kalman gain form of \eqnref{eq:uncorr_KG}:
\begin{equation}
    \frac{\dd \pmb{\Sigma}(t)}{\dt} =  \pmb{F}(t) \pmb{\Sigma}(t) \!+\! \pmb{\Sigma}(t) \, \pmb{F}^{\,\Trans}\!(t)  \!-\!  \pmb{\Sigma}(t) \, \pmb{H}^{\Trans}(t) \pmb{R}^{-1}(t) \pmb{H}(t) \pmb{\Sigma}(t) \!+\! \pmb{G}(t)\pmb{Q}(t)\pmb{G}^\Trans\!(t).
\end{equation}
Similarly, we can derive the  Kalman-Bucy equation from \eqnref{eq:merged_est_eq} by substituting the corresponding first-order approximations:
\begin{align}
    \!\!\est{\pmb{x}}(t_{k+1}) &= \! \left(\I \!+\! \pmb{F}(t_k)\Dt\right)\est{\pmb{x}}(t_{k}) \!+\! \pmb{B}(t_k)\pmb{u}(t_k) \Dt \!+\! \left(\I \!+\! \pmb{F}(t_k)\Dt\right) \pmb{K}(t_k) \Dt \left(\pmb{y}_k \!-\! \pmb{H}(t_k )\est{\pmb{x}}(t_{k})\right)\!,
\end{align}
and then rearranging terms to take the limit $\Dt \to 0$:
\begin{align}
    \frac{\dd \est{\pmb{x}}}{\dt} &= \lim_{\Dt \rightarrow 0} \frac{\est{\pmb{x}}_{k+1}  - \est{\pmb{x}}_{k}}{\Dt} \nonumber \\
    &= \lim_{\Dt \rightarrow 0}  \left[ \pmb{F}(t_k)\est{\pmb{x}}(t_{k}) + \pmb{B}(t_k)\pmb{u}(t_k) + \pmb{K}(t_k) \!\!\left(  \!\frac{1}{\Dt} \! \int_{t_{k\shortminus1}}^{t_k} \!\!\! \pmb{y}(t) \dt - \pmb{H}(t_k )\est{\pmb{x}}(t_{k}) \!\right) \!\right] \nonumber \\
    &= \pmb{F}(t)\est{\pmb{x}}(t) + \pmb{B}(t)\pmb{u}(t) + \pmb{K}(t)\!\left( \lim_{\Dt \rightarrow 0} \!\frac{1}{\Dt} \! \int_{t_{k\shortminus1}}^{t_k} \!\!\! \pmb{y}(t) \dt - \pmb{H}(t)\est{\pmb{x}}(t) \!\right) \nonumber \\
    &=\footnotemark \pmb{F}(t)\est{\pmb{x}}(t) + \pmb{B}(t)\pmb{u}(t) + \pmb{K}(t)\!\left( \pmb{y}(t) - \pmb{H}(t)\est{\pmb{x}}(t) \right).
\end{align}
This yields a differential equation describing the evolution of the continuous-time estimate.
\footnotetext{$\lim_{\Dt \rightarrow 0} \frac{1}{\Dt} \int_{t_k \shortminus \Dt}^{t_k} \pmb{y}(t) \dt = \pmb{y}(t_k)$.}

\begin{table}[h!]
\begin{center}
\renewcommand{\arraystretch}{2.5} 
\setlength{\extrarowheight}{0pt} 
\begin{tabular}{|>{\centering\arraybackslash}m{2.5cm}|>{\centering\arraybackslash}m{12cm}|}
\hline
\textbf{Model} \vspace{1pt}& 
\makecell[c]{$\dot{\pmb{x}}(t) = \pmb{F}(t) \pmb{x}(t) + \pmb{B}(t) \pmb{u}(t) + \pmb{G}(t) \pmb{w}(t), \;\; \pmb{w}(t) \sim \Gauss(0, \pmb{Q}(t))$ \\ $\pmb{y}(t) = \pmb{H}(t) \pmb{x}(t) + \pmb{v}(t), \;\; \pmb{v}(t) \sim \Gauss(0, \pmb{R}(t))$} \vspace{1pt}\\
\hline
\textbf{Initialize} \vspace{1pt}& 
\makecell[c]{$\est{\pmb{x}}(t_0) = \est{\pmb{x}}_{0}$ \\ $\pmb{\Sigma}_{0} = \EE{ (\tilde{\pmb{x}}_{0} - \pmb{x}_0) (\tilde{\pmb{x}}_{0} - \pmb{x}_0)^{\!\Trans} }$} \vspace{1pt} \\
\hline
\textbf{Gain} \vspace{1pt}& 
$\pmb{K}(t) = \pmb{\Sigma}(t) \, \pmb{H}^{\Trans}(t) \pmb{R}^{-1}(t)$ \vspace{1pt}\\
\hline 
\textbf{Covariance} \vspace{1pt}& 
\makecell[c]{$\dot{\pmb{\Sigma}}(t) =  \pmb{F}(t) \pmb{\Sigma}(t) \!+\! \pmb{\Sigma}(t) \, \pmb{F}^\Trans\!(t)  \!-\! \pmb{K}(t) \pmb{H}(t) \pmb{\Sigma}(t) \!+\! \pmb{G}(t)\pmb{Q}(t)\pmb{G}^\Trans\!(t)$} \vspace{1pt}\\
\hline
\textbf{Estimate} \vspace{1pt} & 
\makecell[c]{$\dot{\est{\pmb{x}}}(t) = \pmb{F}(t)\est{\pmb{x}}(t) + \pmb{B}(t)\pmb{u}(t) + \pmb{K}(t)\!\left( \pmb{y}(t) - \pmb{H}(t)\est{\pmb{x}}(t) \right)$} \vspace{1pt}\\
\hline
\end{tabular}
\caption[Summary of the continuous uncorrelated Kalman filter]{Summary of the continuous uncorrelated Kalman filter.}
\label{tab:summary_continuous_uncorrelated_KF}
\end{center}
\end{table}

\section{The continuous correlated Kalman filter} \label{sec:correlated_KF}

In many systems, the measurement noise is uncorrelated with the process noise. However, in quantum systems, the measurement can act back onto the system --- an effect referred to as measurement back-action --- and introduce correlation between the noises. As a result, the standard KF theory has to be extended to properly account for correlated measurement and process noises \cite{crassidis2011optimal}. This requires allowing the zero-mean Gaussian noise processes, $\pmb{w}(t)$ and $\pmb{v}(t)$, present in the LG model of \eqnsref{eq:cont_system_model_L&G_ch1}{eq:cont_meas_model_L&G_ch1} to be cross-correlated:
\begin{align}
    \EE{\pmb{w}(t)\pmb{v}^\Trans(s)} = \pmb{S}(t) \delta(t-s), \label{eq:cont_corr_cov_S}
\end{align}
where the matrix $\pmb{S}(t)$ is not necessarily symmetric. To derive the KF equations for a system with correlated measurement and process noises, we proceed in the same way as we did for the case of uncorrelated noise, with the exception that first we must de-correlate the measurement and process noise \cite{crassidis2011optimal,Tsang2009}. 

In \secref{sec:discretization_cont_model} we established the equivalency between the continuous model in \eqnsref{eq:cont_system_model_L&G_ch1}{eq:cont_meas_model_L&G_ch1} and the following discrete model, which now includes correlated process and measurement noise:
\begin{align}
    &\begin{cases}
        \pmb{x}_k \!=\! \pmb{A}_{k\shortminus1} \, \pmb{x}_{k\shortminus1} \!+\!  \pmb{B}_{k\shortminus1} \, \pmb{u}_{k\shortminus1} \!+\! \pmb{G}_{k\shortminus1} \, \pmb{q}_{k\shortminus1}, \\
        \pmb{y}_k \!=\! \pmb{H}_k \pmb{x}_k \!+\! \pmb{r}_k,
    \end{cases} \label{eq:discrete_process_meas_model}\\
    &\;\;\; \EE{\pmb{q}_{k}\,\pmb{q}_\ell^\Trans} = \pmb{Q}_k \delta_{k\ell}, \\
    &\;\;\; \EE{\pmb{r}_{k}\,\pmb{r}_\ell^\Trans} = \pmb{R}_k \delta_{k\ell}, \\
    &\;\;\; \EE{\pmb{q}_{k\shortminus1}\pmb{r}_\ell^\Trans} = \pmb{S}_k \delta_{k\ell}. 
\end{align}
Starting from this discrete model, we de-correlate the measurement and process equations by introducing the measurement term into the process equation:
\begin{align}
    \pmb{x}_k &= \pmb{A}_{k\shortminus1} \pmb{x}_{k\shortminus1} + \pmb{B}_{k\shortminus1} \pmb{u}_{k\shortminus1} + \pmb{G}_{k\shortminus1} \pmb{q}_{k\shortminus1} + \pmb{D}_{k\shortminus1} \left(\pmb{y}_{k\shortminus1} - \pmb{H}_{k\shortminus1}\pmb{x}_{k\shortminus1} - \pmb{r}_{k\shortminus1}\right)  \\
    &= \left(\pmb{A}_{k\shortminus1} - \pmb{D}_{k\shortminus1} \pmb{H}_{k\shortminus1}\right) \pmb{x}_{k\shortminus1} + \pmb{B}_{k\shortminus1} \pmb{u}_{k\shortminus1} + \pmb{D}_{k\shortminus1} \,\pmb{y}_{k\shortminus1} + \pmb{Z}_{k\shortminus1},
\end{align}
where $\pmb{D}_{k\shortminus1}$ is an arbitrary matrix, since $\pmb{y}_{k\shortminus1} - \pmb{H}_{k\shortminus1}\,\pmb{x}_{k\shortminus1} - \pmb{r}_{k\shortminus1} = 0$. Moreover, we define $\pmb{Z}_{k\shortminus1} \coloneqq \pmb{G}_{k\shortminus1} \, \pmb{q}_{k\shortminus1} - \pmb{D}_{k\shortminus1} \pmb{r}_{k\shortminus1}$ to be the effective process noise. In order to ensure that the correlation between the measurement $\pmb{r}_k$ and the new process noise $\pmb{Z}_{k\shortminus1}$ vanishes, we set
\begin{equation}
    \pmb{D}_k \coloneqq \pmb{G}_k\,\pmb{S}_k\,\pmb{R}_k^{-1}
\end{equation}
such that
\begin{equation}
    \EE{\pmb{Z}_{k\shortminus1} \pmb{r}_k^\Trans} = \left( \pmb{G}_{k\shortminus1} \pmb{S}_{k\shortminus1} - \pmb{D}_{k\shortminus1} \pmb{R}_{k\shortminus1}\right) =0.
\end{equation}
Then, $\pmb{Z}_{k\shortminus1}$ can be rewritten as:
\begin{align}
    \pmb{Z}_{k\shortminus1} = \pmb{G}_{k\shortminus1} \pmb{q}_{k\shortminus1} - \pmb{G}_{k\shortminus1} \pmb{S}_{k\shortminus1} \pmb{R}_{k\shortminus1}^{-1} \pmb{r}_{k\shortminus1} = \pmb{G}_{k\shortminus1} \left(\pmb{q}_{k\shortminus1} - \pmb{S}_{k\shortminus1} \pmb{R}_{k\shortminus1}^{-1} \pmb{r}_{k\shortminus1}\right) = \pmb{G}_{k\shortminus1} \pmb{U}_{k\shortminus1},
\end{align}
with $\pmb{U}_{k\shortminus1} \coloneqq \pmb{q}_{k\shortminus1} - \pmb{S}_{k\shortminus1} \pmb{R}^{-1} \pmb{r}_{k\shortminus1}$. This results in an uncorrelated model:
\begin{align}
    \pmb{x}_k &= \left(\pmb{A}_{k\shortminus1} - \pmb{G}_{k\shortminus1} \pmb{S}_{k\shortminus1} \pmb{R}^{-1}_{k\shortminus1} \pmb{H}_{k\shortminus1}\right) \pmb{x}_{k\shortminus1} + \pmb{B}_{k\shortminus1} \pmb{u}_{k\shortminus1} + \pmb{G}_{k\shortminus1}\pmb{S}_{k\shortminus1} \pmb{R}_{k\shortminus1}^{-1} \pmb{y}_{k\shortminus1} + \pmb{G}_{k\shortminus1} \pmb{U}_{k\shortminus1} \label{eq:de-correlated_process}  \\
    \pmb{y}_k &= \pmb{H}_k \pmb{x}_k + \pmb{r}_k, \label{eq:de-correlated_measurement}
\end{align}
where, in order for the process noise $\pmb{U}_k$ to be uncorrelated w.r.t. the measurement noise, the process must now depend on the measurement $\pmb{y}_{k\shortminus1}$. Additionally, the variance of the process noise $\pmb{U}_k$ now reads as:
\begin{equation}
    \EE{\pmb{U}_k\pmb{U}_k^\Trans} = \pmb{Q}_{k\shortminus1} - \pmb{S}_{k\shortminus1} \pmb{R}_{k\shortminus1}^{-1} \pmb{S}_{k\shortminus1}^\Trans.
\end{equation}

To derive the prediction and update probability densities,  $p(\pmb{x}_k|\pmb{y}_{0:k\shortminus1})$ and $p(\pmb{x}_k|\pmb{y}_{0:k})$, for the system described by \eqnsref{eq:de-correlated_process}{eq:de-correlated_measurement}, we follow the approach of \secref{sec:uncorr_discrete_KF}. As in that derivation, both probabilities turn out to be Gaussian, with means and covariances following predict and update rules. These equations describe the optimal evolution of the estimator and its covariance over time, since the mean of the posterior is the optimal estimator minimizing the MSE. 

Additionally, it is important to note that the de-correlation trick exclusively affects the process equation, i.e. \eqnref{eq:de-correlated_process}. Thus, we expect only the prediction step to be modified, leaving the update step unchanged. In particular, the predict step, i.e. \eqnref{eq:de-correlated_process}, now depends on the measurement outcome $\pmb{y}_{k\shortminus1}$. That might seem counter-intuitive, as measurement outcomes typically only affect the update step, not the prediction. Nonetheless, this is irrelevant for the derivation of the continuous correlated KF, since the equations for the predict and update step have to be merged. 

With this understanding, let us derive an expression for the Gaussian probability density for the prediction step, $p(\pmb{x}_k|\pmb{y}_{0:k\shortminus1}) \coloneqq \Gauss(\pmb{x}_k|\est{\pmb{x}}_{k|k\shortminus1},\pmb{\Sigma}_{k|k\shortminus1})$. This derivation essentially mirrors the one in \eqnref{eq:derivation_predict_PDF} with a slight modification:
\begin{align}
    &p(\pmb{x}_k| \pmb{y}_{0:k\shortminus1}) \underset{\ref{propty:sum_rule}}{=} \int p(\pmb{x}_k,\pmb{x}_{k\shortminus1}| \pmb{y}_{0:k\shortminus1}) \dd \pmb{x}_{k\shortminus1} \nonumber \\
    &\underset{\ref{propty:prod_rule}}{=} \int p(\pmb{x}_k| \pmb{x}_{k\shortminus1}, \pmb{y}_{0:k\shortminus1}) p(\pmb{x}_{k\shortminus1}|\pmb{y}_{0:k\shortminus1}) \dd \pmb{x}_{k\shortminus1} \nonumber \\
    &= \int p(\pmb{x}_k|\pmb{x}_{k\shortminus1},\pmb{y}_{k\shortminus1}) p(\pmb{x}_{k\shortminus1}|\pmb{y}_{0:k\shortminus1}) \dd \pmb{x}_{k\shortminus1}. 
\end{align}
Here, $p(\pmb{x}_k| \pmb{x}_{k\shortminus1}, \pmb{y}_{0:k\shortminus1})$ is simplified to $p(\pmb{x}_k|\pmb{x}_{k\shortminus1},\pmb{y}_{k\shortminus1})$ instead of $p(\pmb{x}_k|\pmb{x}_{k\shortminus1})$, since unlike in \eqnref{eq:derivation_predict_PDF}, the state $\pmb{x}_k$ not only depends on $\pmb{x}_{k\shortminus1}$ but also on $\pmb{y}_{k\shortminus1}$, as evident from \eqnref{eq:de-correlated_process} \cite{Tsang2009}. Fortunately, this still fulfills Markovianity and $p(\pmb{x}_k|\pmb{x}_{k\shortminus1},\pmb{y}_{k\shortminus1})$ is still a Gaussian distribution:
\begin{align}
    &\!p(\pmb{x}_k|\pmb{x}_{k\shortminus1},\pmb{y}_{k\shortminus1}) = \nonumber \\
    &=\Gauss\!(\pmb{x}_k|\pmb{A}^\prime_{k\shortminus1} \pmb{x}_{k\shortminus1} \!\!+\! \pmb{B}_{k\shortminus1} \pmb{u}_{k\shortminus1} \!\!+\! \pmb{G}_{k\shortminus1} \pmb{S}_{k\shortminus1} \pmb{R}_{k\shortminus1}^{-1} \pmb{y}_{k\shortminus1}, \!\pmb{G}_{k\shortminus1} \!\left(\pmb{Q}_{k\shortminus1} \!\shortminus\! \pmb{S}_{k\shortminus1} \pmb{R}_{k\shortminus1}^{-1} \pmb{S}_{k\shortminus1}^\Trans \right) \!\pmb{G}_{k\shortminus1} ),
\end{align}
where $\pmb{A}^\prime_{k\shortminus1} = \pmb{A}_{k\shortminus1} - \pmb{G}_{k\shortminus1} \pmb{S}_{k\shortminus1} \pmb{R}_{k\shortminus1}^{-1} \pmb{H}_{k\shortminus1}$. Assuming that the posterior at the previous time step is Gaussian, as given in  \eqnref{eq:PDF_Gaussian_step_k-1}: $p(\pmb{x}_{k\shortminus1}|\pmb{y}_{0:k\shortminus1}) = \Gauss(\pmb{x}_{k\shortminus1}|\est{\pmb{x}}_{k\shortminus1|k\shortminus1},\pmb{\Sigma}_{k\shortminus1|k\shortminus1})$, we can use \lemref{lem:joint_distribution} to write the joint probability distribution $p(\pmb{x}_k,\pmb{x}_{k\shortminus1}|\pmb{y}_{0:k\shortminus1})$ as:
\begin{align}
    p(\pmb{x}_k,\pmb{x}_{k\shortminus1}|\pmb{y}_{0:k\shortminus1}) = \Gauss\!\left( \! \begin{pmatrix} \pmb{x}_{k\shortminus1} \\ \pmb{x}_k \end{pmatrix} \middle| \,
        \pmb{m}_{k|k\shortminus1} , 
        \pmb{P}_{k|k\shortminus1} \!
        \right),
\end{align}
where
\begin{align}
    \pmb{m}_{k|k\shortminus1} = \begin{pmatrix} \est{\pmb{x}}_{k\shortminus1|k\shortminus1} \\ \pmb{A}^\prime_{k\shortminus1} \est{\pmb{x}}_{k\shortminus1|k\shortminus1} + \pmb{B}_{k\shortminus1} \pmb{u}_{k\shortminus1} + \pmb{G}_{k\shortminus1} \pmb{S}_{k\shortminus1} \pmb{R}^{-1}_{k\shortminus1} \, \pmb{y}_{k\shortminus1}  \end{pmatrix},
\end{align}
and
\begin{align}
    \pmb{P}_{k|k\shortminus1} = \begin{pmatrix} 
        \pmb{\Sigma}_{k\shortminus1|k\shortminus1} & \pmb{\Sigma}_{k\shortminus1|k\shortminus1} \pmb{A}_{k\shortminus1}^{\prime\,\Trans} \\ 
        \pmb{A}_{k\shortminus1}^{\prime} \pmb{\Sigma}_{k\shortminus1|k\shortminus1} & \pmb{A}_{k\shortminus1}^{\prime} \pmb{\Sigma}_{k\shortminus1|k\shortminus1} \, \pmb{A}_{k\shortminus1}^{\prime\,\Trans} + \pmb{G}_{k\shortminus1}\!\left(\pmb{Q}_{k\shortminus1} \!- \pmb{S}_{k\shortminus1}\pmb{R}_{k\shortminus1}^{-1}\pmb{S}^\Trans_{k\shortminus1}\right)\!\pmb{G}_{k\shortminus1}^\Trans
        \end{pmatrix} .
\end{align}

To now obtain the predict probability density $p(\pmb{x}_{k}|\pmb{y}_{0:k\shortminus1})$ from the joint probability $p(\pmb{x}_k,\pmb{x}_{k\shortminus1}|\pmb{y}_{0:k\shortminus1})$, we compute its marginal following \lemref{lem:cond_PDF_Gaussian_var}:
\begin{align}
    p(\pmb{x}_k| \pmb{y}_{0:k\shortminus1}) &= \Gauss(\pmb{x}_k|\est{\pmb{x}}_{k|k\shortminus1},\pmb{\Sigma}_{k|k\shortminus1}), \\
    \est{\pmb{x}}_{k|k\shortminus1} &= \left(\pmb{A}_{k\shortminus1} \!- \pmb{G}_{k\shortminus1} \pmb{S}_{k\shortminus1} \pmb{R}_{k\shortminus1}^{-1} \pmb{H}_{k\shortminus1}\right) \est{\pmb{x}}_{k\shortminus1|k\shortminus1} + \pmb{B}_{k\shortminus1} \pmb{u}_{k\shortminus1} + \pmb{G}_{k\shortminus1} \pmb{S}_{k\shortminus1} \pmb{R}^{-1}_{k\shortminus1} \, \pmb{y}_{k\shortminus1}, \label{eq:predict_est_equation} \\
    \pmb{\Sigma}_{k|k\shortminus1} &= \left(\pmb{A}_{k\shortminus1} \!- \pmb{G}_{k\shortminus1} \pmb{S}_{k\shortminus1} \pmb{R}_{k\shortminus1}^{-1} \pmb{H}_{k\shortminus1}\right) \pmb{\Sigma}_{k\shortminus1|k\shortminus1} \left(\pmb{A}_{k\shortminus1} \!- \pmb{G}_{k\shortminus1} \pmb{S}_{k\shortminus1} \pmb{R}_{k\shortminus1}^{-1} \pmb{H}_{k\shortminus1}\right)^{\Trans} \nonumber \\
    &+ \pmb{G}_{k\shortminus1}\!\left(\pmb{Q}_{k\shortminus1} \!- \pmb{S}_{k\shortminus1}\pmb{R}_{k\shortminus1}^{-1}\pmb{S}^\Trans_{k\shortminus1}\right)\!\pmb{G}_{k\shortminus1}^\Trans, \label{eq:predict_cov_corr_discrete}
\end{align}
whose mean also depends on the measurement outcome $\pmb{y}_{k\shortminus1}$. 

Since the measurement equation remains unchanged when de-correlating the process and measurement noise, the update probability density matches that of the uncorrelated case in \eqnref{eq:update_PDF}:
\begin{align}
    p(\pmb{x}_k|\pmb{y}_{0:k}) &= \Gauss\left(\pmb{x}_k|\est{\pmb{x}}_{k|k},\pmb{\Sigma}_{k|k}\right), \\
    \est{\pmb{x}}_{k|k} &= \est{\pmb{x}}_{k|k\shortminus1} + \pmb{K}^{(u)}_k (\pmb{y}_k - \pmb{H}_k \est{\pmb{x}}_{k|k\shortminus1}), \label{eq:update_est_equation}\\
    \pmb{\Sigma}_{k|k} &= (\I - \pmb{K}^{(u)}_k\pmb{H}_k) \pmb{\Sigma}_{k|k\shortminus1}, \label{eq:update_cov_corr_discrete}\\
    \pmb{K}^{(u)}_k &= \pmb{\Sigma}_{k|k\shortminus1} \pmb{H}_k^\Trans \left(\pmb{H}_k \pmb{\Sigma}_{k|k\shortminus1} \pmb{H}_k^\Trans + \pmb{R}_k\right)^{-1},
\end{align}
where we explicitly denote the Kalman gain of the uncorrelated discrete KF as $\pmb{K}^{(u)}_k$ instead of the $\pmb{K}_k$ used previously (e.g. see \tabref{tab:summary_discrete_uncorrelated_KF}). At this stage, the reason for introducing this separate notation may not be obvious, but it will become clear when we derive the corresponding gain for the correlated case. 

We now merge the predict and update equations for the estimate, namely \eqnref{eq:update_est_equation} and \eqnref{eq:predict_est_equation}, into a single recursive form:
\begin{align}
    \est{\pmb{x}}_{k+1|k} &= \left(\pmb{A}_k - \pmb{G}_k \, \pmb{S}_k \pmb{R}_k^{-1} \pmb{H}_k \right) \est{\pmb{x}}_{k|k} + \pmb{B}_k \pmb{u}_k + \pmb{G}_k \, \pmb{S}_k \, \pmb{R}_k^{-1} \, \pmb{y}_k \nonumber \\
    &= \left(\pmb{A}_k - \pmb{G}_k \, \pmb{S}_k \pmb{R}_k^{-1} \pmb{H}_k \right) \left(\est{\pmb{x}}_{k|k\shortminus1} + \pmb{K}^{(u)}_k (\pmb{y}_k - \pmb{H}_k \est{\pmb{x}}_{k|k\shortminus1})\right) + \pmb{B}_k \pmb{u}_k + \pmb{G}_k \, \pmb{S}_k \, \pmb{R}_k^{-1} \, \pmb{y}_k \nonumber \\
    &= \pmb{A}_k \est{\pmb{x}}_{k|k\shortminus1} + \pmb{B}_k \pmb{u}_k + \left(\pmb{A}_k\pmb{K}^{(u)}_k - \pmb{G}_k\pmb{S}_k\pmb{R}_k^{-1} \left(\pmb{H}_k \pmb{K}^{(u)}_k - \I \right)\right)\left(\pmb{y}_k - \pmb{H}_k \est{\pmb{x}}_{k|k\shortminus1} \right).
\end{align}
Next, we substitute the first-order approximations in $\Dt$ of  \tabref{tab:summary_discretization} into the equation above:
\begin{align}
    \est{\pmb{x}}(t_{k+1}) 
    &=\est{\pmb{x}}(t_k) + \pmb{F}(t_k) \est{\pmb{x}}(t_k) \, \Dt + \pmb{B}(t_k) \pmb{u}(t_k) \Dt + \left(\frac{1}{\Dt} \int_{t_{k\shortminus1}}^{t_k}\pmb{y}(t) \dt - \pmb{H}(t_k) \est{\pmb{x}}(t_{k}) \right) \times \nonumber \\
    &\times \left( \pmb{K}^{(u)}(t_k) - \pmb{G}(t_k)  \pmb{S}(t_k) \pmb{R}^{-1}(t_k) \right) \Dt + \mathcal{O}(\Dt),
\end{align}
which, after dividing by $\Dt$ and taking the limit of $\Dt \rightarrow 0$, yields the Kalman-Bucy equation for a correlated system:
\begin{align}
    \frac{\dd \est{\pmb{x}}}{\dt} &= \lim_{\Dt \rightarrow 0} \frac{\est{\pmb{x}}(t_{k+1}) - \est{\pmb{x}}(t_k)}{\Dt} \nonumber \\
    &= \pmb{F}(t) \est{\pmb{x}}(t) + \pmb{B}(t) \pmb{u}(t) + \pmb{K}^{(c)}(t) \left(\pmb{y}(t) - \pmb{H}(t) \est{\pmb{x}}(t) \right),
\end{align}
with $\pmb{K}^{(c)}(t)$ being the correlated Kalman gain, defined as:
\begin{equation} \label{eq:cont_corr_KG}
    \pmb{K}^{(c)}(t) \coloneqq \pmb{K}^{(u)}(t)  - \pmb{G}(t)  \pmb{S}(t) \pmb{R}^{-1}(t) =  \left( \pmb{\Sigma}(t) \, \pmb{H}^{\Trans}(t)   - \pmb{G}(t)  \pmb{S}(t) \right)\pmb{R}^{-1}(t). 
\end{equation}

Having derived the Kalman–Bucy equation for the state estimate, we now turn to the covariance dynamics. Following the same procedure as before, we combine the update and prediction steps for the covariance by substituting \eqnref{eq:update_cov_corr_discrete} into \eqnref{eq:predict_cov_corr_discrete} in order to obtain a joint recursive equation for the covariance $\pmb{\Sigma}_k$:
\begin{align}
    \pmb{\Sigma}_{k+1|k} &= \left(\pmb{A}_k - \pmb{G}_k\,\pmb{S}_k\,\pmb{R}_k^{-1}\pmb{H}_k\right) \! \left(\I -\pmb{K}^{(u)}_k\,\pmb{H}_k\right) \pmb{\Sigma}_{k|k\shortminus1} \left(\pmb{A}_k - \pmb{G}_k\,\pmb{S}_k\, \pmb{R}_k^{-1}\pmb{H}_k\right)^{\!\Trans} \nonumber \\
    &+ \pmb{G}_k \left(\pmb{Q}_k - \pmb{S}_k\,\pmb{R}_k^{-1}\pmb{S}_k^\Trans\right) \pmb{G}_k^\Trans.
\end{align}
If we now replace each term in this expression with its first-order approximation in $\Dt$ (as summarized in \tabref{tab:summary_discretization}), the recursion can be rewritten in a form suitable for the continuous-time limit:
\begin{align}
    \pmb{\Sigma}(t_{k+1}) &= \left(\I + \,\,\tilde{\!\!\pmb{F}}(t_k)\Dt\right) \pmb{\Sigma}(t_k)\! \left(\I + \,\,\tilde{\!\!\pmb{F}}^{\,\Trans}\!(t_k)\Dt\right) \nonumber \\
    &+ \left(\I + \,\,\tilde{\!\!\pmb{F}}(t_k)\Dt\right) \pmb{\Sigma}(t_k) \pmb{H}^\Trans(t_k) \pmb{R}^{-1}(t_k) \Dt \, \pmb{H}(t_k) \pmb{\Sigma}(t_k) \left(\I + \,\,\tilde{\!\!\pmb{F}}^{\,\Trans}\!(t_k)\Dt\right) \nonumber \\
    &+  \pmb{G}(t_k)\!\left(\pmb{Q}(t_k)\!-\!\pmb{S}(t_k)\pmb{R}^{-1}(t_k)\pmb{S}^\Trans(t_k)\right)\!\pmb{G}^\Trans(t_k) \Dt \nonumber \\
    &= \pmb{\Sigma}(t_k) + \,\,\tilde{\!\!\pmb{F}}(t_k) \pmb{\Sigma}(t_k) \, \Dt + \pmb{\Sigma}(t_k)  \,\,\tilde{\!\!\pmb{F}}^{\,\Trans}\!(t_k)  \Dt \nonumber \\
    &+ \pmb{\Sigma}(t_k) \pmb{H}^\Trans\!(t_k) \pmb{R}^{-1}\!(t_k) \pmb{H}(t_k) \pmb{\Sigma}(t_k) \Dt \nonumber\\
    &+\!  \pmb{G}(t_k)\!\left(\pmb{Q}(t_k)\!-\!\pmb{S}(t_k)\pmb{R}^{-1}(t_k)\pmb{S}^\Trans\!(t_k)\right)\!\pmb{G}^\Trans\!(t_k) \Dt + \mathcal{O}(\Dt),
\end{align}
where $\,\,\tilde{\!\!\pmb{F}}(t_k) = \pmb{F}(t_k) - \pmb{G}(t_k)\pmb{S}(t_k)\pmb{R}^{-1}(t_k)\pmb{H}(t_k)$. By now taking the limit of $\Dt$ approaching zero, we derive a differential equation for the covariance, i.e. the Riccati equation of the correlated Kalman–Bucy filter:
\begin{align}
    \frac{\dd \pmb{\Sigma}}{\dt} &= \lim_{\Dt \rightarrow 0} \frac{\pmb{\Sigma}(t_{k+1})-\pmb{\Sigma}(t_k)}{\Dt} = \left(\pmb{F}(t) - \pmb{G}(t)\pmb{S}(t)\pmb{R}^{-1}(t)\pmb{H}(t)\right) \pmb{\Sigma}(t) \nonumber \\
    &+ \pmb{\Sigma}(t) \left(\pmb{F}(t) - \pmb{G}(t)\pmb{S}(t)\pmb{R}^{-1}(t)\pmb{H}(t)\right)^\Trans + \pmb{\Sigma}(t) \pmb{H}^\Trans\!(t) \pmb{R}^{-1}\!(t) \pmb{H}(t) \pmb{\Sigma}(t) \nonumber\\
    &+\!  \pmb{G}(t)\!\left(\pmb{Q}(t)\!-\!\pmb{S}(t)\pmb{R}^{-1}(t)\pmb{S}^\Trans\!(t)\right)\!\pmb{G}^\Trans\!(t),
\end{align}
which can be rewritten in the more compact form:
\begin{align}
    \frac{\dd\pmb{\Sigma}}{\dt} = \pmb{F}(t) \pmb{\Sigma}(t) + \pmb{\Sigma}(t) \pmb{F}^{\,\Trans}\!(t) - \pmb{K}^{(c)}(t)\pmb{R}(t){\pmb{K}^{(c)}}^\Trans\!(t) + \pmb{G}(t) \pmb{Q}(t) \pmb{G}^\Trans\!(t),
\end{align}
with the correlated Kalman gain being $\pmb{K}^{(c)}(t) = \left( \pmb{\Sigma}(t) \, \pmb{H}^{\Trans}(t)   - \pmb{G}(t)  \pmb{S}(t) \right)\pmb{R}^{-1}(t)$, as defined in \eqnref{eq:cont_corr_KG}. From this point onward, we simply write $\pmb{K}$ for both correlated and uncorrelated Kalman gains. The context will indicate whether we are dealing with the correlated or uncorrelated case, so no superscripts are needed. 

\begin{table}[h!]
\begin{center}
\renewcommand{\arraystretch}{2.5} 
\setlength{\extrarowheight}{0pt} 
\begin{tabular}{|>{\centering\arraybackslash}m{2.5cm}|>{\centering\arraybackslash}m{12cm}|}
\hline
\vspace{0.4cm} \textbf{Model} \vspace{0.5cm} & \vspace{0.4cm} 
\makecell[c]{$\dot{\pmb{x}}(t) = \pmb{F}(t) \pmb{x}(t) \!+\! \pmb{B}(t) \pmb{u}(t) \!+\! \pmb{G}(t) \pmb{w}(t), \;\; \pmb{w}(t) \sim \Gauss(0, \pmb{Q}(t))$ \\ \vspace{-0.3cm} \\$\pmb{y}(t) = \pmb{H}(t) \pmb{x}(t) \!+\! \pmb{v}(t), \;\; \pmb{v}(t) \sim \Gauss(0, \pmb{R}(t))$ \\ \vspace{-0.3cm} \\ $\EE{\pmb{w}(t) \pmb{v}^\Trans(s)} = \pmb{S}(t) \delta(t-s)$} \vspace{0.3cm} \\
\hline
\textbf{Initialize} \vspace{1pt}& 
\makecell[c]{$\est{\pmb{x}}(t_0) = \est{\pmb{x}}_{0}$ \\ $\pmb{\Sigma}_{0} = \EE{ (\tilde{\pmb{x}}_{0} - \pmb{x}_0) (\tilde{\pmb{x}}_{0} - \pmb{x}_0)^{\!\Trans} }$} \vspace{1pt} \\
\hline
\textbf{Gain} \vspace{1pt}& 
$\pmb{K}(t) = \left( \pmb{\Sigma}(t) \, \pmb{H}^{\Trans}(t) \!-\! \pmb{G}(t)  \pmb{S}(t) \right)\pmb{R}^{-1}(t)$ \vspace{1pt}\\
\hline 
\textbf{Covariance} \vspace{1pt}& 
\makecell[c]{$\dot{\pmb{\Sigma}}(t) \!=\!  \pmb{F}(t) \pmb{\Sigma}(t) \!+\! \pmb{\Sigma}(t) \pmb{F}^{\,\Trans}\!(t) \!-\! \pmb{K}(t)\pmb{R}(t)\pmb{K}^\Trans\!(t) \!+\! \pmb{G}(t) \pmb{Q}(t) \pmb{G}^\Trans\!(t)$} \vspace{1pt}\\
\hline
\textbf{Estimate} \vspace{1pt} & 
\makecell[c]{$\dot{\est{\pmb{x}}}(t) \!=\! \pmb{F}(t)\est{\pmb{x}}(t) \!+\! \pmb{B}(t)\pmb{u}(t) \!+\! \pmb{K}(t)\!\left( \pmb{y}(t) \!-\! \pmb{H}(t)\est{\pmb{x}}(t) \right)$} \vspace{1pt}\\
\hline
\end{tabular}
\caption[Summary of the continuous correlated Kalman filter]{Summary of the continuous correlated Kalman filter.}
\label{tab:summary_continuous_correlated_KF}
\end{center}
\end{table}

\subsection{Orthogonality principle}

An important property used later to prove the optimality of the linear-quadratic Gaussian controller is the orthogonality between the MMSE estimate and its error, which also applies to the KF estimate since its the MMSE estimator for LG systems.

\begin{property}[Orthogonality Principle of the MMSE estimate] Let $\pmb{x}(t)$ be the true state of a system and $\est{\pmb{x}}(t)$ its MMSE estimate. The corresponding estimation error is defined as
    \begin{equation}
        \pmb{e}(t) = \est{\pmb{x}}(t) - \pmb{x}(t).
    \end{equation}
    The orthogonality principle states that this error $\pmb{e}(t)$ is uncorrelated with the estimate $\est{\pmb{x}}(t)$ itself. Formally,
    \begin{align} \label{eq:ortho_KF}
        \EE{\est{\pmb{x}}(t) \,\pmb{e}^\Trans\!(t)} = 0,
    \end{align}
    where the expectation value is taken w.r.t. the total probability distribution $p(\pmb{x},\pmb{y}_{t})$. In other words, the cross-correlation between the estimate and its error is zero. 
\end{property}

\begin{myproof}
    First, recall that the optimal estimate minimizing the MSE or quadratic cost function is the mean of the posterior \eqref{eq:mean_of_the_posterior}, i.e. the MMSE estimator:
    \begin{equation}
        \est{\pmb{x}}(t) = \int \pmb{x}(t) \, p(\pmb{x}|\pmb{y}_{t}) \dd \pmb{x} = \E{\pmb{x}(t)}{\pmb{x}|\pmb{y}_{t}},
    \end{equation}
    which is a function of the observations $\pmb{y}_{t}$, i.e. $\est{\pmb{x}}(t) = \est{\pmb{x}}(\pmb{y}_{t})$. We can rewrite the expectation product in \eqnref{eq:ortho_KF} as:
    \begin{align}
        \E{\est{\pmb{x}}(t)\, \pmb{e}^\Trans\!(t)}{\pmb{x},\pmb{y}_{t}} &=  \E{\est{\pmb{x}}(t)\,\est{\pmb{x}}^\Trans\!(t)}{\pmb{y}_{t}} - \E{\est{\pmb{x}}(t) \,\pmb{x}^\Trans\!(t)}{\pmb{x},\pmb{y}_{t}} \\
        &=  \E{\est{\pmb{x}}(t)\,\est{\pmb{x}}^\Trans\!(t)}{\pmb{y}_{t}} - \E{\est{\pmb{x}}(t) \,\E{\pmb{x}^\Trans\!(t)}{\pmb{x}|\pmb{y}_{t}}}{\pmb{y}_{t}} \\
        &= \E{\est{\pmb{x}}(t)\,\est{\pmb{x}}^\Trans\!(t)}{\pmb{y}_{t}} - \E{\est{\pmb{x}}(t)\,\est{\pmb{x}}^\Trans\!(t)}{\pmb{y}_{t}} = 0,
    \end{align}
    where we have used $\E{\est{\pmb{x}}(t)\,\est{\pmb{x}}^\Trans\!(t)}{\pmb{x},\pmb{y}_{t}} = \E{\est{\pmb{x}}(t)\,\est{\pmb{x}}^\Trans\!(t)}{\pmb{y}_{t}}$ since $\est{\pmb{x}}(t)$ is a function only of $\pmb{y}_{t}$.
\end{myproof}

\section{Extended Kalman filter} \label{sec:EKF_theory}

The derivation of the KF so far has relied on the assumption that the model is linear. If that is not the case and the model is nonlinear, a Gaussian input does not necessarily produce a Gaussian output, unlike in the linear scenario. As a result, the linear KF may no longer be the optimal estimator, as there might be nonlinear filters that produce a better estimate. There are a wide range of nonlinear filters one could choose from, but the simplest, most  natural step is to consider nonlinear extensions of the KF, such as the linearized or extended Kalman filter (EKF), even though they are not guaranteed to provide an optimal estimate.

To derive the linearized KF equations, we must first consider the nonlinear model given by a set of coupled nonlinear stochastic equations of the form:
\begin{align}
    \dot{\pmb{x}}(t) &= \pmb{f}\,[\pmb{x}(t),\pmb{u}(t),\pmb{w}(t),t], \label{eq:nonlinear_f} \\
    \pmb{y}(t) &= \pmb{h}[\pmb{x}(t),\pmb{v}(t),t],  \label{eq:nonlinear_h}
\end{align}
where $\pmb{f}$ and $\pmb{h}$ are both continuously differentiable nonlinear functions, $\pmb{x}(t)$ and $\pmb{u}(t)$ are the state and control vectors, and  $\pmb{w}(t)$ and $\pmb{v}(t)$ denote Langevin-noise terms, whose noise covariances fulfill the same relations as the ones specified for a linear and correlated system, detailed in \eqnsref{eq:cov_Q_ch1}{eq:cov_R_ch1} and \eqnref{eq:cont_corr_cov_S}. 

Next, we expand the process and measurement equations around a nominal trajectory:
\begin{equation}
    (\pmb{x}_{\pmb{0}}(t),\pmb{u}_{\pmb{0}}(t),\pmb{w}_{\pmb{0}}(t),\pmb{v}_{\pmb{0}}(t)),
\end{equation}
i.e. nominal values for the state, control and noises. A nominal trajectory is simply \emph{a-priori} guess of what the system trajectory might look like, e.g. either a pre-planned trajectory, such as a flight trajectory, or the actual KF estimate. Crucially, the nominal trajectory should be as close as possible to the real trajectory so the following linearization approximately holds:
\begin{align}
    \pmb{\dot{x}}(t) &\approx \pmb{f}\,[\pmb{x_0}(t),\pmb{u_0}(t),\pmb{w_0}(t),t] + \nabla_{\pmb{x}}\, \pmb{f}\,|_{\pmb{0}} (\pmb{x}(t) - \pmb{x_0}(t))  \nonumber \\
    &+ \nabla_{\pmb{u}}\,\pmb{f}\,|_{\pmb{0}} (\pmb{u}(t) - \pmb{u_0}(t)) + \nabla_{\pmb{w}} \,\pmb{f} \,|_{\pmb{0}} (\pmb{w}(t) - \pmb{w_0}(t)) \nonumber \\
    &=  \pmb{f}\,[\pmb{x_0}(t),\pmb{u_0}(t),\pmb{w_0}(t),t] + \nabla_{\pmb{x}}\, \pmb{f} \, |_{\pmb{0}} \Delta\pmb{x}(t) + \nabla_{\pmb{u}}\, \pmb{f} \, |_{\pmb{0}} \Delta\pmb{u}(t) + \nabla_{\pmb{w}}\, \pmb{f} \, |_{\pmb{0}} \Delta\pmb{w}(t), \label{eq:linearization_x} \\
    \pmb{y}(t) &\approx  \pmb{h}[\pmb{x_0}(t),\pmb{v_0}(t),t] + \nabla_{\pmb{x}} \, \pmb{h} |_{\pmb{0}} (\pmb{x}(t) - \pmb{x_0}(t)) + \nabla_{\pmb{v}} \, \pmb{h} |_{\pmb{0}} (\pmb{v}(t) - \pmb{v_0}(t)) = \nonumber \\
    &= \pmb{h}[\pmb{x_0}(t),\pmb{v_0}(t),t] + \nabla_{\pmb{x}} \, \pmb{h} |_{\pmb{0}} \Delta \pmb{x}(t) + \nabla_{\pmb{v}} \, \pmb{h} |_{\pmb{0}} \Delta \pmb{v}(t), \label{eq:linearization_y}
\end{align}
where $\nabla_{\pmb{x}} \,\pmb{f} \, |_{\pmb{0}}$ denotes the Jacobian of the function $\pmb{f}\,[\pmb{x}(t),\pmb{u}(t),\pmb{w}(t),t]$ with respect to the variable $\pmb{x}$ and evaluated at the nominal values $(\pmb{x_0}(t),\pmb{u_0}(t),\pmb{w_0}(t))$, or for the case of the measurement function $\pmb{h}(\pmb{x}(t),\pmb{v}(t))$, at $(\pmb{x_0}(t),\pmb{v_0}(t))$. Additionally, note that we have introduced the notation $\Delta \pmb{x}(t)$, $\Delta \pmb{u}(t)$, $\Delta \pmb{w}(t),$ and $\Delta \pmb{v}(t)$ to represent the deviation of the real trajectory from the nominal trajectory. Since it is reasonable to assume that the control $\pmb{u}(t)$ is known at all times, we can set the nominal control to the actual value $\pmb{u_0}(t) = \pmb{u}(t)$ s.t. $\Delta \pmb{u}(t) = 0$. Furthermore, we assume that the nominal noise for the process and measurement are both zero at all times: $\pmb{w_0} (t)= 0$ and $\pmb{v_0}(t) = 0$, s.t. $\Delta\pmb{w}(t) = \pmb{w}(t)$ and $\Delta\pmb{v}(t) = \pmb{v}(t)$. Thus, from now on, the nominal values considered are  $(\pmb{x_0}(t),\pmb{u_0}(t),\pmb{w_0}(t),\pmb{v_0}(t)) = (\pmb{x_0}(t),\pmb{u}(t),0,0)$. 

Starting from \eqnsref{eq:linearization_x}{eq:linearization_y}, we can write an equivalent system of SDEs for the deviation of the state from its nominal value, $\Delta \pmb{x}(t)$, as well as the deviation of the measurement, $\Delta \pmb{y}(t)$:
\begin{align}
    \Delta \pmb{\dot{x}}(t) &= \pmb{\dot{x}}(t) - \pmb{\dot{x}_0}(t) = \nabla_{\pmb{x}} \,\pmb{f} \,|_{\pmb{0}} \, \Delta\pmb{x}(t) + \nabla_{\pmb{w}}\, \pmb{f} \,|_{\pmb{0}} \, \pmb{w}(t) = \nabla_{\pmb{x}} \,\pmb{f} \,|_{\pmb{0}} \, \Delta\pmb{x}(t) + \pmb{q}(t), \label{eq:process_Deltax}\\
    \Delta \pmb{y}(t) &= \pmb{y}(t) - \pmb{y_0}(t) = \nabla_{\pmb{x}} \, \pmb{h} |_{\pmb{0}} \, \Delta \pmb{x}(t) + \nabla_{\pmb{v}} \, \pmb{h} |_{\pmb{0}} \, \pmb{v}(t) = \nabla_{\pmb{x}} \, \pmb{h} |_{\pmb{0}} \, \Delta \pmb{x}(t) + \pmb{r}(t), \label{eq:measurement_Deltay}
\end{align}
where we have employed that the nominal trajectory fulfills: $\dot{\pmb{x}}_{\pmb{0}}(t) \!=\! \pmb{f}[\pmb{x_0}(t),\pmb{u_0}(t),\pmb{w_0}(t),t]$ and $\pmb{y_0}(t) = \pmb{h}[\pmb{x_0}(t),\pmb{v_0}(t),t]$. Moreover, both noise terms have been relabeled as:
\begin{align}
    \pmb{q}(t) \coloneqq \nabla_{\pmb{w}}\, \pmb{f} \,|_{\pmb{0}} \, \pmb{w}(t), \quad \text{and} \quad
    \pmb{r}(t) \coloneqq \nabla_{\pmb{v}} \, \pmb{h} |_{\pmb{0}} \, \pmb{v}(t),
\end{align}
with covariances
\begin{align}
    \EE{\pmb{q}(t)\pmb{q}^\Trans\!(s)} &= \tilde{\pmb{Q}}(t) \delta(t-s) \dt ,  &&\text{where} \quad\quad \tilde{\pmb{Q}}(t) \coloneqq \nabla_{\pmb{w}} \,\pmb{f} \,|_{\pmb{0}} \, \pmb{Q}(t) \nabla_{\pmb{w}} \,\pmb{f}^{\,\Trans} |_{\pmb{0}}, \label{eq:q_tilda_EKF}\\
    \EE{\pmb{r}(t)\pmb{r}^\Trans\!(s)} &= \tilde{\pmb{R}}(t) \delta(t-s) \dt ,  &&\text{where} \quad\quad  \tilde{\pmb{R}}(t) \coloneqq \nabla_{\pmb{v}} \, \pmb{h} |_{\pmb{0}} \, \pmb{R}(t)  \nabla_{\pmb{v}} \, \pmb{h}^\Trans |_{\pmb{0}}, \label{eq:r_tilda_EKF} \\
    \EE{\pmb{q}(t)\pmb{r}^\Trans\!(s)} &= \,\tilde{\!\pmb{S}}(t) \delta(t-s) \dt, &&\text{where} \quad\quad  \,\tilde{\!\pmb{S}}(t) \coloneqq \nabla_{\pmb{w}} \,\pmb{f} \,|_{\pmb{0}}  \, \pmb{S}(t)  \nabla_{\pmb{v}} \,\pmb{h}^\Trans |_{\pmb{0}}. \label{eq:s_tilda_EKF}
\end{align}
Crucially, \eqnsref{eq:process_Deltax}{eq:measurement_Deltay} are linear with respect to the state deviation $\Delta\pmb{x}$, which is now the state variable. Therefore, we can use the standard KF to find an estimate for $\Delta \pmb{x}(t)$, i.e. $\Delta\est{\pmb{x}}(t)$:
\begin{align}
    \Delta \dot{\est{\pmb{x}}}(t) &=  \pmb{F}(t) \Delta\est{\pmb{x}} + \pmb{K}(t) \left(\Delta \pmb{y} - \pmb{H}(t) \Delta \est{\pmb{x}} \right), \\
    \dot{\pmb{\Sigma}}(t) &= \pmb{F}(t) \pmb{\Sigma}(t) + \pmb{\Sigma}(t) \pmb{F}^\Trans\!(t) - \pmb{K}(t) \tilde{\pmb{R}}(t) \pmb{K}^\Trans\!(t) + \tilde{\pmb{Q}}(t), \label{eq:cov_ric_linearized}\\
    \pmb{K}(t) &= \left(\pmb{\Sigma}(t) \pmb{H}^\Trans\!(t) -  \,\tilde{\!\pmb{S}}(t)\right)\tilde{\pmb{R}}^{-1}\!(t) \label{eq:KG_linearized},
\end{align}
where $\pmb{\Sigma}(t)$ is the covariance matrix defined as $\pmb{\Sigma}(t) \!=\! \EE{(\Delta\pmb{x}(t) \!\shortminus\! \Delta\est{\pmb{x}}(t))(\Delta\pmb{x}(t) \!\shortminus\! \Delta\est{\pmb{x}}(t))^\Trans}$, and the matrices given by the model which define the KF are:
\begin{align}
    \pmb{F}(t) \coloneqq  \nabla_{\pmb{x}} \, \pmb{f}|_{\pmb{0}}, \quad \pmb{H}(t) \coloneqq  \nabla_{\pmb{x}} \, \pmb{h}|_{\pmb{0}}, \quad \text{and} \quad \pmb{G}(t) \coloneqq  \nabla_{\pmb{w}} \, \pmb{f}|_{\pmb{0}},
\end{align}
with the equations for the covariances $\tilde{\pmb{Q}}(t)$, $\tilde{\pmb{R}}(t)$ and $\tilde{\pmb{S}}(t)$, given in \eqnsref{eq:q_tilda_EKF}{eq:s_tilda_EKF}. It follows from the definition of $\Delta \est{\pmb{x}}$ that the estimate of $\pmb{x}$ is simply the nominal state trajectory plus the deviation of the estimate from this trajectory:
\begin{align}
    \est{\pmb{x}}(t) = \pmb{x_0}(t) + \Delta \est{\pmb{x}}(t).
\end{align}
Therefore, we can combine the nominal state trajectory with the Kalman-Bucy equation to obtain an update rule for the estimate of $\pmb{x}$. Namely,
\begin{align}
    \dot{\est{\pmb{x}}}(t) \!&= \dot{\est{\pmb{x}}}_{\pmb{0}}(t) + \Delta \dot{\est{\pmb{x}}}(t) \nonumber \\
    \!&= \pmb{f}\,[\pmb{x_0}(t),\pmb{u}(t),\!0,t] \!+\! \nabla_{\pmb{x}} \, \pmb{f}\,|_{\pmb{0}} \Delta \est{\pmb{x}}(t) \!+\! \pmb{K}(t) \! \left(\pmb{y}(t) \!-\! \pmb{h}[\pmb{x}_{\pmb{0}}(t),\!0,t] \!-\! \nabla_{\pmb{x}} \, \pmb{h} |_{\pmb{0}} \Delta \est{\pmb{x}}(t) \right)\!, 
\end{align}
where we have used that $\dot{\pmb{x}}_{\pmb{0}}(t) = \pmb{f}[\pmb{x_0}(t),\pmb{u_0}(t),\pmb{w_0}(t),t]$, $\pmb{y_0}(t) = \pmb{h}[\pmb{x_0}(t),\pmb{v_0}(t),t]$, and previous assumptions $\pmb{w}_{\pmb{0}}(t) = \pmb{v}_{\pmb{0}}(t) = 0$ and $\pmb{u}_{\pmb{0}}(t) = \pmb{u}(t)$.

A common issue when employing a linearized KF is that finding a nominal state trajectory is not straightforward. Therefore, a standard workaround is to assume that the KF estimate itself is the trajectory around which we linearize the system, i.e. $\pmb{x_0}(t) \coloneqq \est{\pmb{x}}(t)$. Then, the update rule for the estimate of $\pmb{x}(t)$ becomes:
\begin{align}
    \dot{\est{\pmb{x}}}(t) = \pmb{f}\,[\est{\pmb{x}}(t),\pmb{u}(t),0,t] + \pmb{K}(t) \left(\pmb{y}(t) - \pmb{h}[\est{\pmb{x}}(t),0,t]\right),
\end{align}
since $\Delta \est{\pmb{x}}(t) = 0$, due to our redefinition of the nominal state trajectory as the estimate trajectory. The Kalman gain and the covariance $\pmb{\Sigma}(t) \!=\! \EE{(\pmb{x}(t) - \est{\pmb{x}}(t))(\pmb{x}(t) - \est{\pmb{x}}(t))^{\!\Trans}}$ follow the same equations as in \eqnsref{eq:cov_ric_linearized}{eq:KG_linearized}.

\begin{table}[h!]
\begin{center}
\renewcommand{\arraystretch}{2.5} 
\setlength{\extrarowheight}{0pt} 
\begin{tabular}{|>{\centering\arraybackslash}m{2.5cm}|>{\centering\arraybackslash}m{11cm}|}
\hline
\vspace{0.4cm} \textbf{Model} \vspace{0.5cm} & \vspace{0.4cm} 
\makecell[c]{$\dot{\pmb{x}}(t) = \pmb{f}\,[\pmb{x}(t),\pmb{u}(t),\pmb{w}(t),t], \;\; \pmb{w}(t) \sim \Gauss(0, \pmb{Q}(t))$ \\ \vspace{-0.3cm} \\$\pmb{y}(t) = \pmb{h}[\pmb{x}(t),\pmb{v}(t),t], \;\; \pmb{v}(t) \sim \Gauss(0, \pmb{R}(t))$ \\ \vspace{-0.3cm} \\ $\EE{\pmb{w}(t) \pmb{v}^\Trans(s)} = \pmb{S}(t) \delta(t-s)$} \vspace{0.3cm} \\
\hline
\textbf{Initialize} \vspace{1pt}& 
\makecell[c]{$\est{\pmb{x}}(t_0) = \est{\pmb{x}}_{0}$ \\ $\pmb{\Sigma}_{0} = \EE{ (\pmb{x}_0 - \tilde{\pmb{x}}_{0}) (\pmb{x}_0 - \tilde{\pmb{x}}_{0})^{\!\Trans} }$} \vspace{1pt} \\
\hline
\vspace{0.4cm} \textbf{Gradients} \vspace{0.5cm} & \vspace{0.4cm} 
\makecell[c]{$\pmb{F}(t) \coloneqq \nabla_{\pmb{x}} \, \pmb{f}|_{\pmb{0}}, \quad \quad \pmb{H}(t) \coloneqq  \nabla_{\pmb{x}} \, \pmb{h}|_{\pmb{0}}, \quad \quad \pmb{G}(t) \coloneqq  \nabla_{\pmb{w}} \, \pmb{f}|_{\pmb{0}}$ \\ \vspace{-0.3cm} \\ $\tilde{\pmb{Q}}(t) \coloneqq \nabla_{\pmb{w}} \,\pmb{f}|_{\pmb{0}} \, \pmb{Q}(t) \nabla_{\pmb{w}} \,\pmb{f}^{\,\,\Trans}|_{\pmb{0}}, \quad \tilde{\pmb{R}}(t) = \nabla_{\pmb{v}} \,\pmb{h}|_{\pmb{0}} \, \pmb{R}(t)  \nabla_{\pmb{v}} \,\pmb{h}^\Trans|_{\pmb{0}},$ \\ \vspace{-0.3cm} \\ $\,\tilde{\!\pmb{S}}(t) = \nabla_{\pmb{w}} \,\pmb{f}|_{\pmb{0}} \, \pmb{S}(t)  \nabla_{\pmb{v}} \,\pmb{h}^\Trans|_{\pmb{0}}$} \vspace{0.3cm} \\
\hline
\textbf{Evaluate} \vspace{1pt}& 
\makecell[c]{$\forall \ldots : \;\; \ldots|_{\pmb{0}} \coloneqq \ldots |_{(\est{\pmb{x}}(t),\,\pmb{u}(t),\,\pmb{w} = 0,\,\pmb{v} = 0)}$} \vspace{1pt} \\
\hline
\textbf{Gain} \vspace{1pt}& 
$\pmb{K}(t) =\left(\pmb{\Sigma}(t) \pmb{H}^\Trans(t) -  \,\tilde{\!\pmb{S}}(t)\right)\tilde{\pmb{R}}^{-1}(t)$ \vspace{1pt}\\
\hline 
\textbf{Covariance} \vspace{1pt}& 
\makecell[c]{$\dot{\pmb{\Sigma}}(t) \!=\!  \pmb{F}(t) \pmb{\Sigma}(t) + \pmb{\Sigma}(t) \pmb{F}^\Trans(t) - \pmb{K}(t) \tilde{\pmb{R}}(t) \pmb{K}^\Trans(t) + \tilde{\pmb{Q}}(t)$} \vspace{1pt}\\
\hline
\textbf{Estimate} \vspace{1pt} & 
\makecell[c]{$\dot{\est{\pmb{x}}}(t) \!=\! \pmb{f}\,[\est{\pmb{x}}(t),\pmb{u}(t),0,t] + \pmb{K}(t) \left(\pmb{y}(t) - \pmb{h}[\est{\pmb{x}}(t),0,t]\right)$} \vspace{1pt}\\
\hline
\end{tabular}
\caption[Summary of the continuous correlated extended Kalman filter]{Summary of the continuous correlated extended Kalman filter.}
\label{tab:summary_continuous_correlated_EKF}
\end{center}
\end{table}

\section{Control} \label{sec:control}

So far, we have explored how to estimate time-varying parameters in a Bayesian setting, with a special emphasis on how to update their estimates as more data becomes available. However, in many scenarios we do not only want to estimate the state of a system, we also want to control it.

When the full state is available, the task becomes finding an optimal control law that steers the system toward a desired state. For linear–Gaussian (LG) systems, this optimal control problem is solved by the linear-quadratic regulator (LQR). When only indirect, noisy measurements are available, estimation becomes a prerequisite for control, and the optimal strategy is to combine the LQR with a Kalman filter (KF), which leads to the linear-quadratic-Gaussian (LQG) controller.

While we present relatively direct derivations of LQR and LQG, it is important to note that the rigorous derivations and the proofs of optimality are traditionally obtained using dynamic programming (DP). In that framework, one introduces a value function that quantifies the future cumulative cost of being in a given state, and Bellman's principle of optimality is applied to derive the Hamilton–Jacobi–Bellman (HJB) equation. For LG systems with quadratic costs, assuming a quadratic value function reduces the HJB equation exactly to the Riccati equation that appears later in this chapter. Thus, the Riccati equation we will use is not merely an algebraic convenience: it is the DP/Bellman optimality condition specialized to systems with linear dynamics and quadratic cost.

Since dynamic programming forms a mathematical field of its own, and providing its full treatment would significantly expand the scope of this thesis, we instead rely on more elementary derivations that nevertheless lead to the same results. The reader interested in the formal DP approach can consult \citet{crassidis2011optimal,kolosov2020optimal,Edwards2005} for complete and elegant expositions.

\subsection{The linear-quadratic regulator} \label{sec:LQR}

\begin{figure}[htbp]
\begin{center}
    \begin{tikzpicture}[auto, node distance=2cm,>=latex]
    \tikzstyle{block} = [rectangle, draw, fill=white, 
    text centered, minimum height=2em, minimum width=4em]
    \tikzstyle{input} = [coordinate]
    
    \node [name=y_k] (y_k) {};
    \node [block, right=1cm of y_k, align=center] (controller) {Controller \\ $\pmb{u}(t) = -\pmb{K}_c(t) \, \pmb{x}(t)$};
    \node [block, right=1.5cm of controller, align=center] (system) {System \\ $\dot{\pmb{x}}(t) = \pmb{F}(t)\,\pmb{x}(t) + \pmb{B}(t)\,\pmb{u}(t)$ \\ $\pmb{y}(t) = \pmb{H}(t) \,\pmb{x}(t)$};
    \node [coordinate, right=1.5cm of system] (midpoint) {};
    \node [coordinate, right=1cm of midpoint] (output) {};
    \node [coordinate, below=2cm of y_k] (below_y_k) {};

    \draw [->] (controller) -- node [above, pos = 0.3] {$\pmb{u}(t)$} (system);
    \draw [->] (system) -- node [above, pos = 0.2] {$\pmb{y}(t)$} (output);
    \draw [->] (midpoint) |- (below_y_k) |- (y_k) |- (controller);
\end{tikzpicture}
\end{center}
\caption[A scheme illustrating the feedback loop of a linear-quadratic regulator]{\textbf{A scheme illustrating the feedback loop of a linear-quadratic regulator.} When the full state of the system is accessible, it can be steered using a controller with a control law $\pmb{u}(t)$ proportional to the state $\pmb{x}(t)$ through a control gain $\pmb{K}_c(t)$. The optimal value of $\pmb{K}_c(t)$ is determined by solving the LQR minimization problem. }
\label{fig:LQR_control_scheme}
\end{figure}
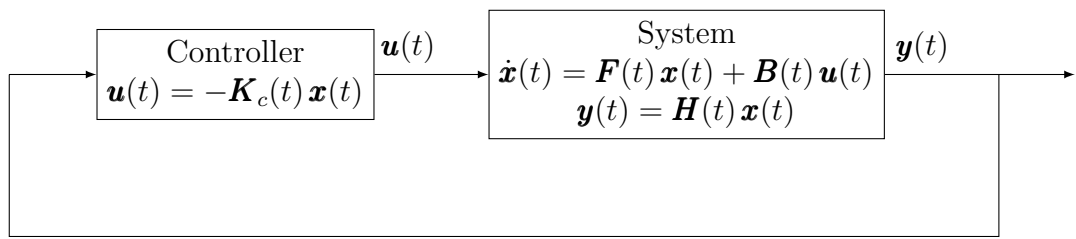

Assume first that the full state $\pmb{x}(t)$ is available. A natural control strategy is to feed back this state to modify the system dynamics:
\begin{align} \label{eq:HJB_sol}
    \pmb{u}(t) = - \pmb{K}_c(t) \pmb{x}(t),
\end{align}
where $\pmb{K}_c(t)$ is a gain matrix. The objective of the LQR is to determine the optimal gain $\pmb{K}_c(t)$ that achieves the best balance between system performance and control effort. The dynamics are assumed linear, 
\begin{align}
    \dot{\pmb{x}}(t) &= \pmb{F}(t) \, \pmb{x}(t) + \pmb{B}(t) \, \pmb{u}(t), \\
    \pmb{y}(t) &= \pmb{H}(t) \, \pmb{x}(t),
\end{align}
and the quality of a control sequence is measured by the quadratic cost:
\begin{align} \label{eq:cost_LQR}
    \pmb{J} = \int_0^\infty \left(\pmb{x}^\Trans\!(t) \,\pmb{\mathcal{P}}(t)\, \pmb{x}(t) + \pmb{u}^\Trans(t) \,\pmb{\mathcal{V}}(t)\,  \pmb{u}(t)\right)  \dt.
\end{align}
Here, the integration is carried out over time from zero to infinity, since we want the controller to continuously operate, i.e. the control horizon is infinite. In the cost function $\,\pmb{J}$, there are two control ``knobs'': the positive semi-definite matrix $\pmb{\mathcal{P}}(t) \geq 0$, which penalizes deviations in the state of the system, and $\pmb{\mathcal{V}}(t) > 0$, a positive definite matrix that penalizes deviations in the control effort. Note that the feedback form in \eqnref{eq:HJB_sol} is actually the solution of the HJB equation for linear systems and quadratic costs.

To find the optimal control law $\pmb{u}(t)$ we must minimize the cost $\;\pmb{J}$ so as to balance two competing objectives: driving the system state as close to zero as possible while minimizing the actuator effort required to achieve this. There are various methods to solve this minimization problem, including learning algorithms like gradient descent. However, we can leverage the constraints of the problem, such as the assumption of linear dynamics and quadratic cost function, to analytically solve it. 

\begin{theorem}[Linear-quadratic regulator] Consider the following minimization problem: 
    \begin{align} \label{eq:minimization_problem_LQR_objective}
        &\pmb{J}_{min} = \underset{\pmb{u}}{\arg \min} \;\; \pmb{J} = \underset{\pmb{u}}{\arg \min}  \int_0^\infty \left(\pmb{x}^\Trans\!(t) \, \pmb{\mathcal{P}}(t) \, \pmb{x}(t) + \pmb{u}^\Trans\!(t) \, \pmb{\mathcal{V}}(t) \, \pmb{u}(t)\right)  \dt \\
        &\text{subject to} \quad \dot{\pmb{x}}(t) = \pmb{F}(t) \,\pmb{x}(t)  + \pmb{B}(t) \, \pmb{u}(t), \label{eq:minimization_problem_LQR_condition}
    \end{align}
    where $\pmb{x}(t)$ is the state vector, $\pmb{\mathcal{P}}(t) \geq 0$, and $\pmb{\mathcal{V}}(t) > 0$. The optimal control law $\pmb{u}(t)$ that solves this optimization problem is known as the LQR:
    \begin{align}
        \pmb{u}(t) &= -\pmb{K}_c(t) \pmb{x}(t), \label{eq:control_law_LQR} \\
        \pmb{K}_c(t) &=  \pmb{\mathcal{V}}^{-1}(t) \pmb{B}^\Trans\!(t) \pmb{\Lambda}(t), \label{eq:control_gain_LQR}\\
        -\dot{\pmb{\Lambda}}(t) &= \pmb{F}^\Trans\!(t) \pmb{\Lambda}(t) \!+\! \pmb{\Lambda}(t) \, \pmb{F}(t) \!+\! \pmb{\mathcal{P}}(t) \!-\! \pmb{\Lambda}(t)\,\pmb{B}(t)\,\pmb{\mathcal{V}}^{-1}(t)\pmb{B}^\Trans\!(t) \pmb{\Lambda}(t), \label{eq:Riccati_eq_control_LQR}
    \end{align}
    where $\pmb{K}_c(t)$ is the control gain, and \eqnref{eq:Riccati_eq_control_LQR} is a Riccati equation with terminal condition $\pmb{\Lambda}(\infty) = 0$ to ensure that the system is continuously controlled over an infinite-time horizon. 
\end{theorem}
\begin{myproof}
To show that the optimal control law given by \eqnsref{eq:control_law_LQR}{eq:Riccati_eq_control_LQR} solves the optimization problem in \eqnsref{eq:minimization_problem_LQR_objective}{eq:minimization_problem_LQR_condition}, let us start by introducing a symmetric matrix $\pmb{\Lambda}(t) = \pmb{\Lambda}^\Trans(t)$ and rewriting the quadratic cost in \eqnref{eq:minimization_problem_LQR_objective} as:
\begin{align}
    \pmb{J} &= \pmb{x}^\Trans\!(0) \pmb{\Lambda}(0) \, \pmb{x}(0) -\pmb{x}^\Trans\!(0) \pmb{\Lambda}(0) \, \pmb{x}(0)  \nonumber \\
    &+ \int_0^\infty \!\!\left(\pmb{x}^\Trans\!(t) \pmb{\mathcal{P}}(t) \,\pmb{x}(t) + \pmb{u}^\Trans\!(t) \pmb{\mathcal{V}}(t) \, \pmb{u}(t)\right)  \dt = \pmb{x}^\Trans\!(0) \pmb{\Lambda}(0)\, \pmb{x}(0) \nonumber \\
    &+ \int_0^\infty \!\!\! \left(\! \frac{\dd}{\dt} \!\left(\pmb{x}^\Trans\!(t) \pmb{\Lambda}(t) \, \pmb{x}(t)\right) \!+\! \pmb{x}^\Trans\!(t) \pmb{\mathcal{P}}(t) \,\pmb{x}(t) \!+\! \pmb{u}^\Trans\!(t) \pmb{\mathcal{V}}(t) \,\pmb{u}(t)\!\!\right) \! \dt  ,\label{eq:J_expanded_1}
\end{align}
where $\pmb{x}(0)$ is the initial state of our system. Furthermore, we assume the state to be stable and go to zero at infinity. Then, we can bring the term $-\pmb{x}(0) \pmb{\Lambda}(0) \,\pmb{x}(0)$ into the integral because $\pmb{x}^\Trans\!(t) \pmb{\Lambda}(t)\, \pmb{x}(t)|_{0}^\infty = 0 - \pmb{x}^\Trans\!(0) \pmb{\Lambda}(0) \,\pmb{x}(0)$. The derivative in \eqnref{eq:J_expanded_1} can be expanded using the state space model as follows:
\begin{align}
    &\frac{\dd}{\dt} \left( \pmb{x}^\Trans\!(t) \pmb{\Lambda}(t) \,\pmb{x}(t) \right) =  \dot{\pmb{x}}^\Trans\!(t) \pmb{\Lambda}(t) \pmb{x}(t) + \pmb{x}^\Trans\!(t)\dot{\pmb{\Lambda}}(t)\pmb{x}(t) + \pmb{x}^\Trans\!(t) \pmb{\Lambda}(t)\dot{\pmb{x}}(t) \\
    & \quad = \pmb{x}^\Trans\!(t)\dot{\pmb{\Lambda}}(t)\pmb{x}(t) +\left(\pmb{F}(t)\pmb{x}(t) + \pmb{B}(t)\pmb{u}(t)\right)^{\!\Trans} \pmb{\Lambda}(t) \,\pmb{x}(t) \nonumber \\
    &\quad \,+ \pmb{x}^\Trans\!(t) \pmb{\Lambda}(t) \left(\pmb{F}(t)\pmb{x}(t)+\pmb{B}(t)\pmb{u}(t)\right).
\end{align}
If now we plug this into \eqnref{eq:J_expanded_1} and group the $\pmb{x}^\Trans\!(t) \, \cdot \; \pmb{x}(t)$ terms, we get:
\begin{align}
    \pmb{J} &= \pmb{x}^\Trans\!(0) \pmb{\Lambda}(0) \,\pmb{x}(0)  + \int_0^\infty \Big(\pmb{x}^\Trans\!(t)\dot{\pmb{\Lambda}}(t)\pmb{x}(t) +  \left(\pmb{F}(t)\pmb{x}(t) + \pmb{B}(t)\pmb{u}(t)\right)^{\!\Trans} \! \pmb{\Lambda}(t) \, \pmb{x}(t) \nonumber \\
    &+ \pmb{x}^\Trans\!(t) \pmb{\Lambda}(t) \left(\pmb{F}(t)\pmb{x}(t)+\pmb{B}(t)\pmb{u}(t)\right) + \pmb{x}^\Trans\!(t) \pmb{\mathcal{P}}(t)\, \pmb{x}(t) + \pmb{u}^\Trans\!(t) \pmb{\mathcal{V}}(t) \,\pmb{u}(t)\Big)  \dt  \\
    &= \pmb{x}^\Trans\!(0) \pmb{\Lambda}(t)\, \pmb{x}(0) + \int_0^\infty \Big( \pmb{x}^\Trans\!(t)\! \left( \dot{\pmb{\Lambda}}(t)+\pmb{F}^\Trans\!(t) \pmb{\Lambda}(t) + \pmb{\Lambda}(t) \pmb{F}(t) + \pmb{\mathcal{P}}(t)\right)\pmb{x}(t)  \nonumber\\
    &+ \pmb{u}^\Trans\!(t) \pmb{\mathcal{V}}(t) \,\pmb{u}(t) + \pmb{x}^\Trans\!(t) \pmb{\Lambda}(t) \pmb{B}(t) \pmb{u}(t) + \pmb{u}^\Trans\!(t) \pmb{B}^\Trans\!(t) \pmb{\Lambda}(t) \, \pmb{x}(t) \Big) \dt. \label{eq:the_last_three_terms}
\end{align}
Recall that the objective of this optimization task is to find the control $\pmb{u}(t)$ that minimizes the cost $\,\pmb{J}$. Therefore, we focus on the last three terms of \eqnref{eq:the_last_three_terms}, since they are the only ones that depend on $\pmb{u}(t)$. In particular, by completing the square, we can rewrite them as follows:
\begin{align}
    &\!\!\!\pmb{u}^{\!\Trans}\!(t) \pmb{\mathcal{V}}(t) \pmb{u}(t) \!+\! \pmb{x}^\Trans\!(t) \pmb{\Lambda}(t) \pmb{B}(t) \pmb{u}(t) \!+\! \pmb{u}^{\!\Trans}\!(t) \pmb{B}^\Trans\!(t) \pmb{\Lambda}(t) \,\pmb{x}(t) \\
    &\quad= \!\left(\pmb{u}(t)\!+\!\pmb{\mathcal{V}}(t)^{-1}\pmb{B}^{\Trans}\!(t)  \pmb{\Lambda}(t)\,\pmb{x}(t)\right)^{\!\!\Trans} \!\pmb{\mathcal{V}}(t) \!\left(\pmb{u}(t)\!+\!\pmb{\mathcal{V}}(t)^{-1}\pmb{B}^\Trans\!(t) \pmb{\Lambda}(t)\,\pmb{x}(t)\right) \nonumber \\
    &\quad - \pmb{x}^\Trans\!(t) \!\!\left(\pmb{\Lambda}(t)\,\pmb{B}(t)\pmb{\mathcal{V}}(t)^{-1}\pmb{B}^\Trans\!(t)  \pmb{\Lambda}(t)\right)\!\pmb{x}(t), 
\end{align}
and substitute them back into the cost function $\;\pmb{J}$:
\begin{align}
    \pmb{J} &= \pmb{x}^\Trans\!(0) \pmb{\Lambda}(0) \, \pmb{x}(0) \nonumber \\
    &+\int_0^\infty \!\!\! \Big(  \pmb{x}^\Trans\!(t) \!\left( \dot{\pmb{\Lambda}}(t) + \pmb{F}^\Trans\!(t) \pmb{\Lambda}(t) \!+\! \pmb{\Lambda}(t) \pmb{F}(t) \!+\! \pmb{\mathcal{P}}(t) \!-\! \pmb{\Lambda}(t)\pmb{B}(t)\pmb{\mathcal{V}}^{-1}(t)\pmb{B}^\Trans\!(t)\pmb{\Lambda}(t)\right)\!\pmb{x}(t) \nonumber \\
    &+ \left(\pmb{u}(t)\!+\!\pmb{\mathcal{V}}^{-1}(t)\pmb{B}^{\Trans}\!(t) \pmb{\Lambda}(t)\pmb{x}(t)\right)^{\!\!\Trans} \!\pmb{\mathcal{V}}(t) \!\left(\pmb{u}(t)\!+\!\pmb{\mathcal{V}}^{-1}(t)\pmb{B}^\Trans \!(t)\pmb{\Lambda}(t)\,\pmb{x}(t)\right) \! \Big) \dt. \label{eq:J_expanded_2}
\end{align}
Now, by inspection of the expression above, we can see that the cost function can be minimized by choosing the control $\pmb{u}(t)$ to be:
\begin{equation}
    \pmb{u}(t) = - \pmb{K}_c(t) \, \pmb{x}(t),
\end{equation}
where 
\begin{equation}
    \pmb{K}_c(t) = \pmb{\mathcal{V}}^{-1}(t) \pmb{B}^\Trans\!(t) \pmb{\Lambda}(t),
\end{equation}
as well as solving for $\pmb{\Lambda}(t)$ such that the first term of \eqnref{eq:J_expanded_2} is also zero, i.e.:
\begin{equation} \label{eq:ARE}
    -\dot{\pmb{\Lambda}}(t) = \pmb{F}^\Trans\!(t) \pmb{\Lambda}(t) \!+\! \pmb{\Lambda}(t) \, \pmb{F}(t) \!+\! \pmb{\mathcal{P}}(t) \!-\! \pmb{\Lambda}(t)\,\pmb{B}(t)\,\pmb{\mathcal{V}}^{-1}(t)\pmb{B}^\Trans\!(t) \pmb{\Lambda}(t).
\end{equation}
In other words, if we can find a matrix $\pmb{\Lambda}(t)$ such that the algebraic Riccati equation in \eqref{eq:ARE} holds, then the optimal control $\pmb{u}(t)$ that minimizes the cost function $\,\pmb{J}$ is a full-state feedback term of the form $\pmb{u}(t) = -\pmb{\mathcal{V}}^{-1}(t) \pmb{B}^\Trans\!(t) \pmb{\Lambda}(t) \, \pmb{x}(t)$.
\end{myproof}

\subsection{The linear-quadratic-Gaussian controller} \label{sec:LQG_controller}

Unfortunately, we often do not have access to the state $\pmb{x}(t)$. While LQR feedback requires full state knowledge, this assumption might unrealistic due to the presence of noise.  In such cases, we need to infer the state of our system from indirect noisy observations. A natural solution is to use the KF to provide state estimates, which can be later used instead of the true state to construct the LQR. For LG systems, the optimal strategy is to combine a KF with a LQR, referred to as the linear-quadratic-Gaussian (LQG) controller \cite{crassidis2011optimal,kolosov2020optimal}. 
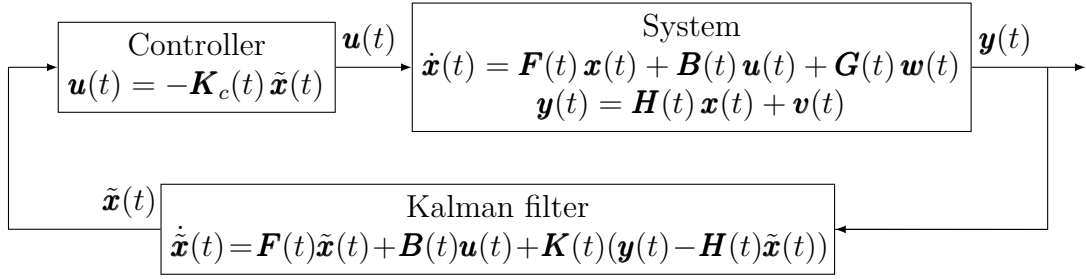
\begin{figure}[htbp]
\begin{center}
    \begin{tikzpicture}[auto, node distance=2cm,>=latex]
    \tikzstyle{block} = [rectangle, draw, fill=white, 
    text centered, minimum height=2em, minimum width=4em]
    \tikzstyle{input} = [coordinate]
    
    \node [name=y_k] (y_k) {};
    \node [block, right=0.5cm of y_k, align=center] (controller) {Controller \\ $\pmb{u}(t) = -\pmb{K}_c(t) \, \est{\pmb{x}}(t)$};
    \node [block, right=1cm of controller, align=center] (system) {System \\ $\dot{\pmb{x}}(t) = \pmb{F}(t)\,\pmb{x}(t) + \pmb{B}(t)\,\pmb{u}(t) + \pmb{G}(t) \, \pmb{w}(t)$ \\ $\pmb{y}(t) = \pmb{H}(t) \,\pmb{x}(t)  + \pmb{v}(t)$};
    \node [coordinate, right=1cm of system] (midpoint) {};
    \node [coordinate, right=0.5cm of midpoint] (output) {};
    \node [coordinate, below=2cm of y_k] (below_y_k) {};
    \node [coordinate, below=2.15cm of midpoint] (below_midpoint) {};
    
    \node [block, right = 2cm of below_y_k, align = center] (KF) {Kalman filter \\ $\dot{\est{\pmb{x}}}(t) \!=\! \pmb{F}(t)\est{\pmb{x}}(t) \!+\! \pmb{B}(t)\pmb{u}(t) \!+\! \pmb{K}(t)\!\left( \pmb{y}(t) \!-\! \pmb{H}(t)\est{\pmb{x}}(t) \right)\!$};
    \draw [->] (below_midpoint) -- (KF);
    \draw [->] (controller) -- node [above, pos = 0.45] {$\pmb{u}(t)$} (system);
    \draw [->] (system) -- node [above, pos = 0.3] {$\pmb{y}(t)$} (output);
    \draw [->] (midpoint) |- (below_midpoint) -- (KF) -- node [above, pos = 0.2] {$\est{\pmb{x}}(t)$} (below_y_k) |- (y_k) |- (controller);
\end{tikzpicture}
\end{center}
\caption[Block diagram of the feedback scheme of a linear-quadratic Gaussian controller]{\textbf{Block diagram of the feedback scheme of a linear-quadratic Gaussian controller.} The controller computes the control input $\pmb{u}(t)$ based on the estimated state $\est{\pmb{x}}(t)$, which is provided by the KF. The system dynamics are influenced by both a process noise $\pmb{w}(t)$ and a measurement noise $\pmb{v}(t)$, preventing full access to the state of the system. For that reason, the KF is needed in order to estimate the state from the measurements $\pmb{y}(t)$. The control law is then generated as $\pmb{u}(t) = -\pmb{K}_c(t) \, \est{\pmb{x}}(t)$ , where the control gain $\pmb{K}_c(t)$ is obtained by solving the LQR problem. }
\label{fig:LQG_control_scheme}
\end{figure}

\begin{theorem}[Linear-quadratic-Gaussian controller]\label{thm:lqg} The objective is to minimize the quadratic cost function:
\begin{align} \label{eq:expected_J}
    \pmb{J} = \EE{\int_{0}^\infty \left(\pmb{x}^\Trans\!(t) \pmb{\mathcal{P}}(t) \pmb{x}(t) + \pmb{u}^\Trans\!(t) \pmb{\mathcal{V}}(t) \pmb{u}(t) \right) \dt }\!,
\end{align}
subject to a linear system driven by white Gaussian noise: 
\begin{align} 
    \dot{\pmb{x}}(t) &= \pmb{F}(t) \, \pmb{x}(t) + \pmb{B}(t) \,\pmb{u}(t) + \pmb{G}(t) \, \pmb{w}(t), \\
    \pmb{y}(t) &= \pmb{H}(t) \, \pmb{x}(t) + \pmb{v}(t), 
\end{align}
where $\pmb{x}(t)$ is the state vector, $\pmb{u}(t)$ is the control input and $\pmb{y}(t)$ is the measurement output. The zero-mean Gaussian noise process $\pmb{w}(t) \sim \Gauss(0,\pmb{Q}(t))$ and $\pmb{v}(t) \sim \Gauss(0,\pmb{R}(t))$ are cross-correlated through the matrix $\pmb{S}(t)$, i.e. $\EE{\pmb{w}(t)\pmb{v}^\Trans(s)} = \pmb{S}(t)\delta(t-s)$. The new cost function is defined as the expected value of the cost function in \eqnref{eq:cost_LQR}, and just as in the case of the LQR, $\pmb{\mathcal{P}}(t) \geq 0$, and $\pmb{\mathcal{V}}(t) > 0$. The control law $\pmb{u}(t)$ that minimizes the cost function is 
\begin{equation} \label{eq:control_LQG}
    \pmb{u}(t) = -\pmb{K}_c(t) \est{\pmb{x}}(t),
\end{equation}
where now the control is not proportional to the state but rather to its estimate, $\est{\pmb{x}}(t)$. The control gain regulating the feedback is given by:
\begin{align} 
    \pmb{K}_c(t) &= \pmb{\mathcal{V}}^{-1}(t) \pmb{B}^\Trans\!(t) \pmb{\Lambda}(t), \\
    \label{eq:ARE_LQG}
    -\dot{\pmb{\Lambda}}(t) &= \pmb{F}^\Trans\!(t) \pmb{\Lambda}(t) \!+\! \pmb{\Lambda}(t) \, \pmb{F}(t) \!+\! \pmb{\mathcal{P}}(t) \!-\! \pmb{\Lambda}(t)\,\pmb{B}(t)\,\pmb{\mathcal{V}}^{-1}(t)\pmb{B}^\Trans\!(t) \pmb{\Lambda}(t), \\
    \pmb{\Lambda}(\infty) &= 0,
\end{align}
where the matrix $\pmb{\Lambda}(t)$ is the solution to a Riccati equation with an infinite control horizon. The estimator of $\pmb{x}(t)$ in \eqnref{eq:control_LQG} is given by the Kalman-Bucy filter:
\begin{align}
    \dot{\est{\pmb{x}}}(t) &= \pmb{F}(t)\est{\pmb{x}}(t) \!+\! \pmb{B}(t)\pmb{u}(t) \!+\! \pmb{K}(t)\!\left( \pmb{y}(t) \!-\! \pmb{H}(t)\est{\pmb{x}}(t) \right)\!, \\
    \est{\pmb{x}}(0) &= \EE{\pmb{x}(0) \, \pmb{x}^\Trans\!(0)}\!, \\
     \pmb{K}(t) &= \left( \pmb{\Sigma}(t) \, \pmb{H}^{\Trans}(t) \!-\! \pmb{G}(t)  \pmb{S}(t) \right)\pmb{R}^{-1}(t), \\
     \dot{\pmb{\Sigma}}(t) &=  \pmb{F}(t) \pmb{\Sigma}(t) \!+\! \pmb{\Sigma}(t) \pmb{F}^{\,\Trans}\!(t) \!-\! \pmb{K}(t)\pmb{R}(t)\pmb{K}^\Trans\!(t) \!+\! \pmb{G}(t) \pmb{Q}(t) \pmb{G}^\Trans\!(t), \nonumber \\
    \pmb{\Sigma}(0) &= \EE{\pmb{x}(0)\,\pmb{x}^\Trans\!(0)}\!,
\end{align}
where to compute the Kalman gain $\pmb{K}(t)$ one also needs to solve the dual (estimation) Riccati problem.
\end{theorem}

\begin{myproof}
First, let us focus on the term $\EE{\pmb{x}^\Trans\!(t) \pmb{\mathcal{P}}(t) \pmb{x}(t)}$ in \eqnref{eq:expected_J} and rewrite it in terms of the estimation error
\begin{equation}
    \pmb{e}(t) = \est{\pmb{x}}(t) - \pmb{x}(t). 
\end{equation}
Namely,
\begin{align}
    \EE{\pmb{x}^\Trans\!(t) \pmb{\mathcal{P}}(t) \pmb{x}(t)} &= \EE{(\est{\pmb{x}}(t) - \pmb{e}(t))^{\!\Trans} \pmb{\mathcal{P}}(t) (\est{\pmb{x}}(t) - \pmb{e}(t))} = \EE{\est{\pmb{x}}^\Trans\!(t) \pmb{\mathcal{P}}(t) \est{\pmb{x}}(t)} \nonumber \\
    &- 2 \EE{\Tr{\left[\pmb{\mathcal{P}}(t) \pmb{e}(t)\est{\pmb{x}}^\Trans\!(t)\right]}} + \EE{\Tr{\left[\pmb{\mathcal{P}}(t) \pmb{e}(t) \pmb{e}^\Trans\!(t)\right]}} \label{eq:exp_xPx_1}
\end{align}
where in the last step we have applied following property of the trace: $\Tr{\left[\pmb{A}\,\pmb{x}\,\pmb{x}^\Trans\right]} = \pmb{x}^\Trans \pmb{A} \pmb{x}$. Note that by applying the orthogonality principle, stated in \eqnref{eq:ortho_KF}, we can simplify the expression above, since
\begin{align}
    \EE{\Tr{\left[\pmb{\mathcal{P}}(t) \pmb{e}(t) \est{\pmb{x}}^\Trans\!(t)\right]}} = \Tr{\left[\pmb{\mathcal{P}}(t)\EE{\pmb{e}(t) \est{\pmb{x}}^\Trans\!(t)}\right]} = 0.
\end{align}
Additionally, by employing the definition of the covariance $\pmb{\Sigma}(t) = \EE{\pmb{e}(t)\pmb{e}^\Trans\!(t)}$, we rewrite \eqnref{eq:exp_xPx_1} as
\begin{align}
    \EE{\pmb{x}^\Trans\!(t) \pmb{\mathcal{P}}(t) \pmb{x}(t)} &= \EE{\est{\pmb{x}}^\Trans\!(t) \pmb{\mathcal{P}}(t) \est{\pmb{x}}(t)} + \Tr{\left[\pmb{\mathcal{P}}(t) \pmb{\Sigma}(t)\right]},
\end{align}
such that then, the cost function can be expressed in terms of the estimator $\est{\pmb{x}}(t)$ and covariance matrix $\pmb{\Sigma}(t)$:
\begin{align}
    \pmb{J} = \EE{\int_{0}^\infty \!\!\!\!  \left(\est{\pmb{x}}^\Trans\!(t) \pmb{\mathcal{P}}(t) \est{\pmb{x}}(t) + \pmb{u}^\Trans\!(t) \pmb{\mathcal{V}}(t) \pmb{u}(t) \right) \dt } + \int_0^\infty \!\!\!\!  \Tr{\left[\pmb{\mathcal{P}}(t) \pmb{\Sigma}(t)\right]} \dt,\label{eq:cost_J_LQG_1}
\end{align}
subject to a new constraint:
\begin{align} \label{eq:KF_LQG}
    \dot{\est{\pmb{x}}}(t) = \pmb{F}(t) \est{\pmb{x}}(t) + \pmb{B}(t) \pmb{u}(t) + \pmb{K}(t) \left(\pmb{y}(t) - \pmb{H}(t) \est{\pmb{x}}\right)\!,
\end{align}
i.e. the Kalman-Bucy filter. The problem of optimizing the cost $\,\pmb{J}$ of \eqnref{eq:cost_J_LQG_1} is equivalent to minimizing the first term of \eqnref{eq:cost_J_LQG_1}, since the second term, $\Tr{\left[\pmb{\mathcal{P}}(t) \pmb{\Sigma}(t)\right]}$, does not depend on $\pmb{u}(t)$. Namely,  
\begin{align}
    &\pmb{J}_{min} = \underset{\pmb{u}}{\arg \min} \;\; \pmb{J} =  \underset{\pmb{u}}{\arg \min} \; \; \pmb{J}^{\,\prime}
\end{align}
where
\begin{align}
    \pmb{J}^{\,\prime} = \EE{\int_{0}^\infty \!\!\!\!  \left(\est{\pmb{x}}^\Trans\!(t) \pmb{\mathcal{P}}(t) \est{\pmb{x}}(t) + \pmb{u}^\Trans\!(t) \pmb{\mathcal{V}}(t) \pmb{u}(t) \right) \dt }\!.
\end{align}
Now, instead of the optimization problem being constrained by a differential equation on the state $\pmb{x}(t)$, it is subjected to the Kalman-Bucy equation of \eqnref{eq:KF_LQG}, since we have rewritten the cost function w.r.t. the KF estimate instead of the state. Then, just like in the case of the LQR, we can use complete-the-squares to find the control $\pmb{u}(t)$ that minimizes the cost,
\begin{align}
    \pmb{J}^{\,\prime} &= \EE{\est{\pmb{x}}^\Trans\!(0)\pmb{\Lambda}(0)\est{\pmb{x}}(0)} \!- \EE{\est{\pmb{x}}^\Trans\!(0)\pmb{\Lambda}(0)\est{\pmb{x}}(0)} \!+\! \EE{\int_{0}^\infty \!\!\!\!\!  \left(\est{\pmb{x}}^\Trans\!(t) \pmb{\mathcal{P}}(t) \est{\pmb{x}}(t) \!+\! \pmb{u}^\Trans\!(t) \pmb{\mathcal{V}}(t) \pmb{u}(t) \right) \!\dt } \nonumber \\
    &= \EE{\est{\pmb{x}}^\Trans\!(0)\pmb{\Lambda}(0)\est{\pmb{x}}(0)} \!+\! \EE{\int_{0}^\infty \!\!\!\!\! \dd \!\left(\est{\pmb{x}}^\Trans\!(t)\pmb{\Lambda}(t)\est{\pmb{x}}(t)\right) \!+\!\! \int_{0}^\infty \!\!\!\!\!  \left(\est{\pmb{x}}^\Trans\!(t) \pmb{\mathcal{P}}(t) \est{\pmb{x}}(t) \!+\! \pmb{u}^\Trans\!(t) \pmb{\mathcal{V}}(t) \pmb{u}(t) \right) \!\dt }  \nonumber\\
    &= \EE{\est{\pmb{x}}^\Trans\!(0)\pmb{\Lambda}(0)\est{\pmb{x}}(0)} +\! \int_{0}^\infty \!\!\! \EE{\dd \!\left(\est{\pmb{x}}^\Trans\!(t)\pmb{\Lambda}(t)\est{\pmb{x}}(t)\right)} \nonumber \\
    &+ \int_{0}^\infty \!\!\!  \left(\EE{\est{\pmb{x}}^\Trans\!(t) \pmb{\mathcal{P}}(t) \est{\pmb{x}}(t)} \!+ \EE{\pmb{u}^\Trans\!(t) \pmb{\mathcal{V}}(t) \pmb{u}(t)} \right) \!\dt .\label{eq:cost_J_LQG_2}
\end{align}

Importantly, the KF estimate is a stochastic process that follows a SDE: the Kalman-Bucy equation. Therefore, to evaluate $\dd \!\left(\est{\pmb{x}}^\Trans\!(t)\pmb{\Lambda}(t)\est{\pmb{x}}(t)\right)$, we have to use It\^{o} calculus:
\begin{align}
    \dd \!\left(\est{\pmb{x}}^\Trans\!(t)\pmb{\Lambda}(t)\est{\pmb{x}}(t)\right) &= \est{\pmb{x}}^\Trans\!(t) \! \left(\dd \pmb{\Lambda}(t)\right)\est{\pmb{x}}(t) \!+\! \dd\est{\pmb{x}}^\Trans\!(t)\pmb{\Lambda}(t)\est{\pmb{x}}(t) \!+\! \est{\pmb{x}}^\Trans\!(t)\pmb{\Lambda}(t)\dd\est{\pmb{x}}(t)  \nonumber \\
    &+ \dd\est{\pmb{x}}^\Trans\!(t)\pmb{\Lambda}(t)\dd\est{\pmb{x}}(t) .\label{eq:xLambdax_expansion_1}
\end{align}
Then, the KF equation can be written in It\^{o} form as
\begin{align}
    \dd \est{\pmb{x}} &= \pmb{F}(t) \est{\pmb{x}}(t) \dt + \pmb{B}(t) \pmb{u}(t) \dt + \pmb{K}(t) \dd \pmb{I}, \\
    \dd \pmb{I} &= \dd \pmb{y} - \pmb{H} \, \est{\pmb{x}} \, \dt = \pmb{H} (\pmb{x} - \est{\pmb{x}}) \dt + \dd \pmb{v} = - \pmb{H} \, \pmb{e} \, \dt+ \dd \pmb{v}
\end{align}
where $\dd \pmb{v} \coloneqq \pmb{v}(t) \dt \sim \Gauss(0,\pmb{R}\dt)$, s.t. $\dd \pmb{I} \sim \Gauss(0,\pmb{R}\dt)$, with
\begin{align}
    \EE{\dd \pmb{I}} &= 0, \\
    \EE{\dd \pmb{I} \dd \pmb{I}^\Trans} &= \pmb{R}\dt, \\
    \EE{\est{\pmb{x}}\, \dd \pmb{I}^\Trans} &= 0.
\end{align}
If we apply these relationships to \eqnref{eq:xLambdax_expansion_1}, we get
\begin{align}
    \dd \!\left(\est{\pmb{x}}^\Trans\!(t)\pmb{\Lambda}(t)\est{\pmb{x}}(t)\right) &= \est{\pmb{x}}^\Trans\!(t) \! \left(\dd \pmb{\Lambda}(t)\right)\!\est{\pmb{x}}(t) \!+\! \left(\est{\pmb{x}}^\Trans\!(t) \pmb{F}^\Trans\!(t) \dt \!+\! \pmb{u}^\Trans\!(t) \pmb{B}^\Trans\!(t) \dt \!+\! \dd \pmb{I}^\Trans \pmb{K}^\Trans\!(t) \right)\!\pmb{\Lambda}(t)\est{\pmb{x}}(t)  \nonumber \\
    &+ \est{\pmb{x}}^\Trans\!(t)\pmb{\Lambda}(t) \!\left(\pmb{F}(t) \est{\pmb{x}}(t) \dt + \pmb{B}(t) \pmb{u}(t) \dt + \pmb{K}(t) \dd \pmb{I}\right) \nonumber \\ 
    &+ \dd\pmb{I}^\Trans\!(t) \pmb{K}^\Trans\!(t)\pmb{\Lambda}(t)\pmb{K}(t)\dd\pmb{I}(t),
\end{align}
the expectation value of which reads as:
\begin{align}
    &\EE{\dd \!\left(\est{\pmb{x}}^\Trans\!(t)\pmb{\Lambda}(t)\est{\pmb{x}}(t)\right)} = \EE{\est{\pmb{x}}^\Trans\!(t) \! \left(\dd \pmb{\Lambda}(t) + \pmb{F}^\Trans\!(t) \pmb{\Lambda}(t) \dt +\pmb{\Lambda}(t) \pmb{F}(t)\dt \right)\!\est{\pmb{x}}(t)} \nonumber \\ 
    &\quad+ \EE{\pmb{u}^\Trans\!(t) \pmb{B}^\Trans\!(t) \pmb{\Lambda}(t)\est{\pmb{x}}(t) \dt} + \EE{\est{\pmb{x}}^\Trans\!(t)\pmb{\Lambda}(t) \pmb{B}(t) \pmb{u}(t) \dt } \nonumber \\
    &\quad+ \EE{\dd \pmb{I}^\Trans \pmb{K}^\Trans\!(t) \pmb{\Lambda}(t)\est{\pmb{x}}(t) + \est{\pmb{x}}^\Trans\!(t)\pmb{\Lambda}(t) \pmb{K}(t) \dd \pmb{I}} + \EE{\dd\pmb{I}^\Trans\!(t) \pmb{K}^\Trans\!(t)\pmb{\Lambda}(t)\pmb{K}(t)\dd\pmb{I}(t)}. \label{eq:xLambdax_expansion_2}
\end{align}
The last two terms of the expansion above can be simplified since $\pmb{x}^\Trans \pmb{A} \pmb{z} = \Tr{[\pmb{A} \pmb{z} \pmb{x}^\Trans]}$ when $\pmb{x}$ and $\pmb{z}$ have the same dimensions, which is the case for $\dd \pmb{I}$ and $\est{\pmb{x}}(t)$. Therefore,
\begin{align}
    &\EE{\dd \pmb{I}^\Trans \pmb{K}^\Trans\!(t) \pmb{\Lambda}(t)\est{\pmb{x}}(t) + \est{\pmb{x}}^\Trans\!(t)\pmb{\Lambda}(t) \pmb{K}(t) \dd \pmb{I}} = \EE{ \Tr{\left[ \pmb{K}^\Trans\!(t) \pmb{\Lambda}(t)\est{\pmb{x}}(t) \dd \pmb{I}^\Trans\right]}} \\
    &\quad+ \EE{ \Tr{\left[\pmb{\Lambda}(t) \pmb{K}(t) \dd \pmb{I} \, \est{\pmb{x}}^\Trans\!(t)\right]}} =  \Tr{\left[ \pmb{K}^\Trans\!(t) \pmb{\Lambda}(t) \EE{\est{\pmb{x}}(t) \dd \pmb{I}^\Trans}\right]} \\
    &\quad+ \Tr{\left[\pmb{\Lambda}(t) \pmb{K}(t) \EE{  \dd \pmb{I} \, \est{\pmb{x}}^\Trans\!(t)}\right]} = 0,
\end{align}
and
\begin{align}
    &\EE{\dd\pmb{I}^\Trans\!(t) \pmb{K}^\Trans\!(t)\pmb{\Lambda}(t)\pmb{K}(t)\dd\pmb{I}(t)} = \EE{\Tr{ \left[\pmb{K}^\Trans\!(t)\pmb{\Lambda}(t)\pmb{K}(t)\dd\pmb{I}(t) \dd\pmb{I}^\Trans\!(t)\right]}} \\
    &\quad= \Tr{ \left[\pmb{K}^\Trans\!(t)\pmb{\Lambda}(t)\pmb{K}(t) \EE{\dd\pmb{I}(t) \dd\pmb{I}^\Trans\!(t)}\right]} = \Tr{ \left[\pmb{K}^\Trans\!(t)\pmb{\Lambda}(t)\pmb{K}(t) \pmb{R}(t) \right]} \dt.
\end{align}
Thus, \eqnref{eq:xLambdax_expansion_2} reduces to
\begin{align}
    &\!\!\!\!\!\!\EE{\dd \!\left(\est{\pmb{x}}^\Trans\!(t)\pmb{\Lambda}(t)\est{\pmb{x}}(t)\right)} = \EE{\est{\pmb{x}}^\Trans\!(t) \! \left(\dd \pmb{\Lambda}(t) + \pmb{F}^\Trans\!(t) \pmb{\Lambda}(t) \dt +\pmb{\Lambda}(t) \pmb{F}(t)\dt \right)\!\est{\pmb{x}}(t)} \nonumber \\ 
    &\!\!+ \EE{\pmb{u}^{\!\Trans}\!(t) \pmb{B}^\Trans\!(t) \pmb{\Lambda}(t)\est{\pmb{x}}(t) \dt \!+\! \est{\pmb{x}}^\Trans\!(t)\pmb{\Lambda}(t) \pmb{B}(t) \pmb{u}(t) \dt } \!\!+\!  \Tr{ \left[\pmb{K}^\Trans\!(t)\pmb{\Lambda}(t)\pmb{K}(t) \pmb{R}(t) \dt \right]} ,
\end{align}
which can finally be inserted back into \eqnref{eq:cost_J_LQG_2}:
\begin{align}
    \pmb{J}^{\,\prime} &= \EE{\est{\pmb{x}}^\Trans\!(0)\pmb{\Lambda}(0)\est{\pmb{x}}(0)} \!+\!\! \int_{0}^\infty \!\!\!\!\! \EE{\est{\pmb{x}}^\Trans\!(t) \! \left(\dd \pmb{\Lambda}(t) \!+\! \pmb{F}^\Trans\!(t) \pmb{\Lambda}(t) \dt \!+\! \pmb{\Lambda}(t) \pmb{F}(t)\dt \!+\! \pmb{\mathcal{P}}(t)\dt \right)\!\est{\pmb{x}}(t)} \nonumber \\
    &+ \int_{0}^\infty \!\!\!  \left(\EE{\pmb{u}^{\!\Trans}\!(t) \pmb{B}^\Trans\!(t) \pmb{\Lambda}(t)\est{\pmb{x}}(t) \dt \!+\! \est{\pmb{x}}^\Trans\!(t)\pmb{\Lambda}(t) \pmb{B}(t) \pmb{u}(t) \dt \!+\! \pmb{u}^\Trans\!(t) \pmb{\mathcal{V}}(t) \pmb{u}(t) \dt} \right) \nonumber \\
    &+\int_{0}^\infty \!\!\! \Tr{ \left[\pmb{K}^\Trans\!(t)\pmb{\Lambda}(t)\pmb{K}(t) \pmb{R}(t) \right]} \dt.
\end{align}
Given that also the last term, $\Tr{ \left[\pmb{K}^\Trans\!(t)\pmb{\Lambda}(t)\pmb{K}(t) \pmb{R}(t) \right]} $, does not depend on $\pmb{u}(t)$, we can equivalently write the minimization problem as 
\begin{align}
    &\pmb{J}_{min} = \underset{\pmb{u}}{\arg \min} \;\; \pmb{J} =  \underset{\pmb{u}}{\arg \min} \; \; \pmb{J}^{\,\prime\prime},
\end{align}
where
\begin{align}
    \pmb{J}^{\,\prime\prime} \! &= \!\! \int_{0}^\infty \!\!\!\!\! \EE{\est{\pmb{x}}^\Trans\!(t) \! \left(\frac{\dd}{\dt} \pmb{\Lambda}\!(t) \!+\! \pmb{F}^\Trans\!(t) \pmb{\Lambda}\!(t) \!+\! \pmb{\Lambda}\!(t) \pmb{F}(t) \!+\! \pmb{\mathcal{P}}(t) \!-\! \pmb{\Lambda}\!(t)\pmb{B}(t)\pmb{\mathcal{V}}^{\shortminus1}\!(t)\pmb{B}^\Trans\!(t)\pmb{\Lambda}\!(t) \!\! \right)\!\est{\pmb{x}}(t)} \! \dt \nonumber \\
    &+ \!\! \int_{0}^\infty \!\!\!\!\!  \EE{\left(\pmb{u}(t)\!+\!\pmb{\mathcal{V}}^{-1}(t)\pmb{B}^{\Trans}\!(t) \pmb{\Lambda}\!(t)\est{\pmb{x}}(t)\right)^{\!\!\Trans} \!\pmb{\mathcal{V}}(t) \!\left(\pmb{u}(t)\!+\!\pmb{\mathcal{V}}^{-1}(t)\pmb{B}^\Trans \!(t)\pmb{\Lambda}\!(t)\,\est{\pmb{x}}(t)\right)}\! \dt .
\end{align}

Then, just like in the LQR case, the control function $\pmb{u}(t)$ that minimizes the cost $\,\pmb{J}$ is 
\begin{align}
    \pmb{u}(t) = -\pmb{K}_c(t) \est{\pmb{x}}(t),
\end{align}
where the control gain is:
\begin{equation}
    \pmb{K}_c(t) = \pmb{\mathcal{V}}^{-1}(t) \pmb{B}^\Trans\!(t) \pmb{\Lambda}(t).
\end{equation}
To determine $\pmb{\Lambda}(t)$, we require the first term of \eqnref{eq:J_expanded_2} to be zero, resulting in the following equation:
\begin{equation} \label{eq:ARE_LQG_proof}
    -\dot{\pmb{\Lambda}}(t) = \pmb{F}^\Trans\!(t) \pmb{\Lambda}(t) \!+\! \pmb{\Lambda}(t) \, \pmb{F}(t) \!+\! \pmb{\mathcal{P}}(t) \!-\! \pmb{\Lambda}(t)\,\pmb{B}(t)\,\pmb{\mathcal{V}}^{-1}(t)\pmb{B}^\Trans\!(t) \pmb{\Lambda}(t),
\end{equation}
where $\pmb{\Lambda}(\infty) = 0$ because the actuator is intended to continuously control the system over an infinite time horizon.
\end{myproof}

\chapter{Continuously monitored quantum systems} \label{chap:cm}

\newthought{To learn something about a quantum system}, like where a particle is, how fast it is moving or how much energy it has, we must perform a measurement. Measurements in quantum mechanics are fundamentally different form those in classical systems: instead of simply observing the system, they interact with it in a way that disturbs its state.


The way quantum measurements are usually introduced in a first Quantum Mechanics course follows the framework of projective measurements developed in the 1930s by Heisenberg, Dirac and von Neumann \cite{Heisenberg1949_translation,Dirac1930,VonNeumann2018_translation}. Von Neumann described measurements using observables, i.e. mathematical objects called self-adjoint operators that live in the Hilbert space of the system. When an observable is measured, the possible outcomes correspond to its eigenvalues, and the quantum state undergoes a discontinuous, probabilistic, and non-unitary transformation, often referred to as ``collapse'' or ``projection'', into an eigenstate associated with the observed eigenvalue. Once this measurement is performed, there is no ambiguity on what the value of the observable is, since the state has been projected onto an eigenstate. While this framework works well for some idealized scenarios: perfect, instantaneous measurements on isolated systems, real-life experiments often deviate from these assumptions. For instance, measurements can be imperfect, i.e. extract only partial information, and non-instantaneous, as any measurement device inevitably requires a finite amount of time to detect and output a measurement value. 


To model non-ideal measurements, Davies \cite{Davies1976} and Kraus \cite{Kraus1983} extended von Neumann’s framework to include positive operator-valued measures (POVMs). Unlike the sharp ``collapse'' described by von Neumann, POVMs allow for imperfect measurements where the state of the system only partially collapses upon measurement.


This framework was sufficient for most quantum experiments until the 1980s, when breakthroughs in laser sources as well as trapping and cooling techniques enabled experimentalists to probe quantum systems in ways that highlighted the need for a continuous description of measurement \cite{Wineland2013_NP,Haroche2013_NP}. In particular, systems like atoms or ions were observed transitioning between discrete energy levels \cite{Nagourney1986,Bergquist1986,Gleyzes2007}. In other words, an atom could ``jump'' between states under continuous observation, and this \emph{quantum jump} could  be influenced by how the system was being measured \cite{Cook1985,Javanainen1986,Zoller1987,Blatt1988}. 

Inspired by the observation of these quantum jumps, a new mathematical theory for \emph{continuous measurements} was developed \cite{Davies1976,Srinivas1981,carmichael_book,Srinivas1996}, describing the evolution of a quantum system under continuous monitoring, as opposed to instantaneous, projective measurements. In classical statistical physics, a system is described by an ensemble of noisy trajectories generated by a set of SDEs. Analogously, in quantum physics, the state of a system conditioned on the measurement record follows a \emph{quantum trajectory} \cite{carmichael_book}. The evolution of such a \emph{conditional state} \cite{carmichael_book} is generated by a stochastic master equation (SME) \cite{Wiseman1993,Wiseman_thesis,wiseman2010quantum}. Furthermore, the trajectory of such a state can be controlled by feeding back the measurement outcomes (or an appropriate function of them) as they are registered \cite{Sayrin2011,Aspelmeyer2014}. A quantum theory of feedback, which naturally arises from and requires a continuous measurement formalism, can be similarly described using SMEs \cite{Wiseman1994_feedback,Wiseman1994_squeezing_feedback}.

These SMEs, central to the theory of continuous measurement, were shown to connect directly to quantum stochastic calculus \cite{Davies1969,Davies1970,Davies1971,Hudson1984,Barchielli1986} and quantum filtering \cite{Belavkin1987,Belavkin1989,Barchielli1991,Belavkin1999}. Quantum stochastic calculus, a non-commutative analogue of It\^{o}’s stochastic calculus, introduces “white noise” Bose fields, $\banihil_t$ and $\bcreat_t$, satisfying canonical commutation relations. In quantum optics, these fields approximate the electromagnetic field and serve as the foundation for a consistent theory of photodetection \cite{Barchielli1986, Gardiner_Collett1985,Gardiner1992}. Using quantum stochastic calculus, Belavkin and Barchielli extended Bayesian filtering into the quantum domain to describe how quantum states are \emph{conditioned} by continuous measurements, thereby establishing the field of quantum filtering \cite{Belavkin1987,Belavkin1989,Belavkin1999}. Turns out that the SME is nothing but the representation of their quantum filter in its adjoint density form \cite{vanHandel2005}. 

\newthought{In this chapter}, we revisit and adapt established derivations from the literature to the context of this thesis. In particular, we rederive the SME for an ensemble of two-level systems continuously monitored through the polarization of a probing light field. To do so, we start by assuming a very general setup: that of a bath (probe) interacting with a system. We first detail in \secref{sec:syst-bath} all the relevant approximations as laid out by Gross et al. (2018) \cite{Gross_2018}. Next, we derive the SME for both photodetection (in \secref{sec:derivation_SME_photo}) and homodyne measurements (in \secref{sec:homodyne_meas}), following the steps of Albarelli et al. (2024) \cite{Albarelli2024}. Then, in \secref{sec:polarimetric_meas}, inspired by the work of Deutsch et al. (2010) \cite{deutsch_quantum_2010}, we explain how a system monitored with polarization spectroscopy can be described by the same SME as one measured by homodyne detection. Last but not least, we also briefly introduce in \secref{sec:mark_meas-based_feedback} Markovian and Bayesian feedback, as they will also be of interest in later chapters. If the reader wants to dive deeper into this topic, many other detailed sources exist, such as the works of Carmichael \cite{carmichael_book}, Gardiner and Zoller \cite{gardiner_zoller2004quantum}, Breuer and Petruccione \cite{Breuer2002}, Wiseman and Milburn \cite{wiseman2010quantum}, and Jacobs \cite{jacobs_book}. Those specifically interested in quantum filtering might find the review by van Handel et al. \cite{vanHandel2005} particularly insightful.

\section{Introducing the system-bath setup} \label{sec:syst-bath}

\begin{figure}[t!]
    \centering
    \includegraphics[width = 0.6\textwidth]{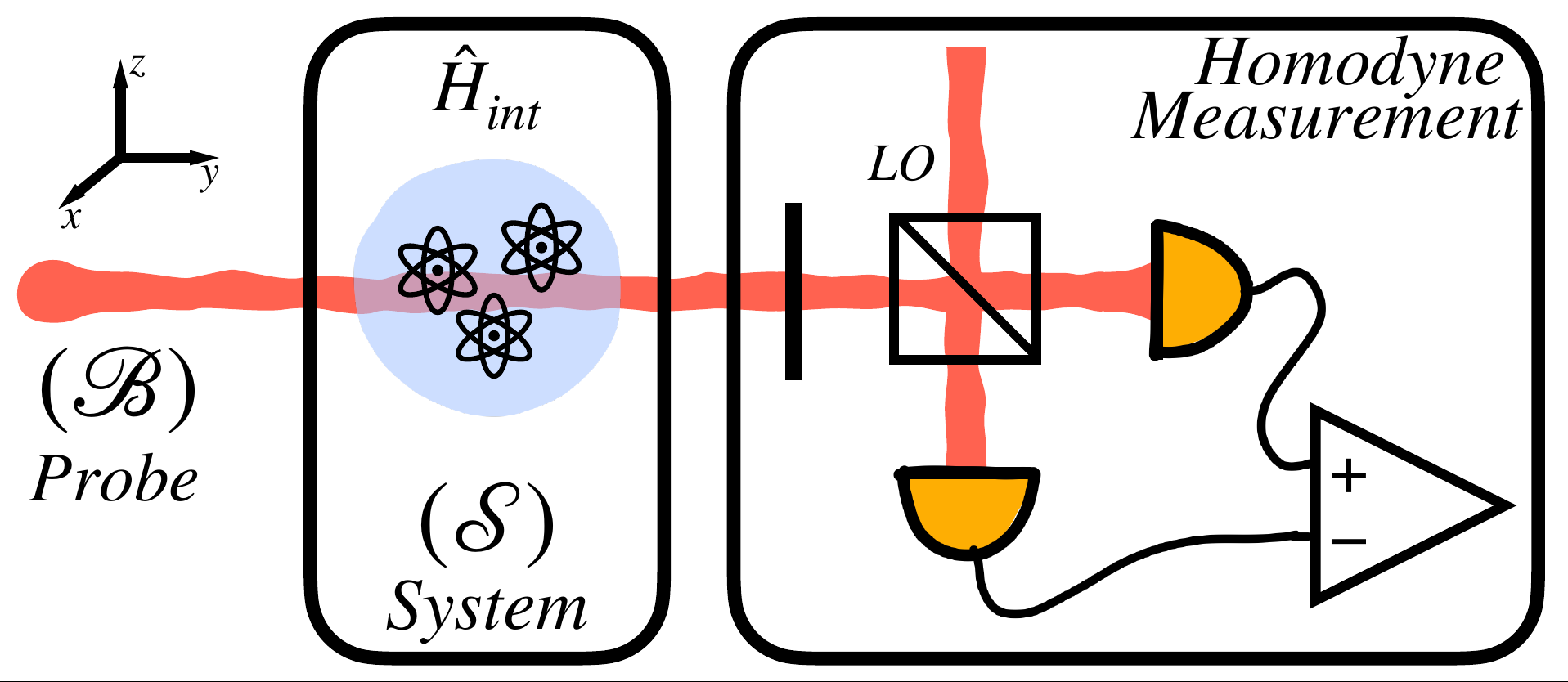}
    \caption[Diagram of the setup]{\textbf{Diagram of the setup.} The setup consists of a main system (labeled with $\mathcal{S}$) continuously monitored by a probe ($\mathcal{B}$) that interacts with the system and is later measured by, in this depiction, a homodyne detector. In this section, though, we consider other types of measurement such as photodetection. The interaction between the system and the probe (also sometimes referred as bath) is governed by the interaction Hamiltonian $\Ham_{\!\interaction}$.}
    \label{fig:system_homodyne}
\end{figure}

Consider the case of a system $\mathcal{S}$ coupled to a bath $\mathcal{B}$ corresponding to a continuum of bosonic modes, as depicted in \figref{fig:system_homodyne}. The total Hamiltonian describing the evolution of the joint system and bath is given by
\begin{equation} \label{eq:total_Ham}
    \Ham = \Ham_{\!\mathcal{S}} + \Ham_{\!\mathcal{B}} + \Ham_{\!\interaction},
\end{equation}
with $\Ham_{\!\mathcal{S}}$ representing the Hamiltonian of the system, and $\Ham_{\!\interaction}$ accounting for the interaction between the system and the bath, with the bath modeled as an infinite collection of bosonic modes whose Hamiltonian $\Ham_{\!\mathcal{B}}$ reads as:
\begin{equation} \label{eq:Ham_bath}
     \Ham_{\!\mathcal{B}} = \int_0^\infty d\omega \, \omega \, \bcreat(\omega) \banihil(\omega), 
\end{equation}
where $\omega$ is the frequency of each bosonic mode, with its corresponding creation and anihilation operators, $\bcreat(\omega)$ and $\banihil(\omega)$, fulfilling the commutation relationships 
\begin{equation} \label{eq:comm_rel_bomega}
    [\banihil(\omega),\bcreat(\omega^\prime)] = 2 \pi \delta(\omega - \omega^\prime) \;\;\;\; \mrm{and} \;\;\;\;  [\banihil(\omega),\banihil(\omega^\prime)] = 0.
\end{equation}

This bath, which we label with $\mathcal{B}$, is in our case the probe that the experimenter wants to use to measure the state of the system. In other words, the probe or bath interacts with the system and it is later measured. Therefore, in this section we will use interchangeably bath, probe and in a lesser extend, environment, all referring to the subspace $\mathcal{B}$.

Additionally, the interaction Hamiltonian is assumed to be linear w.r.t. the bosonic operators $\banihil$:
\begin{equation} \label{eq:interaction_ham_omega_int}
    \Ham_{\!\interaction} = i \int_0^\infty d\omega \, \frac{\sqrt{\kappa(\omega)}}{2\pi} \left(\LinOp \otimes \bcreat(\omega) - \LinOp^\dagger \otimes \banihil(\omega) \right),
\end{equation}
where $\LinOp$ corresponds to a given system operator and $\kappa(\omega)$ is the coupling strength of the system to the bath mode with frequency $\omega$. Throughout this derivation, we assume the system $\mathcal{S}$ to have only one characteristic frequency $\Omega$. Additionally, we will assume that when going into the interaction picture of $\Ham_{\!0} \coloneqq \Ham_{\!\mathcal{S}} + \Ham_{\!\mathcal{B}}$, the system operator acquires a time dependent phase $\LinOp(t) = \LinOp e^{-i\Omega(t-t_0)}$, with $t_0$ denoting the initial time.

Even then, the interaction Hamiltonian is still too complex and further approximations must be performed. Throughout this section, each important approximation will be marked with a dot, for easier identification. The first one relies on the Markovian properties of the bath:
\begin{itemize}
    \item \textit{First Markov approximation:} 
    This approximation relies on the timescale of the system's dynamics being much slower than the timescale of the bath's memory, characterized by its correlation time $\tau_c$. The key idea is that the bath rapidly ``forgets'' its interaction with the system. In other words, any perturbation in the bath caused by its interaction with the system dissipates quickly by propagating away into the bath. From the system's perspective, the excitation it created in the bath has effectively vanished, and the bath has lost all memory of the initial disturbance. This rapid decay of the bath's memory often leads to it being described as ``memoryless''.

    Furthermore, a bath that forgets its past very quickly indicates that its fluctuations are also rapid. In particular, they are much faster than the comparatively slow response of the system, which therefore cannot ``resolve'' them. In other words, the system ``sees'' only the average effect of these fluctuations over longer timescales, i.e., it is no longer sensitive to specific frequencies $\omega$, and therefore, the interaction strength is constant in a (large) frequency bandwidth $\mathcal{W}$. Namely, $\kappa(\omega) = \kappa$ for $\omega \in \mathcal{W} = [\Omega - \theta, \Omega + \theta\,]$, and zero outside of it, where $\Omega$ is the characteristic frequency of the system. 
\end{itemize}

Then, the Hamiltonian becomes:
\begin{equation} \label{eq:interaction_ham_omega}
    \Ham_{\!\interaction} = i \, \sqrt{\kappa} \int_\mathcal{W} d\omega \, \frac{1}{2\pi} \left(\LinOp \otimes \bcreat(\omega) - \LinOp^\dagger \otimes \banihil(\omega) \right).
\end{equation}

We can take this Hamiltonian as the starting point \cite{Gardiner_Collett1985} or derive it from a standard dipole coupling by performing a rotating-wave approximation (RWA) and keeping the slow-varying energy-conserving terms of \eqnref{eq:interaction_ham_omega}. If the latter is performed, the timescales we consider, i.e. $\Dt$, have to be much longer than the characteristic time of the system, $1/\Omega$. Namely, if the RWA is performed, then
\begin{equation}
    \Omega^{-1} \ll \Dt
\end{equation}

Additionally, another assumption intrinsic in the form of \eqnref{eq:interaction_ham_omega} is that the system is small enough, i.e. $\Delta x \ll c\Dt$, such that the spatial integration has been substituted by a point interaction \cite{Gross_2018}.  

If now we go into the interaction frame of $\Ham_0 = \Ham_{\!\mathcal{S}} + \Ham_{\!\mathcal{B}}$, then, the interaction Hamiltonian in the time domain becomes:
\begin{align}
    \Ham_{\!I} (t) &\coloneqq \e^{i \Ham_0(t-t_0)} \Ham_{\!\interaction} \e^{-i \Ham_0(t-t_0)} \nonumber \\
    &= i \frac{\sqrt{\kappa}}{2\pi}\int_\mathcal{W} \dd\omega  \, \e^{i \Ham_0(t-t_0)} \left(\LinOp \otimes \bcreat(\omega) - \LinOp^\dagger \otimes \banihil(\omega) \right) \e^{-i \Ham_0(t-t_0)} \nonumber \\
    &\footnotemark[1] \footnotemark[2]= i \frac{\sqrt{\kappa}}{2\pi} \int_\mathcal{W} \dd\omega \left(\LinOp \e^{-i \Omega(t-t_0)} \otimes \bcreat(\omega) \e^{i \omega (t-t_0)} - \LinOp^\dagger  \e^{i \Omega (t-t_0)} \otimes\banihil(\omega)  \e^{-i \omega (t-t_0)}\right) \nonumber \\
    &= i \frac{\sqrt{\kappa}}{2\pi} \int_\mathcal{W} \dd\omega  \left(\LinOp \otimes \bcreat(\omega) \e^{i (\omega-\Omega) (t-t_0)} - \LinOp^\dagger \otimes \banihil(\omega)  \e^{-i (\omega-\Omega) (t-t_0)}\right) \nonumber \\
    &= i \sqrt{\kappa} \left(\!\LinOp \otimes \frac{1}{2\pi}\!\int_\mathcal{W} \dd\omega \,  \bcreat(\omega) \, \e^{i (\omega-\Omega) (t-t_0)} - \LinOp^\dagger \otimes \frac{1}{2\pi}\!\int_\mathcal{W}\!\! d\omega \, \banihil(\omega)  \e^{-i (\omega-\Omega) (t-t_0)}\right) \nonumber \\ 
    &= i \sqrt{\kappa} \left(\!\LinOp \otimes \bcreat(t) - \LinOp^\dagger \otimes \banihil(t)\right), \label{eq:temporal_int_Ham_derived}
\end{align}
\footnotetext[1]{Given our initial assumption that $\Ham_{\!\mathcal{B}}$ has the form specified in \eqnref{eq:Ham_bath}, by using the Baker-Campbell-Hausdorff formula, it follows that $\e^{i \Ham_{\!\mathcal{B}} (t-t_0)} \banihil(\omega) \e^{-i \Ham_{\!\mathcal{B}} (t-t_0)} = \banihil(\omega) \e^{-i\omega(t-t_0)}$, since $\Ham_{\!\mathcal{B}}$ is time-independent and the modes obey the commutation relationships of \eqref{eq:comm_rel_bomega}. In particular, the key step in this calculation is the derivation of $[\Ham_{\!\mathcal{B}},\banihil(\omega)] = - \omega \banihil(\omega)$. Other work circumvents defining $\Ham_{\!\mathcal{B}}$ and instead require the field operators to acquire a time dependence at the frequency of the mode \cite{Gross_2018}.}
\footnotetext[2]{Earlier in the definition of $\LinOp$ we stated that when going into the interaction picture $\Ham_{\!0} \coloneqq \Ham_{\!\mathcal{S}} + \Ham_{\!\mathcal{B}}$, the system operator acquires a time dependent phase $\LinOp(t)=\LinOp \e^{-i\Omega(t-t_0)}$, with $t_0$ denoting the initial time.}
where in the last step we have made use of the definition of the input modes $\banihil(t)$
\begin{equation} \label{eq:mode_W}
    \banihil(t)= \frac{1}{2\pi}\!\int_\mathcal{W}\!\! d\omega \, \banihil(\omega)  \e^{-i (\omega-\Omega) (t-t_0)} ,
\end{equation}
which one might notice resembles the Fourier transform of the frequency-domain operators but with frequencies restricted to the set $\mathcal{W} = [\Omega - \theta, \Omega + \theta\,]$.

\begin{figure}[t!]
    \centering
    \includegraphics[width=0.8\textwidth]{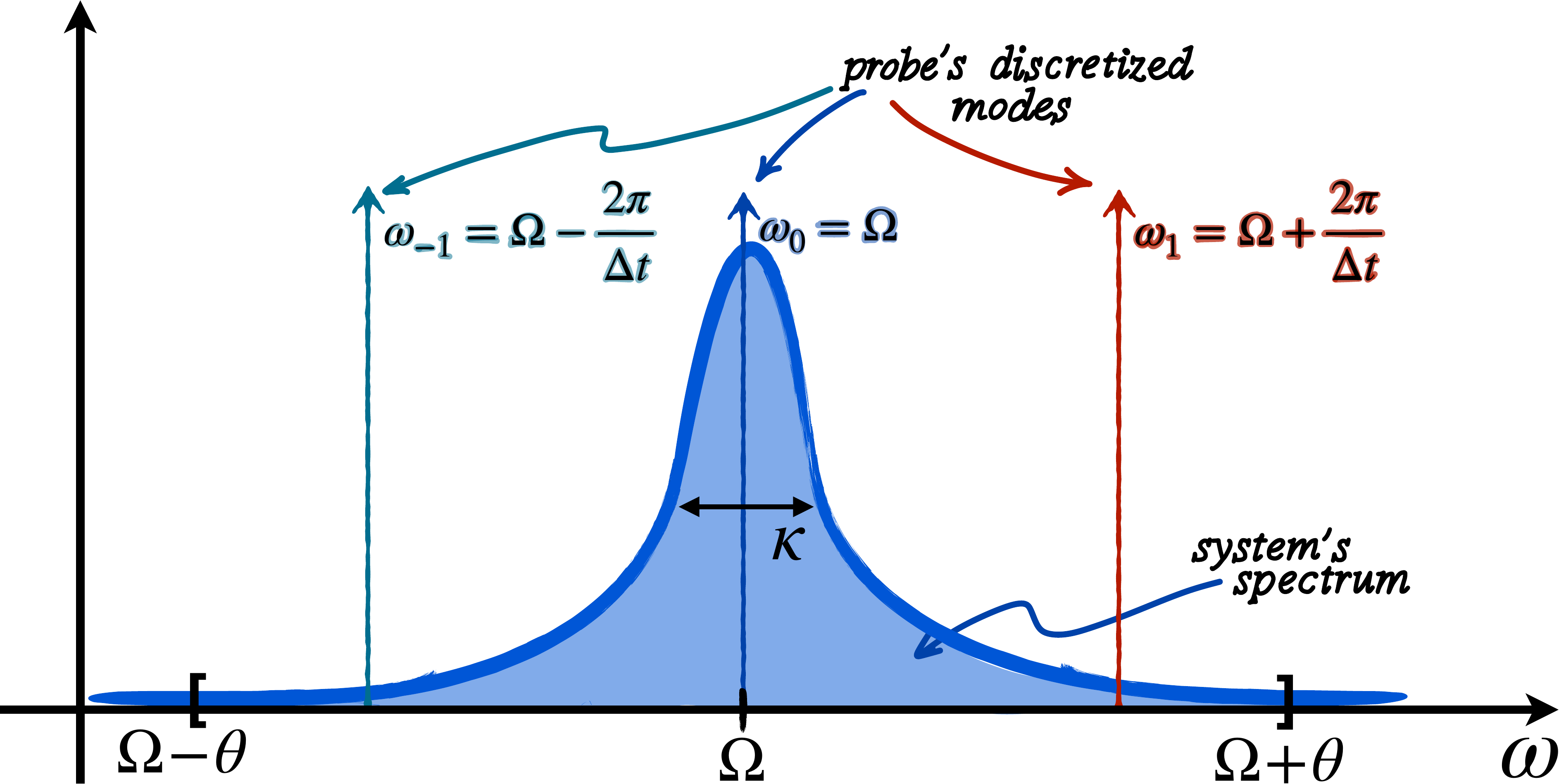}
    \caption[Interaction between the system and the probe in the frequency domain]{\textbf{Interaction between the system and the probe in the frequency domain.} This figure illustrates how the spectrum of the system, centered around $\Omega$ and broadened by a decay $\kappa$, aligns with the discretized frequency modes of the probe, in particular $\omega_0$, $\omega_1$ and $\omega_{-1}$, which are spaced by $\frac{2\pi}{\Dt}$. As detailed in the main text, an important condition when picking the $\Dt$ is that $\Dt \ll \kappa^{-1}$, or equivalently, $1/\Dt \gg \kappa$, ensuring that modes such as $\Omega \pm \frac{2\pi}{\Dt}$ are sufficiently detuned from the spectrum's core to be negligible in system interactions. The parameter $\theta$ defines the interaction bandwidth $\mathcal{W}$ for which we assume constant system-probe interaction, i.e. $\kappa(\omega) = \kappa$ if $\omega \in \mathcal{W} = [\Omega-\theta,\Omega+\theta]$. It is evident also from this depiction why is necessary that $\theta \gg \kappa$. }
    \label{fig:spectrum}
\end{figure}

\begin{itemize}
    \item \textit{Weak coupling:} In our simplified treatment, the system has only one characteristic or transition frequency $\Omega$ with a decay $\kappa$ due to the interaction with the bath. Hence, it stands to reason to expect \footnote{For example, for a signal $S(t) \sim \cos{\Omega t} \, \e^{-k t}$, its power spectral density is a Lorentzian centered around $\Omega$ and with a width of $k$, i.e. $\propto 1/\left((\omega-\Omega)^2+k^2\right)$} its spectrum to be centered at $\Omega$ with a broadening proportional to $\kappa$, with its tails going to zero within a few linewidths $\kappa$ of $\Omega$, as depicted in \figref{fig:spectrum}. Earlier we defined the interaction bandwidth or the effective bath frequencies that the system sees as a set $\mathcal{W} = [\Omega-\theta,\Omega+\theta\,]$. Given that we assume the spectrum of the system to be centered around $\Omega$ with a broadening of $\kappa$, we must then pick $\theta$ to be much larger than $\kappa$, i.e. $\theta \gg \kappa$, to ensure the whole spectrum falls within $\mathcal{W}$. Then, we can extend the integral of \eqnref{eq:mode_W} over the whole real axis $\Real$. However, for the set $\mathcal{W} = [\Omega - \theta, \Omega + \theta\,]$ to be properly defined, the following condition must hold: $\Omega-\theta > 0$. Thus, since $\theta \gg \kappa$, then $\Omega \gg \kappa$. In other words, the system is weakly coupled to the bath, since $\kappa \ll \Omega$.
\end{itemize}

Now, thanks to the weak-coupling approximation, we can extend $\mathcal{W}$ to the whole real axis $\Real$. In this case, the interaction Hamiltonian has the same form as in \eqnref{eq:temporal_int_Ham_derived}:
\begin{equation} \label{eq:interaction_Hamitonian_cont}
    \Ham_{\!I}(t) = i\sqrt{\kappa}  \left(\!\LinOp \otimes \bcreat(t) - \LinOp^\dagger \otimes \banihil(t)\right), 
\end{equation}
but with $\banihil(t)$ and $\bcreat(t)$ becoming instantaneous temporal input modes, i.e., the Fourier transform of the frequency modes $\banihil(\omega)$ and $\bcreat(\omega)$. Namely,
\begin{equation} \label{eq:time_mode}
    \banihil(t)= \frac{1}{2\pi}\! \int_{-\infty}^{\infty}\!\! d\omega \, \banihil(\omega)  \e^{-i (\omega-\Omega) (t-t_0)}, 
\end{equation}
following the standard commutation relationship
\begin{equation} \label{eq:continuous_time_commutation_relationship}
    [\banihil(t),\bcreat(t^\prime)] = \delta(t-t^\prime).
\end{equation}

\begin{figure}[h]
    \centering
    \includegraphics[width = 0.7\textwidth]{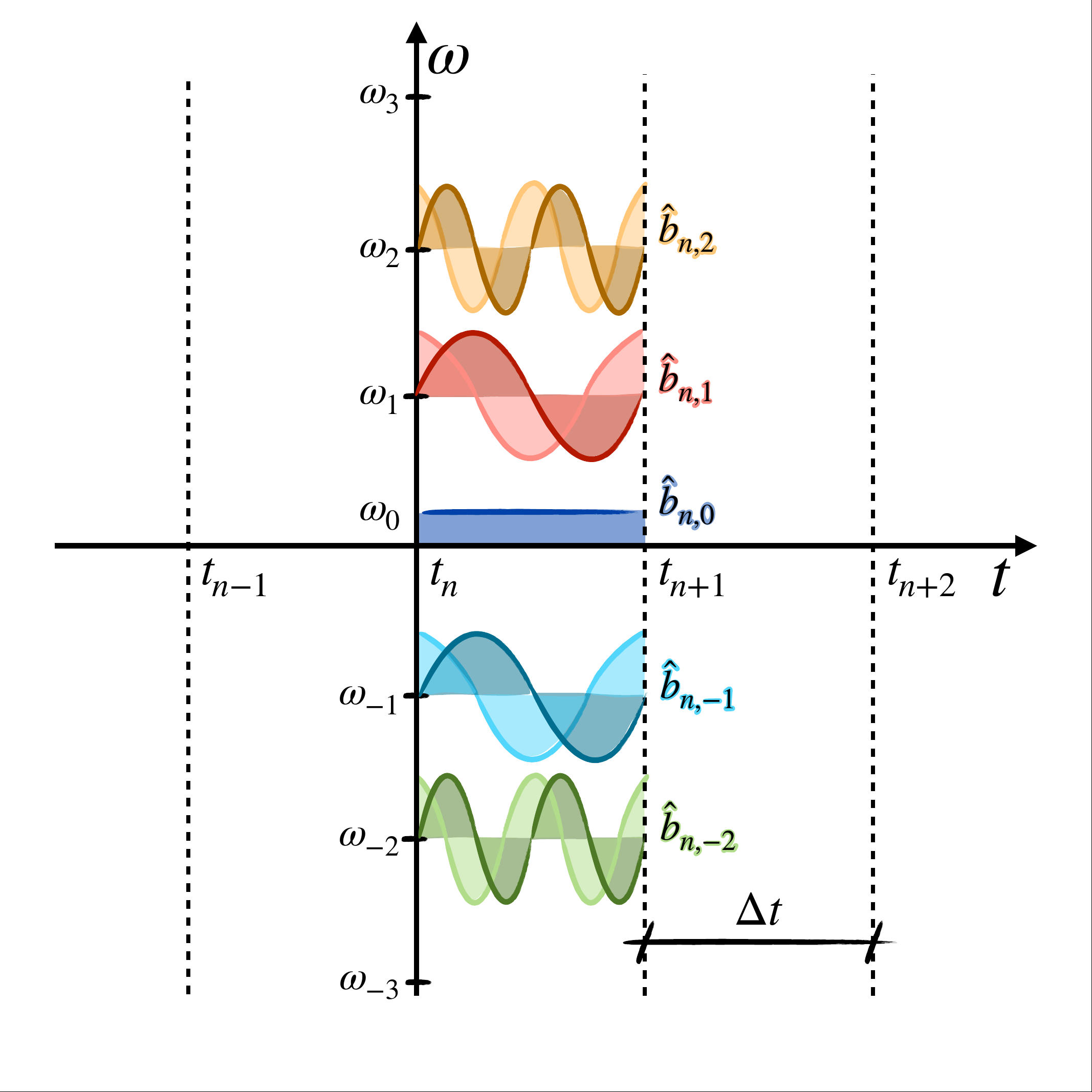}
    \caption[Frequency and time discretization of the probe modes]{\textbf{Frequency and time discretization of the probe modes.} This illustration depicts the time discretization of the probe beam into intervals of $\Dt$, with discretized times labeled as $t_n$ with $n = -\infty,\dots,\infty \in \mathrm{Z}$. At $t_n$, the on-resonant and first four discrete modes are represented, with their frequencies also discretized as $\omega_k = \Omega + \frac{2\pi}{\Dt} k $, with $k = -2,-1,0,1,2$.}
    \label{fig:time_discretization}
\end{figure}

So far we are in the interacting frame and the Hamiltonian governing the dynamics of the joint state of the system and the probe is given in \eqnref{eq:interaction_Hamitonian_cont}, where the probe modes $\banihil(t)$ and $\bcreat(t)$ evolve in time. 

To further simplify this description, we now discretize the probe field into time intervals of duration $\Dt$. The time instances are defined as $t_n = n \Dt$, with $n \in \mathbb{Z}$. Each corresponding segmented mode is labeled as $\banihil_n(t)$, with $n \in \mathbb{Z}$. Consequently, the temporal mode $\banihil(t)$ can be written as a sum of all the discrete time modes $\banihil_n(t)$ as:
\begin{equation} \label{eq:segmented_mode_b}
    \banihil(t) = \sqrt{\Dt} \sum_{n = -\infty}^\infty \banihil_n(t) \Theta(t-t_n), 
\end{equation}
such that $\banihil(t)$ is still continuous but has been structured into (continuous) intervals of size $\Dt$, utilizing the Heaviside function defined as $\Theta(u) = 1$ for $0 \leq u < \Dt$ and zero otherwise. This segmentation ensures that each segment, or ``discrete time mode'', $\banihil_n(t)$, while being a continuous mode within its respective time interval $[t_n,t_n+\Dt)$, is distinctly associated with the time index $n$. This index $n$ highlights that $\banihil_n(t)$ contributes to the overall mode only during its specific interval from $t_n$ to $t_n+\Dt$. 
The factor $\sqrt{\Dt}$ has been added in order to maintain the correct commutation relations among discrete modes, akin to those in continuous modes. This will be demonstrated in detail later in the section. This factor also ensures that the total energy, expressed in continuous time modes as $\int_{-\infty}^\infty \bcreat(t) \banihil(t) \dt$, matches that calculated using discrete modes, $\sum_n \bcreat_n \banihil_n \Dt$. 

Additionally, the segmented time modes $\banihil_n(t)$ can be expressed as a linear combination of discrete frequency modes $\banihil_{n,k}$, with the discretized frequencies ranging as: $\omega_k = \Omega + \frac{2 \pi}{\Dt} k$ with $k = -\infty, \dots, \infty  \in \mathbb{Z}$. In particular, we can use the discrete equivalent of the Fourier transform of \eqnref{eq:time_mode} to expand  $\banihil_n(t)$ as:
\begin{equation} \label{eq:discretized_time_mode_FS}
    \banihil_n(t) = \frac{1}{2\pi} \sum_{k=-\infty}^\infty \banihil_{n,k} \, \e^{-i 2\pi k \, t/\Dt} \Delta \omega = \frac{1}{2\pi} \sum_{k=-\infty}^\infty \banihil_{n,k} \, \e^{-i 2\pi k \, t/\Dt} \frac{2\pi}{\Dt},
\end{equation}
where we have set $t_0 = 0$ for simplicity, and $\Delta \omega = 2 \pi /\Dt$ is the spacing between consecutive frequency samples in the discretized frequency domain. Note that \eqnref{eq:discretized_time_mode_FS} is simply the Fourier series of $\banihil_n(t)$, and $\banihil_{n,k}$ are its Fourier coefficients defined by the integrals:
\begin{equation}
    \banihil_{n,k} = \int_{0}^{\Dt} \banihil_n(t) \, \e^{-\ii 2\pi n \, t /\Dt} \, \dt.
\end{equation}
Note, however, that the integral above defines the Fourier coefficients in terms of $\banihil_n(t)$, i.e. a segmented version of $\banihil(t)$, rather than $\banihil(t)$ itself. 

For practical purposes, we require a direct transformation of the continuous mode $\banihil(t)$ into the discrete modes or Fourier coefficients $\banihil_{n,k}$. This transformation can be achieved by deriving the Fourier coefficients from a \emph{double series expansion} of $\banihil(t)$ w.r.t. time and frequency.

By applying the Fourier expansion of the mode $\banihil_n(t)$ given in \eqnref{eq:discretized_time_mode_FS} to \eqnref{eq:segmented_mode_b}, the continuous mode $\banihil(t)$ can be expressed as a series expansion in both frequency and time as follows \cite{Gross_2018}:
\begin{equation} \label{eq:double_series_expansion}
    \banihil(t) = \frac{1}{\sqrt{\Dt}} \sum_{n=-\infty}^\infty \sum_{k=-\infty}^\infty \banihil_{n,k} \, \Theta(t-t_n) \, \e^{-i 2\pi k \, t/\Dt},
\end{equation}
where the labeling $n$ emphasizes that each frequency mode $\banihil_k$ is associated with a specific time interval $[t_n,t_n+\Dt)$ indexed by $n$. As derived in \propref{prop:ap_fourier_coef_double_series}, the ``Fourier'' or discretized modes $\banihil_{n,k}$ for \eqnref{eq:double_series_expansion} are:
\begin{equation} \label{eq:discrete_modes_bnk}
    \banihil_{n,k} = \frac{1}{\sqrt{\Dt}} \int_{t_n}^{t_n+\Dt} \banihil(t) \e^{i2\pi k t / \Dt} \, \dt,
\end{equation}
which one might notice to highly resemble the Fourier series and the integral form for its coefficients (see \tabref{tab:signal_comparisons}). 

\begin{table}[h]
\centering  
\begin{tabular}{
  |>{\centering\arraybackslash}m{2.5cm}|
  |>{\centering\arraybackslash}m{7.5cm}|
  |>{\centering\arraybackslash}m{4cm}|
  }
\hline
\vspace{0.25cm}\textbf{Type of Series}\vspace{0.25cm} & \vspace{0.25cm}\textbf{Windowed Complex Exponential Series}\vspace{0.25cm} & \vspace{0.25cm}\textbf{Fourier Series}\vspace{0.25cm} \\
\hline
\vspace{0.25cm}\textbf{Time-Domain Signal}\vspace{0.25cm} & \vspace{0.4cm}\( \banihil(t) \!=\!\!\!\! \sum\limits_{n=-\infty}^\infty \sum\limits_{k=-\infty}^\infty \! \frac{1}{\sqrt{\Dt}} \banihil_{n,k} \Theta(t - t_n) \e^{-i2\pi kt/\Dt} \) \vspace{0.3cm}&\vspace{0.1cm} \( f(t) \!=\frac{1}{T} \!\!\! \sum\limits_{n=-\infty}^\infty \!\!\! c_n \, \e^{-i \frac{2\pi n}{T} t} \) \vspace{0.1cm}\\
\hline
\vspace{0.25cm}\textbf{Frequency-Domain Coefficients}\vspace{0.25cm} & \( \banihil_{n,k} \!=\! \frac{1}{\sqrt{\Dt}} \int_{t_n}^{t_n + \Dt} \banihil(t) \e^{i2\pi kt/\Dt} \, \dt \) & \( c_n \!=\! \int_{0}^{T} f(t) \e^{i \frac{2\pi n}{T} t} \, \dt \) \\
\hline
\end{tabular}
\caption[Comparison of Signal Representations]{Comparison of Signal Representations.}
\label{tab:signal_comparisons}
\end{table}

As a sanity check, we make sure that the discretized modes $\banihil_{n,k}$ obey the discrete canonical commutation relation:
\begin{align}
    [\banihil_{n,k},\bcreat_{m,\ell}] &= \frac{1}{\Dt} \int_{t_n}^{t_n+\Dt} \int_{t_m}^{t_m+\Dt} [\banihil(t),\bcreat(t^\prime)] \e^{i2\pi k t/\Dt} \e^{-i 2\pi \ell t^\prime/\Dt} \dt \, \dt^\prime \nonumber \\
    &= \frac{1}{\Dt} \int_{t_n}^{t_n+\Dt} \int_{t_m}^{t_m+\Dt} \!\!\!\! \delta(t-t^\prime) \e^{i2\pi k t/\Dt} \e^{-i 2\pi \ell t^\prime/\Dt} \dt \, \dt^\prime \nonumber \\
    &= \frac{1}{\Dt} \int_{t_n}^{t_n+\Dt} \!\!\!\! \e^{i2\pi k t/\Dt} \!\! 
    \int_{t_m}^{t_m+\Dt} \!\!\!\! \e^{-i 2\pi \ell t^\prime/\Dt}  \delta(t-t^\prime) \, \dt^\prime \, \dt \nonumber \\
    &= \begin{cases} 
    \frac{1}{\Dt} \int_{t_n}^{t_n+\Dt}  \e^{i2\pi k t/\Dt} \e^{-i 2\pi \ell t/\Dt} \, \dt & \text{if} \;\; t_m \le t \le t_m+\Dt , \\
    0 & \text{otherwise},
    \end{cases} \nonumber \\
    &= \frac{1}{\Dt} \delta_{n,m} \int_{t_n}^{t_n+\Dt}  \e^{i2\pi k t/\Dt} \e^{-i 2\pi \ell t/\Dt} \, \dt, \nonumber \\
    &= \frac{1}{\Dt} \delta_{n,m} \, \Dt \, \delta_{k,\ell} = \delta_{n,m} \delta_{k,\ell}, \nonumber
\end{align}
where in the last line we have used the orthogonality condition of integration over the period $\Dt$ given in \eqnref{eq:ap_orthogonality_cond}. Note the analogy with the commutation relationship of \eqnref{eq:continuous_time_commutation_relationship}, which holds due to writing the time expansion of $\banihil(t)$ in \eqnref{eq:segmented_mode_b} with a factor $\Dt$. 

As depicted in \figref{fig:time_discretization}, in each time step, there is an infinite collection of modes $\banihil_{n,k}$ with which the system can interact. Let us perform one more approximation in order to ensure that the system interacts only with the main on-resonant mode $\banihil_{n,0}$. Namely:
\begin{itemize}
    \item \textit{Quasimonochromatic approximation:} Let us pick a time-step $\Dt$ much smaller than the time the state of the system and bath take to ``couple'', i.e. $\Dt \ll 1/\kappa$, so that the system-probe interaction is weak during this time-step $\Dt$. Now, recall that the mode frequencies are discretized as $\omega_k = \Omega + \frac{2\pi}{\Dt} k$, with $k \in \mathrm{Z}$. Since we have just set the condition that $1/\Dt \gg \kappa$, that means that all probe modes except the on-resonant one will be far away from the main part of the spectrum of the system (see \figref{fig:spectrum}).
\end{itemize}

Hence, only the on-resonant mode will interact with the system ($k=0$) and thus we can disregard the rest of the modes ($k\neq0$) from all our calculations moving forward. Namely, the mode $\banihil(t)$ can now be expanded as:
\begin{equation} \label{eq:probe_field_discretized}
    \banihil(t) = \frac{1}{\sqrt{\Dt}} \sum_{n=-\infty}^\infty \banihil_n \Theta(t-t_n) 
\end{equation}
where
\begin{equation} \label{eq:time_discretized_modes}
    \banihil_n \coloneqq \banihil_{n,0} = \frac{1}{\sqrt{\Dt}} \int_{t_n}^{t_n+\Dt} \banihil(t) \, \dt,
\end{equation}
which is obtained by simply setting $k=0$ in \eqnref{eq:discrete_modes_bnk}. Note that so far in this derivation, the mode at time $n$ can be correlated with past or future modes. Thus, to further simplify the treatment, we perform yet another assumption: that the time $\Dt$ is longer than the correlation time of the bath $\tau_c$ such that different segmented modes are uncorrelated with other modes:
\begin{itemize}
    \item \textit{Born-Markov approximation:} As explained in the first Markov approximation, the correlation time $\tau_c$ quantifies how rapidly the bath (a.k.a. the probe) ``forgets'' its interaction with the system. In the setup considered so far, the probe is divided into a series of discrete temporal modes $\banihil_n(t)$ with $n\in\mathbb{Z}$, each interacting with the system over a distinct time interval $[t_n,t_n+\Dt)$. When the condition $\tau_c \ll \Dt$ is satisfied, the memory of the bath decays much faster than the duration of each probe segment. This means that the influence of the system on any given probe segment $\banihil_n(t)$ does not propagate forward to affect subsequent probe segments. 
    Importantly, the state of each probe segment after interacting with the system is not ``lost''. Instead, each segment leaves the interaction region with its new state preserved, i.e. dissipating away from the system, potentially to be measured later. However, the assumption of memorylessness means that when a new probe segment $\banihil_{n+1}(t)$ arrives, it is initialized in the same state as the earlier segments before their interactions, typically $\projector{0}{0}$. This ensures that the joint state of the system and the incoming probe prior to interaction is separable and can be written as  $\rho[n] \otimes \projector{0}{0}$, where $\rho[n]$ describes the state of the system before the interaction with the $n$-th segment.
\end{itemize}

This discretization of the probe, combined with the Born-Markov approximation, yields a ``conveyor belt''-type interaction \cite{Gross_2018,Albarelli2024}: for instance, at time $t_n=0$, modes $\banihil_{n<0}$ have not yet interacted with the system and modes $\banihil_{n\geq0}$ may be correlated with the system but will not interact with it ever again. We can view this from the perspective of Hilbert spaces: each of these discrete modes exist in a distinct Hilbert space $\banihil_n(t) \in \mathcal{H}^{\mathcal{B}}_n$ such that the total subspace of the bath is $\mathcal{H}^{\mathcal{B}} = \bigotimes_{n\in\mathrm{Z}} \mathcal{H}^{\mathcal{B}}_n $. Furthermore, the joint system-bath state will live in an even larger Hilbert space $\mathcal{H}^{\mathcal{S}\mathcal{B}} = \mathcal{H}^{\mathcal{S}} \bigotimes_{n\in \mathrm{Z}} \mathcal{H}^{\mathcal{B}}$. However, at each time $t_n$, it is not necessary to consider the whole Hilbert space but rather the reduced space $ \mathcal{H}^{\mathcal{S}} \otimes \mathcal{H}^\mathcal{B}_n$.

At each time, the Hamiltonian living in $ \mathcal{H}^{\mathcal{S}} \otimes \mathcal{H}^\mathcal{B}_n$ and describing the interaction between system and probe at time $t_n$ will also be labeled with $n$. To derive its expression, we plug in the expression of the segmented probe field given in \eqnref{eq:probe_field_discretized} into \eqnref{eq:interaction_Hamitonian_cont}:
\begin{equation} 
    \Ham_{\!I}(t) = \sum_{n=-\infty}^\infty i\sqrt{\frac{\kappa}{\Dt}} \left(\!  \LinOp \otimes \bcreat_n  - \LinOp^\dagger \otimes \banihil_n \right) \Theta(t-t_n) \coloneqq \sum_{n=-\infty}^\infty \Ham_{\!I}^{(n)} \Theta(t-t_n) \nonumber
\end{equation}
where the interaction Hamiltonian acting from time $t_n$ to time $t_{n+1}$ is:
\begin{equation} \nonumber
    \Ham_{\!I}^{(n)} \coloneqq i\sqrt{\frac{\kappa}{\Dt}} \left(\!  \LinOp \otimes \bcreat_n  - \LinOp^\dagger \otimes \banihil_n \right), 
\end{equation}
which, importantly, is inversely proportional to $\sqrt{\Dt}$. It then follows that the unitary evolution governed by the interaction Hamiltonian from time $t_n$ to $t_n+\Dt$ has the form:
\begin{align} \label{eq:unitary_sqrtDeltat}
    \UnitOp_{\!\!\Dt}^{(n)} = \UnitOp_{\!\!\Dt}(t_n) = \e^{-i \Ham_{\!I}^{(n)} \Dt} &= \exp{\left[ - i \left(i\sqrt{\frac{\kappa}{\Dt}} \left(\!  \LinOp \otimes \bcreat_n  - \LinOp^\dagger \otimes \banihil_n \right)\right)\! \Dt\right]} \nonumber \\
    &= \exp{\left[ \sqrt{\kappa\Dt} \left(\!  \LinOp \otimes \bcreat_n  - \LinOp^\dagger \otimes \banihil_n \right)\right]},
\end{align}
with the exponent proportional to $\sqrt{\Dt}$ instead of $\Dt$. By taking the limit $\Dt \rightarrow 0$, the unitary responsible for the evolution of the joint system and probe from $t$ to $t+\dt$ can be written as:
\begin{equation} \label{eq:unitary_sqrtdt}
    \UnitOp_{\!\!\dt}(t) = \e^{-i \Ham_{I} (t) \dt} = \exp{\left[ \sqrt{\kappa \, \dt } \left(\!  \LinOp  \otimes \bcreat_t  - \LinOp^\dagger \otimes \banihil_t \right)\right]},
\end{equation}
where $\dt$ is an infinitesimally small time increment. Note that the input modes are written as $\banihil_t$ (not as $\banihil(t)$) to highlight that $t$ is a label identifying a particular input operator in the collection $\{\banihil_t\}_{t\in\mathrm{R}}$, rather than a variable. 

The derivation of the unitary in \eqnref{eq:unitary_sqrtdt} is the main result of this section. Now that we understand how the probe interacts with the system through a series of ``conveyor belt''-like interactions, represented by sequential joint unitaries, we will summarize each approximation performed on the way: 
\begin{itemize}
    \item  \emph{Rotating-Wave Approximation (when necessary)}: If the RWA is performed, then it is important that $\Dt \gg 1/\Omega$, so that the interaction of the system with the probe is averaged over many oscillations of the system during the interval $\Dt$.
    \item \emph{The First Markov approximation}, argues that $\kappa(\omega) = \kappa$ for $\omega \in \mathcal{W} = [\Omega - \theta, \Omega + \theta\,]$ and zero elsewhere, since the bath fluctuates so quickly that the system cannot resolve them and sees a uniform coupling. 
    \item \emph{Weak-coupling approximation}: the state of the bath is weakly coupled to the system, $\Omega \gg \kappa$.
    \item \emph{Quasimonochromatic approximation}: If the system interacts weakly enough with the probe during a time interval $\Dt$, i.e., $\Dt \ll 1/\kappa$, then we can ignore all modes with $k\neq0$ since their frequencies would fall outside the spectrum of the system and therefore not interact with it. 
    \item \emph{Born-Markov approximation}: The correlation time $\tau_c$ quantifies how quickly the bath (or probe) "forgets" its interaction with the system. When $\tau_c \ll \Dt$, where $\Dt$ is the duration of each probe segment, the effect of the system on one probe segment does not propagate to subsequent segments. In other words, each discrete temporal mode $\banihil_n(t)$ is uncorrelated from other modes, both in its past and future. Then, each new probe segment starts in the same initial state, typically $\projector{0}{0}$, ensuring that the joint state of the system and probe before interaction is separable: $\rho[n] \otimes \projector{0}{0}$, where $\rho[n]$ represents the state of the system before interacting with the corresponding probe segment at time $t_n$. After interaction, each probe segment retains its state and is not "lost", but simply propagates away from the system, remaining available for future measurements. 
\end{itemize}

To summarize, the conditions that $\Dt$ has to fulfill are \cite{Gross_2018}:
\begin{align}
    \Delta x &\ll c\Dt, \nonumber \\
    \tau_c &\ll \Dt, \nonumber \\
    \frac{1}{\Omega} &\ll \Dt \ll \frac{1}{\kappa}. \nonumber
\end{align}

\section{Derivation of the stochastic master equation} \label{sec:derivation_SME}

Up to this point, we have considered a general system interacting with a probe in a “conveyor belt’’ fashion: the probe is discretized into temporal modes that couple to the system one after another. In other words, at each time step $\Dt$, the $n$-th probe mode, denoted as $\banihil_n$ with $n = t/\Dt$. After the interaction with the system, the probe mode is immediately measured, yielding a classical outcome $y_n$. The combined effect of this interaction and subsequent measurement is captured by the following measurement operator:
\begin{equation} \label{eq:def_POVM_E}
    \measE_{y_n} = \braketop{y_n}{\UnitOp_{\!\!\Dt}^{(n)}}{0},
\end{equation}
which is obtained by evolving the initial probe state, assumed to be $\projector{0}{0}$ \cite{Albarelli2024}, via the unitary interaction of \eqnref{eq:unitary_sqrtDeltat}, and then projected onto the eigenstate associated to the measurement outcome $y_n$.

Moreover, in between two such measurement events, the system also evolves under its own internal dynamics. In the simplest case, this evolution is generated by the system Hamiltonian and is therefore unitary. More generally, if the system also couples to unmonitored degrees of freedom, its evolution is described by a completely positive and trace-preserving (CPTP) map, $\Phi_n$, acting from $t_{n\shortminus1}$ to $t_n$. Thus, the discretized conditional evolution of the state alternates between:
\begin{enumerate}
    \item The internal evolution of the system as described by the CPTP map $\Phi_n$, accounting for both the Hamiltonian dynamics and the coupling to an unmonitored environment
    \item The measurement update via the operator $\measE_{y_n}$ corresponding to a measurement event with a classical outcome $y_n$ obtained from the probe. Collecting these outcomes gives a discretised measurement record $\pmb{y}_{0:n} = \{y_0, y_1, \dots , y_n\}$, with $n = t/ \Dt$, which, in the continuous limit of $n\to\infty$ as $\Dt \to 0$ becomes a continuous measurement record $\pmb{y}_{t}$. Since the measurement update at each time step depends on the specific outcome $y_n$, the evolution of the state is conditional on the measurement record. In other words, different measurement trajectories lead to different state evolutions\footnote{In photodetection, for example, each time step yields either a ``click'' ($y_n = 1$) or ``no click'' ($y_n = 0$), with probabilities that depend on the state of the system prior to the measurement. These discrete outcomes can be modeled as a discrete process that becomes Poissonian in the continuous-time limit. Consequently, the system evolves along a stochastic trajectory, since its state is conditioned on a stochastic sequence of measurement results.}.
\end{enumerate}
Then, the conditioned state of the system after a sequence of interspersed measurements and internal evolutions up to time $t = n \Dt$ can be written in its discretized form as:
\begin{align} \label{eq:rho_n_discrete}
    \rho[n|\pmb{y}_{0:n}] = \frac{\Phi_n\!\!\left[\!\measE_{y_{n}}  \dots \Phi_1\!\!\left[\!\measE_{y_{1}}\, \Phi_0\!\!\left[\!\measE_{y_{0}}\, \rho_{0}\measE_{y_{0}}^{\dagger}\!\right]\!\measE_{y_{1}}^{\dagger}\!\right]\! \dots \measE_{y_{n}}^{\dagger} \! \right]}{\trace{\Phi_n\!\! \left[ \!\measE_{y_{n}}  \dots \Phi_1\!\!\left[\!\measE_{y_{1}}\,\rho_{0}\measE_{y_{1}}^{\dagger}\!\right]\! \dots \measE_{y_{n}}^{\dagger} \right]}},
\end{align}
where the brackets $[\cdot]$ emphasize that this is the \emph{discrete} conditional state, which can also be written as:
\begin{equation} \label{eq:notation_rhoc}
    \rho[n|\pmb{y}_{0:n}] \coloneqq \rho(t_n|\pmb{y}_{t_n}) \coloneqq \rhoc(t_n).
\end{equation}
It follows from \eqnref{eq:rho_n_discrete} that any consecutive conditional states are related as:
\begin{align} \label{eq:iterative_rule_rhoc}
    \rho[n|\pmb{y}_{0:n}] = \frac{\rhoun[n|\pmb{y}_{0:n}] }{\trace{\rhoun[n|\pmb{y}_{0:n}] }} = \frac{\Phi_n \!\!\left[\!\measE_{y_{n}} \, \rho[n\!\shortminus\!1|\pmb{y}_{0:n\shortminus1}] \measE_{y_{n}}^{\dagger} \! \right]}{\trace{\Phi_n \!\!\left[\!\measE_{y_{n}} \, \rho[n\!\shortminus\!1|\pmb{y}_{0:n\shortminus1}] \measE_{y_{n}}^{\dagger} \! \right]}},
\end{align}
where $\rhoun$ denotes the unnormalized state. This iterative rule relating the state at time $n \shortminus 1$ with the one at time $n$ can be further split into two steps: 
\begin{enumerate}
    \item An internal evolution
    \begin{align}
        \rho[n|\pmb{y}_{0:n}]  = \frac{\Phi_n \left[\rho[n\!\shortminus\!1|\pmb{y}_{0:n}] \right]}{\trace{\Phi_n \left[\rho[n\!\shortminus\!1|\pmb{y}_{0:n}] \right]}}.
    \end{align}
    \item A measurement update
    \begin{equation} \label{eq:rho_after_meas}
        \rho[n\!\shortminus\!1|\pmb{y}_{0:n}] = \frac{\rhoun[n\!\shortminus\!1|\pmb{y}_{0:n}]}{\trace{\rhoun[n\!\shortminus\!1|\pmb{y}_{0:n}]}} = \frac{\measE_{y_{n}}\rho[n\!\shortminus\!1|\pmb{y}_{0:n\shortminus1}]\measE_{y_{n}}^\dagger}{\trace{\measE_{y_{n}}\rho[n\!\shortminus\!1|\pmb{y}_{0:n\shortminus1}]\measE_{y_{n}}^\dagger}},
    \end{equation}
    with 
    \begin{equation}
        \trace{\rhoun[n\!\shortminus\!1|\pmb{y}_{0:n}]} = p_{\Dt}(y_n|\pmb{y}_{0:n\shortminus1})
    \end{equation}
    representing the probability of obtaining the outcome $y_n$ given a state evolved according to a previous measurement record $\pmb{y}_{0:n\shortminus1}$.
\end{enumerate}

Due to the analogy with the two-step process of a discrete KF (see \secref{sec:uncorr_discrete_KF}), we label the states above as follows:
\begin{align}
    \text{update : }&\rho[n\!\shortminus\!1|\pmb{y}_{0:n}] \, &= & \quad  \;\; \text{state after measurement update but before internal evolution} \nonumber \\
    \text{predict : }&\rho[n|\pmb{y}_{0:n}] \, &= & \quad  \;\; \text{state after measurement update and internal evolution}.
\end{align}

Having established the general iterative structure of the conditional evolution, we now focus on the measurement update step in more detail. Our aim is to derive the explicit form of the update state from the underlying probe–system interaction. Since the probe couples to the system through a unitary operator over a short interval $\Dt$, the update state can be expressed in terms of this unitary, expanded to first order in $\Dt$. Recalling the definition of the measurement operators $\measE_{y_n}$ in \eqnref{eq:def_POVM_E}, the unnormalized update state can be written as
\begin{align}
    \rhoun[n\!\shortminus\!1|\pmb{y}_{0:n}] &= \measE_{y_{n}}\,\rho[n\!\shortminus\!1|\pmb{y}_{0:n\shortminus1}]\measE_{y_{n}}^\dagger \nonumber \\
    &= \braketop{y_n}{\UnitOp_{\!\!\Dt}^{(n)}}{0} \rho[n\!\shortminus\!1|\pmb{y}_{0:n\shortminus1}] \braketop{0}{\UnitOp_{\!\!\Dt}^{(n) \dagger}}{y_n} \nonumber \\
    &= \bra{y_n}\UnitOp^{(n)}_{\Dt} (\rho[n\!\shortminus\!1|\pmb{y}_{0:n\shortminus1}] \otimes \projector{0}{0}) \UnitOp^{(n) \dagger}_{\Dt} \ket{y_n},
\end{align}
where the joint system–probe state before measurement is separable under the Born–Markov approximation. To proceed, we expand the unitary operator $\UnitOp_{\!\!\Dt}^{(n)}$ of \eqnref{eq:unitary_sqrtDeltat} to first order in $\Dt$:
\begin{align}
    \UnitOp_{\!\!\Dt}^{(n)} &= \I \!\otimes\! \I + \left(\!\LinOp \!\otimes\! \bcreat_n \!-\! \LinOp^\dagger \!\otimes\! \banihil_n \!\right) \!\sqrt{\kappa \, \Dt} \nonumber  \\
    &+ \frac{1}{2} \Big(\!\LinOp^2 \!\otimes\! (\bcreat_n)^2 - \! \LinOp^\dagger \!\LinOp \!\otimes \!\banihil_n\bcreat_n \!-\! \LinOp \LinOp^\dagger \!\otimes \!\bcreat_n \banihil_n \! + \! (\LinOp^\dagger)^2 \!\otimes \!(\banihil_n)^2 \!\Big) \kappa \, \Dt + \littleo(\Dt).
\end{align}
Therefore, the joint state after the probe-system interaction reads as:
\begin{align} \label{eq:joint_state_after_meas}
    \UnitOp^{(n)}_{\Dt} &(\rho[n\!\shortminus\!1|\pmb{y}_{0:n\shortminus1}] \otimes \projector{0}{0}) \UnitOp^{(n) \dagger}_{\Dt}
    =  \rho[n\!\shortminus\!1|\pmb{y}_{0:n\shortminus1}] \!\otimes\! \projector{0}{0} +  \kappa \, \LinOp \rho[n\!\shortminus\!1|\pmb{y}_{0:n\shortminus1}] \LinOp^\dagger  \!\otimes\!   \projector{1}{1}  \, \Dt    \nonumber \\
    &+  \!\left(   \LinOp\rho[n\!\shortminus\!1|\pmb{y}_{0:n\shortminus1}] \!\otimes\!  \projector{1}{0}  +  \rho[n\!\shortminus\!1|\pmb{y}_{0:n\shortminus1}] \LinOp^\dagger  \!\otimes\! \projector{0}{1} \right)  \! \sqrt{ \kappa \, \Dt} \nonumber\\
    &+ \!\frac{\kappa}{2} \! \Big( \!\sqrt{2} \LinOp^2  \rho[n\!\shortminus\!1|\pmb{y}_{0:n\shortminus1}]  \!\otimes\!  \projector{2}{0}  \!-\!  \LinOp^\dagger \LinOp \rho[n\!\shortminus\!1|\pmb{y}_{0:n\shortminus1}] \!\otimes\!  \projector{0}{0} \nonumber \\
    &+\! \sqrt{2} \rho[n\!\shortminus\!1|\pmb{y}_{0:n\shortminus1}] (\!\LinOp^\dagger\!)^2   \!\otimes\!  \projector{0}{2}  \!-\!  \rho[n\!\shortminus\!1|\pmb{y}_{0:n\shortminus1}] \LinOp^\dagger \LinOp \!\otimes\!\projector{0}{0} \!\Big) \Dt + \littleo(\Dt). 
\end{align}
Thus, different SMEs will be obtained depending on our choice of measurement, i.e. the projector $\projector{y_{n}}{y_{n}}$, where $y_n$ can for instance be the number of photons in the case of photodetection, or a Gaussian photocurrent in the case of homodyne measurement.

\subsection{Photodetection} \label{sec:derivation_SME_photo}

Let us consider the case where the probe is continuously measured with a photodetector, i.e., by projecting it on the Fock basis $\projector{n}{n}$. It is obvious from the form of the joint state before the measurement given in \eqnref{eq:joint_state_after_meas} that only when projecting the joint state onto the vacuum state $\projector{0}{0}$ or the single photon state $\projector{1}{1}$, a non-zero result will be obtained. Thus, let us consider each event, no-detection and detection, separately. If no photon is detected, then the unnormalized conditional state is:
\begin{align}
    \rhoun[n\!\shortminus\!1|\{0,\pmb{y}_{0:n\shortminus1}\}] &= \braketop{0}{\UnitOp^{(n)}_{\Dt} (\rho[n\!\shortminus\!1|\pmb{y}_{0:n\shortminus1}] \otimes \projector{0}{0}) \UnitOp^{(n) \dagger}_{\Dt}}{0} \\
    &= \rho[n\!\shortminus\!1|\pmb{y}_{0:n\shortminus1}] - \frac{\kappa}{2} \{ \LinOp^\dagger \LinOp \, , \, \rho[n\!\shortminus\!1|\pmb{y}_{0:n\shortminus1}] \} \, \Dt + \littleo(\Dt),
\end{align}
where $\rho[n\!\shortminus\!1|\pmb{y}_{0:n\shortminus1}]$ is the state conditioned by the measurement at the previous time step. Then, the probability of actually detecting no photons after the state evolves for a time-step $\Dt$ is:
\begin{align}
    p_{\Dt}(0|\pmb{y}_{0:n\shortminus1}) &= \trace{\rhoun[n\!\shortminus\!1|\{0,\pmb{y}_{0:n\shortminus1}\}]} = 1  - \kappa \langle \LinOp^\dagger \LinOp \rangle \Dt + \littleo(\Dt)
\end{align}
where $\langle \LinOp^\dagger \LinOp \rangle = \Tr\{\rho[n \shortminus 1|\pmb{y}_{0:n\shortminus1}] \LinOp^\dagger \LinOp\}$. By now expanding the inverse of $p_{\Dt}(0|\pmb{y}_{0:n\shortminus1})$ to first order in $\Dt$, i.e.
\begin{equation}
    p_{\Dt}(0|\pmb{y}_{0:n\shortminus1})^{-1} = 1 + \kappa \langle \LinOp^\dagger \LinOp \rangle \Dt + \littleo(\Dt),
\end{equation}
we can calculate, also to first order in $\Dt$, the normalized state conditioned on the measurement outcome $0$ w.r.t. the conditional state at the previous time $ \rho[n\!\shortminus\!1|\pmb{y}_{0:n\shortminus1}] \coloneqq \rhoc(t_{n\shortminus1})$:
\begin{align} \label{eq:cond_state_0}
    \rho[n\!\shortminus\!1&|\{0,\pmb{y}_{0:n\shortminus1}\}] = \frac{\rhoun[n\!\shortminus\!1|\{0,\pmb{y}_{0:n\shortminus1}\}]}{p_{\Dt}(0|\pmb{y}_{0:n\shortminus1})} \nonumber \\
    &= \! \left( \rhoc(t_{n\shortminus1}) - \frac{\kappa}{2} \{ \LinOp^\dagger \LinOp \, , \, \rhoc(t_{n\shortminus1}) \} \, \Dt + \littleo(\Dt) \right) \!\!\left(1 + \kappa \langle \LinOp^\dagger \LinOp \rangle \Dt + \littleo(\Dt) \right) \nonumber\\
    &= \rhoc(t_{n\shortminus1}) - \frac{\kappa}{2}\{ \LinOp^\dagger  \! \LinOp \, , \, \rhoc(t_{n\shortminus1}) \} \Dt + \kappa \langle \LinOp^\dagger  \! \LinOp \rangle \rhoc(t_{n\shortminus1}) \Dt + \littleo(\Dt) \nonumber \\
    &= \rhoc(t_{n\shortminus1}) - \frac{\kappa}{2}\{ \LinOp^\dagger  \!  \LinOp \, , \, \rhoc(t_{n\shortminus1}) \} \Dt + \frac{\kappa}{2} \langle  \LinOp^\dagger  \! \LinOp + \LinOp^\dagger  \! \LinOp \rangle \rhoc(t_{n\shortminus1}) \Dt + \littleo(\Dt) \nonumber \\
    &= \rhoc(t_{n\shortminus1}) - \frac{\kappa}{2} \H[\LinOp^\dagger \LinOp] \rhoc(t_{n\shortminus1}) \Dt + \littleo(\Dt),
\end{align}
where we introduce the nonlinear superoperator 
\begin{equation} \label{eq:H_superoperator_def}
    \H[\hat{O}] \bigcdot = \hat{O} \bigcdot + \bigcdot \hat{O}^\dagger - \trace{(\hat{O} + \hat{O}^\dagger)\bigcdot} \bigcdot.
\end{equation}
If instead a photon is detected, the unnormalized conditional state becomes
\begin{equation}
    \rhoun[n\!\shortminus\!1|\{1,\pmb{y}_{0:n\shortminus1}\}] = \braketop{1}{\UnitOp^{(n)}_{\Dt} (\rho[n\!\shortminus\!1|\pmb{y}_{0:n\shortminus1}] \otimes \projector{0}{0}) \UnitOp^{(n) \dagger}_{\Dt}}{1} = \kappa \, \LinOp \rhoc(t_{n\shortminus1}) \LinOp^\dagger \Dt + \littleo(\Dt),
\end{equation}
from which we can calculate the associated probability
\begin{equation}
   p_{\Dt}(1|\pmb{y}_{0:n\shortminus1}) = \trace{\rhoun[n\!\shortminus\!1|\{1,\pmb{y}_{0:n\shortminus1}\}]} = \kappa \langle \LinOp^\dagger \LinOp\rangle \Dt + \littleo(\Dt).
\end{equation}
By dividing the unnormalized updated state by the probability of measuring one photon, we retrieve:
\begin{equation} \label{eq:cond_state_1}
    \rho[n\!\shortminus\!1|\{1,\pmb{y}_{0:n\shortminus1}\}] = \frac{\rhoun[n\!\shortminus\!1|\{1,\pmb{y}_{0:n\shortminus1}\}]}{p_{\Dt}(1|\pmb{y}_{0:n\shortminus1})} = \frac{\LinOp \rhoc(t_{n\shortminus1}) \LinOp^\dagger}{\langle \LinOp^\dagger \LinOp \rangle } + \littleo(\Dt).
\end{equation}
To sum up, at each time step, the photodetector yields either $0$ or $1$ with probabilities:
\begin{align}
    y_n  &= 0 \sim p_{\Dt}(0|\pmb{y}_{0:n\shortminus1}) = 1 -\kappa \langle \LinOp^\dagger \LinOp \rangle \Dt + \littleo(\Dt),\\
    y_n  &= 1 \sim p_{\Dt}(1|\pmb{y}_{0:n\shortminus1}) = \kappa \langle \LinOp^\dagger \LinOp \rangle \Dt  + \littleo(\Dt),
\end{align}
where the rate of the process, $\kappa \langle \LinOp^\dagger \LinOp \rangle $, is time-dependent. Therefore, if one recalls the definition of a non-homogeneous Poisson increment in \defref{def:non-hom_Poisson}, it comes natural to model the $\{0,1\}$ output within the timestep $\Dt$ as a non-homogeneous Poisson increment $\Delta \N \sim \Pois(\kappa \langle \LinOp^\dagger \LinOp \rangle \Dt)$, where as $\Dt \rightarrow 0, \Pr[\Delta \mrm{N} \in \{0,1\}] \rightarrow 1$. Thus, the Poisson increment we introduce has a mean
\begin{equation}
    \EE{\Delta \N} = 0 \cdot p_{\Dt}(0|\pmb{y}_{0:n\shortminus1}) + 1 \cdot p_{\Dt}(1|\pmb{y}_{0:n\shortminus1}) = \kappa \langle \LinOp^\dagger \LinOp \rangle \Dt + \littleo(\Dt)
\end{equation}
which completely describes $\Delta \N$, as discussed in \secref{sec:Poisson_process}. Finally, if we apply the internal evolution map $\Phi_n$, we get the unnormalized conditional state at time $t = n \Dt$:
\begin{align}
    &\text{if} \; y_n = 1 \quad \, \Longrightarrow \quad  \rhoun[n|\{1,\pmb{y}_{0:n\shortminus1}\}] = \Phi_n \left[\rho[n\!\shortminus\!1|\{1,\pmb{y}_{0:n\shortminus1}\}]\right] \\
    &\text{if} \; y_n = 0 \quad \Longrightarrow \quad \rhoun[n|\{0,\pmb{y}_{0:n\shortminus1}\}] = \Phi_n \left[\rho[n\!\shortminus\!1|\{0,\pmb{y}_{0:n\shortminus1}\}]\right]
\end{align}
Let us now consider a measurement-based feedback map $\Phi_n[\;\cdot\;]$, where the Lindbladian governing the evolution of the system depends on the entire history of measurement results $\pmb{y}_{0:n}$. In this setting, we assume the generator $\Lin_{\pmb{y}_{0:n}}$ to be of order $\bigO(1)$, so for sufficiently small time steps the exponential map can always be expanded to first order in $\Dt$:
\begin{align} \label{eq:Phi_map_expansion}
    \Phi_n[\;\cdot\;] = \ee^{\Lin_{\pmb{y}_{0:n}}\Dt}[\;\cdot\;] = \; \cdot \; + \Lin_{\pmb{y}_{0:n}}\; \cdot \;\Dt + \littleo(\Dt).
\end{align}
This assumption does not cover the case of Markovian feedback as introduced by  Wiseman~\cite{Wiseman1994_feedback}, where the feedback Hamiltonian is applied instantaneously and is proportional to the measurement current \cite{Amoros-Binefa2024}. In that scenario, the stochastic part of the generator includes increments that scale as $\sqrt{\Dt}$, so the correct treatment requires the L\'{e}vy-It\^{o} decomposition \cite{Applebaum_2009,Amoros-Binefa2024}\footnote{If the feedback map depends directly on a measurement current modelled as a white Gaussian noise, then the map can be written as $\ee^{\Lin \Delta \W}$, where $\Delta\W \sim \Gauss(0,\Dt)$ denotes a Wiener increment. In this case, the It\^{o} expansion of \lemref{lem:itos_lemma} applies directly.}. Such a general treatment is given in Appendix~C.1. of \refcite{Amoros-Binefa2024}. 

Here, however, we restrict ourselves to the case $\Lin_{\pmb{y}_{0:n}} \sim \bigO(1)$ and combine the expansion in \eqnref{eq:Phi_map_expansion} with the conditional update rules for the measurement outcomes. This yields the following form of the unnormalized conditional state:
\begin{align}
    &\text{if} \; y_n = 1 &\Longrightarrow& \quad \rhoun[n|\{1,\pmb{y}_{0:n\shortminus1}\}] = \Phi_n \!\left[ \frac{\LinOp \rhoc(t_{n\shortminus1}) \LinOp^\dagger}{\langle \LinOp^\dagger \LinOp \rangle } + \littleo(\Dt)\right] \nonumber \\
    & & & \quad\quad= \frac{\LinOp \rhoc(t_{n\shortminus1}) \LinOp^\dagger}{\langle \LinOp^\dagger \LinOp \rangle } + \bigO(\Dt)\\
    &\text{if} \; y_n = 0 &\Longrightarrow& \quad \rhoun[n|\{0,\pmb{y}_{0:n\shortminus1}\}] = \Phi_n \!\left[\rhoc(t_{n\shortminus1}) - \frac{\kappa}{2} \H[\LinOp^\dagger \LinOp] \rhoc(t_{n\shortminus1}) \Dt + \littleo(\Dt) \right] \nonumber \\
    & & & \quad\quad = \rhoc(t_{n\shortminus1}) + \Lin_{\pmb{y}_{0:n}}\,\rhoc(t_{n\shortminus1}) \Dt  - \frac{\kappa}{2} \H[\LinOp^\dagger \LinOp] \rhoc(t_{n\shortminus1}) \Dt + \littleo(\Dt), 
\end{align}
\footnotetext[1]{Big-O notation, $\bigO(\Delta t)$, describes terms that scale at most linearly with $\Delta t$, i.e., they may be proportional to $\Delta t$ or smaller (e.g. $\Delta t^{3/2}$, $\Delta t^2$, $\ldots$). Little-o, $\littleo(\Delta t)$, includes only terms that vanish faster than $\Delta t$ as $\Delta t \to 0$. For instance,  $\Delta t^{3/2} \in \littleo(\Delta t)$, while $\Delta t \in \bigO(\Delta t)$ but not in $\littleo(\Delta t)$.} where for the case of $y_n = 1$, we expand only to zeroth order. Since $\trace{\H[\LinOp^\dagger \LinOp] \rho} = 0$ and $\trace{\Lin \rho} = 0$, the trace of both unnormalized states above is 1. In other words, the states are already normalized:
\begin{align}
    \rho[n|\{1,\pmb{y}_{0:n\shortminus1}\}] &=  \frac{\LinOp \rhoc(t_{n\shortminus1}) \LinOp^\dagger}{\langle \LinOp^\dagger \LinOp \rangle } + \bigO(\Dt) \\
    \rho[n|\{0,\pmb{y}_{0:n\shortminus1}\}] &= \rhoc(t_{n\shortminus1}) + \Lin_{\pmb{y}_{0:n}}\,\rhoc(t_{n\shortminus1}) \Dt  - \frac{\kappa}{2} \H[\LinOp^\dagger \LinOp] \rhoc(t_{n\shortminus1}) \Dt + \littleo(\Dt).
\end{align}
Since $\Delta \N$ indicates whether a particular quantum event within the interval $\Dt$ has occurred, i.e.,
\begin{align}
    &\Delta \N = 1  & &\text{if the event occurs (measurement outcome } 1), \\
    &\Delta \N = 0  & &\text{if the event does not occur (measurement outcome } 0),
\end{align}
then the change on the state when the event occurs v.s. when it does not can be written as
\begin{align}
    &\Delta \N \left(\rho[n|\{1,\pmb{y}_{0:n\shortminus1}\}] - \rho[n\!\shortminus\!1|\pmb{y}_{0:n\shortminus1}] \right)  & &\text{if the event occurs (outcome } 1), \\
    &(1-\Delta \N) \left(\rho[n|\{0,\pmb{y}_{0:n\shortminus1}\}] - \rho[n\!\shortminus\!1|\pmb{y}_{0:n\shortminus1}] \right) & &\text{if the event does not occur (outcome } 0),
\end{align}
where the state $\rho[n\!\shortminus\!1|\pmb{y}_{0:n\shortminus1}] \coloneqq \rhoc(t_{n\shortminus1})$ is the conditional state of the system at the previous step. 
Note that $\Delta \N$ and $1-\Delta \N$ are included when describing the change in the state because they indicate which of the outcomes, i.e. $1$ or $0$, actually occur within the interval $\Dt$. Thus, the overall change in $\rhoc(t_{n\shortminus1})$ can be written as: 
\begin{align} \label{eq:derivation_sme_photondetection_1}
    &\Delta \rhoc(t_n) = \rhoc(t_n) - \rhoc(t_{n\shortminus1})  \\
    &= \Delta \N \! \left(\rho[n|\{1,\pmb{y}_{0:n\shortminus1}\}] - \rhoc(t_{n\shortminus1}) \right) \!+\! (1-\Delta \N) \! \left(\rho[n|\{0,\pmb{y}_{0:n\shortminus1}\}] - \rhoc(t_{n\shortminus1}) \!\right) \!+\! \littleo(\Dt) \nonumber,
\end{align}
where we can disregard any terms $\Dt \Delta \N$ because they represent higher-order infinitesimal contributions that are negligible in this analysis. In particular, for any calculus involving averages or integrations over time, the terms with $\Dt \Delta \N$ become a second-order small term, akin to $\Dt^2$, since for a general Poisson increment with a rate $\lambda>0$:
\begin{equation}
    \EE{\Dt \, \Delta \N} = \Dt \, \EE{\Delta \N} = \Dt \, \lambda \Dt = \lambda \Dt^2.
\end{equation}

Therefore, when substituting \eqnref{eq:cond_state_0} and \eqnref{eq:cond_state_1} into \eqnref{eq:derivation_sme_photondetection_1}, it becomes 
\begin{align}
    &\Delta \rhoc(t_n) \!=\! \Delta \N \!\left(\frac{\LinOp \rhoc(t_{n\shortminus1}) \LinOp^\dagger}{\langle \LinOp^\dagger \LinOp \rangle } -\rhoc(t_{n\shortminus1})\! \right) \!+\! (1\!-\!\Delta \N) \left(\Lin_{\pmb{y}_{0:n}}\,\rhoc(t_{n\shortminus1}) \Dt \! - \frac{\kappa}{2} \H[\LinOp^\dagger \LinOp] \rhoc(t_{n\shortminus1}) \Dt \!\right) \nonumber \\
    &=\Lin_{\pmb{y}_{0:n}}\,\rhoc(t_{n\shortminus1}) \Dt - \frac{\kappa}{2} \H[\LinOp^\dagger \LinOp] \rhoc(t_{n\shortminus1}) \Dt \!+\! \left(\!\frac{\LinOp \rhoc(t_{n\shortminus1}) \LinOp^\dagger}{\trace{\LinOp \rhoc(t_{n\shortminus1}) \LinOp^\dagger} } -\rhoc(t_{n\shortminus1}) \!\right) \!\Delta \N ,
\end{align}
where $\H[\bigcdot]$ is the nonlinear operator defined in \eqnref{eq:H_superoperator_def} and the jumps governed by the Poisson increment $\Delta \N$ are modulated by a nonlinear quantity. As we will also see in the case of the homodyne measurement, these nonlinearities arise as a consequence of the normalization of the conditional state after conditioning due to the measurement. By finally taking the limit of $\Dt \to 0$, we retrieve the SME for photodetection:
\begin{align}
    \dd \rhoc(t) =\Lin_{\pmb{y}_t}\,\rhoc(t) \dt - \frac{\kappa}{2} \H[\LinOp^\dagger \LinOp] \rhoc(t) \dt \!+\! \left(\!\frac{\LinOp \rhoc(t) \LinOp^\dagger}{\trace{\LinOp \rhoc(t) \LinOp^\dagger} } -\rhoc(t) \!\right) \!\dN.
\end{align}

\subsection{Homodyne measurement} \label{sec:homodyne_meas}

Next, we explore monitoring the quantum system $\mathcal{S}$ via a homodyne measurement instead of with photodetection. This method involves the projection of the probe state onto the eigenstates $\{\ket{x^\phi_{n}}\}$ of the general quadrature operator $\Xphi_{n}$ defined as
\begin{equation} \label{eq:def_quadrature_x}
    \Xphi_{n} \coloneqq \frac{\banihil_n \, \ee^{\ii \phi} + \bcreat_n \ee^{-\ii \phi}}{\sqrt{2}},
\end{equation}
which corresponds to the position operator $\position$ defined in \eqnref{eq:pos&mom_a&adagger} when $\phi = 0$, and the momentum operator $\momentum$ when $\phi = \pi/2$.
The set of eigenstates $\{\ket{x^{\phi}_n}\}$ related to $\Xphi_n$ are known as quadrature eigenstates and measure the amplitude or phase quadrature when $\phi = 0$ and $\phi = \pi/2$, respectively. The statistics of the homodyne outcomes $x^\phi_n$ depend on both the state of the probe being measured and the choice of $\phi$. For specific quantum states, such as the vacuum, coherent and squeezed states, the quadrature distribution is Gaussian, characterized by its mean and variance. The mean reflects the displacement of the state, while the variance captures quantum noise, including any squeezing effects. For example, the vacuum state yields a Gaussian distribution centered at zero with variance $1/2$, representing the fundamental quantum noise, or shot noise. 

Since the probe segments are initially assumed to be in the vacuum state (recall \eqnref{eq:def_POVM_E}), performing a homodyne measurement on the probe before its interaction with the system yields a Gaussian distribution:
\begin{align}
    p_0(x^\phi_n|\pmb{x}^{\phi}_{0:n\shortminus1}) &= \trace{\braketop{x^\phi_n}{\UnitOp_0^{(n)}}{0}\rhoc(t_{n\shortminus1})  \braketop{0}{\UnitOp_0^{(n)}}{x^\phi_n} }  \nonumber \\
    &= \trace{\bra{x^\phi_n}(\rhoc(t_{n\shortminus1}) \otimes \projector{0}{0}) \ket{x^\phi_n} }  \nonumber \\
    &= \trace{\rhoc(t_{n\shortminus1})}  \braketsquared{0}{x^\phi_n} = \braketsquared{0}{x_n^\phi} = \frac{1}{\sqrt{\pi}} \e^{-{x_n^\phi}^2},
\end{align}
where in the last step we have used the fact that the ground state wavefunction of the 1D harmonic oscillator -- with its energy eigenstates being the Fock states -- in the (rotated) position representation is a Gaussian function, as derived in \eqnref{eq:rotated_GW_HO}. 

If, instead, we first evolve the joint system-probe state with the interaction unitary before performing the homodyne measurement as described above, the probability distribution of the measurement outcome $x_n^\phi$ will also be Gaussian, but with a shifted mean and potentially modified variance due to the system-probe interaction. To derive this result, we need to consider the trace of the unnormalized state after the measurement, which determines he probability distribution of the measurement outcome:
\begin{align} \label{eq:trace_uncond_hdyn}
    p_{\Dt}(x_n^{\phi}|\pmb{x}_{0:n\shortminus1}^{\phi}) &= \trace{\rhoun[n\!\shortminus\!1|\pmb{x}_{0:n}^\phi]}
\end{align}
where
\begin{align}
    &\rhoun[n\!\shortminus\!1|\pmb{x}_{0:n}^\phi] = \bra{x_n^\phi}\UnitOp^{(n)}_{\Dt} (\rho[n\!\shortminus\!1|\pmb{x}_{0:n\shortminus1}^\phi] \otimes \projector{0}{0}) \UnitOp^{(n) \dagger}_{\Dt} \ket{x_n^\phi} \nonumber \\
    &=\rho[n\!\shortminus\!1|\pmb{x}^\phi_{0:n\shortminus1}]  \braketsquared{0}{x_n^\phi} +  \kappa \, \LinOp \rho[n\!\shortminus\!1|\pmb{x}^\phi_{0:n\shortminus1}] \LinOp^\dagger     \braketsquared{1}{x_n^\phi}  \, \Dt    \nonumber \\
    &+  \!\left(   \LinOp\rho[n\!\shortminus\!1|\pmb{x}^\phi_{0:n\shortminus1}]  \braket{x_n^\phi}{1}\braket{0}{x_n^\phi}  +  \rho[n\!\shortminus\!1|\pmb{x}^\phi_{0:n\shortminus1}] \LinOp^\dagger  \braket{x_n^\phi}{0}\braket{1}{x_n^\phi} \right)  \! \sqrt{ \kappa \, \Dt} \nonumber\\
    &+ \!\frac{\kappa}{2} \! \Big( \!\sqrt{2} \LinOp^2  \rho[n\!\shortminus\!1|\pmb{x}^\phi_{0:n\shortminus1}]   \braket{x_n^\phi}{2}\braket{0}{x_n^\phi}  \!-\!  \LinOp^\dagger \LinOp \rho[n\!\shortminus\!1|\pmb{x}^\phi_{0:n\shortminus1}] \braketsquared{0}{x_n^\phi} \nonumber \\
    &+\! \sqrt{2} \rho[n\!\shortminus\!1|\pmb{x}^\phi_{0:n\shortminus1}] (\!\LinOp^\dagger\!)^2    \braket{x_n^\phi}{0}\braket{2}{x_n^\phi} \!-\!  \rho[n\!\shortminus\!1|\pmb{x}^\phi_{0:n\shortminus1}] \LinOp^\dagger \LinOp \braketsquared{0}{x_n^\phi}\!\Big) \Dt + \littleo(\Dt). \label{eq:uncond_state_after_hdyn_2}
\end{align}
The terms $\braket{x_n^\phi}{1}$ and $\braket{x_n^\phi}{2}$ can be written in terms of $\braket{x_n^\phi}{0}$:
\begin{align}
    &\braket{x_n^\phi}{1} = \braketop{x_n^\phi}{\bcreat_n}{0} = \sqrt{2}  \ee^{\ii\phi} \braketopadj{\!\!x_n^\phi}{ \frac{\banihil_n \, \ee^{\ii \phi} + \bcreat_n \ee^{-\ii \phi}}{\sqrt{2}}}{0\!\!} = \sqrt{2} \ee^{\ii\phi}\braketop{x_n^\phi}{\Xphi_n}{0} \nonumber \\
    &\quad\quad= \sqrt{2} \ee^{\ii\phi} x_n^\phi \, \braket{x_n^\phi}{0}, \label{eq:identity_<x|1>} \\
    &\sqrt{2}\,\braket{x_n^\phi}{2} = (2{x_n^\phi}^2 - 1) \ee^{2\ii \phi} \braket{x_n^\phi}{0}, \label{eq:identity_<x|2>}
\end{align}
which we derived using the fact that the second identity follows from:
\begin{align}
    {x_n^\phi}^2 \, \braket{x_n^\phi}{0} &= \braketop{x_n^\phi}{{\Xphi_n}^2}{0} = \braketopadj{\!\!x_n^\phi}{\frac{\banihil_n \, \ee^{\ii \phi} + \bcreat_n \ee^{-\ii \phi}}{\sqrt{2}}\frac{\banihil_n \, \ee^{\ii \phi} + \bcreat_n \ee^{-\ii \phi}}{\sqrt{2}}}{0\!\!} \nonumber \\
    &= \frac{1}{2} \braketop{x_n^\phi}{(\bcreat_n)^2 \ee^{-2\ii \phi} + \banihil_n \, \bcreat_n}{0} = \frac{1}{2} \!\! \left( \! \sqrt{2} \ee^{-2\ii \phi} \,\braket{x_n^\phi}{2}  + \braket{x_n^\phi}{0} \right)\!\!.
\end{align}

By now substituting \eqnref{eq:identity_<x|1>} and \eqnref{eq:identity_<x|2>} into \eqnref{eq:uncond_state_after_hdyn_2}, and recalling that $\braketsquared{x_n^\phi}{0} = p_0(x_n^\phi|\pmb{x}_{0:n\shortminus1}^\phi)$, then the unnormalized updated state becomes
\begin{align} \label{eq:uncond_state_after_hdyn_3}
    \rhoun[n\!\shortminus\!1|\pmb{x}_{0:n}^\phi]  &= p_0(x_n^\phi|\pmb{x}_{0:n\shortminus1}^\phi) \bigg\{\rhoc(t_{n\shortminus1})\!+\! \big(\LinOp\rhoc(t_{n\shortminus1}) \ee^{\ii \phi} \!+\! \rhoc(t_{n\shortminus1})\LinOp^\dagger \ee^{-\ii \phi}  \big)\sqrt{2\kappa \Dt} \, x_n^\phi  +  \nonumber \\
    &+ \kappa \LinOp \rho \LinOp^\dagger \, 2  \, {x_n^\phi}^2 \Dt + \! \frac{\kappa}{2} \Big(\!\LinOp^2 \rhoc(t_{n\shortminus1}) (2{x_n^\phi}^2 - 1) \ee^{2\ii \phi}  \!-\! \LinOp^\dagger\! \LinOp \rhoc(t_{n\shortminus1})\nonumber \\
    &+\! \rhoc(t_{n\shortminus1})(\!\LinOp^\dagger\!)^2  (2{x_n^\phi}^2 \!\shortminus\! 1) \ee^{-2\ii \phi} \!-\! \rhoc(t_{n\shortminus1})\LinOp^\dagger \LinOp \Big)\Dt \!\bigg\} + \littleo(\Dt) .
\end{align}

By now taking the trace of the unconditional state as described in \eqnref{eq:trace_uncond_hdyn}, we get the probability of obtaining the measurement outcome $x_n^\phi$ after the system and probe interact for a time $\Dt$:
\begin{align} \label{eq:p(t+dt|x)_expansion}
    p_{\Dt}(x_n^\phi|\pmb{x}_{0:n\shortminus1}^\phi) &=  p_0(x_n^\phi|\pmb{x}_{0:n\shortminus1}^\phi) \! \left( \! 1 \!+\! x_n^\phi \braketavg{\LinOp \ee^{\ii \phi}+ \LinOp^\dagger \ee^{-\ii \phi}} \sqrt{2\kappa\Dt} \!+\! \bigO(\Dt)\!\right)\!, 
\end{align}
which, up to order $\sqrt{\Dt}$, follows a Gaussian distribution, since any Gaussian function can be expanded as:
\begin{equation}
    g(t) = A \e^{-(x - \mu \sqrt{t/2})^2} = g(0) \left\{ 1 + x \, \mu \, \sqrt{2 t} + \bigO(t)\right\}.
\end{equation}

Thus, the probability of the homodyne measurement yielding an outcome $x_n^\phi$ at time $t$ is given, up to leading order in $\Dt$, by the following Gaussian distribution:
\begin{align} \label{eq:p(t+dt|x)_gaussian}
    p_{\Dt}(x_n^\phi|\pmb{x}_{0:n\shortminus1}^\phi) &= \frac{1}{\sqrt{\pi}} \exp{\left\{-\left(x_n^\phi - \braketavg{\LinOp \ee^{\ii \phi}+ \LinOp^\dagger\ee^{-\ii \phi}} \sqrt{\frac{\kappa \Dt}{2}}\right)^{\!\!\!2} \right\}} + \bigO(\Dt),
\end{align}
with a mean $\sqrt{\kappa \Dt/2} \braketavg{\LinOp \ee^{\ii \phi}+ \LinOp^\dagger \ee^{-\ii \phi}} = \sqrt{\kappa \Dt/2} \Tr\{(\LinOp \ee^{\ii \phi}+ \LinOp^\dagger \ee^{-\ii \phi})\rhoc(t_{n\shortminus1})\}$ and variance $1/2$.
We can now introduce a new stochastic increment $\Delta y_n$ by multiplying $x_n^\phi$ by $\sqrt{2 \Dt}$:
\begin{equation}
    \Delta y_n \coloneqq x_n^\phi \sqrt{2 \Dt},
\end{equation}
which, therefore, has a mean $\sqrt{\kappa} \braketavg{\LinOp \ee^{\ii \phi}+ \LinOp^\dagger \ee^{-\ii \phi}} \Dt$ and a variance $\Dt$. Namely,
\begin{equation} \label{eq:homodyne_eq_derived}
    \Delta y_n \coloneqq x_n^\phi\sqrt{2\Dt} = \sqrt{\kappa} \braketavg{\LinOp \ee^{\ii \phi} + \LinOp^\dagger \ee^{-\ii \phi}} \, \Dt + \Delta \W,
\end{equation}
where $\Delta \W \sim \mathcal{N}(0,\Dt)$ denotes the Wiener increment introduced in \secref{sec:Wiener_process}. Physically, the derivative $I(t)\coloneqq \lim_{\Dt \to 0}\Delta y_n/\Dt$ corresponds to the stochastically fluctuating photocurrent being measured in real time in a homodyne setup. Next, by using $x_n^\phi\sqrt{2 \Dt} = \Delta y_n$ and $2{x_n^\phi}^2 \Dt = \Delta y_n^2 = \Dt +\littleo(\Dt)$, we can rewrite \eqnref{eq:uncond_state_after_hdyn_3} it into the following, more familiar, form:
\begin{align}
    \rhoun[n\!\shortminus\!1|\pmb{x}_{0:n}^\phi] &= p_0(x_n^\phi|\pmb{x}_{0:n\shortminus1}^\phi) \bigg\{\rhoc(t_{n\shortminus1})\!+\! \sqrt{\kappa} \big(\LinOp\rhoc(t_{n\shortminus1})  \ee^{\ii \phi}  \!+\! \rhoc(t_{n\shortminus1})\LinOp^\dagger \ee^{-
    \ii \phi} \big) \Delta y_n + \kappa \LinOp \rho \LinOp^\dagger \halfsmallspace \Dt   \nonumber \\
    &- \! \frac{\kappa}{2} \Big(\! \LinOp^\dagger\! \LinOp \rhoc(t_{n\shortminus1}) \!+\! \rhoc(t_{n\shortminus1})\LinOp^\dagger \LinOp \Big)\Dt + \littleo(\Dt) \!\bigg\} = p_0(x_n^\phi|\pmb{x}_{0:n\shortminus1}^\phi)  \bigg\{\rhoc(t_{n\shortminus1}) \nonumber \\
    &+  \kappa \D[\LinOp] \rhoc(t_{n\shortminus1}) \Dt + \sqrt{\kappa} \big(\LinOp\rhoc(t_{n\shortminus1})\ee^{\ii \phi} \!+\! \rhoc(t_{n\shortminus1})\LinOp^\dagger\ee^{-\ii \phi}  \big) \Delta y_n + \littleo(\Dt) \!\bigg\}, 
\end{align}
where the superoperator representing the measurement-induced decoherence has the same form as the one defined in \eqnref{eq:dissipation_superoperator}. To normalize this state, we divide it by $p_{\Dt}(x_n^\phi|\pmb{x}_{0:n\shortminus1}^\phi)$ whilst keeping terms to first order in $\Dt$ and lower:
\begin{align}
    \rho[n\!\shortminus\!1|\pmb{x}_{0:n}^\phi] &= \frac{\rhoun[n\!\shortminus\!1|\pmb{x}_{0:n}^\phi]}{p_{\Dt}(x_n^\phi|\pmb{x}_{0:n\shortminus1}^\phi)} = \Big( \rhoc(t_{n\shortminus1}) + \kappa \D[\LinOp] \rhoc(t_{n\shortminus1}) \Dt  \nonumber \\
    &+ \sqrt{\kappa} \big(\LinOp\rhoc(t_{n\shortminus1})  \ee^{\ii \phi}\!+\! \rhoc(t_{n\shortminus1})\LinOp^\dagger \ee^{-\ii \phi}  \big) \Delta y_n  \!\Big) \! \Big(\! 1 \!-\! \sqrt{\kappa} \braketavg{\LinOp  \ee^{\ii \phi}\!+\! \LinOp^\dagger  \ee^{-\ii \phi}} \Delta y_n \nonumber \\
    &+\! \kappa \braketavg{\LinOp  \ee^{\ii \phi} \!+\! \LinOp^\dagger  \ee^{-\ii \phi}}^2 \Dt \! \Big) \!+\! \littleo(\Dt) = \rhoc(t_{n\shortminus1}) + \kappa \D[\LinOp] \rhoc(t_{n\shortminus1}) \Dt \nonumber \\
    & + \sqrt{\kappa} \big(\LinOp\rhoc(t_{n\shortminus1}) \ee^{\ii \phi}  +\! \rhoc(t_{n\shortminus1})\LinOp^\dagger  \ee^{-\ii \phi} \big) \Delta y_n -\sqrt{\kappa} \braketavg{\LinOp  \ee^{\ii \phi}\!+\! \LinOp^\dagger \ee^{-\ii \phi}} \rhoc(t_{n\shortminus1}) \Delta y_n  \nonumber \\
    &-\! \kappa \big(\LinOp\rhoc(t_{n\shortminus1})  \ee^{\ii \phi} \!+\! \rhoc(t_{n\shortminus1})\LinOp^\dagger  \ee^{-\ii \phi} \big) \!\braketavg{\LinOp  \ee^{\ii \phi} \!+\! \LinOp^\dagger  \ee^{-\ii \phi}} \Dt \!+\! \kappa \braketavg{\LinOp  \ee^{\ii \phi} \nonumber \\
    &+\! \LinOp^\dagger  \ee^{-\ii \phi}}^2 \Dt + \littleo(\Dt). \label{eq:state_with_Deltay}
\end{align}
where we have used that the inverse of the conditional probability to leading order $\Dt$ is:
\begin{align}
    &\frac{1}{p_{\Dt}(x_n^\phi|\pmb{x}_{0:n\shortminus1}^\phi)} = \frac{1}{p_{0}(x_n^\phi|\pmb{x}_{0:n\shortminus1}^\phi)} \left( 1 + x_n^\phi  \braketavg{\LinOp  \ee^{\ii \phi} + \LinOp^\dagger  \ee^{-\ii \phi}} \sqrt{2 \kappa \Dt} + \littleo(\Dt) \right)^{-1} \nonumber \\
    &= \frac{1}{p_{0}(x_n^\phi|\pmb{x}_{0:n\shortminus1}^\phi)} \left( 1 + \sqrt{\kappa} \braketavg{\LinOp \ee^{\ii \phi} + \LinOp^\dagger \ee^{-\ii \phi}} \Delta y_n + \littleo(\Dt) \right)^{-1} \nonumber \\
    &= \frac{1}{p_{0}(x_n^\phi|\pmb{x}_{0:n\shortminus1}^\phi)} \left( 1 - \sqrt{\kappa} \braketavg{\LinOp  \ee^{\ii \phi}+ \LinOp^\dagger  \ee^{-\ii \phi}} \Delta y_n + \kappa \braketavg{\LinOp \ee^{\ii \phi} + \LinOp^\dagger  \ee^{-\ii \phi}}^2 \Dt + \littleo(\Dt) \right). 
\end{align}
By now inserting the expression of $\Delta y_n$ defined in \eqnref{eq:homodyne_eq_derived}, we can simplify the state in \eqnref{eq:state_with_Deltay} and get
\begin{align}
    \rho[n\!\shortminus\!1|\pmb{x}_{0:n}^\phi] = \rhoc(t_{n\shortminus1}) + \kappa \D[\LinOp] \rhoc(t_{n\shortminus1}) \Dt + \sqrt{\kappa} \H[\LinOp \ee^{\ii \phi}] \rhoc(t_{n\shortminus1}) \Delta \W + \littleo(\Dt)
\end{align}
with the nonlinear superoperator modulating the stochastic (Gaussian) kicks defined in \eqnref{eq:H_superoperator_def}. Then, if we evolve the state with the internal map $\Phi_n$, we obtain the conditional state at time $t = n\Dt$:
\begin{align}
    \rhoc(t_n) &= \rho[n|\pmb{x}_{0:n}^\phi] = \frac{\rhoun[n|\pmb{x}_{0:n}^\phi]}{\trace{\rhoun[n|\pmb{x}_{0:n}^\phi]}} \nonumber \\
    &= \frac{\Phi_n \! \left[ \rhoc(t_{n\shortminus1}) + \kappa \D[\LinOp] \rhoc(t_{n\shortminus1}) \Dt + \sqrt{\kappa} \H[\LinOp \ee^{\ii \phi}] \rhoc(t_{n\shortminus1}) \Delta \W + \littleo(\Dt)\right]}{\trace{\rhoun[n|\pmb{x}_{0:n}^\phi]}}.
\end{align}
As in \secref{sec:derivation_SME_photo}, we now assume that the CPTP map $\Phi_n[\;\cdot\;]$ admits the same short-time expansion of \eqnref{eq:Phi_map_expansion}, i.e.
\begin{equation} \label{eq:Phi_map_expansion_homodyne}
    \Phi_n[\;\cdot\;] = \ee^{\Lin_{\pmb{y}_{0:n}}\Dt}[\;\cdot\;] = \; \cdot \; + \Lin_{\pmb{y}_{0:n}}\; \cdot \;\Dt + \littleo(\Dt), 
\end{equation}
with the crucial condition that the Lindbladian $\Lin_{\pmb{y}_{0:n}}$ depends only on the measurement record $\pmb{y}_{0:n}$ itself and not on its time derivative, i.e., $I(t) =  \frac{\dd y_{t}}{\dt} = \lim_{\Dt \rightarrow 0} \frac{\Delta y_{n}}{\Dt}$. In other words, the generator remains of order $\bigO(1)$ in $\Dt$. Substituting \eqnref{eq:Phi_map_expansion_homodyne} into the update rule and using $\Tr\{\rhoun[n|\pmb{x}_{0:n}^\phi]\} = 1$, we arrive at the discrete-time stochastic master equation for homodyne detection with feedback:
\begin{align}
    \!\!\!\rhoc(t_n) &\!=\! \rhoc(t_{n\shortminus1}) \!+\! \Lin_{\pmb{y}_{0:n}} \rhoc(t_{n\shortminus1}) \Dt \nonumber \\
&\!+\! \kappa \D[\LinOp] \rhoc(t_{n\shortminus1}) \Dt \!+\! \sqrt{\kappa} \H[\LinOp \ee^{\ii \phi}] \rhoc(t_{n\shortminus1}) \Delta\!\W \!+\! \littleo(\Dt).
\end{align}
Finally, to pass from the discrete update to a continuous description, we interpret the recursion as an Euler-Maruyama approximation of an It\^{o} stochastic differential equation. Thus, by setting $\Delta \rhoc(t_n) = \rhoc(t_n) - \rhoc(t_{n\shortminus1})$, and taking the limit of $\Dt \to 0$, the conditional state satisfies
\begin{align} \label{eq:final_homodyne_SME_eq}
    \dd \rhoc(t) =  \Lin_{\pmb{y}_t} \, \rhoc(t) \dt + \kappa \D[\LinOp] \rhoc(t) \dt + \sqrt{\kappa} \H[\LinOp \ee^{\ii \phi}] \rhoc(t) \dW,
\end{align}
with an associated measurement, obtained from the discrete homodyne increments in \eqnref{eq:homodyne_eq_derived}, as:
\begin{align} \label{eq:final_dy}
    \dd y = \sqrt{\kappa} \braketavg{\LinOp \ee^{\ii \phi} + \LinOp^\dagger \ee^{-\ii \phi}} \dt +\dW.
\end{align}

\subsection{Polarimetric measurement} \label{sec:polarimetric_meas}

Consider a quasimonochromatic probe beam propagating in the $y$-direction, such that a general quantized multi-mode electric field \cite{Kupriyanov2005} can be approximated as a single traveling-wave spacial mode with two orthogonal polarizations $\{\anihil_H,\anihil_V\}$ \cite{deutsch_quantum_2010,Geremia2006}. Then, the positive frequency component for the monochromatic quantized electric field is \cite{deutsch_quantum_2010,Geremia2006}:
\begin{equation}
    \ElectricP = \sqrt{\frac{2\pi \hbar \omega}{Ac\Dt}} \left(\pmb{e}_H \, \anihil_H + \pmb{e}_V \, \anihil_V \right),
\end{equation}
and $\ElectricM = \,\hat{\!\pmb{E}}^{\,(+)\dagger}$, $A$ is the area of the beam and $c$ the speed of light. Thus, to describe the beam it is sufficient with the photon annihilation operators $\{\anihil_H,\anihil_V\}$, associated with horizontal and vertical polarizations, respectively. Similarly as before, for each polarization component, the beam has been divided into a train of modes of duration $\Dt$, each segment interacting for a time $\Dt$ with an atomic cloud of length $\Delta x$ such that $c\Dt \gg \Delta x$ \cite{Geremia2006}. 

\begin{figure}[t]  
\centering 
\begin{tikzpicture}[scale=1, line cap=round, line join=round, >=stealth] 
    \draw[fill=none,thick](0,0) circle (1.5) node [black,yshift=-2.5cm] {};
    \draw[fill=black](0,0) circle (2 pt) node [above] {};
    \draw[thick,->](0,0) -- (2,0) node [right] {$H$};  
    \draw[thick,->](0,0) -- (0,2) node [right] {$V$};  
    \draw[thick,->](0,0) -- (1.41,1.41) node [right] {$H^{\,\prime}$};  
    \draw[thick,->](0,0) -- (-1.41,1.41) node [left] {$V^{\,\prime}$};  
    \draw[thick] (1.5/2,0,0) coordinate (start) arc[start angle=0, end angle=45, x radius=1.5/2, y radius = 1.5/2];
    \node at (1,0.45) {$45^\circ$};
\end{tikzpicture}\hspace{1cm}
\begin{tikzpicture}[3d/install view={phi=110,theta=70},line cap=butt, line join=round,declare function={R=2.5;},c/.style={circle,fill,inner sep=1pt}]
    \path
    (0,0,R)  coordinate (+)
    (0,0,-R)  coordinate (-)
    (R,0,0)  coordinate (H)
    (-R,0,0)  coordinate (V)
    (0,R,0)  coordinate (H')
    (0,-R,0)  coordinate (V')

    ({R*cos(60)}, {R*sin(60)},0)  coordinate (A);
    \node[c] at (+) {};
    \node[below right] at (+) {$\pmb{+}$};
    \node[c] at (-) {};
    \node[below right] at (-) {$\pmb{-}$};
    \node[c] at (H) {};
    \node[below right] at (H) {$H$};
    \node[c] at (V) {};
    \node[below right] at (V) {$V$};
    \node[c] at (H') {};
    \node[below right] at (H') {$H^{\,\prime}$};
    \node[c] at (V') {};
    \node[left] at (V') {$V^{\,\prime}$};
    \draw[3d/screen coords,fill=gray,opacity=0.3] (0,0,0) circle[radius=R]; 
    \path pic{3d/circle on sphere={R=R,C={(0,0,0)}}};
    \path  pic{3d/circle on sphere={R=R,C={(0,0,0)},P={(0,0,0)}, n={(0,1,0)}}}; 

    \draw[3d/hidden] (-) -- (+)  (0,0,0)--(H') (0,0,0)--(H);
    \draw[ultra thick, orange, - latex,dashed] (0,0,0) -- (A);
    \draw[ultra thick, orange, - latex,dashed] (0,0,0) -- (H);

    \draw[thick] (R/2,0,0) coordinate (start) arc[start angle=0, end angle=60, x radius=R/2, y radius = R/2];
    \node at (1.55,1) {$\Theta$};
    \draw[3d/visible, -latex] (R,0,0) -- (R + 4,0,0) node[right]{$\Stokex$};
    \draw[3d/visible, -latex] (0,R,0) -- (0,R + 1,0) node[right]{$\Stokey$};
    \draw[3d/visible, -latex] (0,0,R) -- (0,0,R + 1) node[above]{$\Stokez$};
\end{tikzpicture}
\caption[Stokes vectors on the Pointcar\'{e} sphere]{\textbf{Stokes vectors on the Pointcar\'{e} sphere.} (left) Definition of the $\pm 45^\circ$ polarization basis $\{H^\prime,V^\prime\}$ w.r.t. the linear polarization $\{H,V\}$. (right) Pointcar\'{e} sphere, a graphical representation describing the polarization of light. Points on the sphere correspond to different polarization states, with the equator representing linear polarization, and the poles representing circular polarization. Note that the rotation of linear polarization is confined to the equatorial plane of the Poincaré sphere. A rotation of the polarization by an angle $\Theta/2$ corresponds to a rotation of $\Theta$ in the Poincaré sphere. This can be observed from the state $H^\prime$, which can be generated from $H$ by rotating the polarization by $45$ degrees; as a result, in the Poincaré sphere, $H^\prime$ lies at a 90-degree angle from $H$. } \label{fig:Stokes}  
\end{figure}
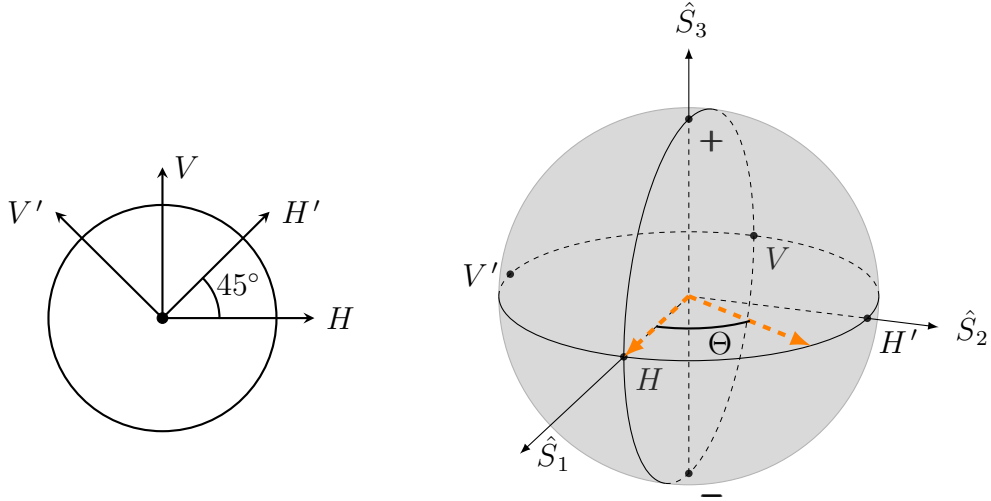

After the segmented electric field at time $t_k = k \Dt$ has interacted with the cloud for a time $\Dt$, we perform a measurement of the transmitted light using a polarimeter: a polarization beam splitter and a differential photo detector \cite{Smith2003,Smith2004} made up of two photodiodes counting the number of horizontal polarized photons $\braketavg{\NOp_H}$ and the number of vertical ones $\braketavg{\NOp_V}$, respectively \cite{Kuzmich1999}. The difference between the number of horizontal photons and the vertical ones is the expected value of the first Stokes vector. The Stokes operators are defined using the polarization modes $\{\anihil_H,\anihil_V\}$, and the position of their vector $\pmb{S} = \left(\Stokex,\Stokey,\Stokez\right)$ on the Poincar\'{e} sphere represents the polarization state of the field. These operators can be introduced from the classical Stokes parameters \cite{Born1999optics} by changing intensities to photon-number operators:
\begin{align}
    & \Stokex = \frac{1}{2} \left(\NOp_H - \NOp_V \right)= \frac{1}{2}\left(\creat_H \, \anihil_H - \creat_V \, \anihil_V\ \right),   \\
    & \Stokey = \frac{1}{2} \left(\NOp_{H^{\,\prime}} - \NOp_{V^{\,\prime}} \right) = \frac{1}{2}\left(\creat_{H^{\,\prime}} \, \anihil_{H^{\,\prime}} - \creat_{V^{\,\prime}} \, \anihil_{V^{\,\prime}}\ \right) = \frac{1}{2}\left(\creat_H \, \anihil_V + \creat_V \, \anihil_H\ \right), \\
    & \Stokez = \frac{1}{2} \left(\NOp_+ - \NOp_- \right)= \frac{1}{2}\left(\creat_+ \, \anihil_+ - \creat_- \, \anihil_-\ \right) = \frac{1}{2i}\left(\creat_H \, \anihil_V - \creat_V \, \anihil_H\ \right) ,
\end{align}
which satisfy the $SU(2)$ algebra $[\hat{S\;}\!_i,\hat{S\;}\!_j] = i\epsilon_{ijk}  \,\hat{S\;}\!_k$. We define the total photon number operator as $\StokeO = \creat_H \, \anihil_H + \creat_V \, \anihil_V$. The Stokes operator $\Stokex$ represents linearly polarized light in either the horizontal or vertical direction, quantified through the annihilation and creation operators $\{\anihil_H,\anihil_V\}$ and  $\{\creat_H,\creat_V\}$. Thus, note how the expected value of $\Stokex$ yields the difference between the number of horizontal v.s. vertical photons. The operator $\Stokey$ measures the difference in population of linearly polarized modes rotated in the $\pm45$ degree direction, and $\Stokez$, of left and right circularly polarized light:
\begin{align}
    \anihil_{H^{\,\prime}} &=  \frac{1}{\sqrt{2}} \left(\anihil_H + \anihil_V\right), \quad
    & \anihil_{V^{\,\prime}} &= \frac{1}{\sqrt{2}} \left(-\anihil_H + \anihil_V \right), \\
    \anihil_{+} &=  \frac{1}{\sqrt{2}} \left(\anihil_H - i \anihil_V\right), \quad
    & \anihil_{-} &= \frac{1}{\sqrt{2}} \left(-\anihil_H - i\anihil_V \right).
\end{align}
Two polarizations perpendicular in real space will be represented by two vectors pointing in the opposite directions in the Poincar\'{e} sphere, and for example, measuring $\ket{H}$ or $\ket{V}$ with $\Stokex$ will yield $+1$ and $-1$, respectively (see \figref{fig:Stokes}).

In general, the Hamiltonian governing the interaction of the atomic cloud with a monochromatic probe detuned off-resonance is given by:
\begin{equation} \label{eq:Tensor_Pol_Ham}
    \Ham_{\!I} = \sum_{g,e} \ElectricM \cdot \frac{\pmb{\alpha}_{g,e}}{\hbar \Delta_{g,e}} \cdot \ElectricP, 
\end{equation}
where $\Delta_{g,e}$ is the detuning of the probe from the $g\rightarrow e$ transition (e.g. for a cloud of Rubidium $\ce{^{87}Rb}$, that would be the transition of the $D_2$ line \cite{de_Echaniz_2005}). The term $\pmb{\alpha}_{g,e}$ is the \emph{atomic polarizability} tensor of that same transition and has the form:
\begin{equation} \label{eq:atom_pol_tens}
    \pmb{\alpha}_{g,e} = \Proj_g \, \pmb{d} \Proj_e \,\pmb{d}^\dagger \Proj_g,
\end{equation}
where $\pmb{d}$ corresponds to the vector of dipole operators, and $\Proj_g$ and $\Proj_e$ are the projectors for the ground and excited states, respectively \cite{deutsch_quantum_2010}. Ultimately, the Hamiltonian in \eqnref{eq:Tensor_Pol_Ham} describes the interaction and eventual transmission of the light through the sample: a photon is annihilated from the probe field through $\ElectricP$, which brings the atom from its ground state to its excited state via the dipole raising operator $\pmb{d}^\dagger$. Then the excited atom returns to a ground state via $\pmb{d}$ by emitting a photon into a transmitted probe mode through $\ElectricM$ \cite{Geremia2006}. 

The atomic polarizability tensor of \eqnref{eq:atom_pol_tens} can be decomposed under the group of rotations into three irreducible components: $\pmb{\alpha}_{g,e} = \pmb{\alpha}_{g,e}^{(0)}+\pmb{\alpha}_{g,e}^{(1)}+\pmb{\alpha}_{g,e}^{(2)}$, and thus, so can the Hamiltonian: 
\begin{align}
    \Ham_{\!I} = \Ham_{\!I}^{(0)} + \Ham_{\!I}^{(1)} + \Ham_{\!I}^{(2)},
\end{align}
where each term corresponds to a scalar, vectorial and tensorial component, respectively \cite{Geremia2006,de_Echaniz_2005,de_echaniz_hamiltonian_2008,deutsch_quantum_2010}. The scalar or rank-0 component is invariant under the group of SO(3) rotations, and since it is state independent, it can be dropped. The vectorial term transforms like a rank-1 vector, and hence, it can be written as a linear combination of components of some vector like the spin $\,\hat{\!\pmb{S}}$. Finally, the tensorial term, a rank-2 tensor, can also be neglected by further increasing the probe detuning \cite{de_Echaniz_2005,de_echaniz_hamiltonian_2008}. Therefore, the Hamiltonian reduces to:
\begin{equation}
    \Ham_{\!I} =  \Ham_{\!I}^{(1)}  = \chi \, \Jy \Stokez,
\end{equation}
where the light, which propagates along the $y$-direction, couples to the $y$ component of the angular momentum of the atoms through the linear Stokes component $\Stokez$, since a rotation of the linear polarization by an angle $\Theta/2$ corresponds to a rotation on the Pointcar\'{e} sphere of an angle $\Theta$ about $\Stokez$ (see \figref{fig:Stokes}). 

Thus, thanks to Hamiltonian engineering \cite{de_echaniz_hamiltonian_2008,Colangelo2013}, the continuous non-demolition measurement required to monitor a system is implemented through a rotation of the polarization of the off-resonant probe beam proportional to the collective angular momentum of the atoms at an angle $\Theta \sim \chi \braketavg{\Jy}(t)$ \cite{Takahashi1999,deutsch_quantum_2010,Kong2020}, a magneto-optical effect known as Faraday rotation \cite{Budker2002RMP,deutsch_quantum_2010}.

Similarly to the approximation of the generalized Bloch sphere as a phase-plane in the linear-Gaussian (or Holstein-Primakoff) approximation, the Poincaré sphere can also be approximated as a plane under the conditions of a large photon number, an initial polarization aligned with $\Stokex$, and small-angle Faraday rotation during transmission through the atomic cloud \cite{deutsch_quantum_2010}. Then, the Stokes vectors can be approximated as:
\begin{equation}
    \Stokex \sim \sqrt{N_{ph}/2}, \;\;\;\;\; \Xph = \Stokey / \sqrt{\frac{N_{ph}}{2}}, \;\;\;\;\; \Pph = \Stokez / \sqrt{\frac{N_{ph}}{2}},
\end{equation}
where $N_{ph}$ is the number of photons in the probe. Now, the interaction Hamiltonian becomes,
\begin{equation}
    \Ham_{\!I} = \chi \sqrt{N_{ph}/2} \, \Jy \, \Pph,
\end{equation}
and a rotation of the polarization along the axis $\Stokez$ becomes a displacement along $\Xph$. Hence, measurement of the Faraday rotation of the probe is fully characterized by the eigenvectors of $\Xph$, i.e. $\ket{\Xph}$. 

Therefore, if now we discretize the system-probe evolution just like in \secref{sec:syst-bath}, the interaction Hamiltonian for a segment $\Dt$ is:
\begin{equation}
    \Ham_{\!I}^{(n)} \coloneqq i \chi \sqrt{\frac{N_{ph}}{4 \Dt}} \; \Jy \left(\bcreat_n - \banihil_n\right), 
\end{equation}
where we have written the momentum operator $\Pph$ in terms of creation and annihilation operators (see \eqnref{eq:pos&mom_a&adagger}) and then discretized the modes according to \eqnref{eq:time_discretized_modes}. Hence, the unitary evolution operator taking the state of the system from $t_n$ to $t_n+\Dt$ reads as 
\begin{equation}
    \UnitOp_{\!\!\Dt}^{(n)} = \e^{-i \Ham_{\!I}^{(n)} \Dt} = \exp{ \left(  \sqrt{\kappa \Dt} \, \Jy \, \left( \bcreat_n - \banihil_n \right)  \right)},
\end{equation}
where $\kappa = \chi^2 N_{ph}/4$. Since angular momentum operators are Hermitian, the unitary form above matches \eqnref{eq:unitary_sqrtDeltat} for $\LinOp \coloneqq \Jy$. Furthermore, since the measurement of the Faraday rotation of the polarization in the Holstein-Primakoff approximation corresponds to a measurement of the displacement along $\Xph$, which is equivalent to \eqnref{eq:def_quadrature_x} but with $\phi = 0$, then this type of polarimetric measurement is equivalent to a homodyne measurement. And thus, we can write the conditional state evolution as:
\begin{align}
    \dd \rhoc(t) =  \Lin_{\pmb{y}_t} \rhoc(t) \dt + \kappa \D[\Jy] \rhoc(t) \dt + \sqrt{\kappa} \H[\Jy] \rhoc(t) \dW,
\end{align}
with a continuous measurement output:
\begin{align} 
    \dd y = 2\sqrt{\kappa} \braketavg{\Jy} \dt +\dW.
\end{align}

\section{Measurement-based feedback} \label{sec:mark_meas-based_feedback}

There are two types of measurement-based feedback one might consider. The first is feedback that utilizes the entire history of measurement outcomes, often referred to as \emph{state-based} or \emph{Bayesian} feedback~\cite{Albarelli2024}. The second is \emph{Markovian} feedback~\cite{Wiseman1994_feedback}, where we feed back in an instantaneous measurement signal, e.g. a photocurrent, modeled at each time step as a stochastic increment.

\subsection{Bayesian feedback} \label{sec:bayesian_feedback}

In the previous sections, we derived the SME for homodyne and photodetection measurements, accounting for the possibility of measurement-based feedback. In particular, we considered a specific class of feedback in which the entire history of measurement outcomes $\pmb{y}_{\leq t}$ enters the dynamics only through terms proportional to $\dt$, such that $\Lin_{\pmb{y}_{\leq t}} \dt = \bigO(\dt)$, and not through stochastic contributions like $\dW$ or the photocurrent $\dd y$. We refer to this subclass of measurement-based feedback as Bayesian feedback following the terminology of \refcite{Albarelli2024}.

Because the GKSL generator appears at order $\dt$, the corresponding feedback-modified evolution map can be expanded as shown in \eqnref{eq:Phi_map_expansion}. This leads directly to the final form of the SME for homodyne detection with Bayesian feedback, given in \eqnref{eq:final_homodyne_SME_eq}. If now we consider the case where the GKSL generator takes the form
\begin{equation}
\Lin_{\pmb{y}_t} \, \rhoc(t) = -\ii \left[\Ham + u(t|\pmb{y}_t) \Fop, \rhoc(t)\right],
\end{equation}
where the control function $u(t|\pmb{y}_t)$ depends on the measurement history $\pmb{y}_t$ but not on its derivative, then the SME becomes
\begin{equation}
    \dd \rhoc(t) =  -\ii \left[\Ham + u(t|\pmb{y}_t) \Fop ,\rhoc(t)\right] \dt + \kappa \D[\LinOp] \rhoc(t) \dt + \sqrt{\kappa} \H[\LinOp]\rhoc(t) \dW. 
\end{equation}

\subsection{Markovian feedback}

We now contrast Bayesian feedback with another common class of measurement-based control: Markovian feedback. In Bayesian control, the feedback appears only at order $\dt$, as part of a Lindbladian contribution $\Lin_{\pmb{y}_{\leq t}} \dt = \bigO(\dt)$. In contrast, Markovian feedback modifies the SME through stochastic terms, with the feedback entering directly through the homodyne signal $\dd y$. As a result, the evolution now includes terms of order $\sqrt{\dt}$ and the feedback cannot be described as a GKSL map at order $\dt$. In other words, to include Markovian feedback, the map in \eqnref{eq:Phi_map_expansion} has to be modified as:
\begin{equation}
    \Phi_n  = \exp{ \{\Lin_0 \Dt + \Lin_{\mrm{Y}} \Delta y_n \}},
\end{equation}
where now the dependence on the measurement outcomes is limited to the term $\Lin_{\mrm{Y}} \Delta y_n$, with $\Delta y_n$ being the discretized photocurrent $\dd y$ at time $n = t/\Dt$ and $\Lin_{\mrm{Y}}$ its associated Lindbladian. The other term, $\Lin_0 \Delta t$, contains the internal dynamics of the atoms, e.g. Larmor precession and dephasing. Expanding the map above yields:
\begin{align}
    \Phi_n  &= \I + \Lin_0 \Dt + \Lin_{\mrm{Y}} \Delta y_n + \frac{1}{2} \left(\Lin_{\mrm{Y}}\right)^2 \left(\Delta y_n\right)^2 + \littleo(\Dt) \nonumber \\
    &= \left(\I + \Lin_0 \Dt + \littleo(\Dt) \right)\left(\I + \Delta y_n \Lin_{\mrm{Y}} + \frac{1}{2} \left(\Delta y_n\right)^2 \left(\Lin_{\mrm{Y}}\right)^2 + \littleo(\Dt) \right) \nonumber \\
    &= \Xi_0 \circ \FF_n,
\end{align}
where the last two expressions correspond to the lowest-order Suzuki–Trotter decomposition into deterministic and stochastic components, with
\begin{equation}
    \Xi_0 \coloneqq \exp{\{ \Lin_0 \Dt \}} \quad \text{and} \quad \FF_n \coloneqq \exp{ \{ \Lin_{\mrm{Y}} \Delta y_n\}}. 
\end{equation}

Thus, similarly to \eqnref{eq:iterative_rule_rhoc}, the conditional state evolves from time $t_{n\shortminus1} = t - \dt$ to $t_n = t$ according to:
\begin{align}
    \rhoc(t_n) = \frac{\FF_n\!\left[  \Xi_0\!\left[ \measE_{y_n} \,\rhoc(t_{n\shortminus1}) \measE^\dagger_{y_n} \right]\right] }{\trace{\! \FF_n\!\left[  \Xi_0\!\left[\measE_{y_n} \,\rhoc(t_{n\shortminus1}) \measE^\dagger_{y_n}\right]\right]}},
\end{align}
where $\measE_{y_n}$ is a general operator representing the weak measurement. In the case of a photocount measurement, the operators read $\measE_{y_n} = \bra{y_n}\UnitOp_{\!\!\Dt}\ket{0}$, where $y_n=0,1$. For a homodyne measurement, the measurement operators are $\measE_{y_n} = \bra{x_n^\phi}\UnitOp_{\!\!\Dt}\ket{0}$, where $\ket{x_n^\phi}$ is the eigenstate of \eqnref{eq:def_quadrature_x}. As shown in \secref{sec:homodyne_meas}, the eigenvalue $x_n^\phi$ is a Gaussian random variable, which defines the homodyne signal $\Delta y_n \propto x_n^\phi$.

In the formulation introduced by Wiseman~\cite{Wiseman1994_feedback,Wiseman1994_squeezing_feedback}, Markovian feedback is realized via a unitary Hamiltonian evolution driven by the instantaneous measurement record:
\begin{equation} \label{eq:Feedback_Ham}
    \FeedHam = \Fop I(t) = \Fop \dd y,
\end{equation}
where $\Fop$ is a fixed Hermitian feedback operator, the homodyne current is defined as  $I(t) = \dd y / \dt$, and $\dd y$ is given by \eqnref{eq:final_dy}. Therefore, the feedback map $\FF_n$ is actually an unitary evolution that reads as:
\begin{equation}
    \FF_n \coloneqq \UnitOp_{\!f} \; \cdot \; \UnitOp_{\!f}^\dagger = \exp{\left\{-\ii \Fop \Delta y_n \right\}} \; \cdot \;  \exp{\left\{\ii \Fop \Delta y_n \right\}},
\end{equation}
where $\Delta y_n$ follows \eqnref{eq:homodyne_eq_derived}. Furthermore, this map can be expanded to first order in $\Dt$ as
\begin{align}
    \FF_n =  \UnitOp_{\!f} \; \cdot \; \UnitOp_{\!f}^\dagger  = \I - \ii \, \Delta y_n \left[ \Fop, \; \cdot \; \right] + \D[\Fop] \, \cdot \, \Dt + \littleo(\dt).
\end{align}
Therefore, to derive a SME which accounts for Markovian feedback, we simply have to apply the feedback unitary to the continuous measurement map, where we have set $\phi = 0$:
\begin{align}
    \rhoc(t_n) &= \UnitOp_{\!f}\, \frac{\Xi_0\!\left[\measE_{y_n} \,\rhoc(t_{n\shortminus1}) \measE^\dagger_{y_n} \right] }{\trace{ \Xi_0\!\left[\measE_{y_n} \,\rhoc(t_{n\shortminus1}) \measE^\dagger_{y_n}\right]}}  \UnitOp_{\!f}^\dagger  \nonumber \\
    &=  \UnitOp_{\!f} \left[\rhoc(t_{n\shortminus1}) + \Lin_{0}\rhoc(t_{n\shortminus1}) \Dt + \kappa \D[\LinOp] \rhoc(t_{n\shortminus1}) \Dt + \sqrt{\kappa} \H[\LinOp] \rhoc(t_{n\shortminus1}) \Delta\!\W \right] \UnitOp_{\!f}^\dagger \nonumber \\
    &= \rhoc(t_{n\shortminus1}) + \Lin_{0}\rhoc(t_{n\shortminus1}) \Dt  + \kappa \D[\LinOp] \rhoc(t_{n\shortminus1}) \Dt+ \D[\Fop] \rhoc(t_{n\shortminus1}) \Dt   - \ii \, \Delta\!\W [ \Fop, \rhoc(t_{n\shortminus1})]  \nonumber \\
    &-\ii \sqrt{\kappa} \left[\!\Fop, \LinOp \rhoc(t_{n\shortminus1}) + \rhoc(t_{n\shortminus1}) \LinOp^\dagger \! \right]\!\Dt \!+\! \sqrt{\kappa} \H[\LinOp] \rhoc(t_{n\shortminus1}) \Delta\!\W + \littleo(\Dt). 
\end{align}
The expression above can be reformulated using the following identities:
\begin{align}
    &\kappa \D[\LinOp] \rhoc(t) + \D[\Fop] \rhoc(t) -\ii \sqrt{\kappa}\left[\!\Fop, \LinOp \rhoc(t) + \rhoc(t) \LinOp^\dagger \! \right]  = \nonumber \\
    &\quad = -\ii \sqrt{\kappa} \left[\frac{\LinOp^\dagger \Fop + \Fop \LinOp}{2},\rhoc(t)\right]+ \D[\sqrt{\kappa}\LinOp - \ii \Fop] \rhoc(t), \\
    &\sqrt{\kappa} \H[\LinOp]\rhoc(t) - \ii \left[\Fop,\rhoc(t)\right] = \H[\sqrt{\kappa} \LinOp - \ii \Fop]  \rhoc(t) .
\end{align}
Namely,
\begin{align}
    \rhoc(t_n) &= \rhoc(t_{n\shortminus1}) + \Lin_{\pmb{y}_{0:n}}\rhoc(t_{n\shortminus1}) \Dt  -\ii \sqrt{\kappa} \left[\frac{\LinOp^\dagger \Fop + \Fop \LinOp}{2},\rhoc(t_{n\shortminus1})\right] \Dt   \nonumber \\
    &+ \D[\sqrt{\kappa}\LinOp - \ii \Fop] \rhoc(t_{n\shortminus1}) \Dt + \H[\sqrt{\kappa} \LinOp - \ii \Fop] \rhoc(t_{n\shortminus1}) \Delta\!\W + \littleo(\dt). \label{eq:markovian_sme_derivation_1}
\end{align}
Finally, if the internal evolution represented by the Lindbladian $\Lin_{0}$ is of the form:
\begin{equation}
    \Lin_{0}\rhoc(t) = -\ii\left[\Ham,\rhoc(t)\right],
\end{equation}
then, \eqnref{eq:markovian_sme_derivation_1} in the limit of $\Dt \to 0$, can be written as
\begin{equation}
    \dd \rhoc(t) = -\ii \left[\Ham_{\text{\!eff}},\rhoc(t)\right] \dt + \D[\LinOp_{\text{eff}}] \rhoc(t) + \H[\LinOp_{\text{eff}}] \rhoc(t) \dW,
\end{equation}
where
\begin{align}
    \Ham_{\text{\!eff}} &= \Ham + \frac{\sqrt{\kappa}}{2} \left( \LinOp^\dagger \Fop + \Fop \LinOp \right),\\
    \LinOp_{\text{eff}} &= \sqrt{\kappa} \LinOp - \ii \Fop.
\end{align}

\chapter{Ultimate precision limits in noisy systems} \label{chap:bounds}

\newthought{In quantum metrology}, finding the fundamental limits to precision is essential for verifying optimal sensing strategies. Thus, this chapter focuses on the derivation and analysis of a lower bound on the BCRB introduced in \secref{sec:BCRB_section} for frequency estimation in the presence of atomic dephasing and field fluctuations\cite{Van-Trees,Fritsche2014}. This bound is referred throughout this thesis as either the classically-simulated (CS) limit or the quantum limit. The term \emph{classically-simulated} stems from its derivation via the decomposition of quantum channels as a convex mixture of unitaries \cite{matsumoto_metric_2010,Demkowicz2012}, while the term \emph{quantum limit} highlights that, given a particular form and strength of noise, no strategy involving any possible quantum effects may surpass it. In that respect, our bound is conceptually related to recent universal-metrology bounds developed for general correlated-noise models \cite{Kurdzialek2025}, which likewise place absolute limits on precision in noisy adaptive protocols \cite{Kurdzialek2023}.

Importantly, as will become clear through the derivation, this bound is entirely independent of the choice of initial quantum state, measurement, or measurement-based feedback. It depends only on the noise model of the system (in our case, local and collective dephasing along the field direction) and the fluctuating strength of the signal we aim to track. As such, attaining this limit would certify that the entire sensing protocol is optimal, i.e. that our particular choice of initial state, measurement, estimator and measurement-based control yields the best possible sensitivity. 

\newthought{This chapter}, which one can view as an extended derivation or proof, is organized as follows: we begin in \secref{sec:discrete-time_pic} by introducing the most general discrete-time evolution of a quantum state, alternating between internal system dynamics and measurement updates, which is connected to the conditional evolution of \chapref{chap:cm}. Next, we specify the types of quantum channels used in our model: the feedback map, and the collective and local dephasing maps. 

In \secref{sec:convex_decomp}, we turn to the central step of this proof: the convex decomposition of the likelihood $p(\pmb{y}_{0:k}|\pmb{\omega}_{0:k})$, which describes the probability of observing measurement outputs $\pmb{y}_{0:k}$ given a signal trajectory $\pmb{\omega}_{0:k}$. This likelihood is rewritten as a mixture of two other conditional probabilities: one that contains the $\omega$-dependence via a classical mixing distribution, and another ``fictitious'' likelihood encoding all the measurement record, independent of the signal. To enable this decomposition, both collective and local dephasing channels must be expressed as Gaussian-weighted integrals over unitary operations (see \secref{sec:coll_decomp} and \secref{sec:local_decomp}). This allows us to reformulate $p(\pmb{y}_{0:k}|\pmb{\omega}_{0:k})$ accordingly, which in turn enables us to upper-bound the FI of the marginal likelihood $p(\pmb{y}_{0:k}|\omega_k)$ in \secref{sec:upperbound_Fisher}. 

This upper-bound on the FI can be analytically derived (see \secref{sec:fisher_ofBigp}), since it ultimately involves computing the FI of several Gaussian distributions, each given by the inverse of their variance. In the case of fluctuating fields, finding such a variance requires solving a recursive relation, which can be done explicitly for the problem at hand. Taking the continuous-time limit of the resulting expression in \secref{sec:cont_time_limit}, yields the CS limit or quantum limit. Finally, in \secref{sec:atom_number_csbound} we extend the analysis to account for scenarios where the number of atoms $N$ fluctuates between experimental runs, and show how the bounds can be appropriately modified. 

\section{Discrete-time picture of measurement-based feedback scheme} \label{sec:discrete-time_pic}

Consider the discrete evolution of a quantum state that alternates between its intrinsic dynamics and measurement updates, as shown in \eqnref{eq:rho_n_discrete}. In this picture, each measurement is modeled by a set of measurement Kraus operators $\measE_{y_k}$, which form a POVM $\{\measE_{y_k}^\dagger \measE_{y_k}\}_{k}$ whose elements fulfill $\sum_{k} \measE_{y_k}^\dagger \measE_{y_k}=\I$ and whose outcome $y_k$ forms part of the discretized measurement record $\pmb{y}_{0:k} = \{y_0,y_1,\dots,y_k\}$. The quantum channel governing the evolution of the state in between measurements is a generic map
\begin{equation}
    \Phi_k \coloneqq \Phi_\Dt(\pmb{y}_{0:k},\omega_k),
\end{equation}
which acts on the state for a time $\Dt$ and depends on a parameter $\omega_k$, as well as, potentially, on all previous measurement records $\pmb{y}_{0:k} = \{y_0,y_1,\dots,y_k\}$. The frequency $\omega_k$ is itself also a time-discretized element from a frequency trajectory $\pmb{\omega}_{0:k} = \{\omega_0,\omega_1,\dots,\omega_k\}$, whose elements will later be assumed to be drawn from a probability distribution at each time-step. Therefore, the state at time $t = k \Dt$, conditional on the measurement outcomes  $\pmb{y}_{0:k} = \{y_0,y_1,\dots,y_k\}$, reads as
\begin{align}
    \!\!&\;
    \rho[k|\pmb{y}_{0:k}] =  
    \frac{\Phi_{k}\!\!\left[\measE_{y_{k}} \Phi_{k\shortminus1}\!\!\left[ \measE_{y_{k\shortminus 1}}   \dots \Phi_{1}\!\!\left[\measE_{y_{1}}\Phi_0\!\left[\measE_{y_{0}} \, \rho_{0} \, \measE_{y_{0}}^\dagger \right]\!\measE_{y_{1}}^{\dagger}\right] \dots \measE_{y_{k\shortminus 1}}^{\dagger} \right]  \! \measE_{y_{k}}^{\dagger} \right]}{p(\pmb{y}_{0:k}|\pmb{\omega}_{0:k})},  \label{eq:cm_discr} 
\end{align}
where we have used a similar notation to \eqnref{eq:notation_rhoc}, $\rho[k|\pmb{y}_{0:k}] \coloneqq \rhoc(k \Dt) = \rho(k\Dt|\pmb{y}_{k\Dt})$, to refer to the discretized conditional state at time $t = k\Dt$. Note that $\rho_{0}$ denotes the initial state of the atoms before any operation is applied, and hence, is different from the state $\rho[0|\pmb{y}_0]$, which is obtained after getting the first outcome $y_0$ and evolving the state with the map $\Phi_0$, dependent on the measurement $y_0$ and the frequency $\omega_0$. The denominator is the discretized version of the likelihood $p(\pmb{y}_t|\pmb{\omega}_t)$, i.e. the probability of measuring $\pmb{y}_{0:k} = \{y_0,y_1,\dots,y_k\}$ given field inputs $\pmb{\omega}_{0:k} = \{\omega_0,\omega_1,\dots,\omega_k\}$. Namely, 
\begin{align}
    p(\pmb{y}_{0:k}|\pmb{\omega}_{0:k}) = \trace{\!\Phi_{k}\!\!\left[\measE_{y_{k}} \Phi_{k\shortminus 1}\!\!\left[ \measE_{y_{k\shortminus 1}}   \dots \Phi_{1}\!\!\left[\measE_{y_{1}}\Phi_0\!\left[\measE_{y_{0}}\,\rho_{0}\,\measE_{y_{0}}^\dagger\right]\!\measE_{y_{1}}^{\dagger}\right]  \!\dots \measE_{y_{k\shortminus 1}}^{\dagger} \right]  \! \measE_{y_{k}}^{\dagger}\right]},
\end{align}
which will later play a crucial role when bounding the aMSE, introduced in \defref{def:aMSE_def}.

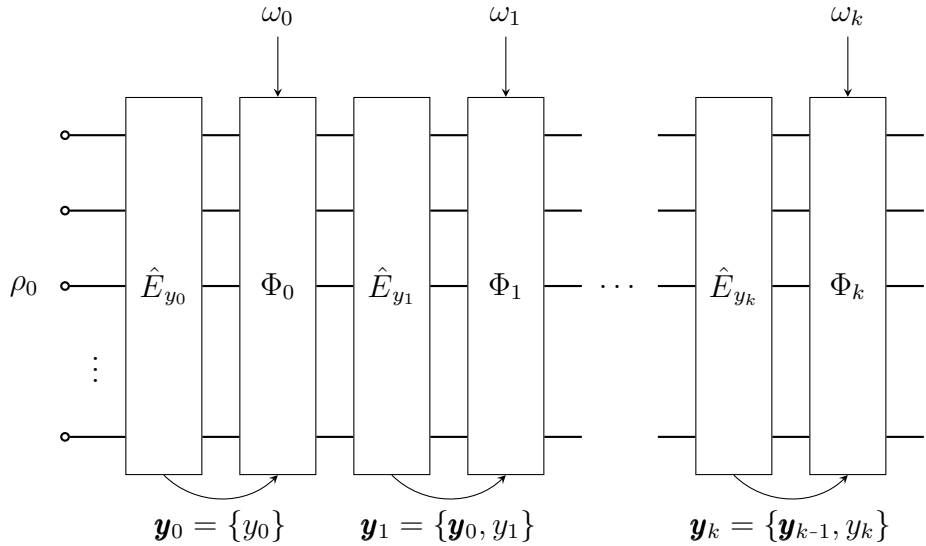
\begin{figure}[htbp]
    \begin{center}
        \begin{tikzpicture}
            \node[circle, draw, thick, minimum size=0.1cm, inner sep=0pt] (rho) at (-1.3,0) {};
            \node[left] at (-1.5,0) {$\rho_0$};
            
            \node[draw, rectangle, minimum width=1cm, minimum height=5cm] (Ey0) at (0,0) {$\measE_{y_0}$};

            \node[draw, rectangle, minimum width=1cm, minimum height=5cm] (Phi0) at (1.5,0) {$\Phi_0$};
            \draw[<-,>=stealth] (Phi0.north) -- ++(0,0.8) node[above] {$\omega_0$};

            \draw[->, bend right=45,>=stealth] (Ey0.south) to node[midway, below] {$\pmb{y}_0 = \{y_0\}$} (Phi0.south);
            
            \node[draw, rectangle, minimum width=1cm, minimum height=5cm] (Ey1) at (3,0) {$\measE_{y_1}$};
            
            \node[draw, rectangle, minimum width=1cm, minimum height=5cm] (Phi1) at (4.5,0) {$\Phi_1$};
            \draw[<-,>=stealth] (Phi1.north) -- ++(0,0.8) node[above] {$\omega_1$};

            \draw[->, bend right=45,>=stealth] (Ey1.south) to node[midway, below] {$\pmb{y}_1 = \{\pmb{y}_0,y_1\}$} (Phi1.south);
            
            \node at (6,0) {$\dots$};
            \node at (-0.9,-1) {$\vdots$};
            
            \node[draw, rectangle, minimum width=1cm, minimum height=5cm] (Eyk) at (7.5,0) {$\measE_{y_k}$};
            
            \node[draw, rectangle, minimum width=1cm, minimum height=5cm] (Phik) at (9,0) {$\Phi_k$};
            \draw[<-,>=stealth] (Phik.north) -- ++(0,0.8) node[above] {$\omega_k$};

            \draw[->, bend right=45,>=stealth] (Eyk.south) to node[midway, below] {$\pmb{y}_k = \{\pmb{y}_{k\shortminus1},y_k\}$} (Phik.south);
            
            \draw[thick] (rho) -- (Ey0.west);
            \draw[thick] (Ey0.east) -- (Phi0.west);
            \draw[thick] (Phi0.east) -- (Ey1.west);
            \draw[thick] (Ey1.east) -- (Phi1.west);
            \draw[thick] (Phi1.east) -- (5.5,0);
            \draw[thick] (6.5,0) -- (Eyk.west);
            \draw[thick] (Eyk.east) -- (Phik.west);
            \draw[thick] (Phik.east) -- (10,0);

            \node[circle, draw, thick, minimum size=0.1cm, inner sep=0pt] (rhoTop) at (-1.3,1) {};
            \draw[thick] (rhoTop) -- ($(Ey0.west) + (0,1)$);
            \draw[thick] ($(Ey0.east) + (0,1)$) -- ($(Phi0.west) + (0,1)$);
            \draw[thick] ($(Phi0.east) + (0,1)$) -- ($(Ey1.west) + (0,1)$);
            \draw[thick] ($(Ey1.east) + (0,1)$) -- ($(Phi1.west) + (0,1)$);
            \draw[thick] ($(Phi1.east) + (0,1)$) -- (5.5,1);
            \draw[thick] (6.5,1) -- ($(Eyk.west) + (0,1)$);
            \draw[thick] ($(Eyk.east) + (0,1)$) -- ($(Phik.west) + (0,1)$);
            \draw[thick] ($(Phik.east) + (0,1)$) -- (10,1);

            \node[circle, draw, thick, minimum size=0.1cm, inner sep=0pt] (rhoTop) at (-1.3,2) {};
            \draw[thick] (rhoTop) -- ($(Ey0.west) + (0,2)$);
            \draw[thick] ($(Ey0.east) + (0,2)$) -- ($(Phi0.west) + (0,2)$);
            \draw[thick] ($(Phi0.east) + (0,2)$) -- ($(Ey1.west) + (0,2)$);
            \draw[thick] ($(Ey1.east) + (0,2)$) -- ($(Phi1.west) + (0,2)$);
            \draw[thick] ($(Phi1.east) + (0,2)$) -- (5.5,2);
            \draw[thick] (6.5,2) -- ($(Eyk.west) + (0,2)$);
            \draw[thick] ($(Eyk.east) + (0,2)$) -- ($(Phik.west) + (0,2)$);
            \draw[thick] ($(Phik.east) + (0,2)$) -- (10,2);

            \node[circle, draw, thick, minimum size=0.1cm, inner sep=0pt] (rhoTop) at (-1.3,-2) {};
            \draw[thick] (rhoTop) -- ($(Ey0.west) + (0,-2)$);
            \draw[thick] ($(Ey0.east) + (0,-2)$) -- ($(Phi0.west) + (0,-2)$);
            \draw[thick] ($(Phi0.east) + (0,-2)$) -- ($(Ey1.west) + (0,-2)$);
            \draw[thick] ($(Ey1.east) + (0,-2)$) -- ($(Phi1.west) + (0,-2)$);
            \draw[thick] ($(Phi1.east) + (0,-2)$) -- (5.5,-2);
            \draw[thick] (6.5,-2) -- ($(Eyk.west) + (0,-2)$);
            \draw[thick] ($(Eyk.east) + (0,-2)$) -- ($(Phik.west) + (0,-2)$);
            \draw[thick] ($(Phik.east) + (0,-2)$) -- (10,-2);
            
        \end{tikzpicture}
    \end{center}
    \caption[Quantum circuit representation of conditional evolution via sequential measurements]{\textbf{Quantum circuit representation of conditional evolution via sequential measurements.} Scheme depicting a quantum circuit representing a sequential measurement on a system of $N$ atoms with an initial density matrix $\rho_0$. Each step involves a POVM $\measE_{y_k}$, which depends on a measurement outcome $y_k$, followed by an evolution through a quantum channel $\Phi_k$. This map depends on an input parameter $\omega_k$ and all prior measurement outcomes $\pmb{y}_{0:k}$, which are progressively collected as the system evolves.  }
\label{fig:class_sim_1}
\end{figure}

If we now focus on a single time interval $\Dt$ in the discrete evolution described above, we can relate the two time-consecutive conditional states, $\rho[k\!\shortminus\!1|\pmb{y}_{0:k\shortminus1}]$ and $\rho[k|\pmb{y}_{0:k}]$ as:
\begin{equation} \label{eq:iteration_rho}
    \rho[k|\pmb{y}_{0:k}] = \frac{\Phi_{k}\!\!\left[\measE_{y_k} \, \rho[k\!\shortminus\!1|\pmb{y}_{0:k\shortminus1}] \measE_{y_k}^\dagger \right]}{\trace{\Phi_{k}\!\!\left[\measE_{y_k} \, \rho[k\!\shortminus\!1|\pmb{y}_{0:k\shortminus1}] \measE_{y_k}^\dagger \right]}},
\end{equation}
with the first time step defined as
\begin{equation}
    \rho[0|\pmb{y}_0] = \frac{\Phi_0 \! \left[\measE_{y_0} \; \rho_{0}\, \measE_{y_0}^\dagger\right] \! \! }{ \trace{ \Phi_0 \! \left[\measE_{y_0} \; \rho_{0}\, \measE_{y_0}^\dagger\right]}}.
\end{equation}

As discussed in \secref{sec:bayesian_feedback}, we focus here on Bayesian feedback\footnote{Nonetheless, our analysis can be straightforwardly generalized to also Markovian feedback or any other form of measurement-based feedback, even with general L\'{e}vy processes \cite{Amoros-Binefa2024}.}, where $\Lin_{\pmb{y}_{0:k}}^\omega \sim \bigO(1)$:
\begin{equation}
	\Phi_k=\ee^{\Lin^\omega_{\pmb{y}_{0:k}} \Dt}.
	\label{eq:semigroup}
\end{equation}
The overall dynamical generator $\Lin^\omega_{\pmb{y}_{0:k}}$ depends on all previous outcomes $\pmb{y}_{0:k}$ through the measurement-based feedback, as well as accounting for $\omega$-encoding but also importantly, decoherence. Nonetheless, it can always be decomposed into two parts: one corresponding to the dynamics generated by the $\omega$-encoding, and the other corresponding to the feedback based on previous measurements by Trotter-Suzuki arguments as $\Dt \to 0$:
\begin{align} \label{eq:splitting_feedback_and_dyn}
	\Phi_k=\ee^{(\Lin_{\omega_k}+\Lin_{\pmb{y}_{0:k}}^f) \Dt}
 &=\ee^{\Lin_{\omega_k}\Dt}\circ\ee^{\Lin_{\pmb{y}_{0:k}}^f\Dt} + \bigO(\Dt^2) \\
 &= \Xi_{\omega_k}\circ\FF_{\pmb{y}_{0:k}} + \bigO(\Dt^2), \nonumber 
\end{align}
where $\Xi_{\omega_k}$ represents the portion of the evolution due to the $\omega$-encoding and decoherence (i.e. the intrinsic, noisy dynamics), while $\FF_{\pmb{y}_{0:k}}$ accounts for the measurement-based feedback. 

As a result, the discrete conditional evolution given in \eqnref{eq:cm_discr} can be equivalently written in a form that explicitly separates the state evolution into three parts: the measurement update, the feedback and the remaining internal dynamics:
\begin{align}
    \!\!&\;
    \rho[k|\pmb{y}_{0:k}]  
    = \frac{\Xi_{\omega_k}\!\!\left[\FF_{\pmb{y}_{0:k}}\!\!\left[\!\hat{E}_{y_{k}} \dots \Xi_{\omega_1}\!\!\left[\FF_{\pmb{y}_{0:1}}\!\!\left[\measE_{y_{1}}\,\Xi_{\omega_0}\!\!\left[\FF_{\pmb{y}_{0}}\!\!\left[\measE_{y_{0}}\,\rho_{0}\measE_{y_{0}}^{\dagger}\right]\right] \! \dots\measE_{y_{1}}^{\dagger}\right]\right] \! \dots  \measE_{y_{k}}^{\dagger} \right]\right]}{p(\pmb{y}_{0:k}|\pmb{\omega}_{0:k})},  \label{eq:cm_discr_wfeed} 
\end{align}
with its likelihood now being:
\begin{align}
    \!\!&\; p(\pmb{y}_{0:k}|\pmb{\omega}_{0:k}) =  \nonumber \\
    &=\trace{\Xi_{\omega_k}\!\!\left[\FF_{\pmb{y}_{0:k}}\!\!\left[\!\hat{E}_{y_{k}} \dots \Xi_{\omega_1}\!\!\left[\FF_{\pmb{y}_{0:1}}\!\!\left[\measE_{y_{1}}\,\Xi_{\omega_0}\!\!\left[\FF_{\pmb{y}_{0}}\!\!\left[\measE_{y_{0}}\,\rho_{0}\measE_{y_{0}}^{\dagger}\right]\right] \! \dots\measE_{y_{1}}^{\dagger}\right]\right] \! \dots  \measE_{y_{k}}^{\dagger} \right]\right]}\!.
\end{align}

Hence, the proof that follows, which is fully based on the form of the map $\Xi_{\omega_k}$ responsible for noisy $\omega$-encoding, applies to any form of measurement-based feedback.

\section{Precision bound for any protocol with local and global noise}

We may further assume that the internal dynamics of the system consists of an $\omega$-encoded unitary evolution, collective dephasing and local dephasing. The map representing this internal evolution, $\Xi_\omega$, can therefore be decomposed into two additional maps,
\begin{equation} \label{eq:splitting_of_dynamics}
	\Xi_\omega=\Omega\circ\Lambda_\omega,
\end{equation}
where $\Omega$ denotes the non-unitary evolution arising in between measurements due to the collective decoherence (of strength $\kcoll$), and the channel $\Lambda_\omega$ accounts for both the unitary frequency-encoding and the non-unitary local decoherence (of strength $\kloc$). Even for non-commuting maps, we can apply the Suzuki-Trotter expansion to first order in $\Dt$ and split up the map $\Omega$ and $\Lambda_\omega$ as required. Then, \eqnref{eq:cm_discr_wfeed} becomes
\begin{align}
    \!\!&\;
    \rho[k|\pmb{y}_{0:k}] = \nonumber \\
    &=\! \frac{\Omega_k\!\!\left[\!\Lambda_{\omega_k} \!\!\left[ \!\FF_{\pmb{y}_{0:k}}\!\!\left[\!\measE_{y_{k}}   \dots \Omega_1\!\!\left[\!\Lambda_{\omega_1} \!\!\left[ \!\FF_{\pmb{y}_{0:1}}\!\!\left[\!\measE_{y_{1}}\Omega_0\!\!\left[\!\Lambda_{\omega_0} \!\!\left[ \!\FF_{\pmb{y}_{0}}\!\!\left[\!\measE_{y_{0}}\,\rho_{0}\measE_{y_{0}}^{\dagger}\!\right]\!\right]\!\right]\measE_{y_{1}}^{\dagger}\!\right]\!\right]\!\right] \! \dots  \measE_{y_{k}}^{\dagger} \!\right]\!\right]\!\right]}{p(\pmb{y}_{0:k}|\pmb{\omega}_{0:k})}, 
\end{align}
such that 
\begin{align}
    \!\!&\; 
    p(\pmb{y}_{0:k}|\pmb{\omega}_{0:k}) \!=\! \nonumber \\
    & = \trace{\!\Omega_k\!\!\left[\!\Lambda_{\omega_k} \!\!\left[ \!\FF_{\pmb{y}_{0:k}}\!\!\left[\!\measE_{y_{k}} \!  \dots \Omega_1\!\!\left[\!\Lambda_{\omega_1} \!\!\left[ \!\FF_{\pmb{y}_{0:1}}\!\!\left[\!\measE_{y_{1}}\Omega_0\!\!\left[\!\Lambda_{\omega_0} \!\!\left[ \!\FF_{\pmb{y}_{0}}\!\!\left[\!\measE_{y_{0}}\,\rho_{0}\measE_{y_{0}}^{\dagger}\!\right]\!\right]\!\right]\measE_{y_{1}}^{\dagger}\!\right]\!\right]\!\right] \!\! \dots  \measE_{y_{k}}^{\dagger} \!\right]\!\right]\!\right]\!} \label{eq:discretized_likelihood}.
\end{align}
 
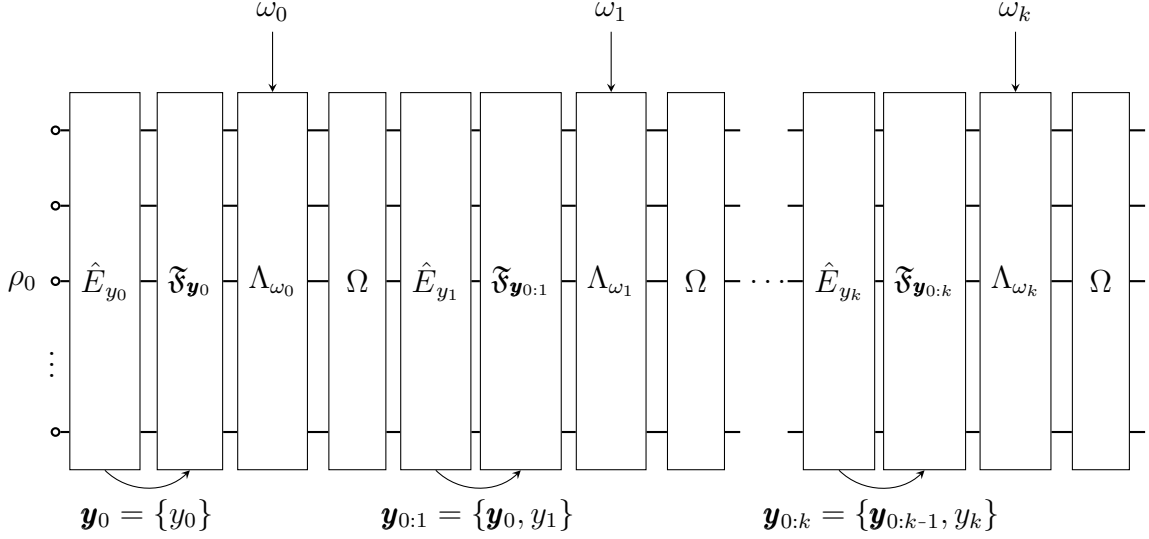
\begin{figure}[htbp]
    \begin{center}
        \begin{tikzpicture}
            \node[circle, draw, thick, minimum size=0.1cm, inner sep=0pt] (rho) at (-0.65,0) {};
            \node[left] at (-0.75,0) {$\!\!\rho_0$};
            
            \node[draw, rectangle, minimum width=0.75cm, minimum height=5cm] (Ey0) at (0,0) {$\measE_{y_0}$};

            \node[draw, rectangle, minimum width=0.75cm, minimum height=5cm] (Fy0) at (Ey0.east) [xshift=0.65cm] {$\FF_{\pmb{y}_0}$};

            \draw[->, bend right=45,>=stealth] (Ey0.south) to node[midway, below] {$\pmb{y}_0 = \{y_0\}$} (Fy0.south);            

            \node[draw, rectangle, minimum width=0.75cm, minimum height=5cm] (Lambda0) at (Fy0.east) [xshift=0.65cm] {$\Lambda_{\omega_0}$};
            \draw[<-,>=stealth] (Lambda0.north) -- ++(0,0.8) node[above] {$\omega_0$};

            \node[draw, rectangle, minimum width=0.75cm, minimum height=5cm] (Omega0) at (Lambda0.east) [xshift=0.65cm] {$\Omega$};

            \node[draw, rectangle, minimum width=0.75cm, minimum height=5cm] (Ey1) at (Omega0.east) [xshift=0.65cm] {$\measE_{y_1}$};

            \node[draw, rectangle, minimum width=0.75cm, minimum height=5cm] (Fy1) at (Ey1.east) [xshift=0.65cm] {$\FF_{\pmb{y}_{0:1}}$};

            \draw[->, bend right=45,>=stealth] (Ey1.south) to node[midway, below] {$\pmb{y}_{0:1} = \{\pmb{y}_0,y_1\}$} (Fy1.south);

            \node[draw, rectangle, minimum width=0.75cm, minimum height=5cm] (Lambda1) at (Fy1.east) [xshift=0.65cm] {$\Lambda_{\omega_1}$};
            \draw[<-,>=stealth] (Lambda1.north) -- ++(0,0.8) node[above] {$\omega_1$};

            \node[draw, rectangle, minimum width=0.75cm, minimum height=5cm] (Omega1) at (Lambda1.east) [xshift=0.65cm] {$\Omega$};
            
            \node at (Omega1.east) [xshift=0.6cm] {$\dots$};
            \node at (-0.7,-1) {$\vdots$};
            
            \node[draw, rectangle, minimum width=0.75cm, minimum height=5cm] (Eyk) at (Omega1.east) [xshift=1.5cm] {$\measE_{y_k}$};

            \node[draw, rectangle, minimum width=0.75cm, minimum height=5cm] (Fyk) at (Eyk.east) [xshift=0.65cm] {$\FF_{\pmb{y}_{0:k}}$};

            \draw[->, bend right=45,>=stealth] (Eyk.south) to node[midway, below] {$\pmb{y}_{0:k} = \{\pmb{y}_{0:k\shortminus1},y_k\}$} (Fyk.south);

            \node[draw, rectangle, minimum width=0.75cm, minimum height=5cm] (Lambdak) at (Fyk.east) [xshift=0.65cm] {$\Lambda_{\omega_k}$};
            \draw[<-,>=stealth] (Lambdak.north) -- ++(0,0.8) node[above] {$\omega_k$};

            \node[draw, rectangle, minimum width=0.75cm, minimum height=5cm] (Omegak) at (Lambdak.east) [xshift=0.65cm] {$\Omega$};

            \draw[thick] (rho) -- (Ey0.west);
            \draw[thick] (Ey0.east) -- (Fy0.west);
            \draw[thick] (Fy0.east) -- (Lambda0.west);
            \draw[thick] (Lambda0.east) -- (Omega0.west);
            \draw[thick] (Omega0.east) -- (Ey1.west);
            \draw[thick] (Ey1.east) -- (Fy1.west);
            \draw[thick] (Fy1.east) -- (Lambda1.west);
            \draw[thick] (Lambda1.east) -- (Omega1.west);
            \draw[thick] (Omega1.east) -- ($(Omega1.east)+(0.2,0)$); 
            \draw[thick] ($(Eyk.west)-(0.2,0)$) -- (Eyk.west);
            \draw[thick] (Eyk.east) -- (Fyk.west);
            \draw[thick] (Fyk.east) -- (Lambdak.west);
            \draw[thick] (Lambdak.east) -- (Omegak.west);
            \draw[thick] (Omegak.east) -- ($(Omegak.east)+(0.2,0)$); 

            \node[circle, draw, thick, minimum size=0.1cm, inner sep=0pt] (rho1) at (-0.65,1) {};
            \draw[thick] (rho1) -- ($(Ey0.west) + (0,1)$);
            \draw[thick] ($(Ey0.east) + (0,1)$) -- ($(Fy0.west) + (0,1)$);
            \draw[thick] ($(Fy0.east) + (0,1)$) -- ($(Lambda0.west) + (0,1)$);
            \draw[thick] ($(Lambda0.east) + (0,1)$) -- ($(Omega0.west) + (0,1)$);
            \draw[thick] ($(Omega0.east) + (0,1)$) -- ($(Ey1.west) + (0,1)$);
            \draw[thick] ($(Ey1.east) + (0,1)$) -- ($(Fy1.west) + (0,1)$);
            \draw[thick] ($(Fy1.east) + (0,1)$) -- ($(Lambda1.west) + (0,1)$);
            \draw[thick] ($(Lambda1.east) + (0,1)$) -- ($(Omega1.west) + (0,1)$);
            \draw[thick] ($(Omega1.east) + (0,1)$) -- ($(Omega1.east)+(0.2,1)$); 
            \draw[thick] ($(Eyk.west)-(0.2,-1)$) -- ($(Eyk.west) + (0,1)$);
            \draw[thick] ($(Eyk.east) + (0,1)$) -- ($(Fyk.west) + (0,1)$);
            \draw[thick] ($(Fyk.east) + (0,1)$) -- ($(Lambdak.west) + (0,1)$);
            \draw[thick] ($(Lambdak.east) + (0,1)$) -- ($(Omegak.west) + (0,1)$);
            \draw[thick] ($(Omegak.east) + (0,1)$) -- ($(Omegak.east)+(0.2,1)$); 

            \node[circle, draw, thick, minimum size=0.1cm, inner sep=0pt] (rho2) at (-0.65,2) {};
            \draw[thick] (rho2) -- ($(Ey0.west) + (0,2)$);
            \draw[thick] ($(Ey0.east) + (0,2)$) -- ($(Fy0.west) + (0,2)$);
            \draw[thick] ($(Fy0.east) + (0,2)$) -- ($(Lambda0.west) + (0,2)$);
            \draw[thick] ($(Lambda0.east) + (0,2)$) -- ($(Omega0.west) + (0,2)$);
            \draw[thick] ($(Omega0.east) + (0,2)$) -- ($(Ey1.west) + (0,2)$);
            \draw[thick] ($(Ey1.east) + (0,2)$) -- ($(Fy1.west) + (0,2)$);
            \draw[thick] ($(Fy1.east) + (0,2)$) -- ($(Lambda1.west) + (0,2)$);
            \draw[thick] ($(Lambda1.east) + (0,2)$) -- ($(Omega1.west) + (0,2)$);
            \draw[thick] ($(Omega1.east) + (0,2)$) -- ($(Omega1.east)+(0.2,2)$); 
            \draw[thick] ($(Eyk.west)-(0.2,-2)$) -- ($(Eyk.west) + (0,2)$);
            \draw[thick] ($(Eyk.east) + (0,2)$) -- ($(Fyk.west) + (0,2)$);
            \draw[thick] ($(Fyk.east) + (0,2)$) -- ($(Lambdak.west) + (0,2)$);
            \draw[thick] ($(Lambdak.east) + (0,2)$) -- ($(Omegak.west) + (0,2)$);
            \draw[thick] ($(Omegak.east) + (0,2)$) -- ($(Omegak.east)+(0.2,2)$);             

            \node[circle, draw, thick, minimum size=0.1cm, inner sep=0pt] (rhomin1) at (-0.65,-2) {};
            \draw[thick] (rhomin1) -- ($(Ey0.west) + (0,-2)$);
            \draw[thick] ($(Ey0.east) + (0,-2)$) -- ($(Fy0.west) + (0,-2)$);
            \draw[thick] ($(Fy0.east) + (0,-2)$) -- ($(Lambda0.west) + (0,-2)$);
            \draw[thick] ($(Lambda0.east) + (0,-2)$) -- ($(Omega0.west) + (0,-2)$);
            \draw[thick] ($(Omega0.east) + (0,-2)$) -- ($(Ey1.west) + (0,-2)$);
            \draw[thick] ($(Ey1.east) + (0,-2)$) -- ($(Fy1.west) + (0,-2)$);
            \draw[thick] ($(Fy1.east) + (0,-2)$) -- ($(Lambda1.west) + (0,-2)$);
            \draw[thick] ($(Lambda1.east) + (0,-2)$) -- ($(Omega1.west) + (0,-2)$);
            \draw[thick] ($(Omega1.east) + (0,-2)$) -- ($(Omega1.east)+(0.2,-2)$); 
            \draw[thick] ($(Eyk.west)-(0.2,2)$) -- ($(Eyk.west) + (0,-2)$);
            \draw[thick] ($(Eyk.east) + (0,-2)$) -- ($(Fyk.west) + (0,-2)$);
            \draw[thick] ($(Fyk.east) + (0,-2)$) -- ($(Lambdak.west) + (0,-2)$);
            \draw[thick] ($(Lambdak.east) + (0,-2)$) -- ($(Omegak.west) + (0,-2)$);
            \draw[thick] ($(Omegak.east) + (0,-2)$) -- ($(Omegak.east)+(0.2,-2)$); 
        \end{tikzpicture}
    \end{center}
    \caption[Scheme illustrating a quantum circuit with sequential measurements, feedback and dephasing on a multi-qubit system]{\textbf{Scheme illustrating a quantum circuit with sequential measurements, feedback and dephasing on a multi-qubit system.} The state evolves step-by step through a repeating sequence of operations: a measurement operator $\measE_{y_j}$, a feedback map $\FF_{\pmb{y}_{0:j}}$ conditioned on past outcomes, and an internal dynamics channels $\Omega$ and $\Lambda_{\omega_j}$. Here, $\Omega$ represents a collective dephasing map, while $\Lambda_{\omega_k}$ encodes both local dephasing and the unitary $\omega_k$-dependent evolution. Measurement outcomes $\pmb{y}_{0:k}$ are collected progressively and used to inform future feedback steps, allowing the system to evolve adaptively under both unitary and non-unitary processes. }
\label{fig:CS_4}
\end{figure}

\subsection{Convex decomposition of the likelihood} \label{sec:convex_decomp}

Next, our goal is to find a convex decomposition of the effective noisy $\omega$-encoding map \cite{Amoros-Binefa2024,Amoros-Binefa2025}, i.e.~$\Omega\left[\Lambda_{\omega}\left[ \; \cdot \; \right]\right]$ in \eqnref{eq:splitting_of_dynamics}, so that the discretised likelihood \eref{eq:discretized_likelihood} can be decomposed as follows:
\begin{align} \label{eq:likelihood_convex_decomp}
    p(\pmb{y}_{0:k}|\pmb{\omega}_{0:k}) = \int \D\pmb{\Zeta}_{0:k} \; q(\pmb{\Zeta}_{0:k}|\pmb{\omega}_{0:k}) \; p(\pmb{y}_{0:k}|\pmb{\Zeta}_{0:k}),
\end{align}
where $\pmb{\Zeta}_{0:k} = \{\pmb{\zeta}_0,\pmb{\zeta}_1,\dots,\pmb{\zeta}_k\}$ is a sequence of sets, each containing $N$ auxiliary frequency-like random variables. For instance, $\pmb{\zeta}_\ell = \{\zeta^{\,(1)}_\ell,\zeta^{\,(2)}_\ell, \dots, \zeta^{\,(N)}_\ell\}$ indicates that within the $\ell$th step, the first probe undergoes the Larmor precession for $\Dt$ with frequency $\zeta^{\,(1)}$, the second probe with $\zeta^{\,(2)}$, etc \cite{Amoros-Binefa2024,Amoros-Binefa2025}. 

While $q(\pmb{\Zeta}_{0:k}|\pmb{\omega}_{0:k})$ in \eqnref{eq:likelihood_convex_decomp} represents the mixing distribution, which contains all the dependence on the frequency trajectory $\pmb{\omega}_{0:k}$, the term $p(y_{0:k}|\pmb{\Zeta}_{0:k})$ can be interpreted as a \emph{fictitious} likelihood of obtaining the measurement record $\pmb{y}_{0:k} = \{y_j\}_{j=0}^k$. Here, the sequence $\pmb{\Zeta}_{0:k}$ specifies a series of unitary maps that encode the frequency information in between discretized measurements and feedback operations:
\begin{align}
    \!\!&\;
    p(\pmb{y}_{0:k}|\pmb{\Zeta}_{0:k}) = \nonumber \\
    &=\trace{\mathcal{U}_{\pmb{\zeta}_k}\!\left[\FF_{\pmb{y}_{0:k}}\!\!\left[ \measE_{y_{k}} \dots \mathcal{U}_{\pmb{\zeta}_1}\!\left[\FF_{\pmb{y}_{0:1}}\!\left[ \measE_{y_{1}} \mathcal{U}_{\pmb{\zeta}_0}\!\!\left[\FF_{\pmb{y}_{0}}\!\!\left[ \measE_{y_{0}} \,\rho_{0}\measE_{y_{0}}^{\dagger}\right]\right]\measE_{y_{1}}^{\dagger}\right]\right]\!\dots\!\measE_{y_{k}}^{\dagger}\! \right]\right]\!}\!. \label{eq:likelihood_Zk_Uk}
\end{align} 

To justify the form of the convex decomposition in \eqref{eq:likelihood_convex_decomp}, we seek to represent the overall noisy encoding channel $\Omega\left[\Lambda_{\omega}\!\left[ \; \cdot \; \right]\right]$ as a probabilistic mixture of unitaries. Rather than decomposing the full channel all at once, we handle its components separately: the collective map $\Omega[\cdot]$, which acts globally, and the local channel, which describes local dynamics and exhibits a tensor product structure with local dephasing and unitary evolution acting independently on each two-level system. 

\subsection{The map $\Omega$ as a convex mixture of unitaries} \label{sec:coll_decomp}

The channel $\Omega$, which represents the evolution of the atomic state under collective dephasing, admits a representation as a convex mixture of unitaries. This follows from the following result established in~\cite{Amoros-Binefa2021}:
\begin{theorem}[Map as a convex mixture of unitaries] \label{thm:mixture_unitaries}
    Given a unitary evolution governed by a Hamiltonian $\xi \Ham$
    \vspace{-10pt}
    \begin{equation}
        \Unitary_{\xi,\tau}[\;\cdot\;] = \ee^{-\ii \, \xi \, \Ham \tau} \; \cdot \; \ee^{\ii \, \xi \, \Ham \tau},
    \end{equation}
    whose scalar encoding $\xi \in \Real$ (frequency) is randomly distributed according to a Gaussian probability density
    \begin{equation} \label{eq:gaussian_mixing_prob}
        \xi \sim p_{\mu,\sigma}(\xi\,) = \mathcal{N}(\mu(\tau),\sigma^2(\tau)) = \frac{1}{\sqrt{2\pi \sigma^2(\tau)}} \exp{\left\{\!\shortminus\frac{(\xi - \mu(\tau))^2}{2 \sigma^2(\tau)}\!\right\}},
    \end{equation}
    then, the quantum map $\Omega$ can be written as a convex mixture of these unitaries as:
    \begin{equation} \label{eq:convex_mixture_of_unitaries}
        \rho(\tau) = \Omega[\rho(0)] = \E{\Unitary_{\xi,\tau}[\rho(0)]}{p(\xi\,)} = \int \dd\xi \; p_{\mu,\sigma}(\xi\,) \, \ee^{-\ii \, \xi \, \Ham \tau} \rho(0) \ee^{\ii \, \xi \, \Ham \tau}, 
    \end{equation}
    if $\rho(\tau)$ corresponds to the solution of the following master equation:
    \begin{align}
        \frac{\dd \rho(\tau)}{\dd \tau} &= -\ii \omega(\tau) [\Ham,\rho(\tau)] + \Gamma(\tau) \left(\Ham \rho(\tau) \Ham - \frac{1}{2} \{ \Ham^2,\rho(\tau) \} \right) \label{eq:form1_master_eq} \\
        &= -\ii \omega(\tau) [\Ham,\rho(\tau)] - \frac{1}{2} \Gamma(\tau) \left[\Ham,[\Ham,\rho(\tau)]\right], \label{eq:form2_master_eq}
    \end{align}
    with the time-dependent frequency and decay parameters being
    \begin{equation}
        \omega(\tau) = \mu(\tau) + \tau \dot{\mu}(\tau) \;\;\;\;\;\; \text{and} \;\;\;\;\;\; \Gamma(\tau) = 2\sigma^2(\tau) \tau \left( 1 + \frac{\dot{\sigma}(\tau)}{\sigma(\tau)} \tau \right).
    \end{equation}
\end{theorem}
\begin{myproof}
    Available in \appref{sec:proofs_ch4}.
\end{myproof}

Now consider a system of $N$ spin$-1/2$ particles evolving for a time $\Dt$ under the dynamics of \eqnref{eq:form2_master_eq}:
\begin{equation}
    \frac{\dd \rho(\Dt)}{\dd (\!\Dt)} = - \frac{1}{2} \kcoll \left[\Jz,\left[\Jz,\rho(\Dt) \right]\right],
\end{equation}
with frequency $\omega(\Dt) = 0$, $\Gamma(\Dt) = \kcoll$ and $\Ham = \Jz$. This particular choice of $\Ham =\Jz$ anticipates the structure of the SME considered in the next chapter, where the atomic sensor undergoes collective dephasing along the $z$-axis.

Then, the effective map $\Omega$ describing the evolution from $\rho(0)$ to $\rho(\Dt)$ can be written as a mixture of unitary channels. Specifically, it takes the form:
\begin{equation}
    \Omega[ \; \cdot \;] = \int \dd\xi \; p_\smallc(\xi\,) \; \ee^{-\ii\xi \; \Jz \Dt} \; \cdot \; \ee^{\ii \xi \; \Jz \Dt},
    \label{eq:coll_overall_map}
\end{equation}
where the mixing probability of \eqnref{eq:gaussian_mixing_prob} is a Gaussian distribution $p_\smallc(\xi) = \mathcal{N}(0,\Vcoll)$ with zero mean and variance:
\begin{equation}
    \Vcoll \coloneqq \kcoll/\Dt. \label{eq:var_coll} 
\end{equation}

\subsection{The map $\Lambda_\omega$ as a convex mixture of unitaries} \label{sec:local_decomp}

On the other hand, the overall map associated with the local dephasing is given by
\begin{equation}
    \Lambda_\omega = \ee^{t \Lin},
    \label{eq:local_overall_map}
\end{equation}
where $\Lin$ is the GKSL generator defined in \eqnref{eq:Linbladian_form}, i.e.:
\begin{align}
    \frac{\dd\rho}{\dt} &= \Lin \, \rho = -\ii [\Ham,\rho(t)] + \sum_{i = 1}^{N} \D{[\hat{L}_i]}\rho(t). 
\end{align}
If we choose the Hamiltonian as $\Ham = \omega \Jz$, with $\Jz = \frac{1}{2}\sum_{i=1}^N \Pauliz^{(i)}$ being the collective angular momentum operator in the $z$-direction, and define the local collapse operators as $\LinOp_i = \sqrt{\kloc/2} \Pauliz^{(i)}$, then the master equation becomes
\begin{align}
    \frac{\dd\rho}{\dt} &= -\ii \omega [\Jz,\rho(t)] +  \frac{\kloc}{2} \sum_{i = 1}^{N} \D{[\Pauliz^{(i)}]}\rho(t)  \nonumber\\
    &=  \left( -\ii \frac{\omega}{2} \sum_{i=1}^N  [ \Pauliz^{(i)} , \; \cdot \; ] +  \frac{\kloc}{2} \sum_{i = 1}^{N} \D{[\Pauliz^{(i)}]} \; \cdot \; \right) \rho(t)  \nonumber \\
    &= \left[ \bigoplus_{i=1}^N \Lin^{(i)} \right] \rho, \label{eq:local_SE}
\end{align}
where the subscript $(i)$ denoting the position of $\Pauliz$ in the tensor-product structure. Just as in the previous section, the choice of $\Ham$ and $\LinOp_i$ reflects our atomic sensor model, where both the field and dephasing occur along the $z$-axis. More details regarding this model will be presented in the next chapter, i.e. \chapref{chap:model}. We then define the local generator for each spin as
\begin{align}
    \Lin^{(i)} = -\ii\frac{\omega}{2}[\Pauliz^{(i)}, \; \cdot \; ] + \frac{\kloc}{2} \D[\Pauliz^{(i)}] \; \cdot \, ,
\end{align}
so that the overall map $\Lin$ is simply the direct sum of the individual contributions $\Lin^{(i)}$. The formal solution to the master equation above is given by the collective map $\Lambda_\omega$, which can be written as a tensor product of individual CPTP maps acting on each atom:
\begin{equation}
    \Lambda_\omega = \ee^{\bigoplus t \Lin^{(i)}} = \bigotimes_{i=1}^N \ee^{t \Lin^{(i)}} = \bigotimes_{i=1}^N \Lambda_\omega^{(i)},
\end{equation}
with the semigroup map $\Lambda_\omega^{(i)} = \ee^{t \Lin^{(i)}}$ defined by the GKSL generator $\Lin^{(i)}$ representing the unconditional evolution of the $i$th atom, i.e.:
\begin{align}
    \frac{\dd\rho_i(t)}{\dt} = -i\omega [\jz^{(i)},\rho_i(t)] + 2\kloc \D[\jz^{(i)}]\rho_i(t) ,
\end{align}
where $\jz = \frac{1}{2} \Pauliz$, and $\rho_i(t) = \Tr_{\forall \neq i}{\{\rho(t)\}}$ is the reduced state of the $i$th atom, i.e. the state after tracing out all atoms except the $i$th one. Applying again the result from~\theoremref{thm:mixture_unitaries}, each local map $\Lambda_\omega^{(i)}$ can be expressed as a convex mixture of unitary channels:
\begin{equation}
    \Lambda_{\omega}^{(i)}[ \; \cdot \; ] = \int \dd\upsilon^{(i)} \, p_{\ell}(\upsilon^{(i)}|\omega) \; \Unitary_{\upsilon^{(i)},\Dt}[\; \cdot \;],
\end{equation}
where the auxiliary variable $\upsilon^{(i)}$ is distributed according to $p_{\ell}(\upsilon^{(i)}|\omega) = \mathcal{N}(\omega,\Vloc)$, i.e. a Gaussian with mean $\omega$ and variance 
\begin{equation}
    \Vloc \coloneqq 2 \kloc/\Dt. \label{eq:var_loc}
\end{equation}

The corresponding unitary channel $\Unitary_{\upsilon^{(i)},\Dt}[\, \cdot \,]$ is also parametrized w.r.t. the auxiliary variable  $\upsilon^{(i)}$:
\begin{equation}
    \Unitary_{\upsilon^{(i)},\Dt} [\; \cdot \;] = \ee^{-\ii \, \upsilon^{(i)} \jz^{(i)} \Dt} \, \cdot \,  \ee^{\ii \, \upsilon^{(i)} \jz^{(i)} \Dt} .
\end{equation}

Putting everything together, the overall local map  $\Lambda_\omega$ defined in \eqnref{eq:local_overall_map} is equivalent to a convex combination of tensor products of unitary maps:
\begin{equation} \label{eq:local_overall_map_as_unitaries}
      \Lambda_\omega [\, \cdot \,]  \! = \bigotimes_{i=1}^N  \! \Lambda_\omega^{(i)} [\, \cdot \,]  \! =  \! \! \int  \! \! \D \pmb{\upsilon} \, \wp_\ell (\pmb{\upsilon}|\omega) \bigotimes_{i = 1}^N \Unitary_{\upsilon^{(i)},\Dt}[\, \cdot \,] =  \! \! \int  \! \! \D \pmb{\upsilon} \, \wp_\ell (\pmb{\upsilon}|\omega) \Unitary_{\pmb{\upsilon}}[ \, \cdot \, ], 
\end{equation}
where $\pmb{\upsilon} = (\upsilon^{(1)}, \dots , \upsilon^{(i)}, \dots ,\upsilon^{(N)})$, the integration measure is $\D\pmb{\upsilon} = \prod_{i=1}^N \dd\upsilon^{(i)}$, and the joint product distribution is $\wp_\ell (\pmb{\upsilon}|\omega) = \prod_{i=1}^N p_{\ell}(\upsilon^{(i)}|\omega)$. Note that since $\exp{\!(A)}\otimes \exp{\!(B)} = \exp{\!(A \oplus B)}$, then the overall unitary channel takes the explicit form:
\begin{align}
    \Unitary_{\pmb{\upsilon}}[ \, \cdot \, ] &\coloneqq \bigotimes_{i = 1}^N \Unitary_{\upsilon^{(i)},\Dt}[\, \cdot \,] = \ee^{-\ii \Dt \sum_{i=1}^N \upsilon^{(i)} \IjzI^{(i)} } \; \cdot \;  \ee^{\ii \Dt \sum_{i=1}^N \upsilon^{(i)} \IjzI^{(i)} },
\end{align}
where $\IjzI^{(i)} = \underbrace{ \I \otimes \dots \otimes \I}_{i-1}  \otimes \, \jz^{(i)} \otimes \underbrace{ \I \otimes \dots \otimes \I}_{N-i}$ denotes the appropriately embedded single-site spin operator $\jz^{(i)} = \frac{1}{2} \Pauliz^{(i)}$ in the full tensor-product Hilbert space.

\subsection{The joint map $\Omega\circ\Lambda_{\omega}$ as a convex mixture of unitaries}

The overall evolution of the system under $\omega$-encoding, including both collective and local decoherence effects, is described by the combined map $\Omega[\Lambda_\omega[\; \cdot \;]]$. Both the collective map $\Omega[\;\cdot\;]$ and the local channel $\Lambda_\omega[\;\cdot\;]$ have already been represented as a convex combination of unitaries in \eqnref{eq:coll_overall_map} and \eqnref{eq:local_overall_map_as_unitaries}, respectively. To now represent the overall map  $\Omega[\Lambda_\omega[\; \cdot \;]]$ as a convex mixture of unitaries, the first step is to combine the collective and local channels as:
\begin{equation} \label{eq:N+1_unitaries}
    \Omega[\Lambda_\omega[\; \cdot \;]] = \int \!\!\dd\xi \, p_{\smallc}(\xi) \int \D \pmb{\upsilon} \, \wp_{\ell}(\pmb{\upsilon}|\omega) \; \Unitary_{\xi,\pmb{\upsilon}}[ \; \cdot \; ],
\end{equation}
where 
\begin{equation} \label{eq:unit_op_overall_map}
    \Unitary_{\xi,\pmb{\upsilon}}[ \; \cdot \; ] =  \ee^{-i \Dt \sum_{i=1}^N (\xi + \upsilon^{(i)}) \IjzI^{(i)} } \; \cdot \;  \ee^{i \Dt \sum_{i=1}^N (\xi + \upsilon^{(i)}) \IjzI^{(i)} }.
\end{equation}
Here, $p_{\smallc}(\xi)$ is the Gaussian probability density associated with the collective dephasing, while $\wp_{\ell}(\pmb{\upsilon}|\omega) = \prod_{i=1}^N p_{\ell}(\upsilon^{(i)}|\omega)$ is the product of local densities.

In the above expression, the unitary operator depends on the sum $\xi + \upsilon^{(i)}$, meaning that the collective parameter $\xi$ and the local variables $\upsilon^{(i)}$ appear together. To recast the overall map into a form that is more amenable to classical simulation, we redefine the integration variables by setting
\begin{equation}
    \upsilon^{(i)} \to \zetai - \xi,
\end{equation}
for each $i$. This change of variables incorporates the contribution of $\xi$ into a new effective variable $\zetai$ for each particle. Consequently, the unitary operator becomes solely a function of $\zetai$, thereby simplifying the overall map into a convex combination of tensor products of local unitaries:
\begin{align}
    &\Omega[\Lambda_\omega[\; \cdot \; ]] = \int \! \D \pmb{\zeta} \! \left[  \! \int \! \dd\xi \, p_{\smallc}(\xi) \prod_{i=1}^N  p_{\ell}(\zetai - \xi|\omega) \right] \! \Unitary_{\pmb{\zeta}}[ \; \cdot \; ]  \nonumber \\
    &=\int \! \! \D\pmb{\zeta} \! \left[ \! \frac{1}{(2\pi \Vcoll)^{1/2}} \frac{1}{(2\pi \Vloc)^{N/2}} \! \! \int \! \!  \dd\xi\, \exp{\left\{-\frac{\xi^{\,2}}{2\Vcoll} \right\}} \exp{\left\{-\!\sum_{i=1}^N \! \! \! \frac{(\zetai - \xi - \omega)^2}{2\Vloc} \right\}} \! \right] \! \Unitary_{\pmb{\zeta}}[ \; \cdot \; ] ,
    \label{eq:collective_channel}
\end{align}
where the vector $\pmb{\zeta} = (\zeta^{(1)}, \dots , \zetai, \dots, \zeta^{(N)})$ collects the $N$ auxiliary frequencies acting on each particle. The corresponding unitary map is defined as:
\begin{align}
    \Unitary_{\pmb{\zeta}}[ \; \cdot \; ] =  \ee^{-\ii \, \Dt \sum_{i=1}^N \zetai \IjzI^{(i)}} \, \cdot \,  \ee^{\ii \, \Dt \sum_{i=1}^N \zetai \IjzI^{(i)}}.
\end{align}

By evaluating the integral in \eqnref{eq:collective_channel} using standard results for Gaussian integrals (see \lemref{lem:ap_integral_N+1_gaussians} in Appendices), we obtain the final expression for the overall map:
\begin{align} \label{eq:unitary_form_joint_map}
        \Omega[\Lambda_\omega[\; \cdot \; ]] &=  \frac{1}{\sqrt{2\pi \left(\Vcoll + \Vloc/N \right)}}\int \! \! \D\pmb{\zeta} \; \fzeta \exp{\left\{\!- \frac{(\avgzeta - \omega)^2}{2(\Vcoll + \Vloc/N)}\right\}}\Unitary_{\pmb{\zeta}}[ \; \cdot \; ] ,
\end{align}
where $\avgzeta$ is the average (mean) of the components of $\pmb{\zeta}$,
\begin{equation}
    \avgzeta \coloneqq \frac{1}{N} \sum_{i = 1}^N \zetai,
\end{equation}
and
\begin{equation}
    \fzeta = \sqrt{\frac{1}{N (2\pi \Vloc)^{N-1}}} \exp{\left\{- \frac{1}{2\Vloc} \left(\sum_{i=1}^N  (\zetai)^{\,2} -  N \avgzeta^{\;2} \right) \right\}}.
\end{equation}

\subsection{Upper-bounding the Fisher information} \label{sec:upperbound_Fisher}

The main goal of this chapter is to find an analytical lower bound for the aMSE of the estimator for $\omega$ at the time-step $t = k \Dt$. To achieve this, we rely on a family of BCRBs~\cite{Van-Trees,Fritsche2014}, and, in particular, choose the one that lower bounds the aMSE of the estimator at the last step $k$~\cite{Fritsche2014}:
\begin{align} \label{eq:BCRB_ch4}
    \EE{\Delta^2\est{\omega}_k} \geq \frac{1}{\Fisher[p(\omega_k)] + \int \dd \omega_k \, p(\omega_k) \Fisher[p(\pmb{y}_{0:k}|\omega_k)]},
\end{align}
where $p(\omega_k)$ represents the prior knowledge about the frequency at time $t = k \Dt$, and $p(\pmb{y}_{0:k}|\omega_k)$ is the probability of observing a measurement trajectory $\pmb{y}_{0:k}$ given that the frequency at time $t = k \dt$ is $\omega_k$. Here, the FI $\Fisher[\,\cdot\,]$ is defined just like in \secref{sec:Fisher_info}:
\begin{align}
    \Fisher[p(\pmb{y}_{0:k}|\omega_k)] &= \E{ \left(\partial_{\omega_k} \log{p(\pmb{y}_{0:k}|\omega_k)} \right)^2}{p(\pmb{y}_{0:k}|\omega_k)} \label{eq:other_def_FI} \\
    &= \E{-\partial^2_{\omega_k} \log{p(\pmb{y}_{0:k}|\omega_k)}}{p(\pmb{y}_{0:k}|\omega_k)}. \label{eq:best_def_FI}
\end{align}
However, for our problem there is no analytical solution for $\Fisher[p(\pmb{y}_{0:k}|\omega_k)]$. Therefore, to avoid the computationally demanding task of calculating $\Fisher[p(\pmb{y}_{0:k}|\omega_k)]$, we opt to derive an upper bound by using another probability for which the FI can be analytically determined. To do so, we begin by rewriting the conditional probability of interest $p(\pmb{y}_{0:k}|\omega_k)$ using the Bayes' rule:
\begin{align} \label{eq:p(y|w)_bayes_rewritten}
    p(\pmb{y}_{0:k}|\omega_k) &= \frac{p(\pmb{y}_{0:k},\omega_k)}{p(\omega_k)} = \frac{1}{p(\omega_k)} \!\! \int \!\! \D \pmb{\omega}_{0:k\shortminus1} \, p(\pmb{y}_{0:k},\pmb{\omega}_{0:k}) \nonumber \\
    &= \frac{1}{p(\omega_k)} \!\! \int \!\! \D \pmb{\omega}_{0:k\shortminus1} \, p(\pmb{\omega}_{0:k}) p(\pmb{y}_{0:k}|\pmb{\omega}_{0:k}),  
\end{align}
which allows us to establish a connection between $p(\pmb{y}_{0:k}|\omega_k)$, i.e. the probability of observing the vector of outcomes $\pmb{y}_{0:k}$ conditioned on the last parameter $\omega_k$, and $p(\pmb{y}_{0:k}|\pmb{\omega}_{0:k})$, i.e. the probability of detecting a measurement trajectory $\pmb{y}_{0:k}$ given that the parameter to estimate has followed a trajectory $\pmb{\omega}_{0:k}$. Then, it is possible to apply \eqnref{eq:likelihood_convex_decomp} to \eqnref{eq:p(y|w)_bayes_rewritten}, which reveals a decomposition analogous to \eqnref{eq:likelihood_convex_decomp} but now for $p(\pmb{y}_{0:k}|\omega_k)$:
\begin{align}
    p(\pmb{y}_{0:k}|\omega_k) &=  \int \D\pmb{\Zeta}_{0:k} \; p(\pmb{y}_{0:k}|\pmb{\Zeta}_{0:k}) \; \left[ \frac{1}{p(\omega_k)} \!\! \int \!\! \D \pmb{\omega}_{0:k\shortminus1} \, p(\pmb{\omega}_{0:k}) q(\pmb{\Zeta}_{0:k}|\pmb{\omega}_{0:k}) \right] \nonumber \\
    &= \int \D\pmb{\Zeta}_{0:k} \; p(\pmb{y}_{0:k}|\pmb{\Zeta}_{0:k}) \; \BigP_{\omega_k}(\pmb{\Zeta}_{0:k}) = \mapS_{\pmb{\Zeta}_{0:k}\to\pmb{y}_{0:k}}\!\left[\BigP_{\omega_k}(\pmb{\Zeta}_{0:k})\right],
\end{align}
where we identify $\mapS_{\pmb{\Zeta}_{0:k}\to\,\pmb{y}_{0:k}}[\; \bigcdot \; ]=\int\! \D\pmb{\Zeta}_{0:k} \,  p(\pmb{y}_{0:k}|\,\pmb{\Zeta}_{0:k})\; \bigcdot \;$ as a stochastic map independent of the parameter $\omega_k$, and the probability distribution $\BigP_{\omega_k}(\pmb{\Zeta}_{0:k})$ as
\begin{equation} \label{eq:def_BigP}
    \BigP_{\omega_k}(\pmb{\Zeta}_{0:k}) = \frac{1}{p(\omega_k)} \!\! \int \!\! \D \pmb{\omega}_{0:k\shortminus1} \, p(\pmb{\omega}_{0:k}) q(\pmb{\Zeta}_{0:k}|\pmb{\omega}_{0:k}), 
\end{equation}
which contains the information on $\omega_k$. As the FI is always nonincreasing under the action of any stochastic map, we can now upper-bound $\Fisher[p(\pmb{y}_{0:k}|\omega_k)]$ as
\begin{align}
    \Fisher[p(\pmb{y}_{0:k}|\omega_k)] = \Fisher[\mapS_{\pmb{\Zeta}_{0:k}\to\pmb{y}_{0:k}}\!\left[\BigP_{\omega_k}(\pmb{\Zeta}_{0:k})\right]] \leq \Fisher[\BigP_{\omega_k}(\pmb{\Zeta}_{0:k})].
\end{align}

Thus, the problem of lower-bounding the BCRB in \eqnref{eq:BCRB_ch4} now reduces to evaluating the FI of $\BigP_{\omega_k}(\pmb{\Zeta}_{0:k})$.

\subsection{Analytical form of $\BigP_{\omega_k}(\pmb{\Zeta}_{0:k})$}

The probability distribution in \eqnref{eq:def_BigP} is made up of three different probability components: the marginal probability distribution $p(\omega_k)$, the prior $p(\pmb{\omega}_{0:k})$ and the CS likelihood or mixing distribution $q(\pmb{\Zeta}_{0:k}|\pmb{\omega}_{0:k})$. To find an analytical expression for \eqnref{eq:def_BigP}, we first need to elaborate on the exact forms of each probability component, which, in turn, depend on the stochastic process governing $\omega(t)$. Specifically, throughout this chapter, we assume that $\omega(t)$ follows an OU process.

\subsubsection{Prior contribution}
As mentioned in the beginning of this section, we are interested in tracking the trajectory of a parameter or frequency, $\pmb{\omega}_{0:k} = \{\omega_0,\omega_1,\dots,\omega_k\}$, where each element $\omega_k$ is drawn from a probability distribution $p(\omega_k)$ defined as
\begin{equation}
    p(\omega_k) = \int \D \pmb{\omega}_{0:k\shortminus1} \; p(\pmb{\omega}_{0:k}).
\end{equation}
In what follows, we choose $\omega_k$ to be the time-discretized version of the process $\omega(t)$ governed by the OU equation (see \secref{sec:OUP_intro}):
\begin{equation} \label{eq:OU_process_ch4}
    \dd \omega(t) = - \chi \omega(t) \dt + \sqrt{q_\omega} \, \dW_{\!\omega},
\end{equation}
where $\chi > 0$ and $q_\omega > 0$ parametrize the decay and volatility of the process, and $\dW_{\!\omega}$ denotes the Wiener differential with mean $\EE{\dW_{\!\omega}} = 0$ and variance $\EE{\dW^2_\omega} = \dt$. The probability of the process transitioning from $\omega_{k\shortminus1}$ at time $(k\!\shortminus\!1)\Dt$ to $\omega_{k}$ at $k\Dt$ is given by
\begin{equation} \label{eq:transition_prob_OUP}
    p(\omega_k|\omega_{k\shortminus1}) = \sqrt{\frac{1}{2\pi \Vp }} \exp{\left\{ - \frac{(\omega_k - \omega_{k\shortminus1} \ee^{-\chi \Dt})^2}{2\Vp}\right\} },
\end{equation}
with the one-step transition variance reading as
\begin{equation} \label{eq:VarP}
    \Vp = \frac{q_\omega}{2\chi} (1-\ee^{-2\chi\Dt}).
\end{equation}

We choose to consider the OU process in \eqnref{eq:OU_process_ch4} instead of the more general process 
\begin{equation}
    \dd \omega(t)~=~-~\chi \left( \omega(t) - \bar{\omega} \right) \dt + \sqrt{q_\omega} \, \dW_{\!\omega},
\end{equation}
because the constant shift $\bar{\omega}$ preserves the aMSE, i.e. $\EE{\Delta^2\est{\nu}(t)} = \EE{\Delta^2\est{\omega}(t)}$ \cite{Amoros-Binefa2025}. Thus, the simplified OU process is sufficient for our purposes.

Since the OU process is a Markov process (see \secref{sec:markov_process}), the probability of the process $\omega(t)$ of following a discrete trajectory $\pmb{\omega}_{0:k} = \{ \omega_0,\omega_1,\dots,\omega_k\}$ is given by
\begin{equation} \label{eq:traj_p(omega_k)_OUP}
    p(\pmb{\omega}_{0:k}) = \prod_{i=1}^k p(\omega_i|\omega_{i-1}) p(\omega_0),
\end{equation}
where we assume $p(\omega_0)$ to be a Gaussian prior with mean zero and variance $\sigma_0^2$, i.e. $p(\omega_0) = \mathcal{N}(0,\sigma_0^2)$. From this expression, we can compute the marginal probability of the frequency taking the value $\omega_k$ at time $k\Dt$, irrespective of the previous values of $\omega$:
\begin{align} \label{eq:p(omega_k)_OUP}
    p(\omega_k) = \frac{1}{\sqrt{2\pi \Vp^{(k)}}}\, \exp\!\left(-\frac{\omega_k^2}{2\Vp^{(k)}}\right),
\end{align}
with the marginal variance at time step $k$ being:
\begin{equation}\label{eq:VarPk}
  \Vp^{(k)} = \sigma_0^2 \ee^{-2k\chi\Dt} + \frac{q_\omega}{2\chi}(1 - \ee^{-2k\chi\Dt}),
\end{equation}
which is derived recursively from the one-step transition variance $\Vp$ in \eqnref{eq:VarP}.

\subsubsection{Classically-simulated contribution}
If now we substitute the CS form of the joint map $\Omega[\Lambda_\omega[\;\cdot\;]]$ of \eqnref{eq:unitary_form_joint_map} into  the likelihood $p(\pmb{y}_{0:k}|\pmb{\omega}_{0:k})$ in \eqnref{eq:discretized_likelihood}, we retrieve the desired decomposition of \eqnref{eq:likelihood_convex_decomp}. Namely,
\begin{align}
    p(\pmb{y}_{0:k}|\pmb{\omega}_{0:k}) &=  \! \int \!\! \D\pmb{\Zeta}_{0:k} \! \left[ \prod_{j=0}^k \fzetaindex{j} \frac{1}{\sqrt{2\pi(\Vcoll+\Vloc/N)}}\,\exp{\left\{- \frac{(\avgzeta_j - \omega_j)^2}{2(\Vcoll + \Vloc/N)}\right\}} \right] \! p(\pmb{y}_{0:k}|\pmb{\Zeta}_{0:k}) \nonumber \\
    &= \int \D\pmb{\Zeta}_{0:k} \; q(\pmb{\Zeta}_{0:k}|\pmb{\omega}_{0:k}) \; p(\pmb{y}_{0:k}|\pmb{\Zeta}_{0:k}),
\end{align}
with $p(\pmb{y}_{0:k}|\pmb{\Zeta}_{0:k})$ consistent with the form given in \eqnref{eq:likelihood_Zk_Uk}.
Thus, we identify the conditional distribution $q(\pmb{\Zeta}_{0:k}|\pmb{\omega}_{0:k})$ as a product of independent likelihood terms:
\begin{align}
    q(\pmb{\Zeta}_{0:k}|\pmb{\omega}_{0:k}) &= \prod_{j=0}^k q(\pmb{\zeta}_j|\omega_j) = \prod_{j=0}^k \fzetaindex{j} \frac{1}{\sqrt{2\pi(\Vcoll+\Vloc/N)}} \exp{\left\{- \frac{(\avgzeta_j - \omega_j)^2}{2(\Vcoll + \Vloc/N)}\right\}} \nonumber \\
    &=\prod_{j=0}^k \fzetaindex{j} \mathcal{Q}(\avgzeta_j|\omega_j) =  f\,(\pmb{\Zeta}_{0:k}) \mathcal{Q}(\pmbavgzeta_{0:k}|\pmb{\omega}_{0:k}),
\end{align}
where we define the prefactor $f\,(\pmb{\Zeta}_{0:k}) = \prod_{j=0}^k \fzetaindex{j}$ as the product of normalization constants (which do not depend on $\pmb{\omega}_{0:k}$), and
\begin{equation}
    \mathcal{Q}(\pmbavgzeta_{0:k}|\pmb{\omega}_{0:k}) = \prod_{j=0}^k \mathcal{Q}(\avgzeta_j|\omega_j) =\prod_{j=0}^k \frac{1}{\sqrt{2\pi(\Vcoll+\Vloc/N)}} \exp{\left\{- \frac{(\avgzeta_j - \omega_j)^2}{2(\Vcoll + \Vloc/N)}\right\}},
\end{equation}
is a product of $k+1$ Gaussians $\mathcal{Q}(\avgzeta_j|\omega_j)$ each centered around $\omega_j$ with a common variance
\begin{equation} \label{eq:defVq}
    \Vq = \Vcoll + \frac{\Vloc}{N} = \frac{\kcoll}{\Dt} + \frac{2\kloc}{N \Dt}, 
\end{equation}
which uses the definitions of the collective and local variances in \eqnref{eq:var_coll} and \eqnref{eq:var_loc}, respectively.

\subsubsection{Integrated form of $\BigP_{\omega_k}(\pmb{\Zeta}_{0:k})$}

Once each contribution to $\BigP_{\omega_k}(\pmb{\Zeta}_{0:k})$ has been established, we can bring them all together and rearrange its integral form in \eqnref{eq:def_BigP} as a set of nested integrals:
\begin{align} \label{eq:rearranging_BigP}
    &\BigP_{\omega_k}(\pmb{\Zeta}_{0:k}) = \frac{1}{p(\omega_k)} \!\! \int \!\! \D \pmb{\omega}_{0:k\shortminus1} \, p(\pmb{\omega}_{0:k}) q(\pmb{\Zeta}_{0:k}|\pmb{\omega}_{0:k}) = \frac{f\,(\pmb{\Zeta}_{0:k})}{p(\omega_k)} \!\! \int \!\! \D \pmb{\omega}_{0:k\shortminus1} \, p(\pmb{\omega}_{0:k}) \mathcal{Q}(\pmbavgzeta_{0:k}|\pmb{\omega}_{0:k}) \nonumber \\
    &= \frac{f\,(\pmb{\Zeta}_{0:k})}{p(\omega_k)} \!\! \int \!\! \D \pmb{\omega}_{0:k\shortminus1} \, \prod_{j=1}^k p(\omega_j|\omega_{j\shortminus1}) \mathcal{Q}(\avgzeta_j|\omega_j) \, p(\omega_0) \mathcal{Q}(\avgzeta_0|\omega_0) \nonumber \\
    &= \frac{f\,(\pmb{\Zeta}_{0:k})}{p(\omega_k)} \mathcal{Q}(\avgzeta_k|\omega_k) \!\!\int\!\! \dd \omega_{k\shortminus1} \, p(\omega_k|\omega_{k\shortminus1}) \mathcal{Q}(\avgzeta_{k\shortminus1}|\omega_{k\shortminus1})  \dots \!\!\int\!\! \dd \omega_0 \,  p(\omega_1|\omega_0) \mathcal{Q}(\avgzeta_0|\omega_0) p(\omega_0).
\end{align}

Since all functions within the integrals are Gaussian, this set of nested integrals admits a closed-form recursive solution, as established in \lemref{lem:recurrence_relation_integral} in Appendices. In particular, by identifying the variances $\Vp$ and $\Vq$ in \eqnref{eq:recursive_relation} with those defined in \eqnref{eq:VarP} and \eqnref{eq:defVq}, respectively, we can directly state that:
\begin{align} \label{eq:closed_form_BigP}
    \BigP_{\omega_k}(\pmb{\Zeta}_{0:k}) &= \frac{f\,(\pmb{\Zeta}_{0:k})}{p(\omega_k)} \mathcal{Q}(\avgzeta_k|\omega_k) \mathcal{P}_k(\omega_k),
\end{align}
where $\mathcal{P}_k(\omega_k)$ is a Gaussian distribution of the form
\begin{equation}
    \mathcal{P}_k(\omega_k) = C_k \exp{\left\{-\left(\dfrac{(\omega_k - \mu_k)^2}{2\mrm{V}_{\mrm{CS}}^{(k)}}\right)\right\}},
\end{equation}
with a variance that evolves recursively in time according to
\begin{equation} \label{eq:variance_recursive_relation}
    \mrm{V}_{\mrm{CS}}^{(k)} = \Vp + \frac{\Vq \mrm{V}_{\mrm{CS}}^{(k\shortminus1)}}{\Vq + \mrm{V}_{\mrm{CS}}^{(k\shortminus1)}},
\end{equation}
starting from an initial value $\mrm{V}_{\mrm{CS}}^{(0)} = \sigma_0^2$.

\subsection{Fisher information of $\BigP_{\omega_k}(\pmb{\Zeta}_{0:k})$} \label{sec:fisher_ofBigp}

Now that we have a closed-form expression for $\BigP_{\omega_k}(\pmb{\Zeta}_{0:k})$, as given in \eqnref{eq:closed_form_BigP}, we can move to computing its FI using the definition provided in \eqnref{eq:best_def_FI}. Namely,
\begin{align}
    \Fisher{[\BigP_{\omega_k}(\pmb{\Zeta}_{0:k})]} &= \int \D \pmb{\Zeta}_{0:k} \, \BigP_{\omega_k}(\pmb{\Zeta}_{0:k}) \left[-\partial^2_{\omega_k} \log{\left( \frac{f\,(\pmb{\Zeta}_{0:k})}{p(\omega_k)} \mathcal{Q}(\avgzeta_k|\omega_k) \mathcal{P}_k(\omega_k) \right)}\right] \nonumber \\
    & = \int \D \pmb{\Zeta}_{0:k}\,  \BigP_{\omega_k}(\pmb{\Zeta}_{0:k}) \left[-\partial^2_{\omega_k} \log{ f\,(\pmb{\Zeta}_{0:k}) } \right] \label{eq:Fisher_of_f} \\
    & - \int \D \pmb{\Zeta}_{0:k}\,  \BigP_{\omega_k}(\pmb{\Zeta}_{0:k}) \left[-\partial^2_{\omega_k} \log{p(\omega_k) } \right]  \label{eq:Fisher_of_p} \\
    & + \int \D \pmb{\Zeta}_{0:k} \, \BigP_{\omega_k}(\pmb{\Zeta}_{0:k}) \left[-\partial^2_{\omega_k} \log{ \mathcal{Q}(\avgzeta_k|\omega_k) } \right]  \label{eq:Fisher_of_curlyQ}\\
    & + \int \D \pmb{\Zeta}_{0:k} \, \BigP_{\omega_k}(\pmb{\Zeta}_{0:k}) \left[-\partial^2_{\omega_k} \log{\mathcal{P}_k(\omega_k)  } \right]  \label{eq:Fisher_of_curlyP} \\
    &= -\frac{1}{\Vp^{(k)}} + \frac{1}{\Vq} + \frac{1}{\mrm{V}_{\mrm{CS}}^{(k)}} \label{eq:Fisher_variances_form},
\end{align}
where in \eqnref{eq:Fisher_of_f} we have used that $f\,(\pmb{\Zeta}_{0:k})$ is independent of $\omega_k$, and therefore, does not contribute to the FI. Meanwhile, the non-zero contributions arising from \eqnsref{eq:Fisher_of_p}{eq:Fisher_of_curlyP} follow from the identity:
\begin{align} \label{eq:Fisher_Gaussian}
    \Fisher[\Gauss(\mu,\mrm{V})] = -\partial_{\omega_k}^2 \log{\exp{\left\{-\frac{(\omega_k - \mu)^2}{2 \mrm{V}}\right\}}} = \mrm{V}^{-1},
\end{align}
where we have used the definition of a Gaussian distribution and the FI of \eqnref{eq:best_def_FI}. While we already have a closed-form expression for both $\Vp^{(k)}$ and $\Vq$, obtaining $\mrm{V}_{\mrm{CS}}^{(k)}$ requires solving the recursive relation in \eqnref{eq:variance_recursive_relation}. Fortunately, despite being somewhat lengthy, an explicit expression for $\mrm{V}_{\mrm{CS}}^{(k)}$ exists:
\begin{align}
    \mrm{V}_{\mrm{CS}}^{(k)} = \frac{W_+ V_+^{\,k} + W_-V_-^{\,k}}{U_-V_-^{\,k} + U_+ V_+^{\,k}},
\end{align}
where the subscript $k$ in $V_\pm^k$ simply denotes $V_{\pm}$ raised to the $k$-th power, where $k$ is the discrete time counter $k = t/Dt$. Moreover, terms $W_\pm$, $V_\pm$, and $U_\pm$ are defined as
\begin{align}
    W_\pm &= \pm 2\Vp\Vq \pm \sigma_0^2 \Vp + \sigma_0^2 \sqrt{\Vp(4\Vq + \Vp)}, \\
    U_\pm &= \mp \Vp \pm 2\sigma_0^2 + \sqrt{\Vp(4\Vq + \Vp)}, \\
    V_\pm &= 2\Vq + \Vp \pm \sqrt{\Vp(4\Vq + \Vp)}, 
\end{align}
where $\Vq$ is the variance given in \eqnref{eq:defVq} and $\Vp$ is the one-step variance of the OU process, i.e. \eqnref{eq:VarP}.

\subsection{The continuous-time limit} \label{sec:cont_time_limit}
If now we take the continuous-time limit of $\Dt \rightarrow 0$, the term $1/\Vq$ in \eqnref{eq:Fisher_variances_form} goes to zero since $\Vq$ is inversely proportional to $\Dt$. The other terms become,
\begin{align}
    \Vp (t) &= \lim_{\Dt \rightarrow 0} \Vp^{(k)} = \sigma_0^2 \ee^{-2\chi t} + \frac{q_\omega}{2\chi}(1 - \ee^{-2\chi t}),
\end{align}
and
\begin{align}
    \mrm{V}_{\mrm{CS}}(t) = \lim_{\Dt \rightarrow 0} \mrm{V}_{\mrm{CS}}^{(k)} = \frac{\sqrt{q_\omega \kappa_Q(N)} \, \sigma_0^2 \cosh{\left(t\sqrt{\frac{q_\omega}{\kappa_Q(N)}}\right)} + q_\omega \kappa_Q(N) \sinh{\left(t\sqrt{\frac{q_\omega}{\kappa_Q(N)}}\right)}}{\sqrt{q_\omega \kappa_Q(N)} \cosh{\left(t\sqrt{\frac{q_\omega}{\kappa_Q(N)}}\right)} + \sigma_0^2 \sinh{\left(t \sqrt{\frac{q_\omega}{\kappa_Q(N)}}\right)}},
\end{align}
where $\kappa_Q(N) = \kcoll + 2\kloc/N$. Therefore, the BCRB of \eqnref{eq:BCRB_ch4} in the continuous-time limit can be now bounded as follows:
\begin{align}
    \EE{\Delta^2\est{\omega}(t)} &\geq \frac{1}{\Fisher[p(\omega(t))] + \int \dd \omega \, p(\omega(t)) \Fisher[p(\pmb{y}_{t}|\omega(t))]} \nonumber \\
    &\geq \frac{1}{\Fisher[p(\omega(t))] + \int \dd \omega \, p(\omega(t)) \Fisher[\BigP_{\omega}(\pmb{\Zeta}_{t})]} \nonumber \\
    &= \dfrac{1}{\dfrac{1}{\Vp (t)} - \dfrac{1}{\Vp (t)} + \dfrac{1}{\mrm{V}_{\mrm{CS}}(t)}} = \mrm{V}_{\mrm{CS}}(t), \label{eq:CSbound_aMSE}
\end{align}
where $\pmb{y}_t = \{y(\tau) \, : \, 0 \leq \tau \leq t \}$ is the realization of the measurement process, and we have used \eqnref{eq:Fisher_Gaussian} to write $\Fisher[p(\omega(t))] = 1/\Vp (t)$. Hence, in its most general form, the aMSE in the continuous limit of $\Dt \to 0$ is lower bounded by
\begin{equation} \label{eq:CSlimit_full}
    \!\EE{\Delta^2\est{\omega}(t)} \! \geq \! \mrm{V}_{\!\mrm{CS}}^{\sigma_0}(t) = \frac{\sqrt{q_\omega \kappa_Q(N)} \, \sigma_0^2 \cosh\!{\left(t\sqrt{\frac{q_\omega}{\kappa_Q(N)}}\right)} \! + q_\omega \kappa_Q(N) \sinh\!{\left(t\sqrt{\frac{q_\omega}{\kappa_Q(N)}}\right)}}{\sqrt{q_\omega \kappa_Q(N)} \cosh\!{\left(t\sqrt{\frac{q_\omega}{\kappa_Q(N)}}\right)} \! + \sigma_0^2 \sinh\!{\left(t \sqrt{\frac{q_\omega}{\kappa_Q(N)}}\right)}},
\end{equation}
which in the limit of $t\to\infty$, i.e. the steady state, simplifies to:
\begin{equation} \label{eq:CSlim_SS}
    \EE{\Delta^2\est{\omega}(t)} \geq \mrm{V}_{\!\mrm{CS}}^{\sigma_0}(t\to\infty) = \sqrt{q_\omega \kappa_Q(N)} = \sqrt{q_\omega \!\left(\kcoll \! + \! \dfrac{2\kloc}{N}\right)},
\end{equation}
since $\coth{x} = 1$ when $x\to\infty$. The general form of \eqnref{eq:CSlimit_full} can be simplified when considering different regimes: (1) an infinitely wide prior, (2) no field fluctuations and (3) a combination of both cases. In particular, when $\sigma_0 \rightarrow \infty$, then \eqnref{eq:CSlimit_full} reduces to 
\begin{align} \label{eq:CSlim_infprior}
    \!\!\EE{\Delta^2\est{\omega}(t)} \geq \mrm{V}_{\!\mrm{CS}}^{\infty}(t) &= \sqrt{q_\omega \, \kappa_Q(N)} \coth\!{\left(\!t \sqrt{\frac{q_\omega}{\kappa_Q(N)}}\right)}  \nonumber \\
    &= \sqrt{q_\omega \left(\kcoll \! + \! \dfrac{2\kloc}{N}\right)} \coth\!{\left(\!t \sqrt{q_\omega \left(\kcoll \! + \! \dfrac{2\kloc}{N}\right)^{\!\!-1}}\right)}. 
\end{align}
If instead we take the limit of $q_\omega \rightarrow 0$ of \eqnref{eq:CSlimit_full}, it becomes
\begin{equation} \label{eq:CSlim_zeroq}
    \EE{\Delta^2\est{\omega}(t)} \geq \mrm{V}_{\!\mrm{CS}}^{\sigma_0}(t,q_\omega\to 0) = \dfrac{1}{\dfrac{1}{\sigma_0^2} + \dfrac{t}{\kappa_Q(N)}} = \dfrac{1}{\dfrac{1}{\sigma_0^2} + \dfrac{t}{\kcoll + \dfrac{2\kloc}{N}}},
\end{equation}
which exhibits the standard quantum limit (SQL) when considering an infinitely wide prior:
\begin{equation} \label{eq:CSlim_zeroq_infprior}
    \EE{\Delta^2\est{\omega}(t)} \geq \mrm{V}_{\!\mrm{CS}}^{\infty}(t,q_\omega\to 0) = \frac{\kappa_Q(N)}{t} = \frac{\kcoll}{t} + \frac{2\kloc}{N t}.
\end{equation}
\begin{figure}
    \centering
    \includegraphics[width=0.75\linewidth]{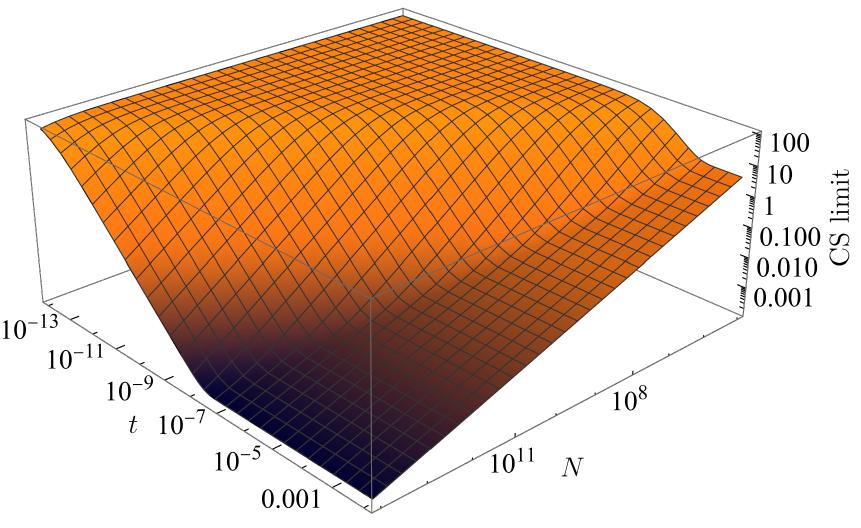}
    \caption[3D Plot of the CS limit with finite $\sigma_0$ w.r.t. $N$ and $t$]{\textbf{3D Plot of the CS limit with finite $\sigma_0$ w.r.t. $N$ and $t$.} Log-log plot of the CS limit in \eqnref{eq:CSlimit_full} as a function of $N$ and $t$, showcasing the behavior of the function over several orders of magnitude. The color gradient indicates the magnitude of the function, transitioning from high (bright) values to low (dark) values. The parameters used to generate this figure: $\sigma_0 = \SI{10}{\radian \second^{-1}}, \; q_\omega = 10^4\SI{}{\radian^2 \second^{-3}}, \; \kcoll = 0, \; \kloc = \SI{100}{\hertz}$. }
    \label{fig:3Dplot_of_CSlimit}
\end{figure}

As briefly hinted at in the introduction, we refer to this lower bound on the BCRB, $\mrm{V}_{\!\mrm{CS}}^{\sigma_0}(t,N)$, as either the CS limit or the quantum limit. After going through its derivation, it is hopefully clear that the term ``classically-simulated'' is used because the bound is derived by expressing the noisy quantum evolution as a convex combination of unitary channels (recall \eqnref{eq:coll_overall_map} or \eqnref{eq:local_overall_map_as_unitaries}), which can be efficiently simulated using classical methods. At the same time, we also call it the ``quantum limit'' because it represents the general bound on sensitivity that cannot be surpassed by any strategy involving any quantum effects. Although this limit sets a lower bound on the aMSE, it is not guaranteed to be tight; that is, there is no guarantee that there exists an estimator can attain this bound. However, the CS limit still disproves the possibility of attaining super-classical scalings of $N^2$ and $t^3$ in the presence of dephasing \cite{Geremia2003,Amoros-Binefa2021,Amoros-Binefa2024}. 

Perhaps most crucially, this quantum limit serves as a powerful benchmark. Since it depends solely on the noise model and field fluctuations and is entirely \emph{independent} on the initial state, measurement and measurement-based feedback, attaining this bound would demonstrate that the chosen sensing strategy is optimal; namely, that the state preparation, measurement, estimation and control all collectively yield the best possible precision in estimating the fluctuating field. 

\subsection{Atom number fluctuations and the convexity of the CS limit} \label{sec:atom_number_csbound}

In practice, the exact number of atoms $N$ in the ensemble may vary from shot to shot. We model this uncertainty by assuming $N \sim p(N) = \Gauss(\barN,\sigma^2)$. As a result, the bound given in \eqnref{eq:CSbound_aMSE} must be averaged over the distribution $p(N)$, yielding 
\begin{align} 
    \E{\EE{\Delta^2\est{\omega}(t)}}{p(N)} &\geq \E{\mrm{V}_{\!\mrm{CS}}^{\sigma_0}(t,N)}{p(N)},
\end{align}
where we emphasize the dependence of $\mrm{V}_{\!\mrm{CS}}^{\sigma_0}(t,N)$ w.r.t $N$. 

\begin{figure}[t]
    \centering
    \includegraphics[width=0.75\linewidth]{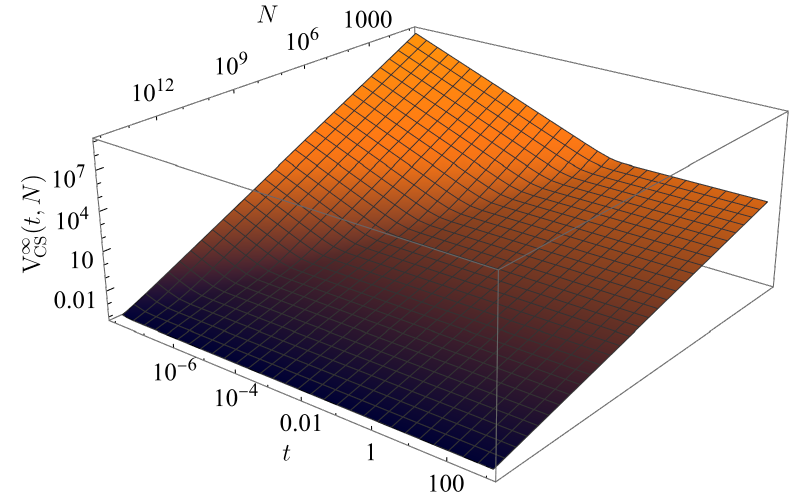}
    \caption[3D Plot of the CS limit with infinitely wide prior w.r.t. $N$ and $t$]{\textbf{3D Plot of the CS limit with infinitely wide prior w.r.t. $N$ and $t$.} Log-log plot of the CS limit $\mrm{V}_{\!\mrm{CS}}^{\infty}$ in \eqnref{eq:CSlim_infprior} as a function of $N$ and $t$ over several orders of magnitude showing its convexity over this specific range. The color gradient indicates the magnitude of the function: bright is high, and dark is low. The parameters used are: $q_\omega = 10^4\SI{}{\radian^2 \second^{-3}}, \; \kcoll = 0, \; \kloc = \SI{100}{\hertz}$. }
    \label{fig:3Dplot_of_CSlimit_infprior}
\end{figure}

Importantly, we focus on the CS limit for an infinitely wide prior (see \eqnref{eq:CSlim_infprior}) such that the influence of the prior is effectively negligible, allowing us to isolate the effects of the noise model on the estimation process. In this case, $\mrm{V}^{\infty}_{\!\mrm{CS}}(t,N)$ is a convex function of $N$, which we verify by evaluating the second derivative of \eqnref{eq:CSlim_infprior} w.r.t. $N$ and checking that it is positive for all $N >0$, $t > 0$, and $q_\omega, \, \kloc, \, \kcoll > 0$\footnote{We omit the second derivative with respect to $N$ in the CS limit with an infinitely wide prior due to its length, but it can be readily computed using a symbolic algebra tool such as \textit{Mathematica}.} (see also \figref{fig:3Dplot_of_CSlimit_infprior} for a more graphical confirmation). Then, by applying Jensen's inequality we obtain
\begin{align} 
    \E{\mrm{V}_{\!\mrm{CS}}^{\infty}(t,N)}{p(N)} \geq \mrm{V}_{\!\mrm{CS}}^{\infty}(t,\E{N}{p(N)}) = \mrm{V}_{\!\mrm{CS}}^{\infty}(t,\barN).
\end{align}
Hence, for an uninformative prior and a system with fluctuating atomic number, the aMSE of the estimate $\omega$ is lower bounded by:
\begin{align}
    \E{\EE{\Delta^2\est{\omega}(t)}}{p(N)} \geq \sqrt{q_\omega \, \kappa_Q(\barN)} \coth\!{\left(\!t \sqrt{\frac{q_\omega}{\kappa_Q(\barN)}}\right)},
\end{align}
where $\kappa_Q(\barN) = \kcoll + 2 \kloc / \barN$.

\chapter{Models for Noisy Real-Time Atomic Magnetometry} \label{chap:model}

\newthought{Tracking a time-varying magnetic field in real time with high sensitivity} is essential for applications ranging from non-invasive medical diagnostics like magnetocardiography~\cite{mcg_paper,Bison2009,jensen_magnetocardiography_2018,Kim2019,Yang2021} or magnetoencephalography~\cite{encephalo_paper,boto_moving_2018}, to navigation in GPS-denied environments~\cite{Canciani2020,Canciani2022}. In these settings, the magnetic field varies continuously, often stochastically, so one cannot rely on repeated runs to lower the estimation error. Instead, the sensor must operate in a single shot and in real time. Besides classical approaches such as boosting signal strength or reducing noise, quantum resources offer another way to enhance sensor performance~\cite{DAriano2001,Giovannetti2004}. In particular, interatomic entanglement in the form of spin squeezing can improve their sensitivity beyond the SQL~\cite{Kuzmich2000,Wasilewski2010,Shah2010,Koschorreck2010,Sewell2012}. 

Much work on this topic has involved step-wise sensing protocols, where state preparation, evolution, detection and estimation are distinct steps \cite{Giovannetti2006_other}. However, practical devices such as optical atomic magnetometers operate continuously, such that those process occur simultaneously. In this continuous-time setting, achieving quantum enhancement requires an accurate open-system description of the quantum dynamics that includes measurement backaction~\cite{Caves1980,Braginsky1996,Belavkin1989,Wiseman1993}, as well as practical estimation and control strategies that operate in real time. 

For Gaussian systems, such models have enabled optimal state estimation and quantum-limited feedback control in mechanical oscillators~\cite{wieczorek_optimal_2015,rossi_observing_2019,Wilson2015,Rossi2018,Magrini2021}. For spin-precession sensors, early theoretical treatments assumed Gaussianity~\cite{Geremia2003,Madsen2004,Molmer2004,Albarelli2017} or resorted to brute-force numerics for low atomic numbers \cite{rossi_noisy_2020}. In experiments, however, spin-precession sensors evolve toward non-Gaussian unconditional states, making Gaussian-only models sufficient only when ignoring~\cite{Jimenez2018} or evading ~\cite{Kong2020,Troullinou2021,Troullinou2023} backaction. 

To describe quantum enhancement in realistic continuous-time sensors, we begin in \secref{sec:atomic_magnetometer_model} by introducing the stochastic master equation (SME) for a continuously monitored atomic spin ensemble, the most rigorous and complete description of its conditional dynamics, as developed in \chapref{chap:cm}. The SME captures the full conditional dynamics under measurement backaction~\cite{Caves1980,Braginsky1996,Belavkin1989,Wiseman1993} as well as environmental decoherence processes that are unavoidable in optical magnetometers~\cite{Kominis2003,Shah2010,Budker2013,Jimenez2018,budker_sensing_2020,Kong2020}. Therefore, the SME offers a complete description of how measurement-induced correlations such as spin squeezing emerge in real time. However, solving the SME exactly becomes computationally intractable for the large ensembles relevant to state-of-the-art devices, where $N \sim 10^{6}-10^{15}$ (see \secref{sec:exact_numerical_sol_SME}). We nevertheless use direct SME simulations for moderate atom numbers ($N \sim 100$) both to benchmark approximations and to verify the presence of conditional and unconditional spin squeezing.

To overcome the scalability barrier while retaining the essential conditional dynamics, we employ the co-moving Gaussian (CoG) approximation~\cite{Amoros-Binefa2024,Amoros-Binefa2025} (see \secref{sec:Co-moving_Gaussian}). This nonlinear model is a 2nd-order cumulant expansion of the total angular momenta that allows us to simulate large ensembles $N \sim 10^{6} \, - \, 10^{15}$. Crucially, the CoG approximation preserves the backaction-driven creation of spin squeezing. We validate this dynamical model by comparing it to the exact SME solution for moderate-sized ensembles ($N \sim 100$), demonstrating its accuracy.

Besides having a reliable and scalable quantum model for the atomic magnetometer, we also address the problem of optimally estimating the field and controlling the sensor in \secref{sec:estim_contr_chap5}. To this end, we propose combining an extended Kalman filter (EKF) with a linear-quadratic regulator (LQR), both introduced in \chapref{chap:bayesian}, which are designed using the CoG nonlinear model. The optimality of the entire sensing protocol, which includes the initial state, measurement, estimate and control, is verified by deriving and attaining bounds on precision applicable to any sensing scheme involving measurement-based feedback, as established in \chapref{chap:bounds}.

We first demonstrate how to reach the quantum limit in the Gaussian regime (i.e., the weak-field regime) in \secref{sec:weak-field_sensing}. We then focus on the more general—and practically important—case of precession-inducing fields. These field can be (1) constant (see \secref{sec:constant_field}), ~(2)~fluctuating stochastically (see \secref{sec:fluctuating_field}), or (3)~determined by a continuously varying waveform, a magnetocardiogram (MCG) (see \secref{sec:nonlinear_MCG_field}), which is distorted by stochastic noise that should be filtered out rather than tracked. The two last signals are tracked by an optical atomic magnetometer with realistic parameters taken from the experimental setting of \refcite{Kong2020} but ignoring spin-exchange atomic collisions~\cite{Happer1977,Savukov2005}. For both constant and fluctuating fields, we show that the magnetometer operates at the quantum limit, and that it generates conditional spin squeezing. 

Finally, in \secref{sec:cond_uncond_spin_squeezing}, we resort back to the exact SME in order to demonstrate that the proposed sensing scheme steers the atomic ensemble into a state that exhibits unconditional spin squeezing~\cite{Pezze2018RMP}, i.e.~the state is entangled even when not recording the measurement data.

\section{Atomic magnetometer model} \label{sec:atomic_magnetometer_model}
\subsection{Setup and its dynamics}

\begin{figure}[t]
	\begin{center}
    \includegraphics[width=0.95\textwidth]{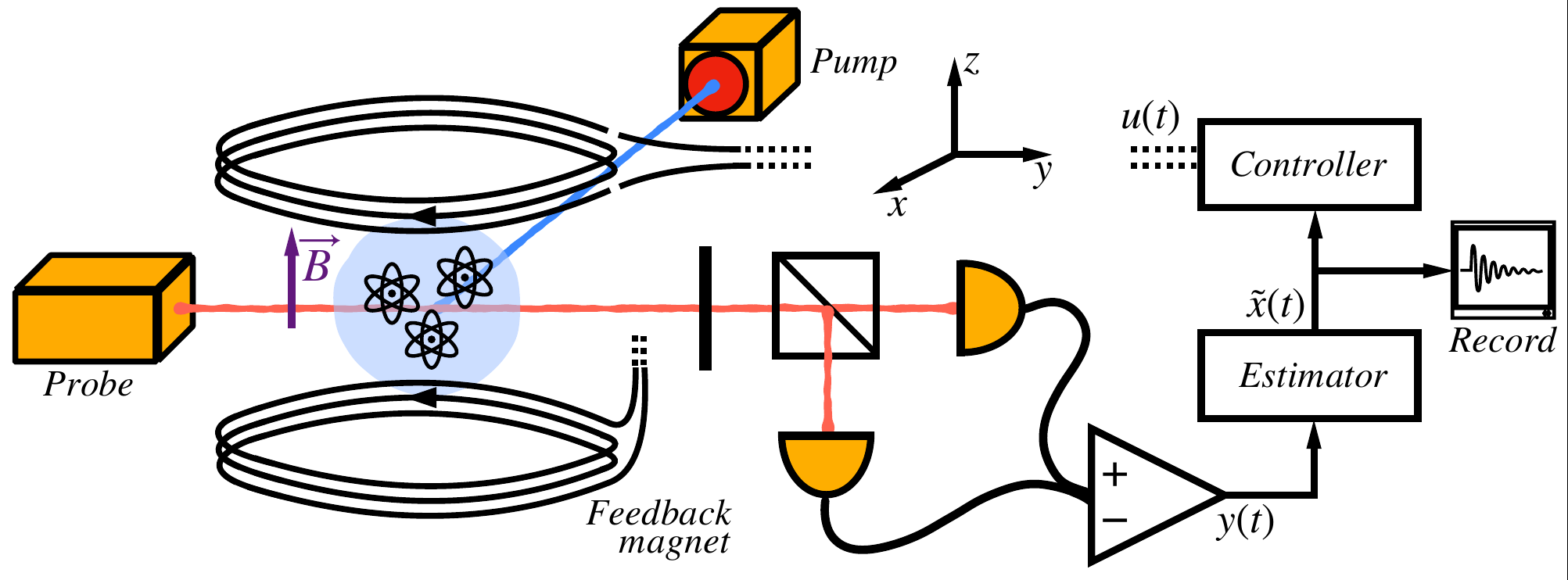}
    \end{center}
    \caption[Geometry of the atomic magnetometer]{
    \textbf{Geometry of the atomic magnetometer.} The magnetometry scheme uses an atomic ensemble optically pumped along the $x$-axis (\emph{blue beam}) into a coherent spin state (CSS). The magnetic field to be measured is oriented along $z$, while a continuous Faraday-rotation measurement is performed with a probe beam propagating along $y$ (\emph{red beam}), producing a photocurrent signal $y(t)$. This signal is fed into an estimator, whose output is then used by a controller to generate and apply a control law to the atoms.}
\label{fig:atomic_ensemble}
\end{figure}

At its core, an optical atomic magnetometer extracts information about a field from light that is scattered after interacting with a collection of $N$ atoms (as depicted in \figref{fig:system_homodyne}). Importantly, these atoms should be sensitive to the time-varying and/or stochastic magnetic field $B(t)$ we aim to sense.

Thus, to monitor the changes in the atomic state caused by the interaction with the magnetic field, the magnetometer must: (1) prepare the atoms into a suitable initial state, (2) allow their state to evolve due to its interaction with the magnetic field, and (3) simultaneously probe that state with a continuous non-demolition measurement \cite{Budker2013}. Additionally, one might devise a control law that (4) steers the atoms towards a state optimized for magnetometry.

\subsubsection{Preparation}
\begin{figure}[t!]
    \centering
    \includegraphics[width=0.8\linewidth]{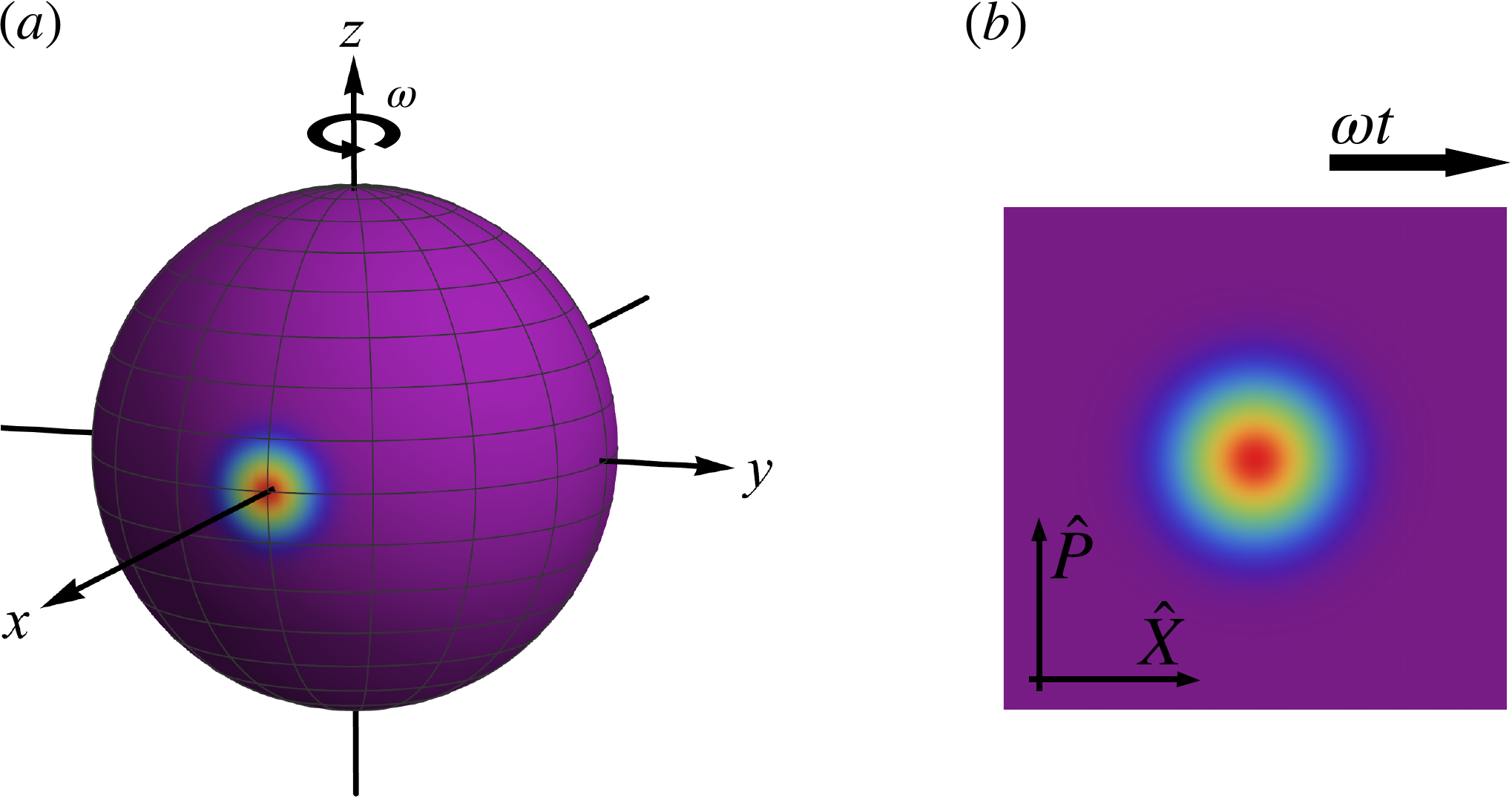}
    \caption[Bloch sphere representation of a CSS and phase representation of a coherent state]{\textbf{Bloch sphere representation of a CSS and phase representation of a coherent state.} (a) Visual representation onto a generalized Bloch sphere of the Wigner function of a CSS of 100 spin-$1/2$. Warmer colors near the center represent higher values of the Wigner distribution, while cooler colors farther away indicate its gradual decay. To generate this figure, we have used the spherical harmonic expansion described in \secref{sec:Wigner_on_Bloch}. (b) Depiction of a coherent state in the phase space. }
    \label{fig:CSS_bloch_plot}
\end{figure}

The ensemble of $N$ atoms is initialized by pumping it with circularly polarized light along the $x$-direction (see \figref{fig:atomic_ensemble}) such that only two energy levels of each atom contributes to the field and probing interactions \cite{happer1972,Budker2002RMP}. Thus, we treat the ensemble as a collection of $N$ spin-$1/2$ particles. Then, the evolution of the total spin can be described through \emph{collective} angular momentum operators:
\begin{equation}
    \hat{\!J}_\alpha = \sum_{i=1}^N \frac{\hat{\sigma}_\alpha^{(i)}}{2}, \quad \trm{with} \quad \alpha = x,y,z,
\end{equation}
defining a vector of collective angular momentum operators $\,\Jqvec{J} = (\Jx,\Jy,\Jz)^\Trans$. Optical pumping of the atoms along the $x$-direction initializes a \emph{coherent spin state} (CSS) along the same direction~\cite{Ma2011} (see also \secref{sec:CSS_main}),
so that the initial mean and variance of the ensemble angular momentum operators, denoted by the vector $\;\Jqvec{J}(t)=(\Jx(t),\Jy(t),\Jz(t))^\Trans$ in the Heisenberg picture, read $\brkt{\,\Jqvec{J}(0)}_\trm{CSS} = (J,0,0)^\Trans$ and $\brkt{\Delta^2\Jqvec{J}(0)}_\trm{CSS} = (0,\,J/2,\,J/2)^\Trans$, respectively, where $J = N/2$.
As shown in \figref{fig:CSS_bloch_plot}a, one may visualize the distribution of the angular momentum of a CSS with the help of the Wigner distribution, which is mapped onto the Bloch sphere (see \secref{sec:Wigner_on_Bloch}).

\subsubsection{Evolution}
Once pumped, the total spin of the polarized atoms starts to precess around the magnetic field axis (see \figref{fig:CSS_bloch_plot}a), assumed hereafter to be the $z$-axis, at a Larmor frequency $\omega(t) = \gmr B(t)$, with $\gmr$ being the effective (constant) gyromagnetic ratio. This translates into a unitary evolution of the state of the atoms:
\begin{equation} \label{eq:field_VonNeumann}
    \dd \rho(t) = -\ii \omega(t) \left[\Jz,\rho(t) \right] \dt,
\end{equation}
where $\rho(t)$ is the density matrix of the atoms. Thanks to this Zeeman term, it becomes clear how the atomic state can be monitored to indirectly track the magnetic field, which is the main goal of any magnetometer. In this thesis, we consider the signal to track, i.e. $B(t)$ (or equivalently, $\omega(t)$), to follow three different profiles:
\begin{enumerate}
    \item \textbf{Constant:} The Larmor frequency does not change in time, i.e. $\omega(t) = \omega$ s.t.,
    \begin{equation}
        \dd \omega(t) = 0.
    \end{equation}    
    \item \textbf{Fluctuating:} The stochastic field we consider is an OUP (see \secref{sec:OUP_intro}), whose evolution is governed by the following SDE: 
    \begin{equation} \label{eq:signal_OUP}
        \dd \omega(t) = -\chi (\omega(t) - \bar{\omega}) \dt + \sqrt{q_\omega} \dW_{\!\omega},
    \end{equation}
    where $\dW_{\!\omega}$ is a Wiener differential, with mean zero and variance $\EE{\dW_{\!\omega}^2} = \dt$. The drift and volatility parameters: $\chi \geq 0$ and $q_\omega \geq 0$, are constant. The long-term mean towards which the process reverts, $\bar{\omega}$, is also positive. 
    \item \textbf{Time-varying:} The last magnetic field profile we aim to track is a time-varying but deterministic signal resembling a heartbeat, i.e. a MCG. This MCG-like signal can be modeled using the dynamics of a \emph{filtered Van der Pol} (VdP) oscillator:
    \begin{align} \label{eq:VdP}
        &\dd \nu(t) = - p \, \omega(t) \,\dt \\
        &\dd \omega(t) = \frac{k}{m} \nu(t) \,\dt + 2 \frac{c}{m} (1-\upsilon(t)) \, \omega(t) \,\dt \\
        &\dd \upsilon(t) = \frac{|\nu(t)|-\nu(t)}{2 T} \,\dt - \frac{\upsilon(t)}{T} \,\dt,
    \end{align}
    where the frequency of interest, $\omega(t)$, is part of a larger state vector describing the signal evolution: $\pmb{x}_s(t) = (\nu(t),\omega(t),\upsilon(t))$. The parameters specifying the MCG-like signal are all positive constants: $p, k, m, c, T > 0$. Additionally, we distort the time-varying signal $\omega(t)$ by adding white noise. This noise is not a feature we aim to track and therefore, should be filtered out without resorting to time-averaging \cite{mcg_paper,ECG_noise_removal}. 
\end{enumerate}

\subsubsection{Measurement} 
In order to track the magnetic field in real time, we must continuously measure the state of the atoms. There are various continuous measurements that can be implemented, but here we consider a continuous polarimetric measurement \cite{Takahashi1999,Kuzmich2000,de_Echaniz_2005,Geremia2006,deutsch_quantum_2010,Kong2020} (see \secref{sec:polarimetric_meas}). 
As depicted in \figref{fig:atomic_ensemble}, a probe beam traveling along $y$ with linearly polarized light is transmitted through an atomic cloud. The interaction with the atoms rotates the polarization of the probe by angle $\Theta \propto \brktc{\Jy(t)}$, due to the Faraday effect \cite{Budker2002RMP}. This change in polarization, and thus, the variation in $\brktc{\Jy(t)}$, is later measured with a polarizing beam splitter and two photodetectors \cite{de_Echaniz_2005}. The effect of such a measurement on the state of the atoms can be described through a SME, as derived in detail in \secref{sec:polarimetric_meas}. Hence, the master equation in \eqnref{eq:field_VonNeumann} becomes a SME of the form: 
\begin{align} \label{eq:VonNeumann_cont_measurement}
    \dd \rhoc(t) = -\ii \omega(t) \left[\Jz,\rhoc(t) \right] \dt  + M \D[\Jy] \rhoc(t) \dt + \sqrt{\eta M} \H[\Jy] \rhoc(t) \dW,
\end{align}
with an associated measurement:
\begin{align} 
    \dd y = 2\eta \sqrt{M}\brktc{\Jy(t)}\dt +\sqrt{\eta }\dW,
\end{align}
where $M$ is the measurement strength, $\eta \in [0,1]$ the detection efficiency,  $\dW$ is the Wiener differential (see \secref{sec:Wiener_process}) and $\brktc{\Jy(t)} \coloneqq \trace{\rhoc(t) \Jy}$. Note that throughout this thesis, the detection efficiency is assumed to be perfect (i.e. $\eta = 1$), and thus, in \secref{sec:derivation_SME} we have not derived the SME for inefficient detection as given in \eqnref{eq:VonNeumann_cont_measurement}. A detailed derivation yielding the SME of \eqnref{eq:VonNeumann_cont_measurement} can be found in \refcite{Albarelli2024}. The sub-$\mrm{(c)}$ notation simply indicates that the density matrix of the atoms is now dependent, i.e. conditional, on the recorded measurement trajectory $\pmb{y}_t = \{y(\tau)\}_{\tau\leq t}$. Namely,
\begin{equation}
    \rhoc(t) \coloneqq \rho(t|\pmb{y}_t),
\end{equation}
which is simply \eqnref{eq:notation_rhoc} but written in continuous time. The two terms in \eqnref{eq:VonNeumann_cont_measurement} describing the measurement backaction arise naturally when discretizing the interaction of the system with a conveyor-belt like probe (see \secref{sec:derivation_SME}). The first term is a dissipative one, with the superoperator $\D[\,\bigcdot\,] \,\bigcdot\,$ given by \eqnref{eq:dissipation_superoperator}. The second term is stochastic as well as nonlinear w.r.t. $\rhoc(t)$, where the superoperator $\H[\,\bigcdot\,] \,\bigcdot\, $ is defined in \eqnref{eq:H_superoperator_def}, and is correlated with the measurement outcomes $\dd y$ through the Wiener differential $\dW$. This last term is the one responsible for the creation of conditional spin squeezing \cite{Geremia2006,Amoros-Binefa2021}.

\subsubsection{Modeling noise}

Noise in quantum systems arise from the interaction of the system with an unmonitored environment. In our work, we incorporate local and global dephasing terms into \eqnref{eq:VonNeumann_cont_measurement}:
\begin{align} \label{eq:VonNeumann_cont_measurement_withnoise}
    \dd \rhoc(t) = &-\ii \omega(t) \left[\Jz,\rhoc(t) \right] \dt  + \frac{\kloc}{2} \sum_{j=1}^N \D[\Pauliz^{(j)}] \rhoc(t) \dt \, + \kcoll \D[\Jz]\rhoc(t) \dt \nonumber \\
    &+ M \D[\Jy] \rhoc(t) \dt + \sqrt{\eta M} \H[\Jy] \rhoc(t) \dW, \\
    \dd y = &\,2\eta \sqrt{M} \braketavg{\Jy}_{\cc}(t) \dt + \sqrt{\eta }\dW,
\end{align}
which model effects such as collisions, stray fields, and laser instabilities occurring along  the $z$-direction, i.e. the direction along which the $B$-field is applied. Local dephasing, as its name suggests, acts independently on each individual atom at a rate $\kloc$ \cite{Huelga1997}. Similar mechanisms affecting the entire ensemble uniformly can instead be modeled with collective dephasing, at a rate $\kcoll$ \cite{Dorner2012,Plankensteiner2016}.

\subsubsection{Feedback}

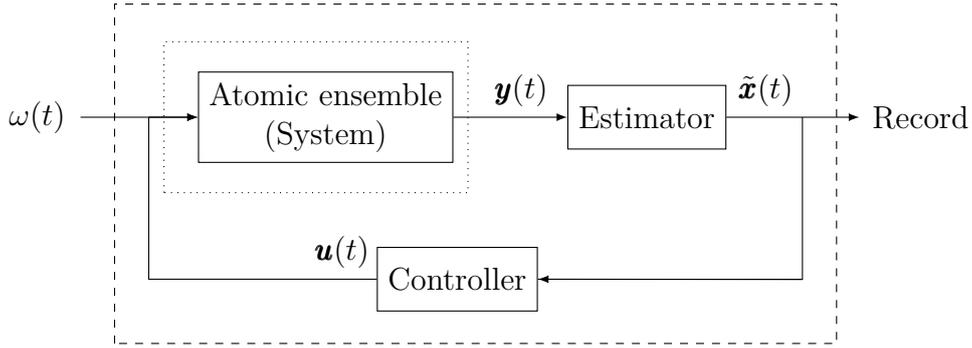
\begin{figure}[h!]
\begin{center}
    \begin{tikzpicture}[auto, node distance=2cm,>=latex]
    \tikzstyle{block} = [rectangle, draw, fill=white, 
    text centered, minimum height=2em, minimum width=4em]
    \tikzstyle{input} = [coordinate]
    
    \node [name=y_k] (y_k) {};
    \node [block, right=0.5cm of y_k, align=center] (system) {Atomic ensemble \\ (System)};
    \node [block, right=1.5cm of system, align=center] (estimator) {Estimator};
    \node [coordinate, right=1cm of estimator] (midpoint) {};
    \node [coordinate, right=0.75cm of midpoint] (output) {};
    \node [coordinate, left=0.75cm of y_k] (bfield_input) {};
    \node [coordinate, below=2cm of y_k] (below_y_k) {};
    \node [coordinate, below=2.15cm of midpoint] (below_midpoint) {};
    
    \node [block, right = 3cm of below_y_k, align = center] (controller) {Controller};
    \draw [->] (bfield_input) node [above, left = 0.05cm] {$\omega(t)$} -- (system);
    \draw [->] (below_midpoint) -- (controller);
    \draw [->] (system) -- node [above, pos = 0.6] {$\pmb{y}(t)$} (estimator);
    \draw [->] (estimator) -- node [above, pos = 0.3] {$\est{\pmb{x}}(t)$} node [above, right = 0.9cm] {Record} (output);
    \draw [->] (midpoint) |- (below_midpoint) -- (controller) -- node [above, pos = 0.15] {$\pmb{u}(t)$} (below_y_k) |- (y_k) |- (system);
    \draw[dotted] (0.2,-1) rectangle ++(4,2);
    \draw[dashed] (-0.45,-3) rectangle ++(9.5,4.5);
\end{tikzpicture}
\end{center}
\caption[Block diagram of the optical atomic magnetometer]{\textbf{Block diagram of the optical atomic magnetometer.} Structure of a closed-loop control feeding back to the atoms a control function $u(t)$ devised from estimators computed from measurement data $\pmb{y}(t)$ provided by a system (in this case, an atomic ensemble). The control law at time $t$ is constructed from the whole measurement history because the estimator at time $t$ depends on the estimator at the previous time step and the measurement at time $t$. The photocurrent $y(t)$ is computed following \eqnref{eq:fullSME_meas}, with the conditional mean of $\Jy$ provided either by the conditional state evolved according to \eqnref{eq:fullSME} or by an approximate dynamical model containing first and second order moments. Assessing the accuracy of such an approximation in replicating the evolution of the system can be done at two different stages: comparing the estimator $\est{\omega}$ w.r.t. the real Larmor frequency $\omega$ (output of dashed box v.s. input), or comparing comparing the first and second moments given by the approximation v.s. the moments provided by the evolution of the conditional density matrix (output of dashed box depending on how we evolve the system). }
\label{fig:closed-loop_atoms}
\end{figure}

Finally, we also consider controlling the system by feeding back estimates constructed from the measurement record through, for instance, coils placed around the atomic ensemble (see \figref{fig:atomic_ensemble}). In particular, by applying the control law $u(t) \coloneqq u(\pmb{y}(t))$ along the direction of the field $B(t)$, the Larmor frequency at time $t$ will be modified to $\omega(t)\to\omega(t)+u(t)$. Importantly, because of the addition of a control field which depends on the whole measurement record $\pmb{y}(t)$ up to time $t$, the dynamics of the atomic ensemble at each time step are now dependent on the \emph{entire} measurement history, instead of the last measurement increment. Since the control is applied as a magnetic field along $z$, it comes up in the SME of \eqnref{eq:VonNeumann_cont_measurement_withnoise} as a unitary term:
\begin{align} \label{eq:fullSME}
    \dd \rhoc(t) = &-\ii \left(\omega(t) + u(t) \right)\left[\Jz,\rhoc(t) \right] \dt  + \frac{\kloc}{2} \sum_{j=1}^N \D[\Pauliz^{(j)}] \rhoc(t) \dt \, + \kcoll \D[\Jz]\rhoc(t) \dt \nonumber \\
    &+ M \D[\Jy] \rhoc(t) \dt + \sqrt{\eta M} \H[\Jy] \rhoc(t) \dW, \\
    y(t) \dt = &\,2\eta \sqrt{M} \brktc{\Jy(t)} \dt + \sqrt{\eta}\dW. \label{eq:fullSME_meas}
\end{align}
This is the final form of the SME that we consider throughout this chapter. One important question that remains to be answered is how to design the control law $u(t)$. As hinted earlier and also depicted in \figref{fig:atomic_ensemble}, the control $u(t)$ will be a function of estimates constructed from the measurement record $\pmb{y}(t) = \{y(\tau)\}_{0\leq\tau\leq t}$, applied back into the system through a closed-loop structure of the form shown in \figref{fig:closed-loop_atoms}. The exact form of the controller and estimator will be discussed later. 

\section{Simulating the system}

In the absence of an experiment providing us with real-time measurements $y(t)$, the next best thing is simulating \eqnsref{eq:fullSME}{eq:fullSME_meas} to generate the measurement trajectories $y(t)$. To explicitly simulate this SME, we adapt recent numerical methods~\cite{Rossi2021} to incorporate estimation and feedback. However, a direct numerical simulation is feasible only for atomic ensembles of small to moderate size. This limitation motivates us to develop an effective model for the dynamics that applies to relevant experiments~\cite{Kuzmich2000,Shah2010,Kong2020}. Moreover, this effective dynamical model can be then used to design the building blocks of an estimation+control scheme. Nevertheless, numerical simulations remain crucial for validating our approach and studying unconditional spin squeezing (see \secref{sec:cond_uncond_spin_squeezing} for more details). In particular, by benchmarking against ``brute-force'' numerics, we ensure that our model can be extrapolated to larger ensemble sizes beyond the reach of direct simulation.

\subsection{Exact numerical solution of the SME} \label{sec:exact_numerical_sol_SME}

Simulating the full ensemble dynamics of a typical optically pumped magnetometer, with sizes ranging in between $N\approx10^{5}\!-\!10^{13}$~\cite{Kuzmich2000,Wasilewski2010,Shah2010,Koschorreck2010,Sewell2012,Kong2020}, is computationally prohibitive, since the dimension of the underlying Hilbert space scales exponentially with $N$, i.e.~as $2^N$ for two-level systems such spin-$1/2$ atoms. Fortunately, the size of the density matrix can be reduced to scale only polynomially with $N$ when the system maintains permutational invariance throughout its evolution, i.e., when any two atoms in the ensemble are indistinguishable. In particular, the complexity of a collection of spin-$1/2$ atoms scales as $\bigO(N^{\,3})$~\cite{Chase2008,Baragiola2010,Shammah2018}, since the density matrix is now a block-diagonal matrix with each block corresponding to a spin-number $j$ ranging from $0$ ($\tfrac{1}{2}$) to $N/2$ for even (odd) $N$. Additionally, if the evolution is exclusively governed by collective operators, which are themselves also permutationally invariant, any state initially living within the totally symmetric subspace (with $j = N/2$), e.g.~CSS, evolves within it, further reducing the complexity to $\bigO(N)$~\cite{Chase2008,Baragiola2010,Shammah2018}. 

Turns out, that the SME in \eqnref{eq:fullSME} preserves the permutational symmetry and, for the case of $\kloc = 0$, it is even sufficient to study the evolution of the density matrix supported by the totally symmetric subspace (with $j = N/2$). Although this symmetry greatly simplifies the simulation of the SME, it still cannot be solved for ensembles with $N\approx10^{5}\!-\!10^{13}$ but rather for ensembles of more moderate sizes of around $N\approx100$ \cite{Shammah2018,rossi_noisy_2020}. For these type of systems, we employ the code of~\citet{Rossi2021} to numerically integrate the SME by exploiting the symmetries described above. In particular, it constructs Kraus operators of the weak measurement at each time-step, while also guaranteeing the positivity of the density matrix~\cite{Rouchon2015,Albarelli2018}. We extend it to perform estimation and control tasks in order to implement the closed-open loop described in \figref{fig:closed-loop_atoms}.

\subsection{Gaussian approximations}

Since the exploitation of permutational symmetry is not sufficient to simulate the system in the experimentally relevant values of $N$, an alternative approach is to derive from \eqnref{eq:fullSME} a system of SDEs of the moments of the collective angular momentum operators $\Jx, \, \Jy$ and $\Jz$. Although the density matrix $\rhoc(t)$ encodes all the statistical information about the system, one may alternatively focus on the evolution of the moments of the system’s observables \cite{carmichael_book,Kubo1963,gardiner_zoller2004quantum}. By taking expectation values and higher-order correlations from the SME, it is possible to derive a set of SDEs for these moments (or cumulants). In general, the evolution of a lower-order moment (such as the mean value of an observable) depends on higher-order moments (such as the variance), resulting in an infinite hierarchy of coupled SDEs. To manage this complexity, one can approximate the higher-order products by neglecting correlations beyond a certain order, effectively truncating the hierarchy \cite{Plankensteiner2022}. In certain special cases, such as when the state of the system is Gaussian, the hierarchy closes; that is, all higher-order moments can be expressed in terms of the first and second moments. Consequently, only a finite number of equations are required to capture the system’s dynamics completely \cite{Geremia2003,jacobs_straightforward_2006}. 

\subsubsection{Linear and Gaussian regime} \label{sec:LG_regime}
\begin{figure}[t!]
    \centering
    \includegraphics[width=0.6\linewidth]{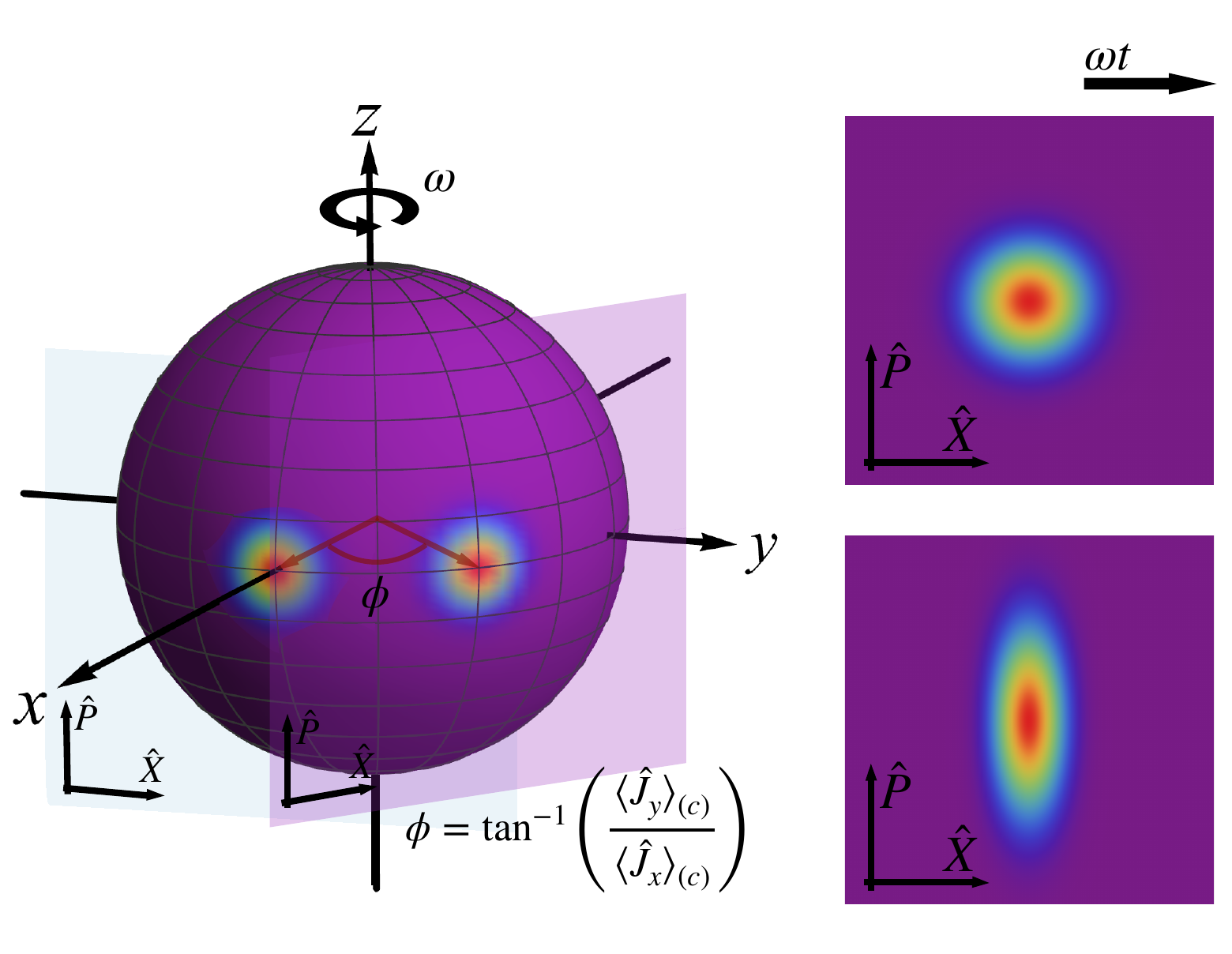}
    \caption[Bloch sphere representation of the atomic state:~Linear and Gaussian and co-moving Gaussian approximations]{\textbf{Bloch sphere representation of the atomic state:~Linear and Gaussian  and co-moving Gaussian approximations.} The 3D-plot on the \emph{left} shows the Wigner function of a CSS aligned along the $x$-axis, represented on a generalized Bloch sphere for $N = 100$ particles. When the atomic ensemble is large ($N\gg1$), the local curvature of the sphere near the peak of the CSS can be approximated by a tangent plane, often referred to as the LG-plane or Holstein-Primakoff plane (in light blue or light pink), in which the effective phase-space quadratures $\position$ and $\momentum$ are defined as in \eqnref{eq:canonical_X&P}. The continuous measurement of the $y$ spin-component induces spin squeezing of the atomic state, which in this planar approximation, corresponds to the (anti)squeezing of the ($\momentum$)$\position$ quadrature (see \emph{right} side). To preserve this Gaussian approximation over longer timescales, the LG-plane is allowed to co-rotate with the spin at the Larmor frequency $\omega$, an approach referred to as the as the CoG approximation.}
    \label{fig:CSS_bloch}
\end{figure}

It is evident from \eqnref{eq:fullSME} that the evolution of the state is not inherently linear, meaning that the atomic state is not guaranteed to remain Gaussian over time. However, under certain approximations, the SME in \eqnref{eq:fullSME} can be reduced to a closed, linear system of SDEs of the first and second moments. Specifically, by neglecting feedback and local noise effects, and assuming the magnetic field $B$ is sufficiently small, then at short enough timescales, $t \lesssim 1/(M + \kcoll)$, we can approximate $\brktc{\Jx(t)}$ with its unconditional average value $\brkt{\Jx(t)}= J \,\ee^{-(M + \kcoll)t/2}$~\cite{Amoros-Binefa2021}, where $J = N/2$. If the $B-$field is not constant and instead follows an OU process like in \eqnref{eq:signal_OUP}, one must ensure that the process obeys the constraints $\chi t \lesssim 1$ and $q_\omega t^3 \lesssim 1$ for any time $t < (M+\kcoll)^{-1}$ (see \appref{sec:UncondJx} for a detailed derivation \footnote{In it we also discuss why the system is unaffected by dephasing acting along the $x$ or $y$ directions}).

Under these approximations, the collective angular momentum vector $\,\Jqvec{J}(t)$ remains predominantly aligned with the $x$-axis, with only small deviations. In that case, and for sufficiently large $N$, the surface of the generalized Bloch sphere can thus be approximated by a two-dimensional phase-plane perpendicular to the vector $\,\Jqvec{J}(t)$~\cite{Geremia2003,Albarelli2017,Amoros-Binefa2021} (see \figref{fig:CSS_bloch} for a depiction). This plane then defines an effective phase space with position and momentum operators given by: 
\begin{equation} \label{eq:canonical_X&P}
    \position \coloneqq \Jy/\sqrt{|\brkt{\Jx(t)}|}, \; \;  \text{and} \; \; \momentum \coloneqq \Jz/\sqrt{|\brkt{\Jx(t)}|},
\end{equation}
which satisfy the canonical commutation relation $[\position,\momentum] \approx i$, as long as $\Jx \approx \left|\brkt{\Jx(t)}\right|\I\,$ for sufficiently large $N$~\cite{Madsen2004,Molmer2004} and $(M+\kcoll) t \lesssim 1$, $q_\omega t^3 \lesssim 1$, and $\chi t \lesssim 1$. With these approximations in place, the SME \eref{eq:fullSME} reduces to a linear set of SDEs for the first and second moments of the quadratures \eref{eq:canonical_X&P}, as well as in the magnetic field $B$~\cite{Madsen2004,Molmer2004}:
\begin{align}
    y(t) \dt &= 2\eta \sqrt{M} \brktc{\Jy(t)} \dt + \sqrt{\eta} \dW, \label{eq:measure_before} \\ 
    \dd \omega(t) &= - \chi \omega(t) \dt + \sqrt{q_\omega} \dW_{\!\omega}, \label{eq:OU_process_before} \\
    \dd \brktc{\Jy(t)} &= \omega(t) \, J \, \ee^{-(M+\kcoll)t/2} \dt + 2\sqrt{\eta M} \Vy(t) \dW, \label{eq:mean_eq_Jy_before} \\
    \dd \Vy(t) &= -4\eta M \Vy(t) \dt + \kcoll \; J^{\,2} \, \ee^{-(M+\kcoll)t} \dt, \label{eq:var_eq_Jy_before}
\end{align}
where we have dropped the sub-index $\trm{(c)}$ because the variance is no longer conditional on the measurement outcomes. In fact, \eqnref{eq:var_eq_Jy_before} is completely decoupled of other state variables and can, therefore, be solved independently (see \appref{sec:VarSol}). Therefore, \eqnsref{eq:measure_before}{eq:mean_eq_Jy_before} define a continuous state-space model that describes both the evolution of the state vector and its relationship to observations (i.e., the measurement vector $\pmb{y}_{\leq t}$. In this case, the state vector is constructed from the variables $\brktc{\Jy(t)}$ and $\omega(t)$ as follows:
\begin{equation}
    \pmb{x}(t) = ( \, \brktc{\Jy(t)} , \omega(t) \, )^\Trans.
\end{equation}
Following the framework of state-space models provided in \secref{sec:state_space_model}, we identify the state and measurement noise vectors as $\dd\pmb{w}(t) = ( \, \dW,\dW_{\!\omega} \, )^\Trans$ and $\dd\pmb{v}(t) = \sqrt{\eta} \, \dW$, respectively. Using these definitions, the system of SDEs given by \eqnsref{eq:measure_before}{eq:mean_eq_Jy_before} can be expressed as:
\begin{align}
  \dd\pmb{x}(t) & = \pmb{F}(t) \, \pmb{x}(t) \, \dt + \pmb{G}(t)  \, \dd\pmb{w}(t),\label{eq:state_system1} \\
  \dd\pmb{y}(t) & = \pmb{H}(t) \, \pmb{x}(t)  \, \dt + \dd\pmb{v}(t), \label{eq:state_system2}
\end{align}
where the matrices are defined as:
\begin{equation}\label{eq:matrices_F_K_B}
  \!\!\pmb{F}(t) = \begin{pmatrix} 0 & J \, \ee^{-r \, t / 2} \\ 0 & -\chi \end{pmatrix}\!, \;\; \pmb{G}(t) = \begin{pmatrix}  2 \sqrt{\eta M} \Vy(t) & 0 \\ 0 & \sqrt{q_\omega} \end{pmatrix}\!, \;\; \pmb{H}(t) =  2\sqrt{\eta M} \begin{pmatrix} 1 & 0 \end{pmatrix}\!,
\end{equation}
with the noise correlations defined as $\EE{\dd\pmb{w}(t) \dd\pmb{w}^\Trans(s)} = \pmb{Q}(t) \, \delta(t-s) \, \dt$, $\EE{\dd\pmb{v}(t) \dd\pmb{v}^\Trans(s)} = \pmb{R}(t) \, \delta(t-s) \, \dt$ and $\EE{\dd\pmb{w}(t) \dd\pmb{v}^\Trans(s)} = \pmb{S}(t) \, \delta(t-s) \, \dt$. Based on the noise vector definitions, the covariance matrix for the system noise is $\pmb{Q}(t) = \I$, the measurement noise has a scalar variance $\pmb{R}(t) = \eta$, and the cross-correlation matrix between the system and measurement is $\pmb{S}(t) = ( \, \sqrt{\eta} \; , \; 0 \, )^\Trans$. Notably, the field and atomic noises are uncorrelated: $\EE{\dW_{\!\omega} \dW} = \EE{\dW \dW_{\!\omega}} = 0$. Furthermore, since $\EE{\dW^2} = \EE{\dW_{\!\omega}^2} =\dt \ge 0$, the covariance matrix $\pmb{Q}(t)$ satisfies semi-positivity (i.e., $\pmb{Q} \ge 0$). Finally, observe that both $\dd\pmb{w}(t)$ and $\dd\pmb{v}(t)$ are partially-correlated Gaussian noises and $\pmb{F}(t) \, \pmb{x}(t)$ and $\pmb{H}(t) \, \pmb{x}(t)$ are both linear functions w.r.t. the state vector. This is consistent with \eqnsref{eq:measure_before}{eq:mean_eq_Jy_before}, which involve Gaussian stochastic terms and are linear with respect to one another. Therefore, a state-space model of the form in \eqnsref{eq:state_system1}{eq:state_system2} (or \eqnsref{eq:measure_before}{eq:var_eq_Jy_before}) is referred to as \emph{linear and Gaussian} (LG)~\cite{Amoros-Binefa2021}. 

\paragraph*{Analytical solution of the spin squeezing parameter in the LG regime}

The dynamics of $\Vy(t)$ can be found analytically by solving \eqnref{eq:var_eq_Jy_before}, since it is a differential equation fully decoupled from the rest. Due to its complexity, the complete analytical solution is presented in detail in \appref{sec:VarSol} \cite{Amoros-Binefa2021}. However, the expression for the variance of $\Jy$ greatly simplifies when dividing it into a short-time and long-time regime:
\begin{subequations}
\begin{numcases}{\Vy(t) =}
     \mrm{V}_{<t^*}(t) =  \frac{J}{2} \, \frac{1+2 J t \kcoll}{1+2 J t M \eta} \ee^{-(M+\kcoll)t/2}, & \text{if\quad $0 \le t \ll t^{*}$}
    \label{eq:varJz_t<<t*}\\
     \trm{V}_{>t^*}(t) =  \frac{J}{2} \, \sqrt{\frac{\kcoll}{\eta M}} \,e^{-(M + \kcoll)t/2}, & \text{if\quad $t \gg t^{*}$}
    \label{eq:varJz_t>>t*}
\end{numcases}
\label{eq:varJz_t}
\end{subequations}
where initially $\Vy(0)=J/2$, and $t^{*} = (2J\sqrt{M\kcoll \eta})^{-1}$ is the transition time between the two regimes. Importantly, note that $t \ll t^*$ implies $2 J t \kcoll \ll \sqrt{\kcoll/(\eta M)}$. Then, if also $\kcoll < \eta M$, we may infer $2 J t \kcoll \ll 1$ and approximate $1 + 2 J t \kcoll \approx 1$ in \eref{eq:varJz_t<<t*}. As a result, we then recover the known noiseless ($\kcoll=0$) solution for the variance within the short-time regime~\cite{Geremia2003}:
\begin{equation}\label{eq:Geremia_VarJz}
  \Vy(t) =
   \frac{J}{2} \, \frac{1+2 J t \kcoll}{1+2 J t M \eta} \ee^{-(M+\kcoll)t/2}
   \approx
   \frac{J}{2+4 J t M \eta},
\end{equation}
despite collective dephasing being present, i.e. $\kcoll>0$. In fact, we prove also in \appref{sec:VarSol} that $\Vy(t)$ is a non-decreasing function at $t\approx0$ if $\kcoll \ge \eta M$. Hence, the noise may be considered negligible at small times \emph{only if} $\kcoll < \eta M$.

Finding an analytical solution for the variance $\Vy(t)$ is crucial because of its close relationship with the spin squeezing parameter of the atomic ensemble~\cite{Ma2011} introduced by~\citet{Wineland1994}. In the experimental setup of \figref{fig:atomic_ensemble}, a continuous measurement performed along $y$ squeezes the variance of $\Jy$, $\Vy(t)$, in detriment of the variance of $\Jz$. Given that we want to use this state to sense small variations in $\omega$, it follows that the state should have maximal polarization along $x$ and minimal variance along $y$ (or maximal squeezing). How closely such a state resembles a CSS pointing along $x$ is therefore quantified by the spin squeezing parameter $\xi_y^{\,2}$~\cite{Kitagaba_Ueda_SSS,Wineland1994,Pezze2018RMP} (see \secref{sec:spin-squeezing_intro}):
\begin{equation}
    \xi^{\,2}_{y}(t) \coloneqq \!\frac{\Vy(t)}{\brkt{\Jx(t)}^2} \left( \frac{\Vy^\CSS}{\brkt{\Jx}_\CSS^2} \right)^{-1}
     = \frac{N\;\Vy(t)}{\brkt{\Jx(t)}^2},
    \label{eq:spin_squeez_par}
\end{equation}
since $\xi^{\,2}_{y}(t) < 1$ indicates a gain in squeezing relative to the CSS~\cite{Wineland1994}, whereas $\xi^{\,2}_{y}(t)~\ge~1$ implies the absence of spin squeezing and, consequently, the inability to certify multi-particle entanglement~\cite{Pezze2018RMP}.  

\begin{figure}[t!]
\centering
\includegraphics[width = \textwidth]{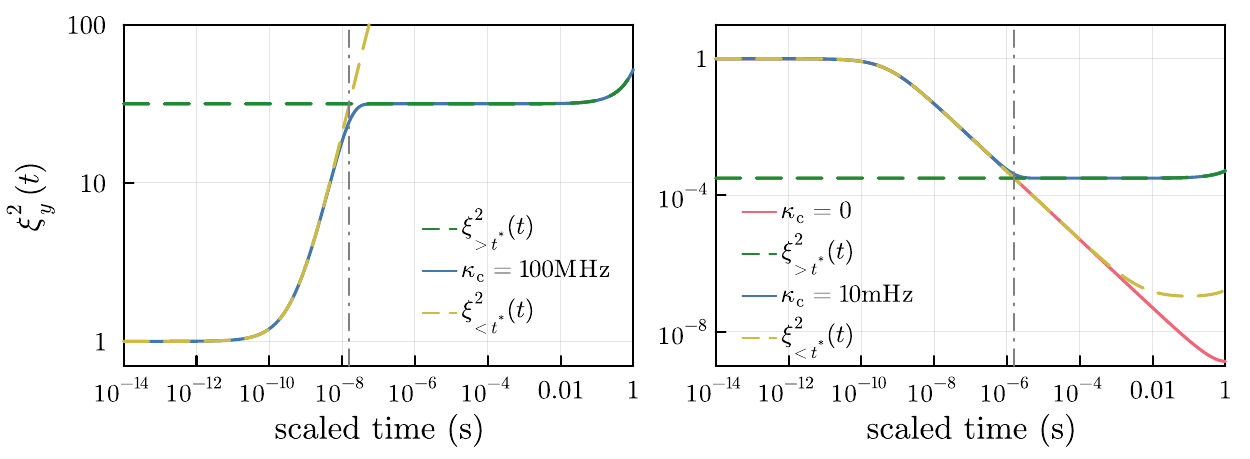}
\caption[Strong and weak dephasing regimes for the squeezing parameter]{\textbf{Strong and weak dephasing regimes for the squeezing parameter.} Both figures plot the \emph{squeezing parameter} $\xi_y^{\,2}(t)$ as defined in equation \eqref{eq:spin_squeez_par} as a function of scaled time $t_s = (M + \kcoll)t$, for the case of $\kcoll = \SI{10}{\milli\hertz} < M = \SI{100}{\kilo\hertz}$ \emph{(right)} and $\kcoll = \SI{100}{\mega\hertz} > M = \SI{100}{\kilo\hertz}$ \emph{(left)}. The exact function $\xi_y^{\,2}(t)$ (solid, blue) is compared with its two different regimes $\xi^{\,2}_{<t^*}(t)$ and $\xi^{\,2}_{>t^*}(t)$ (dashed, yellow and green, respectively), as well as the noiseless solution when $\kcoll < M$ (dashed, red). The gray dash-dot line indicates the inflection point $t^{*}$, where the transition between the two regimes occurs. The other parameters used to generate the plots are $J = 10^9$, $\gmr = \SI{1}{\kilo\hertz/\milli\gaussunit}$, and $\eta = 1$. The time has been rescaled to $t_S = (M + \kcoll) t$ so that its range is therefore limited to $t_S \in [0,1]$, where the LG approximation holds.}
\label{fig:squeezing}
\end{figure}

Next, given that $\Vy(t)$ has two distinct regimes, as presented in  \eqnref{eq:varJz_t}, the spin squeezing parameter can be similarly split as
\begin{subequations}
\label{eq:squeezing_cases}
\begin{numcases}{\xi_y^{\,2}(t)
    \approx\frac{2\Vy(t)}{J \ee^{-(M+\kcoll)t}}
   	=}
    \xi^{\,2}_{<t^*}(t) = \frac{1+2 J t \kcoll}{1+2 J t M \eta} \ee^{(M+\kcoll)t/2}, & \text{if\quad $0 \le t \ll t^{*}$,}\\
    \xi^{\,2}_{>t^*}(t) = \sqrt{\frac{\kcoll}{\eta M}}\ee^{(M+\kcoll)t/2}, & \text{if\quad $t\gg t^{*}$.}
\end{numcases}
\end{subequations}%
In \fref{fig:squeezing} we present explicitly the exact dynamics of the squeezing parameter \eref{eq:spin_squeez_par} in the LG regime for the two important cases:~(\textit{right}), when $\kcoll<\eta M$ and the spin squeezing ($\xi^{\,2}_y(t)<1$) occurs within a finite-time window (see \fref{fig:squeezing} (\textit{right}));~and (\textit{left}), when $\kcoll \geq \eta M$ for which spin squeezing is forbidden (i.e. $\xi^{\,2}_y(t)\geq 1$, as shown in \fref{fig:squeezing} (\textit{left})). In both cases, it is evident that the exact solution for $\xi^{\,2}_y(t)$ very closely follows the two-regime behavior in \eref{eq:squeezing_cases}, with a clear transition at $t\approx t^*$. Moreover, as seen explicitly from the two-regime solution \eref{eq:squeezing_cases} (and the exact solution that can be straightforwardly derived based on the explicit variance evolution of \appref{sec:VarSol}), the dynamics of the squeezing parameter are specified solely by the properties of the continuous measurement ($\eta$ and $M$), collective decoherence ($\kcoll$), and the number of atoms ($\,J=N/2$).

\subsubsection{Co-moving Gaussian approximation} \label{sec:Co-moving_Gaussian}
In practical magnetometers~\cite{Kuzmich2000,Wasilewski2010,Shah2010,Koschorreck2010,Sewell2012,Kong2020}, the atomic spin must precess multiple times during detection to collect a sufficient signal. Since the LG approximation assumes that the angular momentum vector does not precess, it is no longer applicable. To then maintain an approximately Gaussian description of the system at all times, we allow the LG-plane (see \fref{fig:CSS_bloch}) to precess with the mean angular-momentum vector $\brkt{\,\Jqvec{J}(t)}$ at the frequency $\omega$~\cite{Munoz-Arias2020}. This approach is referred to as the \emph{co-moving Gaussian} (CoG) approximation \cite{Amoros-Binefa2024}, and it holds under the following conditions:~(1) the ensemble size is large, i.e., $N \gg 1$; and (2)~the squeezing arising from the continuous measurement is not strong enough to wrap the Wigner function around the Bloch sphere.

Specifically, by analyzing the conditional evolution in the Heisenberg picture for the mean angular momenta $\brktc{\Jindex{\alpha}(t)}, \;\alpha = x,y,z$, and their corresponding covariance matrix $\mathrm{C}^{\cc}_{\alpha \beta}(t) \coloneqq \frac{1}{2} \left(\brktc{\{\Jindex{\alpha}(t),\Jindex{\beta}(t)\}}-2\brktc{\Jindex{\alpha}(t)}\brktc{\Jindex{\beta}(t)} \right)$ with diagonal elements $\mathrm{V}^{\cc}_{\alpha}(t)\coloneqq \mathrm{C}^{\cc}_{\alpha\alpha}(t)$ ($\alpha,\beta = x, y,z$), we derive a set of coupled SDEs based on the SME~\eref{eq:fullSME} (details in \appref{ap:Ito}). Note that this set of coupled SDEs is nothing but a cumulant expansion approximation up to 2nd-order \cite{Kubo1963,Plankensteiner2022}, where the 3rd-order moments in \appref{ap:Ito} are simply the 3rd-order cumulants and $\Cxy$ is a linear combination of the 2nd-order cummulants, i.e. $\Cxy^{\cc} = \frac{1}{2} \left(\brktc{\Jx\Jy} + \brktc{\Jy\Jx}\right)$.

For simplicity, the explicit time dependence of all the quantities is omitted below:
\begin{subequations}
\label{eq:dynamical_model}
\begin{align}  
    \dd\brktc{\Jx} &\!=\!  - (\omega(t) \!+\! u(t))  \brktc{\Jy}  \dt \!- \frac{1}{2}(\kcoll \!+\! 2 \kloc \!+\! M)  \brktc{\Jx} \dt \!+\! 2\sqrt{\eta M}  \Cxy^{\cc} \, \dW \\ 
    \dd\brktc{\Jy} &\!=\! (\omega(t) \!+\! u(t))  \brktc{\Jx} \dt \!- \frac{1}{2}(\kcoll \!+\! 2\kloc)  \brktc{\Jy} \dt \!+\! 2\sqrt{\eta M}  \Vy^{\cc} \, \dW \label{eq:dJy}\\
    \dd \Vx^{\cc} &\! = \! - 2 (\omega(t) \! + \! u(t))  \Cxy^{\cc}\dt \! + \! \kcoll \! \left( \! \Vy^{\cc} \!+\!  \brktc{\Jy}^2 \!-\! \Vx^{\cc} \!\right)\!\dt \!+\! \kloc \!\left( \!\frac{N}{2} \!-\!2 \Vx^{\cc} \!\right)\!\dt  \nonumber \\
    &\!+\! M\! \left( \! \Vz^{\cc} \!-\! \Vx^{\cc} \!-\! 4 \eta {\Cxy^{\cc}}^2 \!\right)\! \dt \label{eq:dVx}\\
    \dd\Vy^{\cc} &\!=\! 2 (\omega(t) \!+\! u(t))  \Cxy^{\cc}\dt \!+\! \kcoll \!\left(\!  \Vx^{\cc} \!+\!  \brktc{\Jx}^2 \!-\!  \Vy^{\cc}\!\right)\! \dt \!+\! \kloc \!\left(\! \frac{N}{2} \!-\!2 \Vy^{\cc} \!\right)\! \dt \nonumber \\
    &\!-\! 4 \eta M  {\Vy^{\cc}}^2\dt \label{eq:dVy}\\
    \dd\Vz^{\cc} &\!=\!  M \!\left( \!\Vx^{\cc} \!+\!  \brktc{\Jx}^2 \!-\! \Vz^{\cc}\! \right)\! \dt \label{eq:dVz}\\
    \dd\Cxy^{\cc} &\!=\! (\omega(t) \!+\! u(t)) \!\left(\!\Vx^{\cc} \!-\!  \Vy^{\cc}\!\right)\! \dt \!-\! \kcoll \!\left(\!2\Cxy^{\cc} \!+\! \brktc{\Jx}\brktc{\Jy}\!\right)\!\dt\!-\! 2\kloc \Cxy^{\cc} \dt \nonumber \\
    &\!-\! \frac{1}{2} M \Cxy^{\cc} \left( 1 + 8\eta \Vy^{\cc}\right)\!\dt \label{eq:dCxy}\\
    \dd\omega &\!=\! -\chi \omega(t) \, \dt + \sqrt{q_\omega} \, \dW_{\!\omega},
\end{align}
\end{subequations}
where we have ignored all the (stochastic) contributions that involve the 3rd-order moments (as discussed in \appref{ap:Ito}). 

This system of SDE gives us a state-space model, a continuous-time version of the discrete-time state-space model introduced in \secref{sec:state_space_model}. If we write \eqnref{eq:dynamical_model} in Langevin form instead of It\^{o}, then:
\begin{align}
    \dot{\pmb{x}}(t) &= \pmb{f}\left(\pmb{x}(t),u(t),\pmb{q}(t),t\right), \label{eq:model_state_eq} \\
    y(t) &= h(\pmb{x}(t),r(t),t), \label{eq:model_measurement_eq}
\end{align}
where $\pmb{x}(t) = ( \brktc{\Jx},\brktc{\Jy},\Vx^{\cc},\Vy^{\cc},\Vz^{\cc},\Cxy^{\cc},\omega )^\TT$ is the state vector, $u(t)$ is the control field, and $\pmb{q}(t)$ and $r(t)$ are the state and measurement noises. The function $\pmb{f}\left(\pmb{x}(t),u(t),\pmb{q}(t),t\right)$ is nonlinear w.r.t. $\pmb{x}(t)$ and given by the SDE system in \eqnref{eq:dynamical_model}:
\begin{align} \label{eq:nonlinear_fun_ch5}
    &\pmb{f}\,(\pmb{x}(t),u(t),\pmb{q}(t),t)  =  \\
    &\left(
    \begin{array}{c}
    - (\omega(t) \!+\! u(t))  \brktc{\Jy}  \!- \frac{1}{2}(\kcoll \!+\! 2 \kloc \!+\! M)  \brktc{\Jx} \!+\! 2\sqrt{\eta M}  \Cxy^{\cc} \, \xi    \\ 
    (\omega(t) \!+\! u(t))  \brktc{\Jx} \!- \frac{1}{2}(\kcoll \!+\! 2\kloc)  \brktc{\Jy} \!+\! 2\sqrt{\eta M}  \Vy^{\cc} \, \xi   \\ 
    - 2 (\omega(t) \! + \! u(t))  \Cxy^{\cc} \! + \! \kcoll \! \left( \! \Vy^{\cc} \!+\!  \brktc{\Jy}^2 \!-\! \Vx^{\cc} \!\right)\! \!+\! \kloc \!\left( \!\frac{N}{2} \!-\!2 \Vx^{\cc} \!\right)\! \!+\! M\! \left( \! \Vz^{\cc} \!-\! \Vx^{\cc} \!-\! 4 \eta {\Cxy^{\cc}}^2 \!\right)\! \\ 
    2 (\omega(t) \!+\! u(t))  \Cxy^{\cc} \!+\! \kcoll \!\left(\!  \Vx^{\cc} \!+\!  \brktc{\Jx}^2 \!-\!  \Vy^{\cc}\!\right)\! \!+\! \kloc \!\left(\! \frac{N}{2} \!-\!2 \Vy^{\cc} \!\right)\!  \!-\! 4 \eta M  {\Vy^{\cc}}^2 \\
     M \!\left( \!\Vx^{\cc} \!+\!  \brktc{\Jx}^2 \!-\! \Vz^{\cc}\! \right)    \\
    (\omega(t) \!+\! u(t)) \!\left(\!\Vx^{\cc} \!-\!  \Vy^{\cc}\!\right)\! \!-\! \kcoll \!\left(\!2\Cxy^{\cc} \!+\! \brktc{\Jx}\brktc{\Jy}\!\right)\!-\! 2\kloc \Cxy^{\cc} \!-\! \frac{1}{2} M \Cxy^{\cc} \left( 1 + 8\eta \Vy^{\cc}\right) \\
     -\chi \omega(t) \,  + \sqrt{q_\omega} \, \xi_\omega
    \end{array}
  \right). \nonumber
\end{align}
On the other hand, $h(\pmb{x}(t),r(t),t)$ is actually a linear function, since the measurement model in \eqnref{eq:fullSME_meas} is linear w.r.t. the state $\pmb{x}(t)$:
\begin{align} \label{eq:meas_model_CGa}
    y(t) = 2 \eta \sqrt{M} \brktc{\Jy(t)} + \sqrt{\eta} \xi = h(\pmb{x}(t),r(t),t) = \pmb{H} \, \pmb{x}(t) + r(t) 
\end{align}
where $\pmb{H} = 2 \eta \sqrt{M}\left(0 \; 1 \; 0 \; 0 \; 0 \; 0 \; 0 \right)$ and the state and measurement noises are:
\begin{align}
    \pmb{q}(t) &= \begin{pmatrix}
        \xi \\
        \xi_\omega
    \end{pmatrix}, & &\trm{and} & r(t) &= \sqrt{\eta}\xi, \label{eq:noises_syst_and_meas}
\end{align}
where $\xi$ and $\xi_\omega$ are Langevin noises defined as $\xi \coloneqq \dW/\dt$ and $\xi_\omega \coloneqq \dW_{\!\omega}/\dt$, respectively. 

In order to validate the CoG approximation, we compare it against the exact SME solution for simulatable ensemble sizes. Our results indicate that while the first and second moments of the system (e.g., $\brktc{\Jx}$, $\brktc{\Jy}$, and $\Vy^{\cc}$) show good agreement between the exact and approximate approaches, the real-time estimation of the Larmor frequency $\est{\omega}(t)$ using the CoG-based dynamics aligns even more closely with the one generated by the exact SME. This accurate generation of an estimate of $\omega(t)$ without having to resort to a full SME simulation, reinforces the usefulness of the proposed CoG approximation, especially in the experimentally relevant regimes of large $N$. A comprehensive error analysis is presented in \appref{sec:verification_CoG}.

\section{Estimation and control} \label{sec:estim_contr_chap5}

So far, we have outlined both the physical dynamics and numerical solution of the system—that is, the atomic ensemble within the optical atomic magnetometer. In particular, we discussed two complementary simulation strategies: one that exploits the permutational symmetry of the ensemble to numerically solve the SME for moderate system sizes, and another based on Gaussian approximations that captures the essential dynamics in larger, experimentally relevant regimes. However, as illustrated in \figref{fig:closed-loop_atoms}, this represents only part of the complete picture. Our next goal is to explore state estimation and control methods that use the real-time measurement data to monitor and steer the system continuously, thereby enhancing the performance of the magnetometer in tracking the magnetic field.

\subsection{Kalman filter} \label{sec:KF_for_weak_field}

For an atomic magnetometer operated in the LG regime, like the one in \secref{sec:LG_regime}, the optimal estimator $\pmb{\est{x}}(t) = ( \, \brktc{\estJy(t)} , \est{\omega}(t) \,)^\Trans$ is given by the \emph{correlated} Kalman filter (KF) \cite{crassidis2011optimal} (see also \secref{sec:correlated_KF}):
\begin{align}
    \dot{\est{\pmb{x}}}(t) &= \pmb{F}(t)\est{\pmb{x}}(t) \!+\! \pmb{K}(t)\!\left( y(t) \!-\! \pmb{H}(t)\est{\pmb{x}}(t) \right), \label{eq:Kalman_Bucy_ch5} \\
    \pmb{K}(t) &= \left( \pmb{\Sigma}(t) \, \pmb{H}^{\Trans}(t) \!-\! \pmb{G}(t)  \pmb{S}(t) \right)\pmb{R}^{-1}(t), \label{eq:Kalman_Gain_ch5}\\
    \dot{\pmb{\Sigma}}(t) &=  \pmb{F}(t) \pmb{\Sigma}(t) \!+\! \pmb{\Sigma}(t) \pmb{F}^{\,\Trans}\!(t) \!-\! \pmb{K}(t)\pmb{R}(t)\pmb{K}^\Trans\!(t) \!+\! \pmb{G}(t) \pmb{Q}(t) \pmb{G}^\Trans\!(t), \label{eq:Riccati_equation_ch5}
\end{align}
where the differential equation for the covariance matrix $\pmb{\Sigma}(t) = \EE{\pmb{x}(t) - \pmb{\est{x}}(t))(\pmb{x}(t) - \pmb{\est{x}}(t))^{\!\Trans}}$, the Riccati equation, yields the (minimal) estimator error of $\pmb{x}(t)$, with initial conditions $\pmb{\Sigma}(0) = \text{Diagonal}\left[0,\sigma_0^2 \right]$, where $\sigma_0^2$ is the variance of the prior distribution $p(\omega_0)$. The rest of the matrices, $\pmb{F}(t),\;\pmb{G}(t),\;\pmb{H}(t),\;\pmb{Q}(t),\;\pmb{S}(t)$ and $\pmb{R}(t)$ are provided by the model in \eqnref{eq:matrices_F_K_B}, and the photocurrent $y(t)$ is given by  \eqnref{eq:state_system2}.

\paragraph*{Solution in the absence of decoherence and field fluctuations}
In the absence of dephasing and field fluctuations ($\kcoll = \kloc = 0$, and $q_\omega = \chi  = 0$), the Ricatti equation in \eqref{eq:Riccati_equation_ch5} can be explicitly solved~\cite{Geremia2003} by redefining the covariance matrix as $\pmb{\Sigma}(t) = \pmb{Y}(t) \pmb{X}(t)^{-1}$, and thus decoupling the Riccati equation as:
\begin{align}
    \label{eq:X2main} \dot{\pmb{X}\,}\!(t) = \ &  \!-\!\Big( \!\pmb{F}(t) -  \pmb{G}(t) \pmb{S}(t) \pmb{R}^{-1}(t) \pmb{H}(t) \Big)^{\!\!\Trans} \pmb{X}(t)  +  \pmb{H}^\Trans\!(t) \pmb{R}^{-1}(t) \pmb{H}(t) \pmb{Y}(t), \\
    \label{eq:Y2main} \dot{\pmb{Y}}(t) = \ &  \Big( \!\pmb{F}(t) \!-\! \pmb{G}(t) \pmb{S}(t) \pmb{R}^{-1}(t) \pmb{H}(t)\!\Big) \!\pmb{Y}(t) \!+\! \pmb{G}(t)\!\Big(\! \pmb{Q}(t) \!-\! \pmb{S}(t) \pmb{R}^{-1}(t) \pmb{S}^\Trans\!(t) \!\Big) \pmb{B}^\Trans\!(t) \pmb{X}(t),
\end{align}
with initial conditions $\pmb{X}(0) = \I$ and $\pmb{Y}(0) = \pmb{\Sigma}(0)$. Since $\kcoll = 0$, the solution to the variance differential equation in \eqnref{eq:var_eq_Jy_before} is 
\begin{equation}
    \Vy(t) = \frac{J}{2+4M\eta \, J \, t},
\end{equation}
as shown in \appref{sec:VarSol}. Moreover, since no fluctuations of the field are being considered, the volatility matrix is now simply $\pmb{G}(t) = \Diag{2\sqrt{\eta M} \Vy(t), 0}$. Under these conditions, the decoupled system of differential equations introduced in \eqnsref{eq:X2main}{eq:Y2main}, given initial conditions $\pmb{X}(0) = \I$ and $\pmb{Y}(0) = \Diag{0,\sigma_0^2}$, can be analytically solved. The solution for $\pmb{\Sigma}_{\omega,\omega}$, which for a LG system matches the aMSE, $\EE{\Delta^2 \est{\omega}(t)}$, is:
\begin{align} \label{eq:Geremia_sol_full}
    \EE{\Delta^2 \est{\omega}^\text{HS}(t,\sigma_0)} = \mrm{A} \frac{(1 + 2\, J \,Mt\eta)}{a(t) \, \ee^{-M t} + 4 \, (1+4\, J \,\eta) \,  \ee^{-Mt/2} + b(t)},
\end{align}
where 
\begin{align}
  \mrm{A} & = M^2 / (16\eta \, J ^2), \\
  a(t) & = -(1 + 2\eta \, J \,(4 + Mt)), \\
  b(t) & = \frac{M^2}{16 \eta \, J^{\,2} \sigma^2_0} + \frac{M^3 t}{8 \, J \, \sigma^2_0} + (Mt-3) + 2 \eta \, J \, (Mt-4) \\
  & = \frac{M^2}{16 \eta \, J^{\,2} \sigma^2_0} + \frac{M^3 t}{8 \, J \, \sigma^2_0} + c(t),
\end{align}
with the subscript $\text{HS}$ highlighting the super-classical scaling of the error in both $t$ and $N$ (Heisenberg scaling). Note that for $\sigma^2_0 \rightarrow \infty$, the functions $b(t)$ and $c(t)$ are equal and $\EE{\Delta^2 \est{\omega}^\text{HS}(t,\infty)}$ matches consistently the solution of \citet{Geremia2003}. In order to highlight its non-classical scaling with $t$ and $N$, the aMSE can be approximated as
\begin{align} \label{eq:HS_approx}
    \EE{\Delta^2 \est{\omega}^\text{HS}(t,\infty)} \approx
    \begin{cases}
             \dfrac{1}{N^{\,2}} \dfrac{3}{\eta M t^3}, & \textrm{for } t\ll (NM)^{-1}, \\
             \dfrac{4}{N^{\,2}} \dfrac{3}{\eta M t^3}, & \textrm{for } (NM)^{-1}\ll t \;(< M^{-1}),
    \end{cases}
\end{align}
when assuming no initial knowledge of the field ($\sigma^2_0 \rightarrow \infty$). The first term in \eref{eq:HS_approx} is derived by Taylor expanding to leading order in time the solution $\EE{\Delta^2 \est{\omega}^\text{HS}(t,\infty)}$. The second term is obtained by expanding $\EE{\Delta^2 \est{\omega}^\text{HS}(t,\infty)}$ to leading order around $(N M t)^{-1}$.

\paragraph*{Steady state solution of the Kalman filter}

The KF of \eqnsref{eq:Kalman_Bucy_ch5}{eq:Riccati_equation_ch5} which tracks an OU process has an analytical steady solution, both in the case of $\chi = 0$ and $\chi \neq 0$. Below we discuss the steady state solution for the case of a pure Wiener noise (i.e. $\chi = 0$), with the general solution for $\chi \neq 0$ presented in \appref{sec:SSS}.

To find the steady state solution of the Riccati differential equation one must set $\dd\pmb{\Sigma}(t) = 0$, which is greatly simplified by noting that the variance of $\Jy$ at $t \gg t^*$  is $\mrm{V}_{>t^*}(t)$ \eqref{eq:varJz_t>>t*}. Then, the steady state solution for $\pmb{\Sigma}_{\omega,\omega}$, which in the case of a KF coincides with the minimal aMSE $\EE{\Delta^2\est{\omega}(t)}$, can be shown to be:
\begin{align} \label{eq:ss_field_sol}
    \EE{\Delta^2\est{\omega}^\mrm{SS}} =
     \left( q_\omega \, \kcoll + \frac{2}{N} \sqrt{\frac{q_\omega^3}{M \eta}} \right)^{1/2},
\end{align}
where second term, which survives in the $\kcoll\to0$ limit, had been derived previously in \refcite{Stockton2004}. By accounting for collective dephasing in our analysis, an additional and important term appears, which dominates for large $N$. Namely, for
\begin{equation}
  N > \frac{2}{\kcoll} \sqrt{\frac{q_\omega}{\eta M}},
\end{equation}
the steady state solution of the Riccati equation simplifies to 
\begin{equation}
    \EE{\Delta^2 \est{\omega}^{\mrm{SS}}} = \sqrt{q_\omega \kcoll} \quad \text{for} \quad N > \frac{2}{\kcoll} \sqrt{\frac{q_\omega}{\eta M}},
\end{equation}
which coincides with the form of the CS limit for an OU process \eqref{eq:CSlim_SS} when $t\gtrsim \sqrt{\kcoll/q_\omega}$, i.e.:
\begin{equation}\label{eq:ss_to_CS}
  \EE{\Delta^2\est{\omega}^\mrm{SS}} = \lim_{t\to\infty} \mrm{V}_{\!\mrm{CS}}^{\sigma_0}(t)  = \sqrt{q_\omega \kcoll}, \;\; \text{for} \; t \gtrsim \sqrt{\frac{\kcoll}{q_\omega}} \; \text{and} \; N \gg \frac{2}{\kcoll}\sqrt{\frac{q_\omega}{\eta M}}.
\end{equation}
Hence, this proves that the chosen set of measurements and initial state, i.e. a continuous homodyne measurement and CSS, give us the best possible precision in the steady state when the atomic ensemble is large (i.e. $N \gg 1$, which is also required for the LG approximation to hold).

\subsection{Extended Kalman filter}

When our system is no longer LG but rather described by a nonlinear dynamical model like the one in \secref{sec:Co-moving_Gaussian}, we need an estimator capable of handing nonlinearities, such as an extended Kalman filter (EKF). As discussed in \secref{sec:EKF_theory}, an EKF estimates the state vector $\pmb{x}(t)$ in real time from the noisy measurement record using a state-space model defined by \emph{nonlinear} state and measurement functions, $\pmb{f}$ and $\pmb{h}$, as well as their associated noise vectors. Namely, at each time $t$, the EKF produces the state estimate $\est{\pmb{x}}(t)$ and its error covariance $\pmb{\Sigma}(t)$ by integrating the differential equations along the photocurrent record $\pmb{y}(t) = \{y(t)\}_{\leq t}$~\cite{crassidis2011optimal}:
\begin{subequations}
\label{eq:EKF_alg}
	\begin{align}
	    \dot{\est{\pmb{x}}}(t) &= \pmb{f}(\est{\pmb{x}}(t),u(t),0,t) + \pmb{K}(t) (y(t)-h(\est{\pmb{x}}(t),0,t)), 
	    \label{eq:EKF_dyn}\\
	    \dot{\pmb{\Sigma}}(t) &= (\pmb{F}(t) - \pmb{G}(t) \pmb{S} \pmb{R}^{-1} \pmb{H})\pmb{\Sigma}(t) + \pmb{\Sigma}(t)(\pmb{F}(t) - \pmb{G}(t) \pmb{S} \pmb{R}^{-1} \pmb{H})^\TT + \nonumber\\
	    & \; \; \; \, + \pmb{G}(t)(\pmb{Q} - \pmb{S} \pmb{R}^{-1} \pmb{S}^\TT) \pmb{G}(t)^\TT - \pmb{\Sigma}(t) \pmb{H}^\TT \pmb{R}^{-1} \pmb{H} \pmb{\Sigma}(t),
	    \label{eq:Sigma_dyn}
	\end{align}
\end{subequations}
where in the state update-predict equation for the estimate \eqref{eq:EKF_dyn}, the term
\begin{align}
    y(t) - h(\est{\pmb{x}}(t),0,t),
\end{align}
represents the innovation, i.e. the difference between the actual measurement (provided by the model in \eqnref{eq:meas_model_CGa}) and the predicted measurement based on the current state estimate. This innovation is scaled by the Kalman gain  
\begin{align}
    \pmb{K}(t) \coloneqq (\pmb{\Sigma}(t) \pmb{H}^\TT - \pmb{G}(t) \pmb{S}) \pmb{R}^{-1},
\end{align}
which couples the state and covariance $\pmb{\Sigma}(t)$ in \eqnref{eq:Sigma_dyn}. The matrices $\pmb{F}(t)$, $\pmb{G}(t)$, and $\pmb{H}$ are the Jacobians of the nonlinear functions, determined by the dynamical model in \eqnref{eq:dynamical_model}. Specifically, we have: 
\begin{align}
    \pmb{F}(t) \coloneqq \nabla_{\pmb{x}} \, \pmb{f} \,|_{(\est{\pmb{x}},u,0)}, \quad \pmb{G}(t) \coloneqq \nabla_\xi \, \pmb{f} \, |_{\est{\pmb{x}}} \quad \text{and} \quad H \coloneqq \nabla_{\pmb{x}} h,
\end{align}
with their exact form provided in \secref{sec:exact_forms_FGH_OUP}. Because the entries of these matrices depend on the current state estimate $\est{\pmb{x}}(t)$, they must be recomputed at each EKF iteration (see \secref{sec:EKF_theory} for an introduction to the EKF). In contrast, the noise matrices are predetermined: 
\begin{align}
    \!\!\!\!\pmb{Q} \coloneqq \EE{\pmb{q}(t) \, \pmb{q}^\TT\!(t)} = \I, \;\; R \coloneqq \EE{r^2(t)} = \eta, \;\; \text{and} \;\; \pmb{S} \coloneqq \EE{\pmb{q}(t) r(t)}= (\sqrt{\eta},\, 0)^\TT.
\end{align}
The initial estimates, $\est{\pmb{x}}(0)$, and its error covariance, $\pmb{\Sigma}(0) = \EE{\Delta^2 \est{\pmb{x}}(0)}$, are chosen based on the assumed prior distribution~\cite{crassidis2011optimal}.

\subsection{Linear-quadratic regulator} \label{sec:LQR_ch5}

A straightforward approach to counteracting atomic precession is to directly compensate the Larmor frequency by setting 
\begin{equation}
    u(t) = -\est{\omega}(t),
\end{equation}
which we refer to as \emph{field compensation}. In principle, this should cancel the precession. In practice, however, the EKF estimate $\est{\omega}(t)$ is only approximate, so residual errors accumulate over time and the control becomes suboptimal, leading to imperfect cancellation that degrades long-term performance. 

To improve on this naive strategy, we take inspiration from the linear-quadratic regulator (LQR) and linear-quadratic Gaussian (LQG) controller introduced in \secref{sec:LQR} and \secref{sec:LQG_controller}, respectively. Our control objective is to stabilize the collective angular momentum along the $x$-axis, perpendicular to the probe direction. If the controller succeeds in doing so, the system enters a LG regime in which linearized dynamics apply. Within that regime the LQR is the optimal feedback law by construction. Strictly speaking, our system does not start in the LG regime, so optimality is not guaranteed. Nevertheless, if there exists a control input that drives the system into this regime, then LQR is the correct controller to maintain it there. This motivates our approach: we apply the LQR control law even outside the LG approximation, relying on its robustness together with EKF state estimates to steer the system towards the $x$-axis and achieve the performance bounds analyzed in \chapref{chap:bounds}.

Assuming LG dynamics hold, the system can be described by the reduced state vector
\begin{equation}
    \pmb{z}(t) = (\brkt{\Jy}, \omega)^\TT,
\end{equation}
with dynamics, given in \eqnsref{eq:state_system1}{eq:matrices_F_K_B}, as~\cite{Geremia2003,Amoros-Binefa2021}:
\begin{align} 
\label{eq:l&g_system}
    \dot{\pmb{z}}(t) = \pmb{A} \, \pmb{z}(t) + \pmb{B} \, u(t) + \pmb{G}(t) \, \pmb{q}(t),
\end{align} 
with
\begin{equation}\label{eq:matrices_A_B_G}
  \!\!\pmb{A}(t) = \begin{pmatrix} 0 & J \\ 0 & -\chi \end{pmatrix}\!, \;\; \pmb{G}(t) = \begin{pmatrix}  2 \sqrt{\eta M} \Vy(t) & 0 \\ 0 & \sqrt{q_\omega} \end{pmatrix}\!, \;\; \pmb{B}(t) =  \begin{pmatrix} J \\ 0 \end{pmatrix}\!,\;\;\pmb{q} = \begin{pmatrix}
      \xi \\ \xi_\omega
  \end{pmatrix},
\end{equation}
where $\xi$ and $\xi_\omega$ are Gaussian white noises. \footnote{Note that in the LG regime, the variance of $\Jy$ is a deterministic function that can be computed analytically~\cite{Stockton2004,Amoros-Binefa2021} (see \appref{sec:VarSol})}. Given the state model above, the LQR design minimizes a \emph{quadratic} cost~\cite{crassidis2011optimal,Stockton2004} (see for more details \secref{sec:LQG_controller}):
\begin{align}
    I(u) &= \EE{\int_0^\infty\!\!\dt\;
    \left[\pmb{z}^\TT(t) \pmb{P} \, \pmb{z}(t) + u(t) V u(t) \right]}
    \label{eq:quad_cost} \\
    &=  \EE{\int_0^\infty \!\!\dt \; 
    \left[p_J \brktc{\Jy}^2 + p_\omega \, \omega^2 + \nu \, u^2(t) \right]},
    \label{eq:quad_cost_ex} 
\end{align}
with the diagonal state penalty $P\ge0$ and scalar control weight $V>0$ defined as~\cite{Stockton2004}:
\begin{align}
    \pmb{P} = \begin{pmatrix}
        p_J & 0 \\ 0 & p_\omega
    \end{pmatrix} \quad \text{and} \quad V = \nu,
\end{align}
with $\nu>0$ as well as $p_J,p_\omega\ge0$, so that deviations in both $\brktc{\Jy}$ and $\omega(t)$ are penalized. As stated in \theoremref{thm:lqg}, the optimal control law that minimizes the cost function is given by:
\begin{align}
    u(t) &= -\pmb{K}_C \, \est{\pmb{z}}(t), \label{eq:LQR_control_law}\\
    \pmb{K}_C &\coloneqq V^{-1} \pmb{B}^\TT \pmb{\Lambda} , \label{eq:gain_matrix}\\
    0 &= \pmb{A}^\TT \pmb{\Lambda} + \pmb{\Lambda} \, \pmb{A} + \pmb{P} - \pmb{\Lambda} \, \pmb{B} V^{-1} \pmb{B}^\TT \pmb{\Lambda}, \label{eq:control_Riccati}
\end{align}
where the optimal control field $u(t)$ is linearly related to the state estimator $\est{\pmb{z}}(t)$ by the gain matrix $\pmb{K}_C$. This gain matrix, defined in \eqnref{eq:gain_matrix}, depends on the matrix $\pmb{\Lambda}$, which is the steady state solution of the algebraic Riccati equation (ARE) in \eqnref{eq:control_Riccati}. We use the algebraic form of the Riccati equation, rather than its differential form, because the control design is intended for the infinite-horizon case.

For the time-independent matrices $\pmb{A}$ and $\pmb{B}$ of \eqnref{eq:l&g_system}, the ARE admits an analytic solution~\cite{Stockton2004}. A detailed derivation in \appref{sec:LQR_derivation} shows that the feedback law then simplifies to
\begin{equation} 
	\label{eq:LQR_final}
    u(t) = - \est{\omega}(t) - \lambda \;\brktc{\estJy(t)},
\end{equation}
where $\lambda \coloneqq \sqrt{p_J/\nu}$ is a tunable parameter functioning as a control ``knob''. Notably, setting $\lambda = 0$ reduces \eqnref{eq:LQR_final} to the naive field-compensation strategy considered earlier. 

In practice, the EKF provides us not with the estimate of the reduced state, $\est{\pmb{z}}(t)$, but actually with the estimate of the full state vector, $\est{\pmb{x}}(t)$:
\begin{equation}
\est{\pmb{x}}(t) = \big( \brktc{\estJx},\brktc{\est{\Jy}},\estVx^{\cc},\estVy^{\cc},\estVz^{\cc},\estCxy^{\cc},\est{\omega} \big)^\TT.
\end{equation}
To embed the reduced-state control into this higher-dimensional setting, we define a projection matrix
\begin{equation}
\pmb{\Xi} \coloneqq
\begin{pmatrix}
0 & 1 & 0 & \dots & 0 & 0 \\
0 & 0 & 0 & \dots & 0 & 1
\end{pmatrix},
\end{equation}
which extracts the $(\brktc{\Jy}, \omega)$ components from $\est{\pmb{x}}(t)$. Then, the generalized control law is
\begin{equation}
u(t) = -\pmb{K}_C \, \pmb{\Xi} \, \est{\pmb{x}}(t).
\end{equation}
In this way, the EKF provides the necessary state estimates, and, as we will show later, the LQR delivers robust feedback even outside the LG regime \cite{Amoros-Binefa2024,Amoros-Binefa2025}.

\section{Results}

In the previous sections, we detailed the model of the optical atomic magnetometer and its numerical simulation, including its atomic ensemble dynamics, simulation methods, and the estimation and control strategies that enhance its performance. We now present our main results, which we organize as follows: in \secref{sec:weak-field_sensing}, we begin by examining the response of the magnetometer under weak-field sensing conditions, where it exhibits LG dynamics. \secref{sec:precessing_field} then details the real-time sensing of a precession-inducing field, with separate analyses for constant, fluctuating, and nonlinear MCG-like field profiles. Finally, \secref{sec:cond_uncond_spin_squeezing} demonstrates the generation of multiparticle entanglement by evaluating conditional as well as unconditional spin squeezing. 

\subsection{Weak-field sensing} \label{sec:weak-field_sensing}

\begin{summary}
In this subsection we focus on large atomic ensembles ($N\gg 1$) sensing weak fields ($B \ll 1$) for times shorter than the coherence time $(M+\kcoll)^{-1}$. In that case, the magnetometer behaves as a LG system and we show:
\begin{itemize}
    \item Numerical solutions of the KF Riccati equation reveal several distinct scaling regimes for the estimation error, including a transient quantum-enhanced scaling of $1/(N^2 t^3)$, consistent with \eqnref{eq:HS_approx}.
    \item The key result is that for sufficiently large $N$, the KF achieves the theoretical minimum error allowed by dephasing and field fluctuations, i.e. the CS limit of $\sqrt{q_\omega \kcoll} \coth{\left(t \sqrt{\frac{q_\omega}{\kcoll}}\right)}$ (derived in \eqnref{eq:CSlim_infprior}).
    \item If the atomic ensemble is not large enough, the KF does not achieve the CS limit and instead reaches the steady state solution given in \eqnref{eq:ss_field_sol}.
    \item This demonstrates that the combination of continuous homodyne measurement and a coherent spin state is optimal in the LG regime of $N\gg 1$, $t < (M+\kcoll)^{-1}$ and $B \ll 1$. 
    \item The analysis also identifies the precise timescales and ensemble sizes at which non-classical scaling transitions into steady state or CS-limited performance.
\end{itemize}
\end{summary}

When employing large ensembles to track weak fields, which is the relevant regime when discussing detection limits,  the system is LG for times within the coherence time of the system and measurement (i.e. $t<(M+\kcoll)^{-1}$ (see \secref{sec:LG_regime}). In this regime, the KF provides an \emph{optimal}, recursive strategy (see \secref{sec:KF_for_weak_field}). By continually incorporating new measurement data, the filter updates both the state estimate and its error covariance, ensuring that the mean-squared error is minimized at every step. Crucially, the KF not only provides a real-time estimate of the magnetic field but also yields a \emph{quantifiable metric of its precision} through the covariance matrix, which for the case of a KF, coincides with the aMSE (see \secref{sec:correlated_KF}). 

\begin{figure}[t!]
    \centering
    \includegraphics[width=\linewidth]{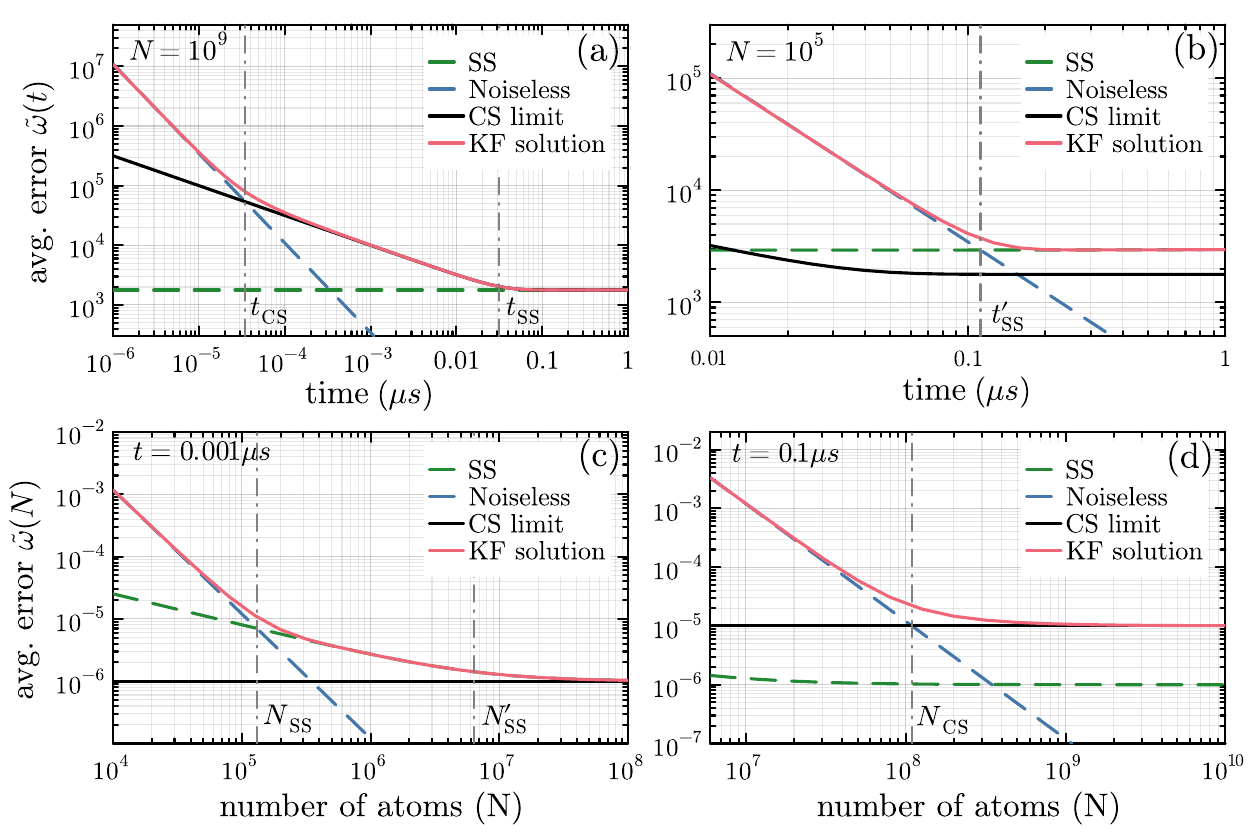}
    \caption[The minimal estimation error as a function of time]{\textbf{The minimal estimation error as a function of time.} Subplots (a) and (b) show the time evolution of the averaged error of estimating $\omega(t)$, $\sqrtaMSEomega$, for large ($N = 10^9$) and small ($N = 10^5$) ensembles, respectively. Subplots (c) and (d) present the dependence of the error on the number of atoms at short ($t = \SI{0.001}{\micro\second}$, (c)) and long $(t = \SI{0.1}{\micro\second}$, (d)) measurement times. Other parameters used to generate this figure are:~$M = \SI{100}{\kilo\hertz}$, $q_\omega = 10^{14}\SI{}{\radian^2 \, \second^{-3}}$, $\kcoll = \SI{100}{\milli\hertz}$, $\chi = 0$ and $\eta = 1$. Colored dashed lines serve as references, representing different scaling behaviors of the Riccati solution: dashed green corresponds to the steady state (SS) solution of the KF, dashed blue represents the analytical solution obtained in an ideal noiseless scenario, and finally, solid black corresponds to the quantum limit imposed by dephasing and field fluctuations, i.e. the CS limit. This quantum limit is attained for large $N$ when $t > t_{\mrm{CS}}$, proving in this regime the optimality of our measurement and initial state. }
    \label{fig:time_plot_LG}
\end{figure}

In \figref{fig:time_plot_LG} we show the minimal averaged error (solid, red) for the estimator of the Larmor frequency (the squared root of the aMSE, $\sqrtaMSEomega = \sqrt{\pmb{\Sigma}_{\omega,\omega}}$). The top plots display the error as a function of time for two atomic ensembles of different sizes: $N = 10^9$ (a) and $N = 10^5$ (b), and the bottom plots show the error as a function of the number of atoms for two different time slices: $t = 10^{-4}\SI{}{\second}$ (c) and $t = 10^{-2}\SI{}{\second}$ (d). In all cases, the error is obtained by numerically solving the Riccati equation of \eqnref{eq:Riccati_equation_ch5}, with the model matrices defined in \eqnref{eq:matrices_F_K_B} and the exact form for the variance of $\Jy$ provided in \eqnref{eq:exactsol}. In \figref{fig:time_plot_LG} we further analyze the different behaviors of the error of the KF (solid, red), by comparing it to the analytical solutions of the KF derived in \secref{sec:KF_for_weak_field} for (1) the steady state (dashed, green) and (2) the noiseless regime (dashed, blue). In particular, for short times or low atomic numbers, the error of $\est{\omega}$ behaves like in the ideal case of negligible dephasing and field fluctuations, revealing a non-classical scaling of the estimation error with both time and the number of atoms:
\begin{equation}
    \EE{\Delta^2 \est{\omega}(t)} \propto \frac{1}{N^{\,2} t^3}, 
\end{equation}
highlighting the quantum-enhanced sensitivity achievable in our setup\cite{Geremia2003}. Over time, however, the precision scaling deteriorates, as shown in the two top plots. This occurs either when the system reaches a steady state (\figref{fig:time_plot_LG} (b)) at time
\begin{equation}
    t_{\mrm{SS}}^\prime  = 3^{1/3} \left( \frac{4}{N^2 \eta M q_\omega}\right),
\end{equation}
or by first attaining the CS limit of \eqnref{eq:CSlim_infprior} (see \figref{fig:time_plot_LG} (a)) at time 
\begin{equation}
    t_{\mrm{CS}} =\frac{2}{N}\sqrt{\frac{3}{\eta M \kcoll}},
\end{equation}
which also later coincides with the KF steady state after $t = t_{\mrm{SS}}$. One might therefore realize that the CS bound in the limit of an infinitely wide prior, as presented in \eqnref{eq:CSlim_infprior}, has two distinct regimes:
\begin{align}
    \mrm{V}_{\!\mrm{CS}}^{\infty}(t) = \sqrt{q_\omega \kcoll} \coth{\left(t \sqrt{\frac{q_\omega}{\kcoll}}\right)} = \begin{cases}
        \dfrac{\kcoll}{t} \quad \text{for} \; t < t_{\mrm{SS}}\\
        \sqrt{q_\omega \kcoll} \quad \text{for} \; t \geq t_{\mrm{SS}}
    \end{cases}
\end{align}
where
\begin{equation}
    t_{\mrm{SS}} = \sqrt{\frac{\kcoll}{q_\omega}}.
\end{equation}
\figref{fig:time_plot_LG} (c) and (d) illustrate that while small ensembles may achieve a precision scaling of $1/N^2$ in estimating $\omega(t)$, this advantage diminishes as the system size increases. As a result, larger atomic ensembles do not provide a net gain in precision. That occurs at 
\begin{align}
    N_{\text{CS}} = \frac{2}{t} \sqrt{\frac{3}{\eta M \kcoll}}, \qquad \text{and} \quad N_{\text{SS}}^{\,\prime} = \frac{2  }{\kcoll} \sqrt{\frac{q_\omega}{\eta M}}.
\end{align}
For a certain range of ensemble sizes, i.e. $[N_{\mrm{SS}},N^{\,\prime}_{\mrm{SS}}]$, where
\begin{equation}
    N_{\text{SS}} = \frac{2 \, 3^{2/3}}{  t^2 \sqrt{\eta M q_\omega}},
\end{equation}
and for short times, the KF error follows the steady state solution (dashed, green) of \eqnref{eq:ss_field_sol}, which scales as $\propto 1/N^{\,1/2}$ (and thus, the error displayed in \figref{fig:time_plot_LG} (c) scales as  $\propto 1/N^{\,1/4}$).

The most important message that \figref{fig:time_plot_LG} aims to convey is that for large ensembles (i.e. $N = 10^9$) the aMSE of the KF attains the lower bound on the error dictated by dephasing and fluctuations, referred throughout this thesis as Classical Simulation (CS) limit. Even though we already know that for a LG system, the KF is the optimal estimator, \figref{fig:time_plot_LG} (a) also confirms that for large atomic ensembles, our choice of measurement and initial state are also optimal, since the CS limit is independent of our choice of measurement and initial state (see \chapref{chap:bounds}). In summary, applying the KF to a system initialized in a CSS and continuously monitored through homodyne measurement not only resolves the estimation error in real time but also establishes an optimal framework for high-precision sensing of weak magnetic fields. 

\subsection{Real-time sensing of a larger frequency} \label{sec:precessing_field}

We now turn our attention to the practical task of real-time sensing of a precession-inducing magnetic field. As before, our main goal is to estimate the Larmor frequency and benchmark the performance of our proposed EKF+LQR strategy against other estimation and control approaches as well as the quantum limit dictated by dephasing and field fluctuations derived in \chapref{chap:bounds}. To explore the role of quantum effects in enhancing estimation, we go beyond analyzing the average error of the frequency estimate, $\sqrtaMSEomega$, and also examine the averaged evolution of the spin squeezing parameter $\mathbb{E}[\xi_y^{\,2}(t)]$ (defined in \secref{sec:spin-squeezing_intro}), along with the averaged ensemble polarization, $\mathbb{E}[\brktc{\Jx(t)}]$.

\subsubsection{Constant field} \label{sec:constant_field}

In what follows, we investigate the real-time tracking of the simplest type of field: a constant magnetic field. First, we do so by using the full SME of \eqnref{eq:fullSME} to model the evolution of the system. Using this ``brute-force'' approach we perform two important tasks: (1) benchmarking of the EKF+LQR strategy against alternative methods and (2) validation of the CoG approximation introduced in \secref{sec:Co-moving_Gaussian}. Once validated, we employ this approximation to study larger atomic ensembles, determining whether our approach reaches the quantum limit imposed by dephasing, and hence, testing the optimality of our magnetometry setup.

\paragraph*{Low atomic numbers}

\begin{summary}
For a constant magnetic field and magnetometers of $N \sim 100$:
\begin{itemize}
    \item Simulations of the full SME for $N = 200$ demonstrate that the EKF+LQR strategy yields convergence of the frequency estimate, maintains substantial spin squeezing, and drives the estimation error down to the CS limit. 
    \item Benchmark comparisons show that alternative strategies (KF with LQR, EKF without control, EKF with frequency compensation) fail to both maintain squeezing and sustain long-term polarization, and they achieve noticeably worse estimation errors.
\end{itemize}

\end{summary}

\begin{figure}[t!]
    \centering
    \includegraphics[width=0.8\columnwidth]{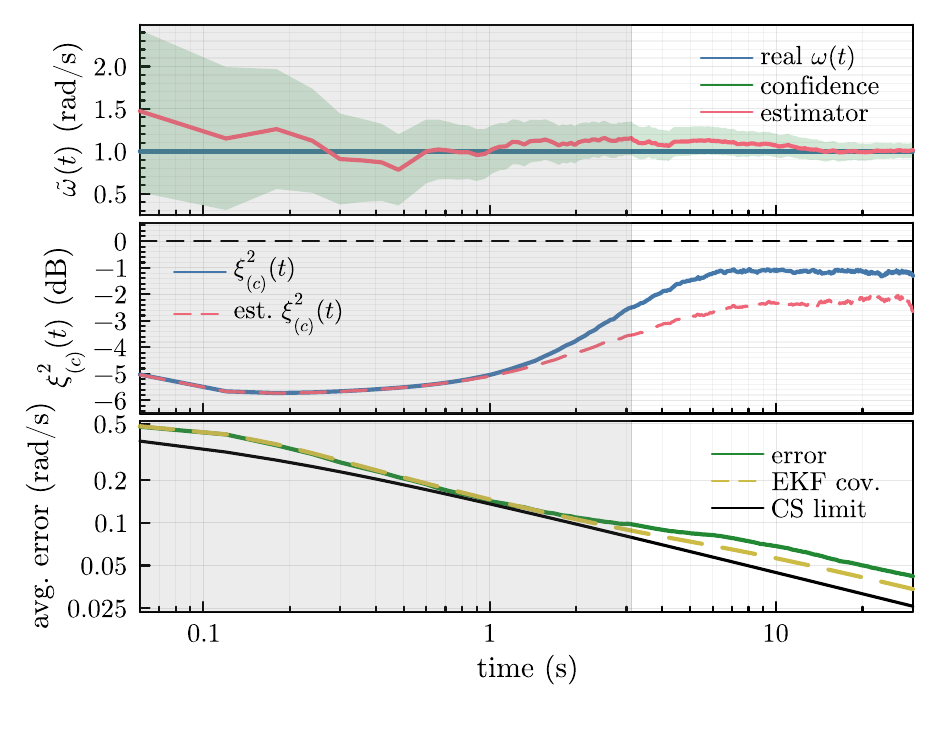}
    \caption[Quantum-enhanced tracking of a constant magnetic field with $N = 200$]{%
    \textbf{Quantum-enhanced tracking of a constant magnetic field with $N = 200$.} \emph{Top}:~Constant field (solid, blue) tracked in real time by the EKF estimate (solid, red). The shaded area represents the confidence band limited by the averaged error, i.e. $\pm 2 \sqrtaMSEomega$.
    \emph{Middle}:~Averaged evolution of the spin squeezing parameter (solid, blue), compared with its real-time EKF prediction (dashed, red). When the parameter is below the dashed black line of $\xi_{y,\cc}^{\,2}(t) = 1$, it indicates squeezing and thus, the presence of multipartite entanglement. 
    \emph{Bottom}:~Evolution of the average error (solid, green) in estimating the fluctuating field, $\sqrtaMSEomega$, which attains the quantum limit imposed by dephasing (solid, black), as correctly predicted by the EKF covariance (dashed, yellow), which matches the averaged error within the LG regime. In all plots, averaging was performed over 1000 atom stochastic trajectories, and the parameters used are:~$N = 200$, $\kcoll = 0.02$, $\kloc = 0$, $M = 0.3$, $\omega = 1$, $\eta = 1$, and $\sigma_0=0.5$.
    }
    \label{fig:fullSME_constant_N200_estimation}
\end{figure}

\figref{fig:fullSME_constant_N200_estimation} \textit{(top)} presents the real-time estimation of a constant magnetic field, simulated via the full SME of \eqnref{eq:fullSME} for a system of $N = 200$ atoms. As time progresses, the continuous incorporation of measurement data reduces the error (green shaded area), while the estimate (solid, red) converges towards the true value (solid, blue). The middle panel illustrates that spin squeezing (solid, blue) quickly emerges and is maintained throughout the experiment, which can be correctly estimated with the EKF (dashed, red) for times within the LG region. Finally, the bottom plot shows that the averaged error approaches the quantum limit dictated by dephasing (i.e., the CS limit, in solid, black):
\begin{equation} \label{eq:CSbound_ch5}
    \EE{\Delta^2 \est{\omega}(t)} \geq \dfrac{1}{\dfrac{1}{\sigma_0^2}+\dfrac{t}{\kcoll}},
\end{equation}
where $\sigma_0$ is the standard deviation of the initial prior of the Larmor frequency. As expected, the EKF covariance (its squared root in dashed yellow) provides a good estimate of the error within the LG regime, although it becomes slightly optimistic beyond the $(M+\kcoll)^{-1}$ mark.

\begin{figure}[t!]
    \centering
    \includegraphics[width=\textwidth]{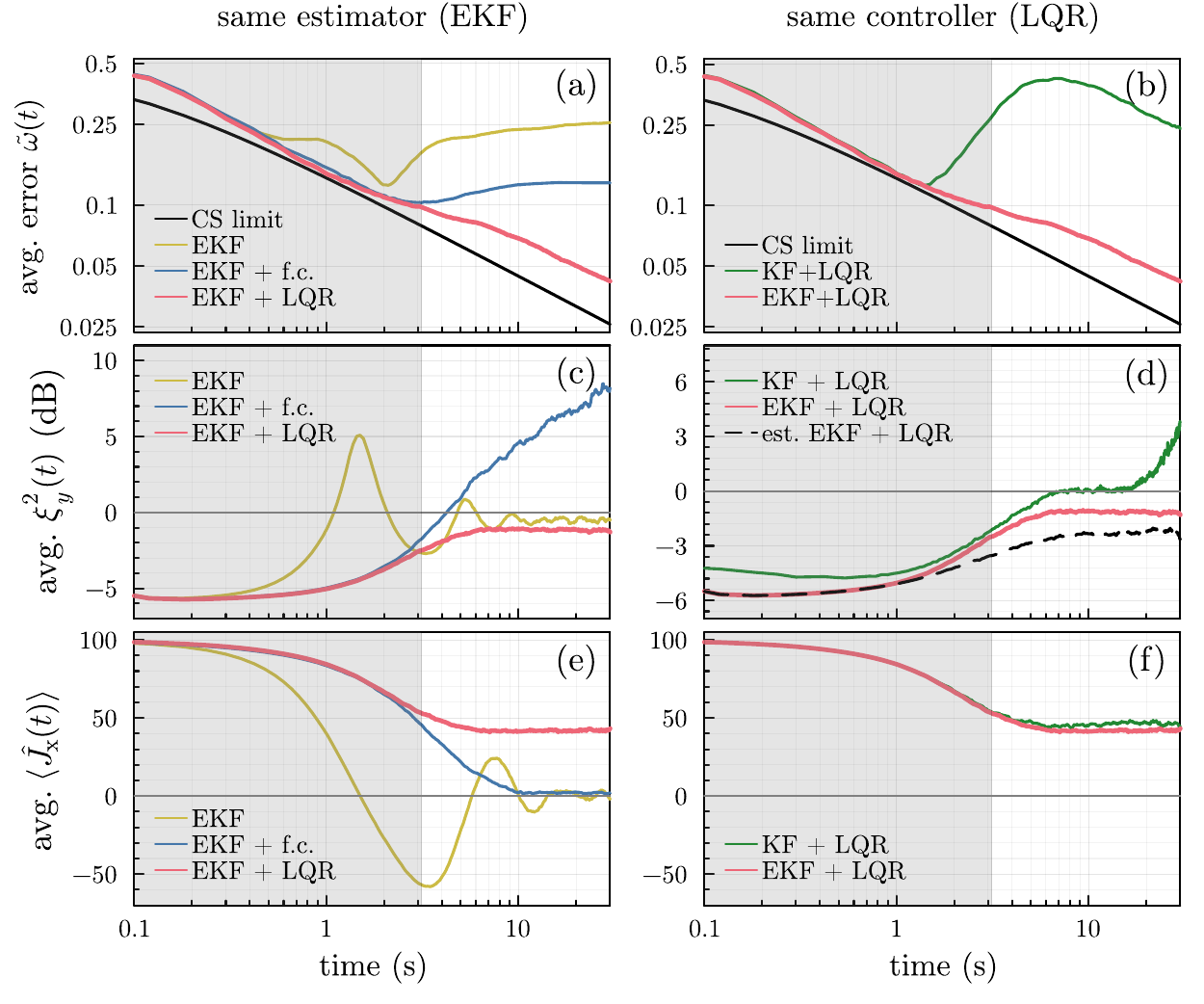}
    \caption[Performance of different estimation and control strategies]{\textbf{Performance of different estimation and control strategies.} The top row shows the averaged estimation error, the middle row displays the spin squeezing parameter (as defined in \eref{eq:spin_squeez_par}), and the bottom row illustrates the ensemble polarization, $\brktc{\Jx}$. In the left column (subplots (a), (c), and (e)), the estimator is fixed as the EKF while three control methods are compared: no control (yellow), field compensation (blue), and LQR (red). In the right column (subplots (b), (d), and (f)), the control method is fixed to LQR while two different estimators are considered: KF (green) and EKF (red). In both subplot (a) and (b), the EKF+LQR strategy (red) is benchmarked against the CS limit (black), which sets the quantum limit imposed by dephasing on the attainable error. Notably, the EKF+LQR approach outperforms the other methods by achieving the lowest estimation error, which continues to decrease even beyond the LG regime (shaded grey). Subplots (c)–(f) demonstrate that only the combination of an EKF with an LQR maintains both spin squeezing and polarization throughout the experiment. Additionally, in subplot (d), the black dashed line indicates the overoptimistic predictions from the EKF, as shown already in \figref{fig:fullSME_constant_N200_estimation}. The parameters used in the SME \eref{eq:fullSME} for simulations are:~$N = 200$, $\kcoll = 0.02$, $\kloc = 0$, $M = 0.3$, $\omega = 1$ and $\eta = 1$. The KF and EKF estimators are initialized with the mean $\Tilde{\pmb{x}}(0) = (N/2,0,0,N/4,N/4,0,\mu_0)^\TT$ and covariance $\pmb{\Sigma}(0) = \Diag{0,0,0,0,0,0,\sigma_{0}^2}$ dictated by the initial CSS state of the atoms, and the Gaussian prior distribution for $\omega\sim\mathcal{N}(\mu_0,\sigma_0^2)$. All results are obtained after averaging over $\nu = 1000$ measurement trajectories, whereas $\omega$-averaging is avoided by choosing its true value $\omega=1$ for a prior with $\mu_0 = \omega + \sigma_0=1.5$ and $\sigma_0=0.5$.}
    \label{fig:err}
\end{figure}

We further compare the performance of the EKF+LQR strategy against less sophisticated methods, including the KF with field compensation (green), the EKF without control (yellow), and the EKF with field compensation (blue). In the left column of \figref{fig:err}, we keep the estimator as the EKF and switch the type of controller: none (yellow), frequency compensation (blue) and LQR (red). In contrast, in the right column we fix the control as LQR and compare different estimators: either KF (green) or EKF (red). This comparison in \figref{fig:err} (a)-(b) under controlled conditions, e.g. no local dephasing such that we can simulate for larger $N$, highlights the superiority of combining an EKF with an LQR, since it yields the lowest average error (red) closest to the quantum limit dictated by dephasing (i.e. the CS limit in black of \eqnref{eq:CSbound_ch5}). The other estimation and control approaches also fail to yield both spin squeezing and a non-vanishing polarization for the full duration of the experiment ($T = \SI{30}{\second} > (M+\kcoll)^{-1} \approx \SI{3}{\second}$), as illustrated in \figref{fig:err} (c)-(d) for squeezing and (e)-(f) for the polarization.

Lastly, simulating the full SME of \eqnref{eq:fullSME} allows us to verify the accuracy of the CoG approximation in simulating our system. As briefly discussed in \secref{sec:Co-moving_Gaussian} and extensively in \appref{sec:verification_CoG}, we compare the Larmor frequency estimates obtained using two methods: one by solving the full SME and the other by solving a system of SDEs given by the CoG approximation (see \secref{sec:Co-moving_Gaussian}). In both cases, the estimation and control is carried out using an EKF combined with an LQR, since \figref{fig:err} established it as the best known strategy.  The results, shown in the first column of \figref{fig:model_vs_exact}, demonstrate that the relative error remains below 1\% at all times and decreases with increasing $N$. This confirms that the CoG approximation is valid and can be confidently used in the subsequent analysis.

\paragraph*{High atomic numbers}

\begin{summary}
For a constant magnetic field, we use the CoG approximation to extend our analysis to large ensembles (here, $N \sim 10^5$):
\begin{itemize}
    \item We find that the EKF+LQR estimator saturates the CS limit for both collective and local decoherence models, confirming optimal performance.
    \item Furthermore, both conditional and unconditional spin squeezing are observed, demonstrating generation of multipartite entanglement even beyond the LG regime. 
    \item The EKF accurately predicts conditional spin squeezing, polarization, and estimation error averages.
\end{itemize}

\end{summary}

For larger ensembles ($N \sim 10^5-10^{15}$ c.f.~\cite{Bohnet2014,hosten_measurement_2016}), simulating the exact SME becomes computationally prohibitive. In this regime, we rely on the CoG approximation, which accurately captures the conditional evolution of the first and second moments of the angular-momentum operators. 

\begin{figure*}[t!]
    \centering
    \includegraphics[width=\textwidth]{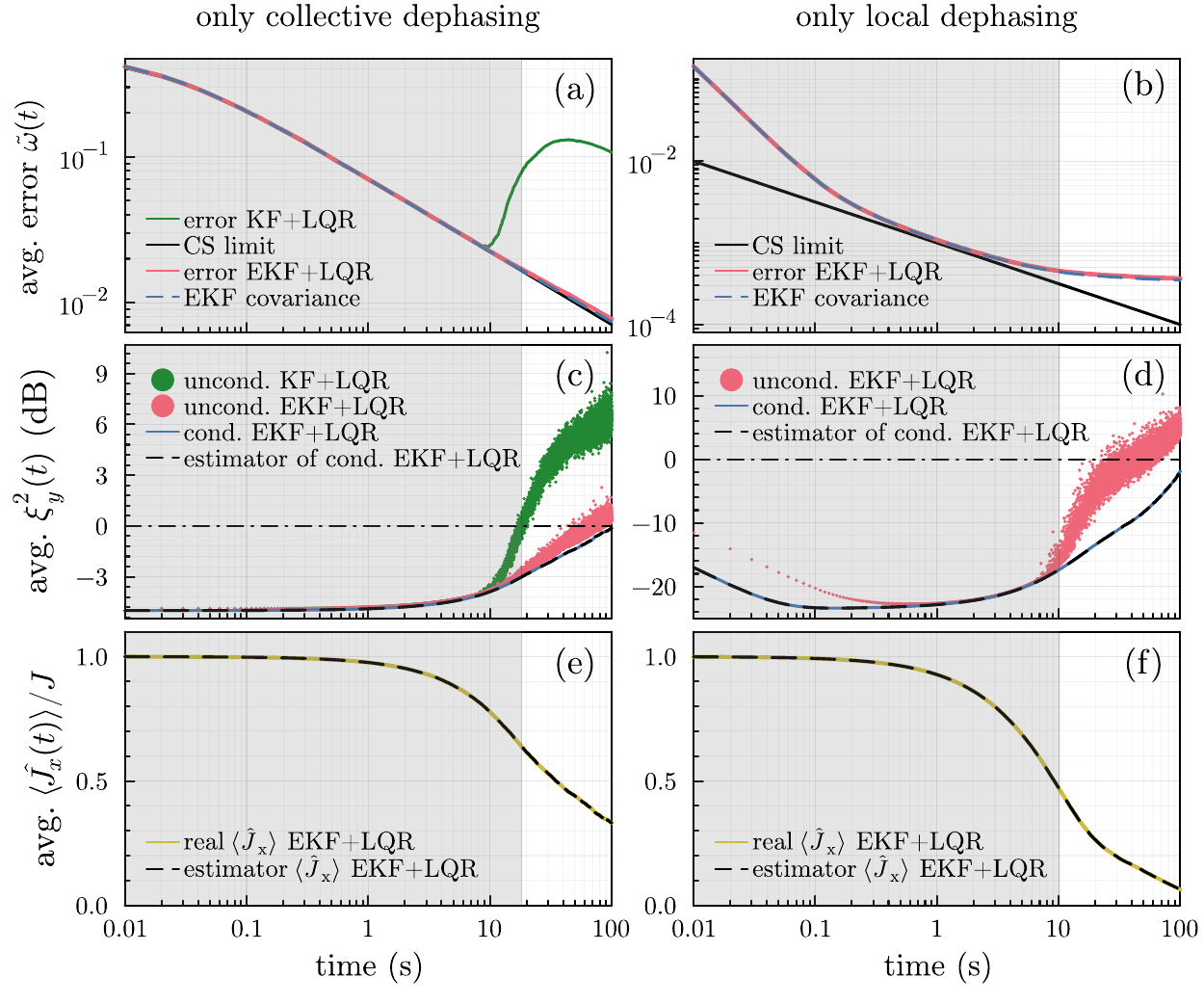}
    \caption[Performance in estimation and spin squeezing extrapolated to large atomic ensembles ($N=10^5$)]{\textbf{Performance in estimation and spin squeezing extrapolated to large atomic ensembles ($N=10^5$).} Subplots (\emph{a}) and (\emph{b}) (\emph{upper row}) depict the case of pure collective decoherence $\kcoll = 0.005$, whereas subplots (\emph{c}) and (\emph{d}) (\emph{lower row}) deal with pure local decoherence $\kloc = 0.05$. \emph{Left column}:~(\emph{a}) and (\emph{c}) show the error (aMSE) attained by the EKF+LQR strategy (red dots) when estimating $\omega$ and its average prediction by the EKF (blue line), $\EE{\pmb{\Sigma}_{\omega\omega}}$, both being lower-bounded by the CS limit \eref{eq:CSbound_ch5} (black line). For collective decoherence, the performance of KF+LQR strategy is also included (grey dots) to emphasize its failure beyond the LG regime (pink shading in all subplots). \emph{Right column}:~(\emph{b}) and (\emph{d}) illustrate the evolution of the spin squeezing parameter \eref{eq:spin_squeez_par} for conditional (blue line) and unconditional (red dots) dynamics, as compared with its classical threshold (horizontal dash-dotted line). The evolution of the ensemble polarization $\brkt{\Jx}=\mathbb{E}[\brktc{\Jx}]$ (green line) is also shown in both cases in extra lower plots. Both the conditional spin squeezing and the polarization in (\emph{b}) and (\emph{d}) are estimated very accurately by the EKF on average (superimposed dashed black lines). The above data is simulated employing the CoG model \eref{eq:dynamical_model} with other parameters set to:~$M = 0.05$, $\omega = 1$ and $\eta = 1$. As in \figref{fig:err}, the KF and EKF estimators are initialized with the mean $\Tilde{\pmb{x}}(0) = (N/2,0,0,N/4,N/4,0,\mu_0)^\TT$ and covariance $\pmb{\Sigma}(0) = \Diag{0,0,0,0,0,0,\sigma_{0}^2}$ dictated by the initial CSS state and the Gaussian prior distribution for $\omega\sim\mathcal{N}(\mu_0,\sigma_0^2)$. All results are obtained after averaging over $\nu = 20000$ measurement trajectories, while $\omega$-averaging is avoided by choosing the prior with $\sigma_0=0.5$ and $\mu_0 = \omega + \sigma_0=1.5$.
    }
    \label{fig:attaining_bounds}
\end{figure*}

The \emph{top row} of \figref{fig:attaining_bounds} demonstrates that, for $N=10^5$ atoms, the EKF+LQR strategy achieves optimal performance when its averaged estimation error (solid, red) reaches the quantum limit imposed by dephasing \eqnref{eq:CSlim_zeroq} (solid, black). This optimal behavior is observed under both collective (a) and local (b) decoherence. Although the quantum limit is reached only briefly within the LG regime under local dephasing, using an EKF is essential, as the KF is not applicable in this scenario (see \secref{sec:LG_regime}). For collective dephasing, the EKF+LQR still outperforms the KF (solid, green), even within the LG regime (shaded grey area). Attaining the quantum limit ensures that the measurement, feedback, and initial state are optimal during that period, thereby addressing the open question posed in \refcite{rossi_noisy_2020}. Moreover, the EKF covariance (its squared root in dashed, blue) closely tracks the averaged estimation error (solid, red), confirming that, despite the nonlinearity of the CoG model \eref{eq:dynamical_model}, the EKF provides reliable trajectory-dependent error estimates.

The \emph{middle row} of \figref{fig:attaining_bounds} shows the averaged spin squeezing parameter in decibels, while the \emph{bottom row} displays the averaged ensemble polarization. In all cases, the EKF predictions (dashed, black) closely match the simulation results: the averaged conditional spin squeezing (solid, blue) and the polarization or averaged $\brktc{\Jx}(t)$ (solid, yellow). This excellent agreement holds as long as the CoG approximation is reliable, and contrasts with our earlier findings in \figref{fig:err}(d), where the EKF predictions overestimated the spin squeezing parameter. Moreover, subplots (c) and (d) confirm that the EKF+LQR strategy not only generates conditional spin squeezing, as already shown in \figref{fig:err}(c) and (d) for $N=200$, but also yields significant unconditional spin squeezing (red dots) above the classical limit (horizontal dash-dot black line). Additionally, subplot (c) demonstrates that the EKF outperforms the KF (green dots) in preserving an unconditional multiparticle entangled state beyond the LG regime, though it also eventually degrades after approximately $\sim\!\SI{50}{\second}$. 

\subsubsection{Fluctuating field} \label{sec:fluctuating_field}
\begin{summary}
    Under realistic experimental parameters, and when the Larmor frequency follows an OU process, we show:
    \begin{itemize}
        \item The EKF+LQR strategy can track field fluctuations within microseconds, while simultaneously generating substantial measurement-induced squeezing (up to $\sim\!\SI{13}{\decibel}$).
        \item Although the CS limit is not reached under current measurement strengths, the EKF accurately predicts the error dynamics, and increasing $M$ would, in principle, enable quantum-limited tracking.
        \item Even in the presence of small collective dephasing, we find that the system operates at the CS limit -- even without squeezing -- and that optimality persists even when the estimator uses mismatched OU parameters.
    \end{itemize}
\end{summary}

\begin{figure}[t!]
    \centering
    \includegraphics[width=0.8\columnwidth]{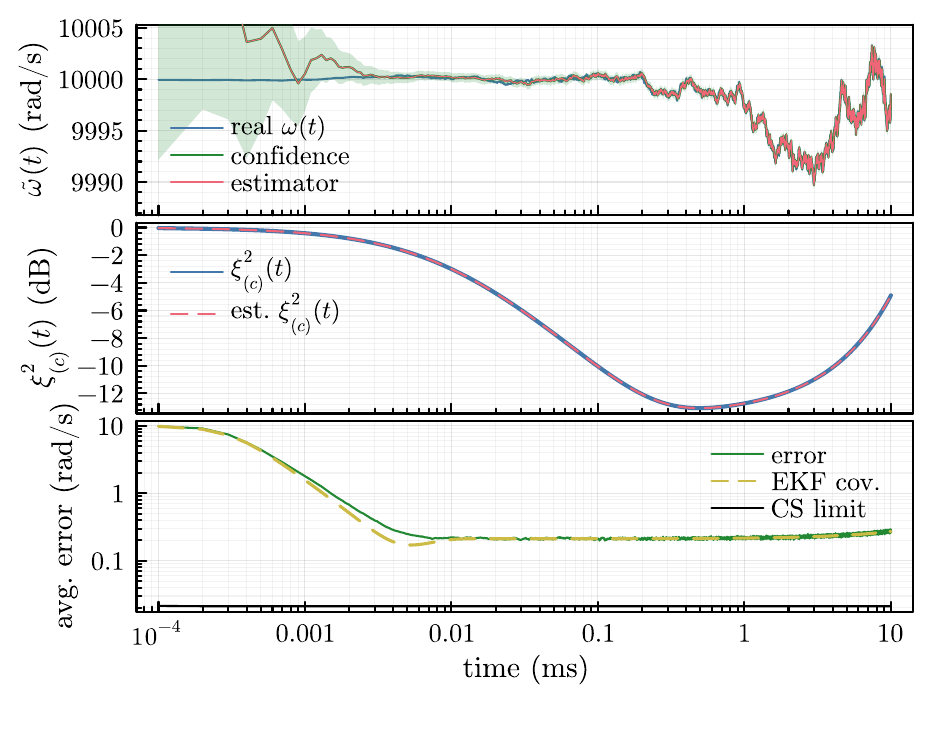}
    \caption[Quantum-enhanced tracking of a fluctuating magnetic field]{%
    \textbf{Quantum-enhanced tracking of a fluctuating magnetic field.} 
    \emph{Top}:~Efficiently tracking in real time of field fluctuations (solid, blue) by its corresponding EKF estimate (solid, red) after gathering only $\approx\!\!\SI{0.01}{\milli\second}$ of photocurrent data. The shaded area represents the confidence band limited by $\pm 2 \sqrtaMSEomega$, i.e.~the error obtained upon averaging.
    \emph{Middle}:~Evolution of the (average) spin squeezing parameter (solid, blue) in $\trm{dB}$, compared with its real-time prediction by the EKF (dashed, red).
    \emph{Bottom}:~The averaged error (solid, green) in estimating the fluctuating field, $\sqrtaMSEomega$, which stabilizes at $\approx \SI{0.21}{\radian \second^{-1}}$, as correctly predicted by the EKF covariance (dashed, yellow). The quantum limit imposed by local dephasing (a.k.a. CS limit, in solid, black) is not attained due to insufficient measurement strength ($M=10^{-8}\SI{}{\hertz}$~\cite{Kong2020}, see also \figref{fig:largerM})~\cite{Amoros-Binefa2024}. Averaging was performed over 1000 field+atom stochastic trajectories. 
    }
    \label{fig:oup_sq_estimation}
\end{figure}

In this subsection, we extend our discussion to the case where the Larmor frequency fluctuates as an Orstein-Uhlenbeck process (OUP)~\cite{Stockton2004,Petersen2006} (see \secref{sec:OUP_intro}):
\begin{equation} \label{eq:oup}
    \dd \omega(t) = -\chi \omega(t) \dt + \sqrt{q_\omega} \dW_{\!\omega},
\end{equation}
where $\dW_{\!\omega}$ is a (new) Wiener increment, $\chi>0$ is the decay rate, and $q_\omega>0$ sets the strength of fluctuations, which are chosen to be $\chi = \SI{0.01}{\second^{-1}}$ and $q_\omega = 10^4\SI{}{\radian^2 \second^{-3}}$ for the simulations presented throughout this section. The results discussed here and in the following sections are obtained using experimentally realistic parameters, chosen to reflect conditions achievable in state-of-the-art setups. These parameter values are inspired by \refcite{Kong2020}, though their adaptation to our setting required interpreting experimental quantities such as the measurement strength, which are not directly provided but must be inferred from various experimental parameters. A detailed explanation is given in \appref{sec:experimental_parameters}. Given these parameters, the standard deviation of the fluctuations over the magnetometer coherence time $T_2$ (here $\SI{10}{\milli\second}$) is approximately $0.1\%$ of the initial Larmor frequency $\omega_0= 10^4 \SI{}{\radian \, \second^{-1}}$ (recall from \eqnref{eq:var_OUP_smallchi} that the variance of an OU process can be approximated as $q_\omega \,t$ at short times $t \lesssim 1/\chi$). \figref{fig:oup_sq_estimation} (\emph{top}) illustrates how the proposed EKF+LQR strategy is capable of tracking the OUP-induced fluctuations in real time for an experiment with realistic parameters (with the EKF equations explicitly written in \appref{sec:exact_forms_FGH_OUP}). Additionally, the EKF also provides an accurate estimate of the atomic spin squeezing parameter~\cite{Ma2011} (\emph{middle}) that reaches $\gtrsim\!\SI{13}{\decibel}$%
~\footnote{
    The value of $\approx\!\SI{2}{\decibel}$ reached in \refcite{Kong2020} should not be directly compared, as therein an unpolarised ensemble in the spin-exchange relaxation-free regime was considered. 
}
at around \SI{0.5}{\milli\second}, being induced purely by the measurement backaction emerging at $\approx\!\SI{0.01}{\milli\second}$. The magnetometer reaches its best resolution at times $\SI{0.01}{\milli\second}\lesssim t \lesssim T_2$, where it tracks field fluctuations in real time with an error of \SI{0.21}{\radian \, \second^{-1}} (\emph{bottom}). The minimal averaged error (solid, green) is correctly predicted by the EKF covariance (its squared root in dashed, yellow). While the quantum limit imposed by local dephasing (\eqnref{eq:CSlimit_full} with $\kcoll=0$) sets a fundamental lower bound of \SI{0.021}{\radian \, \second^{-1}} (solid, black), this limit is not reached under the current measurement strength. However, increasing $M$ can, in principle, allow the system to attain this bound~\cite{Amoros-Binefa2024} (see \figref{fig:largerM}), though the required measurement strengths may not be experimentally feasible.

\begin{figure}[t!]
    \centering
    \includegraphics[width=\textwidth]{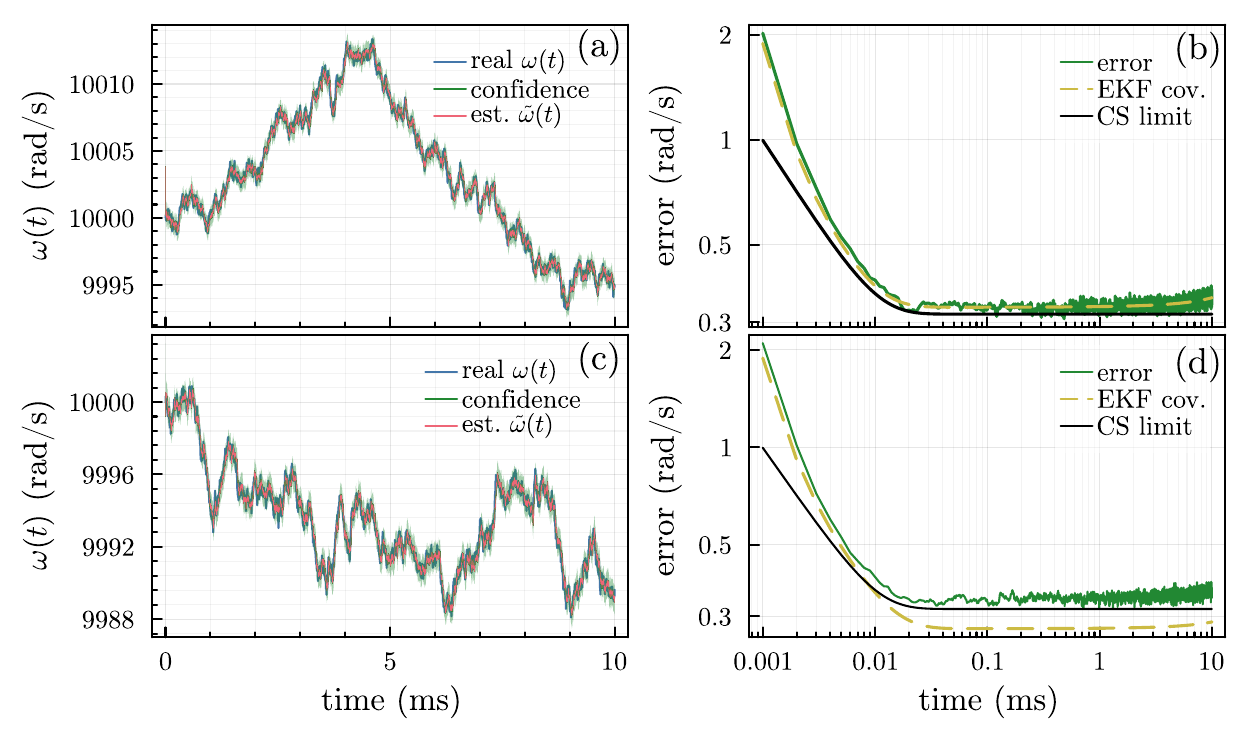}
    \caption[Tracking field fluctuations at the quantum limit]{\textbf{Tracking field fluctuations at the  quantum limit \eref{eq:CSlimit_full}}. As in the bottom plot of \figref{fig:oup_sq_estimation}, the true average error, $\sqrtaMSEomega$ (solid, green), is compared against the error predicted by the EKF (dashed, yellow), i.e. the covariance, and the quantum limit imposed by dephasing, as given in \eqnref{eq:CSlimit_full}, now including collective dephasing $\kcoll = \SI{1}{\micro\hertz}$. Remarkably, the magnetometer operates at the quantum limit regardless of whether the EKF is provided with the exact OUP dynamics \eref{eq:oup} (a-b) or a mismatched version (c-d). In the latter case, despite the EKF assuming a fluctuation strength that is half as large ($q_K = q_\omega/2$) and much faster decay ($\chi_K = 10\chi$), the average error (solid, green) still reaches the quantum limit (solid, black), while the EKF covariance (dashed, yellow) underestimates the error. Both plots were obtained by averaging over 1000 field+atom stochastic trajectories. Additionally, subplots (a) and (c) show a representative field trajectory (blue) alongside its EKF real-time estimate (red), which remains well within the confidence interval $\pm 2\sqrtaMSEomega$ (shaded green).
    }
    \label{fig:oup_estimation}
\end{figure}

Another scenario where the magnetometer operates at the quantum limit, i.e. the CS limit of \eqnref{eq:CSlimit_full}, is when a small amount of collective dephasing is present. In \figref{fig:oup_estimation}, we add $\kcoll = \SI{1}{\micro\hertz}$ ($1/\kcoll \!\approx\!\SI{11}{days}$) alongside the local dephasing of $\kloc=\SI{100}{\hertz}$. As shown in \figref{fig:oup_estimation} (b), while collective dephasing worsens the estimation error and raises the quantum limit, it also leads to a favorable outcome: the averaged estimation error (solid, green) saturates the quantum limit at $\approx\!\SI{0.32}{\radian \, \second^{-1}}$ (solid, black), indicating that the magnetometer operates in an optimal regime. This holds true even though spin squeezing is no longer present, as collective dephasing dominates over measurement-induced correlations  ($\kcoll > \eta M$), preventing the generation of spin-squeezed states~\cite{Amoros-Binefa2021} (see also \figref{fig:squeezing}). Additionally, the quantum limit (solid, black) is not only reached when the EKF is provided with the exact OU field dynamics (subplot (b)), i.e. the parameters defining \eqnref{eq:oup}, but also when the field decay and fluctuating strength are mismatched (subplot (d)). In this later case, the EKF operates assuming that the fluctuation strength is half as strong, $q_K = q_\omega/2$, and the field decay is ten times faster, $\chi_K = 10\chi$. Although under these mismatched conditions the EKF covariance (dashed, yellow) is no longer reliable, the averaged estimation error (solid, green) remains unaffected and still reaches the quantum limit. This shows that even with imperfect knowledge of the OU process parameters, the magnetometer continues to operate optimally: measurement updates keep the estimator accurate, while the covariance becomes inconsistent because it is propagated using incorrect model assumptions.

\subsubsection{nonlinear MCG-like field} \label{sec:nonlinear_MCG_field}

\begin{summary}
In this subsection, we apply the sensing framework developed up to now to nonlinear waveforms resembling magnetocardiography (MCG) signals. The main results are the following:
    \begin{itemize}
        \item We use a Van der Pol oscillator to simulate a MCG-like signal. 
        \item We show that the EKF+LQR strategy can reconstruct full cardiac-like magnetic waveforms, including P-waves and QRS complexes, while filtering noise and maintaining estimation accuracy. 
    \end{itemize}
\end{summary}

Similarly to electrocardiography, atomic magnetocardiography (MCG)~\cite{Bison2009,jensen_magnetocardiography_2018,Kim2019,Yang2021} is a non-invasive technique for imaging the magnetic fields generated by the electrical activity of the heart~\cite{mcg_paper}. The goal in MCG is to reconstruct in real time the full magnetic waveform produced by the heart, capturing the characteristic P-wave, QRS-complex and T-wave~\cite{MCG_signal}, while filtering out unwanted stochastic noise~\cite{ECG_noise_removal} without resorting to extensive time-averaging~\cite{mcg_paper}.

\begin{figure}[t!]
    \centering
    \includegraphics[width=\textwidth]{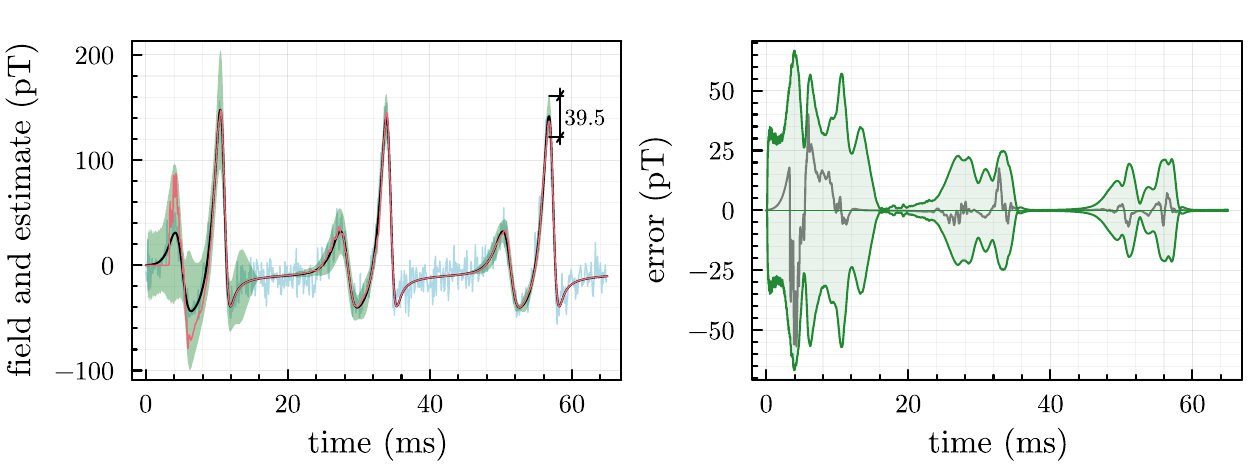} 
    \caption[Tracking a MCG-like signal]{\textbf{Tracking a MCG-like signal.} Under the same conditions as in \figref{fig:oup_estimation}, the magnetometer tracks a MCG signal in the pT range~\cite{Bison2009,Kim2019}. Although the atoms sense the noisy raw signal (blue), our goal is to recover the clean, noise-free waveform (black). The EKF, configured to expect a VdP–type signal~\cite{Kaplan2008,Das2013}, produces an estimate (red) that converges to the clean waveform after one MCG cycle ($\approx \SI{20}{\milli \second}$), as shown in the \textit{left} subplot. In it, the green-shaded area represents the $\pm 3 \sqrtaMSEomega$ confidence band, obtained after averaging over 1000 trajectories. Notably, at the R-wave in the third cycle, this band spans approximately $\SI{40}{\pico\tesla}$, implying an average error of roughly $\SI{6.6}{\pico\tesla}$. The subplot on the \textit{right} displays the difference (gray) between the clean waveform and the EKF estimate, with the same confidence bounds clearly delineating the error magnitude (green). }
    \label{fig:mcg_estimation}
\end{figure}

As in previous sections, we use the EKF+LQR feedback loop to estimate in real time the cardiac magnetic signal from the homodyne photocurrent, since this type of recurrent estimation and control is well suited for tracking such complex, time-varying signals. To simulate a realistic MCG signal, we use a filtered VdP oscillator~\cite{Kaplan2008,Das2013} as defined in \eqnref{eq:VdP}: 
\begin{align} \label{eq:VdP_2}
        &\dd \nu(t) = - p \, \omega(t) \,\dt, \\
        &\dd \omega(t) = \frac{k}{m} \nu(t) \,\dt + 2 \frac{c}{m} (1-\upsilon(t)) \, \omega(t) \,\dt, \\
        &\dd \upsilon(t) = \frac{|\nu(t)|-\nu(t)}{2 T} \,\dt - \frac{\upsilon(t)}{T} \,\dt,
\end{align}
with parameters $p, k, m, c, T > 0$. The VdP model effectively reproduces the P-wave and QRS complex, though it does not fully capture the T-wave. We then superimpose white noise onto the VdP-generated signal to mimic the stochastic disturbances observed in practice. The simulation parameters are adapted from \refcite{Kong2020} for our setting, as discussed in \appref{sec:experimental_parameters}, with the EKF parameters for the VdP estimate matching the ones used to generate the clean VdP signal. Namely, $p = p_{K} = 10^3$, $k =k_K = 1$, $m =m_K = 0.00098$, $c =c_K = 1$ and $T =T_K = 0.003$, with initial values: $\nu(0) = \omega(0) = \upsilon(0) = 0.0045$. The exact expressions for the EKF gradients $\pmb{F}(t)$, $\pmb{G}(t)$ and $\pmb{H}(t)$ needed to construct the EKF, are given in \appref{sec:exact_forms_FGH_VdP}.

\figref{fig:mcg_estimation} \emph{(left)} displays a cyclic MCG-like signal with a period of approximately $\SI{20}{\milli \second}$ and a field $B(t)$ ranging over $[\SI{-14.2}{\pico\tesla},\SI{42.6}{\pico\tesla}]$, compatible with human-heart fields~\cite{Bison2009,Kim2019} (where we have used the Rb-87 ground state hyperfine gyromagnetic ratio,  $2\pi \times \SI{7}{\giga\hertz\per\tesla}$). After an initial transient, the EKF estimate (red) closely follows the true waveform (black), despite the raw signal that the magnetometer actually senses (blue) being contaminated by white noise with a strength of $q_\omega=\SI{2.5e5}{\radian^2 \, \second^{-3}}$. 

In \figref{fig:mcg_estimation} \emph{(right)}, we further assess the tracking performance by plotting the difference between the true Larmor frequency and its EKF estimate (grey). This error is bounded by the squared root of the aMSE times three: $\pm 3\sqrtaMSEomega$ (green) \cite{crassidis2011optimal}, confirming that the estimation remains robust under noisy conditions.

\subsection{Multiparticle entanglement: conditional v.s. unconditional spin squeezing} \label{sec:cond_uncond_spin_squeezing}

\begin{summary}
    In this section:
    \begin{itemize}
        \item We distinguish between conditional and unconditional spin squeezing.
        \item Using the SME to simulate the magnetometer, we show that the EKF+LQR strategy not only generates strong conditional squeezing but also produces unconditional squeezing, even under collective dephasing. 
        \item These exact SME simulations confirm that LQR feedback drives the ensemble into an unconditional spin squeezed state, whereas naive field compensation fails to do so.
        \item Spherical Wigner functions illustrate this process where continuous measurement combined with LQR control squeeze the atoms and keep them pointing along $x$.
    \end{itemize}
\end{summary}

So far, our primary focus has been estimation, where conditional spin squeezing, i.e. squeezing of the atomic state dependent on the measurement record $\rhoc(t) \equiv \rho(t|\pmb{y}_{\leq t})$, has potential for enhancing precision beyond classical limits. However, instead of viewing the magnetometer solely as a sensing device, we can ask whether it can also function as a mechanism for preparing the system in a multipartite entangled state \emph{independent} of the measurement trajectory.

In that case, evaluating the spin squeezing parameter along a specific measurement trajectory (i.e.~\emph{conditional}), is no longer sufficient. Instead, we should quantify the entanglement of the state independent of our observations (i.e.~\emph{unconditional}). While the \emph{conditional} state of the system $\rhoc(t) \equiv \rho(t|\pmb{y}_{\leq t})$ is understood as the one most closely describing the state given a particular measurement record $\pmb{y}_{\leq t}$, an \emph{unconditional} state $\rho(t)$ describes the system when we discard, or do not have access to, the measurement outcomes, what formally corresponds to averaging the conditional state over all the possible past measurement trajectories:
\begin{equation}
    \rho(t) = \E{\rhoc(t)}{p(\pmb{y}_{\leq t})}.
\end{equation}

\begin{figure}[t!]
    \centering
    \includegraphics[width=\textwidth]{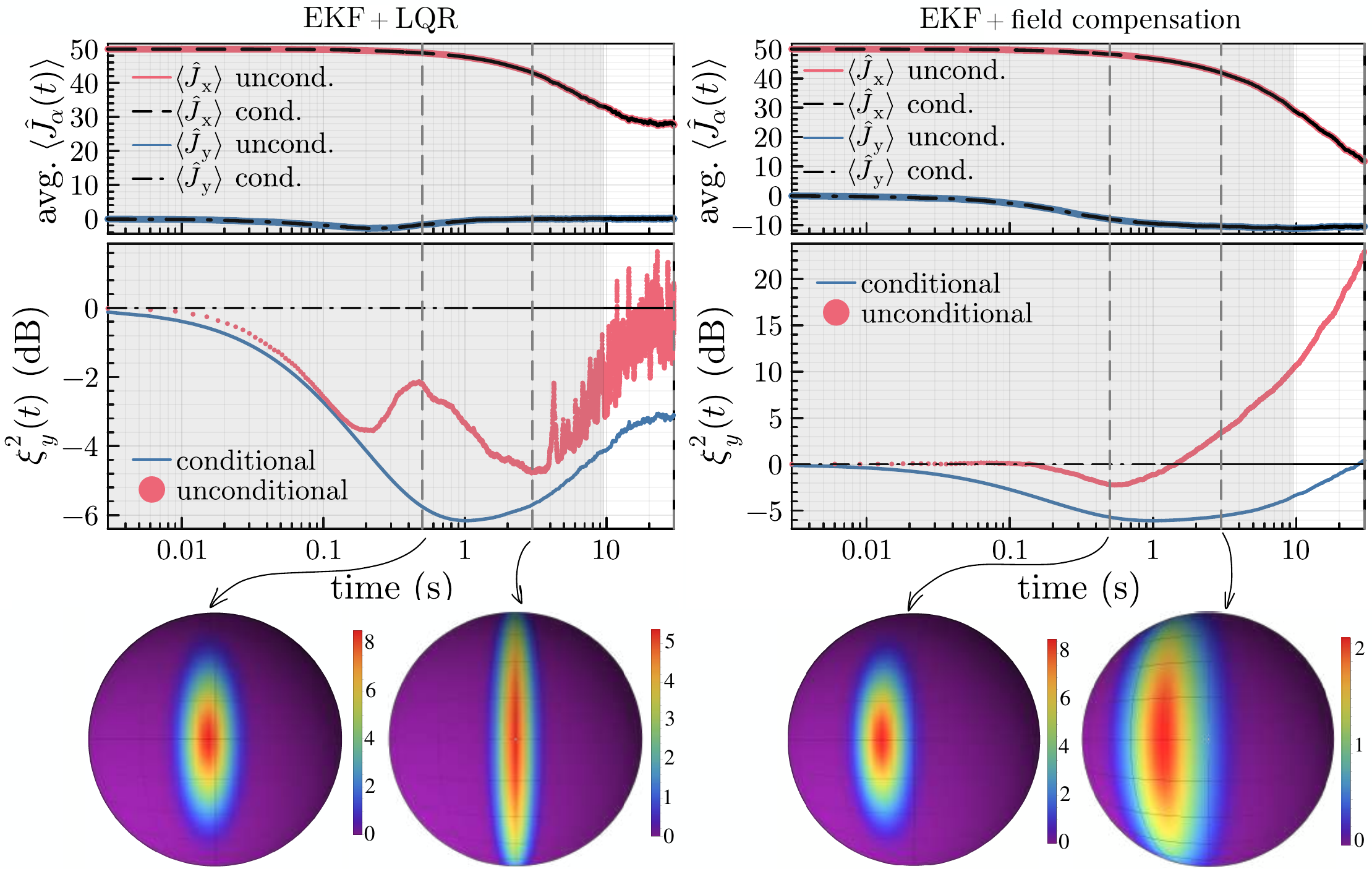}
    \caption[Conditional v.s.~unconditional spin squeezing]{\textbf{Conditional v.s.~unconditional spin squeezing.} Exact spin squeezing dynamics for $N=100$ atoms under collective decoherence and a constant magnetic field, shown for two control strategies:~LQR (\emph{left column}) vs (naive) field compensation (\emph{right column}). \emph{Top row:}
    subplots (\emph{a}) and (\emph{c}) display the evolution of angular momentum components along $x$ (red) and $y$ (blue) directions, where conditional and unconditional means consistently match. \emph{Middle row:} subplots (\emph{b}) and (\emph{d}) compare the average unconditional (red) and conditional (blue) spin squeezing parameters \eref{eq:spin_squeez_par}, also verifying whether they surpass the classical value (horizontal black line). Vertical dashed grey lines mark the relevant times for which we explicitly plot the spherical Wigner functions in the \emph{bottom row}. \emph{Bottom row:} spherical Wigner functions of the unconditional state at times $t = \SI{0.5}{\second}$ and $t= \SI{3}{\second}$. Note that for the LQR control (\emph{left}), even though the width along $y$ of the distribution progressively narrows with time, the amplitude of the Wigner function also decays. The other parameters used to simulate the SME in \eref{eq:fullSME} are:~$\kcoll = 0.005$, $\kloc = 0$, $M = 0.1$, $\omega = 1$, $\eta = 1$. The EKF is initialized with the mean $\Tilde{\pmb{x}}(0) = (N/2,0,0,N/4,N/4,0,\mu_0)^\TT$ and covariance $\Sigma(0) = \Diag{0,0,0,0,0,0,\sigma_{0}^2}$ dictated by the initial CSS state, and the Gaussian prior distribution for $\omega\sim\mathcal{N}(\mu_0,\sigma_0^2)$. All results are obtained after averaging over $\nu = 500$ measurement trajectories, while $\omega$-averaging is avoided by choosing the prior with $\sigma_0=0.5$ and $\mu_0 = \omega + \sigma_0=1.5$.
    }
    \label{fig:sq}
\end{figure}

Without feedback, the only effect experienced by the unconditional state $\rho(t)$ is a collective dephasing induced by the continuous measurement. For example, in the SME of \eqnref{eq:fullSME} with $u(t) = 0$, averaging over measurement records simply introduces an extra \mbox{$M\D[\Jy]$-term}. However, once feedback is turned on, determining an effective master equation for the unconditional dynamics is far less straightforward. In general, feedback-driven evolution relies on Markovianity assumptions~\cite{Wiseman1994_feedback,nurdin2017linear}, which are not satisfied in our LQR-based control scheme (see \secref{sec:LQR_ch5}). Nevertheless, as in Markovian feedback scenarios where the system is unconditionally driven into a spin-squeezed state~\cite{Wiseman1994_feedback,thomsen_continuous_2002}, we demonstrate that this is also the case under the non-Markovian LQR control considered here.

Already in \figref{fig:attaining_bounds} we showed that an EKF combined with a LQR generates unconditional spin squeezing, even in the presence of collective dephasing (red dots in \figref{fig:attaining_bounds}(c)) or local dephasing (red dots in \figref{fig:attaining_bounds}(d)). However, these results rely on the CoG approximation and the solution of the SDE system in \eqnref{eq:dynamical_model} to compute the unconditional moments of angular momentum, and thus, the unconditional spin squeezing. Since the CoG approximation circumvents the need for the full density matrix, it does not allow direct visualization of how continuous measurement and feedback squeeze the atoms. To obtain the Wigner function and map it into the generalized Bloch sphere, a full numerical solution of the SME in \eqnref{eq:fullSME} is required.  For this reason, in \figref{fig:sq} we present results from the exact SME \eref{eq:fullSME} under purely collective decoherence ($N=100$, $\kcoll=0.005$) to compare the EKF+LQR strategy \textit{(left column)} with the naive field compensation \textit{(right column)} in generating conditional as well as unconditional spin squeezing. As in \figref{fig:attaining_bounds}(c-d), \figref{fig:sq} shows the average conditional spin squeezing parameter (solid, blue) alongside the unconditional spin squeezing of the average state (red, dots). Notably, only the EKF+LQR strategy generates an averaged state with unconditional squeezing that consistently breaches the classical value (dash-dot black horizontal line). This is further confirmed by the (spherical) Wigner distribution plots (snapshots at $t=0.5,3$), where the EKF+LQR strategy maintains steady squeezing along the $y$-direction over time. In contrast, under the naive field-compensation strategy, the Wigner distribution begins to lose its structure as early as $t=3$ within the LQ regime.

As the $\omega$-estimate of the EKF is initially set in \figref{fig:sq} to $\est{\omega}(0)>\omega$, the control initially overcompensates for the Larmor precession and rotates the spin in the counter-clockwise direction when viewed along $z$. This is reflected in the spin components, with $\brkt{\Jy}=\mathbb{E}[\brktc{\Jy}]$ acquiring negative values in both \textit{top} plots, and in the corresponding Wigner function being shifted leftward at $\approx t=0.1$. A similar effect would occur when choosing $\est{\omega}(0)<\omega$, in which case the control operation would initially undercompensate the Larmor precession, causing an initial clockwise spin rotation around $z$ (Wigner function shifts to the right). Over time, the LQR control corrects the counter-rotation and stabilizes the spin along $x$ \textit{(left)}, whereas under naive field compensation \textit{(right)}, stability is eventually lost.

\chapter{Conclusions}
\label{chap:conclusions}

In this thesis, we have studied how to track fluctuating or time-varying signals with optical atomic magnetometers at the quantum limit of precision. A key result of our work is the derivation of this quantum limit by establishing a lower bound on the BCRB~\cite{Van-Trees,Fritsche2014,Amoros-Binefa2021}. Notably, the bound scales at best linearly with the number of atoms and sensing time, thus precluding any super-classical scaling~\cite{Amoros-Binefa2024,Amoros-Binefa2025}. Also sometimes referred to as the classically-simulated limit, the quantum limit is independent of the choices of initial state, measurement, estimator, or measurement-based feedback, relying solely on the decoherence model and strength of field fluctuations. Thus, by developing a measurement, estimation, and control strategy that saturates this limit, we demonstrate that the sensing protocol is optimal. 

But how can we develop an optimal sensing protocol? First, we need to be able to simulate the optical atomic magnetometer under typical experimental conditions using a quantum dynamical model that is scalable w.r.t. $N$. This model must rigorously incorporate both the measurement back-action and the environmental decoherence~\cite{Jimenez2018,Troullinou2021,Kong2020}. Second, we must select an estimator and controller that, when combined with the typical measurement (polarization spectroscopy~\cite{Kuzmich1999,Smith2003,Geremia2006,deutsch_quantum_2010}) and initial state (coherent spin state) used in optical atomic magnetometers, will enable us to attain the quantum limit.

The dynamics of an optical atomic magnetometer are rigorously described by a stochastic master equation~\cite{Wiseman1993,Stockton2004,Albarelli2017}, but it is not scalable w.r.t. $N$. To overcome this, we introduce the co-moving Gaussian approximation, which yields a nonlinear dynamical model that scales with $N$ while accurately capturing the effects of measurement backaction and dephasing~\cite{Amoros-Binefa2024}. To validate this approximation, we compared it to the solution of the exact SME, which is numerically tractable for up to $N \approx 100$ atoms~\cite{Amoros-Binefa2024}. Building on this model, we have designed an integrated estimation and control scheme that couples an extended Kalman filter (EKF) with a linear quadratic regulator (LQR)~\cite{Stockton2004,Amoros-Binefa2024,Amoros-Binefa2025}. 

We show that this sensing protocol achieves real-time tracking of constant and fluctuating fields at the quantum limit of precision~\cite{Amoros-Binefa2024,Amoros-Binefa2025}. It also can track more complex waveforms such as those encountered in magneto-cardiography~\cite{Amoros-Binefa2025}. Furthermore, we also demonstrate that the continuous measurement and feedback generate inter-atomic entanglement, manifest as conditional spin-squeezing~\cite{Amoros-Binefa2024,Amoros-Binefa2025}. Remarkably, even if the measurement data is not stored, the sensor is driven by the LQR feedback into an entangled state~\cite{Amoros-Binefa2024}. 

In this work, we have aimed to bridge the gap between mathematical formulations of continuous quantum measurement and estimation theory, and experiments in optical atomic magnetometry. Our results mark a significant step towards real-time quantum-limited metrology, even in the presence of decoherence, by leveraging continuous measurement, feedback, and spin-squeezing. To apply our framework to experiments with orientation-based magnetometers in the Bell-Bloom configuration~\cite{Bell1961,Troullinou2021,Troullinou2023}, or alignment-based magnetometers~\cite{Ledbetter2007,Weis2006,Kozbial2024}, the dynamical model must be extended to incorporate spin-exchange collisions~\cite{Happer1977,Savukov2005} and other relevant sources of atomic decoherence~\cite{Happer2010}. These additions would require more general techniques to bound the Fisher information~\cite{Gorecki2025}, beyond those used in this thesis.

\begin{appendices}
    \chapter{Appendix for \chapref{chap:pre}}
\label{AppendixA}

\section{Proof of $(\dW(t))^{2+N} = 0$} \label{ap:dW^2+N=0_proof}

\begin{prop}[Proof of $(\dW(t))^{2+N} = 0$]
    Consider a standard Wiener process $\W(t)$, then for any $N > 0$, we have:
    \begin{equation}
        (\dW(t))^{2+N} = 0.
    \end{equation}
    To interpret this rigorously, we need to write it in terms of the It\^{o} stochastic integral. Specifically, let $f(t,\mrm{X}(t))$ be a non-anticipating function of time and $\mrm{X}(t)$, a stochastic process. Then, the following holds
    \begin{equation} \label{eq:dW^2+N=0_eq}
        \!\!\int_0^T \! f\left(t,\mrm{X}(t)\right) \dW(t)^{2+N} = \underset{n\to\infty}{\text{ms-lim}} \sum_{i=1}^n \, f\left(t_{i\shortminus1},\mrm{X}(t_{i\shortminus1})\right) \Delta\!\W_i^{2+N} = 0 \;\; \text{for} \;\; N>0.
    \end{equation}
\end{prop}

\begin{myproof}
    To prove \eqnref{eq:dW^2+N=0_eq}, we have to start with the definition of the mean-squared limit and show:
    \begin{equation} \label{eq:ap_def_mslim_1}
        \lim_{n\to\infty} \EE{\left(S_n - S\right)^2} = \lim_{n\to\infty} \EE{S_n^{\,2}} =  0,
    \end{equation}
    where $S = 0$ and
    \begin{equation}
        S_n = \sum_{i=1}^n \, f\left(t_{i\shortminus1},\mrm{X}(t_{i\shortminus1})\right) \Delta\!\W_i^{2+N}. 
    \end{equation}
    Thus, let us start by expanding $S_n^{\,2}$ using \eqnref{eq:identity_sumx2}:
    \begin{align}
        &\lim_{n\to\infty} \EE{S_n^{\,2}} = \lim_{n\to\infty} \EE{\left(\sum_{i=1}^n \, f\left(t_{i\shortminus1},\mrm{X}(t_{i\shortminus1})\right) \Delta\!\W_i^{2+N}\right)^{\!\!\!2}} \nonumber \\
        &\quad = \lim_{n\to\infty} \EE{\sum_{i=1}^n \left(\,f_{i\shortminus1} \Delta\!\W_i^{2+N}\right)^2 + 2\!\sum_{i=1}^n\!\sum_{j=1}^{i-1} \,f_{i\shortminus1} \,f_{j\shortminus1} \Delta\!\W_i^{2+N} \Delta \!\W_j^{2+N}} \nonumber \\
        &\quad = \lim_{n\to\infty} \left\{ \sum_{i=1}^n \EE{f_{i\shortminus1}^{\;2}} \EE{(\Delta\!\W_i^2)^{2+N}} + 2\!\sum_{i=1}^n\!\sum_{j=1}^{i-1} \EE{f_{i\shortminus1}f_{j\shortminus1}} \EE{\Delta\!\W_i^{2+N}} \mrm{E}\big[\Delta\!\W_j^{2+N}\big]\right\}. \nonumber
    \end{align}
    Let us now recall that the higher order moments of a Gaussian random variable can be written as:
    \begin{align}
        &\EE{\left(\Delta\!\W_i^2\right)^{2+N}} = \Dt_i^{2+N}, \\
        &\EE{\Delta\!\W_i^{2+N}} = \begin{cases}
            0 \quad &\text{for} \quad N \; \text{odd},\\
            (N+1)!! \, \Dt_i^{\frac{N}{2}+1} \quad &\text{for} \quad N \; \text{even}.
        \end{cases}
    \end{align}
    
    Note that for any value of $N$, the expectation values of $\Delta\!\W_i^{2+N}$ and $(\Delta\!\W_i^2)^{2+N}$ will be either zero or at least $\Dt_i^2$. Therefore, $\forall N$, in the limit of $n\to\infty$, $\EE{S_n^{\;2}}$ goes to zero:
    \begin{equation}
        \lim_{n\to\infty} \EE{S_n^{\,2}} = 0,
    \end{equation}
    and thus
    \begin{equation}
        \!\!\int_0^T \! f\left(t,\mrm{X}(t)\right) \dW(t)^{2+N} = 0 \;\; \text{for} \;\; N>0,
    \end{equation}
    s.t.
    \begin{equation}
        \dW(t)^{2+N} = 0.
    \end{equation}
\end{myproof}

\section{Proof of $\dW(t) \dt = 0$} \label{ap:dWdt=0_proof}

\begin{prop}[Proof of $\dW(t) \dt = 0$]
    Consider a standard Wiener process $\W(t)$, then we have:
    \begin{equation}
        \dW(t) \dt = 0.
    \end{equation}
    Similarly as before, we write this it in terms of the It\^{o} stochastic integral:
    \begin{equation} \label{eq:dWdt=0_eq}
        \!\!\int_0^T \! f\left(t,\mrm{X}(t)\right) \dW(t) \dt = \underset{n\to\infty}{\text{ms-lim}} \sum_{i=1}^n \, f\left(t_{i\shortminus1},\mrm{X}(t_{i\shortminus1})\right) \Delta\!\W_{\!i} \, \Dt_i = 0,
    \end{equation}
    where $f(t,\mrm{X}(t))$ is a non-anticipating function of time and $\mrm{X}(t)$ is a general a stochastic process. 
\end{prop}

\begin{myproof}
    To prove \eqnref{eq:dWdt=0_eq}, we have to follow the same steps as in \appref{ap:dW^2+N=0_proof}. Since the stochastic integral $S$ is defined as the mean-square limit of the Riemann sum $S_n$, we have to start from the definition of convergence in the mean-squared sense:
    \begin{equation} \label{eq:ap_def_mslim_2}
        \lim_{n\to\infty} \EE{\left(S_n - S\,\right)^{\!2}} = 0,
    \end{equation}
    and identify $S$ and $S_n$. It is straightforward to see that for \eqnref{eq:dWdt=0_eq}, $S = 0$ and
    \begin{equation}
        S_n = \sum_{i=1}^n \, f\left(t_{i\shortminus1},\mrm{X}(t_{i\shortminus1})\right) \Delta\!\W_i \, \Dt_i. 
    \end{equation}
    Then, it follows that \eqnref{eq:ap_def_mslim_2} is simply
    \begin{equation} 
        \lim_{n\to\infty} \EE{S_n^{\,2}} = 0,
    \end{equation}
    which we can easily show by expanding its l.h.s. using \eqnref{eq:identity_sumx2}: 
    \begin{align}
        &\lim_{n\to\infty} \EE{S_n^{\,2}} = \lim_{n\to\infty} \EE{\left(\sum_{i=1}^n \, f\left(t_{i\shortminus1},\mrm{X}(t_{i\shortminus1})\right) \Delta\!\W_i \, \Dt_i\right)^{\!\!\!2}} \nonumber \\
        &\quad = \lim_{n\to\infty} \EE{\sum_{i=1}^n \left(\,f_{i\shortminus1} \Delta\!\W_i \, \Dt_i \right)^2 + 2\!\sum_{i=1}^n\!\sum_{j=1}^{i-1} \,f_{i\shortminus1} \,f_{j\shortminus1} \Delta\!\W_i \, \Dt_i\Delta\!\W_j \, \Dt_j} \nonumber \\
        &\quad = \lim_{n\to\infty} \left\{ \sum_{i=1}^n \EE{f_{i\shortminus1}^{\;2}} \EE{(\Delta\!\W_i^2)} \Dt_i^2+ 2\!\sum_{i=1}^n\!\sum_{j=1}^{i-1} \EE{f_{i\shortminus1}f_{j\shortminus1}} \EE{\Delta\!\W_i} \mrm{E}\big[\Delta\!\W_j\big] \Dt_i \Dt_j \right\} \nonumber \\
        &\quad =  \lim_{n\to\infty} \sum_{i=1}^n \EE{f_{i\shortminus1}^{\;2}} \Dt_i^3 + 0 = 0.
    \end{align}
    Thus
    \begin{equation}
        \!\!\int_0^T \! f\left(t,\mrm{X}(t)\right) \dW(t) \dt = 0,
    \end{equation}
    s.t.
    \begin{equation}
        \dW(t) \dt = 0.
    \end{equation}
\end{myproof}

\section{Solution to forced linear stochastic systems}

\begin{theorem}[General solution of time‑varying inhomogeneous linear stochastic differential equations] \label{theorem:sol_inhom_diff}
    Consider the following time-varying inhomogeneous stochastic differential equation
    \begin{equation}
        \dot{\pmb{x}}(t) = \pmb{F}(t) \, \pmb{x}(t) + \pmb{B}(t) \, \pmb{u}(t) + \pmb{G}(t) \, \pmb{w}(t)
    \end{equation}
    where $\pmb{F}(t)$ is the system matrix, $\pmb{x}(t)$ is the state vector, $\pmb{B}(t)$ is the control matrix, $\pmb{u}(t)$ is the control vector, $\pmb{G}(t)$ is the time-varying matrix modifying the stochastic term, and $\pmb{w}(t)$ is the white noise with mean zero and covariance $\EE{\pmb{w}(t) \pmb{w}(s)^\Trans} = \pmb{Q}(t) \delta(t-s)$. The solution is
    \begin{equation} \label{eq:sol_x_dynamical_system}
        \pmb{x}(t) = \pmb{\Phi}(t,t_0) \, \pmb{x}(t_0) + \int_{t_0}^t \pmb{\Phi}(t,\tau) \pmb{B}(\tau) \pmb{u}(\tau) \dd \tau + \int_{t_0}^t \pmb{\Phi}(t,\tau) \pmb{G}(\tau) \pmb{w}(\tau) \dd \tau,
    \end{equation}
    where $\pmb{\Phi}(t,t_0)$ is the state-transition matrix that satisfies
    \begin{equation} \label{eq:def_phi_transition_matrix}
        \frac{\dd \pmb{\Phi}(t,t_0)}{\dd t} = \pmb{F}(t) \pmb{\Phi}(t,t_0), \;\;\;\;\; \pmb{\Phi}(t_0,t_0) = \I.
    \end{equation}
    and fulfills the following properties for all $t_0 \leq t \leq T$:
    \begin{align}
        \pmb{\Phi}(t_0,t) = \pmb{\Phi}^{-1}(t,t_0) , \label{eq:phi_prop_inv}\\
        \pmb{\Phi}(T,t_0) = \pmb{\Phi}(T,t) \pmb{\Phi}(t,t_0) \label{eq:phi_prop_concat}.
    \end{align}
\end{theorem}

\begin{myproof}
    The standard approach to solving any forced linear system is to first derive the homogeneous solution. Namely, solving the homogeneous system of the form
    \begin{equation} \label{eq:homo_diff_eq}
        \frac{\dd\pmb{x}}{\dd t} = \pmb{F}(t) \, \pmb{x}(t), \;\;\; \pmb{x}(t_0) = \pmb{x}_0.
    \end{equation}
    Let us assume then that the system's response in the absence of external forces is given by
    \begin{equation}
        \pmb{x}(t) = \pmb{\Phi}(t,t_0) \pmb{x}(t_0),
    \end{equation}
    where $\pmb{\Phi}(t,t_0)$ is a so-called transition matrix. By substituting this solution into the homogeneous differential equation in \eref{eq:homo_diff_eq}, we obtain a differential equation for the transition matrix with its corresponding initial conditions:
    \begin{equation}
        \frac{\dd \pmb{\Phi} (t,t_0)}{\dd t} = \pmb{F}(t) \pmb{\Phi}(t,t_0), \;\;\; \pmb{\Phi}(t_0,t_0) = \I.
    \end{equation}
    Let us now consider the full dynamical model
    \begin{equation} \label{eq:model_for_proof}
        \dot{\pmb{x}}(t) = \pmb{F}(t) \, \pmb{x}(t) + \pmb{B}(t) \, \pmb{u}(t) + \pmb{G}(t) \, \pmb{w}(t),
    \end{equation}
    for which, following the method of variation of parameters, we assume a solution of the form:
    \begin{equation} \label{eq:deff_x_nonhom}
        \pmb{x}(t) = \pmb{\Phi}(t,t_0) \pmb{z}(t), \;\;\; \pmb{z}(t_0) = \pmb{x}(t_0),
    \end{equation}
    where $\pmb{z}(t)$ is a vector of unknown functions that modifies the homogeneous solution to account for the non-homogeneous terms in the full differential equation. If now we differentiate \eqnref{eq:deff_x_nonhom} we will get
    \begin{equation} \label{eq:differential_of_x}
        \dot{\pmb{x}}(t) = \pmb{\Phi}(t,t_0) \, \dot{\pmb{z}}(t) + \dot{\pmb{\Phi}}(t,t_0) \, \pmb{z}(t) = \pmb{\Phi}(t,t_0) \, \dot{\pmb{z}}(t) + \pmb{F}(t) \pmb{\Phi}(t,t_0) \, \pmb{z}(t)  .
    \end{equation}
    Let us now substitute \eqnref{eq:deff_x_nonhom} and \eqnref{eq:differential_of_x} in the lhs and rhs of \eqnref{eq:model_for_proof}:
    \begin{align}
        \pmb{\Phi}(t,t_0) \, \dot{\pmb{z}}(t) + \pmb{F}(t) \pmb{\Phi}(t,t_0) \, \pmb{z}(t) = \pmb{F}(t) \, \pmb{\Phi}(t,t_0) \, \pmb{z}(t) + \pmb{B}(t) \, \pmb{u}(t) + \pmb{G}(t) \, \pmb{w}(t)
    \end{align}
    such that
    \begin{align}
        \dot{\pmb{z}}(t) = \pmb{\Phi}^{-1}(t,t_0) \pmb{B}(t) \, \pmb{u}(t) +\pmb{\Phi}^{-1}(t,t_0) \pmb{G}(t) \, \pmb{w}(t).
    \end{align}
    By now integrating this expression and recalling that $\pmb{z}(t_0) = \pmb{x}(t_0)$, we obtain
    \begin{align}
        \int_{t_0}^t \dot{\pmb{z}}(t) &= \pmb{z}(t) - \pmb{z}(t_0) = \pmb{z}(t) - \pmb{x}(t_0) = \pmb{\Phi}^{-1}(\tau,t_0) \, \pmb{x}(t) - \pmb{x}(t_0) \nonumber \\
        &=\int_{t_0}^t  \pmb{\Phi}^{-1}(\tau,t_0) \pmb{B}(\tau) \, \pmb{u}(\tau) \, \dd \tau 
        + \int_{t_0}^t \pmb{\Phi}^{-1}(\tau,t_0) \pmb{G}(\tau) \, \pmb{w}(\tau) \, \dd \tau \label{eq:almost_sol_x(t)}
    \end{align}
    such that
    \begin{align}
        \pmb{x}(t) &= \pmb{\Phi}(t,t_0) \pmb{x}(t_0) + \pmb{\Phi}(t,t_0) \int_{t_0}^t  \pmb{\Phi}^{-1}(\tau,t_0) \pmb{B}(\tau) \, \pmb{u}(\tau) \, \dd \tau  \\
        &+  \pmb{\Phi}(t,t_0) \int_{t_0}^t \pmb{\Phi}^{-1}(\tau,t_0) \pmb{G}(\tau) \, \pmb{w}(\tau) \, \dd \tau.
    \end{align}
    By now using \eqnref{eq:phi_prop_concat}, and first multiplying from the right $\pmb{\Phi}^{-1}(\tau,t_0)$ and then from the left $\pmb{\Phi}^{-1}(t,t_0)$:
    \begin{align}
        &\pmb{\Phi}^{-1}(t,t_0) \pmb{\Phi}(t,t_0) \pmb{\Phi}^{-1}(\tau,t_0) = \pmb{\Phi}^{-1}(t,t_0) \pmb{\Phi}(t,\tau)\pmb{\Phi}(\tau,t_0) \pmb{\Phi}^{-1}(\tau,t_0)
    \end{align}
    we obtain
    \begin{equation}
        \pmb{\Phi}^{-1}(\tau,t_0) = \pmb{\Phi}^{-1}(t,t_0) \pmb{\Phi}(t,\tau)
    \end{equation}
    which when substituted into \eqnref{eq:almost_sol_x(t)} yields the final expression of \eqnref{eq:sol_x_dynamical_system}:
    \begin{equation}
        \pmb{x}(t) = \pmb{\Phi}(t,t_0) \, \pmb{x}(t_0) + \int_{t_0}^t \pmb{\Phi}(t,\tau) \pmb{B}(\tau) \pmb{u}(\tau) \dd \tau + \int_{t_0}^t \pmb{\Phi}(t,\tau) \pmb{G}(\tau) \pmb{w}(\tau) \dd \tau.
    \end{equation}
\end{myproof}

\section{(some) Angular momentum properties}

\begin{prop}[Generation of angular momentum eigenstates via the raising operator] \label{prop:exp_jm_asJpfromj-j}
    The angular momentum eigenstate $\ket{j,m}$ where $j$ is the total angular momentum quantum number and $m$ the magnetic quantum number, can be generated from applying the raising operator $\Jp$ to the lowest eigenstate $\ket{j,-j}$ as follows \cite{Arecchi1972}:
    \begin{equation}
        \ket{j,m} = \frac{1}{(m+j)!} \binom{2j}{m+j}^{\!\!-1/2} \! \Jp^{\,m+j} \ket{j,-j},
    \end{equation}
    where $\Jp$ is the angular momentum raising operator, defined as $\Jp = \Jx + i \Jy$.
\end{prop}

\begin{myproof}
    Recall how $\Jp$ acts on $\ket{j,m}$:
    \begin{align}
        \Jp \ket{j,m} = \sqrt{j(j+1)-m(m+1)} \ket{j,m+1}.
    \end{align}
    Note that the factor inside the square root can be rewritten as:
    \begin{align}
        j(j+1)-m(m+1) = (j+m+1)(j-m).
    \end{align}
    Therefore, applying the operator $\Jp$, $j+m$ times, in order to raise the state $\ket{j,-j}$ to $\ket{j,m}$, as sketched in \tabref{tab:ladder_op_summary}, yields:
    \begin{align}
        \Jp^{\, m+j} \ket{j,-j} = \sqrt{(j+m)! \frac{(2j)!}{(j-m)!}} \ket{j,m},
    \end{align}
    which can be rewritten using a binomial notation, since
    \begin{align}
        (j+m)! \binom{2j}{m+j}^{\!\!\!1/2} \!\!\!= (j+m)! \sqrt{\frac{(2j)!}{(j+m)!(j-m)!}} = \sqrt{(j+m)!\frac{(2j)!}{(j-m)!}}.
    \end{align}
    Hence,
    \begin{equation}
        \ket{j,m} = \frac{1}{(j+m)!} \binom{2j}{m+j}^{\!\!\!-1/2}  \Jp^{\, m+j} \ket{j,-j}.
    \end{equation}
\end{myproof}

\begin{table}[ht]
    \centering
    \tikzmath{real \r; \r=3.1; }
    \begin{tikzpicture}[scale=1]
        
        \node[anchor=north west] at (-2.2, 1) {(a)};
        
        \draw[thick] (0, 0) -- (0, -4);

        \node at (-0.5, 0) {\( \vert j, j \,\rangle \;\;\;\;\;\;\; \)};
        \node at (-0.5, -1.5) {\( \vert j, m \rangle  \;\;\;\;\; \)};
        \node at (-0.5, -4) {\( \vert j, -j \,\rangle  \;\;\;\;  \)};

        \draw[thick] (-0.2, 0) -- (0.2, 0);
        \draw[thick] (-0.2, -1.5) -- (0.2, -1.5);
        \draw[thick] (-0.2, -4) -- (0.2, -4);

        \draw[->, thick, decorate, decoration={snake, amplitude=1.5mm, segment length=5mm}] (0.4, -4) -- (0.4, -1.5) node[midway, right] {\( \,\Jp^{\,j+m} \)}; 
        
        \node[anchor=north west] at (\r - 0.6, 0.3) {(b)};
        \path (\r, -2) node[right, scale=1] {%
            \begin{tabular}{c|c}
              $m$ & $(j+m+1)(j-m)$ \\
              \hline 
               $-j$  & $1 \cdot 2j$ \\
               $-j+1$ & $2(2j-1)$ \\
               $-j+2$ & $3(2j-2)$ \\
               \vdots & \vdots \\
               $m-1$ & $(j+m)(j-m+1)$
            \end{tabular}
        };
      \end{tikzpicture}
      \caption[Raising and lowering operators for the angular momentum]{\textbf{Raising and lowering operators for the angular momentum.} (a) Scheme depicting the transition from state $\ket{j,-j}$ to $\ket{j,m}$ by applying the raising operator $\Jp$ $j+m$ times. (b) Table summarizing the factors that appear every time we apply the $\Jp$ operator to state $\ket{j,-j}$ and subsequent states. }
    \label{tab:ladder_op_summary}
\end{table}

\subsection{Coherent spin state means and variances} \label{sec:CSS_m_v}

\subsubsection{Means}

First, we will compute the term $\braketavg{\Jx}_{\trm{CSS}_x}$. To do so, let us write $\Jx$ in terms of the ladder operators, $\Jx = \frac{1}{2}(\Jp + \Jm)$, such that:
\begin{eqnarray}
    \braketavg{\Jx}_{\trm{CSS}_x} = \braketop{\eta}{\Jx}{\eta} = \frac{1}{2}\big(\braketop{\eta}{\Jp}{\eta} + \braketop{\eta}{\Jm}{\eta}\big).
\end{eqnarray}
Then, the expected value of $\Jp$ w.r.t a CSS can be computed to be $N/4$:
\begin{align}
    &\braketop{\eta}{\Jp}{\eta} = \nonumber \\ 
    &= \frac{1}{2^{N+1}} \!\!\! \sum_{n,m=-\sfrac{N}{2}}^{\sfrac{N}{2}} \! \binom{N}{\frac{N}{2} + n}^{\!\!1/2} \!\! \binom{N}{\frac{N}{2} + m}^{\!\!1/2} \!\!\sqrt{\!\frac{N}{2} \!\left(\!\frac{N}{2} \!+\! 1\!\right) \!-\! m(m+1)}  \braketadj{n,\frac{N}{2}}{\frac{N}{2},m+1} \nonumber \\
    &= \frac{1}{2^{N+1}} \sum_{m=-\sfrac{N}{2}}^{\sfrac{N}{2}} \binom{N}{\frac{N}{2} + m + 1}^{\!\!1/2} \binom{N}{\frac{N}{2} + m}^{\!\!1/2} \sqrt{\frac{N}{2} \left(\frac{N}{2} + 1\right) - m(m+1)}  \nonumber \\
    &\stackrel{\ref{id:id1and2}}{=} \frac{1}{2^{N+1}} \sum_{m=-\sfrac{N}{2}}^{\sfrac{N}{2}} \bigg(\frac{N}{2} - m\bigg) \binom{N}{\frac{N}{2} + m} \stackrel{\ref{id:id3} \text{ \& } \ref{id:id6}}{=}  \frac{1}{2^{N+1}}  N 2^{N-1}  = \frac{N}{4},
\end{align}
and similarly for $\Jm$:
\begin{align}
    &\braketop{\eta}{\Jm}{\eta} =  \nonumber \\
    &= \frac{1}{2^{N+1}} \!\! \sum_{n,m=-\sfrac{N}{2}}^{\sfrac{N}{2}} \!\!\binom{N}{\frac{N}{2} + n}^{\!\!1/2} \!\!\binom{N}{\frac{N}{2} + m}^{\!\!1/2} \!\! \sqrt{\!\frac{N}{2} \!\left(\!\frac{N}{2} \!+\! 1\!\right) \!-\! m(m\!-\!1)} \braketadj{n,\frac{N}{2}}{\frac{N}{2},m-1} \nonumber \\
    &= \frac{1}{2^{N+1}} \sum_{m=-\sfrac{N}{2}}^{\sfrac{N}{2}} \binom{N}{\frac{N}{2} + m - 1}^{\!\!1/2} \binom{N}{\frac{N}{2} + m}^{\!\!1/2} \sqrt{\frac{N}{2} \left(\frac{N}{2} + 1\right) - m(m-1)} \nonumber \\
    &\stackrel{\ref{id:id1and2}}{=} \frac{1}{2^{N+1}} \sum_{m=-\sfrac{N}{2}}^{\sfrac{N}{2}} \bigg(\frac{N}{2} + m\bigg) \binom{N}{\frac{N}{2} - m}  \stackrel{\ref{id:id3} \text{ \& } \ref{id:id6}}{=}  \frac{1}{2^{N+1}}  N 2^{N-1}  = \frac{N}{4}.
\end{align}
Hence,
\begin{equation}
    \braketavg{\Jx}_{\trm{CSS}_x} = \frac{N}{2}.
\end{equation}
The other terms $\braketavg{\Jy}_{\trm{CSS}_x}$ and $\braketavg{\Jz}_{\trm{CSS}_x}$ are straightforward:
\begin{equation}
    \braketavg{\Jy}_{\trm{CSS}_x} = \frac{1}{2i}\big(\braketop{\eta}{\Jp}{\eta} - \braketop{\eta}{\Jm}{\eta}\big) = 0,
\end{equation}
and 
\begin{align}
    &\braketavg{\Jz}_{\trm{CSS}_x} = \braketop{\eta}{\Jz}{\eta} = \frac{1}{2^{N}} \hspace{-.1cm} \sum_{n=-N/2}^{N/2} \sum_{m=-N/2}^{N/2} \hspace{-.1cm} \binom{N}{\frac{N}{2} + n}^{\!\!1/2} \hspace{-.1cm} \binom{N}{\frac{N}{2} + m}^{\!\!1/2} m \braketadj{n,\frac{N}{2}}{\frac{N}{2},m}  \nonumber \\
    &= \frac{1}{2^N} \hspace{-.2cm} \sum_{m=-N/2}^{N/2} \hspace{-.2cm} m \binom{N}{\frac{N}{2} + m} \stackrel{\ref{id:id6}}{=}  0.
\end{align}

\subsubsection{Variances}

Let us start with $\braketavg{\Delta^2\Jx}_{\trm{CSS}_x} = \braketop{\eta}{\Jx^2}{\eta} - \braketop{\eta}{\Jx}{\eta}^2$. Note that,
\begin{equation}
    \braketavg{\Delta^2\Jx}_{\trm{CSS}_x} = \braketop{\eta}{\Jx^2}{\eta} - \braketop{\eta}{\Jx}{\eta}^2 = \braketop{\eta}{\Jx^2}{\eta} - \frac{N^{\,2}}{4}.
\end{equation}
Thus, it only remains to compute the term $\braketop{\eta}{\Jx^2}{\eta}$, i.e.,
\begin{eqnarray}
    \braketop{\eta}{\Jx^2}{\eta} = \frac{1}{4} \big( \braketop{\eta}{\Jp\Jp}{\eta} + \braketop{\eta}{\Jp\Jm}{\eta}+  \braketop{\eta}{\Jm\Jp}{\eta} +  \braketop{\eta}{\Jm\Jm}{\eta} \big).
\end{eqnarray}
To calculate the variances of a CSS polarized along the $x$-axis, the first term, when evaluated for a CSS as defined in \eqnref{eq:CSS_along_x_jmbasis}, reads as:
\begin{equation}
    \braketop{\eta}{\Jp\Jp}{\eta} = \frac{1}{2^N} \!\!\! \sum_{n,m=-\sfrac{N}{2}}^{\sfrac{N}{2}} \!\! \binom{N}{\frac{N}{2} + n}^{\!\!1/2} \binom{N}{\frac{N}{2} + m}^{\!\!1/2} \braketopadj{n,\frac{N}{2}}{\Jp\Jp}{\frac{N}{2},m}.
\end{equation}
Note that $\Jp$ applied to the bra representing the angular state with magnetic quantum number $n$, can be rewritten as:
\begin{align}
    &\bra{n,\frac{N}{2}} \Jp = \bra{n,\frac{N}{2}} \Jm^{\dagger} = \left( \!\Jm \ket{\frac{N}{2},n} \right)^{\!\dagger} \!=\! \left( \! \sqrt{\frac{N}{2} \!\left(\frac{N}{2} \!+\! 1\right) \!-\! n(n\!-\!1)} \ket{\frac{N}{2},n\!-\!1} \!\right)^{\!\!\dagger} \nonumber \\
    &= \bra{n-1,\frac{N}{2}} \sqrt{\frac{N}{2} \left(\frac{N}{2} + 1\right) - n(n-1)}.
\end{align}

Therefore,
\begin{align}
    \braketop{\eta}{\Jp\Jp}{\eta} &= \frac{1}{2^N} \!\! \sum_{n,m=-\sfrac{N}{2}}^{\sfrac{N}{2}} \!\! \binom{N}{\frac{N}{2} + n}^{\!\!1/2} \binom{N}{\frac{N}{2} + m}^{\!\!1/2} \! \braketopadj{n,\frac{N}{2}}{\Jp\Jp}{\frac{N}{2},m} = \nonumber \\
    &= \frac{1}{2^N} \!\! \sum_{n,m=-\sfrac{N}{2}}^{\sfrac{N}{2}} \! \binom{N}{\frac{N}{2} + n}^{\!\!1/2} \! \binom{N}{\frac{N}{2} + m}^{\!\!1/2} \!\! \sqrt{\frac{N}{2} \!\left(\frac{N}{2} \!+\! 1\right) \!-\! n(n-1)} \times \nonumber \\
    &\times \sqrt{\frac{N}{2} \! \left(\frac{N}{2} \!+\! 1\right) \!-\! m(m+1)}  \braketadj{n-1,\frac{N}{2}}{\frac{N}{2},m+1} = \nonumber \\
    & = \frac{1}{2^N} \!\! \sum_{m=-\sfrac{N}{2}}^{\sfrac{N}{2}} \!\! \binom{N}{\frac{N}{2} + m + 2}^{\!\!1/2} \!\! \binom{N}{\frac{N}{2} + m}^{\!\!1/2} \!\! \sqrt{\frac{N}{2}\!\left(\!\frac{N}{2} \!+\! 1 \right) \!-\! (m+2)(m+1)}  \times \nonumber \\
    &\times \sqrt{\frac{N}{2}\bigg(\frac{N}{2} + 1 \bigg) - m(m+1)} = \frac{1}{2^N} \!\! \sum_{m=-\sfrac{N}{2}}^{\sfrac{N}{2}} \frac{N!}{\Big(\frac{N}{2} + m\Big)!\Big(\frac{N}{2}-m - 2\Big)!} = \nonumber \\
    &= \frac{1}{2^N} \sum_{m=-\sfrac{N}{2}}^{\sfrac{N}{2}} \bigg(\frac{N}{2} - m\bigg) \bigg(\frac{N}{2} - m - 1\bigg)\binom{N}{\frac{N}{2} + m}  =\nonumber \\
    &= \frac{1}{2^{N+2}} N (N - 2) \!\! \sum_{m=-\sfrac{N}{2}}^{\sfrac{N}{2}} \binom{N}{\frac{N}{2} + m} + \frac{1}{2^N} \sum_{m=-\sfrac{N}{2}}^{\sfrac{N}{2}} \!\! m^2 \binom{N}{\frac{N}{2} + m}  = \nonumber \\
    &= N(N-2)2^{-2} + N 2^{-2} = N (N - 2 + 1)2^{-2} = \frac{N}{4}(N-1).
\end{align}
The rest of the terms are:
\begin{align}
    \braketop{\eta}{\Jp\Jm}{\eta} &= \frac{1}{2^N} \sum_{n=-N/2}^{N/2} \binom{N}{\frac{N}{2} + n}^{\!\!1/2} \binom{N}{\frac{N}{2} + m}^{\!\!1/2} \sqrt{\frac{N}{2}\bigg(\frac{N}{2} + 1 \bigg) - n(n-1)} \times \nonumber \\ 
    &\times \sqrt{\frac{N}{2}\bigg(\frac{N}{2} + 1 \bigg) - m(m-1)} \braketadj{n-1,\frac{N}{2}}{\frac{N}{2},m-1} = \nonumber \\
    &= \frac{1}{2^N} \sum_{m=-N/2}^{N/2} \binom{N}{\frac{N}{2} + m}\bigg(\frac{N}{2}\bigg(\frac{N}{2} + 1 \bigg) - m(m-1)\bigg) = \nonumber \\
    &= \frac{1}{2^N} N(N+2) \ 2^{N-2} - \frac{1}{2^N} N 2^{N-2} = \frac{N}{4}(N+1),
\end{align}
which yields the same value as:
\begin{align}
    \braketop{\eta}{\Jm\Jp}{\eta} &= \frac{1}{2^N} \sum_{n=-N/2}^{N/2} \binom{N}{\frac{N}{2} + n}^{\!\!1/2} \binom{N}{\frac{N}{2} + m}^{\!\!1/2} \sqrt{\frac{N}{2}\bigg(\frac{N}{2} + 1 \bigg) - n(n+1)} \times \nonumber \\ 
    &\times \sqrt{\frac{N}{2}\bigg(\frac{N}{2} + 1 \bigg) - m(m+1)} \braketadj{n+1,\frac{N}{2}}{\frac{N}{2},m+1} = \nonumber \\
    &= \frac{1}{2^N} \sum_{m=-N/2}^{N/2} \binom{N}{\frac{N}{2} + m}\bigg(\frac{N}{2}\bigg(\frac{N}{2} + 1 \bigg) - m(m+1)\bigg) = \nonumber \\
    &= \frac{N}{4}(N + 1),
\end{align}
with the expected value of $\Jm^2$ being:
\begin{align}
    \braketop{\eta}{\Jm\Jm}{\eta} &= \frac{1}{2^N} \sum_{n=-\sfrac{N}{2}}^{\sfrac{N}{2}} \binom{N}{\frac{N}{2} + n}^{\!\!1/2} \binom{N}{\frac{N}{2} + m}^{\!\!1/2} \sqrt{\frac{N}{2}\bigg(\frac{N}{2} + 1 \bigg) - n(n+1)} \times \nonumber \\ 
    &\times \sqrt{\frac{N}{2}\bigg(\frac{N}{2} + 1 \bigg) - m(m-1)} \braketadj{n+1,\frac{N}{2}}{\frac{N}{2},m-1} = \nonumber \\
    &=\! \frac{1}{2^N} \!\! \sum_{m=-\sfrac{N}{2}}^{\sfrac{N}{2}} \!\! \binom{N}{\frac{N}{2} + m - 2}^{\!\!1/2} \!\! \binom{N}{\frac{N}{2} + m}^{\!\!1/2} \!\! \sqrt{\frac{N}{2}\!\bigg(\!\frac{N}{2} \!+\! 1 \!\bigg) \!-\! (m\!-\!2)(m\!-\!1)}  \times \nonumber \\
    &\times \sqrt{\frac{N}{2}\bigg(\frac{N}{2} + 1 \bigg) - m(m-1)} = \frac{1}{2^N} \sum_{m=-\sfrac{N}{2}}^{\sfrac{N}{2}} \frac{N!}{\Big(\frac{N}{2} + m -2 \Big)!\Big(\frac{N}{2}-m\Big)!} = \nonumber \\
    &= \frac{1}{2^N} \sum_{m=-\sfrac{N}{2}}^{\sfrac{N}{2}} \bigg(\frac{N}{2} + m\bigg) \bigg(\frac{N}{2} + m - 1\bigg)\binom{N}{\frac{N}{2} - m} = \nonumber \\
    &= \frac{1}{2^{N+2}} N (N - 2) \sum_{m=-\sfrac{N}{2}}^{\sfrac{N}{2}} \binom{N}{\frac{N}{2} + m} + \frac{1}{2^N} \sum_{m=-\sfrac{N}{2}}^{\sfrac{N}{2}} m^2 \binom{N}{\frac{N}{2} + m} = \nonumber \\
    &= N(N-2)2^{-2} + N 2^{-2} = N (N - 2 + 1)2^{-2} = \frac{N}{4}(N-1).
\end{align}
Hence, the second moment of $\Jx$ w.r.t. $\ket{\eta}$ is:
\begin{equation}
    \braketop{\eta}{\Jx^2}{\eta} = \frac{1}{4} \Bigg( \frac{N}{2}(N + 1) + \frac{N}{2}(N-1) \Bigg) = \frac{N^{\,2}}{4}.
\end{equation}
We know from previous calculations that $\braketop{\eta}{\Jx}{\eta} = \frac{N}{2}$. Hence,
\begin{equation}
    \braketavg{\Delta^2\Jx}_{\trm{CSS}_x}  = \braketop{\eta}{\Jx^2}{\eta} - \braketop{\eta}{\Jx}{\eta}^2 = \frac{N^2}{4} - \bigg(\frac{N}{2}\bigg)^2 = 0.
\end{equation}
The variances for the rest of the components are,
\begin{align}
    \braketavg{\Delta^2\Jy}_{\trm{CSS}_x} &= \braketop{\eta}{\Jy^2}{\eta} - \braketop{\eta}{\Jy}{\eta}^2 =  \braketop{\eta}{\Jy^2}{\eta} = -\frac{1}{4} \braketop{\eta}{(\Jp - \Jm)(\Jp - \Jm)}{\eta} \nonumber \\
    &= -\frac{1}{4} \big( \braketop{\eta}{\Jp\Jp}{\eta} -  \braketop{\eta}{\Jp\Jm}{\eta} - \braketop{\eta}{\Jm\Jp}{\eta} +  \braketop{\eta}{\Jm\Jm}{\eta} \big) \nonumber \\
    &= -\frac{1}{4} \bigg( \!\frac{N}{4}(N - 1) - \frac{N}{4}(N + 1) - \frac{N}{4}(N + 1) + \frac{N}{4}(N - 1) \! \bigg) = \frac{N}{4},
\end{align}
and,
\begin{align}
    \braketavg{\Delta^2\Jz}_{\trm{CSS}_x} &= \frac{1}{2^N} \sum_{n=-N/2}^{N/2} \sum_{m=-N/2}^{N/2} \binom{N}{\frac{N}{2} + n}^{\!\!1/2} \binom{N}{\frac{N}{2} + m}^{\!\!1/2} \bra{n,\frac{N}{2}} \Jz^2 \ket{m,\frac{N}{2}} = \nonumber \\
    &= \frac{1}{2^N} \sum_{n=-N/2}^{N/2} \sum_{m=-N/2}^{N/2} \binom{N}{\frac{N}{2} + n}^{\!\!1/2} \binom{N}{\frac{N}{2} + m}^{\!\!1/2} n \ m \braketadj{n,\frac{N}{2}}{m,\frac{N}{2}} = \nonumber \\
    &=\frac{1}{2^N} \sum_{m = -N/2}^{N/2} m^2 \binom{N}{\frac{N}{2} + m} = \frac{1}{2^N} N \, 2^{N-2} = \frac{N}{4}.
\end{align}

\section{Binomial identities}

\begin{theorem}[Binomial theorem]
    The expansion of the binomial $x+y$ to the power $n$, where $n$ is any nonnegative integer, is given by
    \begin{equation} \label{eq:binomial}
                (x+y)^{n} = \sum_{k=0}^n \binom{n}{k} x^k y^{n-k},
    \end{equation}
    which in the case of $y = 1$, reads as
    \begin{equation} \label{eq:binomial_y=1}
                (1 + x)^{n} = \sum_{k=0}^n \binom{n}{k} x^k.
    \end{equation}
\end{theorem}

\begin{identity} \label{id:id1and2}
    \begin{align}
        \binom{N}{\frac{N}{2} + m \pm 1}^{1/2} \binom{N}{\frac{N}{2} + m}^{\!\!1/2} = \frac{\big(\frac{N}{2} \mp m\big)}{\sqrt{\frac{N}{2}\left(\frac{N}{2} + 1\right) - m\big(m \pm 1\big)}} \binom{N}{\frac{N}{2} \pm m}
    \end{align}
\end{identity}

\begin{myproof} It is straightforward to show using factorial notation:
    \begin{align}
        &\binom{N}{\frac{N}{2} + m \pm 1}^{\!\!1/2} \binom{N}{\frac{N}{2} + m}^{\!\!1/2} = \nonumber \\
        &=\left(\frac{N!}{\left(\frac{N}{2}+m\pm1\right)!\left(\frac{N}{2}-m\mp 1\right)!}\right)^{\!\!1/2} \left(\frac{N!}{\left(\frac{N}{2}+m\right)!\left(\frac{N}{2}-m\right)!}\right)^{\!\!1/2} \nonumber \\
        &= \frac{1}{\sqrt{\left(\frac{N}{2} \pm m +1\right)\left(\frac{N}{2} \mp m\right)}} \frac{N!}{\left(\frac{N}{2} \pm m\right)!\left(\frac{N}{2} \mp m -1\right)!} \nonumber \\
        &= \frac{\left(\frac{N}{2} \mp m\right)}{\sqrt{\frac{N}{2}\left(\frac{N}{2} + 1\right) - m\left(m \pm 1\right)}} \binom{N}{\frac{N}{2} \pm m} \nonumber
    \end{align}
\end{myproof}

\begin{identity} \label{id:id3}
    Follows straightforwardly from the binomial formula of \eqnref{eq:binomial_y=1} by setting $x=1$ and changing the summand index $k$ to $\frac{N}{2}+m$:
    \begin{equation} \label{eq:id3}
        \sum_{m=-N/2}^{N/2} \binom{N}{\frac{N}{2} + m} = 2^{N}.
    \end{equation}
\end{identity} 

\begin{identity} \label{id:id4and5}
    \begin{equation} \label{eq:id4and5}
         \sum_{m=-N/2}^{N/2} \binom{N}{\frac{N}{2} \pm m} \bigg( \frac{N}{2} \pm m \bigg) = N \, 2^{N-1}.
    \end{equation}
\end{identity}

\begin{myproof}
    The closed forms for the above sums can be found using the derivative of the binomial formula given in \eqnref{eq:binomial_y=1}. Namely,
    \begin{equation}
        \frac{\dd}{\dd x}(x+1)^N = N(x + 1)^{N-1} = \frac{\dd}{\dd x} \sum_{k=0}^N \binom{N}{k} x^k = \sum_{k=0}^N \binom{N}{k} k \, x^{k-1}.
    \end{equation}
    Then, \eqnref{eq:id4and5} is obtained by setting $x = 1$ and changing the summation limits.
\end{myproof}

\begin{identity} \label{id:id6}
     \begin{equation} \label{eq:id6}
        \sum_{m=-N/2}^{N/2} m \binom{N}{\frac{N}{2} + m} = N 2^{N-1}-N 2^{N-1} = 0,
    \end{equation}
    which follows trivially from \idref{id:id3} and \idref{id:id4and5}.
\end{identity}

\begin{identity} \label{id:id7}
    \begin{equation} \label{eq:id7}
        \sum_{k=-N/2}^{N/2} m^2 \binom{N}{\frac{N}{2} + m} = N \, 2^{N-2}.
    \end{equation}
\end{identity}

\begin{myproof}
    To find the close form expression of \idref{id:id7}, we take the second order derivative of the binomial of \eqnref{eq:binomial_y=1} multiplied by x:
    \begin{equation} \label{eq:expression_x_times_binomial}
            x \, (x + 1)^N = \sum_{k = 0}^N \binom{N}{k} x^{k+1}.
    \end{equation}
    To do so, we first compute the second derivative of the l.h.s. is
    \begin{equation}
        \frac{\dd^2}{\dd x^2} \left(x(x+1)^N\right) = 2N(x+1)^{N-1} + N \, x \, (N-1) (x+1)^{N-2},
    \end{equation}
    and of the r.h.s.:
    \begin{equation}
        \frac{\dd^2}{\dd x^2} \sum_{k=0}^N \binom{N}{k} x^{k+1} = \sum_{k=0}^N \binom{N}{k} (k+1) k \, x^{k-1}.
    \end{equation}
    If we now set $x = 1$ and change the summation limits by redefining the index $k$ as $k = \frac{N}{2} + m$, then, \eqnref{eq:expression_x_times_binomial} can be written as
    \begin{align} \label{eq:intermediate_expression_identity_binomial}
        &\sum_{m=-N/2}^{N/2} \left(\frac{N}{2} + m + 1\right) \left(\frac{N}{2} + m\right) \binom{N}{\frac{N}{2} + m} = N2^N + N(N-1) 2^{N-2}.
    \end{align}
    If now we massage the l.h.s. a bit more:
    \begin{align}
        &\sum_{m=-N/2}^{N/2} \left(\frac{N}{2} + m + 1\right) \left(\frac{N}{2} + m\right) \binom{N}{\frac{N}{2} + m} = \nonumber \\
        &= \sum_{m=-N/2}^{N/2} \left(\frac{N}{2}\left(\frac{N}{2} + 1\right)  + m^2 + m(N+1) \right) \binom{N}{\frac{N}{2} + m} \nonumber \\
        &\underset{\ref{id:id6}}{=} N(N+2)2^{-2} \!\! \sum_{m=-N/2}^{N/2} \binom{N}{\frac{N}{2} + m} + \!\!\! \sum_{m=-N/2}^{N/2} \!\! m^2 \binom{N}{\frac{N}{2} + m} \nonumber \\
        &\underset{\ref{id:id3}}{=}N(N+2)2^{N-2} + \sum_{m=-N/2}^{N/2} m^2 \binom{N}{\frac{N}{2} + m}, 
    \end{align}
    we can finally rewrite \eqnref{eq:intermediate_expression_identity_binomial} as
    \begin{align}
        N(N+2)2^{N-2} + \sum_{m=-N/2}^{N/2} m^2 \binom{N}{\frac{N}{2} + m} = N \, 2^N + N(N-1) 2^{N-2} ,
    \end{align}
    and thus, reach the final form of \eqnref{eq:id7}:
    \begin{align}
        &\sum_{m=-N/2}^{N/2} m^2 \binom{N}{\frac{N}{2} + m} = N \, 2^N + N(N-1) 2^{N-2} - N(N+2)2^{N-2} \nonumber \\
        &= N\left(4+(N-1)-(N+2)\right) 2^{N-2} = N 2^{N-2}.
    \end{align}
\end{myproof}

    \chapter{Appendix for \chapref{chap:bayesian}}
\label{AppendixB}

\section{Gaussian properties I}

\begin{property}[Inversion of a 2x2 Block matrix]
    \label{propty:inv_2x2_Block}
    Consider a Block matrix partitioned into four blocks where $\pmb{A}$ and $\pmb{D}$ are square blocks of arbitrary size. Then,
    \begin{align}
        \begin{pmatrix} \pmb{A} & \pmb{B} \\
                        \pmb{C} & \pmb{D}
        \end{pmatrix}^{-1} \!\!\!\!= \begin{pmatrix} 
        \pmb{A}^{-1} + \pmb{A}^{-1} \pmb{B} (\pmb{D} - \pmb{C} \pmb{A}^{-1} \pmb{B})^{-1} \pmb{C} \pmb{A}^{-1} &  -\pmb{A}^{-1}\pmb{B}(\pmb{D}-\pmb{C}\pmb{A}^{-1}\pmb{B})^{-1} \\
        -(\pmb{D} -\pmb{C}\pmb{A}^{-1}\pmb{B})^{-1} \pmb{C} \pmb{A}^{-1} & (\pmb{D} - \pmb{C}\pmb{A}^{-1} \pmb{B})^{-1}
        \end{pmatrix}.
    \end{align}
\end{property}

\begin{property}[Determinant of a 2x2 Block matrix] \label{propty:det_2x2_Block}
    Consider the following Block matrix 
    \begin{equation}
        \begin{pmatrix} \pmb{A} & \pmb{B} \\
                        \pmb{C} & \pmb{D}
        \end{pmatrix}.
    \end{equation}
    If $\pmb{A}$ is invertible, then the determinant is given by:
    \begin{equation}
        \det{\left[\begin{pmatrix} \pmb{A} & \pmb{B} \\
                        \pmb{C} & \pmb{D}
        \end{pmatrix}\right]} = \det{[\pmb{A}]} \det{[\pmb{D}-\pmb{C}\pmb{A}^{-1}\pmb{B}]}
    \end{equation}
\end{property}

\begin{lem}[The joint distribution of Gaussian variables] \label{lem:joint_distribution}
    Let $\pmb{x} \in \Real^n$ and $\pmb{y} \in \Real^m$ be two Gaussian random variables such that:
    \begin{align}
        \pmb{x} &\sim p(\pmb{x}) = \Gauss(\pmb{x}|\,\pmb{\mu},\pmb{\Sigma}), \label{eq:gaussian_x_ap_proof} \\
        \pmb{y}|\pmb{x} &\sim p(\pmb{y}|\pmb{x}) = \Gauss(\pmb{y}|\pmb{A}\pmb{x} + \pmb{B}\pmb{u}, \pmb{G}\pmb{Q}\pmb{G}^\Trans) \label{eq:gaussian_x|y_ap_proof}
    \end{align}
    where $\pmb{A}$ is a transition matrix, $\pmb{B}$ is the control input matrix, $\pmb{u}$ is an external control, and $\pmb{\Sigma}$ and $\pmb{Q}$ are positive-definite covariance matrices. Then, the joint distribution of $(\pmb{x},\pmb{y})$ is a multivariate Gaussian given by:
    \begin{equation} \label{eq:joint_multivariate_gaussian}
        \begin{pmatrix}
            \pmb{x} \\ \pmb{y}
        \end{pmatrix} \sim p(\pmb{x},\pmb{y}) = \Gauss\left( \begin{pmatrix} \pmb{x} \\ \pmb{y} \end{pmatrix} \middle|
        \begin{pmatrix} \pmb{\mu} \\ \pmb{A} \pmb{\mu} + \pmb{B}\pmb{u} \end{pmatrix}, 
        \begin{pmatrix} 
        \pmb{\Sigma} & \pmb{\Sigma} \pmb{A}^\Trans \\ 
        \pmb{A} \pmb{\Sigma} & \pmb{A} \pmb{\Sigma} \pmb{A}^\Trans + \pmb{G}\pmb{Q}\pmb{G}^\Trans
        \end{pmatrix} 
        \right)
    \end{equation}
\end{lem}

\begin{myproof}
    The joint distribution of $(\pmb{x},\pmb{y})$ can be expressed using the product rule (\propertyref{propty:prod_rule}):
    \begin{equation}
        p(\pmb{x},\pmb{y}) = p(\pmb{y}|\pmb{x}) p(\pmb{x}) = \Gauss(\pmb{y}|\pmb{A}\pmb{x}+\pmb{B}\pmb{u}, \pmb{G}\pmb{Q}\pmb{G}^\Trans) \; \Gauss(\pmb{x}|\,\pmb{\mu},\pmb{\Sigma}).
    \end{equation}
    Let us now work backwards by assuming that \eqnref{eq:joint_multivariate_gaussian} holds and see if we can split the joint distribution into the following two explicit forms of the Gaussian distributions of $\pmb{x}$ and $\pmb{y}|\pmb{x}$:
    \begin{align}
        \!\!\Gauss(\pmb{y}|\pmb{A}\pmb{x}, \pmb{G}\pmb{Q}\pmb{G}^\Trans) \!&=\! \frac{1}{\sqrt{(2\pi)^m |\pmb{G}\pmb{Q}\pmb{G}^\Trans|}} \exp\!{\!\left(\!\!\shortminus\frac{1}{2} (\pmb{y} \!\shortminus\! \pmb{A}\pmb{x} \!\shortminus\! \pmb{B}\pmb{u})^\Trans \!(\pmb{G}\pmb{Q}\pmb{G}^\Trans)^{\!\shortminus1} (\pmb{y} \!\shortminus\! \pmb{A}\pmb{x} \!\shortminus\! \pmb{B}\pmb{u}) \!\!\right)}, \\ 
        \Gauss(\pmb{x}|\,\pmb{\mu},\pmb{\Sigma}) \!&=\! \frac{1}{\sqrt{(2\pi)^n|\pmb{\Sigma}|}} \exp\!{\!\left(\!\shortminus\frac{1}{2} (\pmb{x} \!\shortminus\! \pmb{\mu})^\Trans \pmb{\Sigma}^{-1} (\pmb{x} \!\shortminus\! \pmb{\mu}) \!\right)}.
    \end{align}
    Namely, if
    \begin{align} \label{eq:joint_gaussian_exponent_expanded}
        p(\pmb{x},\pmb{y}) &= \Gauss\left( \begin{pmatrix} \pmb{x} \\ \pmb{y} \end{pmatrix} \middle| \,
        \pmb{m}, 
        \pmb{P}
        \right) \\
        &= \Gauss\left( \begin{pmatrix} \pmb{x} \\ \pmb{y} \end{pmatrix} \middle|
        \begin{pmatrix} \pmb{\mu} \\ \pmb{A} \pmb{\mu} + \pmb{B}\pmb{u} \end{pmatrix}, 
        \begin{pmatrix} 
        \pmb{\Sigma} & \pmb{\Sigma} \pmb{A}^\Trans \\ 
        \pmb{A} \pmb{\Sigma} & \pmb{A} \pmb{\Sigma} \pmb{A}^\Trans + \pmb{G}\pmb{Q}\pmb{G}^\Trans
        \end{pmatrix} 
        \right) \\
        &= \frac{1}{\sqrt{(2\pi)^{n+m} |\pmb{\Sigma}||\pmb{G}\pmb{Q}\pmb{G}^\Trans|}} \exp\!{\!\left(\!\!\shortminus\frac{1}{2} \left(\!\!\begin{pmatrix} \pmb{x} \\ \pmb{y} \end{pmatrix} \!\!\shortminus\!\! \begin{pmatrix} \pmb{\mu} \\ \pmb{A} \pmb{\mu} \!+\! \pmb{B}\pmb{u}\end{pmatrix}\!\!\right)^{\!\!\!\!\Trans} \!\!\pmb{P}^{-1}\!\!\left(\!\!\begin{pmatrix} \pmb{x} \\ \pmb{y} \end{pmatrix} \!\!\shortminus\!\! \begin{pmatrix} \pmb{\mu} \\ \pmb{A} \pmb{\mu} \!+\! \pmb{B}\pmb{u}\end{pmatrix}\!\!\right)\!\!\right)} \nonumber
    \end{align}
    where we have used \propertyref{propty:inv_2x2_Block} and \propertyref{propty:det_2x2_Block} to compute the inverse and determinant of $\pmb{P}$ since $\pmb{\Sigma}$ is a positive-definite symmetric matrix and thus, invertible: 
    \begin{align}
        \pmb{P}^{-1} &=\! \begin{pmatrix} 
        \pmb{\Sigma} & \pmb{\Sigma} \pmb{A}^\Trans \\ 
        \pmb{A} \pmb{\Sigma} & \pmb{A} \pmb{\Sigma} \pmb{A}^\Trans + \pmb{G}\pmb{Q}\pmb{G}^\Trans
        \end{pmatrix}^{\!\!-1} \!\!\!\!\!=\!\! \begin{pmatrix}
        \pmb{\Sigma}^{-1} + \pmb{A}^\Trans (\pmb{G}\pmb{Q}\pmb{G}^\Trans)^{-1} \pmb{A} & - \pmb{A}^\Trans (\pmb{G}\pmb{Q}\pmb{G}^\Trans)^{-1}\\ -(\pmb{G}\pmb{Q}\pmb{G}^\Trans)^{-1} \pmb{A} & (\pmb{G}\pmb{Q}\pmb{G}^\Trans)^{-1}
        \end{pmatrix}, \\
        \det{\pmb{P}} &= \det{\left[\begin{pmatrix} 
        \pmb{\Sigma} & \pmb{\Sigma} \pmb{A}^\Trans \\ 
        \pmb{A} \pmb{\Sigma} & \pmb{A} \pmb{\Sigma} \pmb{A}^\Trans + \pmb{G}\pmb{Q}\pmb{G}^\Trans
        \end{pmatrix} \right]} = \det{\left[\pmb{\Sigma}\right]} \det{\left[\pmb{G}\pmb{Q}\pmb{G}^\Trans\right]}. 
    \end{align}
    If now we expand the exponent of \eqnref{eq:joint_gaussian_exponent_expanded}, we get
    \begin{align}
        &\begin{pmatrix}
            \pmb{x} - \pmb{\mu} \\ \pmb{y} - \pmb{A}\pmb{\mu}
        \end{pmatrix}^\Trans  
        \begin{pmatrix}
        \pmb{\Sigma}^{-1} + \pmb{A}^\Trans (\pmb{G}\pmb{Q}\pmb{G}^\Trans)^{-1} \pmb{A} & - \pmb{A}^\Trans (\pmb{G}\pmb{Q}\pmb{G}^\Trans)^{-1}\\ -(\pmb{G}\pmb{Q}\pmb{G}^\Trans)^{-1} \pmb{A} & (\pmb{G}\pmb{Q}\pmb{G}^\Trans)^{-1}
        \end{pmatrix}
        \begin{pmatrix}
            \pmb{x} - \pmb{\mu} \\ \pmb{y} - \pmb{A}\pmb{\mu}
        \end{pmatrix} \nonumber \\
        &= \begin{pmatrix}
            \pmb{x} - \pmb{\mu} \\ \pmb{y} - \pmb{A}\pmb{\mu}
        \end{pmatrix}^\Trans \begin{pmatrix}
            (\pmb{\Sigma}^{-1} + \pmb{A}^\Trans (\pmb{G}\pmb{Q}\pmb{G}^\Trans)^{-1} \pmb{A})(\pmb{x} - \pmb{\mu}) -\pmb{A}^\Trans (\pmb{G}\pmb{Q}\pmb{G}^\Trans)^{-1} (\pmb{y} - \pmb{A}\pmb{\mu}) \\ -(\pmb{G}\pmb{Q}\pmb{G}^\Trans)^{-1}\pmb{A}(\pmb{x} - \pmb{\mu}) + (\pmb{G}\pmb{Q}\pmb{G}^\Trans)^{-1}(\pmb{y} - \pmb{A}\pmb{\mu})
        \end{pmatrix} \nonumber \\
        &= \begin{pmatrix}
            \pmb{x} - \pmb{\mu} \\ \pmb{y} - \pmb{A}\pmb{\mu}
        \end{pmatrix}^\Trans \begin{pmatrix}
            (\pmb{\Sigma}^{-1}(\pmb{x} - \pmb{\mu}) + \pmb{A}^\Trans (\pmb{G}\pmb{Q}\pmb{G}^\Trans)^{-1} (\pmb{A} \pmb{x} - \pmb{y}) \\ -(\pmb{G}\pmb{Q}\pmb{G}^\Trans)^{-1}(\pmb{A}\pmb{x} - \pmb{y} )
            \end{pmatrix} \nonumber \\
        &= (\pmb{x} \!-\! \pmb{\mu})^\Trans \pmb{\Sigma}^{-1} (\pmb{x} \!-\! \pmb{\mu}) +  (\pmb{x} \!-\! \pmb{\mu})^\Trans \pmb{A}^\Trans (\pmb{G}\pmb{Q}\pmb{G}^\Trans)^{-1} (\pmb{A}\pmb{x} \!-\! \pmb{y}) + (\pmb{y} \!-\! \pmb{A}\pmb{\mu})^\Trans(\pmb{G}\pmb{Q}\pmb{G}^\Trans)^{-1}(\pmb{y} \!-\! \pmb{A}\pmb{x}) \nonumber \\
        &= (\pmb{x} \!-\! \pmb{\mu})^\Trans \pmb{\Sigma}^{-1} (\pmb{x} \!-\! \pmb{\mu}) + (\pmb{\mu}^\Trans\pmb{A}^\Trans \!-\! \pmb{x}^\Trans \pmb{A}^\Trans \!+\! \pmb{y}^\Trans \!-\! \pmb{\mu}^\Trans \pmb{A}^T)(\pmb{G}\pmb{Q}\pmb{G}^\Trans)^{-1}(\pmb{y} \!-\! \pmb{A}\pmb{x}) \nonumber \\
        &= (\pmb{x} \!-\! \pmb{\mu})^\Trans \pmb{\Sigma}^{-1} (\pmb{x} \!-\! \pmb{\mu}) + (\pmb{y} \!-\! \pmb{A}\pmb{x})^\Trans(\pmb{G}\pmb{Q}\pmb{G}^\Trans)^{-1}(\pmb{y} \!-\! \pmb{A}\pmb{x}). 
    \end{align}
    It follows then that \eqnref{eq:joint_gaussian_exponent_expanded} can indeed be split into the two Gaussian densities of \eqnref{eq:gaussian_x_ap_proof} and \eqnref{eq:gaussian_x|y_ap_proof}:
    \begin{align}
        p(\pmb{x},\pmb{y}) &=\! \frac{1}{\sqrt{(2\pi)^{n+m} |\pmb{\Sigma}||\pmb{G}\pmb{Q}\pmb{G}^\Trans|}} \exp\!{\!\left(\!\!\shortminus\frac{1}{2} \! \left((\pmb{x} \!-\! \pmb{\mu})^{\!\Trans} \pmb{\Sigma}^{-1} (\pmb{x} \!\shortminus\! \pmb{\mu}) \!+\! (\pmb{y} \!\shortminus\! \pmb{A}\pmb{x})^{\!\Trans}\!(\pmb{G}\pmb{Q}\pmb{G}^\Trans)^{-1}\!(\pmb{y} \!\shortminus\! \pmb{A}\pmb{x})\right)\!\!\right)} \nonumber \\
        &=\! \frac{1}{\sqrt{(2\pi)^{n} |\pmb{\Sigma}|}} \e^{-\frac{1}{2} \left((\pmb{x} - \pmb{\mu})^\Trans \pmb{\Sigma}^{-1} (\pmb{x} - \pmb{\mu}) \right)} \frac{1}{\sqrt{(2\pi)^{m} |\pmb{G}\pmb{Q}\pmb{G}^\Trans|}} \e^{-\frac{1}{2} \left((\pmb{y}- \pmb{A}\pmb{x})^\Trans(\pmb{G}\pmb{Q}\pmb{G}^\Trans)^{-1}(\pmb{y} - \pmb{A}\pmb{x})\right)} \nonumber \\
        &=\! \Gauss(\pmb{x}|\,\pmb{\mu},\pmb{\Sigma}) \Gauss(\pmb{y}|\pmb{A}\pmb{x},\pmb{G}\pmb{Q}\pmb{G}^\Trans).
    \end{align}
\end{myproof}

\begin{lem}[Marginal and conditional distributions of a joint Gaussian probability density] \label{lem:cond_PDF_Gaussian_var} 
    Let two random variables $\pmb{x}$ and $\pmb{y}$ have the joint Gaussian probability density
    \begin{equation}
        \begin{pmatrix}
            \pmb{x} \\
            \pmb{y}
        \end{pmatrix}
        \sim
        \Gauss{\left(
        \begin{pmatrix}
            \pmb{x} \\
            \pmb{y}
        \end{pmatrix}
        \middle|
        \begin{pmatrix}
            \pmb{\mu}_x \\
            \pmb{\mu}_y
        \end{pmatrix},
        \begin{pmatrix}
            \pmb{\Sigma}_{x,x} & \pmb{\Sigma}_{x,y} \\
            \pmb{\Sigma}_{y,x} & \pmb{\Sigma}_{y,y}
        \end{pmatrix}
        \right)}
    \end{equation}
    where $\pmb{\Sigma}_{x,y}^\Trans = \pmb{\Sigma}_{y,x}$. Then, the marginal and conditional probability densities of $\pmb{x}$ and $\pmb{y}$ are
    \begin{align}
        \pmb{x} &\sim p(\pmb{x}) = \Gauss(\pmb{x}|\pmb{\mu}_x,\pmb{\Sigma}_{x,x}), \\
        \pmb{y} &\sim p(\pmb{y}) = \Gauss(\pmb{y}|\pmb{\mu}_y,\pmb{\Sigma}_{y,y}), \\
        \pmb{x}|\pmb{y} &\sim p(\pmb{x}|\pmb{y}) = \Gauss(\pmb{x}|\pmb{\mu}_x+\pmb{\Sigma}_{x,y} \pmb{\Sigma}_{y,y}^{-1} (\pmb{y} - \pmb{\mu}_y), \pmb{\Sigma}_{x,x} - \pmb{\Sigma}_{x,y} \pmb{\Sigma}_{y,y}^{-1}\pmb{\Sigma}_{x,y}^\Trans), \\
        \pmb{y}|\pmb{x} &\sim p(\pmb{y}|\pmb{x}) = \Gauss(\pmb{y}|\pmb{\mu}_y + \pmb{\Sigma}_{x,y}^\Trans \pmb{\Sigma}_{x,x}^{-1}(\pmb{x}-\pmb{\mu}_x),\pmb{\Sigma}_{y,y} - \pmb{\Sigma}_{x,y}^\Trans \pmb{\Sigma}_{x,x}^{-1} \pmb{\Sigma}_{x,y})
    \end{align}
\end{lem}

    \chapter{Appendix for \chapref{chap:cm}}
\label{AppendixC}

\section{Fourier coefficients of a double series expansion}

\begin{prop}[Double time–frequency Fourier series expansion of a continuous mode] \label{prop:ap_fourier_coef_double_series}
    Let $\banihil(t)$ be a continuous mode defined over time $t$. By segmenting time into intervals of duration $\Delta t$ and applying a Fourier expansion, the continuous mode $\banihil(t)$ can be written as a double series expansion in both time and frequency:
    \begin{equation} \label{eq:ap_double_series_exp}
    \banihil(t) = \frac{1}{\sqrt{\Delta t}} \sum_{n=-\infty}^\infty \sum_{k=-\infty}^\infty \banihil_{n,k} \, \Theta(t-t_n) \, \e^{-i 2\pi k \, t/\Delta t},
    \end{equation}
    where $t_n = n \Delta t$ is the discretized time step, $\Theta(t-t_n)$ is the Heaviside function defined as $\Theta(u) = 1$ for $0 \leq u < \Delta t$ and zero otherwise, and $\banihil_{n,k}$ are the ``Fourier'' coefficients of the double series, given by:
    \begin{equation} \label{eq:ap_doubleFourier_coefficients}
        \banihil_{n,k} = \frac{1}{\sqrt{\Delta t}} \int_{t_n}^{t_n+\Delta t} \banihil(t) \e^{i2\pi k \, t / \Delta t} \, \dt.
    \end{equation}
\end{prop}

\begin{myproof}
    Let us start the proof by multiplying both sides of \eqnref{eq:ap_double_series_exp} by $\e^{i 2\pi \ell t/\Delta t}$, as well as integrating them w.r.t. time from $t_m$ to $t_m+\Delta t$:
    \begin{align} 
        \int_{t_m}^{t_m + \Delta t} \!\! \banihil(t) \e^{i 2\pi \ell \, t/\Delta t} \, \dt = \int_{t_m}^{t_m + \Delta t} \!\!\!\!\! \frac{1}{\sqrt{\Delta t}} \! \sum_{n=-\infty}^\infty \sum_{k=-\infty}^\infty \!\! \banihil_{n,k} \, \Theta(t-t_n) \, \e^{-i 2\pi k \, t/\Delta t} \e^{i 2\pi \ell \, t/\Delta t} \, \dt. \label{eq:ap_integral_form_discrete_modes_step1}
    \end{align}
    Since $\Theta(t-t_n)$ is nonzero only for $t_n \leq t< t_n + \Delta t$, and considering that the integration is specifically performed over the interval $t_m$ to $t_m + \Delta t$, $\Theta(t-t_n)$ can be replaced with $1$ within the integration interval $[t_m, t_m+\Delta t)$ and zero otherwise. This allows for the removal of the summation over $n$ in the expression, as the only non-zero contribution comes from the term when $n = m$, i.e., when the interval picked by the Heaviside function coincides with the integration interval. Then, the rhs of \eqnref{eq:ap_integral_form_discrete_modes_step1} reduces to
    \begin{align}
        \int_{t_m}^{t_m + \Delta t} \!\! \banihil(t) \e^{i 2\pi \ell \, t/\Delta t} \, \dt = \sum_{k=-\infty}^\infty \frac{1}{\sqrt{\Delta t}}  \banihil_{m,k} \int_{t_m}^{t_m + \Delta t} \!\!\!\! \e^{-i 2\pi k \, t/\Delta t} \e^{i 2\pi \ell \, t/\Delta t} \, \dt \nonumber.
    \end{align}
    It can be further simplified by using the orthogonality of the exponential functions, i.e., meaning that for terms with $k\neq \ell$, the integral over the interval $\Delta t$ will be zero due to periodicity of the exponents over that time period, and for $k = \ell$, the integral will yield the interval $\Delta t$:
    \begin{equation} \label{eq:ap_orthogonality_cond}
        \int_{t_m}^{t_m+\Delta t}  \!\!\!\! \e^{-i 2\pi k \, t/\Delta t} \e^{i 2\pi \ell \, t/\Delta t} \, \dt = \Delta t \, \delta_{k,\ell},
    \end{equation}
    where $\delta_{k,\ell}$ denotes the Kronecker delta. Therefore,
    \begin{align}
        \int_{t_m}^{t_m + \Delta t} \!\! \banihil(t) \e^{i 2\pi \ell \, t/\Delta t} \, \dt = \sum_{k=-\infty}^\infty \frac{1}{\sqrt{\Delta t}}  \banihil_{m,k} \, \Delta t \, \delta_{k,\ell} = \banihil_{m,\ell} \sqrt{\Delta t}, \nonumber 
    \end{align}
    such that we can easily isolate the discretized modes $\banihil_{n,k}$ as:
    \begin{equation}
        \banihil_{n,k} = \frac{1}{\sqrt{\Delta t}} \int_{t_n}^{t_n+\Delta t} \banihil(t) \e^{i2\pi k \, t / \Delta t} \, \dt.
    \end{equation}
\end{myproof}

    \chapter{Appendix for \chapref{chap:bounds}}
\label{AppendixD}

\section{Writing a quantum map as a convex mixture of unitaries} \label{sec:proofs_ch4}

\begin{theorem}[Quantum map as a convex mixture of unitaries] \label{thm:mixture_unitaries_app}
    Given a unitary evolution governed by a Hamiltonian $\xi \Ham$
    \vspace{-10pt}
    \begin{equation}
        \Unitary_{\xi,\tau}[\;\cdot\;] = \ee^{-\ii \, \xi \, \Ham \tau} \; \cdot \; \ee^{\ii \, \xi \, \Ham \tau},
        \vspace{-15pt}
    \end{equation}
    whose scalar encoding $\xi \in \Real$ (frequency) is randomly distributed according to a Gaussian probability density
    \begin{equation} \label{eq:gaussian_mixing_prob_ap}
        \xi \sim p_{\mu,\sigma}(\xi\,) = \mathcal{N}(\mu(\tau),\sigma^2(\tau)) = \frac{1}{\sqrt{2\pi \sigma^2(\tau)}} \exp{\left\{\!\shortminus\frac{(\xi - \mu(\tau))^2}{2 \sigma^2(\tau)}\!\right\}},
    \end{equation}
    then, the quantum map $\Omega$ can be written as a convex mixture of these unitaries as:
    \begin{equation} \label{eq:convex_mixture_of_unitaries_ap}
        \rho(\tau) = \Omega[\rho(0)] = \E{\Unitary_{\xi,\tau}[\rho(0)]}{p(\xi\,)} = \int d\xi \; p_{\mu,\sigma}(\xi\,) \, \ee^{-\ii \, \xi \, \Ham \tau} \rho(0) \ee^{\ii \, \xi \, \Ham \tau}, 
    \end{equation}
    if $\rho(\tau)$ corresponds to the solution of the following master equation:
    \begin{align}
        \frac{\dd \rho(\tau)}{\dd \tau} &= -\ii \omega(\tau) [\Ham,\rho(\tau)] + \Gamma(\tau) \left(\Ham \rho(\tau) \Ham - \frac{1}{2} \{ \Ham^2,\rho(\tau) \} \right) \label{eq:form1_master_eq_ap} \\
        &= -\ii \omega(\tau) [\Ham,\rho(\tau)] - \frac{1}{2} \Gamma(\tau) \left[\Ham,[\Ham,\rho(\tau)]\right] \label{eq:form2_master_eq_ap}
    \end{align}
    with the time-dependent frequency and decay parameters being
    \begin{equation}
        \omega(\tau) = \mu(\tau) + \tau \dot{\mu}(\tau) \;\;\;\;\;\; \text{and} \;\;\;\;\;\; \Gamma(\tau) = 2\sigma^2(\tau) \tau \left( 1 + \frac{\dot{\sigma}(\tau)}{\sigma(\tau)} \tau \right).
    \end{equation}
\end{theorem}

\begin{myproof}
    The first step is to differentiate $\rho(\tau)$, as defined in \eqnref{eq:convex_mixture_of_unitaries}, with respect to $\tau$:
    \begin{align}
        \frac{\dd \rho(\tau)}{\dd \tau} = \int \dd \xi \left( \frac{\dd}{\dd \tau} \; p_{\mu,\sigma} (\xi\,) \right) \Unitary_{\xi,\tau} [\rho(0)] + \int \dd \xi \, p_{\mu,\sigma} (\xi\,) \frac{\dd}{\dd\tau} \Unitary_{\xi,\tau}[\rho(0)], 
    \end{align}
    which, in turn, requires the differentiation of the Gaussian probability distribution,
    \begin{align}
        \frac{\dd}{\dd \tau} \; &p_{\mu,\sigma} (\xi\,) = \frac{\dd}{\dd \tau} \left[ \frac{1}{\sqrt{2\pi \sigma^2(\tau)}} \exp\!{\left\{\!\shortminus\frac{(\xi \!\shortminus\! \mu(\tau))^2}{2 \sigma^2(\tau)}\!\right\}} \right] \nonumber \\
        &= \frac{\dd}{\dd \tau} \left[ \frac{1}{\sqrt{2\pi \sigma^2(\tau)}} \right] \exp\!{\left\{\!\shortminus\frac{(\xi \!\shortminus\! \mu(\tau))^2}{2 \sigma^2(\tau)}\!\right\}} + \frac{1}{\sqrt{2\pi \sigma^2(\tau)}}  \frac{\dd}{\dd \tau} \left[ \exp\!{\left\{\!\shortminus\frac{(\xi \!\shortminus\! \mu(\tau))^2}{2 \sigma^2(\tau)}\!\right\}} \right] \nonumber \\
        &= - \frac{\dot{\sigma}(\tau)}{\sigma(\tau)} \frac{1}{\!\sqrt{2\pi \sigma^2(\tau)}} \exp\!{\left\{\!\shortminus\frac{(\xi \!\shortminus\! \mu(\tau))^2}{2 \sigma^2(\tau)}\!\right\}} \!-\! \frac{1}{\!\sqrt{2\pi \sigma^2(\tau)}} \exp\!{\left\{\!\shortminus\frac{(\xi \!\shortminus\! \mu(\tau))^2}{2 \sigma^2(\tau)}\!\right\}} \frac{\dd }{\dd \tau} \frac{(\xi \!\shortminus\! \mu(\tau))^2}{2 \sigma^2(\tau)} \nonumber \\
        &= p_{\mu,\sigma} (\xi\,) \left( - \frac{\dot{\sigma}(\tau)}{\sigma(\tau)} - \frac{2 \sigma^2(\tau) \frac{\dd }{\dd \tau}(\xi - \mu(\tau))^2 - (\xi - \mu(\tau))^2 \frac{\dd}{\dd \tau} 2 \sigma^2(\tau) }{4 \sigma^4(\tau)} \right) \nonumber \\
        &= p_{\mu,\sigma} (\xi\,) \left( - \frac{\dot{\sigma}(\tau)}{\sigma(\tau)} - \frac{-4\sigma^2(\tau) (\xi - \mu(\tau)) \dot{\mu}(\tau) - (\xi - \mu(\tau))^2 4 \sigma(\tau) \dot{\sigma}(\tau)}{4\sigma^4(\tau)} \right) \nonumber \\
        &= p_{\mu,\sigma} (\xi\,) \left(\frac{  (\xi-\mu(\tau)) \sigma(\tau) \dot{\mu}(\tau) + \left((\xi - \mu(\tau))^2 - \sigma^2(\tau) \right) \dot{\sigma}(\tau) }{\sigma^3(\tau)} \right),
    \end{align}
    and the unitary map, 
    \begin{align} \label{eq:derivative_U_wrt_tau}
        \frac{\dd}{\dd \tau} \Unitary_{\xi,\tau} [\rho(0)] &= \frac{\dd}{\dd \tau}  \left[ \ee^{-\ii \, \xi \, \Ham \tau} \rho(0) \ee^{\ii \, \xi \, \Ham \tau} \right] = \frac{\dd}{\dd \tau}  \left[ \ee^{-\ii \, \xi \, \Ham \tau} \right] \rho(0) \ee^{\ii \, \xi \, \Ham \tau} +  \ee^{-\ii \, \xi \, \Ham \tau} \rho(0) \frac{\dd}{\dd \tau}  \left[ \ee^{\ii \, \xi \, \Ham \tau} \right] \nonumber \\
        &= (-\ii \xi \Ham) \Unitary_{\xi,\tau} [\rho(0)] + \Unitary_{\xi,\tau} [\rho(0)] (\ii \xi \Ham) = -\ii \left[\Ham,\xi \; \Unitary_{\xi,\tau} [\rho(0)]\right].
    \end{align}
    Therefore, the derivative of the density matrix with respect to $\tau$ becomes,
    \begin{align} \label{eq:first_part_derivation_drho/dt}
        \frac{\dd \rho(\tau)}{\dd \tau} &= \frac{1}{\sigma^3(\tau)} \int \dd \xi \; p_{\mu,\sigma} (\xi\,) \left((\xi-\mu(\tau)) \sigma(\tau) \dot{\mu}(\tau) + \left((\xi - \mu(\tau))^2 - \sigma^2(\tau) \right) \dot{\sigma}(\tau)\right) \Unitary_{\xi,\tau} [\rho(0)] \nonumber \\
        &-\ii \int \dd \xi \; p_{\mu,\sigma} (\xi\,) \left[\Ham,\xi \; \Unitary_{\xi,\tau} [\rho(0)]\right] \nonumber \\
        &= \frac{1}{\sigma^3(\tau)} \E{\left(\xi \, \sigma(\tau) \, \dot{\mu}(\tau)  \!\shortminus\! \mu(\tau) \sigma(\tau) \dot{\mu}(\tau) + (\xi \!\shortminus\! \mu(\tau))^2 \dot{\sigma}(\tau) \!\shortminus\! \sigma^2(\tau) \dot{\sigma}(\tau) \right) \Unitary_{\xi,\tau} [\rho(0)]}{p_{\mu,\sigma} (\xi\,)} \nonumber \\
        &- \ii \left[\Ham,\E{\xi \; \Unitary_{\xi,\tau}[\rho(0)]}{p_{\mu,\sigma} (\xi\,)} \right] = \frac{1}{\sigma^3(\tau)} \E{\xi \, \sigma(\tau) \, \dot{\mu}(\tau) \Unitary_{\xi,\tau} [\rho(0)]}{p_{\mu,\sigma} (\xi\,)} \nonumber \\
        &- \frac{1}{\sigma^3(\tau)} \E{\mu(\tau) \, \sigma(\tau) \, \dot{\mu}(\tau) \Unitary_{\xi,\tau} [\rho(0)]}{p_{\mu,\sigma} (\xi\,)} + \frac{1}{\sigma^3(\tau)} \E{(\xi - \mu(\tau))^2 \dot{\sigma}(\tau) \, \Unitary_{\xi,\tau} [\rho(0)]}{p_{\mu,\sigma} (\xi\,)} \nonumber \\
        &- \frac{1}{\sigma^3(\tau)} \E{\sigma^2(\tau) \dot{\sigma}(\tau)  \, \Unitary_{\xi,\tau} [\rho(0)]}{p_{\mu,\sigma} (\xi\,)} - \ii \left[\Ham,\E{\xi \; \Unitary_{\xi,\tau}[\rho(0)]}{p_{\mu,\sigma} (\xi\,)} \right] \nonumber \\
        &= \frac{\dot{\mu}(\tau)}{\sigma^2(\tau)} \, \E{\xi \, \Unitary_{\xi,\tau} [\rho(0)]}{p_{\mu,\sigma} (\xi\,)} - \frac{\mu(\tau) \, \dot{\mu}(\tau) }{\sigma^2(\tau)}\rho(\tau) + \frac{\dot{\sigma}(\tau)}{\sigma^3(\tau)}  \E{(\xi - \mu(\tau))^2 \, \Unitary_{\xi,\tau} [\rho(0)]}{p_{\mu,\sigma}} \nonumber \\
        &-\frac{\dot{\sigma}(\tau)}{\sigma(\tau)} \rho(\tau) - \ii \left[\Ham,\E{\xi \; \Unitary_{\xi,\tau}[\rho(0)]}{p_{\mu,\sigma} (\xi\,)} \right]. 
    \end{align}
    To continue our proof, we must then explicitly evaluate the averaged expressions $\E{\xi \; \Unitary_{\xi,\tau}[\rho(0)]}{p_{\mu,\sigma} (\xi\,)}$ and $\E{(\xi - \mu(\tau))^2 \, \Unitary_{\xi,\tau} [\rho(0)]}{p_{\mu,\sigma}}$. To do so we will employ the trick of taking the derivative of the Gaussian probability distribution with respect to $\mu(\tau)$. Namely, 
    \begin{align}
        \sigma^2(\tau)\frac{\dd}{\dd \mu(\tau)} p_{\mu,\sigma} (\xi\,) &= \sigma^2(\tau)\frac{\dd}{\dd \mu(\tau)} \left[ \frac{1}{\sqrt{2\pi \sigma^2(\tau)}} \exp\!{\left\{\!\shortminus\frac{(\xi - \mu(\tau))^2}{2 \sigma^2(\tau)}\!\right\}} \right] \nonumber \\
        &=  \xi \, p_{\mu,\sigma} (\xi\,) - \mu(\tau) \, p_{\mu,\sigma} (\xi\,)  \label{eq:first_derivative_p}
    \end{align}
    and
    \begin{align}
        \sigma^4(\tau) \frac{\dd^2}{\dd \mu(\tau)^2} \, p_{\mu,\sigma} (\xi\,) &= \sigma^2(\tau) \frac{\dd}{\dd \mu(\tau)} \left[(\xi-\mu) p_{\mu,\sigma} (\xi\,)\right] \nonumber \\
        &= \sigma^2(\tau) \left( -p_{\mu,\sigma} (\xi\,) + (\xi-\mu(\tau)) \frac{\dd}{\dd \mu(\tau)} p_{\mu,\sigma} (\xi\,) \right) \nonumber \\
        &= -\sigma^2(\tau) \, p_{\mu,\sigma} (\xi\,) + (\xi - \mu(\tau)) \sigma^2(\tau) \frac{\dd}{\dd \mu(\tau)} p_{\mu,\sigma} (\xi\,) \nonumber \\
        &= -\sigma^2(\tau) \, p_{\mu,\sigma} (\xi\,) + (\xi - \mu(\tau))^2 p_{\mu,\sigma} (\xi\,). \label{eq:second_derivative_p}
    \end{align}
    Then, we can move to evaluating the averages, i.e.,
    \begin{align}
        \E{\xi \; \Unitary_{\xi,\tau}[\rho(0)]}{p_{\mu,\sigma} (\xi\,)} &= \int \dd \xi \; \xi \; p_{\mu,\sigma} (\xi\,) \, \Unitary_{\xi,\tau}[\rho(0)] \nonumber \\
        &\stackrel{\eqref{eq:first_derivative_p}}{=}  \int \dd \xi \; \left(\sigma^2(\tau) \frac{\dd}{\dd \mu(\tau)} p_{\mu,\sigma} (\xi\,) + \mu(\tau) \, p_{\mu,\sigma} (\xi\,) \right) \Unitary_{\xi,\tau}[\rho(0)] \nonumber \\
        &= \sigma^2(\tau) \int \dd \xi \;  \frac{\dd}{\dd \mu(\tau)} p_{\mu,\sigma} (\xi\,) \, \Unitary_{\xi,\tau}[\rho(0)] + \mu(\tau) \, \rho(\tau) \nonumber \\
        &\stackrel{\eqref{eq:derivative_of_first_moment}}{=} -\ii \sigma^2(\tau) \tau \left[\Ham, \rho(\tau) \right] + \mu(\tau) \, \rho(\tau), \label{eq:mean_U}
    \end{align}
    and
    \begin{align}
        \E{(\xi - \mu(\tau))^2 \, \Unitary_{\xi,\tau} [\rho(0)]}{p_{\mu,\sigma}} &= \int \dd \xi \; (\xi - \mu(\tau))^2 \; p_{\mu,\sigma} (\xi\,) \, \Unitary_{\xi,\tau}[\rho(0)] \nonumber \\
        &\stackrel{\eqref{eq:second_derivative_p}}{=} \int \dd \xi \left(\sigma^4(\tau) \frac{\dd^2}{\dd \mu(\tau)^2} \, p_{\mu,\sigma} (\xi\,) + \sigma^2(\tau) \, p_{\mu,\sigma} (\xi\,) \right) \Unitary_{\xi,\tau}[\rho(0)] \nonumber \\
        &= \sigma^4(\tau) \int \dd \xi \, \frac{\dd^2}{\dd \mu(\tau)^2} \, p_{\mu,\sigma} (\xi\,) \, \Unitary_{\xi,\tau}[\rho(0)] + \sigma^2(\tau) \, \rho(\tau) \nonumber \\ &\stackrel{\eqref{eq:second_derivative_of_second_moment}}{=} -\sigma^4(\tau) \tau^2 \left[\Ham,\left[\Ham,\rho(\tau)\right]\right] + \sigma^2(\tau) \, \rho(\tau) \label{eq:variance_U}
    \end{align}
    where in the last step of both expressions we have used the change of variable of $\xi \rightarrow \xi + \mu(\tau)$ to calculate
    \begin{align}
        &\frac{\dd}{\dd \mu(\tau)} \int \dd \xi \; p_{\mu,\sigma} (\xi\,) \, \Unitary_{\xi,\tau}[\rho(0)] = \frac{\dd}{\dd \mu(\tau)} \int \dd \xi \; p_{\mu,\sigma} (\xi + \mu(\tau)\,) \, \Unitary_{\xi+\mu(\tau),\tau}[\rho(0)] \nonumber \\
        &=\frac{\dd}{\dd \mu(\tau)} \int \dd \xi \; p_{0,\sigma(\tau)} (\xi\,) \, \Unitary_{\xi+\mu(\tau),\tau}[\rho(0)] = \int \dd \xi \; p_{0,\sigma(\tau)} (\xi\,) \, \frac{\dd}{\dd \mu(\tau)} \Unitary_{\xi+\mu(\tau),\tau}[\rho(0)] \nonumber \\
        &\stackrel{\eqref{eq:derivative_U_wrt_tau}}{=} -\ii \int \dd \xi \, p_{0,\sigma(\tau)} (\xi\,)  \left[\Ham,\tau \; \Unitary_{\xi,\tau} [\rho(0)]\right] = -\ii \tau \left[\Ham, \rho(\tau) \right] ,\label{eq:derivative_of_first_moment}
    \end{align}
    which then implies that
    \begin{align}
        &\frac{\dd^2}{\dd \mu(\tau)^2} \! \int \! \dd \xi \, p_{\mu,\sigma} (\xi\,) \, \Unitary_{\xi,\tau}[\rho(0)] = \frac{\dd}{\dd \mu(\tau)} \left[ \frac{\dd}{\dd \mu(\tau)} \! \int \! \dd \xi \, p_{\mu,\sigma} (\xi\,) \, \Unitary_{\xi,\tau}[\rho(0)] \right] \nonumber \\
        &= \! -\ii \tau \left[\Ham, \frac{\dd}{\dd \mu(\tau)} \rho(\tau) \right] = -\ii \tau \left[\Ham, \frac{\dd}{\dd \mu(\tau)} \int \dd \xi \; p_{\mu,\sigma} (\xi\,) \, \Unitary_{\xi,\tau}[\rho(0)] \right] \nonumber \\
        &= -\tau^2 \left[\Ham,\left[\Ham,\rho(\tau)\right]\right]. \label{eq:second_derivative_of_second_moment}
    \end{align}
    Armed now with the explicit form of the expected values given in \eqnref{eq:mean_U} and \eqnref{eq:variance_U}, we can complete our derivation of \eqnref{eq:first_part_derivation_drho/dt}. Namely, 
    \begin{align}
        \frac{\dd \rho(\tau)}{\dd \tau} &= \frac{\dot{\mu}(\tau)}{\sigma^2(\tau)} \, \E{\xi \, \Unitary_{\xi,\tau} [\rho(0)]}{p_{\mu,\sigma} (\xi\,)} + \frac{\dot{\sigma}(\tau)}{\sigma^3(\tau)}  \E{(\xi - \mu(\tau))^2 \, \Unitary_{\xi,\tau} [\rho(0)]}{p_{\mu,\sigma}}  \nonumber \\
        &-\left(\frac{\mu(\tau) \, \dot{\mu}(\tau) }{\sigma^2(\tau)} + \frac{\dot{\sigma}(\tau)}{\sigma(\tau)}\right)\rho(\tau) - \ii \left[\Ham,\E{\xi \; \Unitary_{\xi,\tau}[\rho(0)]}{p_{\mu,\sigma} (\xi\,)} \right] \nonumber \\
        &= \frac{\dot{\mu}(\tau)}{\sigma^2(\tau)} \, \left(-\ii \sigma^2(\tau) \tau \left[\Ham, \rho(\tau) \right] + \mu(\tau) \, \rho(\tau) \right) -\left(\frac{\mu(\tau) \, \dot{\mu}(\tau) }{\sigma^2(\tau)} + \frac{\dot{\sigma}(\tau)}{\sigma(\tau)}\right)\rho(\tau) \nonumber \\
        &+ \frac{\dot{\sigma}(\tau)}{\sigma^3(\tau)}  \left( -\sigma^4(\tau) \tau^2 \left[\Ham,\left[\Ham,\rho(\tau)\right]\right] + \sigma^2(\tau) \, \rho(\tau) \right) \nonumber \\
        &- \ii \left[\Ham, \left(-\ii \sigma^2(\tau) \tau \left[\Ham, \rho(\tau) \right] + \mu(\tau) \, \rho(\tau)\right) \right]  \nonumber \\
        &= - \ii \, \dot{\mu}(\tau) \tau \left[\Ham,\rho(\tau) \right] - \sigma(\tau) \dot{\sigma}(\tau) \tau^2 \left[\Ham, \left[\Ham, \rho(\tau) \right]\right] \nonumber \\
        &- \sigma^2(\tau) \tau \left[\Ham, \left[\Ham, \rho(\tau) \right] \right] -\ii \mu(\tau) \left[\Ham,\rho(\tau)\right] \nonumber \\
        &= -\ii (\mu(\tau) + \dot{\mu}(\tau) \tau) \left[\Ham, \rho(\tau) \right] - \left(\frac{\dot{\sigma}(\tau)}{\sigma(\tau)} \tau + 1\right) \sigma^2(\tau) \tau \left[\Ham, \left[\Ham, \rho(\tau) \right] \right] 
    \end{align}
    which is then in the desired form of \eqnref{eq:form2_master_eq_ap}.
\end{myproof}

\section{Gaussian properties II}

\begin{lem}[The integral of $N+1$ Gaussians] \label{lem:ap_integral_N+1_gaussians}
    The integral of $N+1$ Gaussians can be written as
    \begin{align} \label{eq:integral_of_gaussians}
        &\frac{1}{(2\pi \Vcoll)^{1/2}} \frac{1}{(2\pi \Vloc)^{N/2}} \! \int \! \dd \xi \; \ee^{-\frac{\xi^2}{2\Vcoll}} \ee^{-\sum_{i=1}^N \frac{(\zetai - \xi - \omega)^2}{2\Vloc}} \nonumber \\
        &\quad \quad=  \frac{1}{\sqrt{2\pi (\Vcoll + \Vloc/N)}} \; \fzeta \exp\!{\left\{\!- \frac{(\avgzeta - \omega)^2}{2(\Vcoll + \Vloc/N)} \right\}} 
    \end{align}
    where $\avgzeta \coloneqq \frac{1}{N} \sum_{i = 1}^N \zetai$ and
    \begin{equation}
        \fzeta = \sqrt{\frac{1}{N (2\pi \Vloc)^{N-1}}} \; \exp\!{\left\{- \frac{1}{2\Vloc} \left(\sum_{i=1}^N  (\zetai)^2 -  N \avgzeta^{\;2} \right) \right\}}
    \end{equation}
\end{lem}

\begin{myproof}
    Let us first start by expanding the exponent of the local term,
    \begin{align}
        (\zetai - (\xi + \omega))^2 = (\zetai)^2 - 2(\xi + \omega) \zetai + (\xi+\omega)^2
    \end{align}
    such that when summing over $i = 1, \dots, N$, we can it as
    \begin{align} \label{eq:ap_sum_over_exponent_1}
        \sum_{i=1}^N (\zetai - (\xi + \omega))^2 &= 
        \sum_{i=1}^N \! \left(\!(\zetai)^2 \!-\! 2(\xi \!+\! \omega) \zetai \!+\! (\xi\!+\!\omega)^2\!\right) \nonumber \\
        &= \sum_{i=1}^N  (\zetai)^2 \!-\! 2(\xi + \omega) \!\sum_{i=1}^N \zetai \!+\! N (\xi\!+\!\omega)^2
    \end{align}
    Next, we introduce the \emph{average} of auxiliary frequencies experienced by $N$ atoms, 
    \begin{equation}
        \avgzeta \coloneqq \frac{1}{N} \sum_{i = 1}^N \zetai
    \end{equation}
    which will be employed to reformulate  \eqnref{eq:ap_sum_over_exponent_1}:
    \begin{align}
        \sum_{i=1}^N (\zetai - (\xi + \omega))^2 &= \sum_{i=1}^N  (\zetai)^2 \!-\! 2(\xi + \omega) N \avgzeta \!+\! N (\xi\!+\!\omega)^2 \nonumber \\
        &= N \avgzeta^{\;2} \!-\! 2(\xi + \omega) N \avgzeta \!+\! N (\xi\!+\!\omega)^2 + \sum_{i=1}^N  (\zetai)^2 -  N \avgzeta^{\;2} \nonumber \\
        &= N (\avgzeta - (\xi + \omega))^2 +  \sum_{i=1}^N  (\zetai)^2 -  N \avgzeta^{\;2} ,
    \end{align}
    where in the penultimate step we have added and subtracted $N\avgzeta$.
    Crucially, the local exponential in \eqnref{eq:integral_of_gaussians} can now be divided into two terms
    \begin{align}
        \ee^{-\sum_{i=1}^N \frac{(\zetai - \xi - \omega)^2}{2\Vloc}} &= \ee^{-\frac{(\avgzeta - (\xi + \omega))^2}{2\Vloc/N}} \ee^{- \frac{1}{2\Vloc} \left(\sum_{i=1}^N  (\zetai)^2 -  N \avgzeta^{\;2} \right)},
    \end{align}
    leaving the first exponential as the only one depending on $\xi$ and $\omega$ and the second term as a function of exclusively $\pmb{\zeta}$.

    Therefore, we have reduced the problem of integrating $N+1$ Gaussian functions with respect to $\xi$, to integrating only two Gaussians, i.e.,
    \begin{align}
         \int \dd \xi \; \ee^{-\frac{\xi^2}{2\Vcoll}} \ee^{-\sum_{i=1}^N \frac{(\zetai - \xi - \omega)^2}{2\Vloc}} =  \ee^{- \frac{1}{2\Vloc} \left(\sum_{i=1}^N  (\zetai)^2 -  N \avgzeta^{\;2} \right)} \int \dd \xi \; \ee^{-\frac{\xi^2}{2\Vcoll}}  \ee^{-\frac{(\avgzeta - (\xi + \omega))^2}{2\Vloc/N}} 
    \end{align}
    which has a known result:
    \begin{align}
         \int \dd \xi \; \ee^{-\frac{\xi^2}{2\Vcoll}}  \ee^{-\frac{(\avgzeta - (\xi + \omega))^2}{2\Vloc/N}}  = \sqrt{2\pi \frac{\Vcoll \Vloc/N}{\Vcoll + \Vloc/N}} \exp\!{\left\{- \frac{(\avgzeta - \omega)^2}{2(\Vcoll + \Vloc/N)} \right\}}.
    \end{align}
    Therefore,
    \begin{align}
        &\frac{1}{(2\pi \Vcoll)^{1/2}} \frac{1}{(2\pi \Vloc)^{N/2}} \int \dd \xi \; \ee^{-\frac{\xi^2}{2\Vcoll}} \ee^{-\sum_{i=1}^N \frac{(\zetai - \xi - \omega)^2}{2\Vloc}} \nonumber \\
        &\quad \quad= \frac{1}{\sqrt{2\pi (\Vcoll + \Vloc/N)}} \; \fzeta \exp\!{\left\{- \frac{(\avgzeta - \omega)^2}{2(\Vcoll + \Vloc/N)} \right\}} 
    \end{align}
    where
    \begin{equation}
        \fzeta = \sqrt{\frac{1}{N(2\pi \Vloc)^{N-1}}} \; \exp\!{\left\{- \frac{1}{2\Vloc} \left(\sum_{i=1}^N  (\zetai)^2 -  N \avgzeta^{\;2} \right) \right\}}
    \end{equation}
\end{myproof}

\begin{lem}[Set of nested integrals of Gaussian functions] \label{lem:recurrence_relation_integral}
    Consider the following recurrence relation defining $\mathcal{P}_j(\omega_j)$ in terms of $\mathcal{P}_{j\shortminus1}(\omega_{j\shortminus1})$ for $j = 0,1,2,\dots$:
    \begin{align} \label{eq:recursive_relation}
        \!\!\!\mathcal{P}_j(\omega_j) \!= \!\!\int\!\! \dd \omega_{j\shortminus1} \frac{1}{\sqrt{2\pi \Vp}} \exp\!{\left\{\!-\frac{(\omega_j - \omega_{j\shortminus1})^2}{2\Vp}\!\right\}} \frac{1}{\sqrt{2\pi \Vq}} \exp\!{\left\{\!-\frac{(\avgzeta_{j\shortminus1}-\omega_{j\shortminus1})^2}{2\Vq}\!\right\}} \mathcal{P}_{j\shortminus1} (\omega_{j\shortminus1})
    \end{align}
    which involves a generalized convolution of Gaussian distributions with variances $\Vp$ and $\Vq$, respectively. The initial condition for the recursive relation is given by a Gaussian of the form
    \begin{align}
        \mathcal{P}_0(\omega_0) = C_0 \exp{\left\{-\frac{(\omega_0 - \mu_0)^2}{2V_0}\right\}},    
    \end{align}
    where $C_0$ is a constant, $V_0\geq 0$ and $\mu_0 \in \mathds{R}$. Then, $\forall j\geq 1$, the solution to the recurrence relation is
    \begin{equation} \label{eq:sol_recursive_relation}
        \mathcal{P}_j(\omega_j) = C_j \exp{\left\{-\frac{(\omega_j - \mu_j)^2}{2V_j}\right\}},
    \end{equation}
    with parameters $C_j$, $\mu_j$ and $V_j$ given in turn by the following coupled recurrence relations:
    \begin{align} 
        \label{eq:recursive_Cj} C_j &= C_{j\shortminus1} \left( 2\pi \left(\Vp + \Vq + \frac{\Vp\Vq}{V_{j\shortminus1}}\right)\right)^{\!\!-1/2} \exp{\left\{-\frac{(\omega_{j\shortminus1} - \mu_{j\shortminus1})^2}{2(\Vq + V_{j\shortminus1})}\right\}} \\
        \label{eq:recursive_muj} \mu_j &= \frac{\Vq \, \mu_{j\shortminus1} + V_{j\shortminus1} \avgzeta_{j\shortminus1}}{\Vq + V_{j\shortminus1}} \\
        \label{eq:recursive_Vj} V_j &= \Vp + \frac{\Vq V_{j\shortminus1}}{\Vq + V_{j\shortminus1}}. 
    \end{align}
\end{lem}

\begin{myproof}
    As with many recursive problem, it is sufficient to use mathematical induction to prove that the proposed solution in \eqnref{eq:sol_recursive_relation} holds. This involves verifying the base case for $j = 0$, and then showing that the relation \eqref{eq:recursive_relation} is fulfilled by \eqnref{eq:sol_recursive_relation} for any $j\geq 1$. The base case is trivially satisfied by the definition. For the inductive step, we assume that $\mathcal{P}_{j\shortminus1}(\omega_{j\shortminus1})$ takes the form given in \eqnref{eq:sol_recursive_relation}, and we show that substituting this into the recurrence relation in \eqnref{eq:recursive_relation} yields the correct form for $\mathcal{P}_j(\omega_j)$. This is done by performing the integral in \eqnref{eq:recursive_relation} explicitly:
    \begin{align} \label{eq:inductive_step_1}
        \mathcal{P}_j(\omega_j) 
        &= \int\! \dd \omega_{j\shortminus1} \frac{1}{\!\sqrt{2\pi \Vp}} \exp\!{\left\{\!\!\shortminus\frac{(\omega_j \! \shortminus \!\omega_{j\shortminus1})^2}{2\Vp}\!\right\}} \nonumber\\
        &\qquad\times \frac{1}{\!\!\sqrt{2\pi \Vq}}  \exp\!{\left\{\!\!\shortminus\frac{(\avgzeta_{j\shortminus1}\! \shortminus \!\omega_{j\shortminus1})^2}{2\Vq}\!\right\}} C_{j\shortminus1} \! \exp\!{\left\{\!\!\shortminus\frac{(\omega_{j\shortminus1} \! \shortminus \! \mu_{j\shortminus1})^2}{2V_{j\shortminus1}}\!\!\right\}} \nonumber \\
        &= \frac{C_{j\shortminus1}}{\sqrt{2\pi \left(\Vq + \Vp + \frac{\Vp\Vq}{V_{j\shortminus1}}\right)}} \exp\!{\left\{\!- \frac{\omega_j^{\,2} - 2\alpha \omega_j + \beta}{2 \left( \Vp + \frac{\Vq V_{j\shortminus1}}{\Vq + V_{j\shortminus1}}\right)} \!\right\}}
    \end{align}
    with $\alpha$ and $\beta$ fulfilling:
    \begin{align}
        \alpha &= \frac{\Vq \, \mu_{j\shortminus1} + V_{j\shortminus1} \avgzeta_{j\shortminus1}}{\Vq + V_{j\shortminus1}} = \mu_j, \\
        \beta &= \frac{\Vq \, \mu_{j\shortminus1}^2 + V_{j\shortminus1} \avgzeta^{\;2}_{j\shortminus1} + \Vp (\avgzeta_{j\shortminus1} - \mu_{j\shortminus1})^2}{\Vq + V_{j\shortminus1}},
    \end{align}
    which are both independent of $\omega_j$ and $\omega_{j\shortminus1}$. If now we `complete the square', i.e., split the exponent of \eqnref{eq:sol_recursive_relation} into $\omega_j^{\,2} - 2\alpha\omega_j + \beta = (\omega_j - \alpha)^2 - \alpha^2 + \beta$, and substitute the expressions for $\alpha$ and $\beta$, we get,
    \begin{align}
        \mathcal{P}_j(\omega_j) = \frac{C_{j\shortminus1} \exp\!{\left\{\!-\frac{-\alpha^2 + \beta}{2\left(\Vp + \frac{\Vq V_{j\shortminus1}}{\Vq + V_{j\shortminus1}}\right)}\!\right\}}}{\sqrt{2\pi \left(\Vq + \Vp + \frac{\Vp\Vq}{V_{j\shortminus1}}\right)}} \exp\!{\left\{ \!- \frac{(\omega_j - \alpha)^2}{2 \left(\Vp + \frac{\Vq V_{j\shortminus1}}{\Vq + V_{j\shortminus1}} \right)}\!\right\}}
    \end{align}
    which, since $\alpha = \mu_j$ and 
    \begin{align}
        -\alpha^2 + \beta &= -\mu_j^{\,2} + \frac{\Vq \, \mu_{j\shortminus1}^2 + V_{j\shortminus1} \avgzeta^{\;2}_{j\shortminus1} + \Vp (\avgzeta_{j\shortminus1} - \mu_{j\shortminus1})^2}{\Vq + V_{j\shortminus1}} \nonumber \\
        &= -\frac{(\Vq \, \mu_{j\shortminus1} + V_{j\shortminus1} \avgzeta_{j\shortminus1})^2}{(\Vq + V_{j\shortminus1})^2} + \frac{\Vq \, \mu_{j\shortminus1}^2 + V_{j\shortminus1} \avgzeta^{\;2}_{j\shortminus1} + \Vp (\avgzeta_{j\shortminus1} - \mu_{j\shortminus1})^2}{\Vq + V_{j\shortminus1}} \nonumber \\
        &= \frac{-\Vq^2 \, \mu_{j\shortminus1}^2 - 2 \Vq \, \mu_{j\shortminus1} V_{j\shortminus1} \avgzeta_{j\shortminus1} - V_{j\shortminus1}^{\,2} \avgzeta_{j\shortminus1}^{\;2} + \Vq^2 \, \mu^2_{j\shortminus1} + V_{j\shortminus1} \Vq \, \mu^2_{j\shortminus1}  }{(\Vq + V_{j\shortminus1})^2} \nonumber \\
        &\quad + \frac{V_{j\shortminus1}^{\,2} \avgzeta^{\;2}_{j\shortminus1} + \Vq V_{j\shortminus1} \avgzeta^{\;2}_{j\shortminus1} + \Vp (\Vq + V_{j\shortminus1}) (\avgzeta^{\;2}_{j\shortminus1} - 2\avgzeta_{j\shortminus1}\mu_{j\shortminus1} + \mu_{j\shortminus1}^2)}{(\Vq + V_{j\shortminus1})^2} \nonumber \\
        &= \frac{- 2 \Vq \, \mu_{j\shortminus1} V_{j\shortminus1} \avgzeta_{j\shortminus1} + V_{j\shortminus1} \Vq \, \mu^2_{j\shortminus1} + \Vq V_{j\shortminus1} \avgzeta^{\;2}_{j\shortminus1} + \Vp\Vq\avgzeta^{\;2}_{j\shortminus1} - 2\Vp\Vq\avgzeta_{j\shortminus1}\mu_{j\shortminus1} }{(\Vq + V_{j\shortminus1})^2} \nonumber \\
        &+ \frac{\Vp\Vq\mu_{j\shortminus1}^2 + \Vp V_{j\shortminus1}\avgzeta^{\;2}_{j\shortminus1} - 2\Vp V_{j\shortminus1}\avgzeta_{j\shortminus1}\mu_{j\shortminus1} + \Vp V_{j\shortminus1}\mu_{j\shortminus1}^2 }{(\Vq + V_{j\shortminus1})^2} \nonumber \\
        &= \frac{(\Vq V_{j\shortminus1} + \Vp \Vq + \Vp V_{j\shortminus1})\avgzeta^{\;2}_{j\shortminus1} - 2 (\Vq V_{j\shortminus1} + \Vp \Vq + \Vp V_{j\shortminus1}) \avgzeta_{j\shortminus1} \mu_{j\shortminus1} }{(\Vq + V_{j\shortminus1})^2} \nonumber \\
        &+ \frac{(\Vq V_{j\shortminus1} + \Vp \Vq + \Vp V_{j\shortminus1}) \mu_{j\shortminus1}^2}{(\Vq + V_{j\shortminus1})^2} = \frac{(\Vq V_{j\shortminus1} + \Vp \Vq + \Vp V_{j\shortminus1})}{(\Vq + V_{j\shortminus1})^2} (\avgzeta_{j\shortminus1} - \mu_{j\shortminus1})^2 \nonumber \\
        &= \left(\Vp + \frac{\Vq V_{j\shortminus1}}{\Vq + V_{j\shortminus1}} \right) \frac{(\avgzeta_{j\shortminus1} - \mu_{j\shortminus1})^2}{\Vq + V_{j\shortminus1}},
    \end{align}
    yields the expected form of \eqnref{eq:sol_recursive_relation}, with $C_j$, $\mu_j$ and $V_j$ as specified by \eqnsref{eq:recursive_Cj}{eq:recursive_Vj}.
\end{myproof}

    \chapter{Appendix for \chapref{chap:model}}
\label{AppendixE}

\section{Unconditional dynamics in the presence of a fluctuating field in the LG regime}
\label{sec:UncondJx}

Consider the most general dephasing evolution possible:
\begin{align}
    \dd \rhoc(t) = &-\ii \, \omega(t) \! \left[\Jz,\rhoc(t) \right] \! \dt  \!+\! \sum_{\alpha = x,y,z} \kappa_\alpha \D[\Jindex{\alpha}]\rhoc(t) \dt \!+\! M \D[\Jy] \rhoc(t) \dt, \label{eq:cond_dyn}
\end{align}
where collective dephasing occurs along the three directions $x, \, y$ and $z$ at rates $\kappa_x, \, \kappa_y$ and $\kappa_z$, respectively. From this master equation, a set of differential equations describing the evolution of the corresponding observables $\Jx, \, \Jy,$ and $\Jz$ can be derived:
\begin{align}
    \dd\omega & = - \chi \omega(t) \dt + \sqrt{q_\omega} \dW_\omega,  \label{eq:uncond_Bt} \\
   \dd\brkt{\Jx} &  = -\omega(t) \brkt{\Jy} \dt - \frac{1}{2}(M + \kappa_y + \kappa_z) \brkt{\Jx} \dt, \label{eq:uncond_Jx} \\
   \dd\brkt{\Jy} &  = \omega(t) \brkt{\Jx} \dt - \frac{1}{2}(\kappa_x + \kappa_z) \brkt{\Jy} \dt, \label{eq:uncond_Jy} \\
   \dd \brkt{\Jz} &= -\frac{1}{2} \left(M + \kappa_y + \kappa_x \right) \brkt{\Jz} \dt, \label{eq:uncond_Jz}
\end{align}
Given that \eqnsref{eq:uncond_Bt}{eq:uncond_Jy} form a closed set of coupled differential equations, they can be numerically solved in order to yield the unconditional evolution of $\brkt{\Jx(t)}$. However, since we would rather find an approximate analytical expression, we focus on short timescales such that $\oline{\omega}(t)\, t\ll1$. Then, by substituting the random variable $\omega(t)$ by its time average $\;\oline{\omega}(t)$ into \eqnsref{eq:uncond_Jx}{eq:uncond_Jz}:
\begin{equation}
    \oline{\omega}(t) = \frac{1}{t} \int_0^t \, \dd\tau \, \omega(\tau),
\end{equation}
the system of differential equations  yields,
\begin{align}
    \brkt{\Jx(t)} = \frac{J}{2\Theta} \, \ee^{- (M + \kappa_x + \kappa_y + 2\kappa_z + \Theta)\,t/4} \left(M \!-\! \kappa_x \!+\! \kappa_y \!+\! \Theta \!-\! \ee^{t\Theta/2} (M \!-\! \kappa_x \!+\! \kappa_y \!-\! \Theta ) \right),
\end{align}
where $\Theta = \sqrt{(M-\kappa_x+\kappa_y)^2 - 16 \; \oline{\omega}^2(t)}$ and the initial conditions set were to the CSS state: $\brkt{\Jx(0)} = J$ and $\brkt{\Jy(0)} = 0$. By expanding the unconditional evolution of $\Jx$ to first order in $\,\oline{\omega}(t)$, we obtain
\begin{align}
    \label{eq:ap_UncondJx_Taylor} \brkt{\Jx(t)} &\approx J \, \ee^{- (M + \kappa_z + \kappa_y)\,t/2} \nonumber \\
    &+ 2 \, J \, \ee^{- (M + \kappa_z + \kappa_y)\,t/2} \, \oline{\omega}^2(t) \, \frac{2-2\ee^{(M-\kappa_x+\kappa_y)t/2}+t(M-\kappa_x+\kappa_y)}{(M-\kappa_x+\kappa_y)^2}.
\end{align}
Hence, he unconditional evolution of $\Jx$ can be approximated as
\begin{align} \label{eq:Jx_approx_ap}
    \brkt{\Jx(t)} \approx J \, \ee^{- (M + \kappa_z + \kappa_y)\,t/2},
\end{align}
when the first term of \ref{eq:ap_UncondJx_Taylor} dominates over the second. In other words, when the time average of $\omega(t)$ fulfills
\begin{align} \label{eq:BBound1}
    |\,\oline{\omega}(t)| \le \frac{\sqrt{2}}{t},
\end{align}
where we have approximated the exponent $\ee^{(M-\kappa_x+\kappa_y)t/2}$ in \eref{eq:ap_UncondJx_Taylor} as its Taylor expansion up to second order. Namely, $\ee^{(M-\kappa_x+\kappa_y)t/2} \approx 1 + \frac{1}{2}(M-\kappa_x+\kappa_y)t + \frac{1}{8}(M-\kappa_x+\kappa_y)^2t^2$.
Additionally, note that within the linear-Gaussian regime, the approximation \eqnref{eq:Jx_approx_ap} is independent of the decoherence rate $\kappa_x$. Therefore, the decoherence rate $\kappa_x$ is redundant and disappears from \eqnsref{eq:measure_before}{eq:var_eq_Jy_before} when performing the Holstein-Primakoff transformation of the SME according to the $\position$ and $\momentum$ quadratures of \eqnref{eq:canonical_X&P}. Mathematically, one can check that the dissipative terms disappear, since $\trace{\momentum \D[\position] \rhoc} = \trace{\position \D[\momentum] \rhoc} = 0$. This is can be intuitively explained---deviations of $\,\Jqvec{J}(t)$ from the $x$-direction are then too small for the collective noise manifested via $\D[\Jx]$ in equation \eqnref{eq:cond_dyn} to have any effect on the quadratures \eqnref{eq:canonical_X&P}. 

\begin{figure}[t!]
        \subfloat{%
            \includegraphics[width=.5\linewidth]{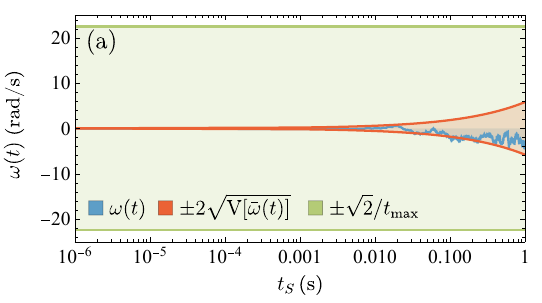}%
        }\hfill
        \subfloat{%
            \includegraphics[width=.5\linewidth]{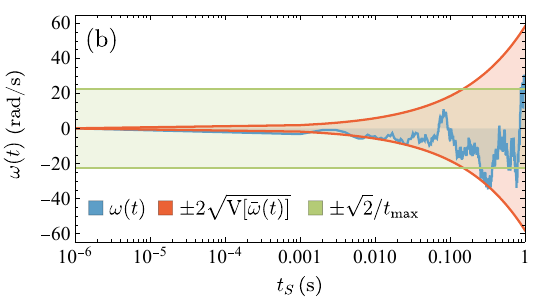}%
        }\\
        \subfloat{%
            \includegraphics[width=.5\linewidth]{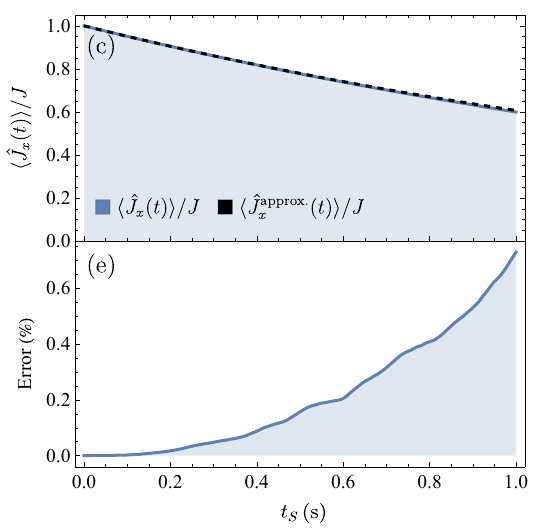}%
        }\hfill
        \subfloat{%
            \includegraphics[width=.5\linewidth]{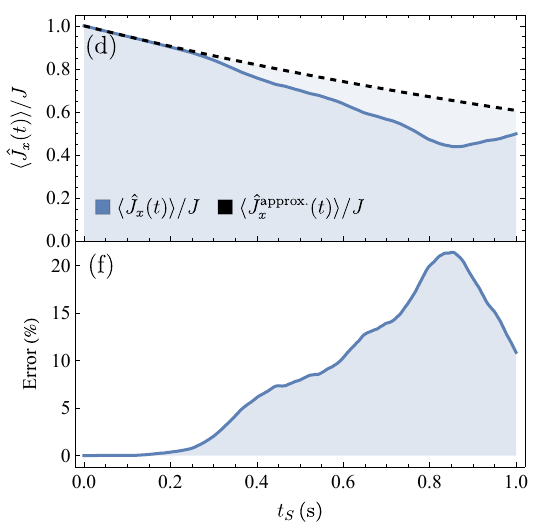}%
        }
         \caption[Validity of the LG approximation for $\brkt{\Jx(t)}$ at different field fluctuation strengths]{\textbf{Validity of the LG approximation for $\brkt{\Jx(t)}$ at different field fluctuation strengths} The parameters used to generate the plots are $M = \SI{100}{\kilo\hertz}$, $\kappa_z = \kcoll = \SI{1}{\hertz}$, $J = 10^7$, $\eta = 1$, and $\chi = 0$, with $t_S = (M + \kcoll)t$ being the scaled time such that $t_S = 1$ when $t = (M + \kcoll)^{-1}$. Plots (a), (c), and (d) (left column)  have been generated with a field fluctuation strength of $q_\omega = 10^{14}\SI{}{\radian \, \second^{-3}}$, and plots (b), (d), (f) (right column) with $q_\omega = 10^{16}\SI{}{\radian \, \second^{-3}}$. The first row (plots (a) and (b)) show the fluctuating field in solid blue juxtaposed with the confidence interval of $\oline{\omega}(t)$ ($\delta$) as well as the upper bound for $|\,\oline{\omega}(t)|$ which stems from the Taylor expansion of $\langle \Jx(t) \rangle$. The plots in the second row (subfigures (c) and (d)), compare the exact solution of $\langle \Jx(t) \rangle$ with its approximation $J \ee^{-(M + \kcoll)t/2}$. Finally, in the bottom row, plots (e) and (f) show the error percentage of the approximation of $\langle \Jx(t) \rangle$.}\label{fig:AppendixA_Figure}
\end{figure}

Next, we would like to find the mean and variance of the time average of $\omega(t)$,
\begin{align}
    \EE{\,\oline{\omega}(t)} = \frac{1}{t} \int_0^t \EE{\omega(\tau)} \,  \d\dd\tau, \\
    \mrm{V}[\, \oline{\omega}(t)] = \EE{(\,\oline{\omega}(t) - \EE{\,\oline{\omega}(t)})^2}.
\end{align}

As discussed in \secref{sec:OUP_intro}, the mean of a random variable $\omega(t)$ driven by an OU process of the form in \eqnref{eq:uncond_Bt} reads
\begin{equation}
    \EE{\omega(t)} = \omega(0) \, \ee^{-\chi t} = 0,
\end{equation}
where we have taken $\omega(0) = 0$, while its covariance (as shown in \eqnref{eq:covariance_OUP}) is:
\begin{align}
    \mrm{C}[\omega(s),\omega(t)] = \EE{\omega(s) \omega(t)} = \frac{q_\omega}{2\chi} \left(e^{-\chi|t-s|} - \ee^{-\chi(t+s)} \right).
\end{align}

Hence, $\EE{\, \oline{\omega}(t)} = 0$ and its variance:
\begin{align}
    &\mrm{V}[\, \oline{\omega}(t)] = \EE{(\,\oline{\omega}(t))^2} = \EE{\frac{1}{t} \! \int_0^t \!\! \dd \tau_1 \, \omega(\tau_1) \; \frac{1}{t} \! \int_0^t \!\! \dd \tau_2 \, \omega(\tau_2) } \\
    &=\frac{1}{t^2} \! \int_0^t \!\! \dd\tau_1 \! \int_0^t \!\! \dd\tau_2 \; \EE{\omega(\tau_1)\omega(\tau_2)} = \frac{1}{t^2} \int_0^t \dd\tau_1 \, \dd\tau_2 \, \frac{q_\omega}{2\chi} \left(\ee^{-\chi |\tau_1 - \tau_2|} - \ee^{-\chi (\tau_1 + \tau_2)}\right)\\
    &= \frac{q_\omega}{2\chi t^2} \left[ \int_0^t \dd\tau_1 \left( \int_0^{\tau_1} \dd\tau_2 \, \ee^{-\chi(\tau_1 - \tau_2)} + \int_{\tau_1}^t \dd\tau_2 \,  \ee^{\chi(\tau_1 - \tau_2)}\right) - \left( \frac{1-\ee^{-\chi t}}{\chi}\right)^2 \right]  \\
    &= \frac{q_\omega}{2\chi t^2} \left[ \frac{1}{\chi} \int_0^t \dd\tau_1 \, \left(1 - \ee^{-\chi \tau_1} - \ee^{\chi (\tau_1-t)} + 1\right) - \left( \frac{1-\ee^{-\chi t}}{\chi}\right)^2 \right]  \\
    &=\frac{q_\omega}{2\chi^3 t^2} (4\ee^{-\chi t} + 2\chi t -\ee^{-2\chi t} - 3).
\end{align}

If $\chi \ll \frac{4}{3\,t}$, we can correctly approximate the variance as:
\begin{align}
    &\mrm{V}[\, \oline{\omega}(t)] = \frac{q_\omega \, t}{3}.
\end{align}
So, using the 68-95-99.7 rule and taking the confidence interval for $\;\oline{\omega}(t)$ to be
\begin{align}
    \left|\EE{\;\oline{\omega}(t)} \pm 2\sqrt{\mrm{V}[\, \oline{\omega}(t)]}\right| = 2 \sqrt{\frac{q_\omega t}{3}},
\end{align}
we can define an inequality that ensures \eref{eq:BBound1} to hold with high (95\%) probability:
\begin{equation} \label{eq:ap_ineq}
    2 \sqrt{\frac{q_\omega t}{3}}\ \lesssim \frac{\sqrt{2}}{t}.
\end{equation}
As shown in figure \fref{fig:AppendixA_Figure}, the inequality in \eref{eq:ap_ineq} correctly assures the approximation $\langle \Jx(t) \rangle \approx J \ee^{-(M+\kcoll)t/2}$ to hold. Moreover, it gives an upper bound to the value of $q_\omega$,
\begin{equation}
    q_\omega \lesssim \frac{3}{2 \, t^3}.
\end{equation}

In summary, the unconditional evolution of $\Jx$ can be taken to be
\begin{align}
    \brkt{\Jx(t)} \approx J \ee^{- (M + \kcoll )\,t/2},
\end{align}
if $\chi \le \frac{4}{3\,t}$ and $q_\omega \lesssim \frac{3}{2  t^3}$. Note that we have renamed $\kappa_z \eqqcolon \kcoll$ and dropped $\kappa_y$, since it is unnecessary. Namely, by transforming the parameters of the continuous measurement in \eqnref{eq:measure_before} as follows:~$M\to M - {\kappa}_y$, $\eta\to\eta M / (M - {\kappa}_y)$ and $y(t) \to y(t) \sqrt{M/(M-{\kappa}_y)}$;~we retrieve the conditional dynamics \eqnsref{eq:measure_before}{eq:mean_eq_Jy_before} with ${\kappa}_y = 0$. Hence, the impact of the collective noise introduced via $\D[\Jy]$ in \eqnref{eq:cond_dyn} can always be interpreted and incorporated into a modified form of the continuous measurement \eqnref{eq:measure_before}.

\section{Conditional dynamics of the variance $\Vy(t)$ in the LG regime}
\label{sec:VarSol}

When taking into account possible decoherence mechanisms along the same direction of the magnetic field, the differential equation for the conditional variance of $\Vy(t)$ is shown to be,
\begin{align} \label{eq:ap_Vardif}
     \dd \Vy(t) & = -4 M \eta \Vy(t) dt + \kcoll \; J^{\,2} \ee^{-(M + \kcoll) t} \dt,
\end{align}
whose solution exists and can be given in terms of modified Bessel functions of first and second kind ($\mathcal{I}_{\beta}[  \cdot  ]$ and $\mathcal{K}_{\beta}[  \cdot  ]$), and regularized confluent hypergeometric functions ($\prescript{}{0}{F}_1[  \cdot  ]$). Namely,
\begin{align} \label{eq:exactsol}
      \Vy(t) = \mrm{V}_\mrm{e}(t) & = J \ee^{-(M+\kcoll)t/2}  \Bigg(\mathcal{I}_{1}\Big[2\beta\Big] \Big(\sqrt{\eta \, \kcoll M}   \mathcal{K}_0[2\alpha] - \kcoll   \mathcal{K}_1[2\alpha]\Big) + \nonumber  \\
    &  + \mathcal{K}_1\Big[2\beta\Big] \Big(\kcoll \, \mathcal{I}_1[2\alpha] + \sqrt{\eta \, \kcoll M}  \, \prescript{}{0}{F}_1[1,\alpha^2] \Big) \Bigg)  \nonumber  \\
    &  \Bigg/ \Bigg( 2\prescript{}{0}{F}_1[1,\beta^2] \Big(\sqrt{\eta \, \kcoll M} \mathcal{K}_1[2\alpha] - M\eta \, \mathcal{K}_0[2\alpha]\Big) \nonumber  \\
    &  +\frac{2 \eta M}{M + \kcoll} \mathcal{K}_0[2\beta] \Big((M+\kcoll)  \prescript{}{0}{F}_1[1,\alpha^2] + 2\kcoll \,J  \prescript{}{0}{F}_1[2,\alpha^2]\Big) \Bigg),
\end{align}
where $\alpha = 2J\sqrt{\eta \, \kcoll M}/(M + \kcoll)$ and $\beta = \alpha   \ee^{-(M + \kcoll)t/2}$. The behavior of the solution in \ref{eq:exactsol} can be better understood when broken down into different regimes.
\begin{table}[ht]
\centering
\begin{tabular}{|p{4cm}|p{4cm}|p{4cm}|}
\hline
\begin{equation*}
\mathcal{I}_1[2\beta] \approx \frac{\ee^{2\beta}}{2\sqrt{\pi \beta}}
\end{equation*} &
\begin{equation*}
\mathcal{K}_0[2\alpha] \approx \frac{1}{2}\sqrt{\frac{\pi}{\alpha}}\ee^{-2\alpha}
\end{equation*} &
\begin{equation*}
\mathcal{K}_1[2\alpha] \approx  \frac{1}{2}\sqrt{\frac{\pi}{\alpha}}\ee^{-2\alpha}
\end{equation*}\\
\hline
\begin{equation*}
\mathcal{K}_1[2\beta] \approx  \frac{1}{2}\sqrt{\frac{\pi}{\beta}}\ee^{-2\beta}
\end{equation*} &
\begin{equation*}
\mathcal{I}_1[2\alpha] \approx \frac{\ee^{2\alpha}}{2\sqrt{\pi \alpha}}
\end{equation*} &
\begin{equation*}
\prescript{}{0}{F}_1[1,\alpha^2] \approx \frac{\ee^{2\alpha}}{2\sqrt{\pi \alpha}}
\end{equation*}\\
\hline
\begin{equation*}
\prescript{}{0}{F}_1[1,\beta^2] \approx \frac{\ee^{2\beta}}{2\sqrt{\pi \beta}}
\end{equation*} &
\begin{equation*}
\mathcal{K}_0[2\beta] \approx \frac{1}{2} \sqrt{\frac{\pi}{\beta}} \ee^{-2\beta}
\end{equation*} &
\begin{equation*}
\prescript{}{0}{F}_1[2,\alpha^2] \approx \frac{\ee^{2\alpha}}{2 \alpha \sqrt{\pi \alpha}}
\end{equation*}\\
\hline
\end{tabular}
\caption[Series expansions of the Bessel functions]{\textbf{Series expansions of the Bessel functions} for $1/\alpha$ and $1/\beta$ around $1/\alpha_0 = 0$ and $1/\beta_0 = 0$, to leading order.}
\label{tab:Bessel_exp}
\end{table}

In order to do so, the first step is to expand the modified Bessel functions and the regularized confluent hypergeometric functions around infinity and only keep the first order, since $\alpha \gg 1$ and $\beta \gg 1$. The relevant expansions are shown in \tabref{tab:Bessel_exp}.

\begin{figure}[t!]
\centering
\begin{subfigure}{.49\textwidth}
  \centering
  \includegraphics[width=\linewidth]{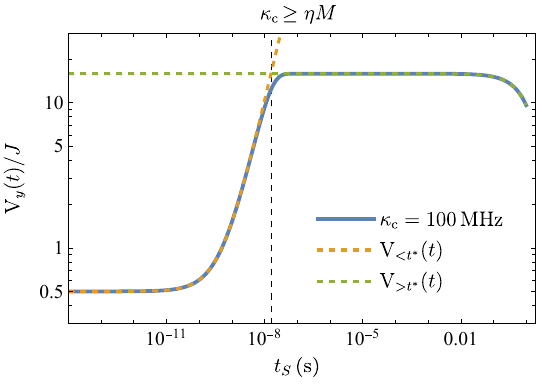}
\end{subfigure} \hfill
\begin{subfigure}{.49\textwidth}
  \centering
  \includegraphics[width=\linewidth]{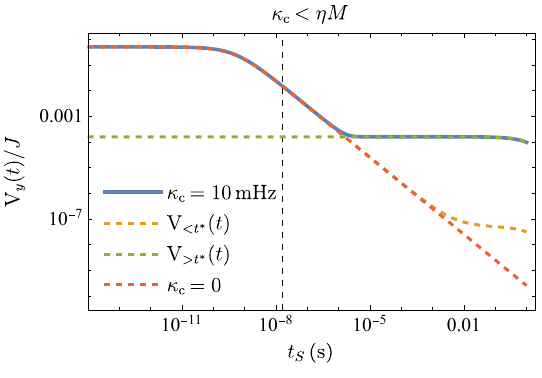}
\end{subfigure}
\caption[Strong and weak dephasing regimes for the variance]{\textbf{Strong and weak dephasing regimes for the variance}. In the subfigure on the left ($\kcoll = \SI{10}{\milli\hertz} < M$), the exact variance solution $\Vy(t)$ is compared to the approximated functions $\mrm{V}_{<t^{*}}(t)$ and $\mrm{V}_{>t^{*}}(t)$ (dashed green and yellow, respectively). The transition time $t^*$ between these two regimes is marked with a dotted black vertical line. In the second graph (on the right, with $\kcoll = \SI{100}{\mega\hertz} > M$), the two different regimes $\mrm{V}_{<t^*}(t)$ (dashed green) and $\mrm{V}_{>t^*}(t)$ (dashed yellow) are superimposed with the exact solution of $\Vy(t)$ (in solid blue). Notation $t_s$ refers to a scaled time, $t_s = t (M + \kcoll)$. All plots have been generated with $M = \SI{100}{\kilo\hertz}$, $\eta = 1$, and $J = 10^9$.}
\label{fig:ap_Variance_plots}
\end{figure}

By then substituting the leading order expansions of \tabref{tab:Bessel_exp} to solution \ref{eq:exactsol} and approximating $\ee^{4\beta}$ as $\ee^{-2\alpha(-2+t(M+\kcoll))}$, the variance of $\Jz(t)$ simplifies to
\begin{align}\label{eq:approx_varJz}
    \Vy(t) \approx \frac{1}{2}\,J\ee^{-(M+\kcoll)t/2} \frac{\sqrt{M\kcoll\eta} \cosh{(2Jt\sqrt{M\kcoll \eta})}+\kcoll \sinh{(2Jt\sqrt{M\kcoll \eta})}}{\sqrt{M\kcoll \eta} \cosh{(2Jt \sqrt{M\kcoll\eta})}+M\eta \sinh{(2Jt\sqrt{M\kcoll\eta})}}.
\end{align}

Note that if $2Jt\sqrt{M\kcoll \eta} \gg 1$, then, $\cosh{(2Jt\sqrt{M\kcoll \eta})}  \approx \frac{1}{2} \ee^{2Jt\sqrt{M\kcoll \eta}}$ and $\sinh{(2Jt\sqrt{M\kcoll \eta})} \approx \frac{1}{2} \ee^{2Jt\sqrt{M\kcoll \eta}}$, such that,
\begin{align}\label{eq:approx_varJz_1}
    \Vy(t) \approx \mrm{V}_{>t^*}(t) = \frac{1}{2}\,J\,\ee^{-(M+\kcoll)t/2} \sqrt{\frac{\kcoll}{\eta M}}.
\end{align}

If $2Jt\sqrt{M\kcoll \eta} \ll 1$, then, $\cosh{(2Jt\sqrt{M\kcoll \eta})} \approx 1$ and $\sinh{(2Jt\sqrt{M\kcoll \eta})} \approx 2Jt\sqrt{M\kcoll \eta}$ such that,
\begin{align}\label{eq:approx_varJz_2}
    \Vy(t) \approx \mrm{V}_{<t^*}(t) = J\ee^{-(M+\kcoll)t/2} \, \frac{(1+2 J t \kcoll)}{2+4 J t M \eta}.
\end{align}
Moreover, note that since $2Jt\sqrt{M\kcoll \eta} \ll 1$, we can then derive the  condition
\begin{align}
  t \ll t^{*} = \frac{1}{2J\sqrt{M\kcoll \eta}} \Rightarrow 2 J t \kcoll \ll \sqrt{\frac{\kcoll}{M \eta}},
\end{align}
for which \ref{eq:approx_varJz_2} holds. From it, it also follows that if $\kcoll < M$, then $(1+2 J t \kcoll) \approx 1$ and,
\begin{equation}
  \Vy(t) \approx \frac{J}{2+4 J t M \eta} \ee^{-(M+\kcoll)t/2} \approx \frac{J}{2+4 J t M \eta}, \label{eq:ap_Geremia_sol}
\end{equation}
where in the last step we used the short-time condition $t \ll t^{*} < (M + \kcoll)^{-1}$.

Next, by showing that $\Vy(t)$ is a non-decreasing function at $t\approx0$ if $\kcoll \ge \eta M$, we can prove that \eref{eq:ap_Geremia_sol} will hold \emph{if and only if} $\kcoll < \eta M$ for $J\gg1$.
Namely, that we can consider the global decoherence $\kcoll$ to be insignificant for small times $t \ll t^*$ only when $\kcoll < \eta M$.
In order to do so, we take the derivative with respect to time of the function \ref{eq:approx_varJz_2} and then compute the limit for time approaching zero:
\begin{equation}
  \lim_{t \rightarrow 0}\frac{\dd}{\dt} \left[\mrm{V}_{<t^*}(t)\right] = -\frac{1}{4} \,J (M + \kcoll -4J\kcoll +4 J M \eta).
\end{equation}
By then setting the solution above equal to zero, we find the value of $\kcoll$ for which the derivative changes signs,
\begin{equation}
  \kcoll = \frac{M + 4 J M \eta}{4J - 1} \approx M \eta + \frac{M(\eta + 1)}{4J} + \frac{M(\eta + 1)}{16 J^{\,2}} + \mathcal{O}\left[\frac{1}{J}\right]^3,
\end{equation}
which can be correctly approximated as $\kcoll = \eta M$ when $J \gg 1$. Hence,
\begin{equation}
  \lim_{t \rightarrow 0} \frac{\dd}{\dt} \left[ \mrm{V}_{<t^*} (t) \right] \geq 0 \, \, \text{if} \, \, \kcoll \geq \eta M,
\end{equation}
and
\begin{equation}
  \lim_{t \rightarrow 0} \frac{\dd}{\dt} \left[ \mrm{V}_{<t^*} (t) \right] < 0 \, \, \text{if} \, \, \kcoll < \eta M.
\end{equation}

Thus proving that the variance of $\Jy$ can be approximated as \eref{eq:ap_Geremia_sol} only when $\kcoll < \eta M$.

\section{Derivation of the CoG dynamical model of \eqnref{eq:dynamical_model}} 
\label{ap:Ito}
The set of stochastic differential equations \eref{eq:dynamical_model} can be derived by carefully applying the rules of It\^{o} calculus, e.g., by noting that the differential of any two functions of time and a stochastic process, $f$ and $g$, reads $\dd (f g) = f \dd g + g \dd f + \dd f \dd g$. In our case, these functions are the means, variances and covariances of some quantum observable $\GenOp$, whose dynamical evolution can then be computed by substituting the conditional dynamics \eref{eq:fullSME} of $\dd\rhoc$ into $\dd\brkt{\GenOp} = \trace{\GenOp \, \dd\rhoc}$. In particular, considering $\Jindex{\alpha}$, $\mathrm{V}^{\cc}_{\alpha}$ and $\mathrm{C}^{\cc}_{\alpha\beta}$ with $\alpha,\beta = x,y,z$ appearing in \eqnref{eq:dynamical_model}, which satisfy 
\begin{subequations}
\label{eq:deriv_model}
\begin{align}  
    &\dd\brktc{\Jindex{\alpha}} = \Tr[\Jindex{\alpha} \, \dd\rho_{\cc}],\\ 
    &\dd \mrm{V}^{\cc}_\alpha = \dd\brktc{\Jindex{\alpha}^2} -  \dd \left(\brktc{\Jindex{\alpha}}^2\right) \\
    &\qquad\, = \dd\brktc{\Jindex{\alpha}^2} - 2\brktc{\Jindex{\alpha}} \dd\left(\brktc{\Jindex{\alpha}}\right) - \dd\brktc{\Jindex{\alpha}} \dd\brktc{\Jindex{\alpha}}, \nonumber \\
    &\dd\mrm{C}^{\cc}_{\alpha\beta} = \frac{1}{2}\dd\brktc{\Jindex{\alpha}\Jindex{\beta}} + \frac{1}{2}\dd\brktc{\Jindex{\beta}\Jindex{\alpha}} - \dd\left(\brktc{\Jindex{\alpha}}\brktc{\Jindex{\beta}}\right)  \nonumber \\
    &\qquad\; = \frac{1}{2}\dd\brktc{\Jindex{\alpha}\Jindex{\beta}} + \frac{1}{2}\dd\brktc{\Jindex{\beta}\Jindex{\alpha}} - \brktc{\Jindex{\beta}}\dd\brktc{\Jindex{\alpha}} \nonumber \\
    &\qquad\qquad - \brktc{\Jindex{\alpha}}\dd\brktc{\Jindex{\beta}} - \dd\brktc{\Jindex{\alpha}}\dd\brktc{\Jindex{\beta}},
\end{align}
\end{subequations}
with the initial conditions set to the mean, variances and co-variances of a CSS along $x$ (see \secref{sec:CSS_main}):
\begin{align}
    \brktc{\;\Jqvec{J} (0)} &= \left(\brktc{\Jx(0)} , \; \brktc{\Jy(0)} , \; \brktc{\Jz(0)}\right)^{\!\!\Trans} = \left(\frac{N}{2},\;0,\;0\right)^{\!\!\!\Trans}, \\
    \brktc{\pmb{\mrm{C}}^{\cc}_J(0)} &= \begin{pmatrix}
        \Vx^{\cc}(0) & \Cxy^{\cc}(0) & \Cxz^{\cc}(0) \\
        \Cxy^{\cc^*}(0) & \Vy^{\cc}(0) & \Czy^{\cc^*}(0) \\
        \Cxz^{\cc^*}(0) & \Czy^{\cc}(0) & \Vz^{\cc}(0)
    \end{pmatrix} = \begin{pmatrix}
        0 & 0 & 0 \\
        0 & N/4 & 0 \\
        0 & 0 & N/4
    \end{pmatrix}.
\end{align}

By carefully working to the relevant order $O(\dt^{3/2})$:
\begin{subequations}
\label{eq:full_dynamical_model}
\begin{align}  
    \dd\brktc{\Jx} &=  - (\omega(t) \!+\! u(t))  \brktc{\Jy} \dt\!- \frac{1}{2}(\kcoll \!+\! 2 \kloc \!+\! M)  \brktc{\Jx}\dt\!+\! 2\sqrt{\eta M}  \Cxy^{\cc} \,\dW \\ 
    \dd\brktc{\Jy} &= (\omega(t) \!+\! u(t))  \brktc{\Jx}\dt\!- \frac{1}{2}(\kcoll \!+\! 2\kloc)  \brktc{\Jy}\dt\!+\! 2\sqrt{\eta M}  \Vy^{\cc} \,\dW \\
    \dd\brktc{\Jz} &= -\frac{1}{2} M \brktc{\Jz}\dt\!+\! 2\sqrt{\eta M} \Czy^{\cc} \,\dW \\
    \dd \Vx^{\cc}  &=  - 2 (\omega(t) \! + \! u(t))  \Cxy^{\cc}\dt\! + \! \kcoll \! \left( \! \Vy^{\cc} \!+\!  \brktc{\Jy}^2 \!-\! \Vx^{\cc} \!\right)\!\dt \!+\! \kloc \!\left( \!\frac{N}{2} \!-\!2 \Vx^{\cc} \!\right)\!\dt  \nonumber\\
    &+\!  M\! \left( \! \Vz^{\cc} \!-\! \Vx^{\cc} \!-\! 4 \eta {\Cxy^{\cc}}^2 \!\right) \!\dt+\! 2\sqrt{\eta M} \!\left(\! \frac{1}{2}\cov_{\cc}(\Jx^2\Jy) \!+\! \frac{1}{2}\cov_{\cc}(\Jy\Jx^2) \!\!\right)\!\dW \\
    \dd\Vy^{\cc} &= 2 (\omega(t) \!+\! u(t))  \Cxy^{\cc}\dt\!+\! \kcoll \!\left(\!  \Vx^{\cc} \!+\!  \brktc{\Jx}^2 \!-\!  \Vy^{\cc}\!\right) \!\dt \!+\! \kloc \!\left(\! \frac{N}{2} \!-\!2 \Vy^{\cc} \!\right) \!\dt\nonumber \\
    &-\! 4 \eta M  {\Vy^{\cc}}^2\dt+ 2 \sqrt{\eta M} \, \cov_{\cc}(\Jy^3) \,\dW  \\
    \dd\Vz^{\cc} &=  M \!\left( \!\Vx^{\cc} \!+\!  \brktc{\Jx}^2 \!-\! \Vz^{\cc}\! \right) \!\dt \nonumber \\
    &+ 2\sqrt{\eta M} \left(\frac{1}{2} \cov_{\cc}(\Jz^2\Jy) + \frac{1}{2} \cov_{\cc}(\Jy\Jz^2) \right) \dW\\
    \dd\Cxy^{\cc} &= (\omega(t) \!+\! u(t)) \!\left(\!\Vx^{\cc} \!-\!  \Vy^{\cc}\!\right) \!\dt \!-\! \kcoll \!\left(\!2\Cxy^{\cc} \!+\! \brktc{\Jx}\brktc{\Jy}\!\right)\!\dt \nonumber \\
    &-\! 2\kloc \Cxy^{\cc}\dt\!-\! \frac{1}{2} M \Cxy^{\cc} \left( 1 + 8\eta \Vy^{\cc}\right)\!\dt \nonumber \\
    &+ 2\sqrt{\eta M} \left(\frac{1}{4}\cov_{\cc}(\Jx \Jy^2) + \frac{1}{2}\cov_{\cc}(\Jy \Jx \Jy) + \frac{1}{4} \cov_{\cc}(\Jy^2 \Jx) \right)\dW \\
    \dd\Czy^{\cc}  &= (\omega(t) \! + \! u(t)) \Cxz^{\cc}\dt - \frac{1}{2}\left(\kcoll  + 2\kloc + M\left(1 + 8\eta\Vy^{\cc}\right)\right) \Czy^{\cc}\dt \nonumber \\
    &+ 2\sqrt{\eta M} \bigg( \frac{1}{4}\cov_{\cc}(\Jz \Jy^2) +  \frac{1}{2}\cov_{\cc}(\Jy \Jz \Jy) + \frac{1}{4} \cov_{\cc}(\Jy^2 \Jz) \bigg)\dW \\
    \dd\Cxz^{\cc}  &= - (\omega(t) \! + \! u(t))\Czy^{\cc}\dt - \frac{1}{2} \left(\kcoll + 2\kloc + 4M \right) \Cxz^{\cc}\dt- M\brktc{\Jz}\brktc{\Jx}\dt \nonumber \\
    &- 4 \eta M \Cxy^{\cc}\Czy^{\cc}\dt + 2\sqrt{\eta M} \bigg(\frac{1}{4} \cov_{\cc}(\Jx \Jz \Jy) + \frac{1}{4} \cov_{\cc}(\Jy \Jx \Jz) \nonumber \\ 
    & + \frac{1}{4} \cov_{\cc}(\Jz \Jx \Jy) + \frac{1}{4} \cov_{\cc}(\Jy \Jz \Jx) \bigg)\dW \\
    \dd \omega &= -\chi \omega(t) \dt + \sqrt{q_\omega} \dW_\omega
\end{align}
\end{subequations}
where for any three operators $\AOp$, $\BOp$ and $\COp$, we define 
\begin{align}
    \cov_{\cc}(\AOp \!\! \BOp \! \COp) &\coloneqq \brktc{\AOp \!\! \BOp \! \COp} \!- \brktc{\AOp\!}\brktc{\BOp \! \COp} - \brktc{\BOp} \brktc{\AOp \!\! \COp} \nonumber \\
    &- \brktc{\COp}\brktc{\AOp \!\! \BOp} + 2 \brktc{\AOp\!}\!\brktc{\BOp}\!\brktc{\COp}.
\end{align}

We simplify the full dynamical model given in \eqnref{eq:full_dynamical_model} by applying a cut-off approximation that discards the third-order moments and higher. This step is crucial for two reasons: it allows us to construct an EKF and provides a self-contained set of stochastic differential equations describing our sensor. Importantly, the impact of neglecting third-order moments on the CoG model is limited, as these moments appear only within the stochastic terms of the second-order moment dynamics.

Additionally, we omit the differential equations for $\brktc{\Jz}$, $\Cxz^{\cc}$, and $\Czy^{\cc}$ in the main text because these quantities remain consistently zero throughout the time evolution. This is due to their initial values being zero (CSS-state conditions: $\brktc{\Jz(0)} = \Cxz^{\cc}(0) = \Czy^{\cc}(0) = 0$) and their exclusively decaying dynamics. By disregarding these irrelevant terms, we arrive at the dynamical equations presented in the main text as \eqnref{eq:dynamical_model}.

Finally, in order to solve the system of SDE numerically, is also convenient to normalize the state and hence the system of SDEs w.r.t. $\sqrt{N}$. Namely,
\begin{align}
    \X &\coloneqq \Jx/\sqrt{N} & \Y &\coloneqq \Jy/\sqrt{N},
\end{align}
with new variances and covariances being:
\begin{align}
    \VX &= \Vx/N & \VY &= \Vy/N & \VZ &= \Vz/N & \CXY &= \Cxy/N.
\end{align}

Then, the system of SDE simulating the system in the co-moving approximation is:
\begin{align}  
    \dd\brktc{\!\X} &\!=\!  - (\omega(t) \!+\! u(t))   \brktc{\!\Y}  \dt \!- \frac{1}{2}(\kcoll \!+\! 2 \kloc \!+\! M) \brktc{\!\X} \dt \nonumber \\
    &\!+\! 2\sqrt{\eta M N} \CXY \dW \\ 
    \dd\brktc{\! \Y} &\!=\! (\omega(t) \!+\! u(t))  \brktc{\!\X} \! \dt \!- \!\frac{1}{2}(\kcoll \!+\! 2\kloc)  \brktc{\!\Y} \! \dt \!+\! 2\sqrt{\eta M N} \VY \! \dW \label{eq:dJy_ap}\\
    \dd \VX &\!=\! - 2 (\omega(t) \! + \! u(t))  \CXY \dt \! + \! \kcoll \! \left( \! \VY \!+\!  \brktc{\!\Y}^2 \!-\! \VX \!\right)\!\dt   \nonumber \\
    &\!+\! \kloc \!\left( \!\frac{1}{2} \!-\!2 \VX \!\!\right)\!\dt\!+\! M\! \left( \!\! \VZ \!\!\!-\!\! \VX \!\!-\! 4 \eta N {\CXY\!\!\!\!}^2 \, \right)\! \dt \label{eq:dVx_ap}\\
    \dd\VY &\!=\! 2 (\omega(t) \!+\! u(t))  \CXY \dt \!+\! \kcoll \!\left(\!  \VX \!+\!  \brktc{\!\X}^2 \!-\!  \VY\!\right)\! \dt  \nonumber \\
    &\!+\! \kloc \!\left(\! \frac{1}{2} \!-\!2 \VY \!\right)\! \dt \!-\! 4 \eta M N {\VY\!\!\!\!}^2 \,\,\dt \label{eq:dVy_ap}\\
    \dd\VZ &\!=\!  M \!\left( \!\VX \!+\!  \brktc{\!\X}^2 \!-\! \VZ\! \right)\! \dt \label{eq:dVz_ap}\\
    \dd\CXY &\!=\! (\omega(t) \!+\! u(t)) \!\left(\!\VX \!-\!  \VY\!\right)\! \dt \!-\! \kcoll \!\left(\!2\CXY \!+\! \brktc{\!\X}\brktc{\!\Y}\!\right)\!\dt  \nonumber \\
    &\!-\! 2\kloc \CXY \dt \!-\! \frac{1}{2} M \CXY \left( 1 + 8\eta N \VY\right)\!\dt \label{eq:dCxy_ap}\\
    \dd\omega &\!=\! -\chi \omega(t) \, \dt + \sqrt{q_\omega} \, \dW_\omega,
\end{align}
with the initial values for the means: $\left(\brktc{\!\X(0)}\!,\brktc{\!\Y(0)}\!\right) = (\sqrt{N}/2,0)$, the variances and co-variances: $\left(\!\VXt{0}\!,\VYt{0}\!,\VZt{0}\!,\CXYt{0}\!\right) = (0,1/4,1/4,0)$ and the Larmor frequency $\omega(0) = \mu$. 

\section{Verification of the CoG approximation} \label{sec:verification_CoG}

In this section, we simulate the exact dynamics of the density matrix for low values of $N$ using the SME of \eqnref{eq:fullSME} in order to verify that the approximate evolution of the lowest moments given by \eqnref{eq:dynamical_model} correctly captures the system behavior for moderate dephasing and measurement-strength parameters. We observe that the agreement between the full model and the approximate equations improves with increasing atomic number at short timescales, and because the experimental regimes involve large $N\approx10^{5}\!-\!10^{13}$~\cite{Kuzmich2000,Wasilewski2010,Shah2010,Koschorreck2010,Sewell2012,Kong2020}, we subsequently use \eqnref{eq:dynamical_model} to simulate the dynamics of the atomic sensor with sufficient accuracy.

\begin{figure}[t!]
    \centering \includegraphics[width = \textwidth]{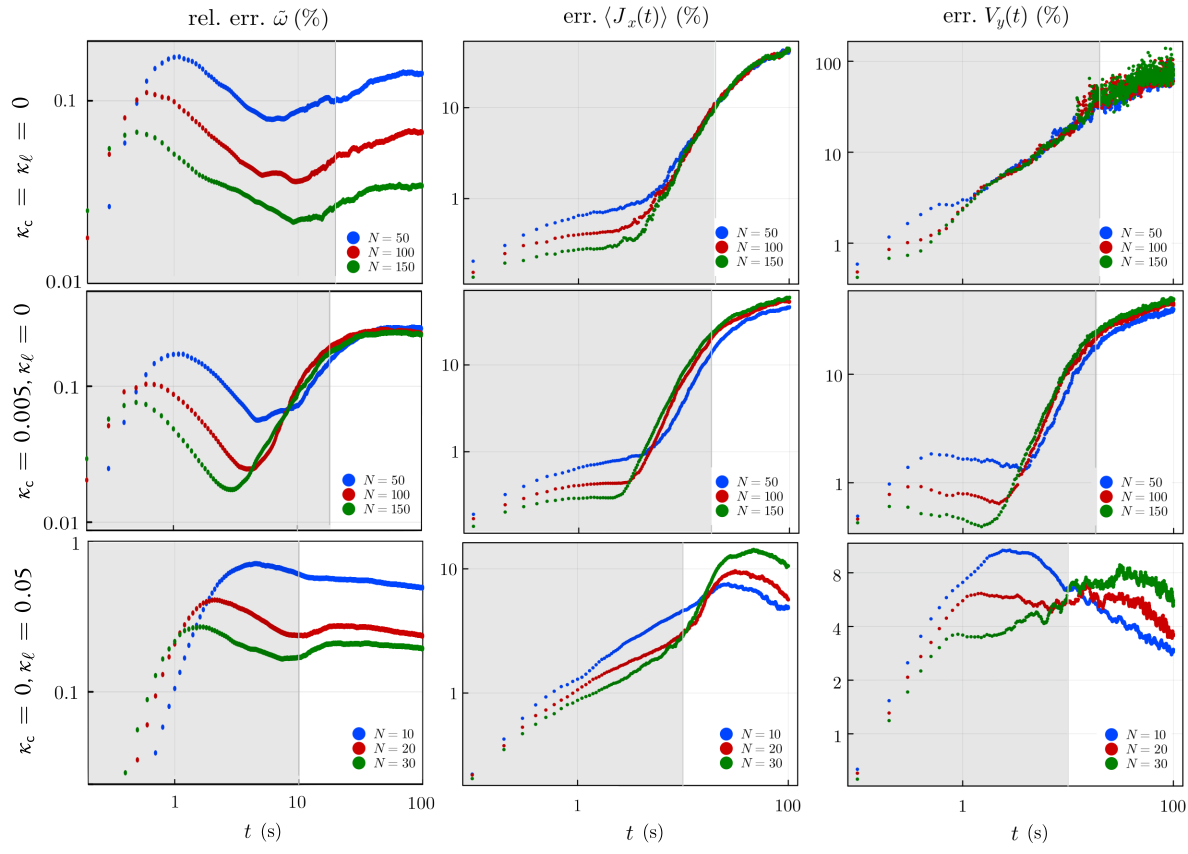}
    \caption[Performance of the CoG approximation: Larmor frequency estimation and moment error analysis]{\textbf{Performance of the CoG approximation: Larmor frequency estimation and moment error analysis.} \textbf{(Left column)} Evolution of the relative error in estimating the Larmor frequency when comparing the exact model (SME, \eqnref{eq:fullSME}) with the approximate CoG model of \eqnref{eq:dynamical_model}. Each graph shows the relative error (in \%) for three different noise scenarios (from top to bottom): dephasing induced solely by continuous measurement; combined measurement-induced and collective dephasing; and combined measurement-induced, collective, and local dephasing. For the first two cases, system sizes $N = 50,\,100,\,150$ (blue, red, and green, respectively) are considered, while for the case including local decoherence, $N = 10,\,20,\,30$ are used. In all cases, the error remains below 1\% and decreases with increasing $N$. \textbf{(Center and right columns)} Comparative error analysis of the moments $\brktc{\Jx(t)}$ and $\Vy^{\cc}(t)$ between the exact SME solution and the CoG model. Here, the relative error (in \%) is defined in \eqnref{eq:error_def}. The analysis is performed for the same three decoherence scenarios: (top row) continuous measurement only ($M=0.05$, $\kcoll=\kloc=0$); (middle row) measurement-induced and collective decoherence ($M=0.05$, $\kcoll=0.005$, $\kloc=0$); and (bottom row) measurement-induced and local decoherence ($M=0.05$, $\kcoll=0$, $\kloc=0.05$). In each plot, increasing system sizes (either $N = 50,\,100,\,150$ or $N = 10,\,20,\,30$ for the local case) demonstrate that the CoG approximation becomes more accurate at short times as $N$ increases. All error values are obtained by averaging over $\nu=1000$ measurement trajectories. Figure adapted from \refcite{Amoros-Binefa2024}.}
    \label{fig:model_vs_exact}
\end{figure}

\figref{fig:closed-loop_atoms} shows the architecture of the feedback loop employed in our atomic magnetometry scheme. In each round, the measurement data $\pmb{y}(t)$ is generated by simulating the ``\emph{Atomic ensemble}'' either exactly -- evolving its full conditional density matrix $\rhoc$ via the SME \eref{eq:fullSME} --- or approximately, through the dynamics of its relevant first and second moments, $\brktc{\Jx}$, $\brktc{\Jy}$ and $\Vy^{\cc}$, according to the CoG model \eref{eq:dynamical_model}. The measurement record generated by the system is then processed by the ``\emph{Estimator}'' (i.e., the EKF), which provides in real time not only an estimate of the Larmor frequency, $\est{\omega}(t)$, but also estimates of the dynamical parameters, $\est{\pmb{x}}(t)$. These estimates are used by the ``\emph{Controller}'' (i.e., the LQR) to adjust the system dynamics on the fly by modifying $u(t)$.

To assess the accuracy of the CoG approximation of \eqnref{eq:dynamical_model} when simulating the system dynamics, we benchmark it against the exact SME solution for moderate atomic numbers, where the exact simulation is computationally feasible. This comparison is carried out at two levels: (1) we focus solely on the estimation task by computing the average error in the real-time estimate $\est{\omega}(t)$ (large box in \figref{fig:closed-loop_atoms}), and (2) we adopt a stricter criterion by requiring that the relevant moments, $\brktc{\Jx}$, $\brktc{\Jy}$, and $\Vy^{\cc}$, are accurately reproduced when compared to their exact values obtained from evolving $\rhoc(t)$ with the SME(smaller box in \figref{fig:closed-loop_atoms}).

In the left column of \figref{fig:model_vs_exact}, we present in percentage the \emph{average relative error} between the real-time estimate $\est{\omega}(t)$ of $\omega$ obtained using the exact model (full SME solution) and the approximate model (CoG), i.e.
\begin{align} 
    \EE{\delta_{\est{\omega}}} (\%) &= 100 \times \EE{\left|\dfrac{\est{\omega}_{\mathrm{SME}}-\est{\omega}_{\mathrm{CoG}}}{\est{\omega}_{\mathrm{SME}}}\right|} \nonumber \\
    &= 100 \times \!\! \int\!\dd\omega\,p(\omega) \int\!\mathcal{D}\pmb{y}_{\leq t}\,p(\pmb{y}_{\leq t}|\omega)\left|\dfrac{\est{\omega}_{\mrm{SME}} - \est{\omega}_{\mrm{CoG}}}{\est{\omega}_{\mrm{SME}}}\right| \label{eq:error_def}
\end{align}
where the expectation is taken over the realizations of the experiment. In the left column of \figref{fig:model_vs_exact}, three plots corresponding to a different noise scenario are showcased. From top to bottom: only measurement decoherence ($\kcoll = \kloc = 0)$, combined measurement and collective decoherence ($M = 0.05$, $\kcoll = 0.005$ and $\kloc = 0$), and combined measurement and local decoherence ($M = \kloc = 0.05$ and $\kcoll = 0.005$). Each plot shows the averaged relative error for increasing system sizes --- specifically, $N = 50,100,150$ for the first two scenarios, and $N = 10,20,30$ for the local case. 
In each case the average relative error decreases as the system size increases, reaching below 1\% error, which indicates that for large ensembles ($N \sim 10^5 - 10^{13}$) the CoG approximation is sufficiently accurate for generating the measurement data used to estimate the Larmor frequency.

We further assess the CoG model by comparing the evolution of key dynamical moments ($\brktc{\Jx}$, $\brktc{\Jy}$, and $\Vy^{\cc}$) to their exact values computed from the full density matrix $\rhoc$ (via the SME, Eq. (1)). In this comparison, the error is quantified as
\begin{equation}
    \EE{\delta_{\mrm{x}}} = 100 \times \frac{\EE{|\mrm{x}_\mrm{SME} - \mrm{x}_\mrm{CoG}|}}{\EE{|\mrm{x}_\mrm{SME}|}} \quad \text{for} \quad \mrm{x} \in \{ \brktc{\Jx}, \brktc{\Jy}, \Vy^{\cc}\},
\end{equation}
where $\mrm{x}$ stands for any of the moments. As illustrated in the center and right columns of \figref{fig:model_vs_exact}, for short times the error in simulating these moments decreases with increasing system size, and remains below approximately 10\% for times $t \lesssim 1/(M+\kcoll+2\kloc)$. This supports the validity of using the CoG approximation to predict quantities such as the spin-squeezing parameter in large atomic ensembles.

\section{Steady-state solution of the Kalman Filter for $\chi \neq 0$ in the LG regime}
\label{sec:SSS}

Let us consider the covariance differential equation introduced in \eref{eq:Riccati_equation_ch5}, with matrices $\pmb{F}(t)$, $\pmb{G}(t)$, and $\pmb{H}(t)$ defined in \eref{eq:matrices_F_K_B} and noise covariance matrices being $\pmb{Q} = \text{Diagonal}[1,q_\omega]$, $\pmb{R}= \eta$, and $\pmb{S} = (   \sqrt{\eta} \; \; 0   )^\Trans$. The model matrices of \eref{eq:matrices_F_K_B} in the steady state read as:
\begin{equation}\label{eq:matrices_F_K_B_steady_state}
  \!\!\pmb{F} = \begin{pmatrix} 0 & J \\ 0 & -\chi \end{pmatrix}\!, \;\; \pmb{G} = \begin{pmatrix} J \, \sqrt{\kcoll} & 0 \\ 0 & \sqrt{q_\omega} \end{pmatrix}\!, \;\; \pmb{H} =  2\sqrt{\eta M} \begin{pmatrix} 1 & 0 \end{pmatrix}\!.
\end{equation}

Next, for simplicity in the upcoming analysis, we will rename the elements of the steady-state covariance matrix $\pmb{\Sigma}$ as
\begin{equation}
    \pmb{\Sigma} = \begin{pmatrix}
        \mathbb{E}[\Delta^2 \estJy] & \mathbb{E}[\Delta \estJy \Delta \est{\omega}] \\ 
        \mathbb{E}[\Delta \est{\omega} \Delta \estJy] & \mathbb{E}[\Delta^2 \est{\omega}]
    \end{pmatrix} = \begin{pmatrix}
        x & y \\
        y & z
    \end{pmatrix}.
\end{equation}
Then, the system of equations in the steady state ($\dd \, \pmb{\Sigma} = 0$) can be written as,
\begin{align}
    -4 \,J \sqrt{\eta M \kcoll}  \, x - 4M\eta \, x^2 - 2 \,  J \, y = 0, \\
    -\chi \, y - 2 \, J \sqrt{\eta M \kcoll} \, y - 4 M\eta \, x \, y -   J \, z = 0, \\
    q_\omega - 4M\eta \, y^2 -2 \chi \, z = 0,
\end{align}
such that the solution for $z(t)$ is
\begin{align} \label{eq:VarB}
    z & = -\kcoll \, \chi - \frac{\chi^3}{4  J^2 M \eta} - \frac{\chi}{J \sqrt{M \eta}} \sqrt{q_\omega  + \kcoll \, \chi^2} \,  \\
    & + \frac{1}{4 J^{\,2} M \eta} \left(\chi^2 + 2 J \sqrt{M \eta(q_\omega  + \kcoll \, \chi^2)} \right) \sqrt{\chi^2 + 4 \kcoll \, J^{\,2} M \eta + 4 J\sqrt{M \eta(q_\omega  + \kcoll \, \chi^2)}}, \nonumber
\end{align}
which, when expanded in powers of $1/J$ around zero and truncated at first order, can be approximated as
\begin{align} \label{eq:approx_VarB}
    \mathbb{E}[\Delta^2\est{\omega}^\text{SS} ]\equiv z \approx -\kcoll \, \chi + \sqrt{\kcoll q_\omega + \kcoll^2 \, \chi^2}.
\end{align}
Moreover, when $\chi = 0$, the term \eqref{eq:VarB} becomes,
\begin{align}
    \mathbb{E}[\Delta^2\est{\omega}^\text{SS}]|_{\chi = 0} = \left( q_\omega \, \kcoll + \frac{1}{  J} \sqrt{\frac{q_\omega^3}{M \eta}} \right)^{1/2},
\end{align}
matching the solution introduced in \eref{eq:ss_field_sol}.

\section{Derivation of the LQG control law} \label{sec:LQR_derivation}

In this section we present a detailed derivation, along the lines of \refcite{Stockton2004}, of the optimal control law used in our LQG design. The final control law is given by
\begin{align}
u(t) =  -\est{\omega}(t) - \lambda\, \brktc{\estJy(t)},
\end{align}
where the constant parameter $\lambda = \sqrt{p_J/\nu}$ is determined by the weights in the quadratic cost function. Our starting point is the linearized system dynamics in the LG regime. In this regime, the state is given by the reduced vector 
\begin{align}
\pmb{z}(t) = \begin{pmatrix}\brktc{\Jy(t)} \\ \omega(t)\end{pmatrix},
\end{align}
which evolves according to
\begin{align}
\dot{\pmb{z}}(t) &= \pmb{A}\,\pmb{z}(t) + \pmb{B}\,u(t) + \pmb{G}(t)\,\pmb{q}(t), \label{eq:lgr_system_app}
\end{align}
with
\begin{equation}
    \pmb{A} = \begin{pmatrix} 0 & J \\ 0 & -\chi \end{pmatrix}, \quad \pmb{B} = \begin{pmatrix} J \\ 0 \end{pmatrix}. \label{eq:AandB}    
\end{equation}
The performance of the controller is measured by the quadratic cost function
\begin{align}
I(u) &= \EE{\int_0^\infty \Bigl[\pmb{z}^\TT(t)\,\pmb{P}\,\pmb{z}(t) + \nu\,u^2(t)\Bigr]\, \dt }, \label{eq:quad_cost_app}
\end{align}
where
$$
\pmb{P} = \begin{pmatrix} p_J & 0 \\ 0 & p_\omega \end{pmatrix} \geq 0, \quad \nu > 0.
$$

For such a LG system with cost (\ref{eq:quad_cost_app}), the optimal feedback control law is:
\begin{align}
u(t) &= -\pmb{K}_C\,\est{\pmb{z}}(t), \label{eq:LQR_control_law_app}
\end{align}
where the gain matrix is given by
\begin{align}
&\pmb{K}_C = \nu^{-1}\,\pmb{B}^\TT\,\pmb{\Lambda}, \label{eq:gain_matrix_app}\\
&\pmb{A}^\TT\,\pmb{\Lambda} + \pmb{\Lambda}\,\pmb{A} + \pmb{P} - \pmb{\Lambda}\,\pmb{B}\,\nu^{-1}\,\pmb{B}^\TT\,\pmb{\Lambda} = \mathbf{0}. \label{eq:control_Riccati_app}
\end{align}
Here, $\pmb{\Lambda}$ is the unique positive semidefinite solution of the algebraic Riccati equation (ARE). We now parameterize $ \pmb{\Lambda} $ as a symmetric $2\times 2$ matrix:
\begin{align}
\pmb{\Lambda} = \begin{pmatrix} \Lambda_{11} & \Lambda_{12} \\ \Lambda_{12} & \Lambda_{22} \end{pmatrix},
\end{align}
and solve \eqnref{eq:control_Riccati_app} component by component. Starting with the $(1,1)$ entry, we find
\begin{align}
0 + p_J - \frac{J^{\,2}}{\nu}\,\Lambda_{11}^2 &= 0 \Longrightarrow \Lambda_{11} = \frac{\sqrt{p_J\,\nu}}{J}. \label{eq:Lambda11}
\end{align}
The $(1,2)$ entry yields
\begin{align}
J\,\Lambda_{11} - \chi\,\Lambda_{12} - \dfrac{J^{\,2}}{\nu}\,\Lambda_{11}\,\Lambda_{12} &= 0 \Longrightarrow \Lambda_{12} = \frac{J\,\Lambda_{11}}{\chi + \dfrac{J^{\,2}}{\nu}\,\Lambda_{11}} = \frac{\sqrt{p_J\,\nu}}{\chi + J\,\sqrt{\dfrac{p_J}{\nu}}}, \label{eq:Lambda12}
\end{align}
where we substituted \eqnref{eq:Lambda11} in the last step. Finally, the $(2,2)$ entry gives
\begin{align}
2\bigl(J\,\Lambda_{12} - \chi\,\Lambda_{22}\bigr) + p_\omega - \frac{J^{\,2}}{\nu}\,\Lambda_{12}^2 &= 0,
\end{align}
which can be solved for
\begin{align}
\Lambda_{22} &= \frac{1}{2\chi}\left(2\,J\,\Lambda_{12} + p_\omega - \frac{J^{\,2}}{\nu}\,\Lambda_{12}^2\right).
\end{align}
This fully specifies $\pmb{\Lambda}$, though $\Lambda_{22}$ is not required for the final control law.

Using \eqref{eq:gain_matrix_app} and the form of $\pmb{B}$ from \eqref{eq:AandB}, the gain matrix becomes
\begin{align}
\pmb{K}_C &= \frac{1}{\nu}\,\pmb{B}^\TT\,\pmb{\Lambda} = \frac{J}{\nu}\, \begin{pmatrix}
    \Lambda_{11} & \Lambda_{12}
\end{pmatrix},
\end{align}
and the feedback law \eqref{eq:LQR_control_law_app} can be written explicitly as
\begin{align}
u(t) &= -\pmb{K}_C\,\est{\pmb{z}}(t) \nonumber \\
     &= -\frac{J\,\Lambda_{11}}{\nu}\,\brktc{\estJy(t)} - \frac{J\,\Lambda_{12}}{\nu}\,\est{\omega}(t). \label{eq:u_intermediate}
\end{align}
Substituting \eqref{eq:Lambda11} and \eqref{eq:Lambda12}, we have
\begin{align}
    \frac{J\,\Lambda_{11}}{\nu} &= \frac{J}{\nu}\cdot \frac{\sqrt{p_J\,\nu}}{J} = \sqrt{\frac{p_J}{\nu}} = \lambda, \\
    \frac{J\,\Lambda_{12}}{\nu} &= \dfrac{1}{1+\dfrac{\chi}{J}\sqrt{\dfrac{\nu}{p_J}}} = \dfrac{1}{1+\dfrac{\chi}{J \lambda}} \approx 1,
\end{align}
where the second term can be approximated to $1$ because $\chi < J$ under typical experimental conditions. Thus, the control law reduces to the simple form:
\begin{align}
u(t) &= -\est{\omega}(t) - \lambda\, \brktc{\estJy(t)}. \label{eq:LQR_final_app}
\end{align}

One might note that the final expression of $u(t)$ depends only on the control parameters $p_J$ and $\nu$. This arises because, in the system of \eqnref{eq:lgr_system_app}, the control input $u(t)$ acts directly only on the first state, $\brkt{\Jy}$, while the second state, $\omega$, is not directly actuated. Recall that the feedback gain is computed as $K_C = \nu^{-1}\,\pmb{B}^\TT\,\pmb{\Lambda}$, meaning it only ``reads'' the parts of $\pmb{\Lambda}$ corresponding to the states that the input $u(t)$ can actually influence. The entries of $\pmb{\Lambda}$ associated with the unactuated state ($\pmb{\Lambda}_{22}$, which depends on $p_\omega$), never enter $K_C$, and therefore never affect the control law. Intuitively, this means that no matter how strongly we penalize the unactuated state in the cost function, the controller can only react to what it can physically influence, so the optimal feedback ends up depending solely on the actuated state.

\section{Experimentally realistic parameters} \label{sec:experimental_parameters}

\begin{figure}[t!]
    \centering
    \includegraphics[width=0.8\linewidth]{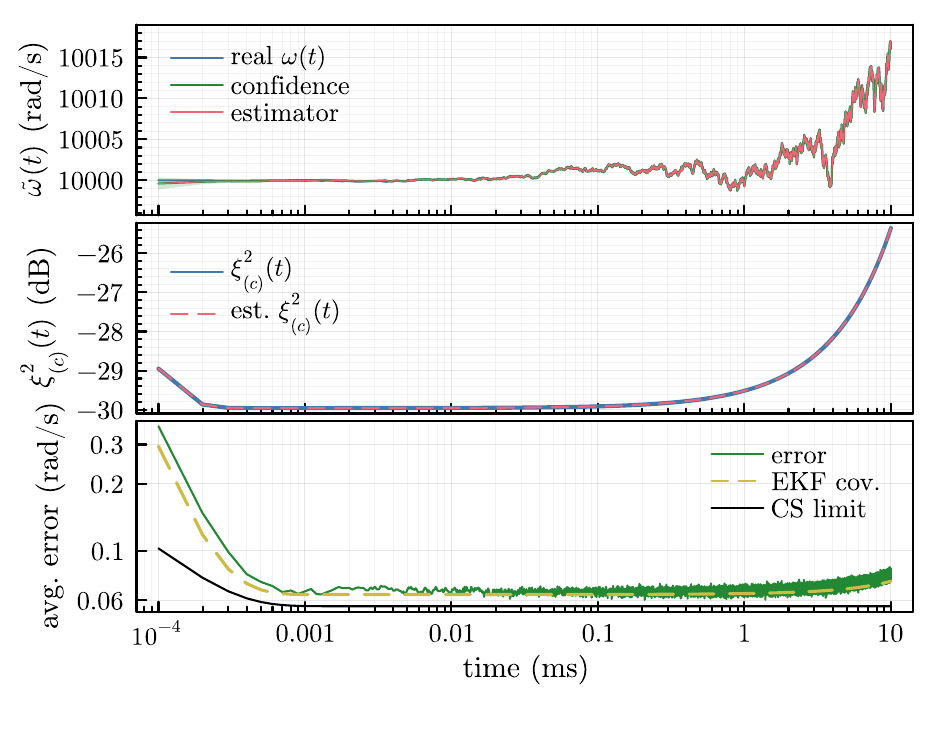}
    \caption[Tracking a fluctuating magnetic field with a strong continuous measurement]{\textbf{Tracking a fluctuating magnetic field with a strong continuous measurement.} The \emph{top plot} illustrates how the EKF estimate (solid, red) closely follows the true field OUP dynamics (solid, blue), staying well within the error bounds of $\pm 2 \sqrt{\EE{\Delta^2 \est{\omega}(t)}}$, which are so small compared to the fluctuating field that they are nearly imperceptible. The \emph{middle plot} shows the conditional spin squeezing (solid, blue) induced by the strong continuous measurement of strength $M = \SI{1}{\milli \hertz} \gg \kcoll = \SI{1}{\nano\hertz}$ and the LQR feedback, along with its real-time estimation by the EKF (dashed red). In the \emph{bottom plot}, the estimation error of $\omega(t)$ (solid, green) reaches a sensitivity of $\sim \SI{0.066}{\radian \, \second^{-1}}$ that matches the square-root of the EKF covariance (dashed, yellow). While the stronger measurement significantly enhances precision, the quantum limit dictated by dephasing (solid, black) at roughly $\SI{0.056}{\radian \, \second^{-1}}$ is not perfectly saturated. A further increase in $M$ could bring the error closer to this optimal limit~\cite{Amoros-Binefa2024}. The results in the bottom two plots are obtained by averaging over 1000 stochastic field-atom trajectories. }
    \label{fig:largerM}
\end{figure}

\secref{sec:fluctuating_field} and \secref{sec:nonlinear_MCG_field} analyze the estimation of a fluctuating and time-varying magnetic field, respectively, in experimentally realistic conditions, using parameters inspired by \refcite{Kong2020}. Most parameters are straightforward to define: we take the atomic ensemble size as $N = 10^{13}$ and set the coherence time to $T_2 = \SI{10}{\milli \second}$, which implies a local dephasing rate of $\kloc = 1/T_2 = \SI{100}{\hertz}$. We further consider frequencies on the order of $\SI{}{\kilo \hertz}$, specifically $\omega_0 = 10^4\SI{}{\radian \, \second^{-1}}$. The one parameter that is trickier to determine is the measurement strength parameter $M$. To establish $M$, we compare the equation for the photocurrent of \eqnref{eq:fullSME_meas} with Equation 18 from \refcite{Kong2020}. Namely, since the Wiener differential in \eqnref{eq:fullSME_meas} has a variance of $\dt$, then Equation 18 from \refcite{Kong2020} should be normalized by $\sqrt{\eta q_e^2 \dot{\!N}\,}$. Expressing it in our notation, the measurement equation of \refcite{Kong2020} can be then rewritten as:
\begin{equation}
    I(t) \dt = \eta \sqrt{g^2 \dot{\!N}\,} \brktc{\Jy(t)} \dt + \sqrt{\eta} \, \dW,
    \label{eq:I(t)}
\end{equation}
where $\EE{\dW^2} = \dt$, and we write $\Jy$ instead of $\Jz$ to account for the different experimental geometry. By directly comparing this to \eqnref{eq:fullSME_meas}, we identify the measurement strength parameter as:
\begin{equation} \label{eq:def_M}
    M = \frac{g^2 \dot{\!N}\,}{4}.
\end{equation}
Here, $\dot{\!N}\,$ represents the photon flux, given by:
\begin{equation}
    \dot{\!N}\, = \frac{P}{2\pi \hbar \, \nu}
\end{equation}
where $P$ is the probe power, varying between \SI{0.5}{\milli\watt} and \SI{2}{\milli\watt}, and $\nu$ is the frequency of the probe light, detuned by $\Delta \nu = \nu_{D_1Rb} - \nu$ from the Rb $D_1$ transition at $\nu_{D_1Rb} = c/\lambda_{D_1Rb}$ of \SI{794.8}{\nano\meter}. The coupling constant $g$ in \eqnref{eq:def_M} is defined in \refcite{Kong2020} as:
\begin{equation}
    g \approx \frac{c \, r_e f_{osc}}{A_{eff}} \frac{1}{\Delta \nu}
\end{equation}
where $c$ is the speed of light, $r_e = \SI{2.82e-13}{\centi\meter}$ is the classical electron radius, $f_{osc} = 0.34$ is the oscillator strength for the Rb $D_1$ transition, and $A_{eff} = \SI{0.0503}{\centi\meter^2}$ is the effective beam area. As a result, $M$ is expected to lie within the range of \SI{1e-10}{\hertz} and \SI{1e-8}{\hertz}, depending on the  probe power $P$ and optical detuning $\Delta \nu$, which can vary from $\Delta\nu \approx \SI{24}{\giga\hertz}$ to \SI{64}{\giga\hertz} when off-resonance.  

Physically, the measurement strength $M$ characterizes the balance between the light-atom interaction to the photon shot-noise in the detection process of \eqnref{eq:I(t)}. However, quantum backaction from continuous measurement unfolds on a timescale dictated by $M'=M\,N$, rather than $M$ alone. This arises because in the quantum model of \eqnref{eq:fullSME}, the variance of the relevant spin operators decays with an effective rate $1/M'$. This behavior has been rigorously established in both decoherence-free cases~\cite{Geremia2003} and scenarios with collective noise~\cite{Amoros-Binefa2021,Amoros-Binefa2024}, and is further corroborated here (see \figref{fig:oup_sq_estimation}), as well as experimentally~\cite{Kong2020}.

\section{Exact forms of gradient matrices $\pmb{F}(t)$, $\pmb{G}$(t) and $\pmb{H}$(t) for an OUP} \label{sec:exact_forms_FGH_OUP}

\begin{sideways}
\centering
\parbox{8in}{
\begin{align}
    &\est{\pmb{x}} = \begin{pmatrix} \est{x}_1 & \est{x}_2 & \est{x}_3 & \est{x}_4 & \est{x}_5 & \est{x}_6 & \est{x}_7 \end{pmatrix} = \begin{pmatrix} \estNormX & \estNormY & \estNormVX & \estNormVY & \estNormVZ & \estNormCXY & \est{\omega} \end{pmatrix}, \nonumber \\
    &\pmb{F}(t) = \nabla_{\!\pmb{x}} \; \pmb{f} \, |_{(\pmb{\Tilde{x}},u,0,t)}  =  \\
    & = \left(
     \scalemath{0.8}{
    \begin{array}{ccccccc}
     -(\kcoll\!+\!2\kloc \!+\! M)/2 & -(\est{x}_7 \!+\! u) & 0 & 0 & 0 & 0 & -\est{x}_2 \\ 
    (\est{x}_7 \!+\! u) & -(\kcoll\!+\!2\kloc)/2 & 0 & 0 & 0 & 0 & \est{x}_1 \\ 
    0 & 2\kcoll \est{x}_2 & -(\kcoll\!+\!2\kloc\!+\!M) & \kcoll & M & -2(\est{x}_7 \!+\! u) \!-\! 8\eta M N \, \est{x}_6 & -2\est{x}_6\\ 
    2\kcoll\est{x}_1 & 0 & \kcoll & -\kcoll \!-\! 2\kloc \!-\! 8\eta M N \, \est{x}_4 & 0 & 2(\est{x}_7 \!+\! u) & 2 \est{x}_6 \\
    2M\est{x}_1 & 0 & M & 0 & -M & 0 & 0 \\
    -\kcoll \est{x}_2 & -\kcoll \est{x}_1 & \est{x}_7 \!+\! u & -(\est{x}_7 \!+\! u) \!-\!4\eta M N \, \est{x}_6 & 0 & -\left(2\kcoll \!+\! 2\kloc \!+\! \frac{M}{2} \right) \!-\! 4\eta M N \, \est{x}_4 & \est{x}_3 \!-\! \est{x}_4\\
    0 & 0 & 0 & 0 & 0 & 0 & -\chi 
    \end{array},
    }
  \right)  \nonumber \\
    &\pmb{G}(t) = \nabla_{\pmb{\xi} \,} \, \pmb{f} \, |_{\pmb{\Tilde{x}}} = 
    \begin{pmatrix} 
    2\sqrt{\eta M N} \, \est{x}_6 & 0 \;\; \\ 
    2\sqrt{\eta M N} \, \est{x}_4 & 0 \;\; \\ 
    0 & 0 \;\; \\ 
    0 & 0 \;\; \\ 
    0 & 0 \;\; \\ 
    0 & 0 \;\; \\ 
    0 & \sqrt{q_K} \;\;
    \end{pmatrix}, \quad \quad \pmb{Q} = \begin{pmatrix}
        1 & 0 \\
        0 & 1
    \end{pmatrix}, \quad \quad \pmb{S} = \begin{pmatrix}
        \sqrt{\eta} \\ 0 
    \end{pmatrix}, \quad \quad \pmb{R} = \eta, \\
    &\pmb{H} = \nabla_{\pmb{x}} \pmb{h} = 2\eta \sqrt{M N} \begin{pmatrix} 0 & 1 & 0 & 0 & 0 & 0 & 0 \end{pmatrix}.
\end{align}
}
\end{sideways}

\section{Exact forms of gradient matrices $\pmb{F}(t)$, $\pmb{G}$(t) and $\pmb{H}$(t) for an VdP oscillator} \label{sec:exact_forms_FGH_VdP}

\begin{sideways}
\centering
\parbox{8in}{
\begin{align}
    &\est{\pmb{x}} = \begin{pmatrix} \est{x}_1 & \est{x}_2 & \est{x}_3 & \est{x}_4 & \est{x}_5 & \est{x}_6 & \est{x}_7 & \est{x}_8 & \est{x}_9 \end{pmatrix} = \begin{pmatrix} \estNormX & \estNormY & \estNormVX & \estNormVY & \estNormVZ & \estNormCXY & \est{\nu} & \est{\omega} & \est{\upsilon} \end{pmatrix}, \nonumber \\
    &\pmb{F}(t) = \nabla_{\!\pmb{x}} \; \pmb{f} \, |_{(\pmb{\Tilde{x}},u,0,t)}  =  \\
    & = \left(
     \scalemath{0.7}{
    \begin{array}{ccccccccc}
     -(\kcoll\!+\!2\kloc \!+\! M)/2 & -(\est{x}_8 \!+\! u) & 0 & 0 & 0 & 0 & 0 & -\est{x}_2 & 0 \\ 
    (\est{x}_8 \!+\! u) & -(\kcoll\!+\!2\kloc)/2 & 0 & 0 & 0 & 0 & 0 & \est{x}_1 & 0 \\ 
    0 & 2\kcoll \est{x}_2 & -(\kcoll\!+\!2\kloc\!+\!M) & \kcoll & M & -2(\est{x}_8 \!+\! u) \!-\! 8\eta M N \, \est{x}_6 & 0 & -2\est{x}_6 & 0\\ 
    2\kcoll\est{x}_1 & 0 & \kcoll & -\kcoll \!-\! 2\kloc \!-\! 8\eta M N \, \est{x}_4 & 0 & 2(\est{x}_8 \!+\! u)& 0 & 2 \est{x}_6 & 0\\
    2M\est{x}_1 & 0 & M & 0 & -M & 0 & 0 & 0 & 0 \\
    -\kcoll \est{x}_2 & -\kcoll \est{x}_1 & \est{x}_8 \!+\! u & -(\est{x}_8\!+\! u) \!-\!4\eta M N \, \est{x}_6 & 0 & -\left(2\kcoll \!+\! 2\kloc \!+\! \frac{M}{2} \right) \!-\! 4\eta M N \, \est{x}_4 & 0 & \est{x}_3 \!-\! \est{x}_4 & 0 \\
    0 & 0 & 0 & 0 & 0 & 0 & 0 & -p_K & 0 \\
    0 & 0 & 0 & 0 & 0 & 0 & k_K/m_K & \dfrac{2c_K(1-\est{x}_9)}{m_K} & -2c_K\est{x}_8/m_K \\
    0 & 0 & 0 & 0 & 0 & 0 & -\dfrac{1+\est{x}_7/|\est{x}_7|}{2T_K} & 0 & -\dfrac{1}{T_K}
    \end{array},
    }
  \right)  \nonumber \\
    &\pmb{G}(t) = \nabla_{\pmb{\xi} \,} \, \pmb{f} \, |_{\pmb{\Tilde{x}}} = 
    \begin{pmatrix} 
    2\sqrt{\eta M N} \, \est{x}_6 & 0 \;\; \\ 
    2\sqrt{\eta M N} \, \est{x}_4 & 0 \;\; \\ 
    0 & 0 \;\; \\ 
    0 & 0 \;\; \\ 
    0 & 0 \;\; \\ 
    0 & 0 \;\; \\ 
    0 & 0 \;\; \\ 
    0 & \sqrt{q_K} \;\; \\
    0 & 0 \;\;
    \end{pmatrix}, \quad \quad \pmb{Q} = \begin{pmatrix}
        1 & 0 \\
        0 & 1
    \end{pmatrix}, \quad \quad \pmb{S} = \begin{pmatrix}
        \sqrt{\eta} \\ 0 
    \end{pmatrix}, \quad \quad \pmb{R} = \eta, \\
    &\pmb{H} = \nabla_{\pmb{x}} \pmb{h} = 2\eta \sqrt{M N} \begin{pmatrix} 0 & 1 & 0 & 0 & 0 & 0 & 0 & 0 & 0 \end{pmatrix}.
\end{align}
}
\end{sideways}

\end{appendices}

\singlespacing

\clearpage
\addcontentsline{toc}{chapter}{References}
\bibliographystyle{myapsrev4-2}
\bibliography{references}

\newpage



\end{document}